\documentclass[preprint, aps, nofootinbib,  10pt]{revtex4-1}
\usepackage{amsmath} \usepackage{graphicx} 
\usepackage{bm} 
\usepackage{amssymb}
\usepackage{pstricks}

\begin{document}

\newlength{\figurewidth}
\setlength{\figurewidth}{\columnwidth}

\newcommand{\prtl}{\partial}
\newcommand{\la}{\left\langle}
\newcommand{\ra}{\right\rangle}
\newcommand{\dla}{\la \! \! \! \la}
\newcommand{\dra}{\ra \! \! \! \ra}
\newcommand{\we}{\widetilde}
\newcommand{\smfp}{{\mbox{\scriptsize mfp}}}
\newcommand{\smp}{{\mbox{\scriptsize mp}}}
\newcommand{\sph}{{\mbox{\scriptsize ph}}}
\newcommand{\sinhom}{{\mbox{\scriptsize inhom}}}
\newcommand{\sneigh}{{\mbox{\scriptsize neigh}}}
\newcommand{\srlxn}{{\mbox{\scriptsize rlxn}}}
\newcommand{\svibr}{{\mbox{\scriptsize vibr}}}
\newcommand{\smicro}{{\mbox{\scriptsize micro}}}
\newcommand{\scoll}{{\mbox{\scriptsize coll}}}
\newcommand{\sattr}{{\mbox{\scriptsize attr}}}
\newcommand{\sth}{{\mbox{\scriptsize th}}}
\newcommand{\sauto}{{\mbox{\scriptsize auto}}}
\newcommand{\seq}{{\mbox{\scriptsize eq}}}
\newcommand{\teq}{{\mbox{\tiny eq}}}
\newcommand{\sinn}{{\mbox{\scriptsize in}}}
\newcommand{\suni}{{\mbox{\scriptsize uni}}}
\newcommand{\tin}{{\mbox{\tiny in}}}
\newcommand{\scr}{{\mbox{\scriptsize cr}}}
\newcommand{\tstring}{{\mbox{\tiny string}}}
\newcommand{\sperc}{{\mbox{\scriptsize perc}}}
\newcommand{\tperc}{{\mbox{\tiny perc}}}
\newcommand{\sstring}{{\mbox{\scriptsize string}}}
\newcommand{\stheor}{{\mbox{\scriptsize theor}}}
\newcommand{\sGS}{{\mbox{\scriptsize GS}}}
\newcommand{\sBP}{{\mbox{\scriptsize BP}}}
\newcommand{\sNMT}{{\mbox{\scriptsize NMT}}}
\newcommand{\sbulk}{{\mbox{\scriptsize bulk}}}
\newcommand{\tbulk}{{\mbox{\tiny bulk}}}
\newcommand{\sXtal}{{\mbox{\scriptsize Xtal}}}
\newcommand{\sliq}{{\text{\tiny liq}}}
\newcommand{\ssol}{{\text{\tiny sol}}}
\newcommand{\snn}{{\text{\tiny nn}}}

\newcommand{\tHS}{\text{\tiny HS}}
\newcommand{\tRCP}{\text{\tiny RCP}}
\newcommand{\tLJ}{\text{\tiny LJ}}

\newcommand{\smin}{\text{min}}
\newcommand{\smax}{\text{max}}

\newcommand{\saX}{\text{\tiny aX}}
\newcommand{\slaX}{\text{l,{\tiny aX}}}

\newcommand{\svap}{{\mbox{\scriptsize vap}}}
\newcommand{\sjam}{J}
\newcommand{\Tm}{T_m}
\newcommand{\sTS}{{\mbox{\scriptsize TS}}}
\newcommand{\sDW}{{\mbox{\tiny DW}}}
\newcommand{\cN}{{\cal N}}
\newcommand{\cB}{{\cal B}}
\newcommand{\br}{\bm r}
\newcommand{\bq}{\bm q}
\newcommand{\bR}{\bm R}
\newcommand{\cH}{{\cal H}}
\newcommand{\cV}{{\cal V}}
\newcommand{\cHlt}{\cH_{\mbox{\scriptsize lat}}}
\newcommand{\sthermo}{{\mbox{\scriptsize thermo}}}
\newcommand{\eU}{U_\text{eff}}

\newcommand{\bu}{\bm u}
\newcommand{\bk}{\bm k}
\newcommand{\bX}{\bm X}
\newcommand{\bY}{\bm Y}
\newcommand{\bA}{\bm A}
\newcommand{\bj}{\bm j}

\newcommand{\drho}{\delta \hspace{-.3mm} \rho} 
\newcommand{\drhot}{\delta \hspace{-.3mm} \tilde{\rho}}

\newcommand{\sloc}{\text{loc}}
\newcommand{\wF}{\widetilde{F}}
\newcommand{\bmu}{\bm \mu}
\newcommand{\bE}{\bm E}
\newcommand{\smol}{{\mbox{\scriptsize mol}}}

\newcommand{\tF}{{\mbox{\tiny $\!F$}}}
\newcommand{\tlambda}{{\mbox{\tiny $\lambda$}}}
\newcommand{\de}{{\dot{\varepsilon}}}
\newcommand{\tmax}{{\mbox{\tiny max}}}
\newcommand{\DV}{\delta V_{12}}
\newcommand{\sout}{{\mbox{\scriptsize out}}}

\newcommand{\tX}{\text{\tiny X}}

\def\Xint#1{\mathchoice
   {\XXint\displaystyle\textstyle{#1}}%
   {\XXint\textstyle\scriptstyle{#1}}%
   {\XXint\scriptstyle\scriptscriptstyle{#1}}%
   {\XXint\scriptscriptstyle\scriptscriptstyle{#1}}%
   \!\int}
\def\XXint#1#2#3{{\setbox0=\hbox{$#1{#2#3}{\int}$}
     \vcenter{\hbox{$#2#3$}}\kern-.5\wd0}}
\def\ddashint{\Xint=}
\def\dashint{\Xint-}

\markboth{Structural Glass Transition}{Structural Glass Transition}


\title{Theory of the Structural Glass Transition: A Pedagogical
  Review}

\author{Vassiliy Lubchenko} \email{vas@uh.edu} \affiliation{Department
  of Chemistry, University of Houston, Houston, TX 77204-5003}
\affiliation{Department of Physics, University of Houston, Houston, TX
  77204-5005}

\date{\today}

\begin{abstract}

  The random first-order transition (RFOT) theory of the structural
  glass transition is reviewed in a pedagogical fashion. The rigidity
  that emerges in crystals and glassy liquids is of the same
  fundamental origin. In both cases, it corresponds with a breaking of
  the translational symmetry; analogies with freezing transitions in
  spin systems can also be made. The common aspect of these seemingly
  distinct phenomena is a spontaneous emergence of the molecular
  field, a venerable and well-understood concept. In crucial
  distinction from periodic crystallisation, the free energy landscape
  of a glassy liquid is vastly degenerate, which gives rise to new
  length and time scales while rendering the emergence of rigidity
  gradual. We obviate the standard notion that to be mechanically
  stable a structure must be essentially unique; instead, we show that
  bulk degeneracy is perfectly allowed but should not exceed a certain
  value. The present microscopic description thus explains both
  crystallisation and the emergence of the landscape regime followed
  by vitrification in a unified, thermodynamics-rooted fashion. The
  article contains a self-contained exposition of the basics of the
  classical density functional theory and liquid theory, which are
  subsequently used to quantitatively estimate, without using
  adjustable parameters, the key attributes of glassy liquids, viz.,
  the relaxation barriers, glass transition temperature, and
  cooperativity size. These results are then used to quantitatively
  discuss many diverse glassy phenomena, including: the intrinsic
  connection between the excess liquid entropy and relaxation rates,
  the non-Arrhenius temperature dependence of $\alpha$-relaxation, the
  dynamic heterogeneity, violations of the fluctuation-dissipation
  theorem, glass ageing and rejuvenation, rheological and mechanical
  anomalies, super-stable glasses, enhanced crystallisation near the
  glass transition, the excess heat capacity and phonon scattering at
  cryogenic temperatures, the Boson peak and plateau in thermal
  conductivity, and the puzzling midgap electronic states in amorphous
  chalcogenides.  \\ 

  Glasses and amorphous solids \\


\end{abstract}

\maketitle

\tableofcontents

\section{Motivation}

Practical use of structural glasses by early hominins---in the form of
tools and weapons---likely goes back to about 2 million years
ago~\cite{liritzis2012obsidian} and thus well predates the appearance
of the anatomically modern human. The lack of crystallite boundaries,
which helped our forefathers to impart sharp and smooth edges to
obsidian rocks, still underlies many uses of structural glasses. For
instance, it results in optical transparency and mechanical sturdiness
of amorphous silicates; the combination of the two makes glasses
uniquely useful in construction and in information
technology. Metallic glasses are exceptionally rigid at room
temperature while being soft and malleable over a rather broad
temperature range near the glass
transition~\cite{PhysTodayMetallicGlasses}.  In contrast,
polycrystalline metals liquefy almost instantly near their melting
temperature and thus have a much narrower processing window. Some of
the applications of glasses are thoroughly modern: The reflectivity
and electrical conductance of chalcogenide alloys depend on whether
the material is in a crystalline or amorphous form, a property
currently utilised in optical drives. In some of these alloys,
crystal-to-glass transition can be induced by electric current or
irradiation, which can be exploited to make non-volatile computer
memory and for other useful applications~\cite{doi:10.1021/cr900040x,
  ISI:000250615400019, ISI:000261127100019, Kolobov2004, Steimer2008,
  Hosseini2014, AIST}.

The relatively gradual onset of rigidity in structural
glassformers---viewed alternatively as a rapid, super-Arrhenius
slowing down of molecular motions with increasing density or lowering
temperature---is as useful to the industrial designer as it is
interesting to the physicist and chemist. For basic symmetry reasons,
liquids freeze into {\em periodic} crystals in a discontinuous fashion
so that shear resistance emerges within a narrow temperature
interval. The mechanical stability of a periodic array of atoms is
intuitive to those familiar with the Debye theory: The
positive-definiteness of the force-constant matrix for a periodic
lattice can be readily shown for a variety of generic force laws
between individual atoms~\cite{Ashcroft}. Even in those difficult
cases where the individual interactions balance each other out in a
delicate fashion---such situations often arise in applications such as
multiferroics---the stability analysis of a {\em periodic} crystal
usually reduces to that for a very small number of normal modes.  In
contrast, the structure and rigidity change continuously on approach
to vitrification, while there is no obvious way to diagonalise the
force-constant matrix for an aperiodic lattice. Another way to look at
this distinction is that glassy liquid and the corresponding crystal,
if any, occupy distinct regions in the phase space that are separated
by a substantial barrier.  The glass transition itself is not even a
phase transition but, rather, signifies that the supercooled liquid
falls out of equilibrium, an expressly kinetic phenomenon.  Glasses
are typically only {\em meta}stable with respect to crystallisation.

Given the above notions, it appears reasonable to question whether the
underlying causes of rigidity in periodic crystals and glasses are
related even as the local interactions in the two types of solids are
very similar, aside from subtle differences in bond lengths and
angles. In such a view, the roles of the cohesive forces and the
thermodynamic driving force for solidification are essentially
gratuitous; the cohesive forces simply prevent the particles from
flying apart and/or fix local coordination on average. To avoid
confusion, we note that in liquids made of rigid particles, there is
no actual bonding and so one speaks of contacts or
collisions. Nevertheless, the thermodynamics of packing-driven
solidification can be put in correspondence with that of
chemically-bonded solids, as will be discussed.

As surprising and unpalatable it may feel, the view of a gratuitous
role of thermodynamics in glassy phenomena would seem to be suggested
by a number of theoretical developments. For instance, one of the
earliest methodologies that yielded an emergence of rigidity in
aperiodic systems was the mode-coupling theory (MCT)~\cite{Leutheuser,
  Gotze_MCT}. Hereby the rigidity arises for expressly kinetic
reasons: At sufficiently high densities, a particle can not keep up
with the feedback it receives from the surrounding particles in
response to its own motions. To reduce the feedback, the particle is
forced to slow down. In the mean-field limit of the MCT, in which
equations become tractable, the slowing down is nothing short of
catastrophic; it leads to a complete kinetic arrest and, hence,
freezing. A thermodynamic signature of this type of freezing, if any,
does not readily transpire in this framework. A variety of models
characterised by complicated kinetic constraints have been conceived
in the past few years~\cite{Keys19032013, PhysRevX.1.021013},
motivated by Palmer et al.'s work on hierarchically constrained
dynamics~\cite{PSAA}. The latter models, like the MCT, were advanced
in the early 80s. These kinetics-based models exhibit a slowing down
of cooperative nature, and so does the MCT. Complicated kinetic
phenomena, which are at least superficially similar to the
non-Arrhenius and non-exponential relaxations observed in actual
glass-formers, can emerge in kinetically constrained models even if
the thermodynamics of the model are
trivial~\cite{PhysRevLett.53.1244}.

A distinct set of models suggesting a somewhat gratuitous role of
thermodynamics in the structural glass transition focus on the
phenomenon of jamming~\cite{Keys2007, LiuNagel1998, SilbertLiuNagel,
  PhysRevLett.103.025701}. During jamming, as epitomised by sand
dunes, the thermal motions are negligible because temperature is
effectively zero compared to the energies involved. In apparent
similarity to glasses, the rigidity of jammed systems appears to form
gradually. Flow occurs through the proliferation of soft, harmonic
modes that arise when particle contacts are removed. Such soft modes
also emerge in random matrix theories~\cite{ParisiBP} and have been
implicated as giving rise to the so called Boson Peak. The Boson Peak
is a set of vibrational-like states in glasses at frequencies of 1 THz
or so, which reveals itself as a ``bump'' in the heat capacity and
excess phonon scattering at the corresponding temperatures, i.e., near
$10^1$~K~\cite{LW_BP, LW_RMP}.

Yet we shall see in the following that one is, in fact, correct in
expectating that thermodynamics are not simply a spectator of the
slowing down that takes place in supercooled liquids when they are
cooled or compressed toward the glass transition. Until vitrification
has taken place, the liquid is in fact equilibrated and thus should
obey detailed balance~\cite{PhysRev.37.405, PhysRev.38.2265}. (The
liquid is equilibrated conditionally in that there is a lower free
energy state, viz., the crystal. The latter, however, is behind a
barrier and is not accessed.)  By detailed balance, cooperative
motions that give to rise to the non-trivial kinetics observed in
supercooled liquids must correspond to a specific set of
microstates. Such states necessarily have a thermodynamic signature in
the form of additional entropy and heat capacity. For instance, the
critical slowing down during a second order transition corresponds to
a non-analyticity in the free energy~\cite{Goldenfeld}, thus leading
to a singularity in the temperature dependence of the heat
capacity. Detailed balance dictates that a complete theory of the
slowing down in supercooled liquids must describe both the
thermodynamics and kinetics in an internally consistent, unified
fashion. The present review is intended as a pedagogical exposition of
a theory that delivers exactly that: a unified, quantitative
description of thermodynamic and kinetic phenomena in supercooled
liquids and glasses.  This theory is called the random first order
transition (RFOT) theory and has been developed since the early-mid
80s, earlier reviews can be found in Refs.~\cite{LW_ARPC, LW_RMP,
  L_JPCL}. The theory has provided a microscopic framework that allows
one to understand the emergence of rigidity in aperiodic systems in
thermodynamic terms and thus connect the structural glass transition
to a better understood---at least conceptually---fields of the
liquid-to-crystal transition and the theory of chemical bonding in
solids~\cite{ZLMicro1}.

Despite showing basic similarity in bonding, supercooled liquids and
glasses differ fundamentally from periodic crystals in that they are
vastly structurally {\em degenerate}, that is, their free energy
exhibits exponentially many minima at a specific value of the free
energy. The thermodynamic signature of the degeneracy is the excess
liquid entropy of the supercooled liquid relative to the corresponding
crystal. Upon vitrification, this excess entropy ceases to change as
the temperature lowers, thus leading to a discontinuity in the
measured heat capacity.  The remarkable variety of relaxations unique
to glassy systems can all be traced to transitions between the
distinct free energy minima.  This microscopic picture gives rise to
quantitative predictions for many signature phenomena that accompany
the structural glass transition and their quantitative
characteristics, without using adjustable parameters. These cardinal
predictions of the theory include the activation barriers for
$\alpha$-relaxation and their relation to the excess liquid
entropy~\cite{KTW, XW} and elastic constants~\cite{RL_sigma0,
  RWLbarrier, LRactivated}, details of the deviations from Arrhenius
behaviour~\cite{XW, LW_soft, StevensonW} and the pressure dependence
of the barriers.~\cite{RL_LJ} In addition, the theory predicts
correlations of thermodynamics with non-exponential
relaxations~\cite{XWbeta}, the cooperativity length~\cite{XW,
  RWLbarrier}, deviations from the Stokes-Einstein
relation~\cite{XWhydro}, ageing dynamics in the glassy
state~\cite{LW_aging}, crossover between activated and collisional
transport in glass-forming liquids~\cite{LW_soft, SSW, RL_Tcr},
decoupling between various types of relaxation~\cite{Lionic},
beta-relaxation~\cite{SWbeta}, shear thinning~\cite{Lthinning},
dynamics near the surface of glasses~\cite{stevenson:234514},
mechanical strength~\cite{Wisitsorasak02102012}, rejuvenation and
front propagation in ultrastable glasses~\cite{2009PNAS..106.1353P,
  PhysRevE.88.022308}, and re-entrant $T$-dependence of the
crystallisation rate~\cite{SWultimateFate}.

The RFOT theory is a {\em microscopic} theory: For simple liquids,
such as hard spheres or Lennard-Jones particles, the theory starts
from the functional form of the interaction and provides a detailed
approach to compute every quantity of interest from scratch. For the
more complicated glass-forming materials of technology, the
interactions must be evaluated by quantum-chemical means. Owing to the
computational complexity of the quantum-chemical problem, detailed
results cannot be expressed in closed form. In these cases, the
microscopic analysis of the RFOT theory delineates a small sufficient
set of system-specific quantities---structural and
thermodynamic---that represent the microscopic input for computations
of the dynamics. These quantities can be extracted by measurement,
such as X-ray scattering, the Brillouin scattering, and the
calorimetry of the liquid-to-crystal transition, while no
phenomenological assumptions are made. The ensuing computations do not
involve adjustable parameters, consistent of course with the
microscopic nature of the description. Importantly, none of those
listed measurements have to do directly with the glass transition per
se or use any kind of dynamic assumptions.

Likewise, the RFOT theory has elucidated in a microscopic fashion
several {\em quantum} phenomena,~\cite{L_JPCL} which play a role at
cryogenic temperatures. Somewhat surprisingly, the low temperature
physics turns out to be intrinsically related to the molecular motions
that froze in at the much higher, glass transition temperature. The
theory ineluctably leads to the result that an equilibrated liquid
must have a specific, rather universal concentration of configurations
that originate from the transition state configurations for
transport. In the equilibrated liquid, these special configurations
are ``domain walls'' of high free energy density separating compact
regions characterised by relatively low free energy density.  The
domain walls quantitatively account for the excess structural states
in cryogenic glasses responsible for the mysterious two-level systems
and the Boson Peak~\cite{LW, LW_BP, LW_RMP, LSWdipole}. More recently,
it has been established that the domain walls have a topological
signature in chalcogenide alloys and host very special midgap
electronic states, responsible for light-induced midgap absorption and
ESR signal~\cite{ZL_JCP, ZLMicro2}. These quantum phenomena underscore
the danger of thinking of crystals merely as some kind of disordered
analogue of periodic crystals. Again, we shall use thermodynamics as
our guiding light in elucidating this important feature of structural
glasses.

Consistent with its being firmly rooted in thermodynamics, the RFOT
theory highlights the symmetry aspects of the structural glass
transition. The basic symmetry that becomes broken en route to the
glass transition is the translational symmetry intrinsic to liquids
and gases. As a result of this symmetry breaking, the one-particle
density profile (gradually) switches from being, in time average,
spatially uniform to a sensibly fixed collection of disparate, sharp
peaks. Similarly to periodic crystals, these peaks indicate the
particles organise themselves into structures that live much longer
than the vibrations.  In contrast with periodic crystals, however, the
structures are not infinitely long-lived but eventually reconfigure,
thus leading to a liquid flow on long times that eventually restores
the translational symmetry.  Only when the activated reconfigurations
become slower than the quenching rate, which depends on glass
preparation, does the system fall out of equilibrium
completely. Importantly, translational symmetry being broken does {\em
  not} imply there is another symmetry, like periodic ordering, that
replaces it.  The symmetry perspective furnishes the requisite
completeness we have come to associate with established physical
theories, such as the theory of second order transitions, which was
originally built on a symmetry-based coarse-grained
functional~\cite{LLstat} and was later complemented by the discovery
of anomalous scaling. The latter is fully determined by the symmetry
and range of the molecular interaction but not its detailed
form~\cite{Goldenfeld}. Likewise, the presence of an underlying
symmetry breaking makes the RFOT description of the structural glass
transition robust with respect to possible ambiguities that inevitably
arise from approximations and incomplete knowledge of detailed
particle-particle interactions; it thus undergirds the applicability
of the theory to rigid particles and chemically-bonded liquids
alike. The symmetry perspective will also allow us to connect the
glass transition with the phenomenon of jamming.

The RFOT theory definitively answers the aforementioned basic question
as to the mechanical stability of an aperiodic array of particles: To
achieve macroscopic mechanical stability on a finite time scale, the
structure does not have to be unique; indeed, even a thermodynamic
degeneracy is perfectly allowed so long as it does not exceed a
certain value. In turn this guarantees that the transitions between
alternative free energy minima are sufficiently slow.  Relics of {\em
  locally} metastable configurations are present in the frozen glass
and persist down to the lowest temperatures measured but do not affect
the {\em macroscopic} stability.


The article is intended to be a rather self-contained source on the
foundations of the RFOT theory and on how to obtain its main results
with a minimum of technical complexity.  The narrative is organised as
follows: Section~\ref{map} describes what the glass transition is from
the viewpoint of macroscopic thermodynamics and explains the relation
between the supercooled-liquid/glass states and the thermodynamically
stable liquid and crystal states.  Section~\ref{Xtal} explains the
thermodynamics of the ordinary liquid-to-periodic-crystal transition
and, along the way, introduces the basic machinery of the classical
density theory and the theory of liquids, which will be our main tools
in discussing things glassy. These tools no longer seem to be part of
standard courses in statistical mechanics; it is hoped that the
present text covers the necessary minimum in a sufficiently
self-contained manner.  Section~\ref{aper} uses the machinery from
Section~\ref{Xtal} to understand the emergence of {\em aperiodic}
solids, from both the thermodynamic and kinetic perspectives. There we
also briefly discuss the connections with several spin models; these
connections have proved to be a source of both insight and some
confusion to many over the years. In Section~\ref{ActTransport}, we
establish in a self-contained manner both the qualitative and
quantitative features of activated transport and the intrinsic,
testable predictions that connect the thermodynamics and kinetics of
glassy liquids. These results will be used in
Sections~\ref{hetero}-\ref{Quantum} to review quantitative predictions
made by the RFOT theory on a great variety of glassy phenomena
mentioned above.  In Section~\ref{concl}, we summarise, briefly review
the formal status of the theory and establish an intrinsic connection
and, at the same time, basic distinction between the glass transition
and jamming.

Last but not least, let us settle a semantic issue that can be
confusing to physicists and chemists alike, in the author's
experience: We will often use the words ``aperiodic crystal'' and
``aperiodic lattice''---or simply ``lattice''---when referring to the
infinite aperiodic array of particles that a glass or a snapshot of a
liquid is. The purpose is to avoid the repeated use of the awkward
``infinite aperiodic array.''  (The author recognises that in their
traditional use, the words ``crystal'' and ``lattice'' usually refer
to {\em periodic} arrays of objects.)

\section{Placing the Glass Transition on the Map, Thermodynamics-wise:
  The Microcanonical Spectrum of Liquid, Supercooled-Liquid, and
  Crystal States. The Definition of the Glass Transition and Ageing}
\label{map}

\begin{figure}[t]
  \includegraphics[width=  \figurewidth]{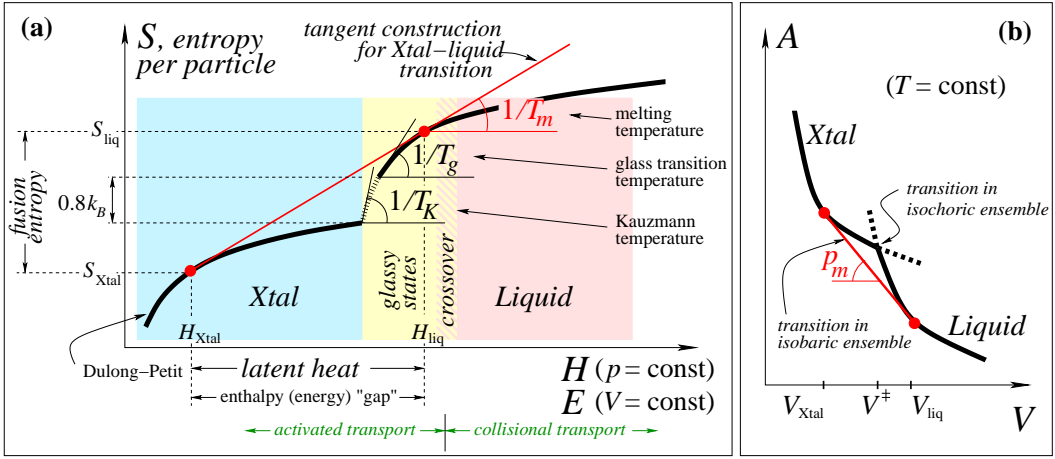}
  \caption{\label{spectrum} {\bf (a)} The ``spectrum'' of a liquid in
    the enthalpy (energy) range of relevance to the liquid-to-crystal
    and the glass transition. The thick, solid black lines depict the
    entropy as a function of enthalpy; the high and low enthalpy
    branches correspond to the liquid and crystal respectively.  The
    states between $H_\sXtal$ and $H_\sliq$ are bypassed during
    crystallisation but are visited, if the liquid can be supercooled
    below the melting temperature. (Some degree of supercooling is
    necessary for crystallisation to proceed anyways because
    crystal-nucleation is subject to a barrier.) The glass transition
    ordinarily occurs at enthalpy values within the enthalpy gap
    $[H_\sXtal, H_\sliq]$, when the liquid excess entropy is $0.8
    \ldots 0.9~k_B$ or so, per rigid molecular unit. The crossover to
    the landscape regime (``glassy states'') could be either above or
    below the melting point $T_m$, the two cases corresponding to
    strong and fragile liquids respectively. {\bf (b)}~The two thick
    solid curves correspond to the Helmholtz free energy of two phases
    characterised by distinct density, such as liquid and crystal. The
    equilibrium transition between the two phases occurs at pressure
    $p=p_m$. It is, in principle, possible to transition between the
    two phases by forcing the system to stay spatially uniform and
    remain on the branch corresponding to the current phase up to the
    crossing point $V^\ddagger$ and then switch to the other phase as
    a whole.  However, the two phases will not be in mechanical
    equilibrium during the transition, $-(\prtl A_\sliq/\prtl
    V)_T|_{V^\ddagger} \ne -(\prtl A_\sXtal/\prtl
    V)_T|_{V^\ddagger}$.}
\end{figure}

We begin by locating the supercooled-liquid state on the energy
landscape of the system, as reflected in its microcanonical
spectrum. Fig.~\ref{spectrum}(a) schematically shows the log-density
of states, or entropy $S$, as a function of energy $E$, if the
experiment is done at constant volume $V$, or enthalpy $H$, if one
employs the more common isobaric conditions $p=\mbox{const}$. Note the
latter is the ensemble of choice for infinitely rigid particles, which
do not undergo phase changes at constant volume because their equation
of state is simply $p/T = f(V)$, where $f(V)$ is a function of volume.
The two (disconnected) thick solid lines correspond to the periodic
crystal states at low enthalpies and the liquid (and gas) states at
high enthalpies.  The slope of each curve is equal to the inverse
temperature at the corresponding value of the enthalpy: $1/T = (\prtl
S/\prtl H)_p = (\prtl S/\prtl E)_V$.

In Fig.~\ref{spectrum}(a), the spectral region bounded by the points
on the curves through which the common tangent passes is special
(these points are shown as large red dots). The states belonging to
this special spectral region are {\em bypassed} during crystallisation
and may thus be said to comprise an enthalpy or energy ``gap'' because
they are inaccessible in true equilibrium. For an enthalpy value
falling within the gap, the system is phase-separated into liquid and
crystal, while the total entropy is a linear function of the enthalpy
and simply reflects the partial quantity of the liquid and crystal, an
instance of the lever rule~\cite{BRR}: $S(H) = x S_\text{liq} + (1-x)
S_\text{Xtal}$, where $H = x H_\text{liq} + (1-x) H_\text{Xtal}$ and
$x$ is the mole fraction of the liquid. The quantities $S_\text{liq}$
and $S_\text{Xtal}$ are the entropies at the the edges of the enthalpy
gap, viz., $H_\text{liq}$ and $H_\text{Xtal}$ respectively. The
numerical value of the width of the gap,
$(H_\text{liq}-H_\text{Xtal})$, which is equal to the latent heat, can
be divided by the melting temperature $T_m$ to obtain the entropy of
fusion $S_m = S_\text{liq} - S_\text{Xtal}$. The latter is generically
about $1.5 k_B$ per particle for ionic compounds~\cite{CRC}; it is
somewhat larger for Lennard-Jones-like substances ($1.68 k_B$ for
Ar~\cite{BRR}) but often less than $k_B$ for covalently bonded
liquids, such as SiO$_2$~\cite{CRC}. The corresponding enthalpy of
fusion is thus comparable to the kinetic energy of the atoms.

If, however, the cooling rate is finite, the liquid must be
supercooled somewhat before it can crystallise, because the nucleation
barrier for crystallisation is infinite strictly at the melting
temperature. Consequently, the liquid states on the right flank of the
enthalpy gap are sampled. These states correspond to a {\em
  supercooled liquid}.  Note the higher the viscosity, the larger the
width of the sampled region, because the prefactor of the nucleation
rate scales roughly inversely with the viscosity.

\begin{figure}[t]
  \centering
  \includegraphics[width= .7 \figurewidth]{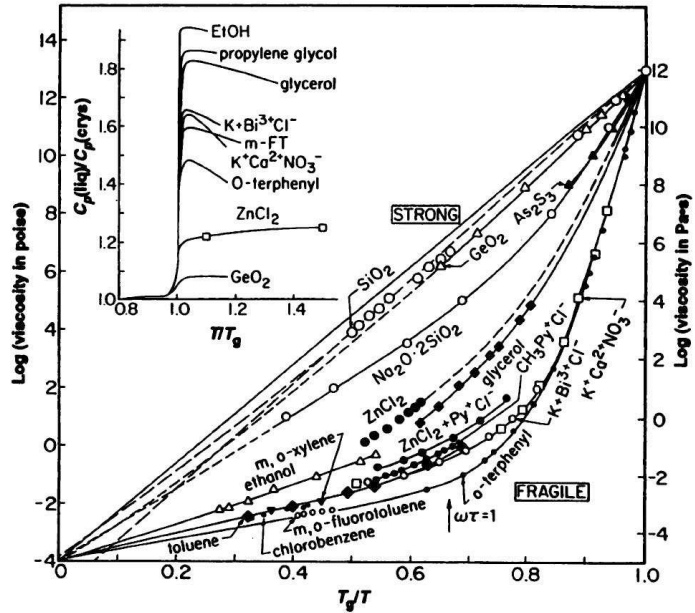}
  \caption{\label{angell} The viscosities of several substances
    plotted as functions of the inverse temperature, the compilation
    and figure due to C.~A.~Angell~\cite{AngellScience1995}. The
    temperature is scaled by the glass transition temperature for each
    substance. The most notable feature of the curves is the deviation
    from the Arrhenius law $e^{\text{const}/T}$. Liquids that deviate
    much or little from the Arrhenius law are often called ``fragile''
    and ``strong'' respectively. The inset shows the temperature
    dependence of the excess liquid heat capacity, relative to the
    corresponding crystal, additionally normalised by the crystal heat
    capacity. There appears to be a correlation between the magnitude
    of the jump in the heat capacity and the liquid's fragility.}
\end{figure}

Now, it is often the case that the viscosity and, thus, the relaxation
times grow rapidly with lowering the temperature, see
Fig.~\ref{angell}; the details of the viscosity growth and the
relation between viscosity and relaxation rates will be discussed in
detail shortly. Given such an increase in the relaxation time, a
liquid is often easy to bring to and maintain in a supercooled
state. A familiar household example of such a supercooled liquid is
glycerol, which is rather difficult to crystallise, as it turns
out. (See, however, Onsager's anecdote about a glycerol factory in
Canada~\cite{PenroseGlycerineAnecdote}.) One can continue to cool down
such a deeply supercooled liquid at a slow rate, with little risk of
crystallisation. Eventually, a liquid cooled at a {\em steady} rate
will fail to reach equilibrium---owing to the rapidly growing
relaxation times. This statement applies at least to the slow rates
realistically achievable in the laboratory; Stevenson and Wolynes have
argued given a slow enough cooling rate, a (periodic-crystal-forming)
liquid will actually crystallise~\cite{SWultimateFate}. The re-entrant
behaviour the glass-to-crystal nucleation, which has been recently
observed in some organic liquids,~\cite{PhysRevB.52.3900,
  doi:10.1021/jp7120577} seems to be consistent with this prediction,
see Section~\ref{fate}.

\begin{figure}[t]
  \begin{tabular*}{\figurewidth} {cc}
    \begin{minipage}{.46 \figurewidth} 
      \begin{center}
        \includegraphics[width= .48 \figurewidth]{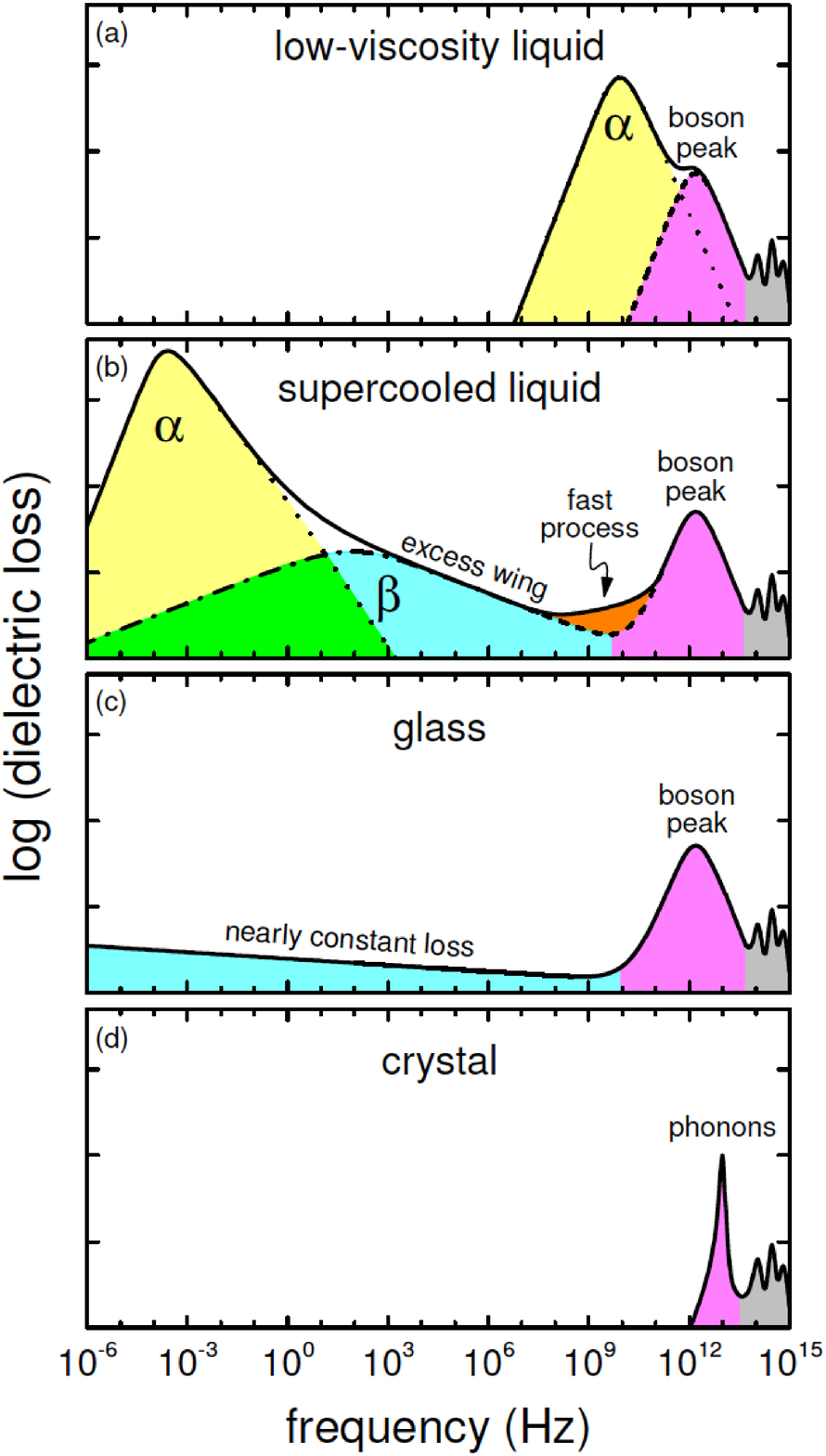}
      \end{center}
    \end{minipage}    
    &
    \begin{minipage}{.48 \figurewidth} 
      \begin{center}
        \includegraphics[width=0.38 \figurewidth]{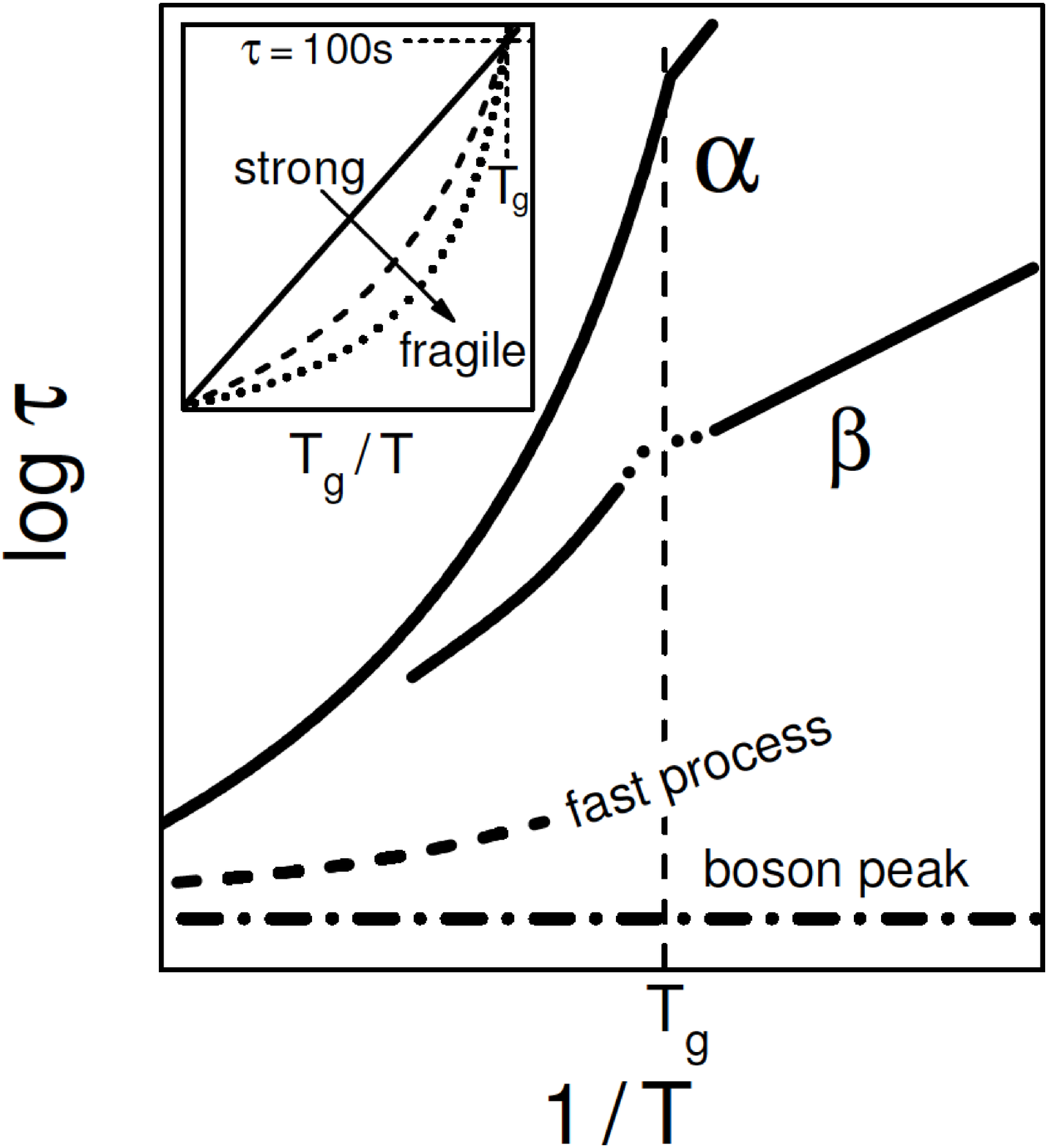}
        \caption{\label{LL} {\bf (Left)} The black solid line depicts
          the imaginary part of the dielectric susceptibility
          $\epsilon''(\omega)$, from Ref.~\cite{LunkenheimerWiley}.
          The coloured peaks correspond to (phenomenologically)
          distinct relaxation processes. The lowest frequency peak is
          labelled as $\alpha$-relaxation. The inverse of the
          corresponding peak frequency matches the structural
          relaxation time as measured by viscosity, see
          Fig.~\ref{angell}. {\bf (Top)} Temperature dependences of
          the relaxation times corresponding to the four (putative)
          relaxation processes from the left panel, subpanel
          (b)~\cite{LunkenheimerWiley}. The precise values of the
          relaxation times depend on the detailed way to de-convolute
          the peaks in the overall spectrum.}
      \end{center}
    \end{minipage}
  \end{tabular*}
\end{figure}

The structural relaxation responsible for the viscous flow can be
readily witnessed in the form of a low-frequency peak in the
dielectric loss spectrum $\epsilon''(\omega)$. This relatively slow
process is traditionally called $\alpha$-relaxation. Other, faster
processes can be argued to take place in addition to main
$\alpha$-relaxation; these faster processes are often non-Arrhenius as
well, see Fig.~\ref{LL}.

Once the liquid that is being cooled fails to equilibrate, we say that
the {\em glass transition}, or {\em vitrification}, has taken place,
at a temperature $T_g$.  Although the system is no longer
equilibrated, the particles continue to move and the material still
relaxes partially, which is called ``ageing.'' These relaxational
motions are even slower than the motions above the glass transition,
whose sluggishness caused the falling out of equilibrium in the first
place; the deeper the quench below the glass transition, the slower
the ageing.

The non-equilibrium, glassy states are no longer identifiable on the
equilibrium spectrum in Fig.~\ref{spectrum}. Rather, they are a
complicated mixture of configurations that are similar to structures
equilibrated in a continuous {\em range} of temperatures; these are
sometimes called ``fictive'' temperatures.  Still, before significant
ageing has taken place, the structure of the glass is very close to
that of the supercooled liquid just above the glass transition, save
for the somewhat reduced magnitude of vibrational motions. Hereby, the
fictive temperature is only weakly distributed and approximately equal
to the glass transition temperature itself. The system is essentially
arrested in the free energy minimum it was occupying during the glass
transition.

The glass transition is not a phase transition, but, instead, is an
instance of kinetic arrest. Still, it can be imparted certain features
of a second order phase transition with enough effort. By employing a
rapid enough quench, one may make ageing largely negligible. Under
these circumstances, the entropy will experience a discontinuity in
its temperature derivative, because the component of the heat capacity
that has to do with the reconfigurational motions of the particles is
zero after the quench.  (The vibrational component of the entropy will
also show a small discontinuity in the temperature derivative because
the $(\prtl V/\prtl T)_p$ derivative will experience a jump.) As a
result, the heat capacity will exhibit a jump at $T_g$. In actuality,
the quench rate is always finite, implying the discontinuity in the
heat capacity is partially smeared out, see the inset of
Fig.~\ref{angell}.

The crystal states within the enthalpy gap are quite distinct from the
supercooled-liquid states. On the left flank, they simply correspond
to vibrational motions of the lattice. However, if one were to
extrapolate the crystal branch toward high enthalpies so that it
overlaps in enthalpy with the liquid branch, things may become more
interesting: Various defect states may now become possible that are
associated with what is called ``mechanical melting.''  Mechanical
melting was proposed early on by Born and others as the cause of the
melting of crystals~\cite{BornHuang}. In this mechanism, the lattice
becomes soft through the proliferation of defects---such as
dislocations---and melts in a relatively smooth fashion, possibly
continuously. A more extreme version of such a defect-based picture of
melting was proposed by Mott and Gurney~\cite{MottGurneyLiquids}, who
visualised a liquid as a polycrystal in a very small crystallite
limit.  Such mechanical melting is hard to observe experimentally,
however, because crystals melt at the surface well before they soften
in the bulk. Since the barrier for surface melting is very low, $k_B
T$ or so~\cite{L_Lindemann}, it is very hard to overheat a crystalline
sample unless its sides are ``clamped'' using a material with a higher
melting point~\cite{Daeges198679, Bilgram}. For these reasons, melting
is usually ``thermodynamic,'' not mechanical, as it takes place near
the temperature $T_m$, where the chemical potentials of the two phases
are equal. Both experiment~\cite{Daeges198679} and
simulation~\cite{MatInterfaces1992} suggest that mechanical melting
would take place at tens to hundreds of degrees above the melting
temperature $T_m$.

It is not particularly conventional to employ the tangent construction
using the enthalpy as the variable and the entropy as the
thermodynamic potential. Yet this is completely analogous to the
tangent construction used to discuss first order transitions
accompanied by volume change in terms of the Helmholtz free energy as
the thermodynamical potential and the specific volume as the variable,
see Fig.~\ref{spectrum}(b). The common tangent in this picture
reflects the mechanical equilibrium between the liquid and crystal
during phase coexistence, analogously to the way the common tangent in
panel (a) reflects the thermal equilibrium between the two
phases. Fig.~\ref{spectrum}(b) also shows that such mechanical
equilibrium can not be achieved in the isochoric ensemble without
compensating externally for the difference in pressures between the
liquid and crystal, by a mechanical partition for instance. The main
drawback of the canonical ensemble exemplified in
Fig.~\ref{spectrum}(b), where one would control the temperature not
enthalpy, is that this ensemble completely misses the supercooled
states, which can be seen explicitly only in a microcanonical
construction, such as in Fig.~\ref{spectrum}(a).

The basic thermodynamic notions discussed above demonstrate that
supercooled liquids and glasses are quite different from the
corresponding crystal in that they belong to a disparate, disconnected
portion of the phase space and so at least some differences in how
rigidity emerges in crystalline and glassy solids may be expected.  We
also directly see that the view of a glass as a defected crystal,
which is often adopted in analyses of cryogenic and optoelectronic
anomalies in glasses, misses the important point that the crystalline
arrangement is not accessible to the atoms and so using it as a
reference state for defect formation is meaningless. (Note the overlap
between the wave-functions of the crystal and glass is negligible.)
Instead, we shall see that one can define such reference states using
mean-field aperiodic free energy minima, while the ``defect''
configurations correspond to interfaces between the minima and are
intrinsic to an equilibrated supercooled liquid; the concentration of
these high free-energy, defective regions depends only logarithmically
on the speed of quenching. In contrast, the quantity of defects in
crystals, such as vacancies, dislocations, twins, etc., is generally
determined by the precise crystal growth setup and the interplay
between heterogeneous and homogeneous nucleation, in addition to a
sensitive dependence on the thermal history of the sample.

\section{Liquid-to-crystal transition as a breaking of translational
  symmetry. Review of the Theory of Liquids and Liquid-to-Solid
  Transition.}
\label{Xtal}

Before we can discuss the thermodynamics of the glass transition, it
is necessary to discuss what drives the {\em ordinary}
liquid-to-periodic-crystal transition. In addition to establishing
commonalities and distinctions between periodic and aperiodic
crystals, this will provide us with a practical reference point in
gauging the quality of our ``understanding'' of the glass transition.

\subsection{What drives crystallisation, why it is a discontinuous
  transition, and why the entropy of fusion is modest}
\label{driving}

\begin{figure}[t]
  \begin{tabular*}{\figurewidth} {ccc}
    \begin{minipage}{.32 \figurewidth} 
      \begin{center}
        \includegraphics[width= .32 \figurewidth]{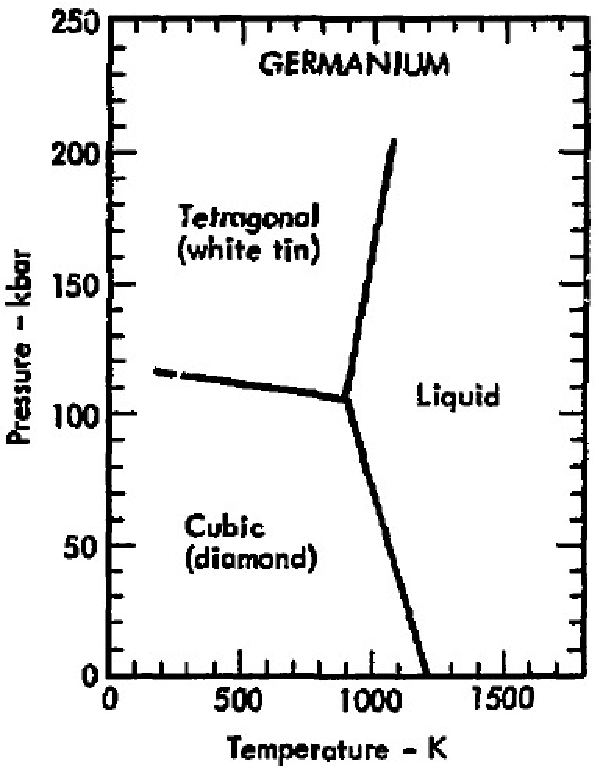}
        \\ {\bf (a)}
      \end{center}
    \end{minipage}    
    &
    \begin{minipage}{.32 \figurewidth} 
      \begin{center}
        \includegraphics[width=0.32 \figurewidth]{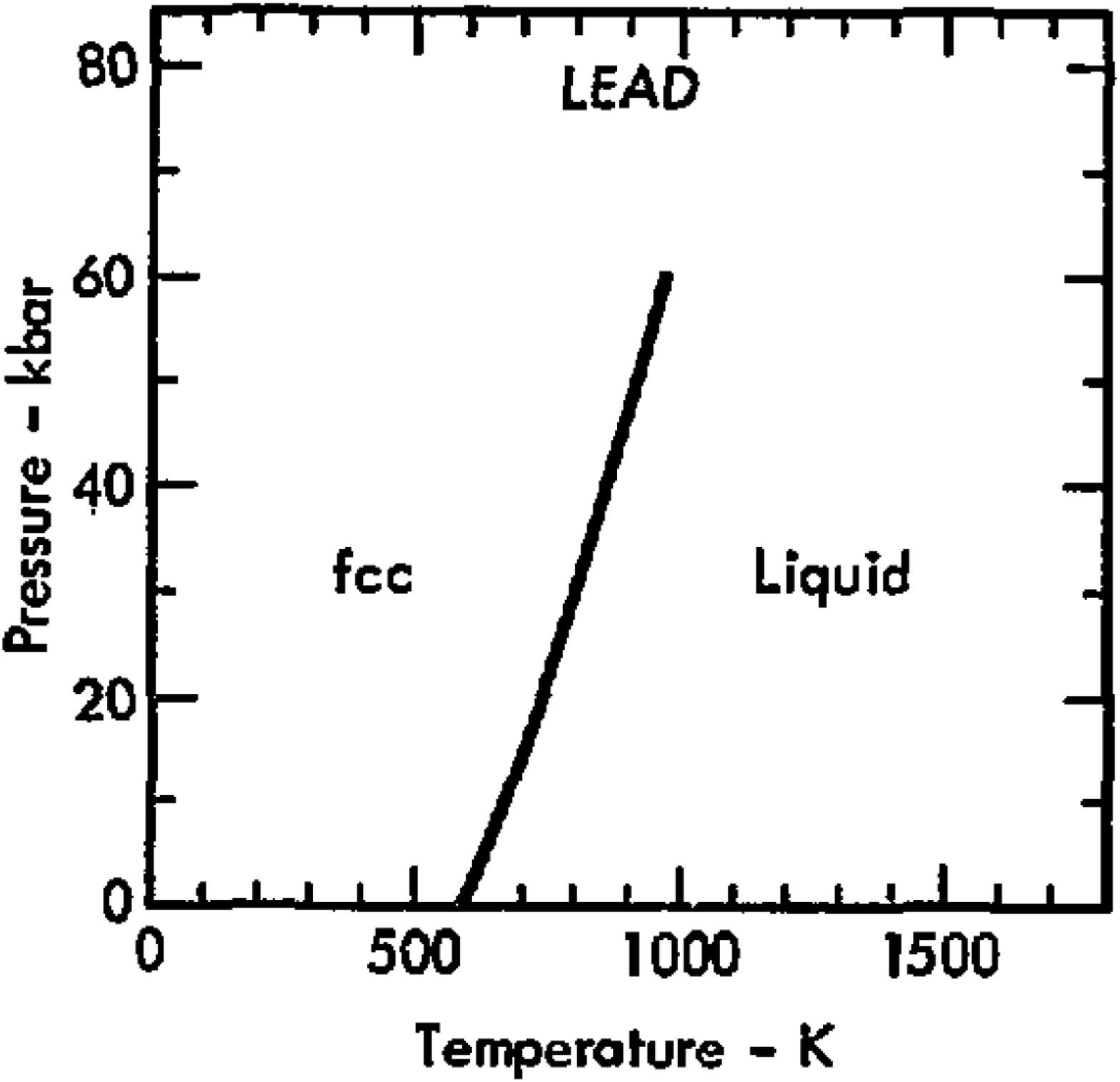}
         \vspace{8mm}
        \\ {\bf (b)} 
      \end{center}
    \end{minipage}
    &
    \begin{minipage}{.32 \figurewidth} 
      \begin{center}
        \includegraphics[width=0.32 \figurewidth]{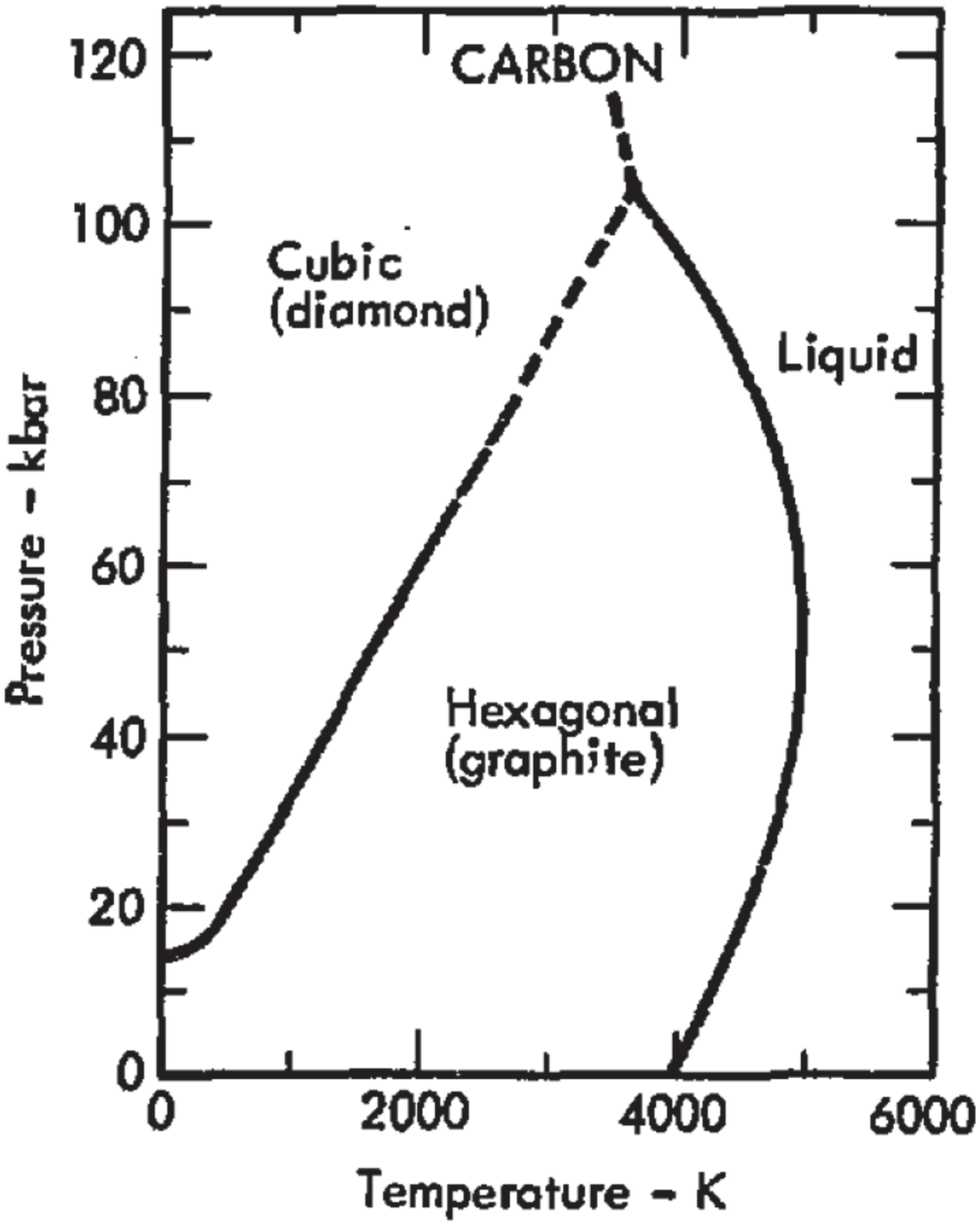}
        \\ {\bf (c)} 
      \end{center}
    \end{minipage}
  \end{tabular*}
  \caption{\label{PhaseDiagrams} Pressure-temperature phase diagrams
    for (a) germanium, (b) lead, and (c) carbon, from
    Ref.~\cite{Young1975}. Note all three elements represent
    closed-shell configurations plus four valence electrons.}
\end{figure}

In an extreme view of solids as very large molecules, it is tempting
to think that crystallisation is driven, thermodynamically, by bond
formation between the atoms. This notion seems to apply particularly
well to crystals with open structures and directional bonding, such as
Si, Ge, and H$_2$O. In the (low pressure) solids of these substances,
atoms are less coordinated than in the corresponding liquids above
melting, as witnessed by a positive volume change following
crystallisation: $\Delta V > 0$, see the phase diagram of germanium in
Fig.~\ref{PhaseDiagrams}(a). One expects that at higher pressures, the
bonding anisotropy becomes progressively subdominant to the steric
repulsion, leading to the more conventional reduction in volume upon
freezing, $\Delta V < 0$, as is the case for lead or high-pressure
germanium, see Fig.~\ref{PhaseDiagrams}. Yet there is generally no
simple correlation between the pressure and the sign of $\Delta V$, as
is exemplified by the phase diagram of carbon shown in
Fig.~\ref{PhaseDiagrams}(c).  Although all three elements in
Fig.~\ref{PhaseDiagrams} represent closed-shell configurations plus
four valence electrons, they show very different phase
behaviours. Clearly, bonding changes play a significant role in the
liquid-to-crystal transition and exhibit remarkable variety even for
seemingly similar electronic configurations.

Yet, while partially correct, the notion of the bonding-driven
crystallisation is potentially misleading.  The majority of enthalpy
change due to bonding in the condensed phase occurs already during the
vapour-to-liquid transition, whose latent heat is typically an order
of magnitude greater than that for the liquid-to-crystal
transition. This fact is reflected in the venerable Trouton's
rule~\cite{BRR}, by which the entropy of condensation at normal
pressure is numerically close to $10^1 k_B$ per particle, compared
with the fusion entropy of about $10^0 k_B$ mentioned earlier.
Indeed, the fusion enthalpy is comparable to the kinetic energy and
thus is much lower than the bond enthalpy, suggesting the crystal
stability is of somewhat subtle origin.

The relatively small entropy change upon freezing can be understood
using the following qualitative, mean-field
argument~\cite{LJDevonshire1937, EyeringLiquid, MottGurneyLiquids}:
Neglecting correlation between particles' movements, the partition
function for the liquid can be roughly estimated as
\begin{equation} \label{Zliq}
  Z_\text{liquid} \sim \frac{1}{N!} \left(\frac{V_f}{\Lambda^3}
  \right)^N,
\end{equation}
where $\Lambda \equiv (2 \pi \hbar^2/m k_B T)^{1/2}$ is the de Broglie
thermal wavelength.  The quantity $V_f$ is the total ``free'' volume,
i.e., the total volume of the system minus the combined volume of the
molecules which we approximate here as relatively rigid, compact
objects.  The combinatorial factor $1/N!$ reflects that the particles
are indistinguishable (Ref.~\cite{LLstat}, Chapter 41). The symmetry
of the Hamiltonian with respect to particle identity is physically
realised by the particles exchanging locations: The defining feature
of an equilibrium liquid/gas is the uniform distribution of a
particle's density on any meaningful time scale. This is just a
technical way to say that the liquid assumes the shape of its
container. In contrast, particles comprising a solid are confined to
``cages'' with specific locations in space, which enables one to
actually {\em label} the particles, even if they are indistinguishable
otherwise.  To estimate the partition function for the corresponding
solid, let us use the Einstein approximation---which {\em also}
neglects particle-particle correlations. Here we simply multiply the
partition functions for individual particles each rattling within its
own cage. The cage volume is the free volume per particle: $V_f/N$,
thus yielding:
\begin{equation}
  Z_\text{solid} \sim \frac{1}{N^N} \left(\frac{V_f}{\Lambda^3}
  \right)^N
\end{equation}
The $N!$ factor is now absent because particles in a solid can be
labelled (according to which lattice sites they are nearest to) and
thus may be regarded as distinguishable, as just mentioned.
Consequently, the excess entropy of the liquid relative to the
corresponding crystal is about $k_B \ln(N^N/N!)  \simeq N k_B$,
roughly consistent with observation. We thus tentatively conclude that
the entropy of fusion is relatively small because the translational
(or ``mixing'') entropy in the uniform liquid only modestly exceeds
the vibrational entropy of particle motions within assigned cages in
the corresponding solid; the interactions enter the analysis through
the free volume and cage shape and contribute toward system-specific
corrections to the simple result $~N k_B$. Note that the modest value
of the translational entropy in gases/liquids is a consequence of an
interaction that is of statistical origin. This interaction is present
even if the particles do not interact in energetic terms: Because two
configurations in which two identical particles are interchanged are
not distinct, the volume statistically available to an individual
particle is not the total free volume $V_f$, but only a tiny portion
of it, i.e., $e V_f/N$ or so, see Eq.~(\ref{Zliq}).

But, in the first place, why should the liquid-to-crystal transition
ordinarily be first order? This is not an entirely trivial question.
For instance, early computer simulations~\cite{HooverRee1968}, which
employed small system-sizes, were ambiguous as to the discontinuous
nature of the transition for hard spheres; particles had to be
confined to individual cells in space to minimise effects of
fluctuations, see discussion in Ref.~\cite{Fixman1969}. An early
constructive argument in favour of a discontinuous nature of the
liquid-to-solid transition is contained in a prescient paper published
by Bernal in 1937~\cite{Bernal1937}, to which we shall return in due
time.  Of the most general applicability is Landau's symmetry-based
argument~\cite{LandauPT1, LandauPT2}, whose publication also dates to
1937 and {\em also} significantly predates the aforementioned liquid
simulations. The main tool in the argument is what is now known as the
Landau-Ginzburg functional~\cite{LLstat, Goldenfeld},
\begin{equation} \label{LG} F = \int d^3 \br [\kappa (\nabla \phi)^2/2
  + V(\phi)].
\end{equation}
We begin from the simplest non-trivial approximation for the bulk free
energy term, viz., in terms of a 4th order polynomial:
\begin{equation} \label{LGbulk} V(\phi) \equiv \frac{A}{2} \phi^2 +
  \frac{B}{3} \phi^3 + \frac{C}{4}\phi^4.
\end{equation}
In the Landau-Ginzburg approach one makes a non-obvious assumption
that both the bulk term and the $(\nabla \phi)$-dependent term---which
we have truncated at the second order---are well-behaved, i.e.,
analytic.

To make analysis constructive, the terms in the expansion
(\ref{LGbulk}) must be traced to physical interactions. Let us do this
first for a familiar system, namely, the Ising spin model with the
energy function 
\begin{equation} \label{EIsing} \cH = - \sum_{i<j} J_{ij} \sigma_i
  \sigma_j, \hspace{20mm} \sigma_i = \pm 1.
\end{equation}
One can formally write down the Helmholtz free energy of the magnet as
a sum of two contributions, both of which are uniquely determined by
the average magnetisation on individual sites:
\begin{equation} \label{IsingF} F(\{ m_i \}) = F_\text{id}(\{ m_i \})
  + F_\text{ex}(\{ m_i \})
\end{equation}
where the ``ideal gas'' contribution:
\begin{equation} \label{IsingID} F_\text{id}(\{ m_i \}) = k_B T \sum_i
  \left( \frac{1+m_i}{2}\ln\frac{1+m_i}{2} +
    \frac{1-m_i}{2}\ln\frac{1-m_i}{2} \right),
\end{equation}
is the free energy of $N$ non-interacting, {\em free} spins and is
simply the sum over the entropies of individual, standalone spins,
times $(-T)$. This expression can be easily derived by noting that the
energy of a free spin is zero while the entropy of a spin with average
magnetisation 
\begin{equation} \label{sigmam} \la \sigma_i \ra = m_i
\end{equation} 
is equal to the log-number of distinct configurations of a macroscopic
number $N$ of spins, of which $N(1+m_i)/2$ point up and $N(1-m_i)/2$
point down (divided by $N$ and multiplied by $k_B$). Thus the entropy
of a free spin with average magnetisation $m_i$ is simply the Gibbs
mixing entropy of two ideal gases with mole fractions $(1+m_i)/2$ and
$(1-m_i)/2$, per particle. Now, the excess term includes all other
contributions to the free energy and is difficult to write down except
in a few cases~\cite{Goldenfeld, baxter:553}. We will content
ourselves with a mean-field approximation, which becomes exact when
each spin interacts infinitely weakly with an infinite number of other
spins. For the Ising magnet from Eq.~(\ref{EIsing}), this would imply
$J_{ij} = J/N$, so that the energy scales linearly with $N$, see
Problem 2-2 of Goldenfeld~\cite{Goldenfeld}. Because each spin is in
``contact'' with $(N-1)$ spins, this situation formally corresponds to
an infinite dimensional space, the exact dimensionality depending on
the type of lattice. For a cubic lattice, the dimensionality would be
$(N-1)/2$.  In the mean-field limit, the correlations between spin
flips can be ignored, since $\la \sigma_i \sigma_j \ra \approx \la
\sigma_i \ra \la \sigma_j \ra + O(1/N) \overset{N \rightarrow
  \infty}{\rightarrow} m_i m_j$. The entropy of such a collection of
uncorrelated spins is equal to that of $N$ free spins, and so
$F_\text{ex}$ is simply the average energy, since $F = E -
TS$. Averaging energy (\ref{EIsing}) in the mean-field limit $\la
\sigma_i \sigma_j \ra = m_i m_j$ readily yields:
\begin{equation} \label{IsingEX} F_\text{ex}(\{ m_i \}) = - \sum_{i<j}
  J_{ij} m_i m_j \hspace{10mm} \text{(mean-field limit).}
\end{equation}
Note that given the knowledge of the free energy as a function of
$m_i$'s, the couplings $J_{ij}$ can be determined by varying the free
energy with respect to the magnetisation on sites $i$ and $j$:
\begin{equation} \label{Jij} J_{ij} = - \frac{\prtl^2
    F_\text{ex}}{\prtl m_i \prtl m_j}.
\end{equation}
$(\prtl^2 F_\text{id}/\prtl m_i \prtl m_j) = 0$, of course.

From now on, assume for concreteness that all of the couplings are
positive: $J_{ij} > 0$. In a translationally invariant system, $J_{ij}
= J(\br_i - \br_j)$, the local magnetisation that minimises the
functional is spatially uniform: $m_i = m$, and is an appropriate
order parameter. In this strictest, spatially-uniform limit of the
mean-field approximation, the Helmholtz free energy of the magnet
reads, per spin:
\begin{equation} \label{IsingMF} F(m)/N = k_B T \left(
    \frac{1+m}{2}\ln\frac{1+m}{2} + \frac{1-m}{2}\ln\frac{1-m}{2}
  \right) - \frac{1}{2}\left(\Sigma_j J_{ij}\right) \, m^2.
\end{equation}
The bulk free energy (\ref{IsingMF}) is the analog of the bulk free
energy $V(\phi)$ from Eq.~(\ref{LGbulk}).

The free energy cost of {\em deviations} from the strictly uniform
configuration in Eq.~(\ref{IsingMF}) can be estimated by still using a
mean-field approximation. The thermally-averaged interaction energy
from Eq.~(\ref{IsingEX}) can be re-written as $- (1/2) \sum_i m_i
\sum_j J_{ij} m_j \simeq - (1/2) \sum_i m_i \sum_j J_{ij} \{1 + (\br_j
- \br_i) \nabla + (1/2)[ (\br_j - \br_i) \nabla]^2\} m(\br)
|_{\br=\br_i}$, where we have truncated the Taylor expansion of $m_j
\equiv m(\br_j)$ around point $\br_i$ at the 2nd order; this suffices
to account for the longest wavelength fluctuations relative to the
average magnetisation. If, on the other hand, we have an
anti-ferromagnet, then an expansion around a finite wave-vector $\bq =
\bq_0$ should be performed, not $\bq = 0$.  Now, after summation in
$j$, the 0th order term in the expansion gives the 2nd term on the
r.h.s. of Eq.~(\ref{IsingMF}), while the 1st order term drops out by
symmetry $(\br_i - \br_{i+j}) = -(\br_i - \br_{i-j})$ and the 2nd
order term simplifies to $- (1/2) [\sum_j J_{ij} (\br_j -
\br_i)^2/(2\cdot 3)] \sum_i m_i \nabla^2 m(\br_i)$. In this
expression, we switch from discrete summation to volume integration
and integrate by parts to obtain, per spin:
\begin{equation} \label{sqGIsing} \frac{1}{2}\left[ \frac{1}{6}
    \left(\Sigma_j r^2_{ij} J_{ij}\right) \right] \, (\nabla m)^2.
\end{equation}
This term represents the square-gradient term $\kappa (\nabla
\phi)^2/2$ in the Landau-Ginzburg expansion (\ref{LG}) for our
ferromagnet. (The one-particle, entropic term from Eq.~(\ref{IsingF})
does not contribute to the square-gradient term.) The coefficient
$\left(\Sigma_j r^2_{ij} J_{ij}\right)$, which corresponds with the
coefficient $\kappa$ in Eq.~(\ref{LG}), follows a general pattern:
Insofar as the interaction $J_{ij}$ possesses a finite range $l$, the
coefficient scales with $l^2$ times an energy scale that characterises
the interaction.

Now, returning to the bulk energy (\ref{IsingMF}), the interaction
term is second order in the order parameter $m$ and is proportional to
the coupling constant $J$, in reflection of its two-body origin.  The
entropic contribution {\em also} has a quadratic contribution, but of
positive sign; it stabilises the symmetric phase, $m=0$, at high
temperatures.  The total coefficient at the second order term reads
as: $A = - (\sum_j J_{ij}) + k_B T$ and vanishes at the critical
(Curie) temperature $T_c = (\sum_j J_{ij})/k_B$ leading to a
ferromagnetic ordering below the Curie point, whereby macroscopic
regions acquire non-zero magnetisation, $m = \pm (-A/C)^{1/2} \ne
0$. (The Curie point is lowered when non-meanfield effects are
included in the treatment~\cite{Goldenfeld}.)  Note that the entropic
part of the free energy limits the possible value of magnetisation:
$|m| < 1$---and thus renders the functional stable. In the low order
expansion (\ref{LGbulk}), such stability is guaranteed by the quartic
term, whereby $C > 0$, however there is no hard constraint on the
magnitude of $m$. (This is reasonable so long as $|m|$ is not too
close to its maximum value of 1.)  The spontaneous magnetisation $m =
\pm (-A/C)^{1/2}$ below the critical point corresponds to a non-zero
(local) field which, in effect, breaks the time-reversal symmetry of
the full Hamiltonians $E(\{ \sigma_i \}) = E( \{- \sigma_i
\})$. Incidentally, because of the time-reversal symmetry, the
coefficient $B$ at the cubic term in the functional is identically
zero for the Ising model, which is an exception rather than the rule,
as we shall see shortly.

To build a free energy functional of the form (\ref{LGbulk}) that is
appropriate for {\em particles}---we first specify that the order
parameter reflect fluctuations of the density around its equilibrium
value $\rho_\seq(\br)$, which we regard for now as
coordinate-dependent, for the sake of generality:
\begin{equation} \label{deltarhoGen} \drho(\br) \equiv \rho(\br) -
  \rho_\seq(\br).
\end{equation}
Analogously to Eq.~(\ref{IsingF}), we present the total Helmholtz free
energy of the liquid (whether uniform of not) as the sum of the
``ideal gas'' and interacting, or ``excess'' free energy
contributions~\cite{Hansen, Evans1979}:
\begin{equation} \label{F} F[\rho(\br)] = F_\text{id}[\rho(\br)]
  +F_\text{ex}[\rho(\br)] 
  ,
\end{equation}
The ``ideal gas'' part is given by the expression
\begin{equation} \label{F_id} F_\text{id}[\rho(\br)] = k_BT\int d^3\br
  \rho(\br ) \left[ \ln\left(\rho (\br ) \Lambda^3 \right)-1\right].
\end{equation}
This equation is the Helmholtz free energy of an ideal gas whose
concentration is not necessarily spatially-uniform. We split space
into elemental volumes $V_i$, each containing $N_i$ gas particles:
$F_\text{id} = \sum_i F_\text{id}^{(i)} = -k_B T \ln Z_\text{id}^{(i)}
= k_B T V_i (N_i/V_i) [\ln(N_i/V_i) -1]$, each of the partial free
energies $F_\text{id}^{(i)}$ corresponds to Eq.~(\ref{Zliq}) with $V_f
= V_i$.  Switching to continuum integration $V_i \rightarrow d^3 \br$
and replacing $N_i/V_i$ with its value $\rho(\br)$ yields
Eq.~(\ref{F_id}). The ideal-gas free energy (\ref{F_id}) is the liquid
analog of the entropic free energy Eq.~(\ref{IsingID}). The only (and
non-essential) difference that it also has an energetic contribution
due to the kinetic energy of individual particles; this contribution
obligingly renders the argument of the logarithm in Eq.~(\ref{F_id})
dimensionless.

Note Eq.~(\ref{F}) is formally exact; it embodies the main idea of the
classical density functional theory (DFT). The DFT builds on the
Hohenberg-Kohn-Mermin theorem~\cite{PhysRev.136.B864,
  PhysRev.137.A1441}, which states that there is a unique free energy
functional $F[ \rho(\br) ]$ that is minimised by the equilibrium
density profile $\rho_\text{eq}(\br)$.  As in the ferromagnet case,
the interacting part $F_\text{ex}$ can be computed in 3D only
approximately. Appropriate approximations will be discussed in due
time. For now, let us formally expand the free energy (\ref{F}) as a
power series in terms of $\drho(\br)$, up to the second order:
\begin{eqnarray} \label{dFgen} F[\rho(\br)] &=& F[\rho_\seq(\br)] +
  k_B T \int d^3 \br \left[ \ln \! \left(\rho_\seq (\br ) \Lambda^3
    \right)
    - c^{(1)}(\br) \right] \drho(\br) \\
  &+& \frac{k_B T}{2} \int d^3 \br_1 d^3 \br_2 \, \drho(\br_1) \,
  \left[\frac{1}{\rho_\seq(\br_1)}\delta(\br_1 - \br_2) -
    c^{(2)}(\br_1,\br_2) \right] \drho(\br_2) \nonumber
\end{eqnarray}
where, by definition, 
\begin{equation} \label{C1r} c^{(1)}(\br) \equiv - \beta \frac{\delta
    F_\text{ex}}{\delta \rho(\br)}.
\end{equation}
and $c^{(2)}(\br_1,\br_2)$ is the standard direct correlation
function:
\begin{equation} \label{c(r)} c^{(2)} (\br_1, \br_2 ) \equiv
  \left. -\beta \frac{\delta^2 F_\text{ex} [\rho(\br)]}{\delta\rho
      (\br_1) \, \delta\rho (\br_2)} \right|_{\rho(\br) =
    \rho_\seq(\br)},
\end{equation}
c.f. Eq.~(\ref{Jij}).  Note that the direct correlation function is
translationally invariant and isotropic for a bulk uniform liquid
\begin{align} \label{uni} \rho_\seq(\br) &= \rho_\sliq \nonumber \\
  c^{(1)}(\br)& = c_\sliq^{(1)} \\
  c^{(2)}(\br_1,\br_2) &= c^{(2)}(|\br_1 - \br_2|), \nonumber
\end{align}
but this assumption is not strictly correct in liquids that are not
uniform---as would be the case in an otherwise homogeneous fluid near
a wall or liquid-vapour interface~\cite{Evans1979} and, of course, in
crystals~\cite{PhysRevE.55.4990}.

\begin{figure}[t]
  \centering
  \includegraphics[width= \figurewidth]{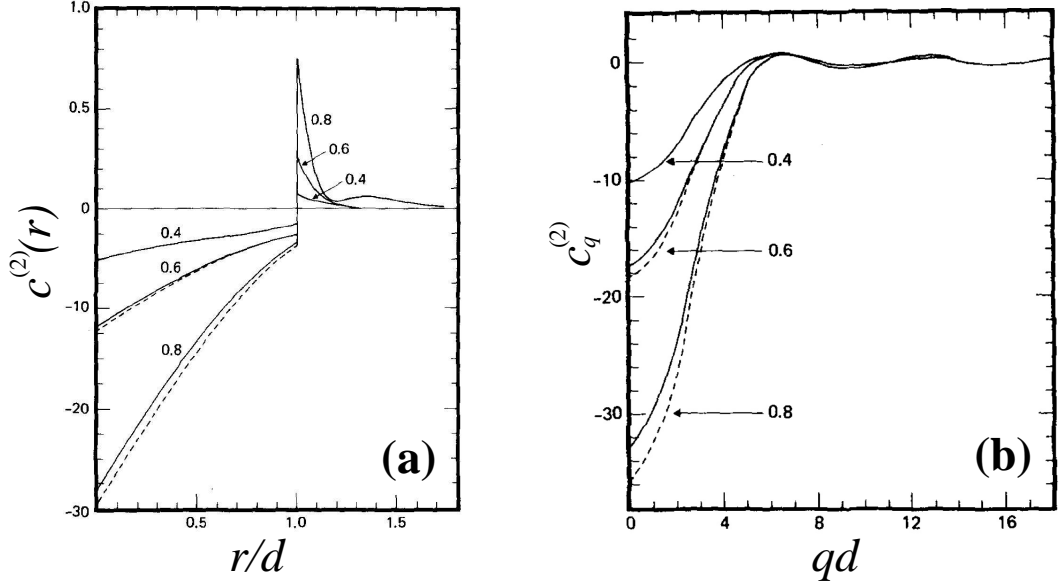}
  \caption{\label{cRHendersonGrundke} We display two specific examples
    of the direct correlation function, corresponding to the
    Percus-Yevick (dashed line) and
    Henderson-Grundke~\cite{HendersonGrundke75} (solid line)
    approximations for the hard sphere liquid. Panels (a) and (b) show
    the function and its Fourier image respectively. Note the
    significantly finer scale on the positive portion of the vertical
    axis in panel (a).}
\end{figure}

As in the ferromagnet case, the second order term in Eq.~(\ref{dFgen})
accounts for the two-body contributions to the free
energy. Appropriately, it can be seen from Eq.~(\ref{dFgen}) that in
the weak-interaction limit~\cite{Hansen}:
\begin{equation} \label{cV} c^{(2)}(\br_1, \br_2) \rightarrow - \beta
  v(\br_1, \br_2), \hspace{5mm} \text{as } \rho \rightarrow 0,
\end{equation}
and so in this limit, the direct correlation function has the same
range as the pairwise interaction $v(\br_1, \br_2)$. Even near
critical points, where the full density-density correlation function
becomes so long-ranged that the compressibility diverges, the direct
correlation function remains integrable (see Eq.~(\ref{OZq}) and
(\ref{SqUni}) below).  The direct correlation function $c^{(2)}(\br_1,
\br_2)$ includes all interactions between these two particles,
including those induced by the rest of the particles. For instance,
$c^{(2)}(r)$ for the hard sphere liquid has a positive---i.e.,
``attractive'' by Eq.~(\ref{cV})---tail around $r = d$, see
Fig.~\ref{cRHendersonGrundke}(a). Two neighbouring particles are
effectively pushed together by being repelled from the surrounding
particles. (This is analogous to the so called ``depletion
interaction'' that can be induced between molecules by adding polymer
to the solution~\cite{AOdepletion}.)  At short separations, $r < d$,
the direct correlation function has a rather different
meaning. Namely, it scales (with the negative sign) with the bulk
modulus of the liquid. This notion will be made precise in a short
while. For now, it is instructive to compare the free energy cost of
quadratic fluctuations in Eq.~(\ref{dFgen}) to the free energy cost of
a weak deformation of an elastic continuum~\cite{LLelast}:
\begin{equation}
  \label{Fuu} F =  \iint d^3\br_1 d^3\br_2 \, D(\br_1-\br_2) 
  \left[ \left( \frac{K}{2} - \frac{\mu}{3} \right)  u_{jj}(\br_1)
    u_{ll}(\br_2) + \mu \, u'_{ij}(\br_1) u'_{ij}(\br_2) \right] 
\end{equation}
where the deformation tensor $u_{ij}$ is defined in the standard
fashion:
\begin{equation} \label{uij}
  u_{ij}=\left(1/2\right)\left( \partial{u_i}/\partial{x_j}
    +\partial{u_j} /\partial{x_i} \right)
\end{equation}
and $u'_{ij}$ stands for its traceless portion $u'_{ij} \equiv u_{ij}
-\frac{1}{3}\delta_{ij}u_{ll}$ that corresponds to pure shear.  The
vector $\bu$ gives a particle's displacement relative to its
equilibrium position, while its divergence $u_{ii}$ yields the
relative volume change of a compact region encompassing a specific
group of atoms, due to uniform contraction or dilation.
Eq.~(\ref{Fuu}) corresponds to a non-local form of the elasticity
theory~\cite{nonlocElast, NTE2, NTE3}, which is reduced to the
classic, ultra-local approximation~\cite{LLelast} by taking the limit
$D(\br) \rightarrow \delta(\br)$. In this continuum limit, the
coefficient $K$ corresponds with the macroscopic bulk modulus:
\begin{equation}\label{kappaT}
  K \equiv - V \left(\frac{\prtl p}{\prtl V} \right)_T ,
\end{equation}
while $\mu$ becomes the standard shear modulus.  This is where we can
make connection with the functional in Eq.~(\ref{dFgen}), since the
$\drho$'s in that equation {\em also} scale with local volume changes.
For a region containing an appreciable number of particles,
$\drho/\rho = - u_{ii}$. Despite this relation, we note that there is
generally no one-to-one correspondence between the deformation tensor
$u_{ij}$ and the local density variation $\drho$ because the former is
a quantity coarse-grained over a mesoscopic region, while the latter
is defined on an arbitrarily small length scale and can change
arbitrarily rapidly in space. For instance, for a stationary particle
at the origin, $\rho(\br) = \delta(\br)$. Now, to connect the
functionals (\ref{dFgen}) and (\ref{Fuu}), we first set $\mu = 0$ as
is appropriate for uniform liquids. Because only long-wavelength
density variations can be compared between the two functionals, we can
adopt the continuum limit of the elasticity: $F = (1/2)\int d^3\br K
u_{jj}^2(\br)$, yielding $K u_{jj}^2(\br)/2$ for the free energy
density at location $\br$. By Eq.~(\ref{dFgen}), the same quantity is
given by $(k_B T/2) \drho^2(\br) \int d^3 \br [1/\rho_\seq -
c^{(2)}(r)]$, where we used that the direct correlation function
decays much faster than the lengthscale for density variations. For
such slow variations, $\drho/\rho = - u_{ii}$, and we obtain:
\begin{equation} \label{sum3} - \rho_\sliq \int d^3 \br c^{(2)}(r) =
  \frac{K}{k_B T \rho_\sliq} - 1,
\end{equation}
c.f. the systematically derived Eq.~(\ref{sum2}).

Now, as a rule of thumb, the bulk modulus is about $(10^1 - 10^2) k_B
T/\rho_\sliq$ for liquids and $10^2 k_B T/\bar{\rho}$ for solids
(consistent with the Lindemann criterion of melting)~\cite{Bilgram}.
The above argument explains why the direct correlation function should
be mostly large and negative at the high densities in question, see
Fig.~\ref{cRHendersonGrundke}(a).  We observe that the direct
correlation function is thus a rather complicated, rich object; it
will be central to our further developments.

The free energy expansion (\ref{dFgen}), when applied to a {\em
  uniform} liquid, Eq.~(\ref{uni}), looks particularly revealing in the
Fourier space as it is simply a sum over independent harmonic
oscillators represented here by the distinct Fourier modes:
\begin{equation} \label{FRYq} F - F(\rho_\sliq) = \frac{k_B T}{2} \int
  \frac{d^3 \bq}{(2 \pi)^3} \left(\frac{1}{\rho_\sliq} -
    \tilde{c}^{(2)}_q \right) |\drhot_{\bq}|^2.
\end{equation}
We have omitted the term linear in $\drho(\br)$, which is equal to
$(\prtl F/\prtl N)_{V, T} (N - \la N \ra) = \mu (N - \la N \ra) $,
where $\mu \equiv (\prtl F/\prtl N)_{V, T}$ is the bulk chemical
potential. This linear term can be set to zero by fixing the total
particle number $N$ at its expectation value $\la N \ra$ and is not
relevant to {\em local} density fluctuations. An example of the
Fourier transform of the direct correlation function is provided in
Fig.~\ref{cRHendersonGrundke}(b).

Consistent with the pattern in Eq.~(\ref{sqGIsing}), the square
gradient term in the Landau-Ginzburg functional corresponding to the
liquid free energy (\ref{F}) is given by~\cite{YangFlemingGibbs}:
\begin{equation} \label{sqGliquid} \frac{1}{2} \left[ \frac{k_B T}{6}
    \int d^3\br \, r^2 c^{(2)}(r) \right] (\nabla \rho)^2,
\end{equation}
which can be seen by expanding $c^{(2)}_q$ up to second order in $q$:
$c^{(2)}_q = \int d^3 \br e^{-i \bq \br} c^{(2)}(r) \approx \int d^3
\br [1- i(\bq \br) -(\bq \br)^2/2] \, c^{(2)}(r) = \int 4 \pi r^2 dr
[1-q^2 r^2/6] \, c^{(2)}(r)$.

\begin{figure}[t]
  \centering
  \includegraphics[width= .8 \figurewidth]{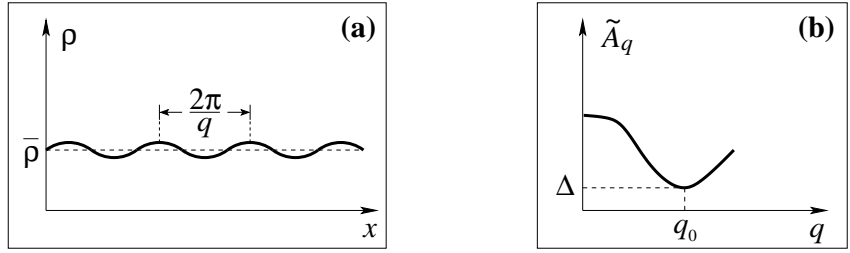}
  \caption{\label{rhoAq} {\bf (a)} The emergence of a solid is
    signalled by the a deviation of the equilibrium density profile
    $\rho(\br)$ from uniformity $\rho(\br) \ne \text{const}$. When the
    solid is periodic, the density profile is, too. {\bf (b)} The
    wavevector dependence of the coefficient $\widetilde{A}_q$ at the
    Fourier component of the second order term in the Landau-Ginzburg
    functional~(\ref{LG}), see Eq.~(\ref{dF}), as originally
    envisioned by Landau~\cite{LandauPT1, LandauPT2}.  Consistent with
    Eq.~(\ref{FRYq}) and Fig.~\ref{cRHendersonGrundke}(b), the second
    order term in Landau-Ginzburg expansion will vanish at a select,
    finite value of the wave-vector $q = q_0$, as $\Delta \to 0$. The
    resultant excitation spectrum is determined by the absolute value
    of $\bq$, but not its direction, and is thus
    quasi-one-dimensional.}
\end{figure}

Now returning to the issue of the liquid-to-crystal transition, the
emergence of a solid will be signalled by the appearance of a finite,
non-uniform component in the density profile, see
Fig.~\ref{rhoAq}(a). Because the density variations of interest for an
ordinary liquid-to-crystal transition are periodic, for that situation
it is convenient to present the variations as a Fourier series in
terms of the vectors $\bq$ of the reciprocal lattice:
\begin{equation} \label{rhoq} \drho(\br) = \sum_{\bq} \drhot_{\bq} \,
  e^{i \bq \br}.
\end{equation}
According to Eq.~(\ref{FRYq}), the second order term in the Landau
expansion can be presented in the form $\sum_{\bq} \, \widetilde{A}_q
|\drhot_{\bq}|^2/2$. To avoid confusion, we note that for a general
form of $\widetilde{A}_q$, the bulk free energy $V(\phi)$ from the
Landau-Ginzburg functional (\ref{LG}) no longer has the ultralocal
form from Eq.~(\ref{LGbulk}). Instead, non-local interactions are now
explicitly included, as in Eqs.~(\ref{dFgen}) and (\ref{Fuu}): $\int
d^3 \br A \phi^2/2 \to \int d^3 \br_1 d^3 \br_2 A(\br_1, \br_2)
\phi(\br_1) \phi(\br_2)/2$, and likewise for higher-order interactions
$\int d^3 \br B \phi^3/3 \to \int d^3 \br_1 d^3 \br_2 d^3 \br_3
B(\br_1, \br_2, \br_3) \phi(\br_1) \phi(\br_2) \phi(\br_3)/3$,
etc. Still, we shall see that when only the dominant Fourier modes are
included, the stability analysis of the functional will simplify to
that of the original form (\ref{LGbulk}).

In the symmetric, uniform phase the coefficients $\widetilde{A}_q$ are
all positive. A (mean-field) instability toward a density wave at a
particular value of $\bq$ such that $|\bq| = q_0$ will reveal itself
through vanishing of the corresponding coefficient
$\widetilde{A}_{q_0}$, see Fig.~\ref{rhoAq}(b).  Landau states that
simultaneous vanishing of $\widetilde{A}_q$ at more than one value of
$q$ is ``unlikely,'' which is consistent with the
$\widetilde{c}^{(2)}_q$ from Fig.~\ref{cRHendersonGrundke}(b). We will
go along with this mean-field view for now and write the second order
term as $\widetilde{A}_{q_0} \, |\delta \widetilde{\rho}_{q_0}|^2/2$
but should recognise that $A_q$ is a smooth function of $q$ and thus
should be rather small in a {\em finite} range of $q$ around the
vicinity of $q_0$. The latter quantity, of course, reflects the
spacing $a$ between the particles.

We can now focus on this most unstable set of modes $|\bq| = q_0$. In
the Fourier sum corresponding to the third-order term in the free
energy, $\sum_{\bq_1} \sum_{\bq_2} \sum_{\bq_3} \,
\widetilde{B}_{\bq_1, \bq_2, \bq_3} \delta \widetilde{\rho}_{\bq_1}
\delta \widetilde{\rho}_{\bq_2} \delta \widetilde{\rho}_{\bq_3}/3$
only the term such that $\bq_1 + \bq_2 + \bq_3 = 0$ survives, if the
coupling $B(\br_1, \br_2, \br_3)$ between density fluctuations is
translationally invariant: $B(\br_1, \br_2, \br_3) = B(\br_1 + \bR,
\br_2 + \bR, \br_3 + \bR)$ for any vector $\bR$. Given that we are
limited to $|\bq_i| = q_0$, the constraint $\bq_1 + \bq_2 + \bq_3 = 0$
implies the three wavevectors form an equilateral triangle with side
$q_0$.  As pointed out by Alexander and
McTague~\cite{PhysRevLett.41.702}, the crystal types whose reciprocal
lattices contain equilateral triangles are easy to enumerate: In 2D,
these are the triangular and hexagonal lattice. (That the triangular
lattice is essentially a sum of three properly phased plane waves
propagating at 60, 180, and 300 degrees is particularly
obvious~\cite{AndersonBasicNotions}.) Note the triangular lattice
maximises the packing density of monodisperse circles in 2D, both
locally and globally~\cite{HalesPOParticle2000}.  In 3D, only the
body-centred cubic (bcc) fits the bill but that alone enables us to
make the crucial conclusion that the coefficient $B$ at the 3rd order
term is generally non-zero for liquids, in contrast with the Ising
magnet for example. This result (of a somewhat torturous argument) is
consistent with the simple intuition that a three-body term---hence, a
three-body contact---is necessary to build a mechanically stable
structure in spatial dimensions 2 and 3. Of course, three-body
interactions are present in actual liquids, as chemical bonds are
generally directional.  Finally note that the regular icosahedron also
consists of the equilateral-triangle motifs but does not tile the
(reciprocal) space owing to its fivefold symmetry. Likewise its
reciprocal (dual) polyhedron, i.e., the regular dodecahedron, does not
tile the direct space. Still note that the regular dodecahedron is the
Voronoi cell for the {\em locally} densest packing of monodisperse
spheres. That such cells do not tile (flat) 3D space, makes the
problem of the densest packing 3D difficult. Indeed, the Kepler
conjecture about the volume fraction of the closest-packed arrays of
monodisperse spheres has been proven only
recently~\cite{HalesPOParticle2000}. To summarise, the third-order
term accounts for the instability of liquids toward {\em
  density}-driven solidification, however given the constraint of
periodicity, this term favours the BCC lattice.

\begin{figure}[t]
  \centering
  \includegraphics[width= .65 \figurewidth]{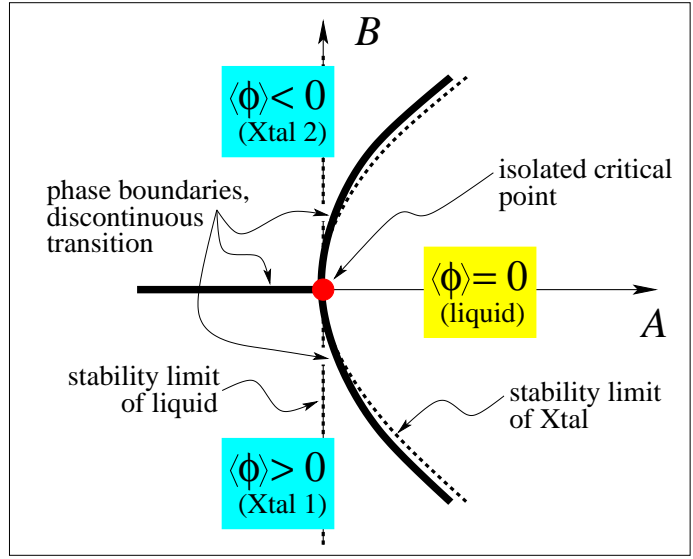}
  \caption{\label{ABphaseDiagram} The mean-field phase diagram of the
    Landau-Ginzburg functional~(\ref{LG}) in the $(A, B)$ plane, $C =
    \text{const} > 0$. The thick solid lines correspond to phase
    boundaries between the symmetric ($\la \phi \ra=0$) and two
    symmetry broken phases, $\la \phi \ra < 0$ and $\la \phi \ra >0$
    respectively. The symmetric phase corresponds to the uniform
    liquid, while the symmetry broken phases to crystalline solids
    with complementary density profiles.  At each point along the
    boundaries, the transition is discontinuous, except at the
    isolated critical point $A=B=0$. In finite dimensions, the
    critical point is avoided thus allowing one to argue that the
    liquid-to-crystal transition is always discontinuous in
    equilibrium, see text for detailed discussion.}
\end{figure}

For the functional (\ref{LGbulk}) to be stable, the coefficient $C$ at
the quartic term must be positive; let us fix it at a certain value
for now and focus solely on the coefficients $A$ and $B$ at the
quadratic and cubic term respectively.  The coefficients $A$ and $B$
are generally linearly independent functions of pressure and
temperature: $A=A(p,T)$, $B=B(p,T)$. As a result, we may consider the
phase diagram of the system in the $(A, B)$ plane, shown in
Fig.~\ref{ABphaseDiagram}, with the understanding that the $(p, T)$
phase diagram can be obtained from that in Fig.~\ref{ABphaseDiagram}
by a simple coordinate transformation $p = p(A, B)$, $T = T(A, B)$.

It is easy to see that there are three distinct phases, as we discuss
in the following and graphically summarise in
Fig.~\ref{ABphaseDiagram}: The symmetric phase $\la \phi \ra = 0$ is
separated from two distinct ordered phases of lower symmetry, $\la
\phi \ra \ne 0$, by a discontinuous transition at $A = 2B^2/9C$,
except in one isolated point, $A = B = 0$, where the transition is
continuous. The higher and lower symmetry phases correspond to the
uniform liquid and crystal states respectively.  The two ordered
phases themselves are separated by a discontinuous transition,
corresponding to the phase boundary at $B=0$, $A < 0$. These two
ordered phases are intimately related: Because the $B$ term switches
sign across their mutual boundary, the density patterns of the two
ordered phases are complementary to each other; to a particle in one
structure, there corresponds a void in the other structure and vice
versa. Finally note that there are two families of curves
corresponding to the mechanical stability limit of the phases
involved: To left of the $A=0$ line, the free energy minimum
corresponding to the symmetric phase vanishes, while in the region $A
> B^2/4C$ the symmetric phase is the sole possible phase.

According to the phase diagram in Fig.~\ref{ABphaseDiagram}, there
cannot be a line of continuous phase transitions separating liquid
from crystal. At most, there is a unique, isolated point at which the
transition could be continuous. It should be understood that in the
direct vicinity of this point, the transitions across the $A =
2B^2/9C$ boundary will be only weakly first order, i.e., they are
accompanied by a small change in the order parameter and a low latent
heat, among other things. Landau notes that whether or not the
critical point of the type in Fig.~\ref{ABphaseDiagram} can be
observed in nature is unclear. Alexander and McTague go further to
argue that, in fact, it {\em cannot}, owing to the
``Brazovsky''~\cite{Brazovskii1975, Brazovskii1975_II} effect: Because
the the free energy contribution of modes $\delta \widetilde{\rho}_q$
depends only on the length of the vectors $\bq$, while
$\widetilde{A}_q$ vanishes at a finite value $q=q_0$, see
Fig.~\ref{rhoAq}(b), the dispersion relation for these modes is
effectively one-dimensional.  Indeed, just above the critical
temperature $T_c$, the free energy cost for spatial density variations
reads (in the quadratic approximation and in $D$ spatial dimensions):
\begin{equation} \label{dF} F - F_\suni = \frac{1}{2} \int \frac{d^D
    \bq}{(2\pi)^D} \, \widetilde{A}_q |\delta \widetilde{\rho}_q|^2
  \propto \int q^{D-1} dq \: \left[ \Delta + (q-q_0)^2 \right] \,
  |\delta \widetilde{\rho}_q|^2,
\end{equation}
c.f. Eqs.(\ref{FRYq}) and Fig.~(\ref{cRHendersonGrundke})(b).
Clearly, the r.h.s. integral above is one-dimensional. On the other
hand, the Landau-Ginzburg bulk energy describes the breaking of a
discrete symmetry, $\phi \leftrightarrow - \phi$, as $B \to 0$.  The
Curie point of an Ising ferromagnet is a good example of such a
discrete symmetry breaking.  Critical points of this type are
generally suppressed by fluctuations in one-dimensional systems at
{\em finite} temperatures~\cite{LLstat}. (This is true unless the
interactions are very long range, $1/r^2$ or
slower~\cite{FrohlichSpencer, 0022-3719-4-5-011}; note excitations in
elastic 3D solids interact at best according to
$1/r^3$~\cite{LLelast}, see also below.) In the present context, these
criticality-destroying fluctuations are motions of domain walls
separating the two symmetry broken phases corresponding to $B>0$ and
$B<0$ and are indeed quasi-one dimensional for interfaces with
sufficiently low curvature.  The Brazovsky effect implies that not
only is the phase boundary between the symmetric and broken-symmetry
phases moved down toward lower values of $A$ owing to
fluctuations---as it generally would---it also dictates that the
continuous transition at $B=0$ will be pushed all the way down to
$T=0$.  Thus, the liquid-to-crystal transition is always first order
in equilibrium.  (Conversely, one may be able to sample some of this
criticality by using rapid quenches, to be discussed in due time.) We
also note that to understand what actually happens at $B = 0$, $A <
0$, i.e., which polymorph will be ultimately chosen by the system,
generally requires knowledge of higher order terms in the
functional~(\ref{LGbulk}).

The erasure of the critical point is not the only non-meanfield effect
we should be mindful of. Recall that the coefficient $\widetilde{A}_q$
is small in a {\em finite} vicinity of the vector $\bq$.  This means
that lattice types other than BCC, such as FCC, can become
stable~\cite{ChaikinLubensky}.  It is these structures in which the
system could settle when the critical point at $A=B=0$ is
avoided. Indeed, elemental solids display a variety of structures,
with bcc being far from prevalent. Still, Alexander and McTague have
argued that even if the BCC structure is not the most stable, it is
likely most kinetically-accessible, especially near weakly
discontinuous liquid-to-solid transitions, consistent with
experiment~\cite{PhysRevLett.41.702}.

The simplest possible approximation (\ref{LGbulk}), in which one
truncates the Landau-Ginzburg expansion at the fourth-order term,
apparently covers the worst-case scenario in the sense that the
presence of negative terms of order higher than three will only act to
further suppress a {\em continuous} liquid-to-crystal
transition. Indeed, suppose the coefficient $C$ at the fourth-order
term in Eq.~(\ref{LGbulk}) is negative, which dictates that we now
expand the free energy up to the {\em sixth} order or higher. A
negative fourth-order term stabilises a broken-symmetry phase, $\phi
\ne 0$, even if the third-order term is strictly zero.  Important
examples of crystal-lattice types for which the fourth or higher order
terms must be non-zero include the graphite, diamond, and simple-cubic
lattices. For instance, to stabilise the graphite lattice, not only
should the bond-angle be fixed at sixty degrees, for which a
third-order term alone would suffice. In addition, the three bonds
emanating from an individual particle must be stabilised in the planar
arrangement. In the case of the diamond lattice, it is likely that the
fifth-order term is non-zero, too: While a fourth-order term alone
could stabilise local pyramidal configuration in which the bond-angles
are at the requisite value of $109.4^\circ$, the resulting energy
function could also favour stacked double-layers giving rise to a
rhombohedral lattice, in addition to the diamond structure. (The
latter stacked structure is exemplified by arsenic, however the angle
is intermediate between $109.4^\circ$ and $90^\circ$.)  Likewise, a
structure with enforced $90^\circ$ bond-angles could, in principle, be
simple-cubic but such a structure is often unstable toward tetragonal
distortion. Interestingly, the only elemental solid that has the
simple-cubic lattice structure at normal pressure is
polonium. (Arsenic, phosphorus, and, of all things, calcium can be
made simple-cubic by applying pressure~\cite{B517777B,
  Olijnyk1984191}.)  Consistent with these notions, we know for a fact
that interactions in actual crystals are truly many-body and are even
affected by (electronic) relativistic
effects~\cite{PhysRevLett.99.016402}. Note the diamond, graphite, and
simple-cubic lattices are some of the most open structures encountered
in actual crystals. (Low density structure with nanovoids can be
readily made~\cite{3538742020081201} but are not uniformly open.)

The following picture emerges from the above discussion: The presence
of a significant, fourth (or higher) order stabilising contribution to
the free energy of the liquid leads to the formation of open
structures with very directional bonding and thus reduces the effects
of steric repulsion on the liquid-to-solid transition.  At the same
time the discontinuity of the liquid-to-crystal transition is
relatively large, see discussion at the end of this Section.  Liquids
that freeze into the diamond lattice and expand alongside illustrate
this situation particularly well. Conversely, if such high order terms
in the free energy expansion are weak, excluded-volume effects become
important that are accounted for by the third-order term.  The smaller
the third order term, the milder the discontinuity of the
liquid-to-solid transition (if the higher-order terms are
non-negative!). Still, the third-order terms cannot completely
disappear in equilibrium because of the steric effects. On the other
hand, quasi-one dimensional fluctuations would destroy the critical
point, even if the third-order terms were very small. We thus conclude
that those quasi-one dimensional fluctuations are {\em also} of steric
origin. We reiterate that despite their importance, the steric effects
are not the only player.  It is essential to remember that the liquid
is equilibrated at a {\em finite} temperature and, thus, is not
jammed; the particles ultimately are allowed to vibrate and exchange
places.

\begin{figure}[t]
  \centering
  \includegraphics[width= .85 \figurewidth]{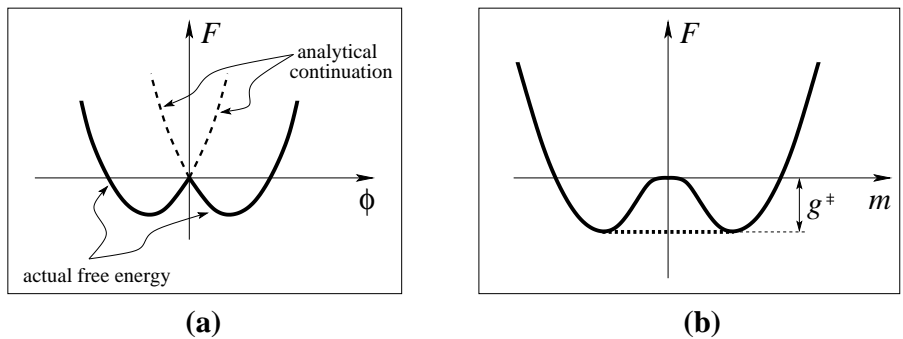}
  \caption{\label{LGfigure} {\bf (a)} This sketch illustrates that
    analytical continuation of the free energy between distinct phases
    is generally ill-defined. {\bf (b)} Thick solid line: The
    Helmholtz free energy $F$ of a spatially {\em uniform} ferromagnet
    below the Curie point, as a function of the average magnetisation
    $m \equiv M/N$ ($M$ is the total magnetisation). The region
    separating the two minima is non-concave; one cannot use the
    Legendre transform to compute the Gibbs free energy $G$: $G = F -
    M h$, $h = (\prtl F/\prtl M)_T$. In equilibrium, the system
    phase-separates, and so the actual, equilibrium Helmholtz free
    energy between the two minima in $F$ is given by the thick dashed
    line. The latter line is entirely analogous to the red line in
    Fig.~\ref{spectrum}(b).}
\end{figure}

The above argument has a somewhat unsatisfactory feature in that it
implies the free energy of a periodic solid can be computed using an
expansion from the uniform state, even though the two states are
separated by a phase transition. The bulk free energy generally
exhibits singularities at phase transitions in finite dimensions and
so expressions for low magnitude fluctuations around metastable
equilibria cannot be analytically-continued across transitions in a
straightforward way, see illustration in
Fig.~\ref{LGfigure}(a). Conversely, there are plenty of examples of
mean-field free energies that are perfectly analytic near a phase
transition, such as the one leading to the familiar van der Waals
equation of state $(V-V_f)[p + c (N/V)^{2}] = Nk_B T$. These potential
issues with the analytical continuation of the free energy do not
invalidate the Landau argument, however.  Indeed, if we assumed, for
the sake of argument, that the transition did {\em not} take place,
the free energy could extrapolate to a value that could only {\em
  exceed} $F(\delta \rho = 0)$. Since it does not do so, in fact, we
then conclude that our assumption that the transition did not take
place was incorrect.  Consistent with this notion, a number of workers
have, in practice, accomplished the program of building a periodic
crystal state starting from the uniform liquid state, starting from
the pioneering work of Kirkwood and Monroe~\cite{KirkwoodMonroe}.
Subsequent, detailed calculations along similar
lines~\cite{RyzhovTareeva, RiceCerjanBagchi1985} explicitly confirmed
the discontinuous nature of the transition. At the same time, they
showed that physical quantities, such as the liquid and crystal
densities at the transition, converge slowly, if at all, with number
of the reciprocal vectors in the expansion
(\ref{rhoq})~\cite{RiceCerjanBagchi1985}.

These difficulties may stem, at least in part, from the large
deviations of the crystalline density profile from a sinusoidal curve.
In fact for a perfectly harmonic crystal, the density profile in a
harmonic solid is a superposition of Gaussian peaks (\cite{LLstat},
Chapter 138):
\begin{equation} \label{rho} \rho(\br, \alpha) = (\alpha/\pi)^{3/2}
  \sum_i e^{-\alpha (\br - \br_i)^2}.
\end{equation}
In contrast with the picture transpiring from Fig.~\ref{rhoAq}(a), the
peaks are rather sharp: Empirically~\cite{Bilgram, GrimvallSjodin,
  PhysRevLett.82.747, L_Lindemann, RL_LJ}, $\alpha \simeq 100/a^2$
nearly universally, see Fig.~\ref{solid}(a). The quantities $\br_i$
stand for the locations of the vertices of the lattice. The
vibrational displacement in crystals goes with temperature roughly as
$T^{1/2}$, above the Debye temperature, and is numerically close to
its value near melting, i.e., $1/\sqrt{\alpha}a \simeq 1/10$ of the
lattice spacing or so. The latter quotient should be nearly universal,
as was argued by Lindemann more than a century ago. It is by no means
intuitively obvious that the vibrational displacement near melting,
which is often called the Lindemann length, should as small as
one-tenths.  Lindemann himself~\cite{Lindemann} thought that this
displacement should be about a half of the particle spacing, which
would be more consistent with the mild density variation in
Fig.~\ref{rhoAq}(a) than with the sharp peaks in
Fig.~\ref{solid}(a). Lindemann's (correct) logic was that the melting
was caused by intense collisions between the atoms. Still, both
experiment and theory indicate the dramatic anharmonicities that
eventually lead to crystal's disintegration should become important
already at $\alpha \simeq 100/a^2$. In Subsection~\ref{RFOTDFT}, we
shall see that for rigid particles, the quantity $\alpha a^2$ is
essentially given by the number of particles in the first coordination
shell (up to a logarithmic correction and a constant factor), which
scales as the dimensionality of space squared, see
Eq.~(\ref{KWalpha1}).

It is straightforward to obtain an explicit expression for
$\alpha$---which is the ``spring constant'' of an effective Einstein
oscillator---given a specific form of the phonon spectrum. For
instance, assuming isotropic elasticity with the isothermal bulk
modulus $K$ and shear modulus $\mu$ and a Debye phonon spectrum with
an ultraviolet cutoff at $|\bk | = \pi/a$, one obtains~\cite{RL_Tcr}:
\begin{equation} \label{alpha1} \alpha^{-1} \equiv \frac{1}{a} \,
  \frac{k_BT}{3\pi\mu} \, \frac{6K + 11\mu}{3K + 4 \mu}.
\end{equation}

It is particularly straightforward to discuss the driving force for
the crystallisation of hard spheres, which do not form actual bonds
and for which the enthalpy change during the transition:
\begin{equation} \label{DH} \Delta H_m = \Delta E_m + p_m \Delta V_m
\end{equation}
is entirely due to the work needed to compress the liquid: $\Delta V_m
< 0$, at constant temperature. The stability of the hard sphere
crystal, relative to the corresponding liquid (gas), stems from the
notion that the densest packing of monodisperse spheres in 3D is a
periodic crystal---face-centred cubic (FCC) or hexagonal close-packed
HCP---which now appears to be rigorously
proven~\cite{HalesPOParticle2000}. This notion has resisted systematic
effort since time of Kepler and is non-trivial because the most
compact Voronoi cell of a particle surrounded by twelve
neighbours---the highest number allowed in 3D---is the regular
dodecahedron.  The latter has a five-fold symmetry and thus does not
tile space. As follows from the proof of the Kepler conjecture, one
can say that thermodynamically, the transition is density-driven or,
equivalently, {\em sterically}-driven. Already the original Landau
argument from 1937, which neglects fluctuations, suffices to show that
the transition in hard spheres is first order: All points on the
liquid-crystal phase boundary in the $(p, T)$ plane are equivalent
because the $p/T$ ratio in hard spheres is determined by the density
exclusively. Thus, it is impossible to have an isolated critical point
on that phase boundary.

To give a partial summary, the liquid-to-crystal transition is a
discontinuous phase transition resulting in a {\em breaking of
  translational symmetry}, upon which individual particles occupy
specific cells in space as opposed the whole space.  The fusion
entropy is essentially the difference between the entropy of
translational motion in the uniform liquid and the vibrational entropy
in the crystal.  Consistent with this, the fusion entropy in open
structures often exceeds that for crystallisation of well-packed
structures, as we shall see explicitly in Subsection~\ref{bead}. For
rigid particles, crystallisation is driven by steric repulsion
alone. The liquid-to-crystal transition breaks a {\em continuous}
symmetry.  Breaking such a symmetry results in the emergence of
Goldstone modes~\cite{Goldenfeld} below the transition, which are
represented in crystals by the phonons~\cite{AndersonBasicNotions}. To
put this in perspective, the symmetry of the Hamiltonian to
simultaneous flips of all spins $\{ \sigma_i \} \leftrightarrow \{-
\sigma_i \}$, which is broken during ferromagnetic ordering, is a
discrete symmetry, see Fig.~\ref{LGfigure}.

\subsection{Emergence of the Molecular Field}

A methodological advantage of studying a monodisperse, hard-sphere
liquid is that one can gain qualitative insight into its
crystallisation while being able to test the corresponding
approximations against simulations. Such tested approximations will
come in handy when studying the emergence of rigidity in {\em
  aperiodic} systems, since monodisperse hard spheres crystallise too
readily thus preventing one from achieving meaningfully deep quenches
in a direct simulation.

Most of the quantitative theories of liquid-to-solid transition can be
traced back to the very old idea of the ``molecular field,'' which was
originally conceived in the context of the vapour-to-liquid transition
by van der Waals~\cite{VdW} and in the context of ferromagnetism by
Curie and Weiss~\cite{Weiss, Ashcroft}. We have already benefited from
thinking about the latter system; let us use it again to illustrate
the concept of the molecular field. For a collection of Ising spins,
Eq.~(\ref{EIsing}), spin $i$ is subject to an instantaneous field
$\sum_j J_{ij} \sigma_j$.  If we, in the mean-field fashion, ignore
correlations in spin flips: $\la \sigma_i \sigma_j \ra \approx \la
\sigma_i \ra \la \sigma_j \ra$, the (thermally-averaged) value of the
molecular field $h_i^\text{mol}$ is simply given by the expression:
$h_i^\text{mol} = \sum_j J_{ij} m_j$.  We observe that when a
ferromagnet spontaneously polarises, so that $m_i > 0$ or $m_i < 0$,
this effective field becomes non-zero thus breaking the time-reversal
symmetry $\{ \sigma_i \} \leftrightarrow \{- \sigma_i \}$ of the
Hamiltonian $\cH = - \sum_{i<j} J_{ij} \sigma_i \sigma_j$.

A systematic way to implement the concept of the effective field is to
employ an additional ensemble. The Helmholtz free energy in
Eq.~(\ref{IsingF}) is the appropriate thermodynamic potential at fixed
magnetisation and is analogous to the canonical construction for
particulate systems, in which one fixes the volume of the
system. Alternatively, one may use the ensemble in which the
magnetisation is allowed to fluctuate while the magnetic field is
fixed. In gases, this is analogous to allow for volume changes at
fixed particle number, while subjecting the gas to external pressure
(the isobaric ensemble); the corresponding thermodynamic potential is
the Gibbs free energy $G = F + pV$.  Alternatively, one may allow the
particle number to change at fixed $V$, while imposing a specific
value of the chemical potential (the grand-canonical ensemble); the
corresponding free energy is $\Omega = F - \mu N = -pV$~\cite{LLstat,
  McQuarrie}.

To evaluate the Gibbs free energy for the magnet, let us formally
write down the partition function for the energy function from
Eq.~(\ref{EIsing}) with an added term $-\sum_i h_i \sigma_i$:
\begin{equation} \label{ZG} Z(\{ h_i \}) = \sum_{\{ \sigma_i = \pm 1
    \} } \exp[- \beta(\cH - \sum_i h_i \sigma_i)].
\end{equation}
Clearly, the average magnetisation $m_i \equiv \la
\sigma_i \ra$ can be computed according to:
\begin{equation} \label{mi} m_i = - \left( \frac{\prtl G(\{ h_i
      \})}{\prtl h_i} \right)_T,
\end{equation}
where $G(\{ h_i \}) \equiv - k_B T \ln Z(\{ h_i \})$ is a function of
the fields $h_i$.  It is easy to convince oneself (without actually
computing $Z$) that one can write out this (Gibbs) free energy as $G =
- \sum_i m_i h_i + E - T S$, where $E \equiv \la \cH \ra$ is the
energy while the quantity $S$ is equal to $ -(\prtl G/\prtl T)_{ \{h_i
  \}}$ and thus must be identified with the entropy. Consequently, the
function $F = G + \sum_i m_i h_i = E - TS$ must be identified with the
Helmholtz free energy, which we have seen can be expressed exclusively
as a function of local magnetisations $m_i$: $F = F(\{ m_i \})$. On
the other hand, $d F = d G + d(\sum_i m_i h_i) = - S dT + \sum_i h_i d
m_i$, and so one has for spin $i$:
\begin{equation} \label{hi} h_i = \left( \frac{\prtl F(\{ m_i
      \})}{\prtl m_i} \right)_T.
\end{equation}
The meaning of Eq.~(\ref{mi}) is that one solves for $h_i$ such that
satisfy the equation for a specified set of magnetisations
$m_i$'s. Likewise, one solves for $m_i$ such that satisfy
Eq.~(\ref{hi}) for a chosen set of on-site fields $h_i$.  It would
appear that the two descriptions in Eq.~(\ref{mi}) and (\ref{hi}) are
equivalent, so that either one can be used. In fact, the Gibbs
ensemble is often more computationally convenient and thus is used
more frequently. However, the two descriptions, which are related by a
Legendre transform: $F = G + \sum_i m_i h_i$, are only equivalent if
both $F$ or $G$ are concave or convex~\cite{Callen}, which is
certainly not the case for the ferromagnet below its Curie point, see
Fig.~\ref{LGfigure}. This figure demonstrates that while to a given
value of magnetisation $m$ there corresponds a unique value of the
field $h = \prtl F/\prtl m$, the converse is not generally true: Up to
three distinct solutions exist for equation $h = \prtl F/\prtl m$ when
the $F(m)$ curve has a convex-up portion. Furthermore, for $h = 0$,
the two stable solutions for the magnetisation are exactly degenerate
in free energy, thus making the choice between distinct solutions of
Eqs.~(\ref{hi}) particularly ambiguous. Furthermore, if one enforces
summation over spin states in the partition function from
Eq.~(\ref{ZG}), at $h_i = 0$, one will recover $m_i = 0$ even below
the Curie point, thus missing the transition. This apparent---but
incorrectly determined---equilibrium value of $m_i = 0$ is an unstable
solution of the free energy functional!  This problematic situation
would not arise in the Helmholtz ensemble, which clearly shows the two
{\em polarised} states as minima of the free energy. These notions
indicate that the Helmholtz ensemble is more basic than the Gibbs
ensemble. In much the same way, the microcanonical ensemble is more
basic than the canonical ensemble in the context of
Fig.~\ref{spectrum}(a); the isochoric ensemble is more basic than the
isobaric ensemble in Fig.~\ref{spectrum}(b).

For an individual spin, however, the $h \leftrightarrow m$
correspondence is one-to-one and we may safely choose to work with
either the magnetisation $m_i$ or the corresponding ``molecular
field'' $h_i^\text{mol}$:
\begin{equation} \label{himol} h_i^\text{mol} = \frac{\prtl
    F_\text{id}}{\prtl m_i} \hspace{5mm} \Rightarrow \hspace{5mm}
  h_i^\text{mol} = k_B T \tanh^{-1}(m_i) \hspace{5mm} \Leftrightarrow
  \hspace{5mm} m_i = \tanh \beta h_i^\text{mol}
\end{equation}

For the field-less energy function in Eq.~(\ref{EIsing}), $h_i = 0$,
and so the equilibrium configuration of $m_i$ simply optimises the
free energy $F$:
\begin{equation} \label{Fm0} \frac{\prtl F}{\prtl m_i} = 0,
\end{equation}
which, together with Eqs.~(\ref{IsingF}) and (\ref{himol}), yields:
\begin{equation} \label{FFm} \frac{\prtl F_\text{id}}{\prtl m_i} +
  \frac{\prtl F_\text{ex}}{\prtl m_i} = 0 \hspace{5mm} \Rightarrow
  \hspace{5mm} h^\text{mol}_i = - \frac{\prtl F_\text{ex}}{\prtl m_i}.
\end{equation}
In the mean-field case (\ref{IsingEX}):
\begin{equation} \label{hmolIsing}
h^\text{mol}_i = \sum_j J_{ij} m_j,
\end{equation}
which is expected since at any given time, spin $i$ is subject to the
instantaneous field $\sum_j J_{ij} \sigma_j$. In the mean-field
approximation, one neglects correlations between spin flips and so, to
compute the expectation value of the instantaneous field, one one may
simply replace $\sigma_j$'s by their average values. 

Whether or not the {\em actual} external field $h_i$ is zero, using
Eqs.~(\ref{hi}) with an arbitrary value of the parameter $h_i$ can be
quite useful because it allows one to analyse configurations other
than the equilibrium ones. Quantifying fluctuations around equilibrium
is necessary for determining various susceptibilities or interactions,
among other things.  For instance, the stiffness of the magnetic
response of an isolated spin is $\prtl h_i/\prtl m_i = k_B T/(1 -
m_i^2)$; it is, of course, the reciprocal of the susceptibility.
Further, $\prtl h_i/\prtl m_j = (\prtl^2 F_\text{ex}/\prtl m_i \prtl
m_j)$ which is equal to $ - J_{ij}$ in the mean-field limit,
c.f. Eq.~(\ref{Jij}). Likewise, one can use Eq.~(\ref{mi}) to compute
the spin-spin correlation function
\begin{equation} \label{Chi} \frac{\prtl m_i}{\prtl (\beta h_j)} = -
  k_B T \frac{\prtl^2 F}{\prtl h_j \prtl h_i} = \la \sigma_i \sigma_j
  \ra - m_i m_j \equiv \chi(\br_{ij}),
\end{equation}
which is a key quantity for detecting second order transitions with a
diverging correlation length.  Note that the two matrices $\prtl
m_i/\prtl h_j$ and $\prtl h_i/\prtl m_j$ are inverse of each other:
$\delta_{ij} = \prtl m_i/\prtl m_j = \sum_k (\prtl m_i/\prtl
h_k)(\prtl h_k/\prtl m_j)$.

\begin{figure}[t]
\begin{tabular*}{\figurewidth} {ll}
\begin{minipage}{.4 \figurewidth} 
  \begin{center}
    \includegraphics[width=0.4
    \figurewidth]{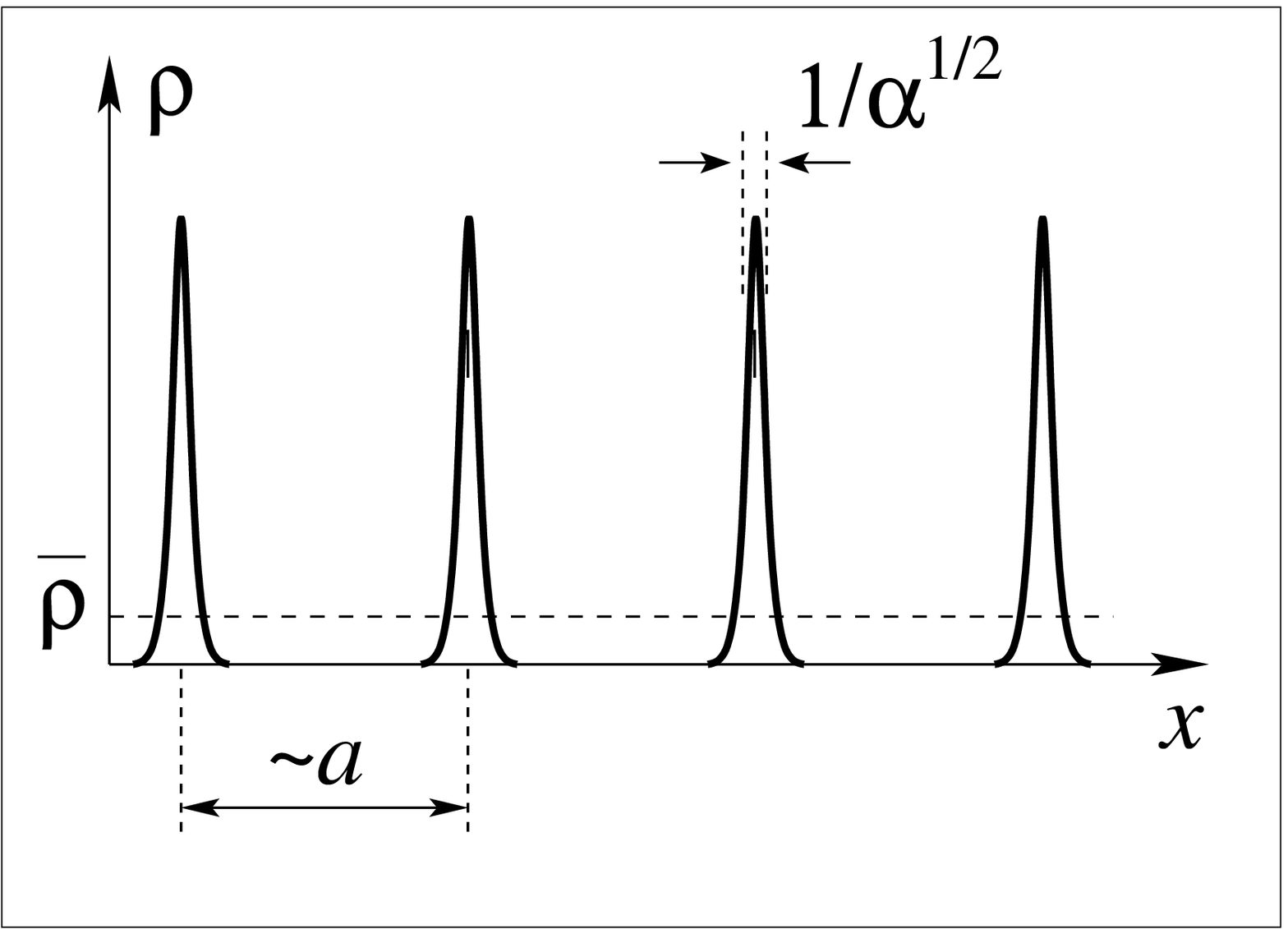} \vspace{7mm}
    \\ {\bf (a)}
   \end{center}
  \end{minipage}
&
\begin{minipage}{.55 \figurewidth} 
  \begin{center}
    \includegraphics[width= .5 \figurewidth]{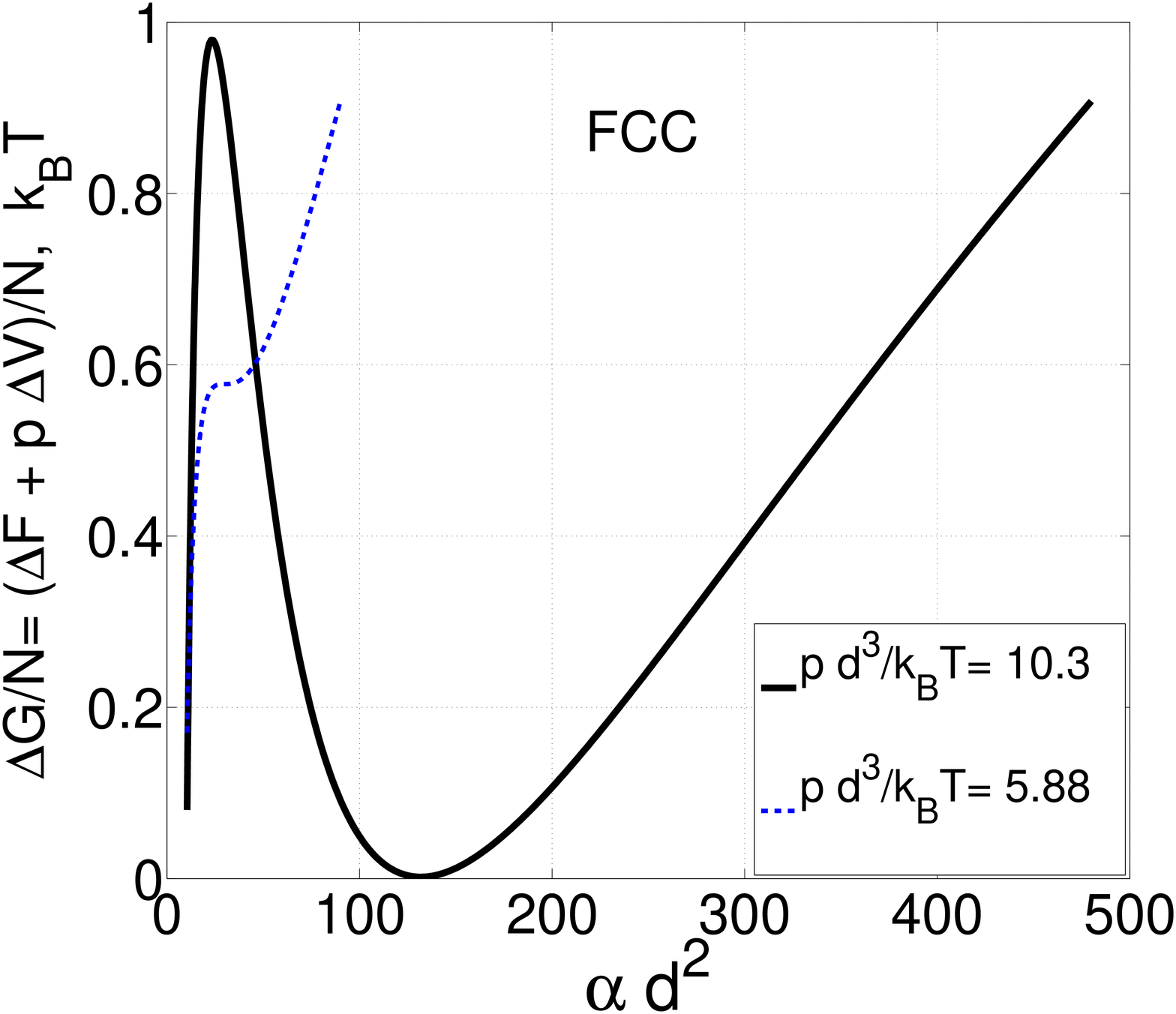}
    \\ {\bf (b)}
  \end{center}
\end{minipage}
\end{tabular*}
\caption{ \label{solid} {\bf (a)} Sketch of the density profile in a
  periodic crystal. In a purely harmonic solid, the profile is a sum
  of gaussians, $(\alpha/\pi)^{3/2} e^{-\alpha (\br-br_i)^2}$, where
  $\br_i$ denotes the average position of particle $i$, see
  Eq.~(\ref{rho}). In important contrast with the density profile in
  Fig.~\ref{rhoAq}(a), the density profile is not simply a periodic
  function of the coordinate, but a superposition of {\em narrow},
  disparate peaks whose width is about one-tenth of the particle
  spacing: $1/\alpha^{1/2} \simeq a/10$.  {\bf (b)} The Gibbs free
  energy of the hard sphere FCC crystal as a function of the force
  constant $\alpha$ of the effective Einstein oscillator from
  Eq.~(\ref{rho}). The thick line corresponds to equilibrium between
  the liquid and crystal. The dashed line corresponds to a liquid that
  is still well above the fusion transition, but the metastable
  minimum corresponding to the crystal just begins to appear. }
\end{figure}

Analogously to how one may view the spontaneous magnetisation as the
emergence of a non-zero molecular field, we may view each peak in
Fig.~\ref{solid}(a) as corresponding to an effective one-particle
potential. Suppose the particles' instantaneous locations are
$\br_i$. In the presence of a one-particle potential $u(\br)$, we must
add the term $\sum_i u(\br_i) = \int d^3 \br \rho_\text{inst}(\br)
u(\br)$ to the {\em energy} function of the liquid, where
\begin{equation} \label{rhoinst} \rho_\text{inst}(\br) \equiv \sum_i
  \delta(\br - \br_i)
\end{equation}
is the instantaneous, highly inhomogeneous density profile of our
liquid, whose thermally {\em averaged} value
(c.f. Eq.~(\ref{sigmam})):
\begin{equation} \label{rhoav}
\rho(\br) \equiv \la \rho_\text{inst} (\br) \ra
\end{equation}
may or may not be spatially uniform. For instance, in a dilute gas,
far from the walls of the container, and in the absence of external
field, the density profile $\rho(\br)$ is spatially uniform. In
contrast, the density profile consists of relatively sharp, disparate
peaks in a mechanically stable solid, as we shall see shortly.  Since
our free energy is defined as a space integral, which is a summation
over fixed (elemental) volumes, this addition to the energy amounts to
switching from the canonical to grand-canonical ensemble, in which the
chemical potential is spatially-varying and equals $-u(\br)$. Call the
corresponding {\em free} energy $\Omega = F + \int d^3 \br \rho(\br)
u(\br)$, so that
\begin{equation} \label{rhor} \rho(\br) = \frac{\delta \Omega}{\delta
    u(\br)},
\end{equation}
c.f. Eq.~(\ref{mi}). Consequently, the appropriate analog of
Eq.~(\ref{hi}) is
\begin{equation} \label{ur} u(\br) = - \frac{\delta F}{\delta
    \rho(\br)}.
\end{equation}
(In the literature, one often uses not the quantity $u(\br)$ but its
negative~\cite{Evans1979}, and also additionally multiplied by
$\beta$~ \cite{LovettMouBuff, YangFlemingGibbs, HaymetOxtoby}.)
Because here the density is defined over continuum space, not a
discrete set over spin sites, we employ functional differentiation in
Eqs.~(\ref{rhor}) and (\ref{ur}). The difference of functional
differentiation from regular differentiation is that in the latter, we
write an increment of a function $f$ defined over a discrete set of
independent variables $g_i$ ($\prtl g_i/\prtl g_j = \delta_{ij}$) as a
discrete sum: $df = \sum_j (\prtl f/\prtl g_j) dg_j$. If, on the other
hand, $g$ is defined over a continuum space, we would like to change
our definition of the derivative so that one can replace the discrete
summation by continuous integration: $d f = \int d^3 \br_j [\delta
f/\delta g(\br_j)] d g(\br_j)$.  This convention does work, but only
if we define $\delta f(\br_i)/\delta f(\br_j) \equiv \delta(\br_i-
\br_j)$ for any function $f(\br)$. As a result, $\frac{\delta}{\delta
  \rho(\br')} \int d^3\br u(\br) \rho(\br) = u(\br')$, if the function
$u$ does not depend on $\rho$ or its derivatives; see, for instance,
Ref.~\cite{Goldenfeld} or \cite{LLmech} for more details. Another
useful notion is $\delta f/\delta h = \int d^3 \br_j [\delta f/\delta
g(\br_j)] [\delta g(\br_j)/\delta h]$. Functional variation of a
quantity with respect to the density, $\delta/\delta \rho({\br})$, is
special in that it yields the (differential) response of the quantity
to adding a particle to the system in the vicinity of site $\br$: On
the one hand, $\int d^3 \br [d \rho(\br)] \frac{\delta}{\delta
  \rho(\br)} = \lim_{dV_i \rightarrow 0} \sum_i [d \rho(\br_i)]
\frac{\prtl}{\prtl \rho(\br_i)}$ by construction. On the other hand,
$\int d^3 \br [d \rho(\br)] \frac{\delta}{\delta \rho(\br)} =
\lim_{dV_i \rightarrow 0} \sum_i dV_i [d \rho(\br_i)]
\frac{\delta}{\delta \rho(\br_i)}$ by definition of the definite
integral. This implies $\frac{\delta}{\delta[\rho(\br_i)]}
\leftrightarrow \frac{\prtl}{\prtl[d V_i \rho(\br_i)]} =
\frac{\prtl}{\prtl[d N_i]}$.

As in the ferromagnet case, the variable $u(\br)$ in Eq.~(\ref{ur})
can be set to an external potential of relevance, thus resulting in an
equation for the density profile. By varying $u(\br)$, we can explore
the density distribution to a desired degree of deviation from
equilibrium.  As the simplest illustration of Eq.~(\ref{ur}), consider
an ideal gas in an external potential $U(\br)$. The Maxwell-Boltzmann
distribution dictates that $\rho(\br) = \Lambda^{-3} e^{-\beta U(\br)}
\Rightarrow U(\br) = - k_B T \ln[\rho(\br) \Lambda^3]$. Substituting
this in $U(\br) = -\delta F_\text{id}/\delta \rho(\br)$ and
integrating in $\rho$ readily yields Eq.~(\ref{F_id}), as expected.

In full correspondence with the ferromagnet case, the descriptions in
Eqs.~(\ref{rhor}) and (\ref{ur}) are generally not equivalent. It is a
corollary of the Hohenberg-Kohn-Mermin theorem~\cite{PhysRev.136.B864,
  PhysRev.137.A1441} that to a given density profile, there
corresponds a unique external potential, and so the solution of
Eq.~(\ref{rhor}) is unique, while the same is not necessarily true of
Eq.~(\ref{ur}). This notion will return in full force in the following
Subsection, when the degeneracy of the solution is not 2, as was the
case for the ferromagnet, but scales {\em exponentially} with the
system size! 

The quantity $\delta F/\delta \rho(\br)$ is the free energy cost of
adding a particle to the system at location $\br$, i.e., the local
value of the chemical potential $\mu$.  In equilibrium and in the
absence of external potential, the free energy cost of adding a
particle to the system should be uniform in space, lest there be
non-zero net particle flux $\bj \propto -\nabla \mu$.  Thus the liquid
analog of Eq.~(\ref{Fm0}) is
\begin{equation} \label{Fopt} \frac{\delta F}{\delta \rho(\br)} = \mu
  = \text{const}.
\end{equation}
In correspondence with equations (\ref{himol}) and (\ref{FFm}), we can
define the molecular field:
\begin{equation} \label{umol} u^\text{mol}(\br) \equiv - \frac{\delta
    F_\text{id}}{\delta \rho(\br)} \hspace{5mm} \Rightarrow
  \hspace{5mm} u^\text{mol}(\br) = - k_B T \ln[\rho(\br) \Lambda^3]
\end{equation}
and establish its relation with the interaction part of the free
energy:
\begin{equation} \label{FFrho} \frac{\delta F_\text{id}}{\delta
    \rho(\br)} + \frac{\delta F_\text{ex}}{\delta \rho(\br)} = \mu
  \hspace{5mm} \Rightarrow \hspace{5mm} -u^\text{mol}(\br) = \mu +
  \beta^{-1} c^{(1)}(\br).
\end{equation}

On the one hand, there is one-to-one correspondence between the
molecular field and the equilibrium density. On the other hand, the
free energy of the liquid is completely determined by the density.
Thus calculation of the free energy amounts to finding the molecular
field. No approximation is made and no generality is lost hereby.

Eq.~(\ref{FFrho}) is trivially satisfied by the uniform liquid, for
which all three terms are spatially-uniform. To see a possibility of
non-trivial solutions, we may present the first-order correlation
function $ c^{(1)}(\br)$ as a power-series expansion in terms of
deviations from the (uniform) liquid density $\rho_l$: $\delta
\rho(\br) = \rho(\br) - \rho_\sliq$, as was done by Ramakrishnan and
Yussouff~\cite{RamakrishnanYussouff}. In the lowest non-trivial order
this expansion reads:
\begin{equation}
  c^{(1)}(\br_ 1) - c^{(1)}_\sliq =
  \int d^3 \br_2 \, [\delta c^{(1)}(\br_1)/\delta \rho(\br_2)] \delta
  \rho(\br_2) = \int d^3 \br_2 \, c^{(2)}(\br_1, \br_2) \delta
  \rho(\br_2), 
\end{equation}
where $c^{(1)}_\sliq$ denotes the value of
$c^{(1)}(\br)$ in the uniform liquid. The quantity $c^{(2)}(\br_1,
\br_2)$ is the familiar direct correlation function:
\begin{equation} \label{c2r} c^{(2)} (\br_1, \br_2)= \frac{\delta
    c^{(1)}(\br_1)}{\delta\rho (\br_2)} = -\beta \frac{\delta^2
    F_\text{ex} [\rho(\br)]}{\delta\rho (\br_1) \, \delta\rho
    (\br_2)}.
\end{equation}
In the uniform liquid state, $c^{(2)} (\br_1, \br_2) = c^{(2)}(\br_1 -
\br_2; \rho_\sliq)$.  Eq.~(\ref{FFrho}) thus becomes:
\begin{equation} \label{RY} 
  \ln[\rho(\br_1)/\rho_\sliq] = \int d^3 \br_2 c^{(2)}(\br_1 - \br_2;
  \rho_\sliq) \delta \rho(\br_2),
\end{equation}
where we had to set chemical potential so that $\beta \mu +
c^{(1)}_\sliq = \ln[\rho_\sliq \Lambda^3]$. In view of
Eqs.~(\ref{umol}) and (\ref{RY}), we observe how a molecular field
could self-consistently arise in the presence of a frozen-in density
wave $\delta \rho(\br) \ne 0$: $u^\text{mol}(\br) = - k_B T \int d^3
\br_1 c^{(2)}(\br_1 - \br_2; \rho_\sliq) \delta \rho(\br_2)$, up to an
additive constant. We can attempt to make a connection with the
ferromagnet case by considering the weak interaction limit, in which
$c^{(2)}(\br) \rightarrow - \beta v(\br)$, as in Eq.~(\ref{cV}). The
resulting expression, $u^\text{mol}(\br) = \int d^3 \br_1 v(\br_1 -
\br_2) \delta \rho(\br_2)$, is quite similar to the expression
$h_i^\text{mol} = \sum_j J_{ij} m_j$ we obtained in the mean-field
limit for the ferromagnet.  We will see in a bit that the two
expressions are in fact identical in structure.

Before that, let us write down an {\em exact} expression that connects
the molecular field with the full density of the solid, which will
turn out to be instructive in other contexts as well.  In the presence
of external potential $U(\br)$ and in equilibrium, we must set
$u(\br)$ = $U(\br)$ in Eq.~(\ref{ur}). As a result, Eqs.~(\ref{F}),
(\ref{F_id}), (\ref{ur}), and (\ref{C1r}) yield $\beta U(\br) = -
\ln[\rho(\br) \Lambda^3] + c^{(1)}(\br)$. Taking the gradient of this
equation results in $\nabla [\beta U(\br) + \ln \rho(\br)] = \nabla
c^{(1)}(\br)$. To evaluate $\nabla c^{(1)}(\br)$, we note that the
dependence of $c^{(1)}(\br)$ on the coordinate arises exclusively
through the density profile $\rho(\br)$, by Eqs.~(\ref{F}) and
(\ref{C1r}).  Consequently, a change in the direct correlation
function upon a small shift $d \br$ in the coordinate $c^{(1)}(\br_1 +
d \br) - c^{(1)}(\br_1) = \int d^3 \br_2 [\delta c^{(1)}(\br_1)/\delta
\rho(\br_2)] \delta \rho(\br_2)$, where $\delta \rho(\br_2)$ is the
(small) change in the density profile resulting from the increment $d
\br$ in the coordinate. This change is equal to $\delta \rho(\br_2) =
\rho(\br_2 + d\br) - \rho(\br_2)$. (Note the change is in the {\em
  argument} of the density, be this argument a dummy variable or not.)
Using this notion and the definition of the direct correlation
function Eq.~(\ref{c2r}), we thus obtain~\cite{LovettMouBuff,
  Wertheim1976}:
\begin{equation} \label{LMBcr} \nabla_1 [\ln \rho(\br_1) + \beta
  U(\br_1)] = \int d^3 \br_2 \, c^{(2)}(\br_1, \br_2) \, \nabla_{\! 2}
  \, \rho(\br_2).
\end{equation}
The equation demonstrates that, on the one hand, the liquid density
obeys the Boltzmann law in the absence of correlations between
particles: $(\rho \rightarrow 0) \Rightarrow (\rho \propto e^{-\beta
  U(\br)} )$. Conversely, it directly shows how the particles produce
an effective force onto each other, via the correlations, since the
r.h.s. enters in the equation additively with external force $- \nabla
U(\br)$.  Note that even given a high quality functional form for the
density profile, such as in Eq.~(\ref{rho}), the equation above cannot
be used to determine the parameters of the profile self-consistently
because it requires the knowledge of the direct correlation function
in the solid. Such knowledge is currently lacking (see however the
insightful analysis by McCarley and Ashcroft~\cite{PhysRevE.55.4990}).

For a homogeneous liquid, $c^{(2)}(\br_1, \br_2) = c^{(2)}(\br_1 -
\br_2)$, the equation above is easily integrated to yield, in the
absence of external potential~\cite{Lovett1977}:
\begin{equation} \label{LMBcrUNI} \ln \rho(\br_1) = \int d^3 \br_2 \,
  c^{(2)}(\br_1-\br_2) \rho(\br_2) + \text{const}.
\end{equation}
In the weak-interaction limit, Eq.~(\ref{cV}), we obtain an equation
that has an identical structure to the simplest type of molecular
field theory: $\ln \rho(\br_1) = - \beta \int d^3 \br_2 \,
v(\br_1-\br_2) \rho(\br_2) + \text{const}$.  Despite its clear
structure, the latter equation is explicitly missing three-body
interactions, which are essential for building a solid, as remarked
earlier. In addition, it is obviously useless for hard spheres, for
which $v(\br)$ is either zero or infinity. Now, the additive constant
in Eq.~(\ref{LMBcrUNI}) can be fixed by noticing that for the uniform
liquid, $\ln \rho_\sliq = \int d^3 \br_2 \, c^{(2)}(\br_1-\br_2)
\rho_\sliq + \text{const}$. Subtracting this from Eq.~(\ref{LMBcrUNI})
yields Eq.~(\ref{RY}). We remind that Eq.~(\ref{LMBcrUNI}) was derived
in the assumption of the liquid's uniformity and so it could be valid
only up to terms of order $(\drho)^2$.

Continuing this line of thought, we realise that Eq.~(\ref{RY}) can be
obtained by varying (w.r.t. to the density) the free energy from
Eq.~(\ref{F}), in which the {\em excess} part of the free energy
$F_\text{ex}$ is expanded as power series in terms of the deviation of
the local density from its value in the uniform liquid, where the
expansion is truncated at the second order. But this expansion is
exactly the expression (\ref{dFgen}) combined with the conditions
(\ref{uni}) to account for the liquid's uniformity. We spell it out
explicitly here for future use:
\begin{eqnarray} \label{FRY} F[\{\rho(\br)\}] &=& k_BT\int d^3\br \rho
  (\br) \left\{ \ln [\rho (\br ) \Lambda^3] -1\right\} \nonumber \\
  &-& \frac{k_B T}{2} \int d^3\br_1 d^3\br_2 \, \delta\rho (\br_1)
  c^{(2)}(\br_2 - \br_1; \rho_\sliq) \delta\rho (\br_2) + F_\text{ex}
  (\rho_\sliq),
\end{eqnarray}
where $F_\text{ex}(\rho_\sliq)$ is the excess free energy of the
uniform liquid at $\rho = \rho_\sliq$. We have again dropped the term
linear in $\delta\rho (\br)$, as in Eq.~(\ref{FRYq}). Functional
(\ref{FRY}) is commonly called the
Ramakrishnan-Yussouff~\cite{RamakrishnanYussouff} (RY) functional, even
though it was first written down in this simple form by Haymet and
Oxtoby~\cite{HaymetOxtoby}, see also Ref.~\cite{YangFlemingGibbs}.

Eqs.~(\ref{RY}) and (\ref{FRY}) can be used to search for crystalline
free energy minima in cases when the direct correlation function is
available, either from a calculation or experiment.  Good
approximations for $c^{(2)} (\br; \rho_\sliq)$ are available for hard
spheres at densities comparable to or below the density at the melting
point. One such approximation is that by Percus and Yevick~\cite{PY},
for which the explicit expression is available~\cite{Thiele1963,
  PhysRevLett.10.321}. Following exactly this route, Ramakrishnan and
Yussouff have recovered a periodic solution of the liquid free energy
in equilibrium with the liquid.  In this method, one expands the
candidate crystal density field in a Fourier series in terms of the
reciprocal lattice vectors $\bq$, see Eq.~(\ref{rhoq}). Note the $\bq
= 0$ component corresponds to the bulk density, which is generally
different between the liquid and solid. In practice, one assumes a
specific lattice type and spacing and uses a modest number of Fourier
components in the expansion. The expansion coefficients are then
determined by solving the (non-linear) Eq.~(\ref{RY}). One can further
compute the free energy of both the uniform liquid and the crystal as
functions of the density and thus compute the pressure $p$ and Gibbs
free energy $F+pV$ of both states.  The equality of the pressure and
chemical potential of the liquid and the crystal signals the two
phases are in equilibrium. Thus, although the procedure does not
predict the lattice type, it does yield the relative stability of
distinct lattice types, which simplifies the search for the true
equilibrium configuration.  Likewise, the procedure predicts the
densities of the phases at ambient conditions and thus enables one to
evaluate the density and entropy jumps during the transition.

On a historical note, the first successful attempt to quantify the
liquid-to-solid transition along the lines of the molecular field
theory is due to Kirkwood and Monroe~\cite{KirkwoodMonroe}, who sought
solid density profiles in the form of periodic waves. These workers
approximated the crystal field by a sum of effective pairwise
potentials and derived a self-consistent equation for the crystal
density distribution similar in structure to Eq.~(\ref{RY}). The
treatment by Kirkwood and Monroe superseded earlier treatments by
Bragg and Williams~\cite{BraggWilliams1934}, Hershfeld, Stevenson, and
Eyring~\cite{EyeringLiquid}, Born~\cite{BornHuang}, and Mott and
Gurney~\cite{MottGurneyLiquids}, to name a few. These treatments
viewed the crystal-to-liquid transition as an order-disorder
transition and were similar in spirit to treatments of the critical
point by Lennard-Jones and Devonshire~\cite{LJDevonshire1937}. Hereby
one regards liquefaction as substitutional disorder in
alloys~\cite{BraggWilliams1934}, or a proliferation of local molecular
rotations~\cite{EyeringLiquid}, dislocations~\cite{BornHuang}, or
domain walls between relatively ordered
crystallites~\cite{MottGurneyLiquids}. While achievable near a
critical point---where density fluctuations are relatively
facile--creation of such defects is too costly near the melting point.
As mentioned in the Introduction, such defect-based treatments
correspond to the {\em mechanical} melting, which takes place in the
bulk, not at the surface; they predict melting points that are too
high. We shall see in the following that the view of supercooled
liquids or glasses as a collection of such ultra-local defects is
similarly flawed.

A more modern approach to building a solid starting from the liquid
state is to note that Eq.~(\ref{Fopt}) is an optimisation problem with
respect to {\em any} parameter on which the density depends; the
optimal values of such parameters are then found by minimisation of
the functional (\ref{FRY}).  The superposition $\rho(\br, \alpha)$ of
gaussians in Eq.~(\ref{rho}) is an excellent ansatz for the crystal
density profile. Exact for a harmonic lattice, it has the flexibility
to effectively include anharmonic effects in a variational
fashion~\cite{feynman1998statistical}. Once the lattice type is
chosen, the free energy of the solid can be computed for any values of
the density and the effective spring constant $\alpha$:
\begin{equation} \label{Far} F(\alpha) \equiv F[\{ \rho(\br, \alpha)
  \} ],
\end{equation}  
where $F[\{ \rho(\br) \} ]$ is the functional from Eq.~(\ref{FRY}).
The $\alpha$-dependence of the free energy, $F(\alpha)$, at fixed
density, is of the type shown in Fig.~\ref{solid}(b). This dependence
is particular noteworthy since it allows one to directly see that the
vibrational amplitude in the crystal is determined by a competition
between two factors. On the one hand, the entropic cost of
localisation---as reflected by the ideal free energy of the crystal
$F_\text{id}$---grows approximately logarithmically with increasing
$\alpha$:
\begin{equation} \label{F_id1} F_\text{id} \approx
  k_BTN\left\{\ln\left[
      \Lambda^3\left(\frac{\alpha}{\pi}\right)^\frac{3}{2}\right]-
    \frac{5}{2}\right\}.
\end{equation}
The approximation is quantitatively adequate for not too small
$\alpha$, namely, $\alpha$ exceeding 20 or so.  On the other hand, the
increasing localisation mitigates the mutual collision rate and,
hence, the pressure that the particles exert on each other, thus
providing progressive stabilisation in a certain range of
$\alpha$'s. Formally this stabilisation comes about because the direct
correlation function is large and negative at small separations owing
to the repulsion. As the peaks in Fig.~\ref{solid}(a) and
Eq.~(\ref{rho}) become progressively narrower, at a given density, the
excess free energy decreases.  The resulting interplay of the
localisation entropy and stabilisation due to reduced collisions
results in a minimum in the $F(\alpha)$ curve at a finite value of
$\alpha$.  The equilibrium solid corresponds to the bottom of the
minimum, of course.  We note that an analogous sterically-driven
symmetry breaking takes place during nematic ordering in liquid
crystals, as elucidated by Onsager a long time
ago~\cite{OnsagerNematic}. In the liquid crystal case, it is the
rotational symmetry that gets broken when the molecules spontaneously
align, locally. (The translational symmetry remains preserved in
liquid crystals, as particles can still exchange places readily.)

When the direct correlation function is not available, we can take
advantage of the intrinsic relation between $c^{(2)} (\br)$ of a
uniform liquid and the liquid structure factor, which is
experimentally measurable. Indeed, let us rewrite $\delta
u(\br_1)/\delta u(\br_2) = \delta(\br_1 - \br_2)$
as \begin{equation} \label{inv}\int d^3 \br_3 \frac{\delta
    u(\br_1)}{\delta \rho(\br_3)} \frac{\delta \rho(\br_3)}{\delta
    u(\br_2)} = \delta(\br_1 - \br_2).
\end{equation}
Eqs.~(\ref{F}), (\ref{F_id}), (\ref{ur}), and (\ref{C1r}) yield
\begin{equation} \label{ur1} \beta u(\br) = - \ln[\rho(\br) \Lambda^3]
  + c^{(1)}(\br)
\end{equation}
and, with the help of Eq.~(\ref{c2r}),
\begin{equation} \label{urrhor} \beta \frac{\delta u(\br_1)}{\delta
    \rho(\br_2)} = - \frac{1}{\rho(\br_1)} \delta(\br_1 - \br_2) +
  c^{(2)} (\br_1, \br_2).
\end{equation}

On the other hand,
\begin{equation} \label{rhorur} \beta^{-1} \frac{\delta
    \rho(\br_1)}{\delta u(\br_2)} = \la \rho_\text{inst}(\br_1)
  \rho_\text{inst} (\br_2) \ra - \rho(\br_1) \rho(\br_2),
\end{equation}
yields the density-density correlation function, c.f. Eq.~(\ref{Chi}).
The thermally averaged quantity on the r.h.s. is closely related to
the pairwise correlation function
\begin{equation} \label{rho2}
  \rho^{(2)}(\br_1, \br_2) \equiv  \la \sum_{i \ne j} \delta(\br_1 -
  \br_i) \delta(\br_2 - \br_j) \ra
\end{equation}
that reflects the probability of finding a particle at location
$\br_2$, if there already is a particle at location $\br_1$. With the
help of Eq.~(\ref{rhoinst}), one obtains
\begin{equation} \label{rhorho} \la \rho_\text{inst}(\br_1)
  \rho_\text{inst} (\br_2) \ra = \rho^{(2)}(\br_1, \br_2) +
  \rho(\br_1) \delta(\br_1 - \br_2).
\end{equation}
We have used the equality $\la \sum_{i} \delta(\br_1 - \br_i)
\delta(\br_2 - \br_i) \ra = \rho(\br_1) \delta(\br_1 - \br_2)$, which
can be established by checking that this latter sum is non-zero only
for $\br_1 = \br_2$, while the multiplicative factor $\rho(\br_1)$ is
verified by integrating the equality in $\br_1$ or $\br_2$.  

The connection with experiment is made by realising that (the coherent
part of) the intensity of light scattered off a sample at wave-vector
$\bq$ is proportional to $\la |\sum_i e^{- i \bq \br_i}|^2 \ra$, as
light scattering is much faster than the thermal
motions~\cite{Hansen}. This quantity, up to a multiplicative factor,
is commonly known as the structure factor $S_{\bq}$:
\begin{eqnarray} \label{Sq} S_{\bq} \equiv \frac{1}{N} \la \sum_{ij}
  e^{- i \bq (\br_i - \br_j) } \ra &=& \frac{1}{N} \int d^3 \br_1 d^3
  \br_2 \, e^{- i \bq (\br_1 - \br_2) } \la \rho_\text{inst}(\br_1)
  \rho_\text{inst} (\br_2) \ra \nonumber \\ &=& 1 + \frac{1}{N} \int
  d^3 \br_1 d^3 \br_2 \, e^{- i \bq (\br_1 - \br_2) }
  \rho^{(2)}(\br_1,\br_2).
\end{eqnarray}

Combining Eqs.~(\ref{inv}), (\ref{urrhor}), (\ref{rhorur}), and
(\ref{rhorho}) yields~\cite{LebowitzPercus1963, Evans1979}:
\begin{eqnarray} \label{genOZ}
  \rho^{(2)}(\br_1,\br_2)-\rho(\br_1)\rho(\br_2)&=&
  \rho(\br_1)\rho(\br_2) c^{(2)}(\br_1,\br_2)  \\ &+&
  \rho(\br_2) \int d \br_3
  [\rho^{(2)}(\br_1,\br_3)-\rho(\br_1)\rho(\br_3)] 
  c^{(2)}(\br_3,\br_2). \nonumber
\end{eqnarray}
The quantity on the l.h.s. and inside the brackets in the integrand is
important. By Eqs.~(\ref{rhorho}) and (\ref{rhoinst}), its integral is
equal to 
\begin{equation} \label{int2rho2} \int d^3 \br_1 d^3 \br_2 \,
  [\rho^{(2)}(\br_1,\br_2)-\rho(\br_1)\rho(\br_2)] = \la N^2 \ra - \la
  N \ra^2 - \la N \ra.
\end{equation}
At the same time, the mean square deviation of the particle number
from its average value for {\em uniform} liquids is directly related
to the isothermal compressibility:
\begin{equation} \label{chiT} \chi_T \equiv -\frac{1}{V} \left(
    \frac{\prtl V}{\prtl p} \right)_T \equiv K^{-1}.
\end{equation}
via $\chi_T = \frac{\la (\delta N)^2 \ra}{N}
\frac{V}{T}$~\cite{LLstat}. The quantity $K$ is the isothermal bulk
modulus.  In addition, the l.h.s. of Eq.~(\ref{int2rho2}) can be also
readily evaluated for a harmonic solid with isotropic elasticity to
yield~\cite{stillinger:3983, RL_sigma0}: $[k_B T \bar{\rho}(K + 4
\mu/3)^{-1} - 1] \bar{\rho} V$, where $\mu$ is the shear modulus and
$\bar{\rho}$ is the density.  We thus obtain an important sum rule
that, among other things, discriminates between uniform liquids and
solids:
\begin{equation} \label{sum1} \frac{1}{\bar{\rho} V} \int d^3 \br_1
  d^3 \br_2 \, [\rho^{(2)}(\br_1,\br_2)-\rho(\br_1)\rho(\br_2)] + 1 =
  \left\{\begin{array}{ll} \frac{k_B T \bar{\rho}}{K}, &
      \text{liquid} \\ \\
      \frac{k_B T \bar{\rho}}{K + 4 \mu/3}, \hspace{3mm} &
      \text{solid}
    \end{array}
  \right.
\end{equation}
Interestingly, if we had the {\em adiabatic} values for the elastic
coefficients on the r.h.s. of Eq.~(\ref{sum1}), the r.h.s. would yield
the same expression in terms of the speed $c_l$ of longitudinal sound,
since for liquids, $K_S = \rho_m c_l^2$, while for solids, $(K_S +
4\mu_S/3) = \rho_m c_l^2$, where $\rho_m$ is the mass density. (The
adiabatic and isothermal shear moduli are strictly
equal~\cite{LLelast}, while the bulk moduli are usually close
numerically.) The distinction of the sum rule between the liquid in
solid is not widely known but is not too surprising, if one thinks of
the density fluctuations not in terms of particle number fluctuations
at constant volume, as in Eq.~(\ref{int2rho2}), but in terms of volume
fluctuations at fixed particle number. Because of a finite shear
modulus, the fluctuations of two subvolumes of a solid are coupled,
however large the subvolumes are. The only way to decouple such
fluctuations is to clamp the sides of the subvolumes. The restoring
force for deformation of an elastic body with clamped sides is greater
than that for a free solid and this force corresponds precisely to a
modulus $(K + 4\mu/3)$, see Chapter 5 of Ref.~\cite{LLelast}.

\begin{figure}[t]
  \begin{tabular*}{\figurewidth} {ll}
    \begin{minipage}{.48 \figurewidth} 
      \begin{center}
        \includegraphics[width=0.55 \figurewidth]{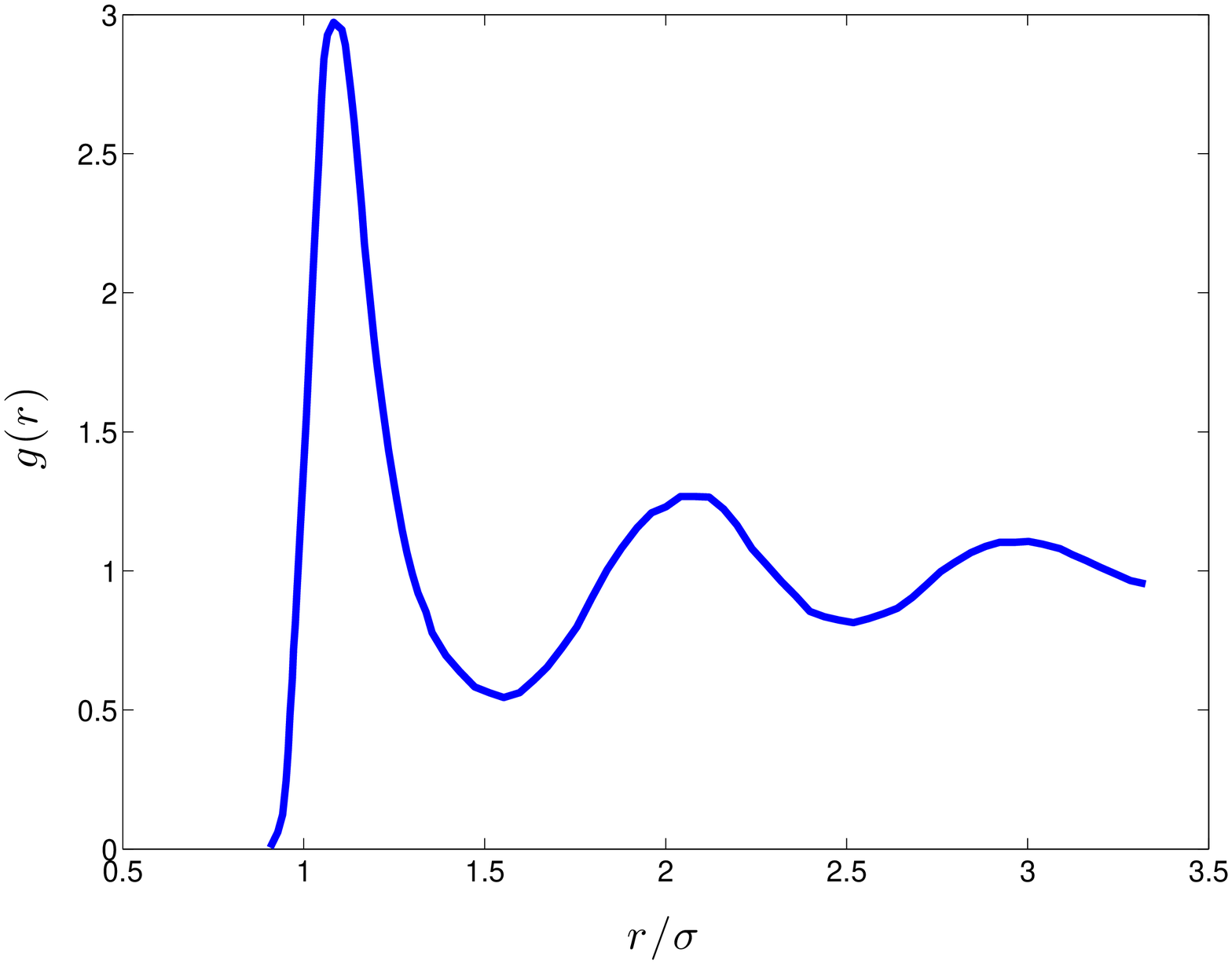} \\
        {\bf (a)}
      \end{center}
    \end{minipage}
    &
    \begin{minipage}{.48 \figurewidth} 
      \begin{center}
        \includegraphics[width=0.35 \figurewidth]{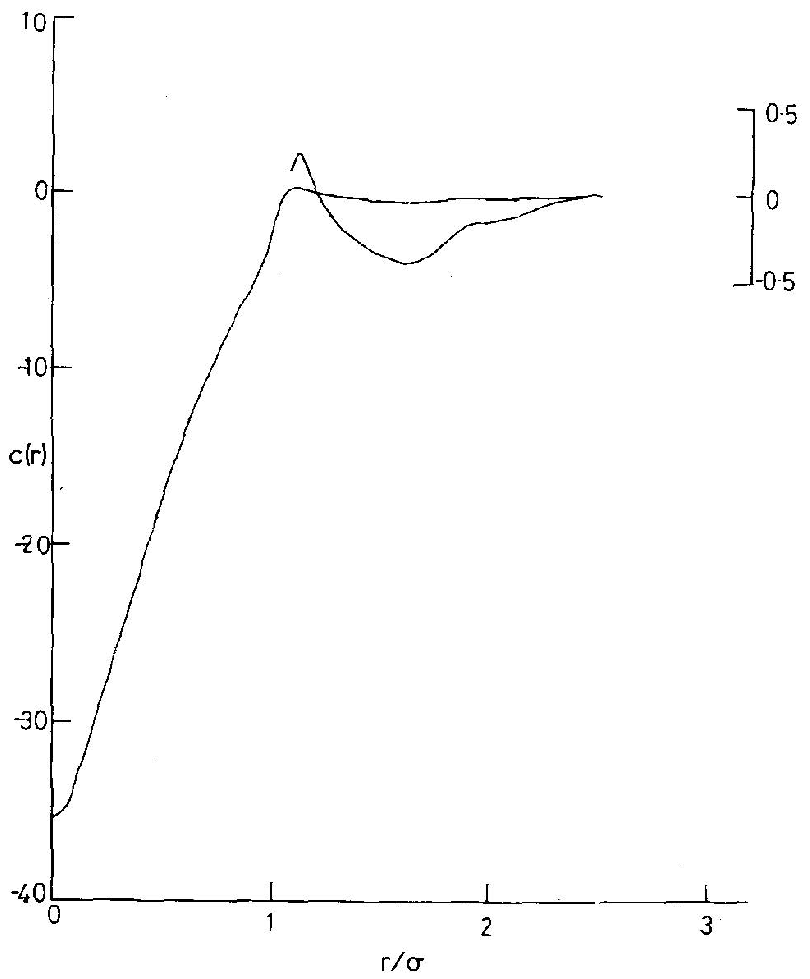}
        \\ 
        {\bf (b)}
      \end{center}
    \end{minipage}
  \end{tabular*}
  \caption{\label{gRcR} {\bf (a)} Pair-distribution function $g(r)$
    for the Lennard-Jones liquid near its triple point, data from
    Allen and Tildesley~\cite{AllenTildesleley1989}. {\bf (b)} Direct
    correlation function $c^{(2)}(r)$ for the Lennard-Jones
    liquid. Magnified version of the attractive tail at large
    separation is shown as the dashed line. From Dixon and
    Hutchinson~\cite{DixonHutchinson1977}.}
\end{figure}

For an equilibrium uniform liquid, both the density-density and the
direct correlation functions are translationally invariant and
isotropic: $\rho^{(2)}(\br_1,\br_2) = \rho^{(2)}(|\br_1- \br_2|)$,
$c^{(2)}(\br_1,\br_2) = c^{(2)}(|\br_1- \br_2|)$. After introducing
the standard, dimensionless pair-correlation function $g(r)$:
\begin{equation} \label{g(r)} g(r) \equiv \frac{1}{\rho_\sliq^2}
  \rho^{(2)}(r),
\end{equation}
Eq.~(\ref{genOZ}) reduces to the familiar Ornstein-Zernike
equation~\cite{OrnsteinZernike, Evans1979, McQuarrie}:
\begin{equation} \label{OZ} g(r) - 1 = c^{(2)}(r) + \rho_\sliq \int
  d^3 \br' [g(r') -1] c^{(2)}(|\br - \br'|).
\end{equation}
Upon defining a new function
\begin{equation} \label{hr} h(r) \equiv g(r) -1,
\end{equation}
Eq.~(\ref{OZ}) looks particularly simple in the Fourier space:
\begin{equation} \label{OZq} \tilde{h}_q = \tilde{c}^{(2)}_q +
  \rho_\sliq \tilde{h}_q \, \tilde{c}^{(2)}_q,
\end{equation}
while the structure factor becomes (for a uniform liquid!):
\begin{equation} \label{SqUni} S_q = 1 + \rho_\sliq \, \tilde{g}_q.
\end{equation}
Eqs.~(\ref{hr})-(\ref{SqUni}) form the sought connection between the
experimentally determined structure factor and the direct correlation
function. The functions $g(r)$ and $c^{(2)}(r)$ are exemplified in
Fig.~\ref{gRcR}. The pair-correlation $g(r)$ looks intuitive: It
reflects the steric repulsion between the particles at short
distances, while its oscillating nature at larger separations accounts
for the short-range order in the liquid. The interpretation of the
direct correlation function is more complicated. The (very weak)
``attractive'' tail at large distances is consistent with
Eq.~(\ref{cV}). What is the meaning of the much larger, and negative,
portion at small $r$? By Eqs.~(\ref{sum1}), (\ref{g(r)}), and
(\ref{OZ}), we obtain:
\begin{equation} \label{sum2} - \rho_\sliq \int d^3 \br c^{(2)}(r) =
  \frac{K}{k_B T \rho_\sliq} - 1.
\end{equation}
Note also that, by Eqs.~(\ref{sum1}) and (\ref{hr}-\ref{SqUni}),
\begin{equation} \label{Sq0} \lim_{q \rightarrow 0} S_q = k_B T
  \rho_\sliq \chi_T ,
\end{equation}
where $\chi_T = K^{-1}$ is the usual compressibility,
Eq.~(\ref{chiT}). This equation helps one mitigate the (usually
significant) uncertainty in the measured structure factor at small
wavevectors. Note $q$ should remain strictly positive in the limit
above, which is usually the case in experiment anyway because the
$q=0$ component includes the incident light.

Given the shape of the direct correlation function in Fig.~\ref{gRcR},
the aforementioned rule of thumb $K \simeq (10^1 - 10^2) k_B T
\rho_\sliq$ along with Eq.~(\ref{sum2}) explains the large and
negative value of the direct correlation function at the
origin. Furthermore this notion implies that upon compression, the
direct correlation function becomes increasingly negative at the
origin since both the pressure and its derivative $\prtl p/\prtl V$
will increase in magnitude with density.

Like the Landau expansion that we used to demonstrate the
discontinuous nature of the liquid-to-crystal transition, the
approximation in Eq.~(\ref{RY}) amounts to analytically continuing the
free energy from the liquid state to the solid state. Although
Eq.~(\ref{RY}) is derived by a systematic expansion---which can be, in
principle carried out to higher orders---one is left wondering whether
the direct correlation function at the density $\rho_\sliq$ can
adequately describe correlations in the solid. Despite the somewhat
decreased entropy, the crystal becomes more stable than the liquid
above certain liquid densities because the particles in the crystal do
not collide with each other as much, which is another way to say that
crystallisation of hard spheres is sterically-driven. At a first
glance, this decrease in the collision rate may sound
counter-intuitive given that the close-packed crystal is, of course,
denser than the liquid.  On the other hand, the coordination in the
crystal exceeds that in the liquid, suggesting the particles do not
have to spend as much time near each other.  It turns out that
appropriate quantitative tools to address this, somewhat subtle aspect
of crystallisation are {\em also} supplied by the classical density
functional theory.

The excess free energy of a hard sphere liquid has the form $k_B T
f_\sliq(\rho_\sliq)$ because there is no finite energy scale in the
problem other than the temperature. As mentioned, good approximations
for this free energy, such as Percus-Yevick, are available. Because
the particles in the crystal occupy disparate, well defined cells in
space, an individual particle is subject to collisions with many fewer
molecules than in the uniform liquid. We thus anticipate that
correlation functions in the crystal, for short particle-particle
separations, should be similar to those in a {\em uniform} liquid at
an effective density which is {\em lower} that the actual density of
the liquid in equilibrium with the solid. Specifically, in the
modified-weighted density approximation (MWDA) due to Denton and
Ashcroft~\cite{MWDA}:
\begin{equation} \label{F_HS} F_\text{ex}\left[\rho\right]= k_BT N
  f_\sliq \left(\hat{\rho}\right),
\end{equation}
where the effective---or ``weighted''---density $\hat{\rho}$ is
defined by the equation:
\begin{equation} \label{rhow}
  \hat{\rho}\left[\rho\left(\br\right)\right] \equiv N^{-1}\int\int
  \tilde{w}\left(\br'-\br;\hat{\rho}\right)\rho\left(\br\right)
  \rho\left(\br'\right)d^3\br d^3\br'.
\end{equation} 
This quantity scales linearly with the actual, non-uniform density but
with a weight. The weight function $\tilde{w}$ is chosen so that in
the uniform limit, the exact density is recovered:
\begin{equation} \label{wint} \int
  \tilde{w}(\br'-\br;\hat{\rho}) d^3\br' =1
\end{equation}
and, likewise, so that the direct correlation function is reproduced:
\begin{equation} \label{c} c^{(2)} \left(\br-\br';\rho\right)=
  -\beta\lim_{\rho(\br)\rightarrow\rho} \frac{\delta^2F_\text{ex}
    \left[\rho\left(\br\right)\right]}{\delta\rho\left(\br\right)
    \delta\rho\left(\br'\right)}
\end{equation}
Multiplying the equation above by $\rho(\br) \rho(\br')$, integrating
over $\br$ and $\br'$, and using Eqs.~(\ref{rhow})-(\ref{wint}) and
the density ansatz~(\ref{rho}) yield an expression that can be used to
determine the weighted density $\hat{\rho}$
self-consistently~\cite{Lowen}:
\begin{align} \label{rhoh_eqn} 2\hat{\rho} \left. \frac{
      f_\sliq}{\prtl \rho}\right|_{\rho=\hat{\rho}} + \rho \hat{\rho}
  \left. \frac{ \prtl^2 f_\sliq}{\prtl \rho^2}
  \right|_{\rho=\hat{\rho}} &=-\left(\frac{\alpha}{\pi}\right)^3\int
  d^3\br\int d^3\br'\int d^3\!\bR \:
  c^{(2)}_\tHS (|\br-\br'|;\hat{\rho} ) \nonumber \\
  &\times\exp\left\{-\alpha\left[\left(\br-\bR\right)^2 +
      \br'^2\right]\right\}\left[\delta\left(\bR\right)+\rho
    g\left(\bR\right)\right].
\end{align}
We remind that the function $f_\sliq (\rho)$ must be specified for
this equation to be useful. Good approximations, such as
Percus-Yevyck, are available for this function at not too high
densities.  Note that to evaluate the effective density $\hat{\rho}$,
determining an explicit form for $\tilde{w}$ is not required.  (The
weighting function can be readily computed and is related to the
direct correlation function~\cite{MWDA}.)  The quantity
$g\left(\bR\right)$ is the site-site correlation function of the
lattice: $g(\bR) \equiv (1/N) \sum_{i \neq j} \delta[\bR-(\bR_i -
\bR_j)]$. A density ansatz other than the superposition of Gaussians
from Eq.~(\ref{rho}) could be used if desired but would lead to a
different equation for $\hat{\rho}$. The above expression can be
formally applied to any lattice. If the latter is periodic, the double
integral can be conveniently recast as a Fourier sum~\cite{MWDA}.

One can now use Eqs.~(\ref{F_HS}), (\ref{rhoh_eqn}), and (\ref{F_id1})
to determine the Helmholtz free energy $F$ of the solid as a function
of the density---which is specified automatically given a
lattice---and the effective spring constant $\alpha$. This free energy
is further optimised with respect to $\alpha$ thus giving an
approximation for the free energy of the solid as a function of the
density, thus allowing one to compute the pressure.  A solid at a
density such that its pressure and chemical potential are equal to
their counterparts in the uniform liquid, is in equilibrium with the
liquid (at the same temperature of course). To illustrate this, we
show in Fig.~\ref{solid}(b) the Gibbs free energy difference between
crystal and liquid, $\Delta F(\alpha) + p \Delta V$, as a function of
the order parameter $\alpha$. The solid line corresponds to
$F(\alpha)$ at the density and pressure such that the solid and liquid
would be in equilibrium for the optimal value of $\alpha$. The dashed
line in Fig.~\ref{solid}(b) shows the {\em spinodal} of the solid with
respect to the liquid. It corresponds to the density at which a
metastable minimum in $F(\alpha)$ just begins to appear. The
respective pressure is, of course, considerably lower than the
pressure at which the two phase would coexist. Note that
Fig.~\ref{solid}(b) is a {\em prima facie} evidence of the
discontinuous nature of the liquid-to-crystal transition since the
uniform liquid is {\em also} described by the density ansatz
(\ref{rho}) if one sets $\alpha = 0$. The transition thus corresponds
to a discontinuous jump from $\alpha = 0$ to $\alpha \simeq 10^2/a^2$.
Note the latter value is in agreement with the phenomenological
criterion of melting due to Lindemann~\cite{Lindemann, Bilgram,
  GrimvallSjodin}, which has been rationalised relatively recently
based on the surface character of crystal melting~\cite{L_Lindemann}.

It turns out that whenever there exists a crystal solution to the free
energy functional, the weighted density $\hat{\rho}$ is always lower
than the actual density. This is consistent with our earlier
expectation that the effective local density, which determines the
collision rate, is lowered in the solid compared with the uniform
liquid, even though the actual density has increased.  Because of the
relative smallness of the effective density, it is adequate to use the
Percus-Yevick (PY) expressions~\cite{Hansen, McQuarrie} for the free
energy $f_\sliq$ and the translationally-invariant portion of the
direct correlation function $c^{(2)}$ in Eq.~(\ref{rhoh_eqn}) in the
solid.  Importantly, the weighted-density approximation alleviates our
earlier concerns about using the direct correlation function at the
liquid density $\rho_\sliq$ in a (solid) phase separated from the
uniform liquid by a free energy barrier. Here we explicitly obtain
that the correlations do experience a discrete jump during the
transition.

\begin{table}[t]
\centering
\includegraphics[width=  .7 \figurewidth]{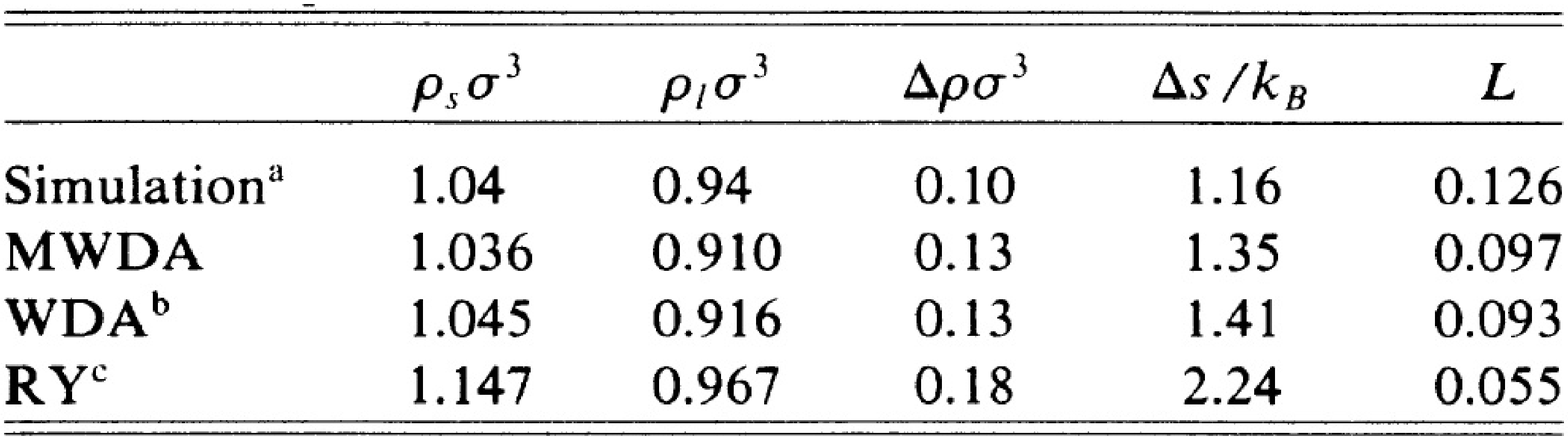}
\caption{\label{DentonAshcroftTable} Densities of hard sphere solid 
  ($\rho_s$) and liquid ($\rho_l$) at liquid-crystal equilibrium.
  $\Delta \rho \equiv \rho_s - \rho_l$. The quantity $\Delta s$ 
  is the fusion entropy per particle and $L$ is the Lindemann 
  ratio from Eq.~(\ref{Lratio}). Letter references: 
  (a): Ref.~\cite{HooverRee1968}; (b): see Ref.~\cite{MWDA}; 
  (c): Ref.~\cite{BarratHansenPastoreWaisman1987} and also 
  see Ref.~\cite{MWDA}.}
\end{table}

\begin{figure}[t]
  \centering
  \includegraphics[width= .5 \figurewidth]{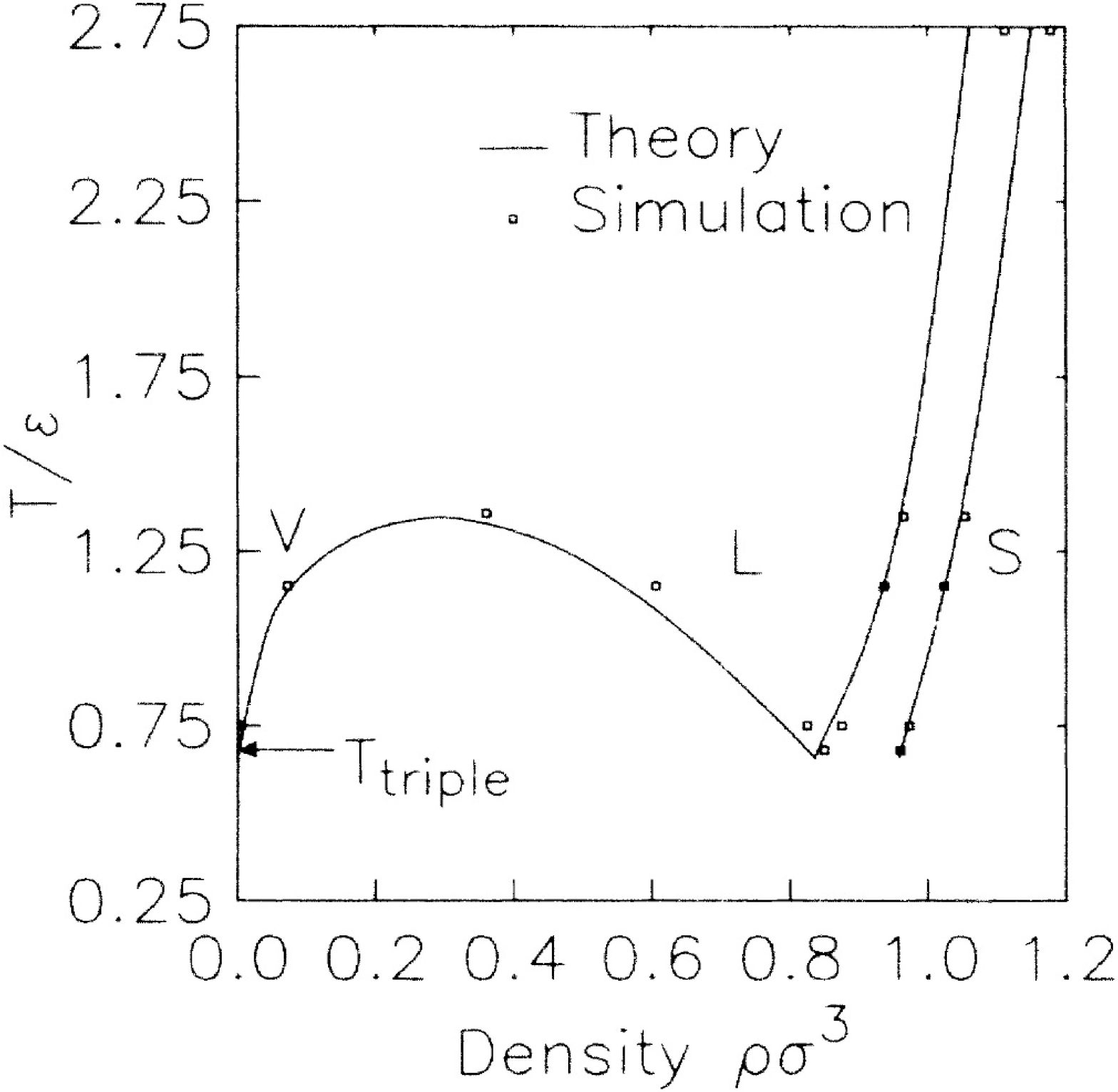}
  \caption{\label{LJphaseDiagram} The phase diagram of the
    Lennard-Jones system in the density-temperature plane. The
    continuous lines are obtained using the classical DFT in the
    modified weighted density
    approximation~\cite{PhysRevLett.56.2775}. The symbols correspond
    to simulation studies~\cite{PhysRev.184.151}.}
\end{figure}

Because of the self-consistency requirement on the weighted density,
the MWDA is a non-perturbative approximation; the corresponding free
energy contains an infinite subset of exact high order terms in the
corresponding perturbative expansion~\cite{MWDA}. This is in addition
to the PY approximation itself being non-perturbative in the first
place. The PY approximation is of excellent quality at liquid
densities in question, judging by comparison with
simulations~\cite{McQuarrie}. Perhaps for these reasons, the MWDA
yields predictions for the transition entropy and densities during the
transition, and the vibrational particle displacement in the crystal
that are in good agreement with simulations, see
Table~\ref{DentonAshcroftTable}. Furthermore, the MWDA can be extended
to non-rigid and weakly interacting systems, such as Lennard-Jones
particles~\cite{PhysRevLett.56.2775}. Here the long range interaction
can be included as a perturbation, while the now soft repulsion at
small separations can be handled by using an effective hard-core
repulsion. Hereby the effective hard-sphere diameter can be determined
using, for instance, the Barker-Henderson prescription~\cite{Hansen}:
\begin{equation} \label{d} d\left(T\right)=
  \int^{\sigma}_{0}dr [1-e^{- V_\tLJ(r)/k_BT}],
\end{equation}
and
\begin{equation} V_\tLJ(r) \equiv 4\epsilon [ (\sigma/r )^{12} -
  (\sigma/r )^{6} ]
\end{equation}
is the Lennard-Jones interaction potential. The resulting prediction
for the phase diagram is in remarkable agreement with simulations,
with respect to the equilibrium between all three phases of the
system~\cite{PhysRevLett.56.2775}, see Fig.~\ref{LJphaseDiagram}.
This approximation yields the following estimates for several key
characteristic of the transition close to the triple point (result of
simulation~\cite{Hansen} in brackets): $\rho_\sliq d^3 = 0.855
(0.875)$, $\rho_\sXtal d^3 = 0.970 (0.973)$, $p d^3/\epsilon = 0.970
(0.973)$, $T \Delta S/\epsilon = 1.1 (1.31)$, and for the Lindemann
displacement~\cite{PhysRevLett.56.2775}:
\begin{equation} \label{Lratio} L \equiv \la (\delta
  r)^2\ra^{1/2}/r_\snn ,
\end{equation}
$L = 0.127 (0.145)$. Here, $r_\snn$ is the nearest-neighbour spacing.

A rather different---and early---line of attack on the problem of
crystallisation of hard spheres, which will be also relevant later, is
due to Fixman~\cite{Fixman1969}. He has expanded the singular
interaction potential between the rigid particles in terms of Hermite
polynomials. In the lowest order, his method amounts to variationally
finding the best value of the effective spring constant $\alpha$ from
Eq.~(\ref{rho}) that approximates the effect of the hard repulsion
between particles when combined with their thermal motion. This
method, which can be systematically improved, produced excellent
estimates for the pressure and free energy. It is often referred to as
the ``self-consistent phonon theory.''

\subsection{Transferability of DFT results from model liquids to
  actual compounds}
\label{bead}

We finish this Section by remarking on whether the results for hard
spheres and Lennard-Jones systems are transferable to actual,
non-model liquids.  Except for argon, which is a Lennard-Jones liquid
to a good accuracy, most substances exhibit much more complicated
interactions. Already in molecular liquids, such as organic compounds,
molecules have complicated shapes.  Covalently bonded substances are
exemplified by various chalcogenides, such As$_2$Se$_3$ and elemental
compounds such Ge and Si, while ZnCl$_2$ is a good example of an ionic
compound. Hydrogen bonding is also very common, as in water and
various alcohols. In contrast, the model systems we have considered so
far exhibit only isotropic steric repulsion and a weak (isotropic)
attraction, which can be treated perturbatively. Clearly, the cohesive
forces in germanium cannot be treated as a perturbation to the steric
repulsion since germanium {\em expands} upon freezing. The
significance of bonding during the latter process is also witnessed by
germanium's fusion entropy, which is $\sim 3.5 k_B$ per atom and thus
significantly exceeds the fusion entropy of Ar, which is $\sim 1.68
k_B$. Generally, the work contribution $p\Delta V$ in the fusion
enthalpy (\ref{DH}) is very small---$10^{-4}$ or so---compared to the
energy contribution at ordinary pressures.

One reason it is difficult to apply the DFT to actual substances is
our lack of knowledge of the functional form of the interactions.
However, even if available, such functional forms would probably be
too complicated to allow for a tractable analytical
approach. Incidentally, while Quantum Chemistry, combined with various
evolutionary algorithms, has made impressive progress in predicting
structures of solids and assessing relative stability of distinct
polymorphs, adequate sampling of the {\em liquid} state for actual
substances near melting temperature and below is still out of reach
for direct molecular modelling. In developing a
computationally-tractable description of the glass transition, our
best bet may be to find a minimal simplified model for the actual
interactions that can reproduce only select, but key features of their
crystallisation or glass transition. An example of such a simplified
interaction is the pairwise potential for SiO$_2$ due to Beest,
Kramer, and Santen~\cite{PhysRevLett.64.1955}.

\begin{table} 
  \centering
{\small
\begin{tabular}{|c|c|c|}
  \hline
  substance & $s_m/k_B$, per atom & $s_m/k_B$, per group \\
  \hline \hline \hline

  HS & 1.19 & \\
  \hline
  LJ & 1.75 &  \\
  \hline \hline \hline

  Ar & 1.68 & \\
  \hline
  Pb & 0.96 & \\
  \hline  \hline \hline

  SiO$_2$ &  & 0.58 (SiO$_{4/2}$) \\
  \hline
  ZnCl$_2$ & 0.69  & 2.07 (ZnCl$_{4/2}$) \\
  \hline   \hline \hline

  NaCl & 1.58 &  \\
  \hline
  CsCl & 1.34 &  \\
  \hline  \hline \hline

  As & 2.70 &  \\
  \hline
  Se & 1.62 &  \\
  \hline
  As$_2$Se$_3$ & 1.51 & 3.78 (AsSe$_{3/2}$) \\
  \hline \hline \hline

  TNB &  & 1.54 (ring) \\
  \hline
  OTP &  & 2.1 (ring) \\
  \hline  \hline  \hline

  C & 2.97 & \\
  \hline  
  Ge & 3.67 & \\
  \hline
  H$_2$O &   & 2.65 (OH$_2$) \\  
  \hline

\end{tabular}
}

\caption{\label{FentropyTable} Fusion entropy, $s_m = h_m/T_m$, 
  for a variety   of substances,  elemental, compound, and model 
  such as the monodisperse  hard sphere (HS) and Lennard-Jones
  (LJ) system, the latter near the triple   point~\cite{Hansen}. 
  Enthalpy of fusion $h_m$ and melting   temperature $T_m$ for
  actual substances are from CRC Tables~\cite{CRC}, 
  except  TNB~\cite{MagillTNB} and Ar~\cite{BRR}.  
  The  polymorphs are as follows: SiO$_2$ (crystobalite), Se (grey), 
  rhombohedral As (grey). 
  OTP (ortho-terphenyl) consists of  three aromatic rings, 
  TNB (tris-naphthyl benzene) of seven rings.} 
\end{table}

A distinct---and equally phenomenological---approach is to ask whether
we can make a correspondence between the actual substances and {\em
  isotropically} interacting particles that exhibit repulsion at short
distances and modest attraction at long distances, such as argon. Such
a description is motivated by our good knowledge of the thermodynamics
and packing properties of isotropically interacting objects, and their
being amenable to semi-analytical treatments such as the classical
density functional theory. The key dimensionless number characterising
crystallisation of such particles is the fusion entropy per
particle. 

In assessing the possibility of such an effective description, we list
the fusion entropies for a small set of model liquids and actual
substances that cover a broad range of bonding patters in
Table~\ref{FentropyTable}. We observe that the fusion entropy per
particle or rigid molecular unit is quite consistent between these
distinct systems; it is numerically close to its value in model
systems, represented here by hard spheres and Lennard-Jones particles,
and some actual systems, such as argon and lead, both of which
crystallise into close-packed structures.

Among the materials listed in Table~\ref{FentropyTable}, stand out the
very open structures represented by the three substances at the bottom
and arsenic. All four lose significantly more translational entropy
per particle, upon freezing, than the rest of the substances. Of these
four, solid carbon (graphite) consists of weakly bonded, flat graphene
sheets; germanium crystallises into the diamond lattice, in which each
atom is bonded equally strongly to four atoms located at the corners
of a regular tetrahedron; the normal, hexagonal ice is, in a sense, an
intermediate case between the graphite and diamond lattice: First of
all, we need to focus on the oxygens since the protons are mobile and
probably contribute comparably to the entropies of the liquid and
solid state. Each oxygen atom is bonded to three oxygens (within a
sheet) while being bonded to one atom from the adjacent sheet, which
is shifted accordingly sideways. Each sheet is in reality a puckered,
double layer consisting of hexamers in the chair conformation; if
flattened out, each double layer would be just like a graphene
sheet~\cite{ABW}. The structure of rhombohedral (grey) arsenic is also
a stack of sheets.  Each arsenic atom is strongly bonded to three
atoms within its sheet and only weakly bonded to three atoms across
the inter-sheet gap.  Note all four substances are poor glassformers
and all, except for As, {\em expand} upon freezing. (Note that
antimony and bismuth, which are in the same group as arsenic, {\em do}
expand upon freezing. These two elements are even worse glassformers
and better electric conductors than As.) To avoid confusion we note
that having relatively little entropy per atom does not guarantee
being a good glass-former, as can be seen by comparing NaCl and CsCl
(poor glass-formers forming a simple-cubic-like and BCC-like
structures respectively) and ZnCl$_2$ and As$_2$Se$_3$, the latter two
being good glass-formers. Zinc chloride consists of relatively rigid,
corner-sharing ZnCl$_{4/2}$ tetrahedra (but apparently not as rigid as
in SiO$_{4/2}$) while As$_2$Se$_3$ consists of relatively rigid
AsSe$_{3/2}$ pyramids co-joined through the Se corners and forming
puckered sheets like those of black phosphorus but with
vacancies~\cite{ZLMicro1}.

To appreciate just how open the diamond structure is, and to
rationalise its high fusion entropy, we first consider the FCC
lattice, which is a close-packed structure. The FCC structure has two
types of cavities: two octahedral (cornered by the six face-centred
vertices) per three tetrahedral. The diamond lattice can be produced
from the FCC lattice by placing a particle in every other tetrahedral
cavity in a (3D) checker pattern. Alternatively, to generate the
diamond lattice, one can superimpose two identical FCC lattices, with
lattice constant 1, which are shifted by $\sqrt{3}/4$ along the main
diagonal. On the other hand, the simple-cubic lattice can be obtained
by superimposing two identical FCC lattices, with lattice constant 1,
but which are shifted by $\sqrt{3}/2$ along the main
diagonal. Consequently, two diamond lattices complement each other to
form a BCC lattice.  Thus roughly, the entropy of freezing of the
diamond lattice is the total entropy due to the ordering of an equal
measure of particles and vacancies into a BCC lattice. Per {\em
  particle}, we get about twice as much fusion entropy as for a BCC
lattice made of two distinct particles. (We should be mindful of the
mixing entropy, too.) Indeed, the fusion entropy of CsCl per atom is
at least twice less than that of germanium or water.  The high fusion
entropy of open structures is in full harmony with our earlier
analysis of the discontinuous nature of the liquid-to-crystal
transition. Per that discussion, materials with open structures freeze
well before the steric effects---which are signalled by relatively low
values of the fusion entropy---become important. It is instructive to
note that the filling fraction in the diamond lattice made of
touching, identical spheres is only $\pi \sqrt{3}/16 \approx 0.34$, to
be compared with the filling fraction of the FCC lattice, $\pi/3
\sqrt{2} \approx 0.74$, or the random close-packed structure, viz.,
$\approx 0.64$~\cite{RevModPhys.82.789}. Conversely, many example of
pressure-induced amorphisation of relatively open structures are
known~\cite{Sharma1996, Brazhkin2003, Ponyatovsky1992, Hemley1988}.

We thus tentatively conclude that with the exception of these very
open structures, steric interactions contribute significantly to the
thermodynamics of freezing. The degree of ``openness'' is positively
correlated---but not without exception---with the sign of the volume
change during freezing and the value of the fusion entropy; it is {\em
  negatively} correlated with the glass-forming ability. We will be
able to rationalise this negative correlation in the following
Section, where we show that {\em equilibrium} aperiodic structures are
stabilised by steric effects.  Now, the glass forming ability is also
decreased when a relatively well-packed structure is agreeable with
the stoichiometry, even if the fusion entropy is not too high. (The
rock salt structure is, in fact, rather well-packed for NaCl, because
of the disparity in the ion sizes and despite the relatively low
coordination number of six.) Thus it appears that at least for
structures that are not too open, an effective description in terms
isotropically interacting particles is possible.  Certain layered
compounds, such as As$_2$Se$_3$, are good glass-formers despite having
a relatively large entropy of fusion.  We shall see that determining
the effective particle size for such compounds is difficult, in
contrast with compounds that exhibit less propensity for local
ordering.

To avoid confusion, we note that substances with very {\em isotropic}
bonding, such as monodisperse hard-spheres or Lennard-Jones particles,
are {\em also} poor glassformers. These fail to vitrify readily
despite the prominence of steric effects because monodisperse spheres
easily find close-packed arrangements. We will return to this topic at
the end of the article.

\section{Emergence of Aperiodic Crystal and Activated Transport, as a
  Breaking of Translational Symmetry}
\label{aper}

The density functional theory (DFT) provides us with reliable tools
to determine the free energy of the uniform liquid and, at the same
time, the free energy of specific crystalline arrangements.  As such,
the DFT enables us to assess the stability of distinct crystalline
polymorphs relative to each other and to the uniform liquid state in
the first place. Likewise, it will enable us to assess the stability
of {\em aperiodic} structures. Until further notice, we will focus on
rigid systems exemplified by hard spheres and also with added weak
attraction as in Lennard-Jones liquids. These results will be
generalised for actual substances in due time.

\subsection{The Random First Order Transition (RFOT)}
\label{RFOTDFT}

Let us now consider a supercooled liquid just above its glass
transition, so that its structure relaxes on a timescale comparable to
one hour. The latter time scale is 16 orders of magnitude longer than
the vibrational relaxation time, which is numerically a picosecond or
so. In other words, the supercooled liquid, despite being able to flow
on very long times, is a solid on mesoscopic length scales and below.
Frozen glasses, if quenched considerably below the glass transition,
not only fail to flow but are often even more rigid than the
corresponding crystal. Given their remarkable mechanical stability, it
is natural to enquire whether aperiodic structures such as those
pertaining to supercooled liquids or glasses are free energy minima.

This question was answered affirmatively by Stoessel and Wolynes in
1984~\cite{dens_F2}, who used the self-consistent phonon theory and
assumed an aperiodic structure characterised by the pair-correlation
function $g(r)$ of a {\em uniform} liquid. These authors have
determined self-consistently the force constant $\alpha$ from the
Gaussian density profile (\ref{rho}), where the lattice is now {\em
  aperiodic}. The self-consistent phonon theory also predicts the
liquid density above which the aperiodic free energy minima begin to
exist.

Soon afterwards, Singh, Stoessel, Wolynes~\cite{dens_F1} (SSW)
reported their stability analysis of aperiodic structures using the
Ramakrishnan-Yussouff density functional (\ref{FRY}), in which the free
energy was determined explicitly as a function of $\alpha$,
analogously to the periodic-crystal calculation in
Fig.~\ref{solid}(b). There are notable differences between how the
calculations are set up in the periodic and aperiodic case, in
addition to the use of an aperiodic lattice. (SSW employed a lattice
generated using the Bennett algorithm~\cite{Bennett}.) In the regular
liquid-to-crystal transition, the two phases occupy distinct parts of
the space. The two phases can coexist for an indefinite time, if the
temperature, pressure, and chemical potential are uniform throughout
the whole system. The appropriate free energy for analysing the
relative stability of the two phases is thus the Gibbs free energy.
One may still consider a transient coexistence between the two phases
when one of them is {\em metastable}, and so only temperatures and
pressures are equal between the two phases, while the metastable free
energy is now in excess of the Gibbs free energy. (The metastable
phase will eventually convert into the stable phase, subject to
pertinent kinetics.) In contrast, there is no phase separation in the
aperiodic case. The aperiodic minimum at finite $\alpha$ is built with
the very same particles comprising the liquid and thus {\em replaces}
the uniform liquid, it does not coexist with the uniform liquid; there
is no spatial interface involved. Nor is there volume change and so
the appropriate free energy is the Helmholtz free energy from
Eq.~(\ref{F}).

\begin{figure}[t]
  \centering
  \includegraphics[width=  \figurewidth]{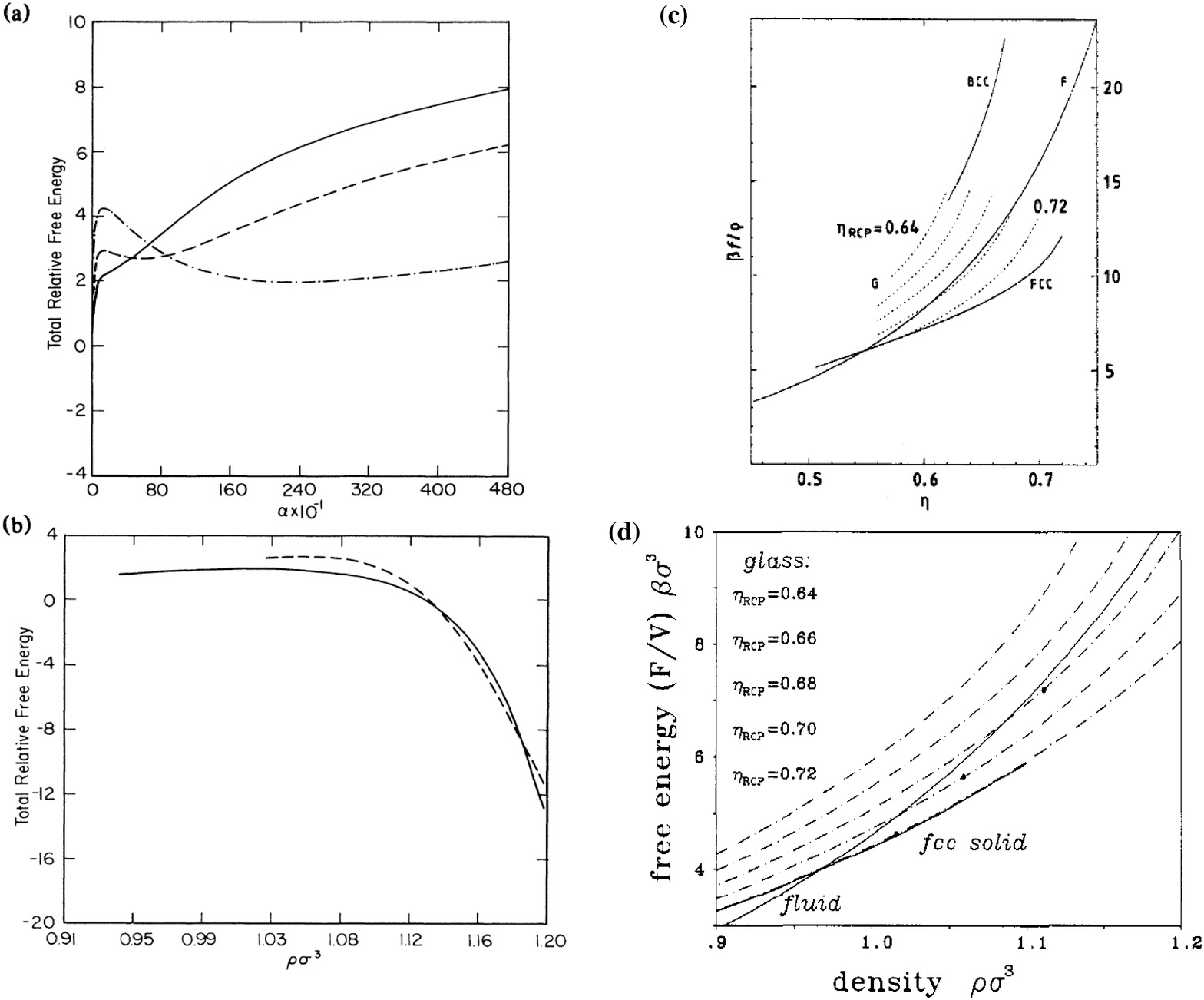}
  \caption{\label{SSW} {\bf (a)} The $\alpha$ dependence of the
    relative free energy $F(\alpha) - F_\suni$ from Eq.~(\ref{FRY}) as
    computed by Singh, Stoessel, and Wolynes~\cite{dens_F1} for an
    {\em aperiodic}, Bennett~\cite{Bennett} lattice, for three values
    of density: $\rho \sigma =1$ (solid line), $\rho \sigma =1.05$
    (dashed line), $\rho \sigma =1.1$ (dash-dotted line).  {\bf (b)}
    Same relative free energy as in panel (a), but as a function of
    density at constant $\alpha$. The solid and dashed curves
    correspond to two distinct lattices, Bennet and
    hypothetical-icosahedral, respectively. {\bf (c)} The free energy
    of an aperiodic (Bennett) structure as a function of filling
    fraction $\eta$, at several values of local coordination, in
    comparison with the free energy of the FCC and BCC structures, and
    the uniform liquid. From Ref.~\cite{BausColot}. {\bf (d)} The free
    energy of an aperiodic (Bennett) structure as a function of
    filling fraction $\eta$, at several values of local coordination,
    in comparison with the free energy of the FCC structure and
    uniform liquid. From Ref.~\cite{Lowen}.}
\end{figure}

The result of the SSW calculation of the free energy $F(\alpha)$ for
an aperiodic lattice, relative to the uniform liquid is shown in
Fig.~\ref{SSW}(a).  Here we observe that in complete analogy with the
regular liquid-to-crystal transition, a metastable minimum develops at
a {\em finite} value of $\alpha \simeq 10^1-10^2$. This metastable
minimum thus corresponds to an assembly of particles localised to
certain locations in space and vibrating about those locations.  The
locations themselves may still move about---the present article is
largely about those movements!---but the movements and much slower
than the vibrational oscillations. The metastable minimum in
Fig.~\ref{SSW} thus corresponds to a solid in the sense that the
vibrationally averaged coordinates of the particles move on
significantly longer timescales than the vibrational relaxation time.

The emergence of a minimum at a finite $\alpha \gg 1/a^2$ implies a
discontinuous transition accompanied by a breaking of the
translational symmetry, in complete analogy with the emergence of
periodic solutions of the free energy considered in the preceding
Section. Before we discuss the thermodynamic significance of the
metastable minimum, let us review its properties.  As in the periodic
case, the metastable minimum appears at a certain threshold
density. The thus emerged aperiodic-crystal phase is further
stabilised with increasing density, see Fig.~\ref{SSW}(b). However the
latter phase reaches the uniform liquid in stability only if the
attractive tail is included in the direct correlation function
discussed earlier; the specific form tail used by SSW is due to
Henderson and Grundke~\cite{HendersonGrundke75}, see
Fig.~\ref{cRHendersonGrundke}. Baus and Colot~\cite{BausColot}
subsequently showed that the stability of the periodic phase is quite
robust. These authors have generalised the SSW treatment to account
for the circumstance that the lattice not only shrinks uniformly with
pressure/density but also that the coordination must {\em increase}
alongside. The specific device they employed to show this is the
following ansatz for the site-site correlation function (which
corresponds to the pairwise correlation function $g(r)$ for an
equilibrium structure):
\begin{equation} \label{gRBC} g\left(\bR\right)= g_B
  [(\eta/\eta_\tRCP)^{1/3}R ],
\end{equation} 
where $g_B(R)$ is the site-site correlation function for Bennett's
lattice of hard spheres with diameter $d$~\cite{Bennett}, and $\eta$
is the packing fraction. The quantity $\eta_\tRCP$ is a parameter that
enables one to emulate, to a degree, changes in local
coordination. For instance, the Bennett function $g_B(x)$ has a sharp
peak at $x=d$ corresponding to particles in immediate contact, if
$\eta = \eta_\tRCP$. For such immediate neighbours at distance $R$,
$(\eta/\eta_\tRCP)^{1/3} R = d$, or $\eta = \eta_\tRCP (d/R)^3$. Thus
raising $\eta_\tRCP$ at fixed distances between particles and their
sizes emulates increasing the volume fraction $\eta$---and hence
coordination---and vice versa for smaller $\eta_\tRCP$. On the other
hand, it is straightforward to see that this effective change in
volume fraction is local, since modifying $\eta_\tRCP$ actually does
not change the average density, by virtue of the relation $\rho \int
d^3 \br g(r) = (N-1)$. (To avoid confusion, we note the latter formula
is valid in the canonical ensemble, in contrast with Eq.~(\ref{sum1}),
which applies in the grand-canonical ensemble.)

\begin{figure}[t]
\begin{tabular*}{\figurewidth} {ll}
\begin{minipage}{.48 \figurewidth} 
  \begin{center}
    \includegraphics[width=0.45 \figurewidth]{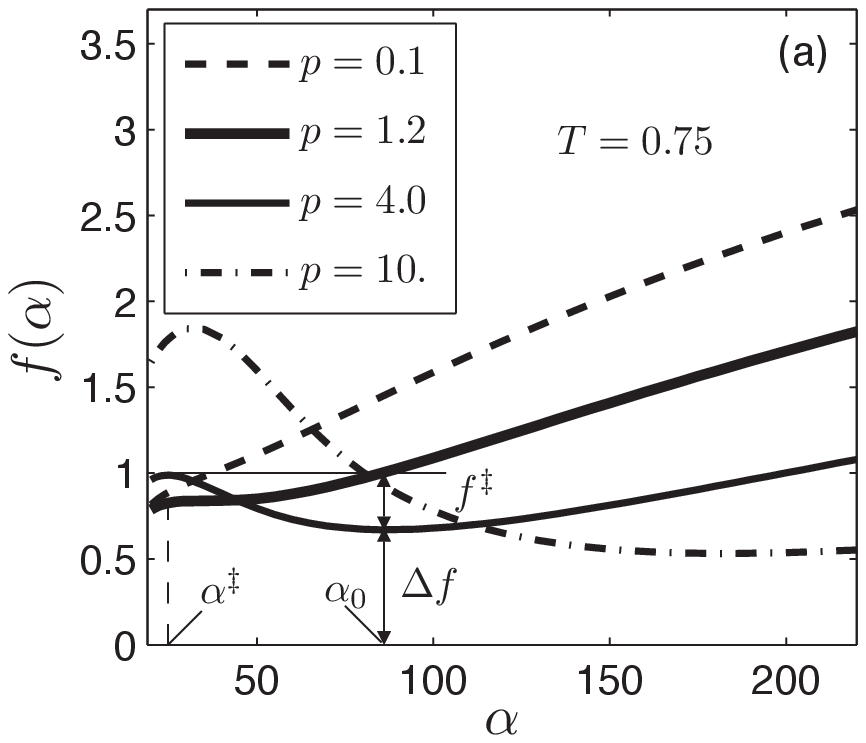} \\
    {\bf (a)}
   \end{center}
  \end{minipage}
&
\begin{minipage}{.48 \figurewidth} 
  \begin{center}
    \includegraphics[width=0.45 \figurewidth]{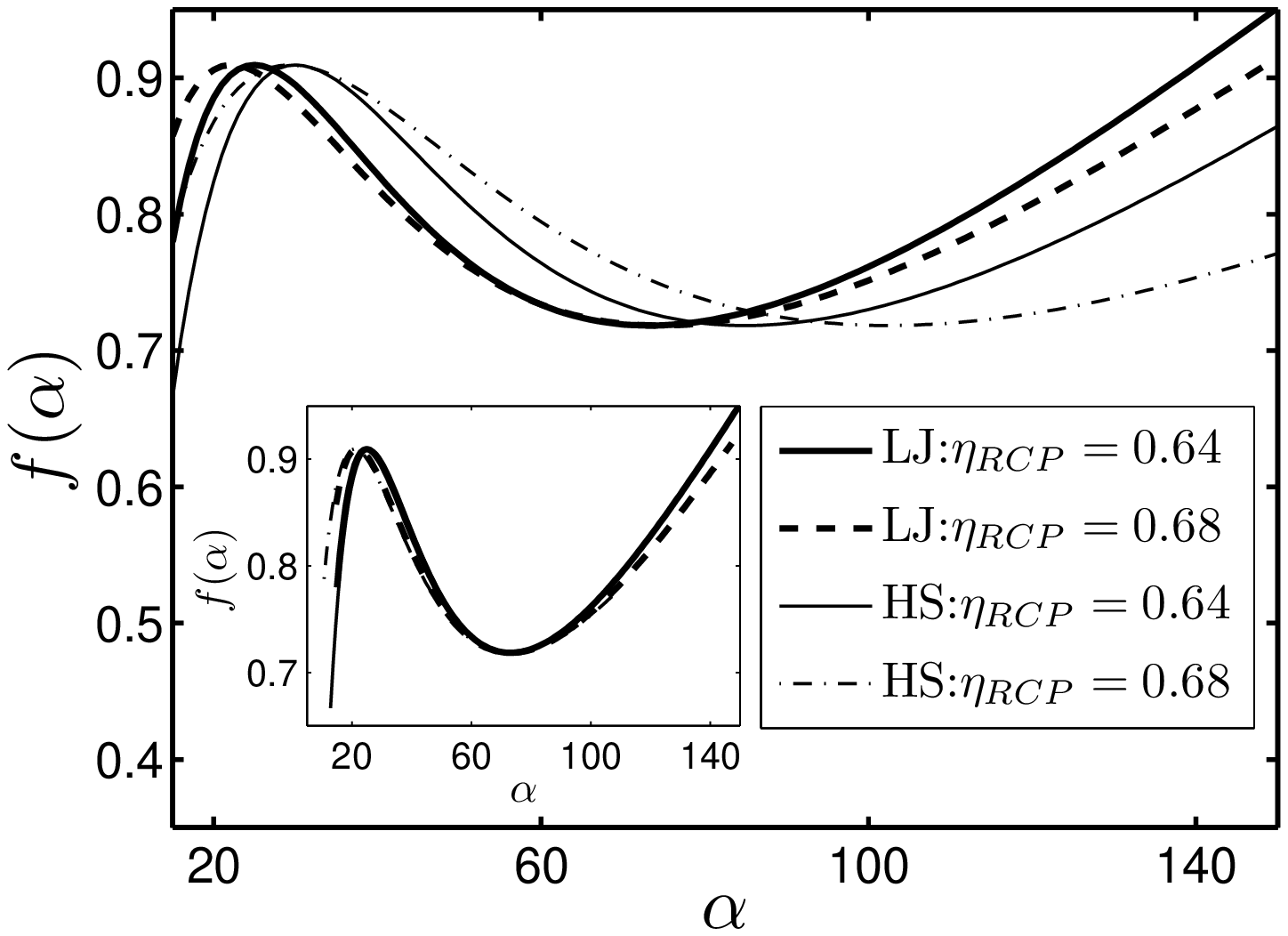}
    \\ 
    {\bf (b)}
  \end{center}
\end{minipage}
\end{tabular*}
\caption{\label{RLFalpha} {\bf (a)} The free energy $F(\alpha)$ of an
  aperiodic crystal of Lennard-Jones particles as a function of the
  force constant $\alpha$ of the effective Einstein oscillator from
  Eq.~(\ref{rho}). The temperature is fixed near the triple
  point. Four values of pressure are represented. The liquid is seen
  to undergo the RFOT at sufficiently high pressure (density). {\bf
    (b)} The $F(\alpha)$ plotted for hard spheres and Lennard-Jones
  particles, each at two distinct values of the parameter $\eta_\tRCP$
  that can be used to effectively vary local coordination.  The inset
  illustrates that given the same barrier height, the shape of the
  curve is robust with respect to detailed interactions and
  coordination. Hereby all four curves are rescaled along the
  horizontal axis and translated vertically, if necessary, so that the
  locations of the metastable minima coincide. Both graphs are from
  Ref.~\cite{RL_LJ}.}
\end{figure}

Baus and Colot~\cite{BausColot} thus showed that although the
aperiodic crystal with the Bennett structure does not ever reach the
uniform liquid in stability---if the attractive tail in $c(r)$ is not
included---increasing the local coordination can yield a structure as
stable as or more stable than the uniform liquid, see
Fig.~\ref{SSW}(c). This result was later confirmed by
Lowen~\cite{Lowen}, see Fig.~\ref{SSW}(d) who used the more reliable
MWDA approximation which is less sensitive to the large $r$ form of
the direct correlation function. (To avoid confusion, we repeat the
``large'' $r$ corresponds to separations just exceeding the particle
diameter, see Figs.~\ref{cRHendersonGrundke} and \ref{gRcR}(b).) As in
the periodic crystal case, the effective, weighted density is {\em
  lowered} when the liquid transitions to the metastable minimum,
implying the collision rate in the aperiodic solid is lowered compared
with the uniform liquid. Rabochiy and Lubchenko~\cite{RL_LJ} have
extended the results of Lowen to Lennard-Jones systems, see
Fig.~\ref{RLFalpha}. The graph is instructive in that it confirms that
a high-quality free energy function---which successfully reproduces
the phase diagram of the LJ system,
Fig.~\ref{LJphaseDiagram}---robustly yields the RFOT transition. It
also explicitly demonstrates that one can vary either density or
temperature to cause the crossover, in systems other than fully rigid
particles.

Aside from some uncertainty as to the quality of the aperiodic
lattice, we observe that an aperiodic arrangement of particles is in
fact a minimum of the free energy and, furthermore, could be made more
stable that the uniform liquid while remaining less stable than the
FCC structure. This observation provides the thermodynamic basis for
our understanding of the stability of supercooled liquids and glasses.

Additional support for the density-functional framework comes from the
mean-field calculation of Kirkpatrick and Wolynes~\cite{MCT} (KW). We
have seen an example of numerically evaluated free energy $F(\alpha)$,
which can be used determine the equilibrium value of $\alpha$ in the
aperiodic crystal.  One may ask, is there a closed form expression one
could use to evaluate the force constant $\alpha$ self-consistently,
similarly to how Eqs.~(\ref{himol}) and (\ref{hmolIsing}) can be used
to determine the spontaneously generated magnetisation below the Curie
point?  KW start from the Ramakrishnan-Yussouff functional (\ref{FRY})
and substitute the gaussian density ansatz
(\ref{rho}). Differentiating the resulting expression with respect to
$\alpha$ yields a self-consistent equation for $\alpha$ (in $D$
spatial dimensions):
\begin{equation} \label{KWalpha} \alpha = \frac{1}{6} \int \frac{d^D
    \bq}{(2\pi)^D} q^2 \tilde{c}_q^{(2)} \widetilde{S}_0(q)
  e^{-q^2/2\alpha},
\end{equation}
where $\widetilde{S}_0(q)$ is the structure factor of the aperiodic
lattice $\widetilde{S}_0(q) = \frac{1}{N} \sum_{ij} e^{- i \bq (\br_i
  - \br_j) }$, c.f. Eq.~(\ref{Sq}). In the mean-field, $D \to \infty$
limit, only the quadratic term in the free energy expansion survives
and so both the Ramakrishnan-Yussouff and the Eq.~(\ref{KWalpha})
become exact. Also in this limit, the functional form of the direct
correlation function simplifies. Thus Eq.~(\ref{KWalpha})
yields~\cite{MCT}:
\begin{equation} \label{KWalpha1} \alpha \sigma^2 = \frac{n^{*2}}{8
    \pi} e^{-D^2/2\alpha \sigma^2},
\end{equation}
where $\sigma$ is the diameter of the hard (hyper)sphere in $D$
dimensions and the quantity $n^* \equiv \bar{\rho} \sigma^D
\pi^{D/2}/\Gamma(1+D/2)$ is closely related to the packing fraction
since the volume of a $D$-dimensional hypersphere is equal to
$\sigma^D \pi^{D/2}/2^D \Gamma(1+D/2)$. For large $D$, $n^* \sim
O(D)$. Eq.~(\ref{KWalpha1}) is the liquid analog of the Weiss equation
for the spontaneous magnetisation of a uniform ferromagnet: $m =
\tanh[\beta (\sum_j J_{ij}) m]$. Like the latter equation,
Eq.~(\ref{KWalpha1}) always has the $\alpha=0$ solution corresponding
to the uniform phase. At sufficiently high densities, two new
solutions at {\em finite} values of $\alpha$ emerge that correspond to
the metastable minimum and the saddle point on the $F(\alpha)$ curve
in Figs.~\ref{SSW}(a) and \ref{RLFalpha}. (Exactly at the spinodal,
there is only one finite-$\alpha$ solution.) Thus in contrast with the
Ising magnet, the liquid-to-solid transition is discontinuous, be the
solid periodic or otherwise. By Eq.~(\ref{KWalpha1}), $\alpha \simeq
D^2/\sigma^2$ and thus simply reflects the number of particles in the
first coordination shell. The $\alpha \simeq D^2/\sigma^2$ scaling
implies the transition is the more discontinuous the closer the system
is to the strict mean-field limit.

Yet to fully appreciate the thermodynamic significance of the
aperiodic minimum in $F(\alpha)$, we must ask ourselves how many
distinct amorphous structures could represent such a minimum. It
appears \`{a} priori likely that when generating a sufficiently large
sample, the variety of such structures scales exponentially with the
sample size. One way to appreciate this is to recall our earlier
discussion of the 3rd order term in the Landau expansion
(\ref{LGbulk}) for the liquid. There we noted that the regular
icosahedron is one of the polyhedra that contains equilateral
triangles made of reciprocal lattice vectors. Alexander and
McTague~\cite{PhysRevLett.41.702} estimate the coefficient at the 3rd
order term due to icosahedra is about 0.63 of that for the bcc
lattice. They however note that because icosahedra do not tile space,
we can dismiss their contribution to the 3rd order term. Yet we should
recognise that on the one hand, the tiling does not have to be
perfect, since the second order term is still quite small even for
$q$'s that are not strictly equal to $q_0$, see
Fig.~\ref{rhoAq}(b). On the other hand, the number of ways to put
together such an imperfect lattice using nearly-equilateral triangle
motifs scales exponentially with the lattice size: An individual
``tile'' does not fit in perfectly; there are more than one way to
insert it in the matrix with a comparable degree of mismatch. The full
multiplicity is the configuration multiplicity for an individual tile,
taken to power $N$.  This multiplicity of alternative tilings will
make a bulk contribution to the stability of the 3rd order term.

Returning to the direct space, we quickly recognise that the
multiplicity of alternative structures in equilibrated liquids {\em
  is} in fact exponential in the system size: Assuming the vibrational
entropy of the periodic and aperiodic crystals are similar, the
multiplicity of the dissimilar aperiodic structures is reflected in
the excess entropy of a supercooled liquid relative to the
corresponding {\em periodic} crystal~\cite{Bernal1937}. At melting,
the excess entropy is the same as the fusion entropy; representative
values for the latter are listed in Table~\ref{FentropyTable}. To
summarise, the metastable minimum on the $F(\alpha)$ curve corresponds
to an exponentially large number of actual free energy minima. The
$F(\alpha)$ curve is thus a one-dimensional projection of the full
free energy surface where the order parameter is the localisation
parameter $\alpha$.

\begin{figure}[t]
\begin{tabular*}{\figurewidth} {ll}
\begin{minipage}{.4 \figurewidth} 
  \begin{center}
    \includegraphics[width=0.4
    \figurewidth]{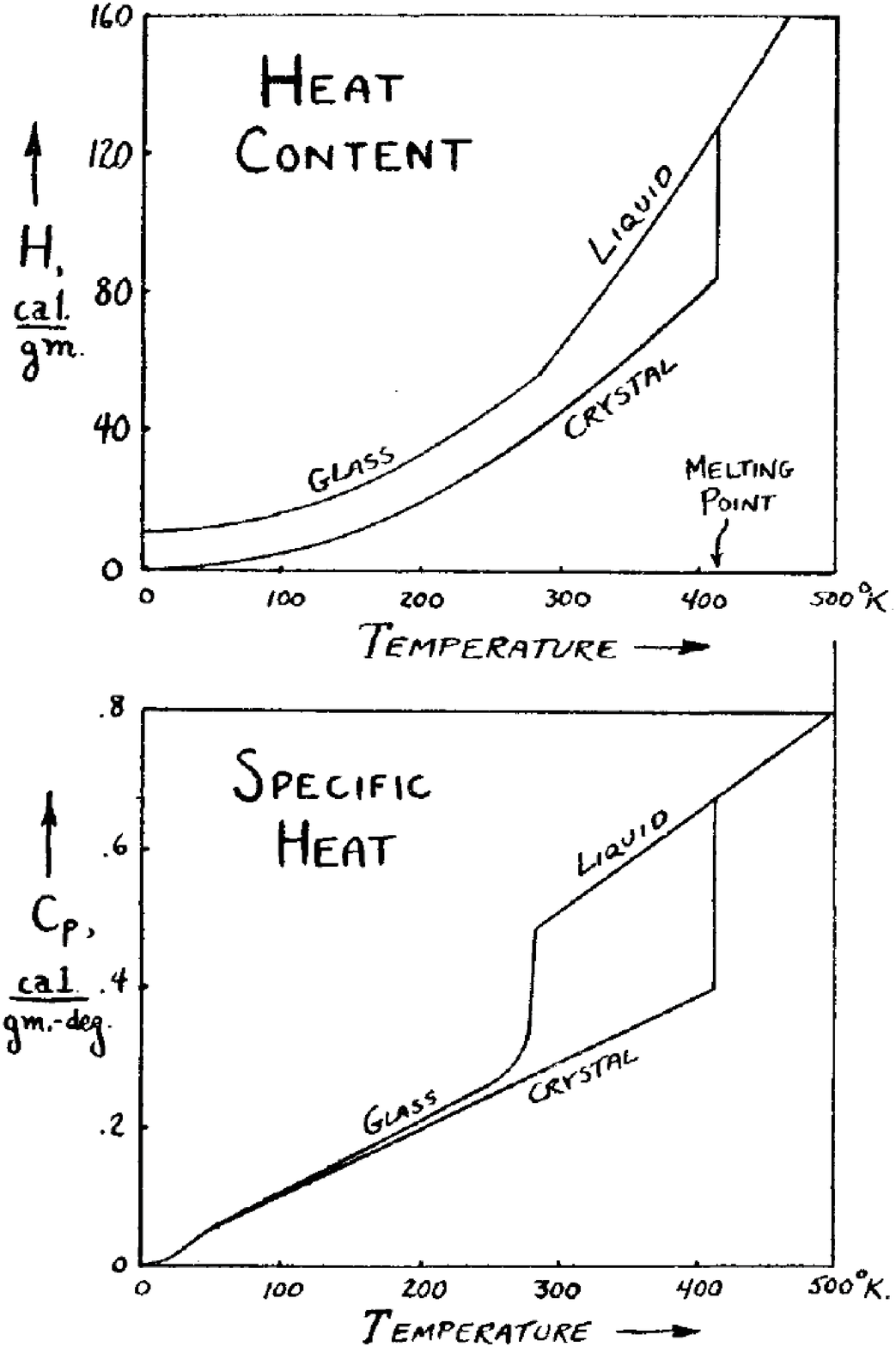} \\
    {\bf (a)}
   \end{center}
  \end{minipage}
&
\begin{minipage}{.55 \figurewidth} 
  \begin{center}
    \includegraphics[width=0.55
    \figurewidth]{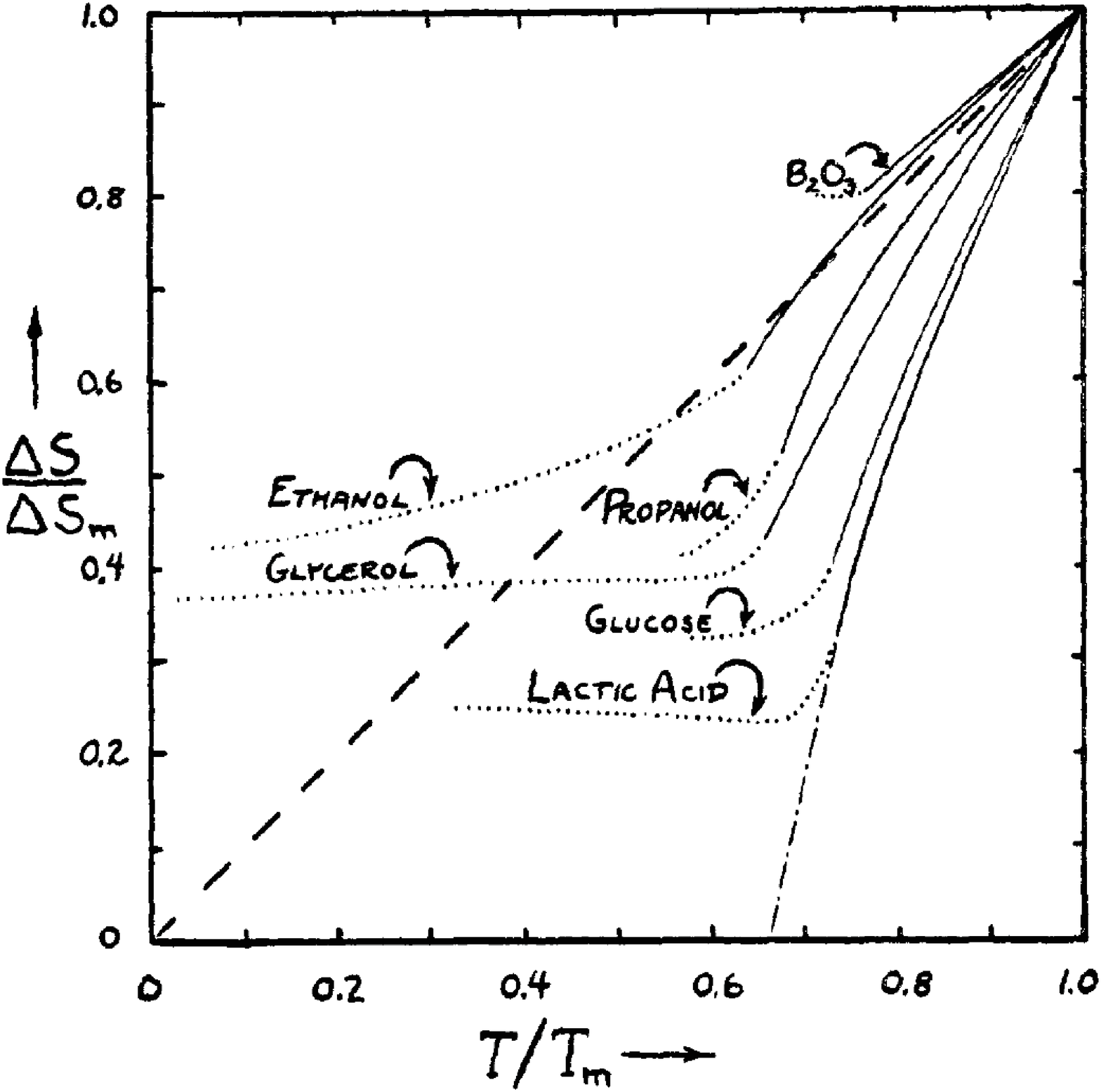} \\ \vspace{7mm}
    {\bf (b)}
  \end{center}
\end{minipage}
\end{tabular*}
\caption{\label{Kauzmann} {\bf (a)} Top: The enthalpies of the liquid
  and the corresponding crystal, as inferred by integrating the heat
  capacity data (bottom)~\cite{Kauzmann}. {\bf (b)} Excess liquid
  entropy $\Delta S$, relative to the corresponding crystal (thin sold
  line) as a function of temperature, for a number of specific
  substances. Extrapolation of this excess entropy below the glass
  transition temperature, due to Kauzmann~\cite{Kauzmann}. $\Delta
  S_m$ and $T_m$ stand for the fusion entropy and the melting
  temperature respectively.}
\end{figure}

Since the vibrational entropy of an individual minimum is numerically
close to that of the crystal, the exponentially large multiplicity of
the free energy minima results in an excess contribution to the
entropy of the liquid, relative to the crystal. This contribution is
directly seen by calorimetry, Fig.~\ref{Kauzmann}, in the form of an
excess heat capacity. The latter can be integrated in temperature to
infer both the excess liquid enthalpy and entropy, by virtue of $C_p =
(\prtl H/\prtl T)_p = T (\prtl S/\prtl T)_p$. Furthermore, since
molecular translations freeze out below the glass transition---apart
from some ageing---the excess entropy is approximately
temperature-independent below $T_g$. This leads to an observable jump
in the heat capacity at the glass transition:
\begin{equation} \label{DCp} \Delta c_p(T_g) = T \left(\frac{\prtl
      s_c}{\prtl T} \right)_{\!\! p, \: T = T_g^+} - T
  \left(\frac{\prtl s_c}{\prtl T} \right)_{\!\! p, \: T = T_g^-}
  \approx T_g \left(\frac{\prtl s_c}{\prtl T} \right)_{\!\! p, \: T =
    T_g^+},
\end{equation}
see Fig.~\ref{Kauzmann} and the inset of Fig.~\ref{angell}.

To determine the microscopic consequences of this multiplicity of
minima it is instructive to assume first that fluctuations around the
free energy minima are such that the system is allowed to transition
between minima only as a whole, but not locally. This mean-field
constraint---which will be lifted shortly---concerns the mutual
transitions between the individual distinct aperiodic free energy
minima but also the transitions between the uniform liquid state and
individual aperiodic minima.  Let us denote the free energy of an
individual minimum as $F_i$. This quantity contains the energy proper
$E_i$ of the configuration plus the vibrational free energy:
\begin{equation} F_i = E_i - T S_{\svibr, i}.
\end{equation}
Suppose the multiplicity of the minima, or ``configurations'', at a
given value of $F_i$ is $\Omega(F_i)$. We define the {\em
  configurational entropy} as the logarithm of this multiplicity times
$k_B$. Per particle,
\begin{equation} S_c = k_B \ln \Omega(F_i).
\end{equation}
Apart from the the contribution of the periodic crystal to the
thermodynamic ensemble, the partition function can be written as
\begin{equation} \label{ZliqaXtal} Z = e^{-\beta F_\sliq} +
  \Omega(F_i) e^{-\beta F_i} = e^{-\beta F_\sliq} + e^{-\beta [F_i - T
    S_c(F_i)]},
\end{equation}
where $F_\sliq$ is the free energy of the uniform-liquid
state. Clearly, when
\begin{equation} \label{aXtalCondition} F_i < F_\sliq + T S_c,
\end{equation}
the aperiodic crystal state will be more stable than the uniform
liquid. More precisely, in the assumption of spatial homogeneity of
transitions between the distinct aperiodic states, exactly one will be
realised at a time. Owing to the vast superiority of the second term
on the r.h.s. of Eq.~(\ref{ZliqaXtal})---the exponents all scale
linearly with the particle number---the system will spend a vanishing
fraction of time in the uniform liquid state if the condition
(\ref{aXtalCondition}) is satisfied.

\begin{figure}[t]
  \begin{tabular*}{\figurewidth} {cc}
    \begin{minipage}{.46 \figurewidth} 
      \begin{center}
        \includegraphics[width= .35 \figurewidth, angle =
        -90]{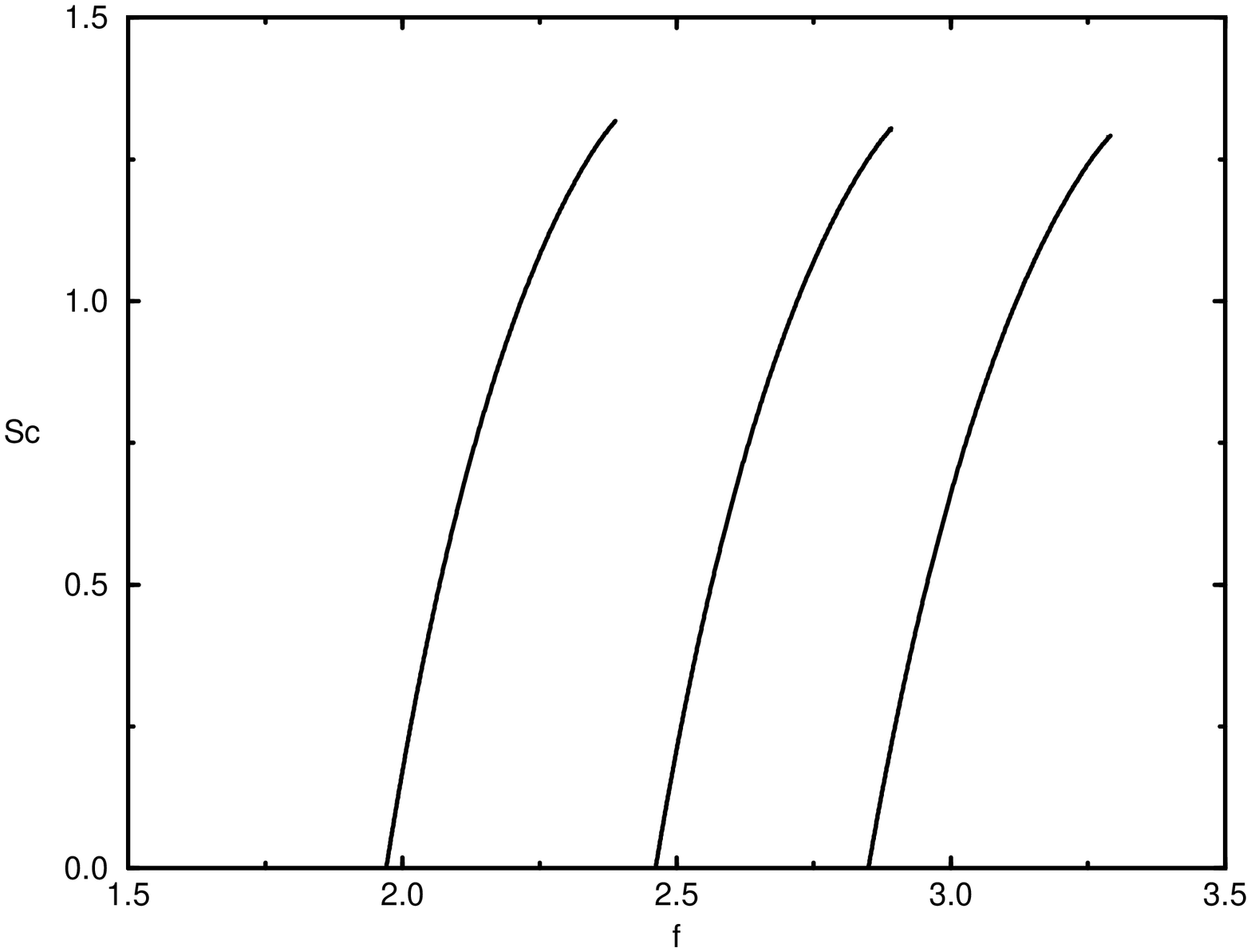}
         \\ {\bf (a)}
      \end{center}
    \end{minipage}    
    &
    \begin{minipage}{.48 \figurewidth} 
      \begin{center}
        \includegraphics[width= 0.35 \figurewidth, angle =
        -90]{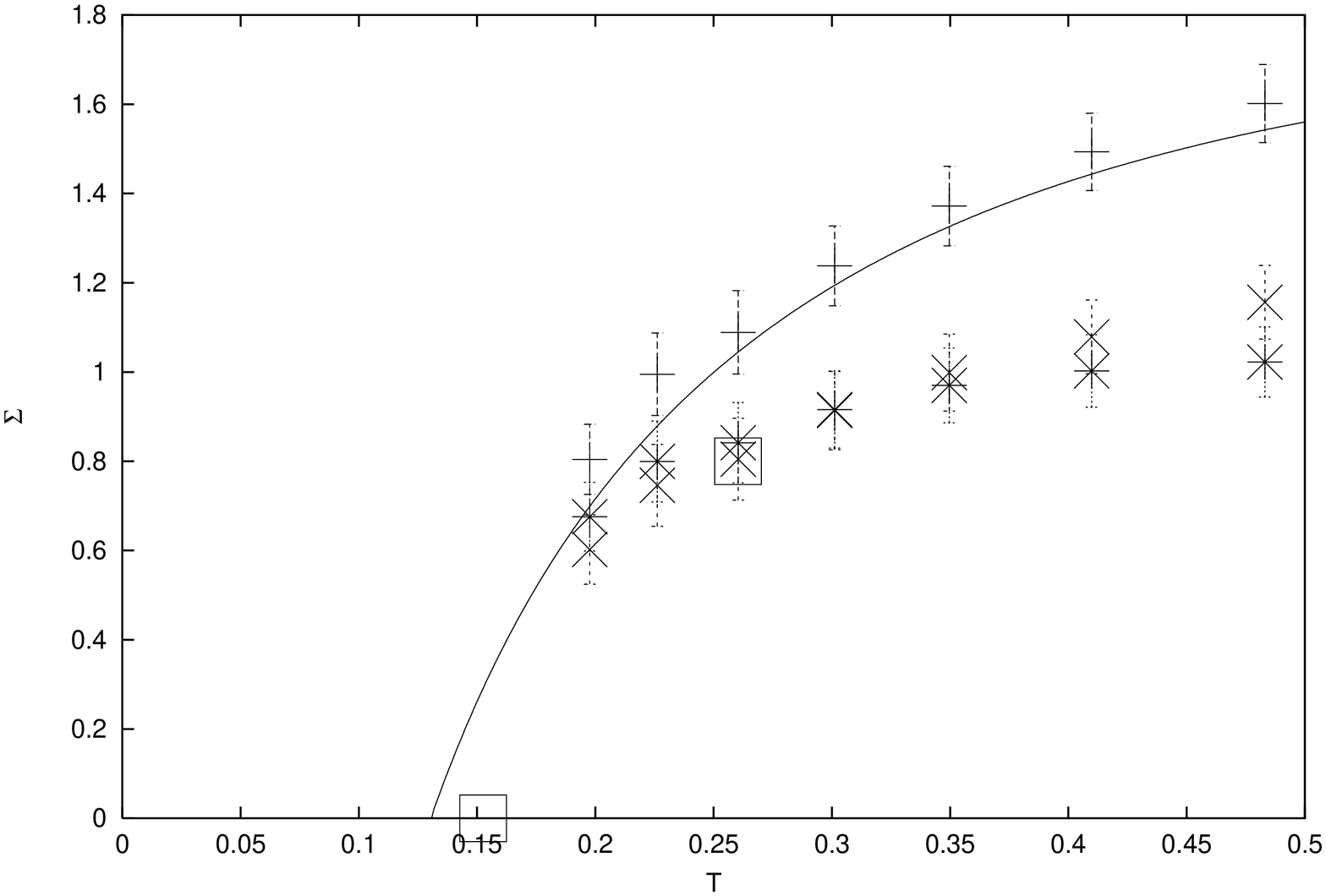}
        \\ {\bf (b)}
      \end{center}
    \end{minipage}
  \end{tabular*}
  \caption{\label{Sc_f} {\bf (a)} The configurational entropy of hard
    spheres computed as a function of the free energy, per particle
    and $k_BT$, at three distinct values of
    temperature~\cite{mezard:1076}. {\bf (b)} The configurational
    entropy for a binary mixture of soft spheres as a function of
    temperature according to a replica calculation(---), alongside
    with results of alternative numerical procedures including direct
    Monte-Carlo (*). From Ref.~\cite{MP_Wiley}.}
\end{figure}

Is this condition satisfied in actual liquids? According to
Fig.~\ref{SSW}, the altitude of the aperiodic minimum, relative to the
uniform liquid, is about $1 - 1.5 k_B T$ at liquid densities near
melting, see Table~\ref{DentonAshcroftTable} for numerical values of
those densities.  If one accounts for the increase in coordination
with density, the excess free energy of an individual aperiodic
minimum relative to the uniform liquid could be significantly
lower. According to Table~\ref{FentropyTable}, the excess liquid
entropy per rigid molecular unit does, in fact, have similar values
near melting.  Of particular value in the present context are results
of M\'ezard and Parisi~\cite{mezard:1076, MP_Wiley}, who used a
replica methodology to estimate the exact type of the mean-field
configurational entropy that enters in
Eq.~(\ref{aXtalCondition}). Their results are shown in Fig.~\ref{Sc_f}
and indicate that indeed the configurational entropy is sufficiently
large to stabilise the aperiodic phase at sufficiently high
density. In the strict mean-field limit, the transition from the
uniform liquid to the aperiodic solid occurs at a sharply defined
temperature, which we will denote with $T_A$. The transition itself is
called the Random First Order transition (RFOT), to reflect that the
liquid freezes into a random lattice, while the freezing itself is a
discontinuous transition. It would not be obvious from the preceding
discussion whether the RFOT would take place at the same temperature
at which the metastable minimum in the $F(\alpha)$ curve just begins
to develop or below that temperature. We shall see shortly, in
Subsection~\ref{MCT}, that the mean-field spinodal does in fact take
place exactly at the temperature $T_A$ at which the (degenerate)
aperiodic crystal becomes thermodynamically stable.  We shall also
observe in Subsection~\ref{spinConnections} that this is a rather
general pattern that also covers spin systems.

\begin{figure}[t] \centering
  \includegraphics[width= .6\figurewidth]{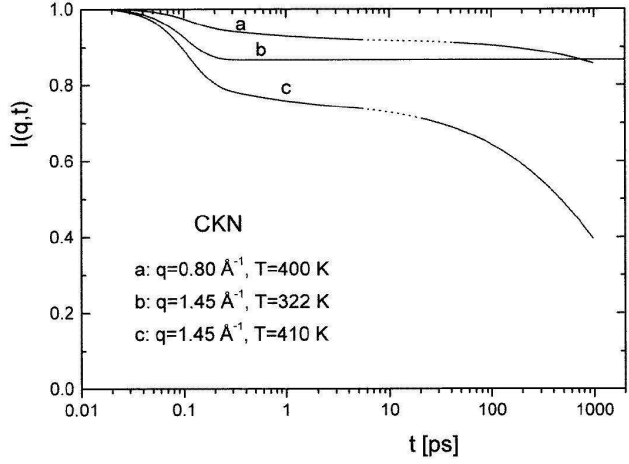}
  \caption{\label{mezei} Intermediate structure factor as determined
    by neutron scattering~\cite{MezeiRussina}. Note curve (b) was
    measured below the glass transition.}
\end{figure}

The formation of long-lived aperiodic structures predicted by the RFOT
theory is consistent with neutron scattering data~\cite{MezeiRussina},
see Fig.~\ref{mezei}, which clearly shows that below a certain
temperature, each particle becomes trapped in a cage made of the
surrounding molecules.  The time needed for relaxation of these
long-lived structures is considerable already above the glass
transition and becomes even longer below the transition, see
Fig.~\ref{mezei}.  We postpone the calculation of the structural
relaxation rate until Subsection~\ref{mosaic}, where we go beyond the
mean-field limit.

The presence of the random first order transition imply that the fluid
is no longer represented by a unique free energy minimum; the
exponentially many free energy minima all correspond to non-uniform
density profiles. Conversely, one may ask what happens to the free
energy minimum corresponding to the uniform liquid. This minimum
becomes {\em metastable} below the RFOT.  One may further ask under
which conditions the uniform minimum disappears, which would
correspond to the mechanical stability limit of the uniform liquid.
The appearance of a marginally stable mode can be detected by standard
stability analysis~\cite{Lovett1977}. First note that a density
response to a weak, coordinate-dependent external field $U(\br)$ can
be written generally as:
\begin{equation} \label{drhodU} \drho(\br_1) = \int d^3 \br_2 \,
  \chi(\br_1, \br_2) \, U(\br_2).
\end{equation}
Where the susceptibility:
\begin{equation} \chi(\br_1, \br_2) \equiv
  \left. \frac{\drho(\br_1)}{\delta U(\br_2)} \right|_{U=0},
\end{equation}
can be expressed through the density-density correlation function
$\rho^{(2)}(\br_1, \br_2)$ and the density, according to
Eqs.~(\ref{rhorur}) and (\ref{rhorho}). For the uniform liquid, which
is translationally invariant, Eq.~(\ref{drhodU}) can be rewritten as
simple equations for individual Fourier components of the functions in
question:
\begin{equation}\label{rhoqUq}
  \drhot_{\bq} = \widetilde{\chi}_q \widetilde{U}_{\bq},
\end{equation}
while each Fourier component of the response function is
straightforwardly related to the that of the direct correlation
function, by Eqs.~(\ref{g(r)})-(\ref{OZq}):
\begin{equation} \label{chiq} \chi_q = \frac{\rho_\sliq k_B T}{1 -
    \rho_\sliq \tilde{c}^{(2)}_q}.
\end{equation}
This expression is consistent with Eq.~(\ref{FRYq}), since the latter
implies $(k_B T/2) V (\rho_\sliq^{-1} - \tilde{c}^{(2)}_q) \la
|\drhot_q|^2 \ra = k_B T/2$ by virtue of the equipartition theorem.
On the other hand, $\chi_q = \la |\delta \tilde{\rho}_q|^2 \ra/V$ by
Eq.~(\ref{rhorur}).  Note also that the $q \rightarrow 0$ limit of
Eq.~(\ref{chiq}) could have been inferred from Eqs.~(\ref{sum1}),
(\ref{g(r)}) and (\ref{sum2}).

According to Eq.~(\ref{chiq})---and in full correspondence with our
earlier analysis of the possibility of a continuous liquid-to-crystal
transition, Eq.~(\ref{FRYq})---the first liquid modes to reach the
spinodal are spatially-varying patterns with $|\bq| = q_0$ such that
\begin{equation} \label{unstable} 1 - \rho_\sliq \tilde{c}^{(2)}_{q_0}
  = 0.
\end{equation}
Lovett~\cite{Lovett1977} points out that at least within the
Percus-Yevick approximation, the uniform liquid state remains
(meta)stable at densities such that the pressure is finite. Whether
these conclusions applies to actual liquids remains an open question.

\begin{figure}[t] \centering
  \includegraphics[width=0.9 \figurewidth]{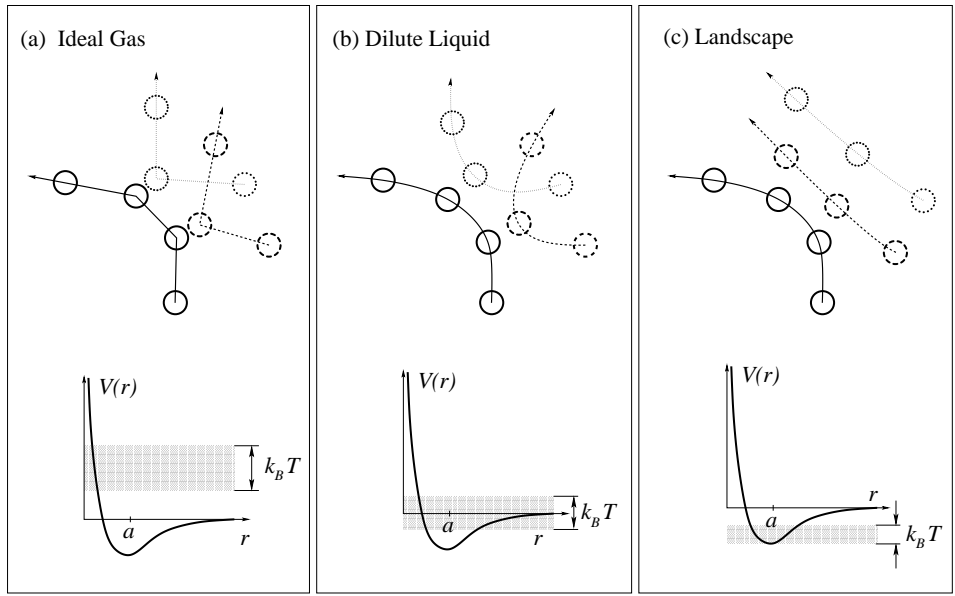}
  \caption{\label{flights} The three relatively distinct kinetic
    regimes of an equilibrium liquid, or a conditionally equilibrium
    liquid, as the case would be below the melting temperature
    $T_m$. See text for explanation. The bottom portions illustrate
    the region of the two-particle potentials explored in the
    corresponding regimes. (a) Nearly ideal gas, (b) dilute liquid,
    and (c) landscape.}
\end{figure}

We conclude by reiterating the central features of the random first
order transition in liquids, see Fig.~\ref{flights}: It is a
transition from a truly uniform liquid characterised by a single free
energy minimum, Fig.~\ref{flights}(a) and (b), into a symmetry-broken
state characterised by an exponentially large number of free energy
minima, Fig.~\ref{flights}(c). It is the translational symmetry that
becomes broken, whereby the uniform density profile for individual
particles no longer minimises the free energy. Instead, the optimum
density profile consists of a collection of disparate narrow
peaks. The translational symmetry is eventually restored by activated
transitions between distinct metastable configurations. The transition
is sharp in the mean-field limit but becomes a soft crossover in
finite dimensions because distinct metastable configurations have a
finite lifetime. As a result, the translational degrees of freedom
freeze out gradually, starting from the highest frequency
motions. Despite the gradual character of the crossover, one may
define a temperature, $T_\scr$, that corresponds to the mid-point of
the crossover~\cite{LW_soft}. In the mean-field limit, the crossover
would be a sharp (first-order) transition at a well-defined
temperature $T_A$.  Although it is customary to assign a {\em
  temperature} to the RFOT crossover, we must bear in mind that
crossover is better thought of as {\em density}-driven, as is clear
from the argument. In actual substances characterised by finite
interactions, compactification is realised most readily by cooling,
hence our use of a temperature $T_\scr$. The notion of the steric
origin of the RFOT transition is consistent with the fact that
substances that are characterised by open structures and very
directional bonding crystallise {\em before} the steric effects become
significant---in ways other than dictating the nearest-neighbour
distance---and thus are poor glassformers, as discussed in
Subsection~\ref{bead}. Conversely, pressurising such open structures
leads to their vitrification~\cite{Sharma1996, Brazhkin2003,
  Ponyatovsky1992, Hemley1988}.

Finally we make an important remark, terminology-wise, that the
crossover from the collisional to activated transport could occur
either above or below the melting temperature, the former and latter
situations corresponding to strong and fragile substances
respectively, see Section~\ref{crossover}. As a result, a strong
liquid can be in the activated transport regime while not being
technically supercooled.  At the same time, a very fragile substance
can be supercooled even though its molecules still move in a largely
collisional manner.  To avoid ambiguity in this regard, we will term
liquids below the crossover {\em glassy} or will say the liquid is in
the {\em landscape} regime.

\subsection{Configurational Entropy}
\label{sc}

We just saw that the magnitude of the configurational entropy is key
to whether the emerging aperiodic structures can compete, free-energy
wise, with the uniform liquid state. We shall also see shortly that
knowing the configurational entropy is central to computing the
activation barrier for liquid relaxations below the crossover. The
configurational entropy has been evaluated from first principles only
for the simplest systems, even if approximately, see
Fig.~\ref{Sc_f}. Such calculations for actual substances are very
difficult, especially given that the explicit form of the actual
quantum-chemical interactions is complicated and not known in closed
form. On the other hand, the classical DFT calculations for aperiodic
lattices, from Subsection~\ref{RFOTDFT}, strictly apply only to
isotropically interacting particles that repel at short distances and
attract modestly at large distances. It is thus imperative to be able
to take advantage of the {\em measured} excess liquid entropy of
actual liquids, which, however, must be calibrated in terms the
entropy content of such a model liquid that is amenable to the DFT
analysis.

Before we discuss the calibration, we provide in
Fig.~\ref{Kauzmann}(b) a graph of the excess liquid entropy, as a
function of temperature, from the seminal paper of
Kauzmann~\cite{Kauzmann}. Kauzmann emphasised, following
Simon~\cite{Simon}, that the excess entropy would vanish at a {\em
  finite} temperature, $T_K > 0$, if extrapolated beyond the glass
transition. Approximately, one may write:
\begin{equation} \label{scT} s_c \approx \Delta c_p(T_K) (T-T_K)/T_K.
\end{equation}
In practice, one often uses the following functional form for the
temperature dependence of the configurational
entropy~\cite{PhysRevLett.64.1549, RichertAngell}:
\begin{equation}\label{scTRA} s_c \simeq
  \Delta c_p(T_g) T_g(1/T_K - 1/T),
\end{equation}
c.f. Eq.~(\ref{DCp}), which works very well for many substances.  

\begin{figure}[t]
  \begin{center}
    \includegraphics[width= .41 \figurewidth]{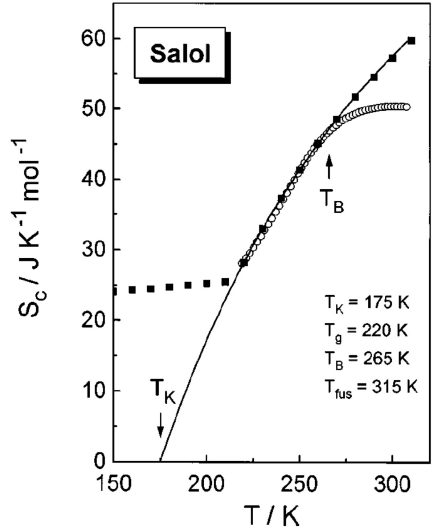} 
  \end{center}
  \caption{\label{RichertAngell} Extrapolation of the configurational
    entropy for a number of specific substances below the glass
    transition temperature for salol (symbols) superimposed on inverse
    log-relaxation time data (solid line), showing a clear correlation
    between the decrease in configurational entropy and the increase
    in the relaxation time~\cite{RichertAngell}.  }
\end{figure}

To calibrate the excess liquid entropy of an actual substance we first
consider a useful, extreme limit of the hard sphere: The entropy of
monodisperse spheres depends only on the density. Likewise, the fusion
entropy is also temperature independent and is about $1.19 k_B$ per
particle (Table~\ref{DentonAshcroftTable}).  Furthermore, we also
understand fairly well how hard spheres pack: For instance, the
highest filling fraction for monodisperse spheres is $\pi/3\sqrt{2}
\approx 0.74$. We know accurately the density of the liquid and
crystal in equilibrium (Table~\ref{DentonAshcroftTable}), and have a
good idea as to the filling fraction of the random close-packed
structure, $0.64$ or so.  A somewhat better reference system can be
obtained by starting with hard spheres and then including a weakly
attractive tail to the particle-particle interaction. The next step is
to find a naturally occurring system that shows strong---but not
infinitely strong---repulsion at short distances and weak attraction
otherwise. The finiteness of the energy scale associated with both the
repulsive and attractive portion of the interaction affords one more
flexibility in calibrating actual interactions.  (We have already seen
an example of this in Section~\ref{Xtal}, see
Fig.~\ref{LJphaseDiagram}.)  Argon is a good choice: On the one hand,
Ar is a closed-shell atom and so two argon atoms repel at short
distances owing to the Pauli exclusion principle while being attracted
through weak, dispersive forces at larger separations. On the other
hand, Ar is not too light so quantum effects are not large, and it is
not too heavy, so that polarizability is not too high.

In all known glassformers---in contrast with argon---the interactions
are {\em directional}, to a lesser or greater degree. For instance,
suppose the magnitude of the fusion entropy per atom in an actual
glass-former is smaller than in argon, see
Table~\ref{FentropyTable}. This means that fewer degrees of freedom
freeze out at the liquid-to-crystal transition, {\em per atom}, than
for isotropically interacting particles: Some of the degrees that
would be available to isotropically interacting particles are, in
fact, constrained by the directionality.  One may thus think of the
solidification of a substance with directional bonding as the freezing
of weakly attractive, isotropically-interacting particles, but with an
effective size {\em exceeding} the average atomic size. These
qualitative conclusions apply equally to solidification into a periodic
or aperiodic crystal.

Thus, we will usually calibrate the configurational entropy content of
actual substances in the following way. We divide the substance's
fusion entropy per molecule by the fusion entropy of argon per
atom. This gives us the number $N_b$ of rigid molecular units, or
``beads,'' per molecule~\cite{LW_soft, StevensonW}:
\begin{equation} \label{Nb} N_b= \frac{\Delta H_m}{s_\text{bead} k_B
    T_m},
\end{equation}
where $\Delta H_m$ and $T_m$ are the fusion enthalpy and temperature
respectively. The quantity $s_\text{bead}$ is the entropy content per
bead in units $k_B$ in a reference liquid. If the latter is Ar (which
is usually the case), $s_\text{bead} = 1.68$.  As a result, the
volumetric bead size is given by
\begin{equation} \label{a3} a^3 = v^3_m/N_b,
\end{equation}
where $v_m$ is the specific volume of the substance.  Clearly, the
calorimetric way to determine the bead size is phenomenological. Yet
it turns out to produce bead counts that are chemically
sensible~\cite{LW_soft}. In those cases where the bead size does
affect the final answer, one should be mindful of potential
ambiguities, as will be discussed in due time.  In any event, many of
the RFOT-based quantitative predictions are universal in that they are
independent of the bead size.

In view of the potential numerical uncertainly in the bead size, it is
reassuring that there is another sense in which this size can be
defined, viz., the ultraviolet cut-off of the theory~\cite{BL_6Spin,
  BLelast}. In this elasticity theory-based view, all vibrational
motions whose wave vector exceeds $\pi/a$ do not affect the structural
reconfigurations and are not included in the phonon sums, see
Section~\ref{crossover}.

To finish the discussion of calibration of the excess liquid entropy
we note that the experimental values of this entropy, when calibrated
according to the bead-counting procedure above, are in fact consistent
with the magnitude of the configurational entropy requisite for the
thermodynamical stability of the (degenerate) aperiodic-crystal state,
relative to the uniform liquid state. These calculations are described
in Section~\ref{crossover}.

Now, the vanishing of the (extrapolated) configurational entropy is in
apparent contradiction with the Nernst law and suggests that there is
something special about the translational degrees of freedom in
liquids. It may appear that the question of whether the entropy crisis
would actually take place is somewhat of an academic nature: All known
liquids become too slow to equilibrate on the laboratory scale well
before the configurational entropy vanishes: Empirically, the liquid
relaxation time diverges {\em exponentially} according to an empirical
relation
\begin{equation} \label{VFT} \tau = \tau_0 e^{D T_0/(T-T_0)},
\end{equation}
known as the Vogel-Fulcher-Tammann (VFT) law.  The coefficient $D$ is
called the fragility.  Despite being somewhat academic, the question
of the vanishing of the configurational entropy is fundamentally
interesting and, ultimately, must be confronted if one were to
reliably estimate the configurational entropy for actual substances
from first principles.

\begin{figure}[t]
  \centering
  \includegraphics[width= .6 \figurewidth]{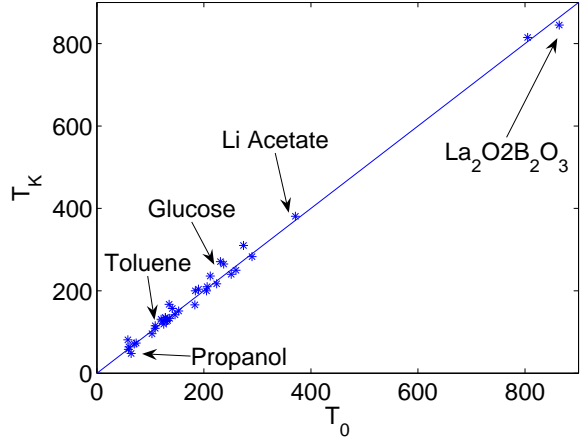}
  \caption{\label{Tk_Vs_T0} The Kauzmann temperature $T_K$ plotted
    versus the temperature $T_0$. At $T_K$, the configurational
    entropy extrapolated below the glass transition vanishes. At $T_0$
    the viscosity extrapolated below the glass transition diverges.}
\end{figure}

Early on, Adam and Gibbs~\cite{AdamGibbs} argued that liquid transport
above the glass transition is activated with a barrier scaling
inversely proportionally to the configurational entropy. Motivated by
these ideas, one may ask whether, in fact, the putative vanishing of
the configurational entropy and divergence of the viscosity would take
place at the same temperature.  Richert and
Angell~\cite{RichertAngell} have carefully analysed uncertainties in
the extrapolation beyond the glass transition temperature. This
analysis demonstrates that the temperature dependences of the
configurational entropy and relaxation data clearly correlate, see
Fig.~\ref{RichertAngell}. Consistent with this detailed analysis,
extrapolation of fitted kinetic and thermodynamic data for many more
substances indicate a very tight correlation, if not downright
coincidence of the temperatures $T_0$ and $T_K$, see
Fig.~\ref{Tk_Vs_T0}.

The putative Kauzmann entropy crisis adds another layer of complexity
to the already intriguing behaviour of glassy liquids. In addition to
freezing into one of the exponentially many aperiodic minima---in the
mean-field limit---an equilibrium liquid appears to be able to reach,
as least in principle, a state in which the log-number of those minima
scales {\em sublinearly} with the system size.  In other words, there
is essentially a unique aperiodic configuration that a liquid could
presumably reach if sufficiently pressurised or cooled in a
quasi-equilibrium fashion. Because this state is unique, it is also
mechanically stable, as is a periodic crystal! To avoid confusion, we
emphasise that this mechanical stability would be achieved even in
{\em finite} dimensions. (In contrast, the mean-field system would
become stable already below the temperature $T_A$.)  Kauzmann himself
suggested that the entropy crisis would be avoided if the liquid
reached its mechanical stability limit and crystallised before $T_K$
is reached.  Stevenson and Wolynes~\cite{SWultimateFate} suggest that
the actual story about the ultimate fate of molecular liquids is more
complicated, to be discussed in due time.

\begin{figure}[t]
  \begin{tabular*}{\figurewidth} {cc}
    \begin{minipage}{.46 \figurewidth} 
      \begin{center} 
        \includegraphics[width=0.46 \figurewidth]{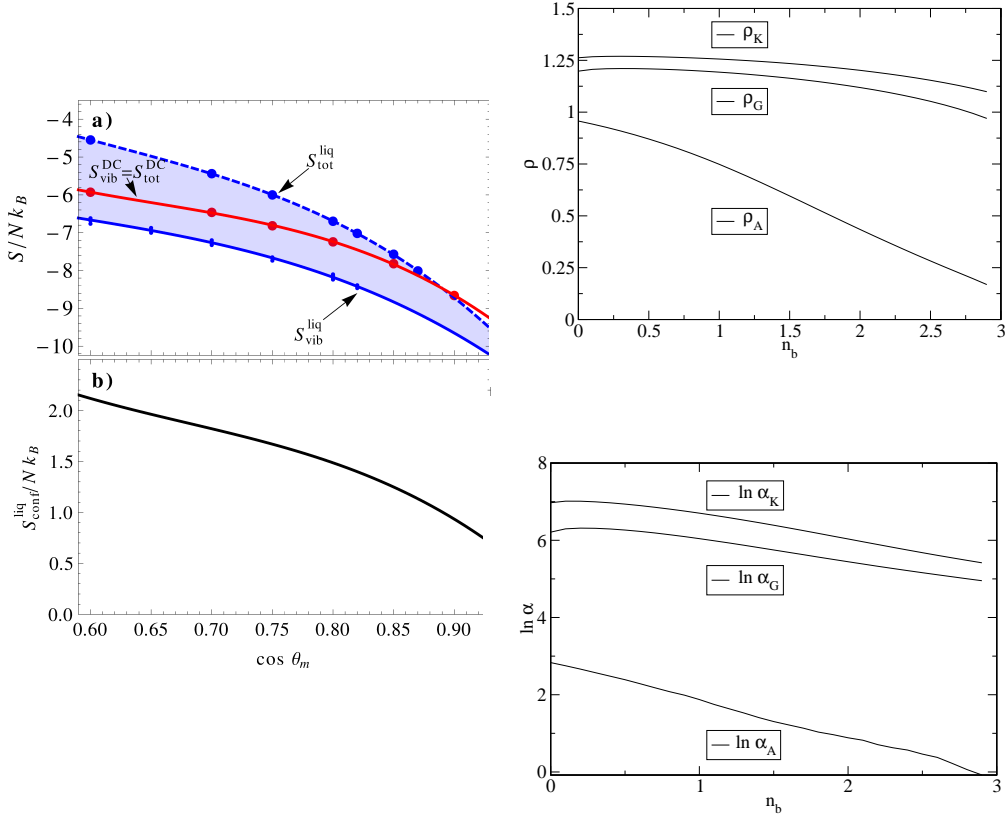}
      \end{center}
    \end{minipage}
    &
    \begin{minipage}{.5 \figurewidth} 
      {\bf (c)} \begin{center}
        \includegraphics[width=0.46 \figurewidth]{HallW1.eps}
      \end{center}
      \vspace{2mm} 
      {\bf (d)} \begin{center} \includegraphics[width=0.46
        \figurewidth]{HallW2.eps}
      \vspace{0mm} 
      \end{center}
    \end{minipage}
  \end{tabular*}
  \caption{\label{Sciortino} {\bf (a)} The (extrapolated) $T=0$
    entropy of the liquid and diamond crystal (DC) made of patchy
    colloidal particles as a function of the angular width $\theta_m$
    of the patch. To large values of $\cos \theta_m$, there correspond
    smaller patches. Both the vibrational and full entropy of the
    liquid are shown. The density is fixed at $\rho \sigma = 0.57$,
    which corresponds to the filling fraction $\approx 0.30$. Note
    this value is below the filling fraction of the diamond lattice,
    viz., $\approx 0.34$.  {\bf (b)} The configurational entropy of
    the patchy colloid, computed by subtracting the vibrational
    entropy of the liquid from the full liquid entropy, extrapolated
    to zero temperature. Due to Smallenburg and
    Sciortino~\cite{9004795520130901}. {\bf (c)} Dependence of the
    densities at the Kauzmann ($\rho_K$), glass transition ($\rho_g$),
    and mean-field RFOT ($\rho_A$) transitions on the network
    connectivity in the Hall-Wolynes model~\cite{HallWolynes}. {\bf
      (d)} Dependence of the force constant of the effective Einstein
    oscillator $\alpha$ from Eq.~(\ref{rho}), according to
    Ref.~\cite{HallWolynes}.}
\end{figure}

\begin{figure}[t]
  \centering
  \includegraphics[width= .7 \figurewidth]{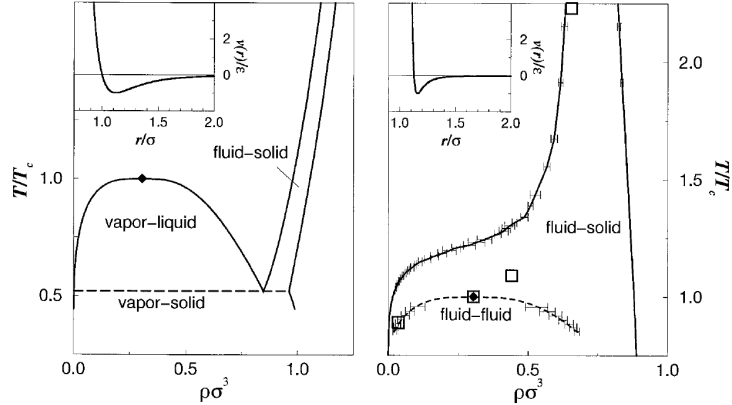}
  \caption{\label{tenWoldeFrenkel} Concentration-temperature phase
    diagrams for isotropically interacting particles. The l.h.s. and
    r.h.s panels show two distinct situations in which the width of
    the attractive well in the interaction potential is comparable to
    or significantly less than the particle size, see
    insets~\cite{WoldeFrenkel}. The former situation is typical in
    ordinary liquids while the latter is characteristic of protein or
    colloidal solutions, as in Ref.~\cite{9004795520130901}. Given a
    sufficiently narrow attractive well, the vapour-liquid coexistence
    line is moved under the liquidus. As a result the liquid can
    exists only as a metastable phase, while in equilibrium, the
    vapour transitions directly to the crystal phase. Since colloids
    are suspended in a solution, the vapour-liquid coexistence is
    observed as a coexistence between a relatively dilute and
    concentrated solution, and is often called ``liquid-liquid''
    coexistence.}
\end{figure}

In this regard, it is interesting to mention a relatively recent
simulation study by Smallenburg and Sciortino~\cite{9004795520130901}
on patchy colloids with highly anisotropic interactions, see
Fig.~\ref{Sciortino}(a) and (b). The model particles are rigid spheres
with four added attractive patches in a tetrahedral arrangement, each
patch of angular width $\theta$. The patches are attractive, but only
in a rather narrow distance range $\delta = 0.12 \sigma$; one contact
per patch is enforced.  The (extrapolated) configurational entropy of
this patchy colloid, {\em at constant volume}, remains positive down
to absolute temperature, implying a negative Kauzmann temperature!
This entropy is so large, in fact, that the liquid remains more stable
than the diamond lattice that the particles can form at the density in
question. The large configurational entropy should not be too
surprising in light of our discussion in Subsection~\ref{bead}; fixing
the density at a value below that of the very open, diamond structure
minimises the steric effects that lead to the crossover and,
eventually, to the (putative) Kauzmann crisis. Indeed, according to
Fig.~1 of Ref.~\cite{9004795520130901}, the ground state of the
colloid at constant pressure is close-packed. The short-range
character of the attraction in the model is also important. Because of
the narrow width of the attractive minimum, relative to the particle
size, the gas-liquid coexistence region of the system is ``hidden''
under the liquidus, see Fig.~\ref{tenWoldeFrenkel}.  This phenomenon
is well known---and often irksome---to X-ray crystallographers, and
has been elucidated by Wolde and Frenkel some time
ago~\cite{WoldeFrenkel}.  In equilibrium, such a gas condenses into
the crystal while bypassing the liquid state. (Since protein solutions
are often very hard to equilibrate, structural biologists often
observe a liquid-liquid separation that may lead to gelation and other
types of aggregation and preventing the protein crystal from forming.)
This type of gas-crystal coexistence can indeed be seen in Fig.~1(a)
of Ref.~\cite{9004795520130901}. The notion that we are dealing with a
gas here buttresses our earlier statements that steric repulsion is
key to understanding the temperature dependence of the configurational
entropy and the glass-forming ability of actual substances.

It would be fair to say that at present, we do not have a reliable way
to evaluate the configurational entropy of actual glass-formers from
first principles. Aside from approximations and the model nature of
the liquid, the M\'ezard-Parisi calculation~\cite{mezard:1076} is
still subject to the assumption that the entropy of the {\em uniform}
liquid can be extrapolated to $T_K$. This may introduce some numerical
uncertainty, if the Kauzmann state is {\em below} the mechanical
stability limit of the uniform liquid.  Regardless of various
technical difficulties, one can establish useful connections between
the bonding patterns in the liquid and its configurational entropy
using a microscopically-motivated model. Hall and
Wolynes~\cite{HallWolynes} (HW) used the self-consistent phonon theory
to analyse a hard sphere liquid with randomly added bonds between
particles, so that a particle is bonded on average to $n_b$
neighbours. Adding bonds stabilises contacts energetically but
destabilises them entropically. As expected, increasing the number of
bonds dramatically decreases the density $\rho_A$ at which metastable
structure begin to form while affecting the density $\rho_K$ of the
Kauzmann state only modestly, Fig.~\ref{Sciortino}(c).  The rapid
decrease of $\rho_A$ with the number of rigid bonds is consistent with
the results from Fig.~\ref{Sciortino}(a) and (b) in that increasing
directionality makes local bonding patterns more open. Beyond a
certain critical value of bonds a rigid percolation takes place in the
network, as signalled by $\alpha^{-1/2}$ approaching the particle size
$a$, see Fig.~\ref{Sciortino}(d). The rigidity percolation at $n_b
\simeq 2.8$ is not inconsistent with independent estimates by Thorpe
and coworkers~\cite{PhysRevE.59.2084, PhysRevE.64.041503}, who obtain
$n_b \simeq 2.4$.  Similarly to the behaviour of the density, adding
bonds also affects the temperature $T_A$ much more than $T_K$ (both
increase), so that the $T_K/T_A$ ratio {\em decreases}. We shall see
in Section~\ref{crossover} that the laboratory glass transition occurs
at a nearly universal value of the configurational entropy. This
notion, in combination with Eq.~(\ref{scTRA}), implies that a smaller
value of the $T_K/T_A$ corresponds with the smaller value of the heat
capacity jump at the glass transition $\Delta c_p(T_g)$.  This is
qualitatively consistent with observation~\cite{AngellScience1995},
see also the inset in Fig.~\ref{angell}.

One should be cautious in interpreting the results in
Figs.~\ref{Sciortino}(c) and (d) at values of $\alpha$ significantly
below $10^2/a^2$. While these graphs give correct qualitative trends
for large to medium values of $\alpha$, they could not correspond to
an equilibrated liquid for the smallest values of $\alpha$, certainly
not as small as $1/a^2$. Indeed, we know that the vibrational
displacement near the mechanical stability limit in directionally
bonded materials, such as silicon, is quite similar in magnitude to
that in well-packed materials, see Ref.~\cite{GrimvallSjodin}. In
other words, the model liquid in Figs.~\ref{Sciortino}(c) and (d), if
equilibrated, should melt well before the percolation transition at
$\alpha \sim 1/a^2$ can be reached.  Consequently, the liquid is not
in the landscape regime on the low end of the investigated range of
$\alpha$. Conversely, given the apparent metastability of the
network-like states with $\alpha \sim 1/a^2$, one may infer from the
Hall and Wolynes model that such small values of $\alpha$ could be
still be reached, but only {\em off-equilibrium}, i.e., by preparing
the glass through rapid quenching or vapour deposition.  This notion
will be of use toward the end of the article.

Finally, given the complexity of the interactions in actual materials,
it is reasonable to ask whether one could evaluate the configurational
entropy semi-phenomenologically. This way, those interactions enter in
the form of the measured values of materials constants such as elastic
moduli and thermal expansion coefficient, etc. Bevzenko and
Lubchenko~\cite{BL_6Spin, BLelast} have, in part, accomplished this
program, to be discussed in Subsection~\ref{spinConnections}.

We reiterate that it is imperative to recognise that whether or not
the configurational entropy strictly vanishes at some low temperature
is not directly relevant to the translational symmetry breaking that
results from the random first order transition. In fact, it is a
common misconception that the existence of the entropy crisis at $T_K$
is the essential feature of the the RFOT. It is not. The only
prerequisite for the translational symmetry breaking at $T_A$ is that
condition~(\ref{aXtalCondition}) be satisfied. We note also that there
is a rigorous sense, in which the Kauzmann crisis can be defined for a
liquid that would not ordinarily have one in a {\em macroscopic}
sample. This crisis can be achieved when a {\em finite} sample runs
out of configurations, to be discussed in Section~\ref{quant}.

\subsection{Qualitative discussion of the transition at $T_A$ as a
  kinetic arrest, by way of mode-mode coupling. Connection between
  kinetic and thermodynamic views on the transition at $T_A$. Short
  discussion on colloids, binary and metallic mixtures, and ionic
  liquids.}
\label{MCT}

Studying the thermodynamic stability of aperiodic solids is certainly
not the only way to approach the problem of the glass transition.
Much effort toward understanding the dramatic slowing-down in
supercooled liquids, Fig.~\ref{angell}, was undertaken in the early
1980s starting from a purely kinetic prospective.  This work
culminated in the creation of the mode coupling
theory~\cite{Leutheuser, Gotze_MCT, GotzeSjogren} (MCT), which
predicts that under certain conditions, the traditional view of the
dynamics of a liquid as being that of a dense gas fails. This theory
builds on early theories of collisional transport~\cite{Hirschfelder}
and arrives at the conclusion that already in a uniform liquid, the
view of particle transport as a memory-less, Langevin process
eventually becomes internally-inconsistent as the density increases.
Given the difficulty of summing high-order terms in the mode-mode
coupling expansion~\cite{WuCao2005}, one has to resort to a mean-field
approximation. In this mean-field limit, one discovers that at a high
enough density, a particle's memory extends indefinitely implying that
it no longer moves about but, instead, forever vibrates around a
certain location in space.  This behaviour is often called ``caging.''
As the cages form, the viscosity diverges, leading to a kinetic
catastrophe. These ``cages'' turn out to correspond to the metastable
structures that emerge during the RFOT. Indeed, Kirkpatrick and
Wolynes~\cite{MCT} have shown that in the mean-field limit, the
kinetic catastrophe of the MCT theory and the RFOT theory are
equivalent. In this approach, the response of the liquid becomes
elastic while the equation that self-consistently specifies the force
constant $\alpha$ of the effective Einstein oscillator from
Eq.~(\ref{rho}) has the same structure as the DFT-based
Eq.~(\ref{KWalpha1}) for determining $\alpha$ self-consistently:
\begin{equation} \label{KWalpha3} \alpha = \frac{\rho}{6} \int
  \frac{d^D \bq}{(2\pi)^D} q^2 \tilde{c}^{(2)} \tilde{h}(q)
  e^{-q^2/2\alpha},
\end{equation}
where the function $h$ is defined in Eq.~(\ref{hr}). It turns out that
this equation identically coincides with Eq.~(\ref{KWalpha}) in the
mean-field limit $D \to \infty$. Indeed, $\widetilde{S}_q \to
\bar{\rho} \tilde{g}_q$ in this limit (since the $i=j$ contribution in
Eq.~(\ref{Sq}) becomes vanishingly small), and so $\bar{\rho}
\tilde{h}_q \to \widetilde{S}_q$ (at $q > 0$), by Eq.~(\ref{hr}). On
the other hand, it can be shown~\cite{MCT} that $\alpha \propto D^2$
as $D \to \infty$, c.f. Eq.~(\ref{KWalpha1}); thus $\widetilde{S}_0(q)
\to \widetilde{S}(q)$. As a result, the kinetic arrest of the MCT
exactly coincides with the inception of the spinodal in $F(\alpha)$,
in the mean-field limit. Furthermore, since the system is equilibrated
and must obey detailed balance, it must be true that the set of states
corresponding to the finite value of $\alpha$ at the spinodal are
thermodynamically stable. Otherwise, the system would not have been
arrested in a state characterised by a finite $\alpha$ but, instead,
would remain in the uniform state $\alpha = 0$. In other words, the
temperature $T_A$ of the RFOT transition strictly coincides with the
mean-field spinodal.  The above argument is somewhat subtle because in
the strictest mean-field limit, the system could, in principle, get
arrested even in a {\em metastable} free energy minimum because the
barrier for escape would be infinite. To make the argument work, we
must assume a finite $D$ at the onset, to make the escape barrier
finite and take advantage of detailed balance. At finite $D$, the
temperature of the spinodal may or may not strictly coincide with the
temperature $T_A$ at which the aperiodic-crystal state becomes
thermodynamically stable. In the $D \to \infty$ limit, however, the
two temperatures become equal. We shall observe a similar pattern for
mean-field spin models in the next Subsection.

In finite dimensions, the cages do not live forever and eventually
disintegrate via activated processes, as will be discussed in detail
in Section~\ref{ActTransport}. Reproducing the activated processes
formally within the the MCT framework appears to be difficult and has
not been accomplished to date, to the author's knowledge and
judgement~\cite{BBMR}. Adding activated processes on top of the MCT
effects can be done phenomenologically and results in good fits of
temperature dependences of relaxation times~\cite{PhysRevE.72.031509,
  BBW2008}. At any rate, the correspondence between the
thermodynamics-based RFOT treatment and the MCT survives in finite
dimensions in the sense that the sharp RFOT transition becomes a soft
crossover, while the kinetic catastrophe becomes a gradual rise in
viscosity.

Even in its simplest version, the MCT is rather formal in that it
discusses correlations in liquid in terms of coupling between
hydrodynamic modes, not molecular motions themselves; discussing the
technical details of the MCT would be beyond the scope of the
article. Extensive reviews of the MCT can be found
elsewhere~\cite{Goetze2008complex, Das_RMP}.  Here, instead, we
discuss the kinetic perspective on the RFOT transition in a
qualitative, phenomenological fashion, yet with relatively explicit
reference to molecular motions.

Suppose first that particle motions have no memory whatsoever. Under
these circumstances, we can connect the (thus frequency-independent)
self-diffusivity $D$ with the viscosity $\eta$ via the Stokes-Einstein
relation: $D = k_B T/(6 \pi (a/2) \eta)$, where we set the
hydrodynamic radius of the particle at one-half of its volumetric
size. By Einstein's formula, the typical travelled distance squared
goes with time $t$ as $\la r^2 \ra = 6 D t$. Thus the time is takes
for a particle to diffuse its own size is given by:
\begin{equation} \label{tauex} \tau_\text{ex} = \frac{\pi \eta a^3}{2
    k_B T}.
\end{equation}
Note that this size represents a secure upper bound on the time it
takes for two particles to exchange identities and thus locally
establish configurational equilibrium.

Now suppose that transient structures could, in principle, form in the
same system, whether as a result of a phase transition or not.
According to the phenomenological Maxwell relation (which can be
derived constructively, see Section~\ref{rheo}), the lifetime of the
metastable structures in such a liquid is intrinsically related to its
viscosity and elastic constants:
\begin{equation} \label{Maxwell1} \tau \simeq \frac{\eta}{K}
\end{equation}

The liquid may be regarded as uniform only insofar as we cannot
distinguish the particles by their location.  Consequently, when the
viscosity is so high that the time a particle ``sticks around''
becomes comparable to the lifetime of a transient structure from
Eq.~(\ref{Maxwell1}), we must conclude that the assumption of the
liquid's uniformity is internally-inconsistent and that transient
structures do, in fact, form.  This is the essence of the symmetry
breaking that takes place during the crossover from collisional to
activated transport, if expressed in kinetic terms. Equating the times
in Eqs.~(\ref{tauex}) and (\ref{Maxwell1}) yields $\eta=
\infty$. (There is no $\eta = 0$ solution because because the
viscosity of hard spheres is finite and concentration-independent in
the infinite dilution limit.)  The $\eta = \infty$ solution is
internally consistent and corresponds with the kinetic catastrophe of
the MCT, by which the liquid would be completely arrested, in the
mean-field limit.

If one attempts to equate the timescales from Eqs.~(\ref{tauex}) and
(\ref{Maxwell1}), with the aim to estimate the viscosity at which the
crossover would actually occur, one quickly discovers a potential
issue. These two equations, in combination with the Lindemann
criterion, $K a^3 \simeq \alpha a^2 k_B T \simeq 10^2 k_B T$, would
seem to indicate that the exchange time is {\em always} longer than
the structural relaxation time $\tau$, whenever the transient
structures could exist. (Some of this excess is intrinsic and some is
due to a likely overestimate in Eq.~(\ref{tauex}).) Yet there is not
necessarily a contradiction here. The condition that $\tau_\text{ex} >
\tau$ simply implies that the emergence of the solid is driven {\em
  thermodynamically}, not kinetically. The finite difference between
$\tau_\text{ex}$ and $\tau$ is inevitable because the kinetic arrest
is due to a solidification transition and is thus a {\em
  discontinuous} transition, as we saw in Section~\ref{Xtal} for the
regular liquid-to-crystal transition and in Subsection~\ref{RFOTDFT}
for the random-first order transition; this implies that the
viscosities in ``pure'' uniform liquid and aperiodic crystal should
differ by a finite amount. The viscosity of actual substances
interpolates between the values given in Eqs.~(\ref{tauex}) and
(\ref{Maxwell1}) and thus changes continuously through the crossover.
It seems worthwhile to recall that the finite (and large) value of the
inverse Lindemann length squared $\alpha$, at a liquid-to-solid
transition, is crucial in avoiding the critical point in
Fig.~\ref{ABphaseDiagram}. Likewise, the present, crude argument
suggests that a {\em continuous} kinetic arrest would also be avoided,
in equilibrium, owing to the finite value of $\alpha$ in a solid, be
it periodic or aperiodic.
 
Still, is there a simple way to estimate the viscosity at which the
crossover takes place based on the simplistic reasoning above? To do
so, we must use a measure of liquid's memory that would be more
appropriate at the crossover than that afforded by Eq.~(\ref{tauex}).
The latter equation clearly does not apply when nearby particles'
movements are tightly correlated, as in Fig.~\ref{flights}(b). In this
regime, the configurational and velocity equilibration processes are
{\em coupled} and so one expects the corresponding equilibration times
to be mutually tied. Indeed, the hopping time of vibrational packets
is tied to the particle hopping time in that the latter is an upper
bound for the former. On the other hand, the vibrational hopping time
is directly connected with the velocity equilibration.  To put this in
perspective, the configurational equilibration is much faster in
dilute gases, Fig.~\ref{flights}(a), and vice versa in very dense
liquids or solids, in which collisions are very frequent,
Fig.~\ref{flights}(c).  In both of these cases configurational and
velocity equilibration processes are decoupled. Velocities equilibrate
on the times comparable to auto-correlation time $\tau_\sauto$:
\begin{equation} \label{tauto} \tau_\sauto = \frac{m}{\zeta} \simeq
  \frac{a^2 \rho}{3 \pi \eta},
\end{equation}
where $m = \rho a^3$ is the particle mass. Thus, {\em near the
  crossover}, we may associate the duration of a particle's memory
with the time $\tau_\sauto$ multiplied by a dimensionless number $C$
that signifies how many collisions the particle must undergo before
the memory of its location is completely erased: $\tau_\text{memory}
\sim C \tau_\sauto$. Substitution of typical values of the density,
particle size and viscosity in Eq.~(\ref{tauto}) shows that
$\tau_\sauto$ is very short, shorter than molecular vibrations. (This
is expected as each molecule directly interacts with molecules from
several coordination shells.)  The velocity equilibration time should
be about two orders of magnitude greater than $\tau_\sauto$, while the
configurational equilibration is even slower.  Consequently, the
numerical factor $C$ is, very roughly, $10^3$ or greater.  We
reiterate that $\tau_\sauto$ reflects the configurational
equilibration only in a {\em narrow} density range; note the inverse
scaling of $\tau_\sauto$ with the viscosity, which is the opposite of
what is expected for the configurational equilibration
timescale. Equating $\tau_\text{memory}$ with $\tau$ from
Eq.~(\ref{Maxwell1}) yields the following estimate for the viscosity
at the crossover:
\begin{equation} \label{viscCrossover} \eta \sim K (a/c_s) \, C^{1/2},
\end{equation}
where $c_s$ is the speed of sound and so $a/c_s$ scales with but
exceeds a typical vibrational period of a molecule. Quantitative
criteria for crossover will be presented in Section~\ref{crossover},
where we shall see that at $T_\scr$, $\eta \simeq K \tau_\svibr \times
10^3$, where $\tau_\svibr$ stands for the vibrational relaxation time.

Although lacking the sophistication of the mode-coupling theory, the
qualitative discussion above on the kinetic aspects of the emergence
of the landscape makes one realise that in liquids made of
indistinguishable particles, particle collisions and configurational
equilibration are intrinsically connected: Each particle collides
exclusively with particles it must exchange places with for the liquid
to equilibrate. Conversely, the viscous drag on a given particle is
solely due to particles that are identical to that given particle.
Colloidal suspensions, among other systems, present a radically
different situation. The viscous drag on a colloidal particle is now
largely due to the solvent. Thus by varying the solvent or the
particle size, one can vary the timescale for configurational
equilibration at a fixed filling fraction. Alternatively said, the
bulk viscosity and structural relaxation are largely {\em decoupled}
in colloidal suspensions.  Owing to this decoupling, one can make
colloidal particles arbitrarily sluggish---by increasing their
size---without them ever entering the landscape regime.  To quantify
these notions, we note the dynamic range accessible to an ordinary,
molecular liquid between the melting point ($\tau \simeq 10^{-12}$)
and the crossover to the landscape regime ($\tau \simeq 10^{-9} \ldots
10^{-8}$) is about 3-4 order of magnitude.  By Eq.~(\ref{tauex}), the
exchange time $\tau_\text{ex}$ for micron-sized colloidal particles
exceeds that for ordinary liquid easily by ten orders of magnitude, at
the same value of the bulk viscosity.  This implies that on ordinary
laboratory timescales, such mesoscopic colloids only {\em begin} to
sample the landscape regime when their arrest arrest becomes
macroscopically apparent.

Likewise, one expects some decoupling between collisions and
structural relaxations in viscous metallic mixtures and heterodisperse
mixtures of model particles employed in simulational studies of slow
liquids. In such mixtures, the mole fractions of the ingredients are
comparable and so the degree of decoupling is not nearly as large as
in colloids. An alternative, thermodynamic way to see this is as
follows: One often employs eutectic mixtures of elements to
destabilise the crystal state, relative to the liquid, at the same
value of viscosity. By the same token, employing carefully chosen size
ratios or molar fractions in mixtures, one may lower the temperature
$T_A$, while not affecting the liquid's viscosity. Again, this will
lead to an increased dynamic range of the pre-landscape regime.  We
shall see additional indications in Subsection~\ref{TLS} that metallic
glasses freeze just below the crossover.

Room temperature ionic liquids represent an intermediate case between
colloidal suspensions and metallic mixtures. Like the metals, such
liquids are mixtures of molecules that do not have obvious crystalline
ground states. On the other hand, the decoupling between collisional
processes and structural relaxation is significantly greater than in
those mixtures. Because of the very long range of the Coulomb
interaction, each molecule directly collides with a very large number
of molecules, not just its nearest neighbours. Indeed, although the
intensity of an individual collision decays as $1/r$ with distance,
the number of molecules grows as $r^2$ with distance. (The author does
not have a simple argument to determine the collisional range.) We
thus conclude ionic liquids are at least as slow and are likely slower
than metallic mixtures and thus are further away from the landscape
regime, at a given value of viscosity. To avoid confusion we note that
the above logic does not apply to traditional ionic liquids such as
ZnCl$_2$, since these exhibit a great deal of covalent bonding.  Such
liquids are more appropriately though of as dipolar.

\subsection{Connection with spin models}
\label{spinConnections}

The formal status of the RFOT theory as of the late 80s--mid 90s was
somewhat uncertain for a number of reasons. Given the inherent
approximations of the density-functional theory, the mean-field
character of the DFT-MCT connection, and the technical complexity of
the MCT theory~\cite{Goetze2008complex}, many regarded the picture
advanced by the RFOT theory of the structural glass transition as
lacking in formal foundation or were unaware of the theory in the
first place.  The replica-based calculation of the configurational
entropy by M\'ezard and Parisi~\cite{mezard:1076} was not available
until 1999. Although constructive, that replica calculation, again, is
approximate and relies on the assumption that the mechanical stability
limit of the uniform liquid is not reached above the (putative)
temperature $T_K$.  On the other hand, calorimetry data for {\em
  actual} supercooled liquids yield the magnitude of the
configurational entropy requisite for the stability of the aperiodic
crystal.  Perhaps the most forceful proof of the activated nature of
the liquid transport near the glass transition is that the rate of the
transport is Arrhenius-like below the glass transition, as emphasised
in Ref.~\cite{LW_Wiley}. Thus, although the premise of the RFOT theory
is clearly consistent with observation---and so are its many
predictions!---some of the aspects of the theory itself could be
perceived as having been established only phenomenologically.  To fill
this perceived formal gap, much technical effort has been directed at
finding solvable models that exhibit features reflecting those of the
structural glass transition.

By the late 1980s, a very interesting disordered model having a
glass-like transition had already been worked out, viz., the
mean-field Sherrington-Kirkpatrick (SK)
model~\cite{PhysRevLett.35.1792, SpinGlassBeyond}. This model exhibits
a proliferation of metastable minima below a certain temperature,
similar to what happens during the RFOT.  The model is an Ising magnet
from Eq.~(\ref{EIsing}), but with randomly distributed couplings
$J_{ij}$.  The distribution's mean is zero. In the mean-field limit,
the couplings should scale not as $1/N$, but $1/\sqrt{N}$, in contrast
with the Ising ferromagnet. Most commonly, the distribution is assumed
to be Gaussian:
\begin{equation} \label{pJ} p(J_{ij}) = \frac{1}{\sqrt{2 \pi J^2/N}}
  e^{-\frac{J_{ij}^2}{2 J^2/N}}
\end{equation}
In an interesting contrast with the regular Ising magnet, we must
retain the second order term in the expression for the molecular field
even in the strict mean-field limit:
\begin{equation} \label{hmolSK} h^\text{mol}_i = \sum_j J_{ij} m_j -
  m_i \sum_j \beta J_{ij}^2 (1 - m_j^2),
\end{equation}
c.f. Eq.~(\ref{hmolIsing}). The reader will recognise $\beta
(1-m_j^2)$ at the susceptibility of a spin with magnetisation $m_j$
discussed following Eq.~(\ref{hmolIsing}). Thus the second order term
on the r.h.s.  gives the effective field spin $i$ exerts on itself
through interactions with the rest of the spins, consistent with this
term being proportional to $m_i$ itself. This term corresponds to the
Onsager cavity field in electrodynamics~\cite{Onsager1936} and turns
out to make a finite contribution to the molecular field in disordered
magnets even in the mean-field limit; this contribution is in fact
numerically equal to the first term at the glass transition and thus
must be retained in the free energy expansion~\cite{SpinGlassBeyond,
  PWA_LesHouches}.

Combined with Eq.~(\ref{himol}), which connects the molecular field
with local magnetisation, Eq.~(\ref{hmolSK}), forms a closed system of
equations, often called the Thouless-Anderson-Palmer~\cite{TAP} (TAP)
equations that can be used, in principle, to determine the equilibrium
magnetisation in the SK magnet. The corresponding free energy reads:
\begin{eqnarray} \label{FSK} F(\{ m_i \}) &=& k_B T \sum_i \left(
    \frac{1+m_i}{2}\ln\frac{1+m_i}{2} +
    \frac{1-m_i}{2}\ln\frac{1-m_i}{2} \right) \nonumber \\ &-&
  \sum_{i<j} J_{ij} m_i m_j - \frac{1}{2} \sum_{ij} \beta J_{ij}^2 (1
  - m_i^2)(1 - m_j^2),
\end{eqnarray}
c.f. Eqs.~(\ref{IsingF}), (\ref{IsingID}), and (\ref{IsingEX}).

The mean-field system freezes into a ``spin-glass'' state below a
certain, sharply defined temperature, whereby the number of possible
states to freeze into scales with the system size, in contrast, for
instance, with the ordinary Ising magnet, in which the number of such
distinct states is just two. All free energy minima are automatically
aperiodic---owing to the disorder in the couplings---similarly to the
liquid below the RFOT. Yet the ergodicity-breaking transition in the
SK model is {\em continuous}, in contrast with the liquid.  The
continuity of the transition is consistent with the discrete symmetry
of the time-inversion symmetry of the SK hamiltonian, $\{ \sigma_i \}
\leftrightarrow \{- \sigma_i \}$, by which the cubic term and the rest
of the odd terms in the Landau-Ginzburg expansion are expressly
forbidden by symmetry.

An equally important piece of distinction between the SK model and the
liquid case is that in the former, the disorder is built-in, or {\em
  quenched}. The multiplicity of the solution is due to the vast
number of configurations available to a magnet with the disordered
couplings from Eq.~(\ref{pJ}) and a lack of a unique stable
state. This type of frustration can be seen already with an Ising {\em
  anti}ferromagnet on the 2D triangular lattice: $E = - J (\sigma_1
\sigma_2 + \sigma_1 \sigma_3 + \sigma_2 \sigma_3)$ with $J < 0$. The
ground state of this system is six-fold degenerate ($E = J$) and the
excited state is two-fold degenerate ($E = - 3J$).  In contrast with
the SK model, there is no built-in disorder in liquids. Instead, the
interactions between molecules are perfectly translationally
invariant. The disorder in structural glass is thus entirely {\em
  self-generated}. Although the latter fact is often cited as a most
enigmatic feature of the structural glass transition, it is not hard
to see how frustration could, in principle, arise in 3D liquids. For
instance, the Voronoi cell corresponding to the locally-densest
packing in 3D is the regular dodecahedron, which has a 5-fold symmetry
axis and thus does not tile space.

It turns out that the solutions of the free energy (\ref{FSK}) have an
incredibly rich structure. The free energy minima are organised into a
multi-tiered hierarchy, by which distinct solutions can be classified
according to the degree of mutual similarity. It is convenient to
think of these solutions as distinct replicas in the Gibbs ensemble
that happened to freeze into distinct free energy minima, below the
glass transition, the same way extensive portions of an Ising
ferromagnet could polarise up or down below the Curie point. For the
mean-field SK model, the similarity of distinct solutions---or replica
overlap---is continuously distributed. This multi-tiered ergodicity
breaking was elucidated by Parisi~\cite{SpinGlassBeyond}, among
others, in the early 1980s and is often called {\em full} replica
symmetry breaking (RSB).

Already in their 1985 paper, Singh, Stoessel, and
Wolynes~\cite{dens_F1} point out that the free energy landscape of an
aperiodic crystal can bear similarities to the Parisi solution of the
SK model, as the distinct replicas are also aperiodic upon ergodicity
breaking. Still, a meaningful connection between the two systems, if
any, was difficult to establish given their distinct symmetries, even
setting aside the lack of quenched disorder in liquids.

Spin models that afford such a connection must not have the
time-inversion symmetry of the Ising model. One such model is the so
called $p$-spin model with an odd $p$:
\begin{equation} \label{pspin} E = - \sum_{i_1 < i_2 < \ldots < i_p}
  J_{i_1 i_2 \ldots i_p} \sigma_{i_1} \sigma_{i_2} \ldots \sigma_{i_p}, 
\end{equation}
where the couplings are distributed according to
\begin{equation} \label{pJpspin} p(J_{i_1 i_2 \ldots i_p} ) =
  \frac{1}{\sqrt{p! \pi J^2/N}} e^{-\frac{J_{i_1 i_2 \ldots i_p}^2}{p!
      J^2/N}}.
\end{equation}
This model exhibits a one-stage replica symmetry breaking (RSB), for
$p>2$~\cite{MezardMontanariBook2009}.  (In the limit $p \to \infty$,
the $p$-spin model reduces to the so called random energy model
(REM)~\cite{GrossMezard, Derrida} while the calculations simplify
significantly.) That all the replica overlaps are the same implies
that all free energy minima are equivalent.

In contrast with the SK model (which corresponds to $p=2$), the
ergodicity breaking in the $p$-spin model is {\em discontinuous} for
$p > 2$~\cite{GrossMezard}.  Motivated by this notion and the
mean-field equivalence between the kinetic catastrophe of the MCT and
the RFOT~\cite{MCT} discussed in Subsection~\ref{MCT}, Kirkpatrick and
Thirumalai~\cite{KT_PRL87, KT_PRB87} have analysed, upon Wolynes's
suggestion, a dynamic version of the $p$-spin model. One way to impart
the model with dynamics is to treat the length of an individual spin
as a particle with a finite mass, whose motion is subject to a soft,
bistable potential. Aside from some potential uncertainty stemming
from approximations, these workers discovered that the dynamic
$p$-spin model undergoes, in mean-field, a kinetic catastrophe at a
temperature $T_g$ {\em above} the temperature $T_g'$ at which the
replica symmetry is broken.

In 1987, Kirkpatrick and Wolynes~\cite{MCT1} (KW) definitively
rationalised those intriguing but enigmatic findings by obtaining a
complete solution of the mean-field version of the so called Potts
glass model, using a state-counting strategy employed by Thouless,
Anderson, and Palmer~\cite{TAP}. A formally exact calculation of the
free energy of the mean-field Potts glass had been furnished two years
prior, by Gross, Kanter, and Sompolinsky~\cite{GrossKanterSomp}.  The
Potts model is similar to the SK model, but generally does not exhibit
the time-inversion symmetry and may exhibit a discontinuous
ergodicity-breaking transition under certain circumstances, similarly
to the $p$-spin model. In its most basic form~\cite{Baxter}, the Potts
model is a generalisation of the Ising model to an energy function of
the form $E = - \sum_{i < j} J_{ij} \sigma(s_i, s_j)$, where the
(generally vectorial) spin variables $s_i$ are allowed to have a
chosen, discrete set of values. The test-function $\sigma(s_i, s_j) =
\delta_{s_i, s_j}$ singles out only the configurations in which the
spin variables on sites $i$ and $j$ have the same value. More
generally, one may use a test function that assigns a non-zero weight
to $s_i \ne s_j$ configurations also; this weight, however, must be
different from that of the identical configuration, by definition. A
common test function of this kind is simply the scalar product of the
vectors, as would be natural in a Heisenberg model: $\sigma(s_i, s_j)
= s_i s_j$, however in contrast with Heisenberg-like models, the spins
are not continuous but are allowed to point only in specified
directions. Specifically one often dictates that the spins point
toward the vertices of a $p$-cornered hypertetrahedron in a flat
$(p-1)$-dimensional space~\cite{0305-4470-8-9-019}. Here we set the
length of each vector at $\sqrt{p-1}$, by definition. For $p=2$, we
recover the usual Ising spins; the corresponding test function $s_i
s_j$ is equal to either $1$ ($s_i= s_j$) or $-1$ ($s_i \ne s_j$).
When $p=3$, the spins point toward the corners of an equilateral
triangle with the side $\sqrt{6}$. The test function can have the
value $2$ ($s_i= s_j$) or $-1$ ($s_i \ne s_j$). For $p=4$, the spins
could point toward the corners of the regular 3D tetrahedron, which
implies four distinct orientations, and so on.  KW considered exactly
this version of the Potts model (for an arbitrary value of $p$):
\begin{equation} \label{HPotts} E = - \sum_{i < j} J_{ij} \, s_i s_j,
\end{equation}
where the couplings $J_{ij}$ are random and obey the distribution
(\ref{pJ}).

For $p>4$, the mean-field Potts glass exhibits a discontinuous
transition at a temperature $T_A$, in which an {\em exponential}
number of solutions emerge, each solution corresponding to a free
energy minimum. The transition exhibits itself as a spinodal at a
finite value of the replica-overlap order parameter, call it $q$. The
free energy just below the spinodal, as a function of $q$, looks like
the $F(\alpha)$ curve in Fig.~\ref{SSW}(a) corresponding to $\rho
\sigma = 1.05$.  It turns out that each individual solution is higher
in free energy than the symmetric phase exactly by $T S_c$, where
$S_c$ is the log-number of those solutions times $k_B$. Despite the
spectacular symmetry breaking at $T_A$, the {\em full} free energy or
entropy of the mean-field Potts glass does not experience a
singularity of any sort. In further contrast with the SK model, the
replica-symmetry breaking at $T_A$ is {\em one}-stage, not
infinite-stage.

Furthermore, the configurational entropy $S_c$ is found to vanish at a
finite temperature $T_K < T_A$. Note that in the mean-field model from
Eq.~(\ref{HPotts}), the barriers separating the distinct minima
emerging below $T_A$ are strictly infinite. This means that the system
is already completely arrested at a temperature $T_A > T_K$ even as
its entropy is finite!  KW~\cite{MCT1} pointed out that in finite
dimensions, however, the barriers separating the distinct free energy
would be, in fact, finite and so activated transitions between the
minima would be allowed. Accounting for spatial fluctuations already
in the long-wavelength approximation, dictates that there be
correlation lengths diverging both at $T_A$, owing to critical-like
fluctuations at the spinodal, and diverging at $T_K$, to be discussed
in detail in Section~\ref{ActTransport}. The infinite-range
correlations at $T_A$ would be destroyed by the aforementioned
activated transitions.  Complemented by these notions, the exact
solution of the mean-field Potts glass model was important in that it
showed that the RFOT-advanced picture of the glass transition as a
kinetic arrest into a {\em degenerate} aperiodic crystal is not unique
from a purely formal standpoint. As an added bonus, the Potts glass
also exhibits a Kauzmann state.

Because of these apparent similarities between the thermodynamics and
kinetics of glassy liquids and the disordered Potts model, it is still
very common to see statements in the literature that the RFOT picture
of the structural glass transition is simply an analogy with the
(mean-field) Potts glass. It is not. As we have already remarked, the
emergence of the aperiodic solid in the form of the RFOT has been
shown independently and, in fact, prior to the solution of the
mean-field Potts glass model.

More recently, the applicability of mean-field Potts models to liquids
has been questioned~\cite{0034-4885-77-4-042501}. Additionally, some
renormalisation-group analysis of $p$-spin-like models that exhibit
both the kinetic and thermodynamic catastrophes suggests, with the aid
of a perturbative argument, that the Kauzmann crisis disappears in
finite dimensions and so do the corresponding long-range
correlations~\cite{MooreNOdiverge, PhysRevE.86.052501,
  PhysRevB.85.100405}. On the other hand, two recent works have shown
that finite-dimensional Potts-like models could in fact exhibit the
RFOT, given sufficient frustration~\cite{2014arXiv1407.7393C,
  2014arXiv1408.1495T}.  These notions are, perhaps, not surprising in
light of an earlier discussion by Eastwood and
Wolynes~\cite{EastwoodW} who argued that spin systems are ``softer''
than liquids, in the following sense: The surface tension between
distinct free energy minima is significantly lower in the former than
in the latter.  This softness is expected to result in a significant
rounding of the RFOT in finite-dimensional (!) spin systems.

We next mention a separate set of works that directly map the dynamics
of glassy liquids onto those of spin or spine-like systems. The
earliest such work, to the author's knowledge, is due to Stevenson et
al.~\cite{stevenson:194505}, who have used a combination of the
replica methodology~\cite{PhysRevLett.75.2847} and classical
density-functional theory to map liquid dynamics onto an Ising model
with randomly distributed couplings {\em and} on-site fields. Such
random on-site fields will turn out to be of special significance in
the next Section.

\begin{figure}[t]
  \begin{tabular*}{\figurewidth} {ccc}
    \begin{minipage}{.25 \figurewidth} 
      \begin{center}
        \includegraphics[width= .25 \figurewidth]{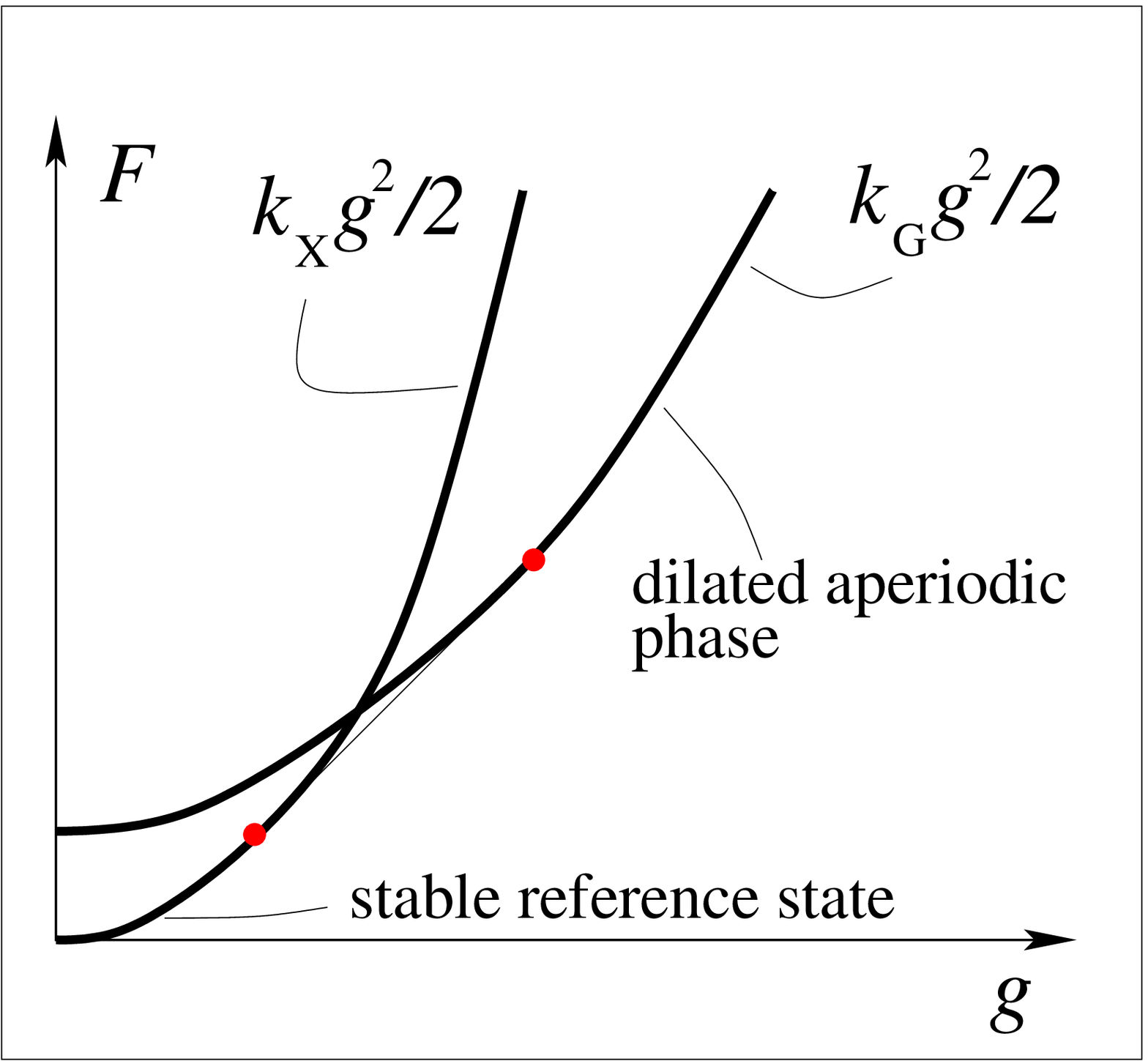}
        \vspace{2mm} 
        \\ {\bf (a)}
      \end{center}
    \end{minipage}    
    &
    \begin{minipage}{.3 \figurewidth} 
      \begin{center}
        \includegraphics[width=0.3 \figurewidth]{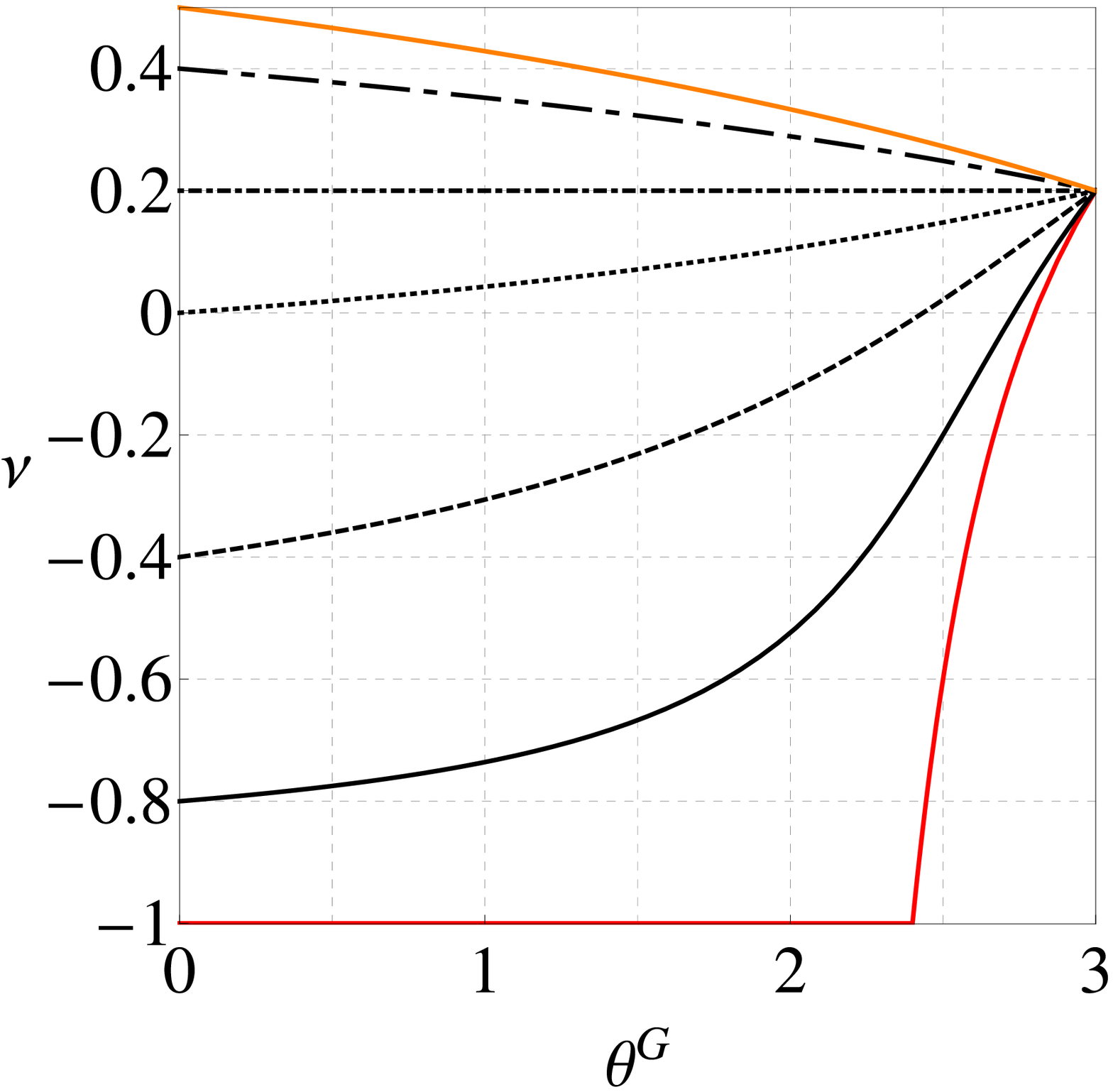}
        \\ {\bf (b)} 
      \end{center}
    \end{minipage}
    &
    \begin{minipage}{.32 \figurewidth} 
      \begin{center}
        \includegraphics[width=0.35 \figurewidth]{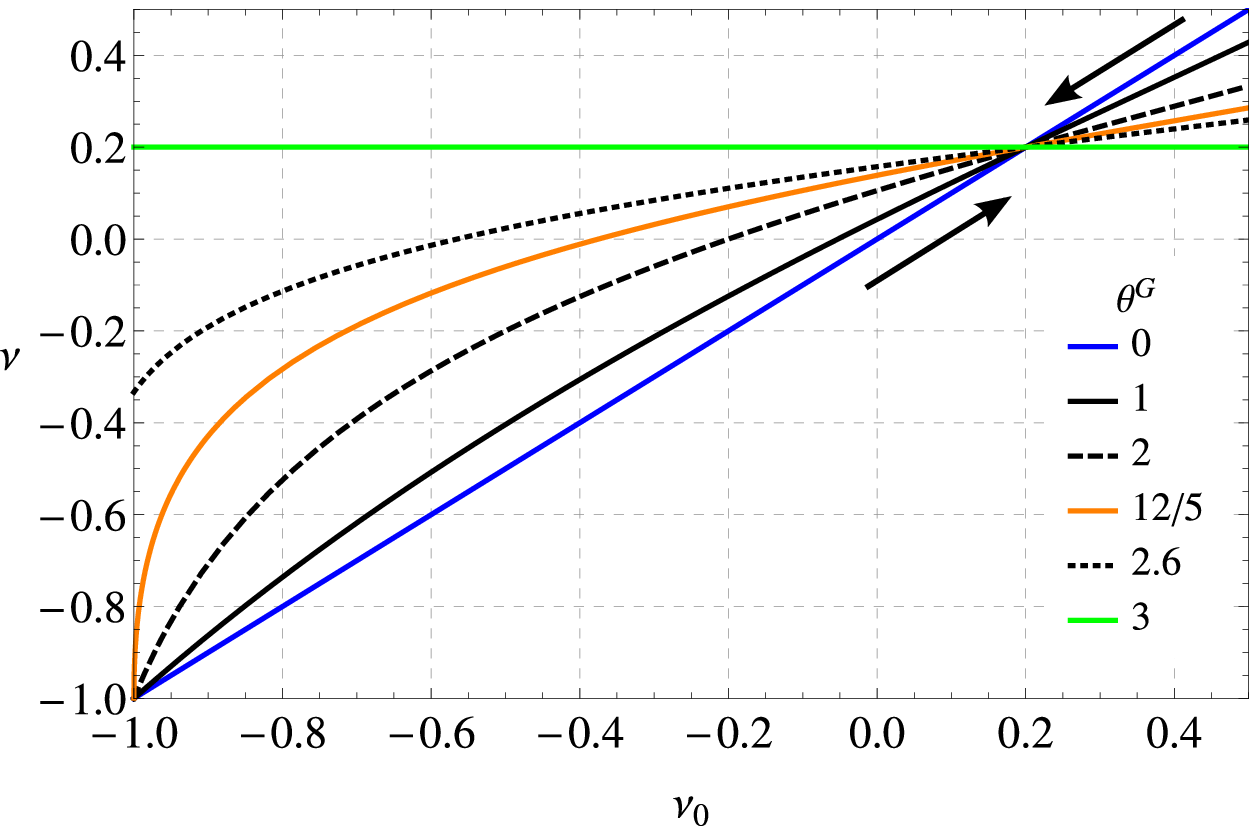}
        \vspace{3mm}
        \\ {\bf (c)} 
      \end{center}
    \end{minipage}
  \end{tabular*}
  \caption{\label{BLfigure} {\bf (a)} The two parabolas correspond to
    the free energy of a reference, mechanically stable state (``X'')
    and a metastable, aperiodic state (``G''). The tangent
    construction demonstrates the excess free energy at the contact
    between the two states, after Ref.~\cite{BL_6Spin}. {\bf (b)} The
    renormalisation of the Poisson ratio $\nu$ in an equilibrium,
    aperiodic solid, as a function of the built-in stress
    $\theta^G$~\cite{BLelast}. {\bf (c)} Renormalised value of the
    Poisson ratio $\nu$ plotted versus its bare value $\nu_0$, for
    several values of the built-in stress $\theta^G$ (expressed in
    terms of temperature). Note the attractive fixed point at $\nu =
    1/5$. The discontinuities at $\nu = 1/2$ ($\mu = 0$) and $\nu =
    -1$ ($K = 0$) correspond to the discontinuous transitions
    (uniform-liquid $\leftrightarrow$ aperiodic-solid) and
    (stable-solid $\leftrightarrow$ aperiodic-solid)
    respectively~\cite{BLelast}.}
\end{figure}

Bevzenko and Lubchenko~\cite{BL_6Spin, BLelast} (BL) have arrived at a
spin-like description of glassy liquids from a very different, purely
elastic perspective. In this view, a glassy liquid---which is an
equilibrated, degenerate aperiodic crystal---is obtained by expanding
a crystal, then letting the atoms settle into one of the many
aperiodic arrangements, and then letting the volume relax. According
to a heuristic construction in Fig.~\ref{BLfigure}(a), the free
energies of the periodic crystal and the degenerate aperiodic crystal
correspond to two distinct terms, as functions of deformation. The two
terms must cross since the glass term starts at a higher value of the
free energy, but has a lower curvature because the bonding in the
aperiodic sample is softer. This construction explicitly demonstrates
there is a mismatch penalty between the crystal and glass, which
corresponds to the excess free energy above the common
tangent~\cite{RowlinsonWidom, Bray}. To make the glassy state
metastable, the aperiodic crystal must be sufficiently stabilised
entropically by the multiplicity of distinct aperiodic arrangements
and by steric repulsion between ill-fitting molecular fragments.  The
built-in stress pattern resulting from this repulsion is of shorter
wavelength than the regular elastic waves. Thus in the full free
energy of an aperiodic crystal, $\int dV u_{ij} \Lambda_{ijkl}
u_{kl}/2$, the full displacement $u_{ij}$ can be profitably presented
as a sum of a short-wavelength contribution $u_{ij}^>$ and
long-wavelength contribution $u_{ij}^<$. The former and latter take
care of the built-in and regular elastic stress, respectively. The the
elastic-moduli tensor $\Lambda_{ijkl}$ looks particularly simple for
isotropic elasticity~\cite{LLelast}: $\Lambda_{ijkl} = (K - 2\mu/3)
\delta_{ij} \delta_{kl} + 2 \mu (\delta_{ik} \delta_{jl} + \delta_{il}
\delta_{jk} ) $, c.~f. Eq.~(\ref{Fuu}). The model is made non-linear
by fixing the magnitude of the local built-in stress; three specific
ways to do so are described in Ref.~\cite{BLelast}.

The component of the deformation corresponding to the built-in stress
violates the so-called Saint-Venant compatibility condition, whereby a
certain combination of the derivatives of the strain (or stress)
tensor, called the ``incompatibility'', is non-zero:
\begin{equation} \label{eq:SaintVenant} \text{inc}(u^>)_{ij} \equiv
  -\epsilon_{ikl} \epsilon_{jmn} \prtl u^>_{ln}/\prtl x_k \prtl x_m
  \ne 0.
\end{equation}
Setting the incompatibility to zero in the ordinary elasticity theory
allows one to obtain the displacement field $\bm{u}$ from the tensor
of the derivatives $u_{ij}$ from Eq.~(\ref{uij}) unambiguously; this
way the result is independent of the integration
contour~\cite{Teodosiu1982, lovett91}. In physical terms the
Saint-Venant compatibility condition guarantees, among other
possibilities, that no bonds are broken. To illustrate this point with
a more familiar analogy, one imposes the rotor-free constraint on the
electric field in electrodynamics: $\bm{\nabla}\times\bm{E} = 0$ so
that the electric field can be expressed as the gradient of a
single-valued, scalar field, viz., the electrostatic potential. In
turn, this implies the energy of an electric charge subject to
electric field is a well defined, single-valued function of the
coordinate.  Note that the existence of a unique reference state in
continuum mechanics is analogous to stipulating that vacuum be unique
in electrodynamics; it allows one to unambiguously match the particles
in the ground and any excited vibrational state.  In contrast, a
glassy liquid is a spatial superposition of many distinct vacua, which
thus invalidates the premise of the ordinary elasticity
theory. BL~\cite{BLelast} have shown that as long as the degenerate
solid is equilibrated, finite-frequency elastic moduli can still be
defined and, in fact, are renormalised versions of the elastic
constants in the individual vibrational ``vacua,'' as is briefly
described below.

Integrating out the regular vibrations, $u^<$ in the partition
function $\text{Tr} \exp(-\beta \int dV u_{ij} \Lambda_{ijkl}
u_{kl}/2)$, leads to a Hamiltonian of the type $E = - \sum_{i < j} s_i
J_{ij} s_j$, c.~f.  Eq.~(\ref{HPotts}), where the variables $s_i$ are
now {\em six}-component vectors that are allowed to point in any
direction and $J_{ij}$ is a six-by-six matrix. The number six stems
from the number of independent entries in the elastic strain (stress)
tensor. (The variables $u^>_{ij}$ and $s$ are linearly related.) The
coupling $J_{ij}$ is rather complicated and spatially anisotropic, see
the explicit expression in Ref.~\cite{BLelast}, but scales with the
distance and elastic constants as $1/\mu(1-\nu)r^3$ ($\nu$ is the
Poisson ratio).  In the BL approach, there is no built-in
disorder. The frustration comes about exclusively because of the
anisotropy of the interaction between local structural
excitations. This anisotropy is similar to, but more complicated than
that in the electric dipole-dipole interaction.

In the strict mean-field limit~\cite{BL_6Spin}, the BL model yields a
second order transition between two regimes corresponding to two types
of frozen-in stress, namely, mostly shear and uniform
dilation/contraction respectively. The transition takes place at a
specific value of the Poisson ratio $\nu = 1/5$. BL argued the two
regimes typically correspond to strong and fragile liquids
respectively. At the next level of approximation~\cite{BLelast}, the
Onsager cavity term was included, as in the last term in
Eqs.~(\ref{hmolSK}) and (\ref{FSK}). This results in much richer
thermodynamics compared with the strictest mean-field limit. In
addition, the effective (finite-frequency) elastic constants of the
degenerate aperiodic solid can be determined self-consistently,
similarly to how Onsager determined the dielectric constant of a polar
liquid based on the molecular dipole moment. The role of the dipole
moment is played here by the magnitude of the built-in stress, which
determines the length of the 6-spin.  The transition at $\nu = 1/5$
now becomes an {\em attractive} fixed point, see
Fig.~\ref{BLfigure}(b) and (c). In addition, one recovers two more
fixed points: one at the uniform liquid ($\mu/K = 0$) and the other at
the infinitely compressible liquid ($K = 0$). The former transition is
self-consistently determined to be first order, consistent with the
RFOT theory. The transition at $K = 0$---which corresponds to an
infinite compressibility---was speculated to correspond to a
mechanical instability that occurs during pressure-induced
amorphisation.

Recently, Yan, D\"uring, and Wyart~\cite{Yan16042013} have put forth
an elasticity-based model for glassy-liquid dynamics in which local
groups of particles are postulated to possess multiple alternative
states; the associated particle motions are coupled via elastic force
fields.

\section{Quantitative Theory of Activated transport in Glassy Liquids}
\label{ActTransport}

We begin by remarking that the mean-field limit---which is often a
good starting point for analysing phase transitions---could be
somewhat confusing in the case of the random first order
transition. On the one hand, the aperiodic crystal state becomes
thermodynamically stable partially owing to its degeneracy, which in
itself is somewhat counter-intuitive since ordinarily, transitions
driven by lowering temperature are accompanied by a {\em decrease} in
the (total) entropy.  (No discontinuity in the total entropy occurs at
the RFOT transition.)  In any event, because the multiplicity of the
aperiodic states is a key part of the stabilisation, it is essential
that the system be able to sample all of those aperiodic structures
without exception. Yet, there are no transitions between the
alternative aperiodic minima in the mean-field limit.  This seeming
paradox is resolved by noting that the transitions between the
distinct aperiodic are {\em not} subject to infinite barriers in
finite dimensions. As already mentioned, Kirkpatrick and
Wolynes\cite{MCT1} discussed a mechanism by which individual aperiodic
minima would locally interconvert by means of activated transitions
already in 1987.  The driving force for the transitions to occur is
{\em precisely} the multiplicity of the distinct aperiodic minima.
According to this early argument, the barriers for the activated
transport would scale inversely with the square of the configurational
entropy. While the experimental viscosity curves could be fitted with
such a dependence, the inverse {\em linear} scaling matches the
empirical Vogel-Fulcher-Tammann law (\ref{VFT}) better. In the
following, we discuss $\alpha$-relaxation in great detail.

Throughout this Section, we will assume the liquid is already in the
landscape regime, i.e., below the temperature $T_\scr$ of the
crossover between mainly collisional and activated transport, so that
the locally-stable aperiodic structures live significantly longer than
the vibrational relaxation time. In this limit, counting the
metastable free energy minima, each of which corresponds to individual
mechanically-metastable structures, becomes unambiguous. Consequently,
the configurational entropy, too, is well defined, while the
structural relaxation can be quantitatively described as rare,
activated processes, with the help of the transition state
theory~\cite{FW, Hanggi_RMP}.

\subsection{Glassy liquid as a mosaic of entropic droplets}
\label{mosaic}

We are used to systems whose free energy surface is essentially
independent of the system size: For instance, the free energy surface
of a macroscopic Ising ferromagnet below its Curie point has two
distinct free energy minima that can be distinguished by their average
magnetisation, which can be taken to be up or down respectively. If
the system is made thrice bigger, the free energy is simply multiplied
by a factor of three; that is, while the number of states within each
minimum increases exponentially, the number of minima themselves
remains the same, i.e., two.  In contrast, the number of free energy
minima in a glassy liquid scales exponentially with the system size
$N$: $e^{s_c N}$, as already mentioned. Under these circumstances, the
system will break up into separate, contiguous regions that are
relatively stabilised; the contiguous regions are separated by
relatively strained interfaces characterised by a higher free energy
density. To see this, suppose the opposite were true and the free
energy density were uniform throughout. Owing to the multiplicity of
distinct free energy minima, the nucleation rate for another
relatively stabilised configuration is finite, as we will see
shortly. As a result, the original configuration will be {\em locally}
replaced by another configuration while the boundary of the replaced
region will be relatively strained because of a mismatch between the
new structure and its environment.  Thus in equilibrium, there is a
steady-state concentration of the strained regions; local
reconfigurations take place at a steady rate between distinct
aperiodic structures.  The concentration of the strained regions and
the escape rate from the current liquid configuration can be
determined self-consistently, as discussed by Kirkpatrick, Thirumalai,
and Wolynes~\cite{KTW}, Xia and Wolynes~\cite{XW}, and Lubchenko and
Wolynes~\cite{LW_aging}.

First we provide a somewhat more explicit, ``microcanonical'' version
of that argument, as detailed recently in Ref.~\cite{LRactivated}. In
a standard fashion, thermodynamic quantities are Gaussianly
distributed in a sufficiently large system.  Thus the partition
function of a thermodynamic system in contact with a thermal bath at
temperature $T \equiv 1/k_B \beta$ and pressure $p$ can be expressed
as a Gaussian integral over the fluctuating value of the Gibbs free
energy $G$:
\begin{equation} \label{Z3} Z = \int \frac{dG}{\sqrt{ 2 \pi \delta G^2
    }} e^{-\beta \overline{G}} \, e^{-(G - \overline{G})^2/2 \delta
    G^2}.
\end{equation}
Here $\overline{G}$ is the most probable value of the Gibbs free
energy and $\delta G = \la (G - \overline{G})^2 \ra^{1/2}$ is the
corresponding standard deviation. It is easy to
show~\cite{LRactivated} that for a region containing $N$ particles,
\begin{equation} \label{dG} \delta G = N^{1/2} \left[ \frac{k_B T
      K}{\bar{\rho}} + (K \alpha_t - \tilde{s})^2 \frac{k_B
      T^2}{\bar{\rho} \tilde{c}_v} \right]^{1/2},
\end{equation}
where $K \equiv -V (\prtl p/\prtl V)_T$ is the bulk modulus, and
$\alpha_t \equiv (1/V) (\prtl V/\prtl T)_p$ the thermal expansion
coefficient. The quantities $\tilde{c}_v$ and $\tilde{s}$ are,
respectively, the heat capacity at constant volume and entropy, both
per unit volume. The average particle density $\bar{\rho}$ can be
rewritten in terms of the volumetric particle size $a$:
\begin{equation} \label{rhoa}
  \bar{\rho} \equiv 1/a^3.
\end{equation}
There is some freedom in choosing the identity and/or size of the
effective particle of the theory. Often one chooses actual atoms as
effective particles so that the particle-particle interaction can be
directly estimated using quantum-chemistry calculations.  It is also
common to use a coarse-grained description, in which the particle
contains several atoms or even a non-integer number of atoms. For
instance, when treating a molecular substance, it is often most
convenient to use computed from scratch or suitably parametrised
potentials, the most common example of such a parametrised interaction
is the Lennard-Jones potential. (Incidentally, evaluating such
intermolecular interactions {\em ab initio} is not necessarily an easy
task even for small molecules.) Here we assign the effective particle
of the theory as the bead defined in Eq.~(\ref{Nb}). This way, we
shall be able to take advantage of several quantitative predictions of
the DFT theory, as discussed in Subsection~\ref{sc}.

Let us now consider a liquid below the crossover to activated
transport but above the glass transition; the liquid is thus
equilibrated. Below the crossover, the reconfigurations are rare
events compared with vibrational motions, which amounts to a
well-developed time scale separation between net translations and
local vibrations. This time-scale separation takes place in ordinary
liquids at viscosities of order $10$~Ps.~\cite{LW_soft, LW_Wiley}
Because of it, the entropy of the liquid can be written as a sum of
distinct contributions:
\begin{eqnarray} \overline{G} &=& \overline{H}_i - T
  \overline{S}_{\svibr, \, i} - T S_c \\ \label{gequil} &\equiv&
  \overline{G}_i - T S_c(\overline{G}_i), 
\end{eqnarray}
where $H_i$ is the enthalpy of an individual aperiodic state, while
the total entropy is presented here as the sum of the vibrational and
configurational contributions, the configurational contribution taking
care of particle translations. The subscript ``$i$'' refers to
``individual'' metastable aperiodic states. The quantity
$\overline{G}_i \equiv \overline{H}_i - T \overline{S}_{\svibr, \, i}$
would be the Gibbs free energy of the sample if the particles were not
allowed to reconfigure but were allowed to vibrate only.  Despite an
entropic contribution due to the vibrations the quantity $G_i$ is an
enthalpy-like quantity as far as the configurational equilibration is
concerned. Although superficially similar in structure to the standard
canonical free energy, the ``microstates'' characterised by the
``enthalpy'' $G_i$ are much different from the microstates from the
canonical ensemble in that the configurational entropy numbers {\em
  long-lived} states. In contrast, the conversion between microstates
in the conventional canonical ensemble is assumed to be faster than
any meaningful observation time.

It is interesting that the free energy of liquid was written in the
form similar to Eq.~(\ref{gequil}) already in 1937 by
Bernal,~\cite{Bernal1937} who apparently assumed that molecular
vibrations and translations were distinct motions at {\em any}
temperature, even though this notion is well-justified only below the
crossover. Bernal used his formulation of the liquid entropy to argue
that liquid-to-crystal transitions are always discontinuous, owing to
the non-zero configurational component of the liquid entropy, which,
then, effects a non-zero latent heat.

Of direct interest is the distribution not of the full free energy $G$
but that of the free energies $G_i$ of individual metastable states:
\begin{equation} \label{Z3a} Z = \int \frac{dG_i}{\sqrt{ 2 \pi \delta
      G_i^2 }} e^{S_c(\overline{G}_i)/k_B-\beta \overline{G}_i} \,
  e^{-(G_i - \overline{G}_i)^2/2 \delta G_i^2}.
\end{equation}
The corresponding width of the distribution, $\delta G _i \equiv \la
(G_i - \overline{G}_i)^2 \ra^{1/2}$, can be evaluated similarly to
$\delta G$ from Eq.~(\ref{dG})~\cite{LRactivated}:
\begin{equation} \label{dGi} \delta G_i = N^{1/2} \left\{ \left[ K - T
      \left(\frac{\prtl S_c}{\prtl V} \right)_T \right]^2 \frac{k_B
      T}{K \bar{\rho}} + [K \alpha_t + ( \Delta \tilde{c}_v -
    \tilde{s}_\svibr) ]^2 \frac{k_B T^2}{\bar{\rho} \tilde{c}_v}
  \right\}^{1/2},
\end{equation}
where $\Delta \tilde{c}_v \equiv T(\prtl \tilde{s}_c/\prtl T)_V$ is
the configurational heat capacity at constant volume and
$\tilde{s}_\svibr$ the vibrational entropy, both per unit volume.

It is convenient, for the present purposes, to shift the energy
reference so that $\overline{G}_i = 0$:
\begin{equation} \label{Z4} Z = \int \frac{dG_i}{\sqrt{ 2 \pi \delta
      G_i^2 }} e^{S_c/k_B} e^{-G_i^2/2 \delta G_i^2}.
\end{equation}
This way, the partition function gives exactly the number
$e^{S_c/k_B}$ of the (thermally available) states that is not weighted
by the Boltzmann factor $e^{-\beta \overline{G}_i}$.

Consider now a local region that is currently {\em not} undergoing a
structural reconfiguration.  Because the region is certainly known not
to be reconfiguring, its free energy---up to finite-size
corrections---is equal to $G_i$, which is typically higher than the
equilibrium free energy $\overline{G}$ from Eq.~(\ref{gequil}). The
free energy difference $\overline{G} - \overline{G}_i = - T S_c < 0$
is the driving force for the eventual escape from the current
structure, and, hence, relaxation toward equilibrium.  Next we
estimate the actual rate of escape and the typical region size that
will have reconfigured as a result of the escape event.

We specifically consider escape events that are local. Therefore, the
environment of a chosen compact region is static, up to vibrations.
Consider the partition function for a compact region of size $N$
surrounded by such a static, aperiodic lattice. The vast majority of
the configurations do not fit the region's boundary as well as the
original configuration, and so there is a free energy penalty
$\Gamma_i > 0$ due to the mismatch between the static boundary and any
configuration of the region other than the original configuration.  We
anticipate that since local replacement of a structure amounts to a
legitimate fluctuation, $\Gamma$ and $\delta G_i$ should be
intrinsically related, which will indeed turn out to be the case.

In the presence of the mismatch penalty, the density of states can be
obtained by replacing $\overline{G}_i \to \overline{G}_i + \Gamma$
under the integral in Eq.~(\ref{Z3a}), where $\Gamma \equiv
\overline{\Gamma}_i$ is the typical value of the mismatch. The latter
generally scales with the region size:
\begin{equation} \label{Gamma1} \Gamma = \gamma N^x,
\end{equation}
but in a sub-thermodynamic fashion: $x < 1$, where the coefficient
$\gamma(N \to \infty) = \text{const}$.  Thus we obtain for the total
number of thermally available states for a region embedded in a static
lattice:
\begin{equation} \label{Z5} Z = \int \frac{dG_i}{\sqrt{ 2 \pi \delta
      G_i^2 }} e^{S_c/k_B - \beta \Gamma} e^{-(G_i-\Gamma)^2/2 \delta
    G_i^2},
\end{equation}
where we set the expectation value of the free energy in the {\em
  absence} of the penalty at zero, as before. (The expectation value
of $G_i$ corresponding to Eq.~(\ref{Z5}) is {\em not} zero at $N >
0$.) Note the argument of the first exponential on the r.h.s. is
independent of $G_i$ but does depend on the region size $N$, and so
does the total number of thermally available states $Z$:
\begin{equation} \label{Z6} Z(N) = e^{s_c N/k_B - \beta \gamma N^x},
\end{equation}
where $s_c \equiv S_c/N$ is the configurational entropy per particle.

Because of the sub-linear $N$-dependence of the mismatch penalty, the
number of thermally available states $Z(N)$ depends non-monotonically
on the region size.  For small values of $N$, this number {\em
  decreases} with the region size, which is expected since the region
is stable with respect to weak deformation such as movement of a few
particles. At the value $N^\ddagger$ such that $(\prtl Z/\prtl
N)_{N^\ddagger} = 0$, the number of available stats reaches its
smallest value and increases with $N$ for all $N > N^\ddagger$. This
critical size $N^\ddagger$:
\begin{equation} \label{Ncr1} N^\ddagger = \left(\frac{x \gamma}{T
      s_c}\right)^{1/(1-x)},
\end{equation}
corresponds to the least likely size of a rearranging region, and thus
corresponds to a bottleneck configuration for the escape event:
Indeed, any state at $N < N^\ddagger$ is less likely than the initial
state and so cannot be a final state upon a reconfiguration; such
final state must thus be at $N > N^\ddagger$. On the other hand, to
move any number of particles $N$ in excess of $N^\ddagger$, one must
have moved $N^\ddagger$ particles as an intermediate step.

The size $N^* > N^\ddagger$ such that
\begin{equation} \label{Z*} Z(N^*) = 1
\end{equation}
is special in that the region of this size is guaranteed to have a
thermally available configuration, distinct from the original one,
even though the boundary is fixed. By construction, this configuration
is mechanically (meta)stable. This implies that a region of size
$N^*$:
\begin{equation} \label{Nstar1} N^* = \left(\frac{\gamma}{T
      s_c}\right)^{1/(1-x)},
\end{equation}
can always reconfigure. What happens physically is that the centre of
the free energy distribution from Eq.~(\ref{Z5}) moves to the right
with $N$ according to $\gamma N^x$ because the mismatch typically
increases with the interface area. This alone would lead to a
depletion of states that are degenerate with the original state, which
is typically at $G_i = 0$.  Yet as $N$ increases, the free energy
distribution also {\em grows} in terms of the overall area, height,
and, importantly, width, as more states become available. For a
sufficiently large size $N^*$, the distribution is so broad that the
region is guaranteed to sample a state at $G_i = 0$ even though the
distribution centre is shifted to the right by $\Gamma$. One may say
that in such a state, a negative fluctuation in the free energy
exactly compensates the mismatch penalty. For this to be typically
true, we must have
\begin{equation} \label{GammadG} \Gamma(N) = \delta G_i(N)
  \hspace{5mm} \text{ at } \hspace{5mm} N = N^*,
\end{equation}
where we have emphasised that both $\Gamma$ and $\delta G$ depend on
$N$.

Finally note that the physical extent $\xi$ of the reconfiguring
region:
\begin{equation} \label{xi} \left(\frac{\xi}{a} \right)^3 \equiv
  \frac{4 \pi}{3} \left(\frac{R^*}{a} \right)^3 \equiv N^*
\end{equation}
yields the volumetric cooperativity length for the reconfigurations.

\begin{figure}[t] \centering
 \includegraphics[width=0.75 \figurewidth]{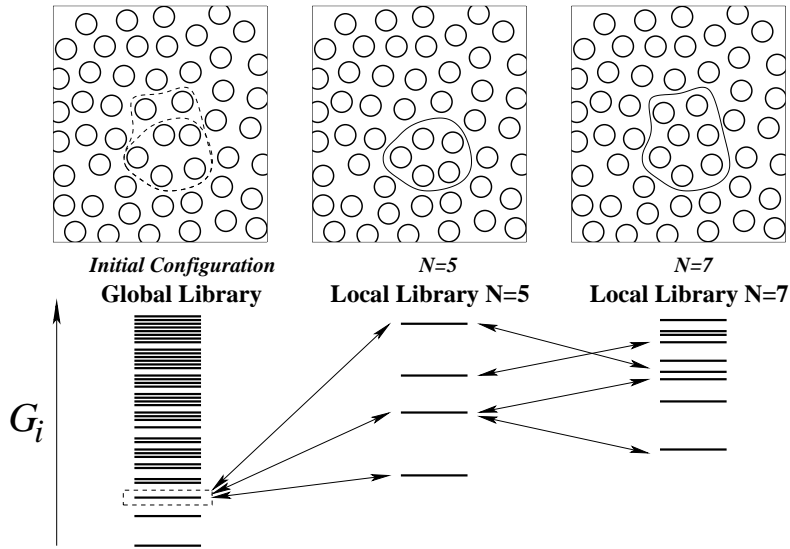}
 \caption{\label{library} Illustration of the library construction of
   aperiodic states~\cite{LW_aging}. On the left, we start out with
   some metastable structure. The density of the horizontal bars
   reflects the increase of the density of states (DoS)
   $e^{S_c(G_i)/k_B}$ with $G_i$. This DoS is distinct from the
   probability distribution in Eq.~(\ref{Z5}), which also includes the
   Boltzmann factor $e^{-\beta G_i}$.  In the centre and right panels,
   5 and 7 particles have been moved. The density of states pertaining
   to the corresponding local regions are much lower than the global
   density of states.  In addition, the majority of the thus obtained
   configurations are higher in free energy than the original
   configuration, owing to the mismatch with the environment. As the
   region size grows, the distribution of free energies $G_i$ of
   individual structures is determined by a competition between a
   depletion due to the mismatch and an entropically-driven increase
   in the DoS. For a large enough $N = N^*$, there will be a
   configuration whose free energy $G_i$ is comparable to that of the
   original structure.}
\end{figure} 

The above notions can be discussed explicitly in terms of particle
movements, within the library construction of aperiodic
states~\cite{LW_aging}, graphically summarised in Fig.~\ref{library}.
We start out with the original state, which is some state from the
full set of states available to the system. We then draw a surface
encompassing a compact region containing $N$ particles and consider
all possible configurations of the particles inside while the
environment is static up to local elastic deformations, i.e.,
vibration. Most of the resulting configurations are, of course, very
high in energy because of steric repulsion between the particles
comprising the region itself and between the particles on the opposite
sides of the the boundary. Only few configurations will contribute
appreciably to the actual ensemble of states, and even those are
offset upwards, free energy-wise, by the mismatch penalty. As the
chosen region is made progressively bigger, three things happen at the
same time: (a) the mismatch penalty typically increases as $\gamma
N^x$; (b) the log-number of available states first decreases but then
begins to increase, asymptotically as $\propto N$; and (c) the
spectrum of states becomes broader, the width going as $\propto
\sqrt{N}$. Think of (a) as a density of states $e^{S(G_i)/k_B}$ that
moves up according to $\gamma N^x$ with the region size (see
Fig.~\ref{library}), thus leading to a ``depletion'' of the density of
states at low $G_i$. Items (b) and (c), on the other hand, mean the
density of states increases in magnitude roughly as $e^{s_c N -\Gamma
  - (G_i-\Gamma)^2/2 \delta G^2}$, at fixed $G_i$ ($\delta G \propto
\sqrt{N}$ and $\Gamma \propto N^x$). For a large enough $N$, this
growth of the density of states, at fixed $G_i$, dominates the
depletion due to the mismatch penalty and so the free energy of the
substituted configuration eventually stops growing and begins to
decrease with the region size $N$, after the latter reaches a certain
critical value $N^\ddagger$. Eventually, at size $N^*$, one will
typically find an available state that is mechanically stable.

The discussion of the statistical notions embodied in
Eqs.~(\ref{Gamma1})-(\ref{Nstar1}) in terms of particle motions helps
one to recognise that the full set of configurations in some range
$[0, N_\smax]$ can be sorted out into (overlapping) subsets according
to the following protocol: (a) within each subset, every region size
is represented at least once and (b) two configurations characterised
by sizes $N$ and $(N+1)$ differ by the motion of exactly one
particle. The subsets thus correspond to dynamically-connected paths,
along each of which particles join the reconfiguring region {\em one
  at a time}. Along each of the dynamically connected paths, one could
thus think of the reconfiguration as a {\em droplet} growth.  A
certain path will dominate the ensemble of the paths given a
particular final configuration.  This dominating path is the one that
maximises the number of states $Z(N)$ from Eq.~(\ref{Z6}) (including
all intermediate values of $N$, of course). This is entirely analogous
to the Second Law, whereby the equilibrium configurations are those
the maximise the density of states (subject to appropriate
constraints).

One may take the notion of the dynamic connectivity even further by
noting that consecutive movements must be directly adjacent in space
because the nucleus grows is a sequence of individual bead movements.
Such movements---whether belonging to the same or distinct
reconfigurations---interact, the interaction scaling with distance $r$
as $1/r^3$, similarly to the electric dipole-dipole
interaction~\cite{BL_6Spin}. The nucleus growth in Eq.~(\ref{FN1})
differs, for instance, from nucleation of liquid inside vapour in that
in the latter, gas particles can join the nucleus from any side, while
in the former, consecutive bead movements are relatively likely to be
neighbours in physical space, in which case the movements form a
contiguous chain.  Now, how extensive are these individual bead
displacements? At the crossover, the individual free energy minima are
marginally stable with respect to transitions between each other and
with respect to the decay into the uniform liquid state
$\alpha=0$. Xia and Wolynes~\cite{XW} thus concluded that the
vibrational displacements within individual free energy minima should
be numerically close to the value prescribed by the Lindemann
criterion of melting, i.e., one-tenth of the volumetric particle size
or so:
\begin{equation} \label{dLa10} d_L \simeq a/10.
\end{equation}
Conversely, displacements just beyond this length will result in a
transition between distinct free energy minima. An important corollary
of the result in Eq.~(\ref{dLa10}) is that individual bead motions are
nearly harmonic, that is, {\em no bonds are broken}.  The total
combination of the individual moves is, of course, an anharmonic
process since both the initial and final state are metastable and are
thus separated by a barrier. A well-known example of such anharmonic
processes consisting of individual harmonic motions are rotations of
rigid SiO$_{4/2}$ tetrahedra in silica~\cite{Trachenko}. Another
explicit example will be considered in Subsection~\ref{midgap}, in
which a covalent bond does not break but, instead, becomes a weaker,
secondary bond. The energy cost of an individual bead movement is
modest.

One may question whether the most likely bottle-neck configuration in
the set of all dynamically connected paths leading to the final state
at $N^*$ is, in fact, as likely as what is prescribed by
$Z(N^\ddagger)$ with $Z(N)$ from Eq.~(\ref{Z6}). The answer is yes
because sampling of all possible shapes and locations for a region of
size $N$ is implied in the summation in Eq.~(\ref{Z5}). By
backtracking individual dynamically-connected trajectories from $N =
N^*$ to $N = 0$, we can determine the precise reconfigured region that
produces the most likely bottle-neck configuration with probability
$Z(N^\ddagger)$.

Now, for region sizes in excess of $N^*$, the number $Z(N)$ of
available states exceeds one, implying that, for instance, {\em two}
distinct metastable configurations are available to a region of size
$N$ such that $Z(N) = 2$. According to the above discussion, the
trajectories leading to these two states are generally distinct,
though the probabilities of the respective bottle-neck configurations
and the corresponding critical sizes $N^\ddagger$ should be
comparable.

Structural reconfiguration can be equally well discussed not in a
microcanonical-like fashion, through the number of states $Z(N)$, but,
instead, in terms of the corresponding free energy $F(N) = - k_B T \ln
Z(N)$, as was originally done by Kirkpatrick, Thirumalai, and
Wolynes~\cite{KTW}. This yields the following activation profile for
the reconfiguration:
\begin{equation} \label{FN1} F(N) = \Gamma - T s_c N \equiv \gamma N^x
  - T s_c N,
\end{equation}
Since we are considering all possible configurations for escape,
subject to the appropriate Boltzmann weight, this amounts to locally
replacing the original configuration encompassing $N$ particles by the
{\em equilibrated} liquid while the particles in the surrounding are
denied any motion other than vibration.  Upon the replacement, the
local {\em bulk} free energy is typically lowered by $\overline{G} -
\overline{G}_i = - T s_c N$, hence the driving term $- T s_c N$ in
Eq.~(\ref{FN1}).

The equilibrated liquid is a Boltzmann-weighted average of
alternative, metastable aperiodic structures that are mutually
distinct and are also generally distinct from the initial
configuration.  Another way of saying two structures are distinct is
that the particles belonging to the structures inside and outside do
not fit as snugly---at the interface between the structures---as they
do within the respective basis structures. This is quite analogous to
the mismatch between two distinct crystalline polymorphs in contact,
such as must occur during a first order transition between the
polymorphs. In contrast with a polymorphic transition, the scaling of
the mismatch penalty with the area of the interface will turn out to
be somewhat complicated, viz, the exponent $x$ need not be $(D-1)/D$.

Combining the free energy view with the notion of dynamically
connected trajectories, due to the library construction, we conclude
that the activation profile in Eq.~(\ref{FN1}) is also a {\em
  nucleation} profile.  Naturally, the bottle-neck configuration
corresponding to $N = N^\ddagger$ from Eq.~(\ref{Ncr1}) thus
corresponds to the critical nucleus size.  The corresponding barrier
is equal to
\begin{equation} \label{F1} F^\ddagger \equiv F(N^\ddagger) = \gamma
  \left(\frac{x\gamma}{T s_c } \right)^{x/(1-x)} (1-x) = \gamma (1-x)
  \, (N^\ddagger)^x .
\end{equation}
This directly shows that the escape rate from a specific aperiodic
state is indeed finite. Note also that the cooperativity size is
always equal to the critical size $N^\ddagger$ times an $x$-dependent
numerical coefficient:
\begin{equation} \label{NNdagger} N^* = N^\ddagger \, x^{-1/(1-x)} >
  N^\ddagger.
\end{equation}

Yet there is more to the activation profile in Eq.~(\ref{FN1}). In
ordinary theories of nucleation, the nucleus continues to grow
indefinitely once it exceeds the critical size, unless it collides
with other growing nuclei, as happens during crystallisation, or, for
instance, when the supply of the contents for the minority phase runs
out, as happens when a fog forms. This essentially unrestricted growth
takes place because this way, the system can minimise its free energy
by fully converting to the minority phase. Such a view is adequate
when there are only two free energy minima to speak of and the system
converts between those two minima.

However in the presence of an exponentially large number of free
energy minima, we must think about the meaning of the $F(N)$ curve
more carefully. We must recognise that both the initial and {\em
  final} state for the escape event are individual aperiodic states
that are, on average, equally likely. In fact, because we have chosen
$G_i = 0$ as our free energy reference, $F(N)$ gives exactly the
log-number (times $-k_B T$) of thermally available states to the
selected region.  As a result, that the free energy $F(N)$ reaches its
initial value of zero indicates that a mechanically metastable state
has become available to escape to.  One is accustomed to situations in
which the initial and final state for a barrier-crossing event are
{\em minima} of the free energy, which does not seem to be the case in
the above argument.  There is no paradox here, however. The quantity
$F(N)$ is not the actual free energy of the system. Instead, by
construction, it is the free energy under the constraint that the
outside of the selected compact region is not able to relax in the
usual matter, but, instead, is forced to be in a specific, metastable
aperiodic minimum.  The monotonic decrease of $F(N)$ at $N=N^*$ is
trivial in that it simply says the surrounding of the droplet will
eventually proceed to reconfigure again and again, as it should in
equilibrium. As we have already emphasised, the state to which the
initial configuration has escaped is perfectly meta{\em stable}.

We now shift our attention to the {\em energy}. Suppose the liquid is
composed of particles that are not completely rigid, and so the
mismatch penalty has an energetic component. Furthermore, it is
instructive to suppose that the penalty is {\em mostly} energetic,
which is probably the case for covalently bonded substances such as
silica or the chalcogenides.~\cite{ZLMicro1} At a first glance, the
energy of the system appears to grow with each nucleation event, since
the driving force in Eq.~(\ref{FN1}) is exclusively entropic, at
equilibrium. Such unfettered energy growth is, of course, impossible
in equilibrium. On the contrary, the configurations before and after a
reconfiguration are typical and the energy must be conserved, on
average. The energy change following a transition must be within the
typical fluctuation range, which reflects the heat capacity $C_V$ at
constant volume and the bulk modulus $K$:~\cite{LLstat}
\begin{equation} \label{deltaE} \delta E = \{k_B C_v T^2 - V [T(\prtl
  p/\prtl T)_V -p]^2T/K \}^{1/2}.
\end{equation}
Note both $C_V$ and $V$ pertain to a single cooperative region. The
conservation of energy, on average, means that since one new interface
appears following an escape event, an equivalent of one interface must
have been {\em subsumed} during an event, as emphasised in
Ref.~\cite{ZLMicro2}.

We thus conclude that the equilibrium concentration of the interface
configurations is given by $1/\xi^3$ with $\xi$ from Eq.~(\ref{xi}),
and so a glassy liquid is a {\em mosaic} of aperiodic
structures,~\cite{XW} each of which is characterised by a relatively
low free energy density, while the interfaces separating the mosaic
cells are relatively stressed regions characterised by excess free
energy density due to the mismatch between stabilised regions. This
stress pattern is not static, but relaxes at a steady pace so that a
region of size $\xi$ reconfigures once per time $\tau$, on average:
\begin{equation} \label{tauF1}
\tau = \tau_0 e^{F^\ddagger/k_B T},
\end{equation}
where the pre-exponent $\tau_0$ corresponds to the vibrational
relaxation time. It is the same vibrational-hopping time we
encountered in Subsection~\ref{MCT}. Since the nucleation is driven by
the multiplicity of distinct aperiodic configurations, i.e., by the
configurational entropy, a glassy liquid may be said to be a {\em
  mosaic} of entropic droplets~\cite{XW}.

Note that the total free energy stored in the strained regions
corresponding to the domain walls is equal to $\Gamma (N/N^*) = T s_c
N$, i.e. the enthalpy difference between the liquid and the
corresponding crystal at the temperature in question, up to possible
differences in the vibrational entropy between the crystal and an
individual aperiodic structure.

Note that every metastable configuration---which contains both the
relatively relaxed and strained regions---is a true free energy
minimum of the liquid. This is in contrast with the mean-field view we
usually take of {\em macroscopic} phase coexistence, by which
extensive portions of the sample are occupied by structures
corresponding to true minima of the bulk free energy density, such as
the two minima in Fig.~\ref{LGfigure}.  Appropriately, the physical
boundary between those macroscopic regions then corresponds to the
saddle-point in the bulk free energy in Fig.~\ref{LGfigure} {\em and}
to a saddle point solution of the free energy of a nucleating
droplet. Thus only in the mean-field limit do the relatively
stabilised regions in glassy liquids correspond with true free energy
minima. (Their multiplicity is still given by the configurational
entropy!) In finite dimensions, there is a steady-state, uniform
density of both relatively stabilised and relatively strained regions,
whose spatial density can be determined self-consistently, as
explained above.

We finish this Subsection by describing a somewhat distinct way to
think about the mosaic, due to Bouchaud and
Biroli~\cite{BouchaudBiroli}. In this view, the ensemble of all states
of a compact region of size $\xi$ consists of a contribution from the
current state and contributions of the full, exponentially large set
of alternative structures.  As in the library construction, the
surrounding of the chosen region is constrained to be static up to
vibrational displacement. What sets apart the current state from all
the alternative states is that it fits the environment better.  The
partition function for the region thus goes roughly as:
\begin{equation} \label{ZBB} Z_\text{BB} \simeq e^{-\beta(-\gamma
    N^x)} + e^{N s_c/k_B}.
\end{equation}
The mismatch free energy $\gamma N^x$ scales with the region size $N$
sublinearly, $x < 1$, while the log-number of alternative states
scales linearly. Thus the stability of sufficiently {\em small}
regions---smaller than $\xi$---can be understood thermodynamically in
a straightforward manner: The energetic advantage of being in the
current state, due to the matching boundary, outweighs the
multiplicity of poorer matching, higher energy states. This is not
unlike the stability of a crystal relative to the liquid below
freezing. Indeed, at liquid-crystal equilibrium, the partition
function of the system can be written as $Z = e^{-\beta (-\Delta H)} +
e^{\Delta S/k_B}$, where $\Delta H > 0$ and $\Delta S > 0$ are the
fusion enthalpy and entropy respectively.

The just listed aspects of the Bouchaud-Biroli picture are essentially
equivalent to the library construction, and, in particular, with
regard to the entropic nature of the driving force for the activated
transport.  In contrast with the KTW~\cite{KTW} and library
construction~\cite{LW_aging}, however, the BB scenario is agnostic as
to the concrete mechanism of mutual reconfiguration between
alternative aperiodic states, other that the reconfigurations must be
rare, activated events.  In the absence of such a concrete mechanism,
BB posited a generic scaling relation between the cooperativity size
and the relaxation barrier that is similar to that transpiring from
Eqs.~(\ref{F1}) and (\ref{NNdagger}). In this view, a region {\em
  larger} than the size $N^*$ can still reconfigure via a {\em single}
activated event but would do so typically {\em more slowly} than the
region of size $N^*$. Combining this notion with the lower bound on
the cooperativity size obtained above one concludes that the
cooperativity size is in fact $N^*$ and one does not face the subtlety
stemming from the downhill decrease of the free energy profile $F(N)$
from Eq.~(\ref{FN1}). The present picture contrasts with the
Bouchaud-Biroli approach in that it specifically prescribes that the
activated reconfigurations take place through a process akin to {\em
  nucleation}.

Now, according to Eqs.~(\ref{Ncr1})-(\ref{xi}), we need to evaluate
the exponent $x$ and the coefficient $\gamma$ for the mismatch
penalty, to estimate the escape rate and the cooperativity size for
the activated reconfigurations, to which we proceed next.

\subsection{Mismatch Penalty between Dissimilar Aperiodic Structures:
  Renormalisation of the surface tension coefficient }
\label{mismatch}

The mismatch penalty between two ill-fitting structures can be
evaluated with the help of the free energy functional. This evaluation
is particularly straightforward in the long wavelength approximation
represented by the Landau-Ginzburg functional from Eq.~(\ref{LG}),
which is often called the Cahn-Hiilliard~\cite{CahnHilliard}
functional in the context of nucleation.  Further simplification is
achieved when the nucleus is very large~\cite{Bray, RowlinsonWidom},
provided that the interface remains of finite width. Hereby the
interface can be effectively regarded as flat while the total mismatch
penalty scales asymptotically linearly with the interface {\em
  area}. Under these circumstances, the characteristics of the
interface are determined by two parameters. One parameter is the
coefficient $\kappa$ at the square gradient term in Eq.~(\ref{LG}),
the other is the barrier height $g^\ddagger$ in the bulk free energy
density $V(\phi)$, see Fig.~\ref{LGfigure}.  The expressions for the
width $l$ and the surface tension coefficient $\sigma$ read
\begin{equation} \label{lintf1} l \sim \sqrt{\kappa/g^\ddagger},
\end{equation}
and
\begin{equation} \label{sigmaFlat} \sigma \sim \sqrt{\kappa g^\ddagger}
  \sim g^\ddagger l,
\end{equation}
respectively; both equalities are within factors of order one, which
depend on the specific form of the free energy functional. The total
mismatch penalty goes as $4 \pi r^2 g^\ddagger l$, consistent with
simple dimensional analysis. Roughly speaking, the excess free energy
$g^\ddagger l$ per unit area reflects the free energy penalty due to
unsatisfied bonds at the interface, see also Eq.~(\ref{sqGIsing}). In
the case of rigid particles, bonds are not defined, but analogous
expressions involving the direct correlation function instead of the
pair-interaction potential can be written, as in
Eq.~(\ref{sqGliquid}).
One recognises that the interface width $l$ reflects the interaction
(or direct correlation) range---according to Eqs.~(\ref{sqGIsing}) and
(\ref{sqGliquid})---and thus is not directly tied to the molecular
size, even though the two are often numerically similar.

Because the states on both sides of our interface are aperiodic, the
degree of mismatch is distributed~\cite{DSW2005, Franz}.  Thus in some
places the two structures may fit quite well and so the scaling of the
surface energy term $\Gamma$ from Eq.~(\ref{FN1}) with the droplet
size $N$ may be weaker than the $N^{(D-1)/D}$ scaling expected for
interfaces separating periodic or spatially uniform phases.  The
mechanism of this partial lowering of the mismatch penalty is as
follows: The number of distinct aperiodic structures available to a
sufficiently large region, we remind, scales exponentially with the
region size. The free energies $G_i$ of individual structures from
Eq.~(\ref{gequil}) are {\em distributed}; they are equal on average
but differ by a finite amount for any specific pair of aperiodic
states.  Fluctuations of extensive quantities scale with $\sqrt{N}$ as
functions of size $N$.~\cite{LLstat} (The size $N$ at which the
$\sqrt{N}$ scaling sets in can be rather small in the absence of
long-range correlations, such as those typical of a critical point.)
Thus the free energy difference between the configurations outside and
inside scales as $\sqrt{N}$, for two regions of the same size $N$, and
could be of either sign.  Suppose now, for concreteness, that the
configuration on the outer side of the domain wall happens to be lower
in free energy than the adjacent region on the inside. Imagine
distorting the domain wall so as to replace a small portion of the
inside configuration by that from the outside.  It turns out the free
energy stabilisation due to the replacement outweighs the
destabilisation due to the now increased area of the interface, as we
shall see shortly.

Before we proceed with this analysis, it is instructive to discuss why
such surface renormalisation and the consequent stabilisation would
{\em not} take place during regular discontinuous transitions when one
phase characterised by a {\em single} free energy minimum nucleates
within another phase also characterised by a {\em single} free energy
minimum.  After all, both phases represent superpositions of
microstates whose energies are {\em also} distributed. Furthermore,
there seems to be a direct correspondence between, say, the regular
canonical ensemble and the situation described in Eq.~(\ref{gequil}).
Hereby, the free energies $G_i$ in Eq.~(\ref{gequil}) seem to
correspond to the energies of the microstates, while the
configurational entropy $S_c$ seems to correspond to the full entropy
in the canonical ensemble.  One difference between the situation in
Eq.~(\ref{gequil}) and the canonical ensemble is that in the latter,
transitions between the microstates within individual phases occur on
times much shorter than the observation time or mutual nucleation and
nucleus growth of the two phases. As a result, the energies of the
phases on the opposite sides of the interface are always equal to
their {\em average} values. In contrast, the distinct aperiodic states
from Eq.~(\ref{gequil}) are long-lived. In fact, the fastest way to
inter-convert between those states is via creation of the very
interface we are discussing! In the canonical ensemble analogy, this
would correspond to having {\em individual} microstates on the
opposite sides of the interface as opposed to ensembles resulting from
averaging over all microstates (with corresponding Boltzmann
weights). Conversely, the situation in Eq.~(\ref{gequil}) would be
analogous to the canonical ensemble only at sufficiently long times
that much exceed the nucleation time from Eq.~(\ref{tauF1}). Note that
at such long times, we have identical, equilibrated liquid on both
sides and so there is no surface tension in the first place.

Now, the situation where the system can reside in long-lived states
whose free energies are distributed in a Gaussian fashion can be
equivalently thought of as a perfectly ergodic, equilibrated system in
the presence of a {\em static}, externally-imposed random field whose
fluctuations scale in the Gaussian fashion.  In the absence of this
additional random field, the mismatch penalty between such two regular
phases would be perfectly uniform along an interface with spatially
uniform curvature. The simplest system one can think of, in which this
situation is realised, is the random field Ising model:
\begin{equation} \label{RFIM} \cH = - J \sum_{i<j} \sigma_i \sigma_j -
  \sum_i h_i \sigma_i , \hspace{10mm} \sigma_i = \pm 1,
\end{equation}
where $J > 0$ while the Zeeman splittings $h_i$'s are random,
Gaussianly distributed variables. In the Hamiltonian above, if one
were to impose a smooth interface between two macroscopic domains with
spins up and down, the domain wall would distort some to optimise the
Zeeman energy. However, the amount of distortion is also subject to
the tension of the interface between the spin-up and spin-down
domains. The overall lowering of the free energy, due to the interface
distortion, corresponds to the optimal compromise between these two
competing factors.  Likewise, a smooth interface between two distinct
aperiodic states will distort to optimise with respect to local bulk
free energy, which is distributed. The energy compensation will scale,
again, as the square root of the variation of the volume swept by the
interface during the distortion.  The final shape of the interface
will be determined by the competition between this stabilisation and
the cost of increasing the area of the interface.

The mapping between the random field Ising model (RFIM) and large
scale fluctuations of the interface between aperiodic liquid
structures was exploited by Kirkpatrick, Thirumalai, and Wolynes (KTW)
\cite{KTW}, who used Villain's argument \cite{Villain} for the
renormalisation of the surface in RFIM to deduce how the droplet
interface tension scales asymptotically with the droplet size.  The
mapping relies crucially on the condition that the undistorted
interface must not be too thick, as it would be near a critical
point. This assumption turns out to be correct since the width of the
undistorted interface is on the order of the molecular length
$a$~\cite{RL_sigma0}.

Let us now review a variation on the KTW-Villain argument concerning
the surface tension renormalisation. This argument produces the
scaling relation we seek for the mismatch penalty but some of its
steps are only accurate up to factors of order one, and so the latter
will be dropped in the calculation. All lengthscales will be expressed
in terms of the molecular length $a$, which simply sets the units of
length. Now, consider two dissimilar aperiodic states in contact, and
assume we have already coarse-grained over all length-scales less than
$r$, while explicitly forbidding interface fluctuations on greater
lengthscales. The interface is thus {\em taut}. Further, consider
spatial variations in the shape of the interface on lengthscales
limited to a narrow interval $[r, r(1+ \Delta)]$. The dimensionless
increment
\begin{equation} \label{increment}
  \Delta = d \ln r
\end{equation}
is the increment of the running argument for our real-space
coarse-graining transformation $r \to r(1+ \Delta)$.  (Ultimately, $r$
will be set at the droplet radius).  We may assume, without loss of
generality, that the mismatch penalty may be written in the following
form, in $D$ spatial dimensions:
\begin{equation} \label{Fsr} \Gamma = \sigma(r) r^{D-1},
\end{equation}
The quantity $\sigma(r)$ may be thought of as a renormalised surface
tension coefficient, where the amount of renormalisation generally
depends on the wavelength, which is distributed in the (narrow) range
between $r$ and $r(1+\Delta)$. Our task is to determine under which
condition such renormalisation takes place, if any.

\begin{figure}[t]
  \centering
  \includegraphics[width= 0.5 \figurewidth]{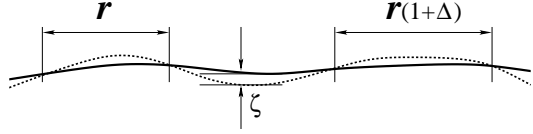}
  \caption{\label{bump} Illustration of a step in the coarse-graining
    procedure of an interface distorted because of frozen-in free
    energy fluctuations, \`a la Villain~\cite{Villain}. The interface
    has already been coarse-grained over lengthscales less than
    $r$. At each spectral interval $[r, r(1+\Delta)]$, the optimal
    value of the distortion $\zeta$ and the resulting free energy
    stabilization are determined using Eq.~(\ref{minF}).}
\end{figure}

To do this, let us deform the interface so as to create a bump of
(small) height $\zeta$ and lateral extent $r$, see sketch in
Fig.~\ref{bump}.  Because the interface is taut, the area will
increase quadratically with $\zeta$. The resulting increase in the
interface area will incur a free energy cost
\begin{equation}
  \delta F_s \sim \sigma(r) r^{D-1} (\zeta/r)^2 \Delta,
\end{equation}
when $\zeta \ll r$.  It will turn out to be instructive to use a more
general form
\begin{equation} \delta F_s \sim \sigma(r) r^{D-1} (\zeta/r)^z \Delta.
\end{equation}
This generalised form is convenient because (a) rough interfaces may
exhibit a $z$ other than $2$, 
(b) the scaling of the interface tension with $r$ will turn out be
independent of $z$ at the end of the calculation. Thus the obtained
$\sigma$ vs. $r$ scaling can be argued to still apply even to
situations when $\zeta/r$ is not necessarily small. (Which it will not
be!)  Now, as already mentioned, one can always flip a region (of size
$N$) at the interface so a to lower its bulk free energy by $\sim h
\sqrt{N}$. The resulting bulk free energy gain is thus:
\begin{equation} \label{Fb} \delta F_b \sim - h \sqrt{N} \Delta \sim -
  h (r^{D-1} \zeta)^{1/2} \Delta,
\end{equation}
where the constant $h$ is straightforward to estimate in light of our
earlier discussion that the bulk stabilisation above is the result of
fluctuations of the Gibbs free energy.  Thus,
\begin{equation} \label{h2} h \sim \delta G_i /\sqrt{N},
\end{equation}
with $\delta G_i$ from Eq.~(\ref{dGi}).  Properly, we should have
written $h^2 = 2 (\delta G_i)^2/N$ in Eq.~(\ref{h2}) because the bulk
free energy stabilisation is a {\em difference} between two random
Gaussian variables, whose distribution widths are $\delta G_i$ each,
but we have agreed to drop factors of order one in the derivation, see
also below.

Next we find the value of $\zeta$ that optimises the total free energy
stabilisation: $\prtl (\delta F_s + \delta F_b)/\prtl \zeta = 0$,
which yields:
\begin{equation} \label{zeta} \zeta \sim (h/\sigma)^{2/(2z-1)}
  r^{(2z-D+1)/(2z-1)}.
\end{equation}
The resulting energy gain per unit area,
\begin{align} \label{minF}
  \min_\zeta\{& \delta F_s + \delta F_b\}/r^{D-1} \sim \nonumber \\
  &\sim - (h^z/\sigma^{1/2})^{1/(z-1/2)} r^{-z(D-2)/(2z-1)} \Delta,
\end{align}
thus represents the renormalisation $\delta \sigma(r)$ of the
$r$-dependent ``surface tension coefficient'' that resulted from
integrating out degrees of freedom in the $k$-vector range between
$1/r$ and $1/(1+\Delta) r$.

The energy gain per unit area from Eq.~(\ref{minF}), due to the
real-space renormalisation in the wavelength range $[r, r(1+\Delta)]$,
can be viewed as an iterative relation, by Eq.~(\ref{increment}):
\begin{equation} \label{dsdr}
  d \sigma \sim - (h^z/\sigma^{1/2})^{1/(z-1/2)} r^{-z(D-2)/(2z-1)} 
  d \ln r.
\end{equation}
A quick inspection of this differential equation shows that the
surface tension coefficient {\em decreases} with $r$. To determine the
actual $r$-dependence of $\sigma$, we must decide on the boundary
condition $\sigma(r=\infty) \equiv \sigma_\infty$. Suppose for a
moment that $\sigma_\infty > 0$, which implies that at sufficiently
large distances, the interface width tends to some {\em finite} value
$l_\infty$, however large, given by Eq.~(\ref{lintf1}). At the same
time, the free energy excess per unit volume $g^\ddagger$ tends to a
finite value $g^\ddagger_\infty$. This is because in the $r \to
\infty$ limit, a steady value of $\sigma$ implies the interface
becomes truly flat and none of its parameters could depend on
$r$. Since the free energy excess per unit volume $g^\ddagger$ tends
to a finite value $g^\ddagger_\infty > 0$ for a flat interface, so
should $l_\infty$, if $\sigma_\infty$ is finite, by
Eq.~(\ref{sigmaFlat}).  The surface tension coefficient is thus given
by the expression
\begin{equation*} \sigma^{2z/(z-1)}(r) = h^{2z/(2z-1)}
  r^{-z(D-2)/(2z-1)} + \sigma^{2z/(z-1)}_\infty.
\end{equation*}
Inserting the above formula in expression (\ref{zeta}) yields:
\begin{equation} \label{zeta1} \zeta \sim \frac{r^{(2z-D+1)/(2z-1)}}
  {[r^{-z(D-2)/(2z-1)}+ (\sigma_\infty/h)^{2z/(2z-1)} ]^{1/z}}.
\end{equation}
The above formula indicates that although incremental changes in the
interface curvature following the renormalisation are small, the
compound increase in the interface thickness---due to the curvature
changes in the broad wavelength range spanned by the coarse-graining
procedure---is not necessarily so.

Eq.~(\ref{zeta1}) indicates that there are two internally-consistent
options regarding the value of the surface tension coefficient
$\sigma_\infty$. In the conventional case of zero random field, $h =
0$, $\sigma_\infty$ is finite while $\zeta = 0$, and so no
renormalisation takes place while the interface width tends to a
steady value $l_\infty$ at diverging droplet radii. If, on the other
hand, the random field is present, the only remaining option is
$\sigma_\infty = 0$. Indeed, by Eq.~(\ref{zeta1}), the interface width
$\zeta$ diverges as $r \to \infty$, when $h > 0$, implying the
supposition of a finite $\sigma_\infty$ and, hence, finite $l_\infty$
was internally inconsistent.  For this argument to be valid, the
renormalised interface width $\zeta$ should exceed the width $l$ of
the original, flat interface. Condition $\zeta > l$ and
Eq.~(\ref{zeta1}), combined with $\sigma_\infty = 0$, yield
\begin{equation} r > l,
\end{equation}
which happens to coincide with the criterion of validity of the thin
interface approximation. Note that in their analysis of barrier
softening effects near the crossover, Lubchenko and
Wolynes~\cite{LW_soft} self-consistently arrived at a similar
criterion, viz., $r > a$, for when interface tension renormalisation
would take place.

It follows that an arbitrarily weak, but finite random field $h$ makes
an interface with a sufficiently low curvature unstable with respect
to distortion and lowering of the effective surface tension:
\begin{equation} \label{sr} \sigma(r) \sim \frac{h}{a^{D-1}}
  (r/a)^{-(D-2)/2},
\end{equation}
independent of $z$, apart from a proportionality constant of order
one, giving us confidence in the result even when the undulation size
$\zeta$ is not very small. Notice we have restored the units of length
for clarity. 

Eq.~(\ref{sr}) yields that the renormalised mismatch energy $\Gamma$
from Eq.~(\ref{Fsr}) scales with the droplet size $N$ in a way that is
independent of the space dimensionality, namely $\sqrt{N}$:
\begin{equation} \label{Gr} \Gamma \sim \sigma(r) r^{D-1} \sim h
  (r/a)^{D/2} \sim h \sqrt{N},
\end{equation}
which is ultimately the consequence of the Gaussian distribution of
the free energy.  Because of the lack of a fixed length scale in the
problem---other than the trivial molecular size $a$, which sets the
units---it should not be surprising that the interface width $\zeta$
scales with the radius $r$ itself:
\begin{equation} \label{zetar}
  \zeta \sim r,
\end{equation}
again independent of $z$. The numerical constant in the above equation
is of order one, as is easily checked, and so $\zeta/r$ is not small
generally.

The large effective interface width can be thought of as a result of
the distortion of the original thin interface where the extent of the
distortion is not determined by a fixed length, but the curvature of
the interface itself. In other words, this interface is a {\em
  fractal} object. Because of this fractality, the structure at the
interface is not possible to characterise as either of the aperiodic
structures on the opposite sides of the original flat interface before
the renormalisation. We could thus informally think of this fractal
interface as the original thin interface {\em wetted}~\cite{XW} by
other structures that interpolate, in an optimal way, between the two
original aperiodic structures. While we are not aware of direct
molecular studies with regard to the fractality of cooperative regions
in non-polymeric liquids, such studies of polymer melts do suggest the
mobile regions have a fractal character~\cite{Starr2014}.

\begin{figure}[t]
  \begin{tabular*}{\figurewidth} {cc}
    \begin{minipage}{.48 \figurewidth} 
      \flushleft
      \begin{center} 
        \includegraphics[width=0.46 \figurewidth]{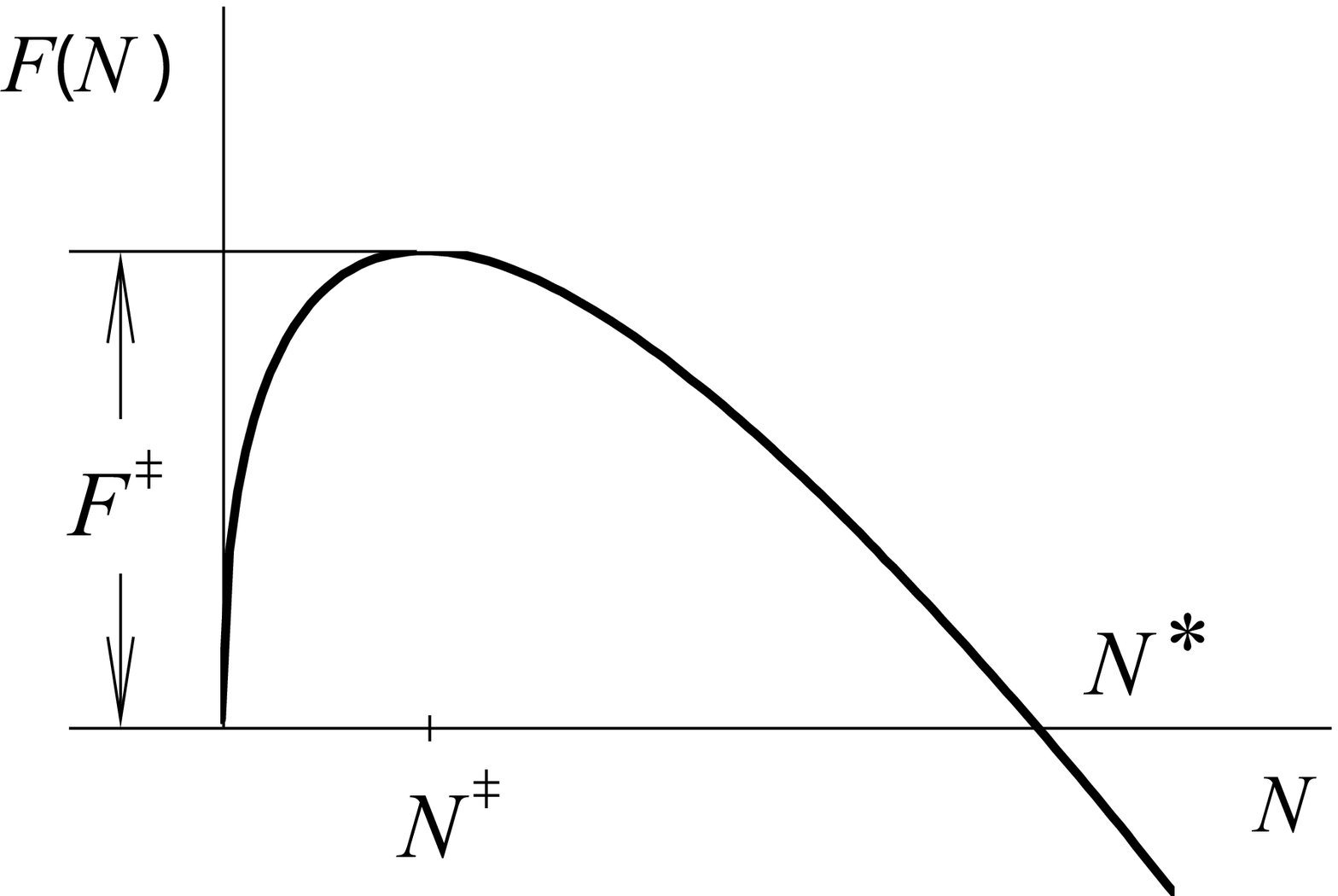}
        \\ {\bf (a)}
      \end{center}
    \end{minipage}
    &
    \begin{minipage}{.48 \figurewidth} 
      \begin{center}
        \includegraphics[width= .46 \figurewidth]{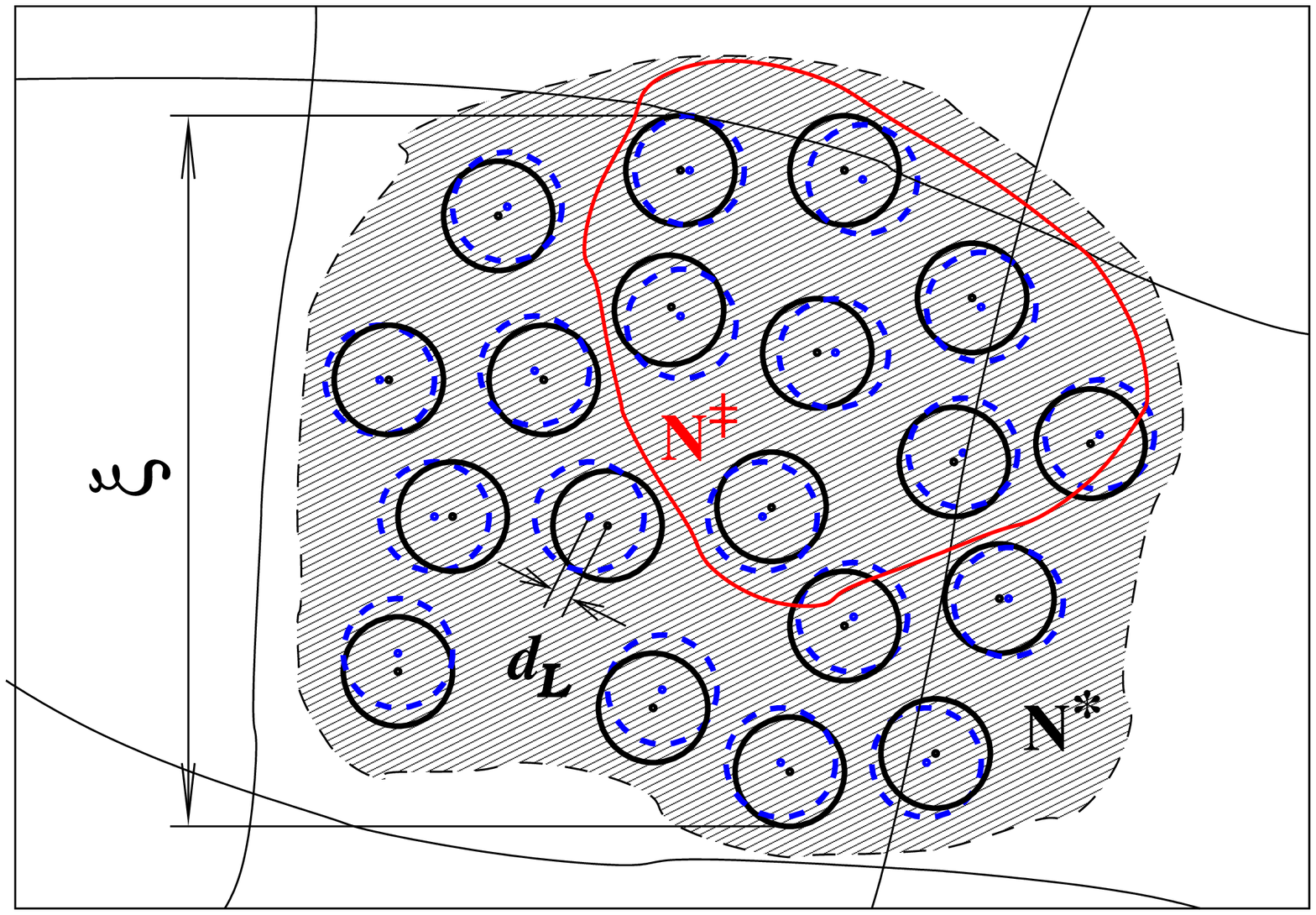} \\
        {\bf (b)}
      \end{center}
    \end{minipage}
  \end{tabular*}
  \caption{\label{FNgraph} {\bf (a)} The free energy nucleation
    profile for structural reconfiguration in a glassy liquid from
    Eq.~(\ref{FN12}). Indicated are the critical size $N^\ddagger$,
    cooperativity size $N^*$, and the barrier $F^\ddagger$. {\bf (b)}
    Cartoon illustrating a cooperative reconfiguration. The latter
    becomes downhill typically past the critical size $N^\ddagger$ and
    is completed when $N^* \equiv (\xi/a)^3$ particles
    moved. Individual particle displacements are typically $d_L \simeq
    a/10$, but decrease toward the edge of the reconfiguring
    region~\cite{LW, LW_RMP}.}
\end{figure}

According to Eq.~(\ref{Gr}), the scaling exponent $x$ is equal to
$1/2$. Thus the matching condition in Eq.~(\ref{GammadG}) is valid at
all values of $N$ and so we arrive at a central result~\cite{KTW, XW,
  LRactivated}:
\begin{equation} \label{Gamma} \Gamma = \gamma \sqrt{N}.
\end{equation}
where
\begin{equation} \label{gammadG}
\gamma = \delta G_i/\sqrt{N} = \text{const}.
\end{equation}
The resulting free energy nucleation profile:
\begin{equation} \label{FN12} F(N) = \gamma \sqrt{N} - T s_c N,
\end{equation}
is shown in Fig.~\ref{FNgraph}(a).

In retrospect, the square-root scaling in Eq.~(\ref{Gamma}) is
natural: In view of Eq.~(\ref{Gamma1}), the $(G_i-\Gamma)^2/2 \delta
G_i^2$ term under the second exponential in Eq.~(\ref{Z5}) scales
asymptotically with $N$ according to $N^{2x-1}$, see also
Ref.~\cite{capillary}.  For any $x$ other than $1/2$, this would
result in an anomalous scaling~\cite{Goldenfeld} of the density of
states with the system size that would be hard to rationalise given
the apparent lack of criticality in actual liquids between the glass
transition and fusion temperatures.

We thus obtain for the nucleation barrier from Eq.~(\ref{F1}):
\begin{equation} \label{F2} F^\ddagger = \frac{\gamma^2}{4T s_c },
\end{equation}
see Fig.~\ref{FNgraph}(a). In view of Eq.~(\ref{scT}), the scaling of
the expression is consistent with the VFT law, Eq.~(\ref{VFT}),
provided the coefficient $\gamma$ does not exhibit a singularity at
the temperature at which the configurational entropy is extrapolated
to vanish.

\subsection{Quantitative estimates of the surface tension, the
  activation barrier for liquid transport, and the cooperativity size}
\label{quant}

Lubchenko and Rabochiy~\cite{LRactivated} (LR) have argued for a
direct identification between the mismatch penalty at the
cooperativity size $N^*$ and the typical value of the free energy
fluctuation at that size, Eq.~(\ref{GammadG}), as was explained in
Subsection~\ref{mosaic}. Except for this identification, the arguments
in Subsections~\ref{mosaic} and \ref{mismatch} represent an expanded
version of the original argument of Kirkpatrick, Thirumalai, and
Wolynes~\cite{KTW} (KTW), with additional clarification due to the
library construction~\cite{LW_aging}. Already the original KTW
argument yields that $\Gamma = \delta G(N)$ up to a factor of order
one. LR~\cite{LRactivated} have, in a sense, completed the KTW
programme and directly estimated the surface tension coefficient
$\gamma$ based on the notion that the renormalisation of the surface
tension is driven by local fluctuations of the free energy. According
to Eqs.~(\ref{dGi}) and (\ref{gammadG}), we obtain:
\begin{equation} \label{gammaFull} \gamma = \left\{ \left[ K - T
      \left(\frac{\prtl S_c}{\prtl V} \right)_T \right]^2 \frac{k_B
      T}{K \bar{\rho}} + [K \alpha_t + (
    \Delta \tilde{c}_v - \tilde{s}_\svibr) ]^2 \frac{k_B
      T^2}{\bar{\rho} \tilde{c}_v} \right\}^{1/2}.
\end{equation}

Next we estimate the $\gamma$ from Eq.~(\ref{gammaFull}). As a rule of
thumb, the bulk modulus is about $(10^1 - 10^2) k_B T/\bar{\rho}$ for
liquids and $10^2 k_B T/\bar{\rho}$ for solids near the melting
temperature $T_m$~\cite{Bilgram} (consistent with the Lindemann
criterion of melting~\cite{Lindemann, L_Lindemann}).  The rate of
change of the configurational entropy with volume is not known but can
be crudely estimated based on the observation that upon freezing, the
hard sphere liquid loses $\approx 1.2 k_B$ worth of entropy per
particle while its volume reduces by about
$10$\%.~\cite{HooverRee1968, Hansen} Assuming our liquid will run out
of configurational entropy at about the same rate---though
gradually---we obtain $(\prtl S/\prtl V)_T \sim 10^1 k_B/a^3$. This is
consistent with the theoretical prediction for this quantity in
Lennard-Jones systems by Rabochiy and Lubchenko, see Fig.~10 of
Ref.~\cite{RL_LJ}.  Further, $\tilde{s}$ is about $10^0 k_B/a^3$.  The
dimensionless expansivity $\alpha_t T$ is generically $10^{-1}$,
although could be much smaller for strong substances, see Fig.~12 of
Ref.~\cite{RL_LJ}. $\Delta \tilde{c}_v$ and $\tilde{s}_c$ are both
$\sim 10^0 k_B/a^3$, while $\tilde{c}_v \sim 10^1 k_B/a^3$. As a
result, we conclude that the volume contribution to the free energy
fluctuation in Eq.~(\ref{dGi}) generically exceeds the temperature
contribution by one-two orders of magnitude, thus yielding:
\begin{equation} \label{gamma1} \gamma^2 \approx \left[ K - T
    \left(\frac{\prtl S_c}{\prtl V} \right)_T \right]^2 \frac{k_B T}{K
    \bar{\rho}},
\end{equation}
since $\bar{\rho} \equiv a^{-3}$.  Consequently,
\begin{equation} \label{FKsc} F^\ddagger \approx \left[ K - T
    \left(\frac{\prtl S_c}{\prtl V} \right)_T \right]^2 \frac{k_B}{4 K
    \tilde{s}_c},
\end{equation}
where $\tilde{s}_c$ is the configurational entropy per unit volume.

According to the above estimate of the $(\prtl S_c/\prtl V)_T$ term,
it is likely that at least for rigid, weakly attractive systems, the
second term in the square brackets is an order of magnitude smaller
than the first term.  We thus expect the following, simple expression
for the surface tension coefficient $\gamma$ to be of comparable
accuracy to Eq.~(\ref{gamma1}):
\begin{equation} \label{gamma2} \gamma \approx \left( \frac{K k_B T}{
      \bar{\rho}} \right)^{1/2} \equiv \sqrt{K a^3 \, k_B T}.
\end{equation}
It is interesting that the coefficient $\gamma$ above, which reflects
coupling of structural fluctuations to its environment, has exactly
the same form as the coupling between the structural reconfigurations
corresponding to the two-level systems (TLS) in cryogenic glasses and
the phonons~\cite{LW, LW_RMP} to be discussed in Subsection~\ref{TLS}.

The simplified form in Eq.~(\ref{gamma2}) implies for the nucleation
barrier:
\begin{equation} \label{FKsc1}
  F^\ddagger \simeq \frac{K}{4 ( \tilde{s}_c/k_B) }.
\end{equation}
Given that the temperature dependence of the bulk modulus is usually
rather weak, Eq.~(\ref{FKsc1}) yields to a good approximation the
venerable Adam-Gibbs functional relation~\cite{AdamGibbs}. The
relation in Eq.~(\ref{FKsc1}) is a central result of the theory. Note
that it allows one to compute the reconfiguration barrier, an
expressly kinetic quantity, using thermodynamic quantities. All these
quantities can be experimentally determined. We will compare the
predictions due to Eq.~(\ref{FKsc1}) with observation shortly, after
we discuss alternative approximations for the mismatch penalty.

A welcome feature of the expression (\ref{gammaFull}) for the
coefficient $\gamma$ is that it yields, when combined with
Eq.~(\ref{F2}), an expression for the barrier that is expressly
independent of the bead size. This notion, of course, also applies to
the simplified expressions in Eqs.~(\ref{FKsc}) and
(\ref{FKsc1}). Thus, in the end, these results do not rely on the
phenomenological assumption of an effective particle of the theory.

The expression for the cooperativity size $\xi \equiv a (N^*)^{1/3} =
(\gamma/T s_c)^{2/3}$ corresponding to the approximation in
Eq.~(\ref{FKsc1}) reads:
\begin{equation} \label{xiKsc} \xi \simeq \left[ \frac{K}{k_B T
      (\tilde{s}_c/k_B)^2 } \right]^{1/3}
\end{equation}
This formula can be rewritten in a convenient form for the
cooperativity volume:
\begin{equation} \label{xiK} \xi^3 \equiv N^* a^3 = [4
  \ln(\tau/\tau_0)]^2 \frac{k_B T}{K}.
\end{equation}
Given the rule of thumb that $K a^3/k_B T \simeq 10^2$, we obtain that
at the glass transition, where $\ln(\tau/\tau_0) \approx 35$, the
cooperativity size is about $10^2$ beads.

Note that neither of the expressions (\ref{FKsc}), (\ref{FKsc1}) and
(\ref{xiKsc}) depends explicitly on the molecular length scale $a
\equiv \bar{\rho}^{-1/3}$. In this sense, these expressions are truly
{\em scale-free}, consistent with the scale-free character of the
ordinary, gaussian fluctuations of thermodynamic quantities. In
contrast, the presence of anomalous scaling generally requires that a
molecular length scale be present explicitly~\cite{Goldenfeld}, see
also our earlier comment on the square-root scaling of the mismatch
penalty following Eq.~(\ref{gammadG}). The simple result in
Eq.~(\ref{FKsc1}) has another notable feature with regard to
scaling~\cite{LRactivated}. It is the only expression of units energy
one could write down using the bulk modulus and the configurational
entropy per unit volume that does not involve temperature. The
proportionality of the barrier to the bulk modulus is a direct
indication of the activated nature of the reconfigurations and implies
that the latter must involve bond stretching.

On the other hand, the configurational entropy in the denominator of
Eq.~(\ref{FKsc1}) reflects the progressively smaller number of degrees
of freedom available to the liquid at lower temperatures (or higher
pressures), which thus leads to fewer possibilities to find an
alternative metastable state and a downhill trajectory to reach that
state. This is also reflected in the entropy dependence of the
cooperativity length (\ref{xiKsc}): Given the decreasing log-number of
states per unit volume, at lower temperatures, searching through
larger regions is required to find an alternative structural state.

We have already remarked that the square-root scaling of the mismatch
penalty is natural in the absence of criticality because otherwise the
free energy would exhibit anomalous scaling.  The square root scaling
is due to the Gaussian nature of the fluctuations that lead to
reconfigurations. But why would generic Gaussian fluctuations lead to
non-generic consequences in glassy liquids?  The answers lies in the
exponential multiplicity of distinct free energy minima, which
introduces new physics in the problem. Following Bouchaud and
Biroli~\cite{BouchaudBiroli}, one may think of the long-lived
structures as residing in traps that are enthalpically stabilised
relatively to a state in which such traps are only marginally stable
and in which the liquid rearrangements are thus nearly
barrierless. This is similar to the way a crystal in equilibrium with
the corresponding liquid can be thought of as a trap which is lower in
enthalpy than the liquid whereby the enthalpic stabilisation due to
the trapping exactly matches the excess liquid entropy times
temperature. A similar situation takes place when a folded protein
molecule is in equilibrium with the unfolded chain. Now, in the
presence of a thermodynamically large number of free energy minima,
fluctuations become possible that enthalpically stabilise local
regions.  The extent of such a special fluctuation is not arbitrary;
its value can be determined self-consistently, in equilibrium, by
matching the (enthalpic) stabilisation due to the fluctuation with the
log-multiplicity of the minima: $\delta G_i = T s_c$. (To avoid
confusion we repeat that the quantity $G_i$ is enthalpy-like with
respect to the configurational degrees of freedom even though it
includes entropic contributions due to vibrations.) Combined with
Eq.~(\ref{gammadG}), this yields $\gamma \sqrt{N^*} = T s_c N^*$ at
the cooperativity size.  No such local trapping is possible when the
multiplicity of the minima is sub-thermodynamic, i.e., when the system
is not in the landscape regime. Indeed, this equation has no
solutions, if the log-number of the minima per particle, $s_c/k_B$, is
identically set to zero.

This simple result above would be modified in the presence of
long-range correlations such as those near critical points. Yet we
have seen that a liquid-to-solid is an {\em avoided} critical point,
the degree of separation from criticality reflected in the substantial
value of the force constant of the effective Einstein oscillator
$\alpha \sim K a/k_B T \sim 10^2/a^2$, see Eq.~(\ref{alpha1}).
Consistent with this notion, the energetics of structural fluctuations
are dominated by the elastic modulus $K$, see Eqs.~(\ref{gammadG}) and
(\ref{gamma2}), which greatly exceeds the thermal energy scale $k_B
T/a^3$.  On the other hand, the appearance of the bulk modulus in
Eqs.~(\ref{gamma2}) and (\ref{FKsc1}) is consistent with our earlier
notion that no bonds are broken during the structural
reconfigurations.

It is quite possible that in addition to the trivial, Gaussian
fluctuations, smaller-scale fluctuations of distinct nature are also
present. For instance, in their recent replica-based work, Biroli and
Cammarota~\cite{2014arXiv1411.4566B} argue there is a ``wandering''
length scale associated with fluctuations of the domain wall
shape. This length diverges with lowering the temperature as
$1/s_c^{1/2}$, as opposed to the stronger divergence $\xi \propto
1/s_c^{2/3}$ from Eq.~(\ref{xiKsc}). Such wandering could be important
at higher temperatures, see Section~\ref{crossover}.

The preceding discussion pertains exclusively to glassy liquids.  It
seems instructive to consider now an apparent counterexample in the
form of an amorphous silicon film. When quenched sufficiently below
the glass transition, a glassy liquid forms an amorphous solid. Just
below the glass transition, the glass can still undergo activated
reconfigurations of the type discussed above, but somewhat modified
for the initial state being off-equilibrium; this will be explained in
detail in Subsection~\ref{aging}. In contrast, silicon is a very poor
glass-former and so its amorphous form must be prepared by means other
than quenching, for instance, by sputtering on a cold substrate.  The
poor glass-forming ability of silicon is consistent with our earlier
notion that this substance would not crossover into the landscape
regime; instead it crystallises at relatively low densities such that
the steric effects are still relatively unimportant. Still, according
to the present discussion it would seem that amorphous silicon could
undergo structural reconfigurations similar of glassy liquids.  After
all, silicon films are aperiodic and low-density thus resulting in
vast structural degeneracy. Indeed, the density of such films is at
least 10\% less than that of the crystal (p. 1004 of
Ref.~\cite{Pohl_review}) let alone the liquid. The liquid excess
entropy of {\em equilibrated} silicon at this density much exceeds the
configurational entropy below the crossover.  To be fair, the
landscape of amorphous silicon is largely energetic, not
free-energetic; it is well above the equilibrium energy at the ambient
conditions. But this would not seem to make a huge difference as far
as the barriers between the (metastable) energy minima are
concerned. Are these barriers given by an expression of the type in
Eq.~(\ref{FKsc1})? First note that in contrast with glassy liquids, we
may no longer use the expression (\ref{gammaFull}) for the mismatch
penalty, since it was derived assuming equilibrium.  Furthermore,
since the enthalpy of the film is dominated by the (highly
anisotropic) bonding, it is likely that silicon will rearrange by {\em
  breaking bonds}; the mismatch penalty will thus reflect the
corresponding, higher energy scale.  Additional complications arise
from the fact that the films do not correspond to a structure that was
equilibrated at any temperature, see Subsection~\ref{aging}.  Another
indicator that reconfigurations in amorphous silicon films must
involve bond breaking is that such films host bulk quantities of
dangling bonds~\cite{Pohl_review}, something of relevance in the
context of photo-voltaic applications of the material. In addition to
being efficient scatterers of charge carriers, the dangling bonds are
directly witnessed by a substantial ESR signal; in contrast, a proper
bond is formed by a filled molecular orbital. Reconfigurations, if
any, clearly fail to ``heal'' such dangling bonds in silicon
films. Conversely, glasses made by quenching equilibrated liquids do
not exhibit significant numbers of dangling bonds.  Thus based on the
high energy cost of rearrangements and the relationship between the
barrier and the cooperativity size from Eq.~(\ref{F1}), we tentatively
conclude that the cooperativity size in amorphous silicon likely
exceeds that in glasses made by quenching. We will continue this
discussion later on.

Now, the linear scaling of the activation barrier in Eq.~(\ref{FKsc1})
with the elastic modulus hearkens back to earlier, enthalpy based
approaches to activated dynamics by Hall and
Wolynes~\cite{ISI:A1987G269600055} and the so called ``shoving
model''~\cite{PhysRevB.53.2171}. Some of the particular realisations
of the shoving model~\cite{dyre:224108, klieber:12A544} posit that the
dominant contribution to the temperature dependence of the barrier is
due to the temperature dependence of the elastic moduli. More
specifically, the shoving model postulates that the barrier goes as $K
V_c$, where $V_c$ stands for the volume of a rearranging region which
is prescribed by system specific interactions and is temperature
independent. In contrast, the RFOT theory demonstrates that the
configurational entropy is the leading contributor to the
$T$-dependence of the barrier, and increasingly so at lower
temperatures~\cite{RWLbarrier}, see also below. Appropriately, the
cooperativity volume must increase with decreasing entropy, see
Eq.~(\ref{xiKsc}).

In view of the intrinsic relation between the elastic constants and
the localisation parameter $\alpha$~\cite{RL_Tcr}, Eq.~(\ref{alpha1}),
the above argument connects the mismatch penalty $\gamma$ with the
``localisation'' of the particles upon the crossover at $T_\scr$. Yet
this connection is only indirect. The original calculation of
$\gamma$, due to Xia and Wolynes~\cite{XW}, {\em explicitly} connects
the mismatch penalty to the formation of the metastable structures in
which the particles vibrate around steady-state average locations.  It
is this original calculation, dating from 2000, that enabled the
majority of quantitative predictions by the RFOT theory. This
calculation will be reviewed next.

The activation profile from Eq.~(\ref{FN12}) can be rewritten in terms
of the droplet radius (assuming a spherical geometry) as:
\begin{equation} \label{Fr} F(r) = 4 \pi r^2 \sigma_0 (a/r)^{1/2} - (4
  \pi/3) (r/a)^3 T s_c,
\end{equation}
using the following connection between the particle number contained
within the droplet and the droplet radius:
\begin{equation} N \equiv \frac{4 \pi}{3} (r/a)^3.
\end{equation}
This yields for the reconfiguration barrier:
\begin{equation} \label{FXW} \frac{F^\ddagger}{k_B T} =
  \frac{3\pi(\sigma_0 a^2/k_B T)^2}{s_c/k_B}.
\end{equation}
and the cooperativity length:
\begin{equation} \label{xiXW} \xi/a = (4\pi/3)^{1/3} \left( \frac{3
      \sigma_0 a^2}{T s_c} \right)^{2/3}.
\end{equation}

To estimate the surface tension coefficient $\sigma_0$, which
corresponds to the mismatch penalty at the molecular scale: $\sigma =
\sigma(r=a)$, we begin with the density profile from Eq.~(\ref{rho})
in which the lattice sites $\br_i$ are arranged in an aperiodic
fashion, as in Subsection~\ref{RFOTDFT}.  This profile is an excellent
approximation to the actual density distribution, even though it
ignores the possibility that the force constant $\alpha$ of the
effective Einstein oscillator may be somewhat spatially distributed.
The aperiodic crystal forms despite the {\em one-particle} entropic
cost $\Delta f_\text{loc} > 0$ of the localisation of a particle
around a certain location in space. The reason is that this
localisation also offers a free energy gain, due to multi-particle
effects, as the collisions are now less frequent, as we have already
discussed following Eq.~(\ref{F_id1}).  In the presence of chemical
interactions, there is also an enthalpy gain due to bond
formation. Both gains contribute to the free energy in formally
equivalent ways (through the direct correlation function) and thus can
be formally termed ``bonding;'' denote the corresponding free energy
difference $\Delta f_\text{bond} < 0$. Additional stabilisation comes
from the multiplicity of aperiodic states: $-T s_c < 0$.

The mismatch between two degenerate states can be thought of as
partial lack of ``bonds''---a half or so---for a particle at the
interface, leading to about $(-\Delta f_\text{bond})/2$ per particle
in missing free energy. Estimating the bonding free energy appears
generally difficult, however obtaining a formal lower bound on
$-\Delta f_\text{bond}$ is not: Below the crossover, the aperiodic
crystal is stable, implying $\Delta f_\text{loc} + \Delta
f_\text{bond} - T s_c \le 0$ and, hence, $ -\Delta f_\text{bond} \ge
\Delta f_\text{loc} - T s_c$, the equality achieved strictly at the
crossover.  The one-particle localisation penalty is computed in the
standard fashion:
\begin{equation} \label{Df_loc} \Delta f_\sloc/k_BT = N^{-1} \int
  \rho(\br) \left[ \ln\rho(\br) -1 \right] d^3\br - (\ln\bar{\rho} -
  1) \simeq (3/2) \ln(\alpha a^2/\pi e),
\end{equation}
where $\bar{\rho} \equiv 1/a^3$ is the number density of beads,
Eq.~(\ref{rhoa}). The density profile $\rho(\br)$ is from
Eq.(\ref{rho}). We thus arrive at
\begin{equation} \label{sigmaLower} \sigma_0 \ge [(3/2) \ln(\alpha
  a^2/\pi e) - s_c/k_B] \, k_B T/2a^2.
\end{equation}
To simplify the expression in the square brackets we compare the
$T$-dependences and magnitudes of the first and second terms: First
off, $\alpha$ increases, while $s_c$ decreases upon lowering the
temperature. At the crossover, Rabochiy and Lubchenko's (RL)
estimates~\cite{RL_LJ} yield $\alpha$ ranging from 71 to 104, hence
$(3/2) \ln(\alpha a^2/\pi e) \simeq 3.2 \ldots 3.7$. For $s_c$ at the
crossover, Stevenson et al.~\cite{SSW} predict $\approx 1.2 k_B$,
while RL have argued it could be as high as $1.75 k_B$. On the other
hand, at $T_g$, $s_c \simeq 0.8 \ldots 0.9 k_B$.~\cite{XW, LW_soft}
Judging from its high-density trend (Fig.~3 of Ref.~\cite{RL_LJ}),
$\alpha (T_g)$ is probably 200 or so, yielding $(3/2) \ln(\alpha
a^2/\pi e) \simeq 4.7$. Xia and Wolynes argued a good estimate for
$\sigma_0$ would be obtained by taking the smallest value of $\alpha$,
i.e. that achieved at the crossover, and disregarding $s_c$; the
latter we have seen becomes progressively smaller than the
localisation term as the temperature is lowered. Note the
$T$-dependence of $s_c$ may well be immaterial to the $T$-dependence
of $\sigma_0$, since Eq.~(\ref{sigmaLower}) is stated as an
inequality. Note also that at temperatures at which $s_c$ is not
small, it is the barrier softening effects that determine $\sigma_0$,
in the first place~\cite{LW_soft}, see Section~\ref{crossover}.  These
effects can be thought of as resulting from fluctuations of the
parameter $\alpha$ near the crossover, which leads to
``short-circuiting'' of the activated transitions by partial melting
of the lattice. (This is another way of saying that near the
crossover, the usual collisional transport becomes
important~\cite{LW_Wiley}.) Finally we note that the linear
temperature dependence of $\sigma_0$ is obvious for rigid systems but
makes just as much sense for strongly interacting systems: The free
energy cost of localisation is determined by the {\em kinetic}
pressure, which is determined by the ambient temperature (and is
usually much greater than the ambient pressure). We thus obtain
\begin{equation} \label{sigmaXW} \sigma_0 = \frac{3}{4} (k_B T/a^2)
  \ln(\alpha a^2/\pi e) \approx 1.85 \, k_B T/a^2,
\end{equation}
if one adopts the generic value for the vibrational displacement, $a^2
\alpha = 100$, c.f. Figs.~\ref{solid} and \ref{SSW}.  In terms of
$\gamma$:
\begin{equation} \label{gammaXW} \gamma = \frac{3}{2} \sqrt{3 \pi} k_B
  T \ln(\alpha_L a^2/\pi e) \approx 11.3 \, k_B T.
\end{equation}

Eqs.~(\ref{sigmaXW}) and (\ref{FXW}) yield an extremely simple
expression for the activation barrier~\cite{XW}:
\begin{equation} \label{XWbarrier} \frac{F^\ddagger}{k_B T} \simeq
  \frac{32.}{s_c}.
\end{equation}
Note that the configurational entropy is per {\em bead}, not per unit
volume, in contrast with Eq.~(\ref{FKsc1}). In the XW~\cite{XW}
approximation for the mismatch penalty, it is the Arrhenius exponent
that does not explicitly depend on the temperature, not the barrier
itself. Similarly simple to Eq.~(\ref{XWbarrier}) is the formula for
the cooperativity size that can be easily obtained using
Eq.~(\ref{F1}):
\begin{equation} \label{NXW} N^* \equiv (\xi/a)^3 \simeq
  [\ln(\tau/\tau_0)/2.83]^2 = [(F^\ddagger/k_B T)/2.83]^2 \approx
  130/s_c^2.
\end{equation}

Notwithstanding its approximate nature, the argument leading to
Eq.~(\ref{XWbarrier}) makes explicit use of the physical picture
emerging from the RFOT theory in that it directly traces the origin of
the mismatch penalty to the confinement of the particle following the
breaking of the translational symmetry and the general inability of
arbitrary aperiodic arrays of such localised particles to mutually
fit. The approximations boil down to a few simple notions: On the one
hand, we have neglected the configurational entropy, which would be
strictly valid only at $T_K$. On the other hand, we have replaced the
inequality in Eq.~(\ref{sigmaLower}) by the equality, which would be
strictly valid at the crossover temperature $T_\scr$. In addition,
there is an uncertainty in the value of $\alpha$, which, however, is
only logarithmic. As a result, the energy scale in Eq.~(\ref{sigmaXW})
could, in principle, vary between $k_B T_\scr$ and $k_B T_K$, but not
much beyond that. In any event, the temperature dependence of the
numerator in Eq.~(\ref{XWbarrier}) is much weaker than that of the
configurational entropy. According to this equation, the barrier
should, in fact, diverge if the configurational entropy vanishes. This
prediction is, arguably, the most important result of the RFOT
theory. It shows that the kinetic catastrophe and the thermodynamic
singularity, which are the landmarks of the glass transition, are
intrinsically related.  The kinetic catastrophe manifests itself
through the rapid growth of the viscosity above the glass transition
that implies the viscosity would diverge at some putative temperature
$T_0$, Fig.~\ref{angell}, if one had the ability to equilibrate the
liquid on the correspondingly increasing time scales. The
thermodynamic singularity is the vanishing of the configurational
entropy at the (putative) Kauzmann temperature $T_K$, as in
Fig.~\ref{RichertAngell}. According to Eq.~(\ref{XWbarrier}), the
temperatures $T_0$ and $T_K$ must coincide, which is consistent with
available extrapolations of kinetic and calorimetric data, see
Fig.~\ref{Tk_Vs_T0}.

\begin{figure}[t]
  \begin{tabular*}{\figurewidth} {cc}
    \begin{minipage}{.62 \figurewidth} 
      \flushleft
      \begin{center} 
        \includegraphics[width=0.62 \figurewidth]{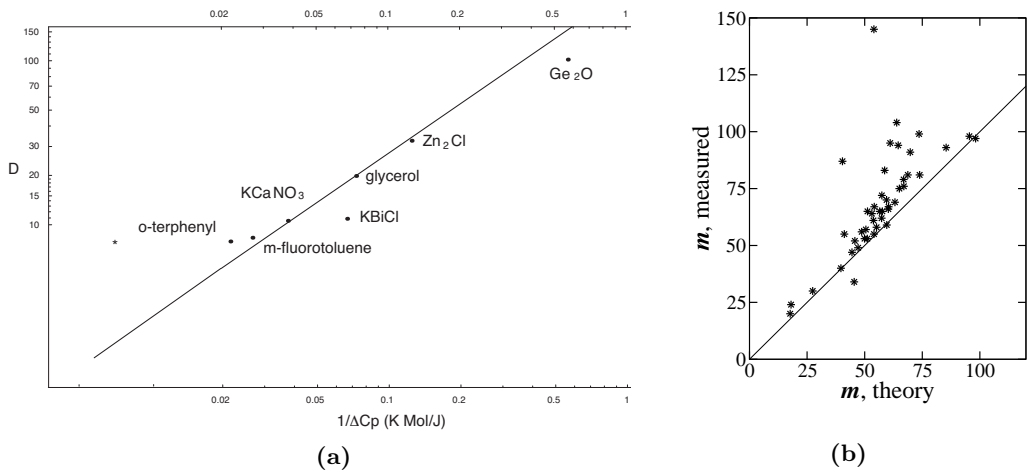}
        \\ {\bf (a)}
      \end{center}
    \end{minipage}
    &
    \begin{minipage}{.34 \figurewidth} 
      \begin{flushright}
        \includegraphics[width= .32 \figurewidth]{m_m.eps} 
      \end{flushright}
      \begin{center}        {\bf (b)}
      \end{center}
    \end{minipage}
  \end{tabular*}
  \caption{\label{frTests} Tests of the Xia-Wolynes relation between
    the barrier for activated transport and the configurational
    entropy per bead (\ref{XWbarrier}), from Ref.~\cite{XW}. {\bf (a)}
    The relation (\ref{DvsDCpEq}) between the fragility from
    (\ref{VFT}) and heat capacity jump per bead, is shown with the
    solid line. The experimental points are indicated with
    symbols. The bead count is determined based on chemical
    considerations. {\bf (b)} The straight line corresponds to the
    Lubchenko-Wolynes~\cite{LW_soft} relation (\ref{mLW}). The
    experimental points, shown with the symbols, were compiled by
    Stevenson and Wolynes~\cite{StevensonW}. The bead count is from
    Eq.~(\ref{Nb}). No adjustable constants used in either estimate.}
\end{figure}

In addition to prescribing that $T_0 = T_K$, Eq.~(\ref{XWbarrier})
predicts that the {\em rate} of change of the barrier with temperature
is determined by the rate of change of the configurational entropy.
Both of these rates have a clear physical meaning and value in
themselves: The slope of the $T$-dependence of the log-viscosity
actually reflects the width of the temperature window in which the
supercooled liquid will be in a certain viscosity interval of
convenience for processing. The broader the window, the less need for
maintaining a constant temperature during the processing. The
corresponding materials are called ``long'' glasses.  Conversely
substances with a narrow processing window are called ``short''
glasses. In a more modern parlance, due to
Angell~\cite{AngellScience1995}, the former and latter and often
called ``strong'' and ``fragile,'' see Fig.~\ref{angell}. Accordingly,
the coefficient $D$ from the VFT law in Eq.~(\ref{VFT}) is called the
fragility. On the other hand, the rate of change of the
configurational entropy with temperature is directly related to the
excess heat capacity associated with the translational degrees of
freedom, see Eq.~(\ref{DCp}). Using Eqs.~(\ref{VFT}),
(\ref{XWbarrier}), and (\ref{scT}), Xia and Wolynes (XW)
obtain~\cite{XW}:
\begin{equation} \label{DvsDCpEq} D \simeq \frac{32.}{\Delta c_p},
\end{equation}
where $\Delta c_p$ is the heat capacity jump at the glass transition,
{\em per bead}. XW tested their formula using the
chemically-reasonable bead size, whereby the bead was identified with
a chemically rigid unit, such as CH$_3$. The results are certainly
very encouraging, see Fig.~\ref{frTests}(a), especially considering
that no adjustable constants are involved. In other words, the
straight line in Fig.~\ref{frTests}(a) is not a fit but a prediction.

To determine the fragility $D$, one must extrapolate kinetic data
below the glass transition temperature.  One may reasonably object to
such extrapolation into regions that are inaccessible in experiment,
both on formal grounds and because such extrapolations are likely to
result in some numerical uncertainty. In view of these potential
ambiguities, an alternative way to test the connection between the
kinetics of the activated transport and the driving force behind the
transport, as predicted by relation (\ref{XWbarrier}) is to compare
the {\em slope} of the temperature dependence of the relaxation time
(or its logarithm) with the rate of change of the configurational
entropy at some standard temperature. The most obvious choice for such
standard temperature is the glass transition temperature $T_g$ itself
since the time scale for the latter usually does not vary much between
different labs or substances. It is convenient, for the sake of
comparing different substances, to work with a dimensionless slope,
called the {\em fragility coefficient}:
\begin{equation} \label{mdef} m = \frac{\partial
    \log_{10}\tau}{\partial \left(T_g/T\right)}\Bigg|_{T=T_g}.
\end{equation} 
Low values of the fragility coefficient correspond to strong
substances, while high values correspond to fragile substances, see
Fig.~\ref{angell}.

With the help of the calorimetric bead count, Eq.~(\ref{Nb}),
Lubchenko and Wolynes~\cite{LW_soft} (LW) have written down the
following simple relation
\begin{equation} \label{mLW} m = \frac{T_g \Delta c_p(T_g)}{\Delta
    H_m} \left\{s_\text{bead} (\log_{10} e) \frac{32.}{s_c^2(T_g)}
    \frac{T_m}{T_g} \right\}.
\end{equation}
Generically, the ratio of the melting and glass transition
temperatures $T_m/T_g$ is about $3/2$ (the actual figure seems to vary
between 1.2 and 1.6 or so). Also, by virtue of Eq.~(\ref{XWbarrier}),
the value of the configurational entropy per bead, at the glass
transition on timescale of $10^5$~sec, is $s_c(T_g) =
32./\ln(10^{5}/1e^{-12}) \simeq 0.82$. Using these generic figures and
$s_\text{bead} = 1.68$ (see discussion of Eq.~(\ref{Nb})), LW obtain
$m = 52 T_g \Delta C_p/\Delta H_m$, which is rather close numerically
to the empirical relation $m = 56 T_g \Delta C_p/\Delta H_m$ first
noticed by Wang and Angell~\cite{WangAngell}. Stevenson and Wolynes
went further and used the actual measured values of $T_m$ and $T_g$ to
compare the value of the fragility coefficient determined in kinetic
measurements with the one determined using the thermodynamic
quantities according to Eq.~(\ref{mLW}). The results of this
comparison are shown in Fig.~\ref{frTests}(b); again, no adjustable
parameters are involved.  We observe a clear correlation between the
kinetics and thermodynamics on approach to the glass transition,
although the precise form of the correlation seems to deviate somewhat
from the simple expression (\ref{mLW}) for the more fragile
substances. One possible source of the deviation are the
barrier-softening effects~\cite{LW_soft}, to be discussed in
Section~\ref{crossover}.
 
Although it directly reflects microscopic aspects of the RFOT, the XW
argument leading to Eq.~(\ref{XWbarrier}) is an estimate, not a fully
microscopically-based calculation of the penalty for bringing two
dissimilar aperiodic states in contact. Rabochiy and
Lubchenko~\cite{RL_sigma0} attempted to perform such a calculation
using standard methods of the classical density-functional
theory. They employed the
Landau-Ginzburg-Cahn-Hillard~\cite{CahnHilliard, RowlinsonWidom}
functional, c.f. Eq.~(\ref{LG}):
\begin{equation} \label{FCH} F\left[\eta(\br)\right] = \int
  \left[\frac{\kappa(\eta)}{2} (\nabla\eta)^2 +f(\eta)\right]d^3\br,
\end{equation}
where the order parameter $\eta$ ($-1 < \eta < 1$) allows one to
consider arbitrary spatial superpositions of two distinct structures
with density profiles $\rho_1(\br)$ and $\rho_2(\br)$ respectively:
\begin{equation} \label{eta} 
 \rho(\br)= \rho_1(\br)\frac{1 -\eta}{2} + \rho_2(\br)\frac{1 +\eta}{2}.
\end{equation}

\begin{figure}[t]
  \begin{tabular*}{\figurewidth} {cc}
    \begin{minipage}{.48 \figurewidth} 
      \flushleft
      \begin{center} 
        \includegraphics[width=0.46 \figurewidth]{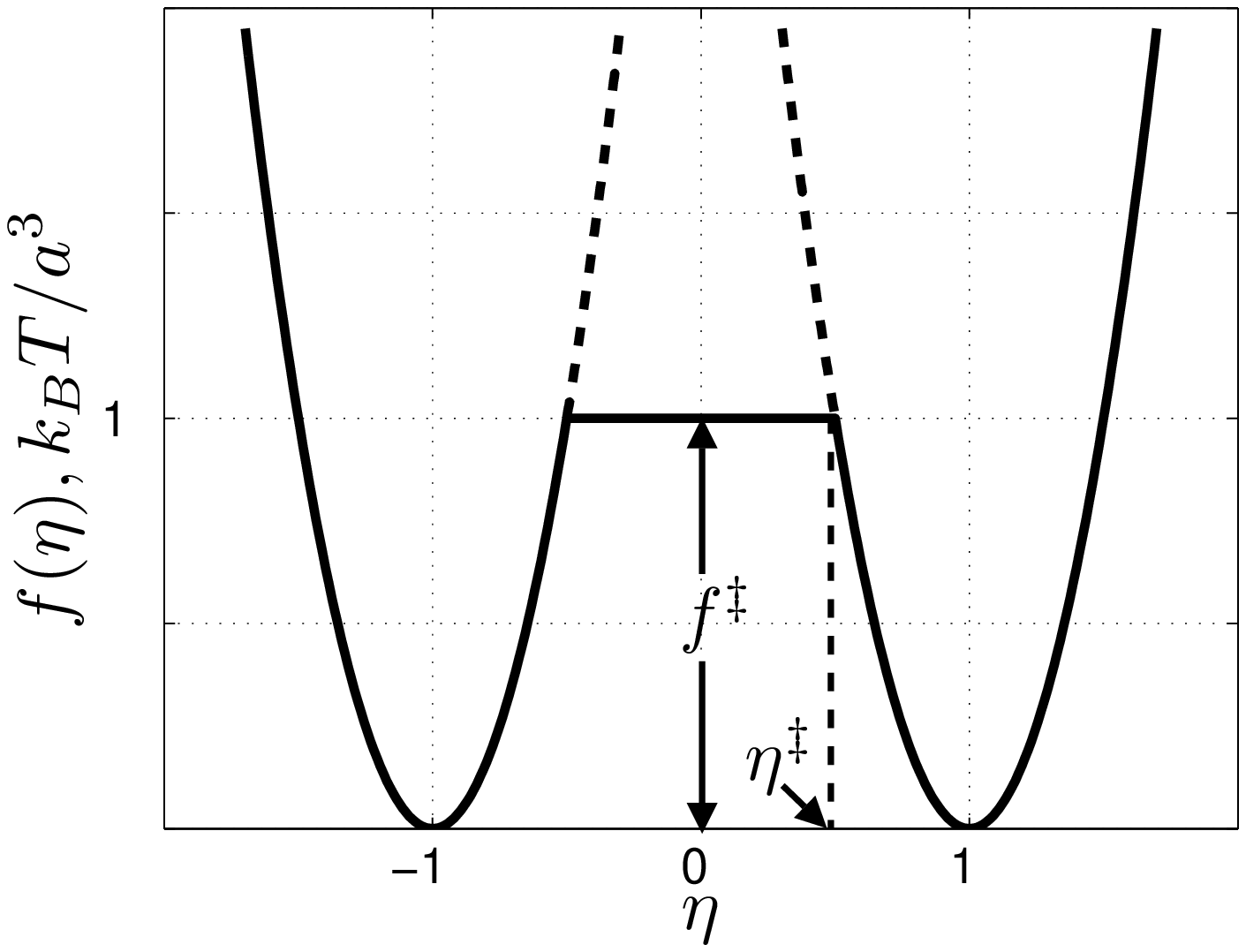}
        \\ {\bf (a)}
      \end{center}
    \end{minipage}
    &
    \begin{minipage}{.48 \figurewidth} 
      \begin{center}
        \includegraphics[width= .46 \figurewidth]{mVSsigmaLabelled.eps} \\
        {\bf (b)}
      \end{center}
    \end{minipage}
  \end{tabular*}
  \caption{\label{sigmaRL} {\bf (a)} The bulk free energy density
    $f(\eta)$ from the Landau-Ginzburg-Cahn-Hilliard density
    functional (\ref{FCH})~\cite{RL_sigma0}. {\bf (b)} The resulting
    predictions for the mismatch penalty $\sigma_0$ at the molecular
    scale~\cite{RL_sigma0}. There are three distinct values for each
    substance that correspond to three distinct ways to determine the
    bead size, see text. The substances can be distinguished by the
    value of the fragility coefficient $m$, except SiO$_2$ and
    GeO$_2$, whose $m$-values are numerically equal. See
    Ref.~\cite{RL_sigma0} for the tabulated values of $\sigma_0$.}
\end{figure}

A reasonable approximation for the bulk free energy density $f(\eta)$,
see Fig.~\ref{sigmaRL}(a), can be obtained by first noticing that the
two aperiodic packings are mechanically stable. Thus for small
deviations of the order parameter from its values $\eta = \pm 1$, the
free energy is quadratic, denote the corresponding curvature by
$m$:
\begin{equation} \label{mLGdef} f(\eta) \approx \frac{m_i(\eta -
    \eta_i)}{2},
\end{equation}
where $\eta_i = \pm 1$.  Already using the quadratic approximation in
the full range of the order parameter---a parabola per each
minima---could suffice so long as the two parabolas cross at a value
below $k_B T/a^3$. Otherwise, the potential $f(\eta)$ must be
corrected to account for the fact the one-particle barrier for a
transition between the crystal and liquid in mutual equilibrium equals
$k_B T$~\cite{L_Lindemann}, which sets the upper bound for the barrier
height in $f(\eta)$. Under these circumstances, it is sufficient to
``cut off'' the portion of the potential in excess of $k_B T/a^3$, see
Fig.~\ref{sigmaRL}(a).  Interestingly, computations of the
coefficients $m$ for actual substances show that the two parabolas
cross at an altitude $f^\ddagger$ that always turns out to be
numerically close to $k_B T/a^3$, at least near $T_g$.

Straightforward calculation shows that~\cite{RL_sigma0}:
\begin{equation}\label{m_fnl}
  m_i = -\frac{k_B T}{4V} \iint  d^3\br_1 d^3\br_2 c_i^{(2)}(\br_1,\br_2) 
  \left[\tilde{\Delta} \rho_1(\br_1) \tilde{\Delta} 
    \rho_1 (\br_2) + \tilde{\Delta}  \rho_2(\br_1) \tilde{\Delta} 
    \rho_2 (\br_2)\right]
\end{equation}
and
\begin{equation} \label{kappa_fnl} \kappa_i = \frac{k_B T}{8V} \iint
  d^3\br_1 d^3\br_2 z^2_{12} c_i^{(2)}(\br_1,\br_2)
  \left[\tilde{\Delta} \rho_1(\br_1) \tilde{\Delta} \rho_1 (\br_2) +
    \tilde{\Delta} \rho_2(\br_1) \tilde{\Delta} \rho_2 (\br_2)\right],
\end{equation}
where $\Delta\rho(\br)\equiv\rho_1(\br)-\rho_2(\br)$,
c.f. Eq.~(\ref{sqGliquid}). As expected, the coefficients at the
quadratic terms in the density functional couple density fluctuations
that correspond to two-body interactions; only fluctuations within
individual states are coupled. The inter-state fluctuations drop out
because the structures are uncorrelated by construction.  Since both
states 1 and 2 are typical, we have $m_1=m_2=m$ and
$\kappa_1=\kappa_2=\kappa$.

Next, one may repeat the analysis analogously to that following
Eq.~(\ref{Fuu}), but also including the shear component of the elastic
free energy, in computing the pairwise correlation function
$\rho^{(2)}(\br_1, \br_2)$. Note that individual aperiodic structures
are mechanically (meta)stable and so $\mu > 0$.  This analysis results
in the sum rule for the solid listed in Eq.~(\ref{sum1}). Combined
with the long-wavelength, Debye approximation for the vibrational
spectrum of our aperiodic solid and the generalised Ornstein-Zernike
relation (\ref{genOZ}), this allows one to write down the following
expression for the Fourier image of the direct correlation function
from Eq.~(\ref{c(r)}):
\begin{equation} \label{c2kk} \hat{c}^{(2)}(\bk_1,\bk_2) \simeq
  -(2\pi)^3\delta(\bk_1+\bk_2) \theta(\pi/a-k) \frac{M c_l^2}{k_B T
    \bar{\rho} N_b},
\end{equation}
Note in this long-wavelength approximation, the direct correlation
function turns out to be translationally-invariant: $c^{(2)}(\br_1,
\br_2) = c^{(2)}(\br_1 -\br_2)$, which is generally not the case for
solids~\cite{PhysRevE.55.4990}. Note that in writing down the equation
above, we have ignored the difference between the isothermal and
adiabatic speeds of longitudinal sound; this difference is not
substantial in solids, see our earlier comments following
Eq.~(\ref{sum1}).

The density-density correlations within each phase are
straightforwardly related to the structure factor, by
Eq.~(\ref{Sq}). Combining this with Eqs.~(\ref{rhoa}), (\ref{c2kk}),
(\ref{m_fnl}), and (\ref{kappa_fnl}) yields the following simple
expressions for the coefficients $\kappa$ and $m$~\cite{RL_sigma0}:
\begin{equation} \label{kappa} \kappa=
  \frac{Mc_l^2a^{-2}S'(\pi/a)}{24N_b},
\end{equation}
and 
\begin{equation} \label{m}
 m= \frac{Mc_l^2}{4\pi^2N_b}\int_0^{\frac{\pi}{a}} S(k)k^2 dk.
\end{equation}
where $S'(k)=dS/dk$. 

Finally, we note that the surface tension coefficient $\sigma_0$
corresponds to the mismatch penalty at the molecular lengthscale,
implying no surface tension renormalisation takes place, as if the
interface were entirely flat.  Combining the standard formula for the
surface tension coefficient for a flat interface~\cite{Bray}, $\sigma
= \int_{-1}^{1} d\eta \left[ 2 \kappa f(\eta)\right]^{1/2}$, and the
two equations above one can write down an explicit expression for the
surface coefficient $\sigma_0$ in terms of measured quantities:
\begin{align} \label{sigma1} & \sigma_0= \frac{k_BT}{a^2} \left[
    \frac{Mc_l^2a^{-1}S'(\pi/a)}{12N_bk_BT} \right]^{1/2} \left\{2 -
    \left[ \frac{Mc_l^2}{8 \pi^2 k_B T N_b} \int_0^{\frac{\pi}{a}}
      S(k)k^2 dk \right]^{-1/2} \right\}, \text{ if } m \ge
  2f^\ddagger \\ \label{sigma2} & \sigma_0=
  \frac{Mc_l^2a^{-1}}{4\sqrt{6} N_b} \left[S'(\pi/a)
    \int_0^{\frac{\pi}{a}} S(k)k^2 dk \right]^{1/2}, \text{ if } m <
  2f^\ddagger
\end{align}
Note that the value of the factor in the curly brackets varies between
1 and 2. 

Despite its relatively computational complexity and requiring the
knowledge of many experimental inputs, the RL formalism leading to
Eqs.~(\ref{kappa}) and (\ref{m}) has a side dividend of allowing one
to compute the correlation length in the landscape regime. By
Eqs.~(\ref{FCH}) and (\ref{mLGdef}), the correlation length is simply
$\sqrt{\kappa/m}$. According to RL's estimates for actual
glass-formers, this length is of molecular dimensions, consistent with
our earlier conclusions that glassy liquids are far away from any sort
of criticality.

The values of $\sigma_0$ predicted by Eqs.~(\ref{sigma1}) and
(\ref{sigma2}) computed using experimentally determined values of the
input parameters are plotted in Fig.~\ref{sigmaRL}(b), where they are
compared with the XW-predicted value $1.85 k_B T/a^2$ from
Eq.~(\ref{sigmaXW}). The filled circles correspond to the calculation
that employs the calorimetric bead count from Eq.~(\ref{Nb}).  As
noted in Ref.~\cite{RL_sigma0}, in instances where the $\sigma_0$
value differed particularly significantly from the XW prediction, the
value of the configurational entropy at $T_g$ also turned out to
differ quite a bit from the XW prediction that $s_c(T_g) \approx
32/\ln(\tau(T_g)/\tau_0)$, see Eq.~(\ref{XWbarrier}). (Such
differences seem to come up often for substances with a significant
degree of local ordering, such as the chalcogenide
alloys~\cite{ZLMicro2}.)  This suggests that the bead count may be off
for these specific substances. To test for the significance of the
precise bead count, Rabochiy and Lubchenko also determined the bead
count self-consistently by requiring that the barrier from
Eq.~(\ref{FXW}) yield exactly $k_B T \ln(3600/10^{-12})$, which
corresponds to the glass transition on 1 hr scale. The resulting bead
count depends on whether the temperature $T_0$ or $T_K$ is used for
the location of the putative Kauzmann state. The values of $\sigma_0$
corresponding to both ways to determine the bead count
self-consistently are shown in Fig.~\ref{sigmaRL}(b) with squares
($T_K$) and triangles ($T_0$). The self-consistently determined bead
count turned out to be chemically reasonable in all cases, except for
OTP.

The general agreement between the $\sigma_0$ values for different
substances and the XW-estimate, even for the calorimetrically
determined bead size, is notable. On the one hand, there would appear
to be little {\em \`a priori} reason for a complicated combination of
several material constants to be so consistent among
chemically-distinct substances.  On the other hand, one expects that
the several approximations involved and some uncertainty in the
experimental input parameters, especially the structure factor, would
introduce a potential source of error in the final result, see
detailed discussion in Ref.~\cite{RL_sigma0}. In any event, it is an
important corollary of the XW calculation that despite apparent
differences in detailed interactions among distinct substances, the
value of $\sigma_0$ is expected to be relatively universal in that it
can be expressed approximately through very few materials constants.

Elastic constants enter the expressions (\ref{sigma1}) and
(\ref{sigma2}), similarly to Eq.~(\ref{gammaFull}) but in a more
complicated way. Still, in Eq.(\ref{sigma1}), which applies when $m
\ge 2f^\ddagger$, i.e., when the ``ledge'' in the $f(\eta)$ is present
(Fig.~\ref{sigmaRL}(a)), the resulting barrier clearly has a
contribution that is linear in the bulk modulus.

\begin{figure}[t]
  \begin{tabular*}{\figurewidth} {cc}
    \begin{minipage}{.48 \figurewidth} 
      \flushleft
      \begin{center} 
        \includegraphics[width=0.48 \figurewidth]{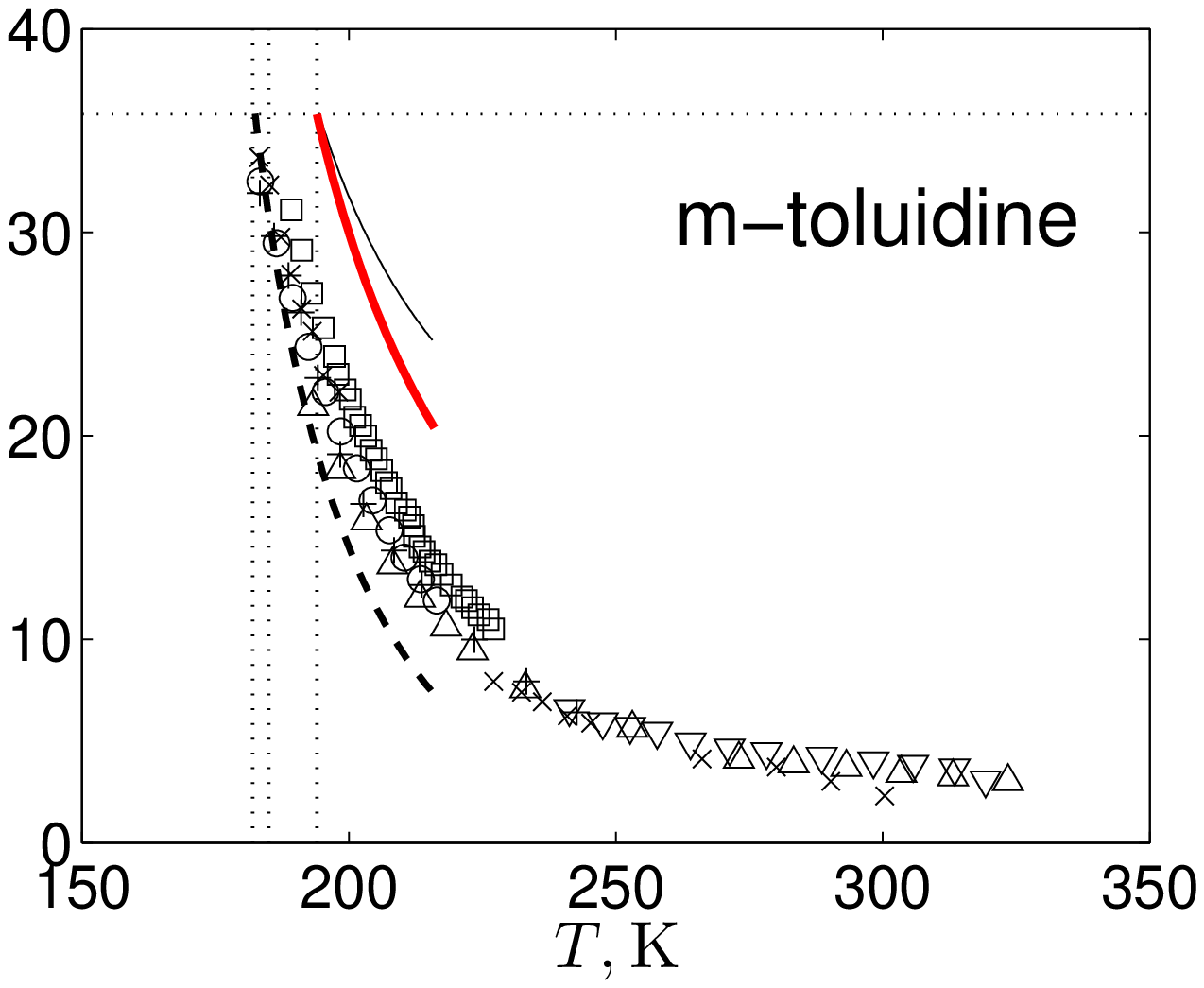}
        \\ {\bf (a)}
      \end{center}
    \end{minipage}
    &
    \begin{minipage}{.48 \figurewidth} 
      \begin{center}
        \includegraphics[width= .48 \figurewidth]{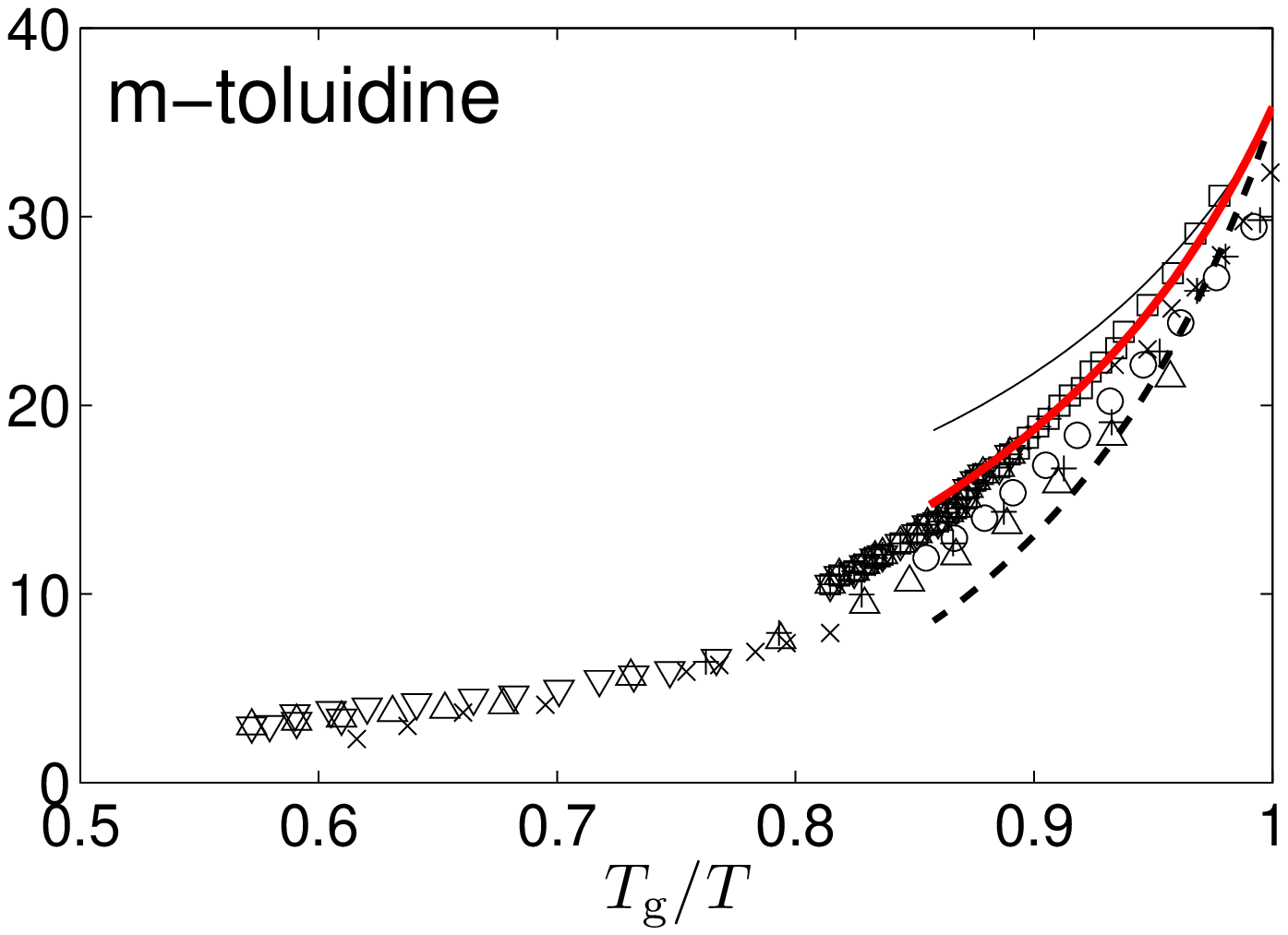} \\
        {\bf (b)}
      \end{center}
    \end{minipage}
  \end{tabular*}
  \caption{\label{barrierT} {\bf (a)} $\alpha$-relaxation barrier
    (divided by $k_B T$) as a function of temperature, computed using
    three distinct approximations for the mismatch penalty, due to Xia
    and Wolynes~\cite{XW} (XW), Rabochiy and
    Lubchenko~\cite{RL_sigma0} (RL), and Lubchenko and
    Rabochiy~\cite{LRactivated} (LR).  {\bf (b)} Same as (a), but as a
    function of the inverse temperature scaled by $T_g$ and adjusted
    so that the barrier at $T_g$ corresponds to 1 hr relaxation time
    (Eq.~(\ref{tauF1}) with $\tau = 3600$~sec and $\tau_0 =
    1$~psec). Both figures are from Ref.~\cite{RWLbarrier}.}
\end{figure}

We have thus discussed three distinct approximations for the mismatch
penalty in this Subsection.  The Xia-Wolynes~\cite{XW} (XW)
approximation (\ref{sigmaXW}) and Rabochiy-Lubchenko~\cite{RL_sigma0}
(RL) approximation (\ref{sigma1})-(\ref{sigma2}) for $\sigma_0$ can be
combined with Eq.~(\ref{FXW}) to compute the free energy barrier for
activated transport in glassy liquids in an extended temperature
range. This was accomplished recently by Rabochiy, Wolynes, and
Lubchenko~\cite{RWLbarrier} (RWL) for eight specific
substances. Likewise, the barrier can be computed, in principle, using
the Lubchenko-Rabochiy~\cite{LRactivated} (LR) approximation for the
coefficient $\gamma$, Eq.~(\ref{gammaFull}), combined with
Eq.~(\ref{F2}). The simplified form (\ref{FKsc1}) only requires the
knowledge of two experimental quantities, which has enabled LR to
estimate the barrier for seven actual substances. In
Fig.~\ref{barrierT}(a) we display the values for the barrier, as a
function of temperature, as predicted by all three approximations for
a specific substance (m-toluidine). We plot the barriers within the
dynamical range representative of actual liquids,
viz. $\ln(\tau/\tau_0) \le 35.7$, which corresponds to a glass
transition on 1 hr time scale and $\tau_0 = 1$~ps. This way, the error
of the approximation exhibits itself through an error in the
temperature corresponding to a particular value of the relaxation
time. As already mentioned, the RL and XW approximations require the
knowledge of the bead size. In addition, the RL approximation requires
the knowledge of the structure factor in an extended temperature
range.  For the lack of this knowledge, the $S(k)$ at a fixed
temperature (usually around $T_g$) were used by RWL, which introduces
another source of uncertainty. Given the aforementioned potential
sources of uncertainty, it is interesting to rescale the computed
barriers by a constant so that the barrier at $T_g$ matches its known
value. This way, one may better judge the error of the approximation
as regards the {\em slope} of the temperature dependence of the
barrier.  We show the so rescaled barriers in
Fig.~\ref{barrierT}(b). The LR approximation does not rely on the bead
size, and so any discrepancy with experiment is due to the error of
the approximation itself and the uncertainty in the experimental
values of the elastic constants and the configurational entropy. The
rescaled LR-based barrier is {\em also} shown in
Fig.~\ref{barrierT}(b). The results in Fig.~\ref{barrierT}(b) are
representative of the results for the rest of the substances analysed
in Refs.~\cite{RWLbarrier} and \cite{LRactivated} in that the XW
approximation tends to underestimate the fragility coefficient in the
{\em extended} temperature range, while the RL-based estimate tends
overestimate the fragility coefficient. (Note the fragility $D$ and
fragility coefficient $m$ anti-correlate!)  None of the three
approximations account for the barrier softening effects
(Section~\ref{crossover}), although the RL and LR approximation may
partially include those effects through the temperature dependence of
the bulk modulus.  The LR-based values appear to fare better than the
other two approximations as far as the slope is concerned, but not the
absolute value of the barrier. Overall, the predictions due to all
three approximations, Fig.~\ref{barrierT}(a), leave room for
improvement but agree with experiment reasonably well given that no
adjustable parameters are used.

\begin{figure}[t] \centering
  \includegraphics[width= .6 \figurewidth]{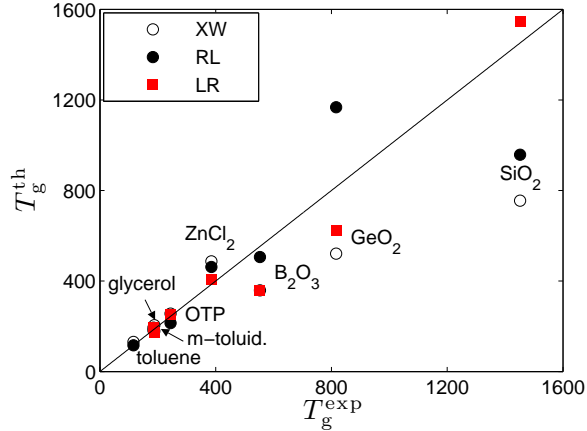}
  \caption{\label{TgTg} The glass transition temperature computed
    using the three distinct approximations for the mismatch penalty,
    due to Xia and Wolynes~\cite{XW} (XW), Rabochiy and
    Lubchenko~\cite{RL_sigma0} (RL), and Lubchenko and
    Rabochiy~\cite{LRactivated} (LR). These theoretically predicted
    glass transition temperatures are plotted against their
    experimental values. }
\end{figure}

It is encouraging that the results of three, rather distinct-looking
approximations for the mismatch penalty yield results that are
numerically similar. These results also imply that there is an
intrinsic connection between the material constants and structure
factor entering in Eqs.~(\ref{gamma1}), (\ref{sigmaXW}-\ref{gammaXW}),
and (\ref{sigma1}-\ref{sigma2}).  Such a connection should have been
expected in view of Eqs.~(\ref{KWalpha}), (\ref{KWalpha1}), and
(\ref{KWalpha3}).

The results displayed in Fig.~\ref{barrierT}(a) are of very special
significance for testing the theory because they do not use any
adjustable parameters. Because no adjustable parameters are involved,
these calculations allow one to {\em predict} the glass transition
temperature, a kinetic quantity, based on measured quantities that are
entirely static. We compile the predictions for the glass transition
temperatures due to the XW and RL approximations, for eight
substances, and due to the LR approximation, for seven substances, all
in Fig.~\ref{TgTg}. We note the larger degree of deviation from
experiment for the stronger substances. This may well be related to
the greater uncertainty in determining the putative Kauzmann
temperature $T_K$ for strong substances, for which $T_K$ differ from
$T_g$ easily by a factor of two. Note that in addition to the
uncertainties inherent in the approximations and in the experimental
input parameters used to compute the barrier, there is also some
uncertainty in the experimentally determined glass transition itself,
as the latter depends on the cooling protocol and sample purity.

\begin{figure}[t]
  \begin{tabular*}{\figurewidth} {cc}
    \begin{minipage}{.48 \figurewidth} 
      \flushleft
      \begin{center} 
        \includegraphics[width=0.48 \figurewidth]{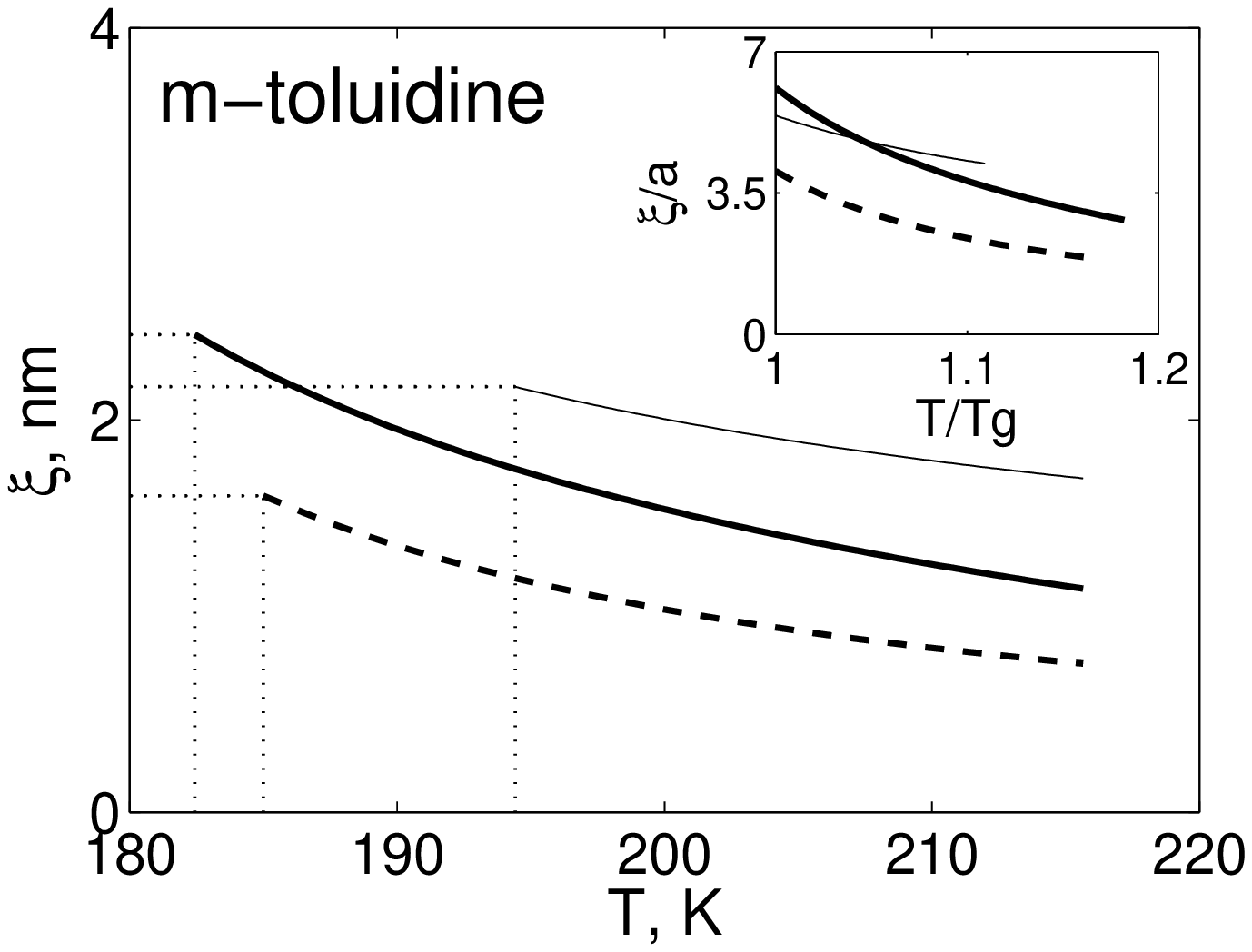}
        \\ {\bf (a)}
      \end{center}
    \end{minipage}
    &
    \begin{minipage}{.48 \figurewidth} 
      \begin{center}
        \includegraphics[width= .46 \figurewidth]{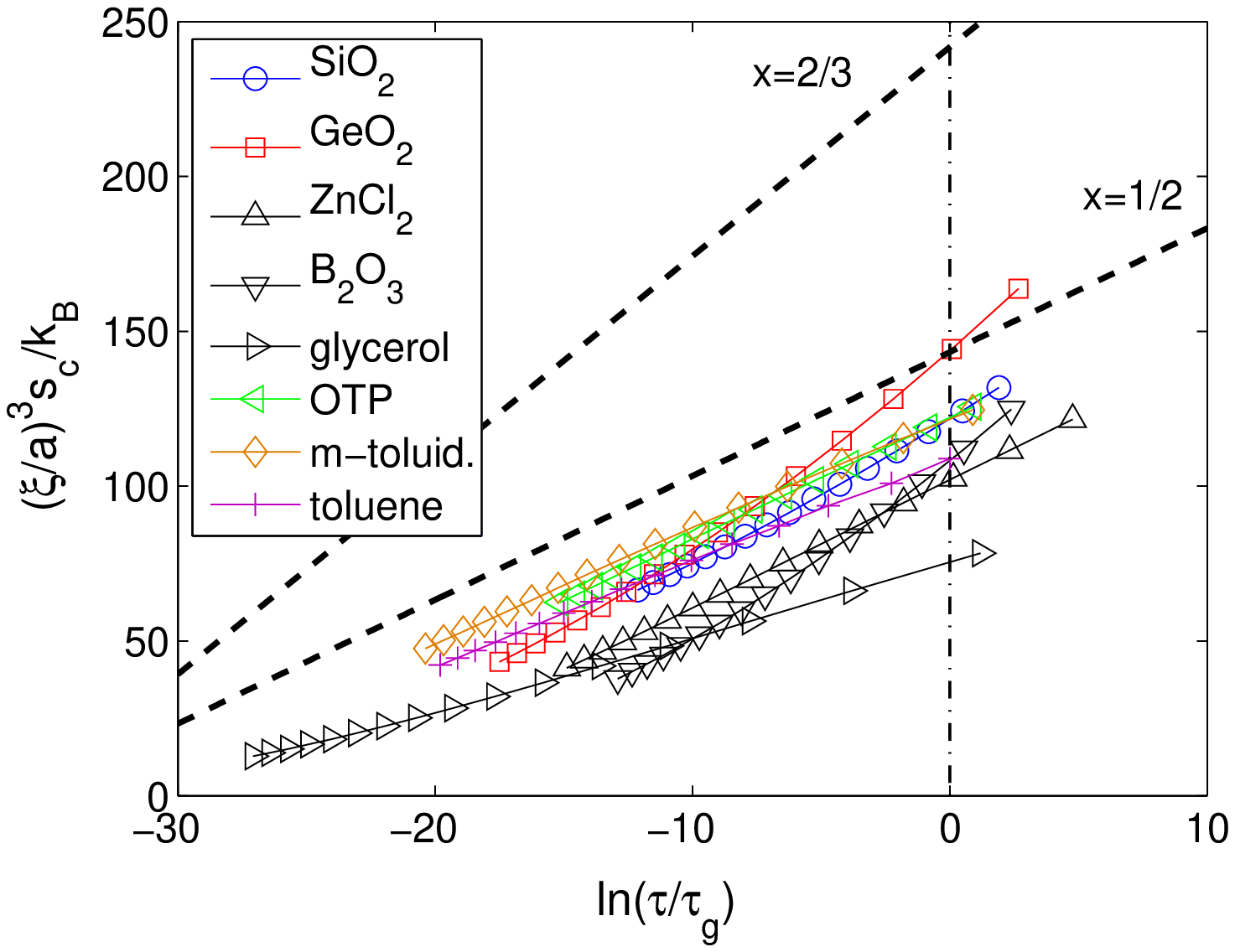} \\
        {\bf (b)}
      \end{center}
    \end{minipage}
  \end{tabular*}
  \caption{\label{xiT} {\bf (a)} The cooperativity length $\xi$ as a
    function of temperature computed using Eq.~(\ref{xiXW}) with the
    value of the surface tension coefficient $\sigma_0$ estimated
    using the XW (thin solid line) and RL (thick solid line)
    approximation. The dashed line corresponds to the cooperativity
    length $\xi$ estimated as in Ref.\cite{Capaccioli}, according to
    the procedure by Berthier et al.~\cite{Berthier}. {\bf (b)} The
    complexity of a rearranging region $s_c (\xi/a)^3$ is plotted as a
    function of $\ln (\tau/\tau_g)$, where $\tau_g$ is the relaxation
    time $\tau$ at $T_g$.  The cooperativity length $\xi$ in this
    graph is estimated as in Ref.\cite{Capaccioli}, except here we use
    the Xia-Wolynes-Lubchenko expression (\ref{XWLbeta}) for the
    temperature dependence of the stretching exponent $\beta$
    normalised so that it matches its experimental value at $T_g$.}
\end{figure}

Also of particular significance is the ability of these RFOT-based
methodologies to predict the cooperativity length $\xi$. In
Fig.~\ref{xiT}(a), we show the temperature dependence of $\xi$ for
m-toluidine, as computed according to Eq.~(\ref{xiXW}) with the help
of the XW and RL approximations for $\sigma_0$ (thin and thick solid
lines respectively). The dashed line shows the dynamical correlation
length $\xi$ computed according to the procedure of Berthier et
al.\cite{Berthier, Capaccioli}:
\begin{equation} \label{xiBerthier} \xi_\text{B}/a = \left\{
    \frac{1}{\pi} \left[ \frac{\beta}{e} \frac{\prtl \ln \tau}{\prtl
        \ln T} \right]^2 \frac{k_B}{\Delta c_p} \right\}^{1/3},
\end{equation}
where $\beta$ is the stretching exponent in the stretched exponential
relaxation profile: $e^{-(t/\tau)^\beta}$.  (The quantity $\beta$, not
to be confused with the inverse temperature $\beta \equiv 1/k_B T$,
will be discussed in Section~\ref{hetero}). The dynamical correlation
length from Eq.~(\ref{xiBerthier}) is an experimentally-inferred lower
bound on the cooperativity length. As before, $\Delta c_p$ is the
(temperature dependent) excess liquid heat capacity.

First we note that the cooperativity length is on the order of a
nanometre, consistent with experimental determinations using
non-linear spectroscopy~\cite{Spiess, RusselIsraeloff,
  CiceroneEdiger}, and also {\em direct} observation of cooperative
reconfigurations on the surface of metallic
glasses~\cite{GruebeleSurface}. Furthermore, the temperature
dependence of $\xi$ is also clearly consistent with that determined
according to the Berthier et al. procedure~\cite{Berthier}.  The
dimensionless measure of the spatial extent of cooperativity $\xi/a$
is shown in the inset of Fig.~\ref{xiT}(a). According to
Eq.~(\ref{NXW}), the cooperativity size at the glass transition
depends only the quench speed and only logarithmically at that.  Thus
it is expected to be quite consistent between different
substances. For the glass transition on 1 hr timescale, the
cooperativity size $N^*$ is about $[\ln(3600/10^{-12})/2.83]^2 \simeq
160$ beads. For the glass transition on the timescale of $10^5$~sec,
this size is $190$ or so.

Finally we calculate the so called {\em complexity} of a rearranging
region, which is defined as the amount of configurational entropy
contained within the region, i.e, $N^* s_c$. This quantity is of
interest partially because it is independent of the bead count.
Indeed, dividing Eq.~(\ref{FXW}) by (\ref{xiXW}), one obtains that the
complexity of a rearranging region is simply the barrier, in units of
$k_B T$, times a factor of four:
\begin{equation} \label{compl} N^* s_c/k_B \equiv (\xi/a)^3 s_c /k_B =
  4 \, (F^\ddagger/k_B T) = 4 \ln(\tau/\tau_0)
\end{equation}
It turns out that similarly universal relationship between the
complexity and the reconfiguration barrier exists for all values of
the exponent $x$ that gives the scaling of the mismatch penalty with
the region size, Eq.~(\ref{Gamma1}).  Indeed, Eqs.~(\ref{Nstar1}) and
(\ref{F1}) yield:
\begin{equation} \label{complX} N^* s_c/k_B = (F^\ddagger/k_B T)
  \left[ x^{x/(1-x)}(1-x) \right]^{-1}.
\end{equation}
The relationships in Eqs.~(\ref{compl}) and (\ref{complX}) can be
interpreted as saying that rearrangement involves searching through
all the states of a fixed fraction of the rearranging
region~\cite{WolynesNIST}. Note the numerical coefficient in
Eq.~(\ref{complX}) does {\em not} depend on the magnitude of surface
tension coefficient $\sigma_0$, but is quite sensitive to the precise
scaling of the mismatch penalty with the droplet size.

The relationship in Eq.~(\ref{compl}) was tested by Capaccioli,
Ruocco, and Zamponi~\cite{Capaccioli} for a large number of actual
substances using the cooperativity length from
Eq.~(\ref{xiBerthier}). Here we show the complexities for the eight
substances from Fig.~\ref{TgTg} as functions of $\ln(\tau/\tau_g)$,
where $\tau_g \equiv \tau(T_g)$, see Fig.~\ref{xiT}(b). In contrast
with Ref.~\cite{Capaccioli}, here we employ a temperature dependent
stretching exponent $\beta$, Eq.~(\ref{XWLbeta}).  The dashed line in
Fig.~\ref{xiT}(b) corresponds to the universal prediction by the RFOT
theory, Eq.~(\ref{compl}). This result does not depend on substance.

The data in Fig.~\ref{xiT}(b) demonstrate that the RFOT-based
estimates of the complexity are consistent with experiment; we remind
the reader that the length $\xi_\text{B}$ is a lower bound. The
complexity graph also allows one to test the RFOT-predicted value of
the scaling exponent $x$ for the mismatch penalty from
Eq.~(\ref{Gamma1}). Indeed, if the exponent were equal to its
conventional value $2/3$, the numerical coefficient $\left[
  x^{x/(1-x)}(1-x) \right]^{-1}$ in Eq.~(\ref{complX}) would be equal
to $6.75$, which significantly exceeds its value of $4$ for $x=1/2$,
see Fig.~\ref{xiT}(b). The exponent $1/2$ is clearly much more
consistent with observation.

The complexity trends exhibited by actual liquids directly contradict
the basic premise of the Adam-Gibbs (AG) argument~\cite{AdamGibbs} on
the connection between the configurational entropy and relaxation
rates in liquids. In short, Adam and Gibbs posited that the
reconfiguration barrier is determined by the enthalpic cost of
deforming a region of a specific size. AG argued the size should be
such that the liquid has exactly two structural states available to
the region: $N^* s_c = \ln 2$.  Such a view would be probably adequate
at high temperatures, however clearly fails for actual substances:
First, given that $s_c$ is of order $k_B$ per particle, the AG
argument predicts the dynamical size that is too small. Second, the
total entropy content of a reconfiguring region, $N^* s_c$, clearly
depends on the temperature, in contrast with the premise of Adam and
Gibbs.  Aside from these shortcomings, the AG ideas must be credited
for bringing attention to the possibility of an intrinsic connection
between the thermodynamics and kinetics in glassy liquids and for
spurring much experimental work that has studied this connection
empirically.  In any event, it should be clear to the reader that the
RFOT theory is not a variation on the phenomenological Adam-Gibbs
scenario, as is often mistakenly stated in the literature, but is a
constructive argument.

On a somewhat related note, the data in Fig.~\ref{xiT} can be used to
refute the premise of the so called ``shoving model,'' which
postulates that the reconfiguration barrier is determined by the
product of an elastic modulus and the volume of the rearranging
region, which is assumed to be fixed~\cite{PhysRevB.53.2171,
  dyre:224108, klieber:12A544}, in direct contradiction with
Eq.~(\ref{NXW}), which predicts the region's volume changes by at
least factor of a few hundred, within the dynamical range of a typical
experiment. (The quantity $\ln(\tau/\tau_0)$ changes between 0 and 35
or so, in Eq.~(\ref{NXW}). Alternatively, the configurational entropy
per bead changes between 1.5 and 0.8 or so.)

\begin{figure}[t]
\begin{tabular*}{\figurewidth} {ll}
\begin{minipage}{.46 \figurewidth} 
  \begin{center}
    \includegraphics[width= .5 \figurewidth]{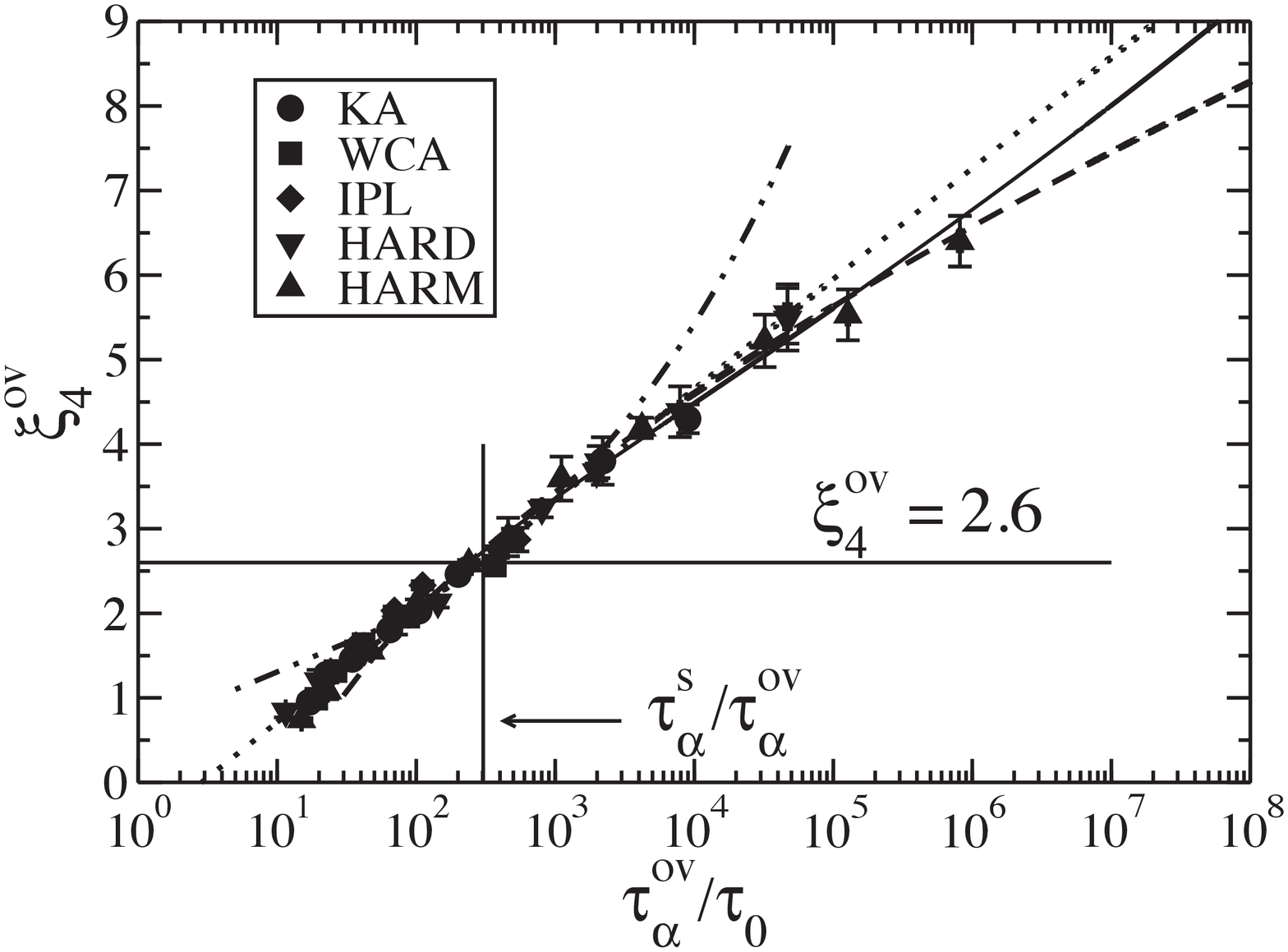}
    \\ {\bf (a)}
   \end{center}
  \end{minipage}
&
\begin{minipage}{.46 \figurewidth} 
  \begin{center}
    \includegraphics[width=0.5 \figurewidth]{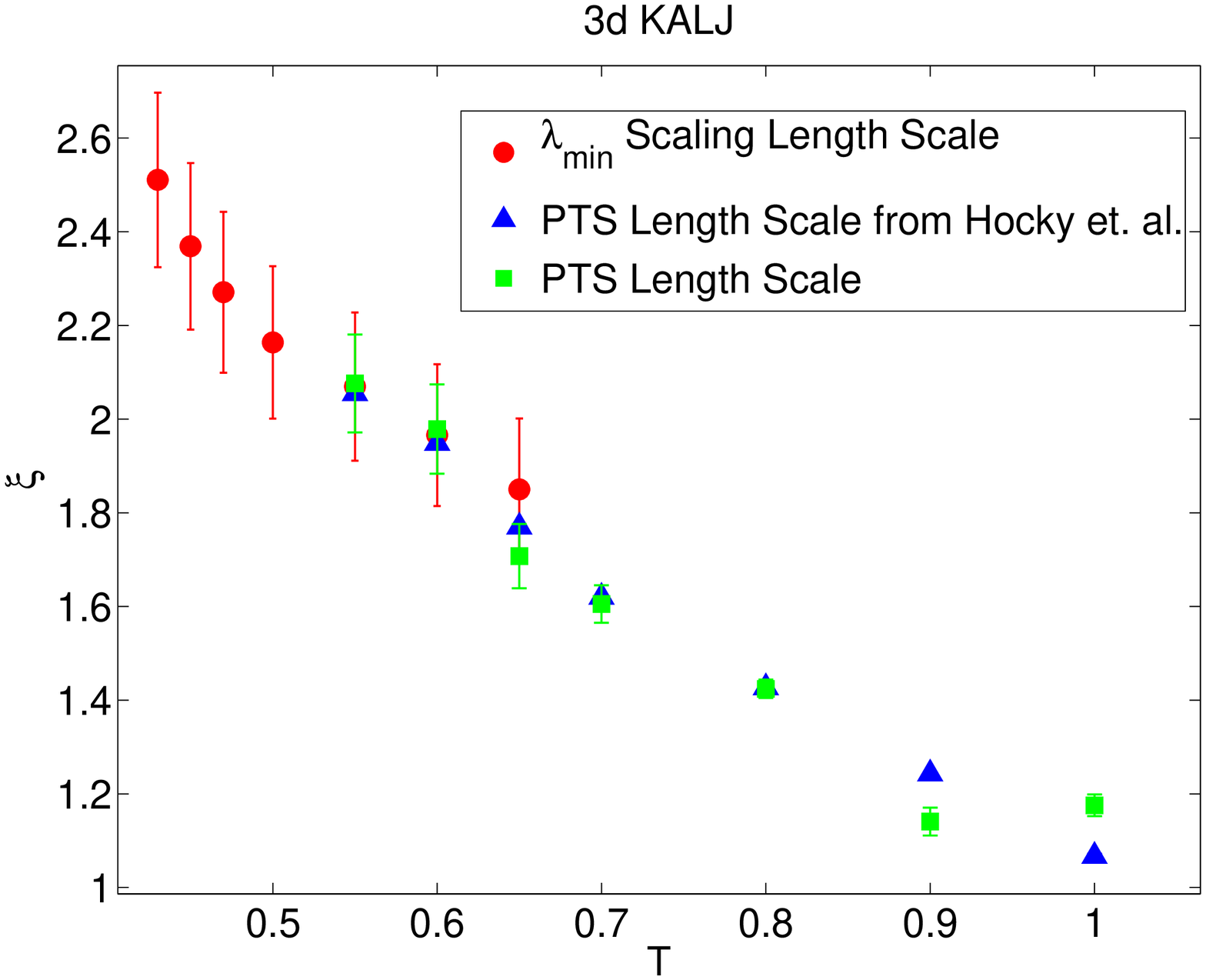}
    \\ {\bf (b)}
  \end{center}
\end{minipage}
\end{tabular*}
\caption{\label{szamel} {\bf (a)} The dependence of the four-point
  correlation length on the relaxation time, according to
  computational studies by Flenner, Staley, and
  Szamel~\cite{PhysRevLett.112.097801}. The fitted solid line has
  slope $1$, the dashed line slope $2/3$. {\bf (b)} Comparison between
  the point-to-set length and the extent $\lambda_\smin$ of marginally
  stable harmonic modes in the inherent structures derived from
  samples equilibrated at temperature $T$, from Biroli, Karmakar, and
  Procaccia~\cite{PhysRevLett.111.165701}.  The blue triangles give
  data due to Hocky, Markland, and
  Reichman~\cite{PhysRevLett.108.225506}.}
\end{figure}

Despite system-specific deviations of the predicted values for the
barrier and the cooperativity size from experiment, it is clear that
the predictions of the RFOT theory are consistent with observation
both with regard to the absolute value of the parameters and the key
scaling relationships. Given this consistency with measurement on
actual substances, the reader may reasonably wonder if similar trends
have been observed in computer simulations of glassy liquids.  It
appears that at present, simulations are largely in the mode-coupling
regime and are only beginning to reach into the activated regime.
Recent simulations on a variety of model binary mixtures by Flenner et
al.~\cite{PhysRevLett.112.097801}, indicate that the dependence of the
dynamic correlation length $\xi_4$ on $\log \tau$ can be well fit by a
power law with the exponent $1$ over the full temperature
range. Still, the low-temperature flank of the data are more
consistent with the exponent $2/3$, see Fig.~\ref{szamel}. This is
exactly what the RFOT theory predicts for the scaling of the
cooperativity length $\xi$ in the activated transport regime, see
Eq.~(\ref{NXW}). Note that for the conventional scaling of the
mismatch with the region size $x=2/3$, the exponent on the Flenner et
al. plot would be equal $1/2$, by Eqs.~(\ref{F1}) and (\ref{xi}). The
four-point correlation length $\xi_4$ from Fig.~\ref{szamel} is
readily evaluated in simulations and, at the same time, is naturally
computed in field-theoretical approaches such as the mode-coupling
theory. The latter, however, has not been extended to the landscape
regime characterised by an exponential multiplicity of free energy
minima. In meanfield, the correlation length $\xi_4$ of the MCT
diverges at the temperature $T_A$ thus signalling the emergence of
macroscopic rigidity. As already mentioned and will be discussed in
more detail in Section~\ref{crossover}, the establishment of such
macroscopic rigidity is avoided, owing to the activated
reconfigurations, because the metastable structures do not extend
beyond the reconfiguration size $\xi$. Consequently, while the length
$\xi$ represents a trivial lower bound on $\xi_4$ above the crossover,
it is a non-trivial {\em upper} bound on $\xi_4$ sufficiently {\em
  below} the crossover. A semi-phenomenological way to estimate
$\xi_4$ within the RFOT theory, through spatial correlations in the
mobility field, has been discussed by Wisitsorasak and
Wolynes~\cite{doi:10.1021/jp4125777} (see their Eqs.~(14)-(16)); it
appears from the latter work that the lengths $\xi_4$ and $\xi$ are
asymptotically equal as $T$ tends to $T_K$.  The mobility field that
will be discussed in the context of glass rejuvenation, see
Section~\ref{rejuvenation}.

One way to quantify the spatial extent of the cooperativity---in a
simulation---that is the closest in spirit to the present formalism
comes about in the library~\cite{LW_aging} or the Bouchaud-Biroli
construction~\cite{BouchaudBiroli}, but goes back to the concept of
the entropic droplet developed by Wolynes and coworkers in the
1980s~\cite{MCT1, KTW}. Take a typical configuration and fix particles
outside a compact region, up to vibration. Now perform this procedure
for regions of various size. The size at which exactly two states are
available corresponds with the cooperativity length of the library
construction and is often called the point-to-set length
$\xi_\text{PS}$ . Alternatively, one may fix particles outside of a
compact region not at positions corresponding to an equilibrated
liquid but, instead, at some generic positions. This length can be
shown to be equivalent to $\xi_\text{PS}$ for a large enough
region~\cite{PhysRevLett.111.107801}. Equilibration of the thus
confined liquid is computationally difficult, and so only rather small
system sizes have been investigated. At any rate, the cooperativity
length does increase with lowering temperature, see
Fig.~\ref{szamel}(b). This confirms the basic notion that the lower
the temperature, the fewer states are available to a liquid region and
the larger region must rearrange to find an alternative structural
state. Incidentally, the relative difficulty of equilibrating a region
in a confined geometry, as compared with periodic boundary conditions,
is consistent with the earlier comment that the present day
simulations are only beginning to enter the landscape regime. Indeed,
subjecting a liquid to a rough wall is equivalent to imposing a
quasi-static field similar to that imposed by a long-lived structure.

We thus observe that a finite-size system with pinned surroundings can
in fact run out of configurational degrees of freedom---thus
effectively undergoing a Kauzmann-like crisis---at a {\em finite}
value of the {\em bulk} configurational entropy. According to the
library (or Bouchaud-Biroli) construction, this finite value of the
configurational entropy, $s_c^*$, can be evaluated with the help of
Eq.~(\ref{Nstar1}) by setting $N^*$ at the system size $N$, thus
yielding $s_c^* = \gamma/T N^{1-x}$. Whether or not the true Kauzmann
crisis, at $N \to \infty$, could truly take place is probably
system-specific. We have seen that such crises do take place in
mean-field models. It appears likely that chemically homogeneous
systems without an obvious periodic ground state, such as atactic
polymers, may reach states with very low values of $s_c$. Otherwise,
some local ordering likely takes place before the thermodynamic
temperature $T_K$ can be reached, in principle.

A cautionary remark concerning common model glass-formers employed in
simulational studies is due here.  Such model liquids are usually
heterodisperse mixtures with a specially chosen size mismatch so as to
avoid crystallisation. Such specially chosen mixtures exhibit various,
fascinating types of local ordering~\cite{Royall2013,
  2014arXiv1405.5691R, PhysRevE.90.032311, 4874755520100401,
  Hirata2013}. The slow dynamics in such mixtures can thus be viewed
as motions of twin-like objects.  Naturally, such motions are subject
to high barriers. These notions are consistent with our earlier
remarks on destabilisation of the landscape regime, relative to the
uniform liquid, that takes place in eutectic-like mixtures while
little changes in the viscosity take place, see
Subsection~\ref{MCT}. It is an interesting question as to how much
local ordering of this type, if any at all, takes place in actual
glassformers. While little crystallites are constantly being formed in
liquids below the melting temperature, they should be very
short-lived. In addition, in order for a well-ordered local structure
to rearrange, bonds may have to deform to a greater degree than is
prescribed by Eq.~(\ref{dLa10}).  Now, the tendency to form local
patterns with very distinct orientational ordering in binary mixtures
stems from the discrete nature of particles, of course. The precise
relation between this discrete phenomenon, which may or may not occur
in actual glass-formers, and the essentially continuum nature of the
RFOT theory has not been elucidated to date. In this regard, early
work on icosahedral order comes to mind~\cite{PhysRevB.32.1480}.

We finish this Section by summarising the main features of the
activated transport in glassy liquids.  The transport is realised
through local activated events. These events occur in a
nucleation-like fashion and are cooperative; they involve between a
few tens and hundreds of particles for realistic quenching
schedules---bigger cooperativity sizes are expected for slower
quenches. Reconfigurations are strongly anharmonic events, according
to the free energy profile in Fig.~\ref{FNgraph}(a), yet local
deformations are harmonic as individual atoms move relatively little
during the reconfiguration, just in excess of the typical vibrational
amplitude, see Fig.~\ref{FNgraph}(b). Last but not least, the simple
expressions Eqs.~(\ref{XWbarrier}) and (\ref{FKsc1}), which quantify
the escape barrier from the current structure, reveal the essential
feature of the free energy landscape of a glassy liquid. The lower the
temperature, the fewer ``valleys'' the landscape exhibits, while their
depth becomes greater.


\section{Dynamic heterogeneity}
\label{hetero}

The central feature of the random first order transition is the
multiplicity of the distinct aperiodic arrangements available to the
liquid below the crossover to the activated transport.  This
multiplicity constitutes the thermodynamic driving force for the
activated transport and directly enters the expression for the
activation barrier. The {\em distribution} of the bulk free energy of
distinct stable structures also has observable consequences.  One of
these is the relatively slow scaling of the mismatch penalty between
distinct aperiodic state with the region size.  We have seen that the
long-lived glassy structures can be thought of as stabilising free
energy fluctuations.  Such fluctuations naturally produce not one, but
a variety of structures. The resulting structural heterogeneity in
glassy liquids is subtle in that there seems to be no static length
associated with it~\cite{0295-5075-98-3-36005}, consistent with the
scale-free character of the generic Gaussian fluctuations. The
heterogeneity is {\em dynamic} in that it can be detected by
monitoring the activated reconfigurations in real time, which requires
either direct observation~\cite{GruebeleSurface} or non-linear
spectroscopy~\cite{Spiess, RusselIsraeloff, CiceroneEdiger}.  This
rather subtle type of dynamic heterogeneity is distinct from but is
often confused with a simpler kind of dynamic heterogeneity, which is
{\em also} intrinsic to glassy liquids and leads to observable
consequences in terms of spatially distributed relaxation rates. This
we discuss next.

\subsection{Correlation between non-exponentiality of 
liquid relaxation and fragility}
\label{betaDcorr}

According to Eqs.~(\ref{F1}) and (\ref{FXW}), the reconfiguration
barrier is determined by the entropy content of the local region. If
fewer states are available, the escape time from the current
configurations is longer. This is straightforward to rationalise using
the landscape picture of the free energy surface of glassy liquids:
Because the entropy is a monotonically increasing function of
enthalpy, entropy-poor states are also likely to be correspond to
deeper ``valleys'' in the landscape, thus implying a higher barrier
must be surmounted in order to escape from the valley. Thus, by
Eqs.~(\ref{F1}) and (\ref{FXW}), the escape barrier---and hence
relaxation rate---will fluctuate in reflection of local fluctuations
in the configurational entropy. Xia and Wolynes~\cite{XWbeta}
exploited this idea to estimate a lower bound on the barrier
fluctuation in glassy liquids. (Local fluctuations in the mismatch
penalty, if any, will also contribute to the barrier fluctuation.) By
this argument, the relative value of local fluctuation in the escape
barrier is given by:
\begin{equation} \label{relf} \frac{\delta
    F^\ddagger}{F_\smp^\ddagger} = \frac{\delta S_c^*}{S_c^*} =
  \frac{1}{2 \sqrt{D}},
\end{equation}
where $S_c^* \equiv s_c N^*$ is the configurational entropy of a
single cooperative region and the index ``mp'' signifies ``most
probable.''  In the first equality, we assume that $S_c^*$ is
Gaussianly distributed. The second equality gives the relative
fluctuation of the configurational entropy for a region of size $\xi$
and is seen as follows: On the one hand, $\delta S_c^* = \sqrt{ k_B
  \Delta C_p^*} = \sqrt{k_B \Delta c_p N^* }$, based on the general
expression for entropy fluctuations: $\delta S = \sqrt{k_B
  C_p}$~\cite{LLstat} and remembering that $\Delta C_p$ corresponds to
the excess liquid heat capacity relative to the crystal. On the other
hand, $s_c \sqrt{N^*} = (\gamma/T)$ by Eq.~(\ref{Nstar1}) with
$x=1/2$. Combined with Eqs.~(\ref{scT}) and (\ref{VFT}), this yields
the second equality in Eq.~(\ref{relf}).  It is interesting that the
value $x=1/2$ of the scaling exponent for the mismatch penalty is
unique in that it is the only value at which the relative value of
entropy fluctuations of a cooperative region is independent of its
size: $\delta S_c^*/S_c^* \propto (T-T_K)^{(2x-1)}$. (For the latter
estimate, one has to assume the ``generalised'' form of the VFT law
$\tau = \tau_0 \exp[D/(T/T_K -1)^{x/(1-x)}]$, see Eq.~(\ref{F1})).

\begin{figure}[t]
  \begin{tabular*}{\figurewidth} {ll}
    \begin{minipage}{.46 \figurewidth} 
      \begin{center}
        \includegraphics[width= .46 \figurewidth]{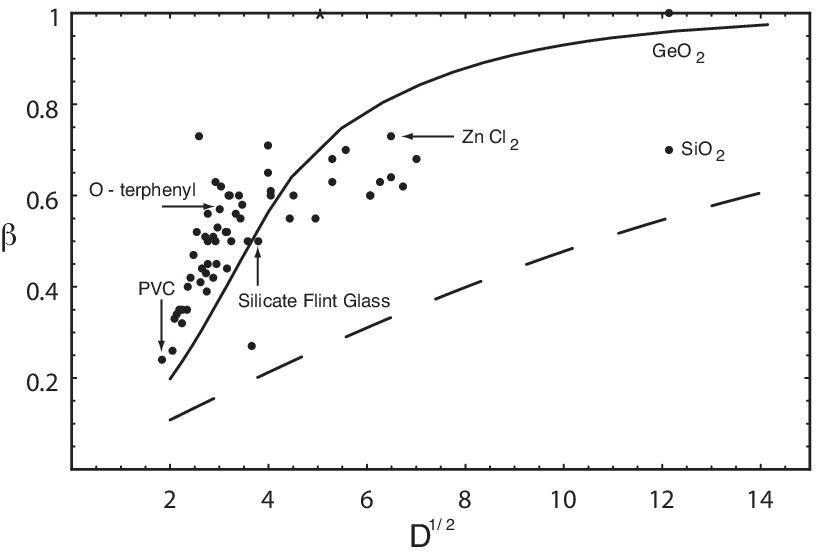}
        \\ {\bf (a)}
      \end{center}
    \end{minipage}
    &
    \begin{minipage}{.46 \figurewidth} 
      \begin{center}
        \includegraphics[width=0.46 \figurewidth]{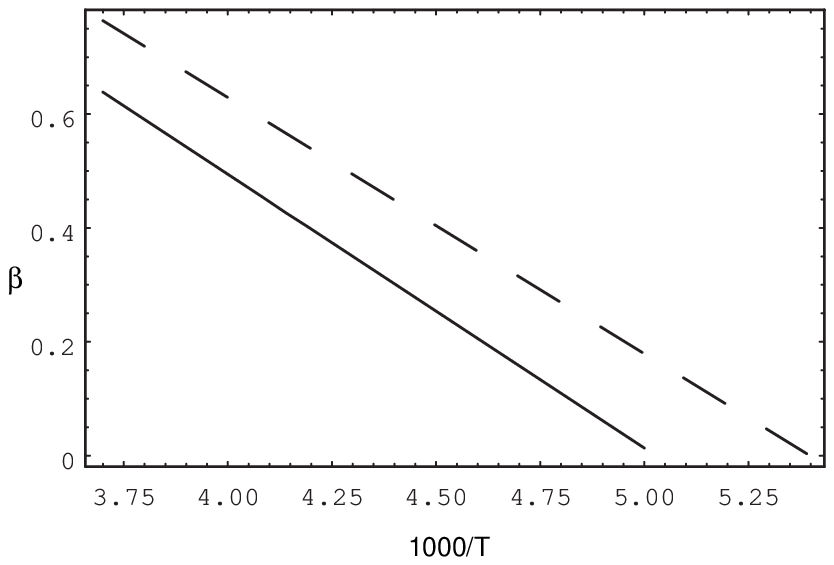}
        \\ {\bf (b)}
      \end{center}
    \end{minipage}
  \end{tabular*}
  \caption{ {\bf (a)} \label{XWbetaD} The exponent $\beta$ from the
    stretched-exponential relaxation profile $e^{-(t/\tau)^\beta}$
    characteristic of glassy liquids plotted against the square root of
    fragility $D$, as predicted by the RFOT theory~\cite{XWbeta}. The
    dashed line corresponds to the simplest assumption of the
    Gaussianly distributed barrier, while the solid line is obtained
    using the more complicated form in Eq.~(\ref{pF}) that takes into
    account the ``facilitation'' effect, by which a very slow region
    is unlikely to be exclusively surrounded by comparably slow
    regions and thus will relax not slower than a typical region. {\bf
      (b)} Solid line: RFOT-predicted temperature dependence of
    $\beta$ for a fragile glass-former~\cite{XWbeta}. Dashed line:
    experimental measurements of Dixon and Nagel and their
    extrapolation to lower temperatures for
    o-terphenyl~\cite{DixonNagel}. }
\end{figure}

Assuming the barrier distribution is Gaussian, it is straightforward
to show~\cite{XWbeta} that the exponential relaxation profile averaged
over the barrier distribution yields approximately a
stretched-exponential:
\begin{equation} \la e^{-t/\tau} \ra \approx e^{-(t/\tau_\smp)^\beta},
\end{equation}
where the stretching exponent is determined by the width of the
barrier distribution and temperature: $\beta = [1 + (\delta
F^\ddagger/k_B T)^2]^{-1/2}$ thus yielding
\begin{equation} \label{betaD} \beta = \left[ 1 + \left(
      \frac{F^\ddagger_\smp/k_BT}{2 \sqrt{D}} \right)^2
  \right]^{-1/2}.
\end{equation}
In view of the near universality of the $F^\ddagger_\smp/k_BT$ ratio
near the glass transition, the above equation implies the stretching
exponent $\beta$ is a simple function of the fragility $D$. The
function is particularly simple for fragile substances, for which $D$
is relatively small: $\beta \propto \sqrt{D}$.  The predictions due to
Eq.~(\ref{betaD}) for the glass transition on $10^4$~sec timescale,
$F^\ddagger_\smp/k_BT \simeq \ln(10^4/10^{-12}) \approx 37$, are shown
in Fig.~\ref{XWbetaD}(a) with the dashed line. The agreement is only
qualitative; the lack of quantitative agreement can be explained by
the neglect of the following, ``facilitation''-like effect. XW argued
that a region can be considered unrelaxed insofar as the particles in
both the region {\em and} its immediate environment have not moved;
this is non-meanfield effect. A given region may be very slow, but the
probability that the immediate surrounding is {\em also} slow is
small. For the same reason in a coin-flipping experiment, the
probability of the persistence of a particular pattern, such as many
tails in row, is small.  XW thus argued that using a Gaussian
distribution for the barriers in excess of the most probable value of
the distribution is too simplistic. The simplest way to fix this is to
replace the r.h.s. of the barrier distribution by a delta function
with the area equal to $1/2$ and centred at $F^\ddagger_\smp$:
\begin{equation} \label{pF} p_1(\wF) = \frac{e^{-(1/\wF-1)^2/2\delta
      \wF^2}}{\sqrt{2 \pi (\delta \wF)^2} \wF^2} +
  \frac{1}{2}\delta(\wF-1),
\end{equation}
where $\wF \equiv F/F_\smp < 1^+$, and we took advantage of the
temperature-independence of the relative width in Eq.~(\ref{relf}).

The resulting prediction for the $\beta$ vs. $D$ relation is shown as
the solid line in Fig.~\ref{XWbetaD}(a). The agreement is now much
improved. Lubchenko~\cite{Lionic} has pointed out, in a different
context, that an exponential barrier distribution yields a comparable
agreement with experiment, while also agreeing with the
frequency-dependent dielectric response $\epsilon''(\omega)$. The
width of this distribution is approximately $\delta F^\ddagger/4$,
i.e.  the average between the l.h.s. of the original Gaussian and the
delta function:
\begin{equation} \label{pFm} p(\wF) = \left\{ \begin{array}{ll}
      \frac{c_1}{\wF^2} e^{-(1/\wF-1)^2/2\delta \wF^2}, & \wF \le \wF_e\\
      \frac{c_2}{\wF^2} e^{\wF/(\delta \wF/4)}, & \wF_e < \wF \le 1,
    \end{array}
  \right.
\end{equation}
This leads to a simple formula that works as well as the more
complicated XW form consisting of a half-gaussian and a delta
function:
\begin{equation} \label{XWLbeta} \beta = \left[ 1 + \left(
      \frac{F^\ddagger_\smp/k_BT}{8 \sqrt{D}} \right)^2
  \right]^{-1/2}.
\end{equation}

Regardless of the size of the facilitation effects and how they are
treated, the RFOT theory unambiguously predicts that given the same
value of the relaxation time (or viscosity), the stretching exponent
$\beta$ should be smaller for more fragile substances. At the same
time, relaxations in the uniform liquid are exponential. This implies
that the temperature dependence of $\beta$ will the more pronounced
the more fragile the substance.  We note that the questions of the
determination of both the stretching exponent $\beta$ and its
temperature dependence are not without controversy. Xia and Wolynes
predictions for the temperature dependence of $\beta$ for a very
fragile substance are shown in Fig.~\ref{XWbetaD}(b) alongside
experimental data for OTP~\cite{DixonNagel}. These results are also
consistent with large-scale simulations of OTP by Eastwood et
al.~\cite{doi:10.1021/jp402102w}. Now, the temperature dependence of
$\beta$ in many substances seems to deviate
significantly~\cite{Richert_Wiley} from the steady dependence seen in
Fig.~\ref{XWbetaD}(b).  This can be rationalised by the presence of
beta-relaxations which overlap frequency-wise with alpha-relaxation,
see Fig.~\ref{LL}. Indeed, in the absence of intervening processes,
the quantity $\beta$ can be inferred by either fitting the relaxation
profiles in direct time or can be extracted from the slope of the
high-frequency wing of the $\alpha$-relaxation peak in the Fourier
transform of $\epsilon''(\omega)$, whereby $\epsilon''(\omega) \sim
\omega^{-\beta}$~\cite{Lionic}. Likewise, there will be one-to-one
correspondence between the width of the peak and the value of $\beta$.
Clearly, the relatively high-frequency $\beta$-relaxation, if present,
will affect the apparent shape of the $\alpha$-peak.  (Note the
``beta'' in beta-relaxation has nothing to do with the stretching
exponent $\beta$ from Eq.~(\ref{XWLbeta}).) In contrast, the estimates
of Xia and Wolynes~\cite{XWbeta} pertain exclusively to alpha
relaxation and do not extend to other processes. Another, early source
of confusion on the temperature dependence of $\beta$ has been
discussed by the present author~\cite{Lionic}. When measured in even
mildly conducting liquids, the $\alpha$-relaxation peak in
$\epsilon''(\omega)$ is largely masked by a divergence at zero
frequency. At the same time, the {\em inverse} of the dielectric
susceptibility, called the dielectric modulus $M(\omega)$, remains
perfectly finite. There have been attempts to extract the value of
$\beta$ from the width of the $\alpha$-peak in $M''(\omega)$. This
procedure often yields a temperature dependence of the quantity
$\beta$ opposite from that in Fig.~\ref{XWbetaD}(b). This opposite
$T$-dependence is an artifact of the finite conductivity and does not
pertain to the actual $\alpha$-relaxation.

\subsection{Violation of the Stokes-Einstein relation and decoupling
  of various processes}
\label{decoupling}

The diffusivity $D$ (not to be confused with the fragility $D$ from
Eq.~(\ref{VFT})) reflects the efficiency of particle transport, in
response to density fluctuations:
\begin{equation} \label{jDn} {\bm j} = -D {\bm \nabla} n,
\end{equation}
where ${\bm j}$ is the particle flux.

Diffusion of small molecules in a glassy liquid in the activated
regime is qualitatively different from largely collisional
transport. Indeed, the motion of an individual bead, when in the
landscape regime, is completely slaved to the structural
reconfigurations.  A bead typically moves a distance $d_L$ during a
reconfiguration, Eq.~(\ref{dLa10}), on average once per
$\alpha$-relaxation time $\tau_\sloc$. Here the index ``loc''
signifies that the waiting time is distributed and generally differs
from location to location, by Eq.~(\ref{pF}) or its like. Suppose for
a moment the quantity $\tau_\sloc$ is not distributed. The relation
between the particle displacement $d_L$ and the waiting time
$\tau_\sloc$ then implies the particle transport obeys the diffusion
equation with a diffusivity:
\begin{equation} \label{Dactivated} D_\sloc = d_L^2/6 \tau_\sloc.
\end{equation}
Now averaging Eq.~(\ref{jDn}) with respect to the barrier
distribution, we obtain the actual diffusivity:
\begin{equation} \label{DactivatedAv} D = (d_L^2/6) \la \tau^{-1} \ra,
\end{equation}
where we have dropped the label ``local.'' We observe that the
diffusivity---and, thus, any other type of mobility coefficient---is
determined by the average of the inverse relaxation time. This is not
surprising since mobility has to do with the {\em rate} of particle
transfer.  According to Eq.~(\ref{DactivatedAv}), the transport is
dominated by faster regions, if the relaxation rates are distributed.
Because in this case, $\la \tau^{-1} \ra^{-1} < \la \tau \ra$.

In contrast with the diffusivity, the viscosity $\eta$ reflects the
efficiency of {\em momentum} transfer, when the velocity profile is
spatially non-uniform. In the simplest case of an incompressible
liquid and in the limit of low velocity gradients, the non-hydrostatic
portion of the stress tensor $\sigma_{ij}$ can be approximated via a
Fick-like law:
\begin{equation} \label{viscDef} \sigma_{ij} = \eta (\prtl v_i/\prtl
  x_j + \prtl v_j/\prtl x_i).
\end{equation}
This quantity reflects the transfer rate of the $i$th component of the
momentum along the direction $j$.  

The efficiency of momentum transfer can be related to the transport
properties of individual particles via the Stokes formula. For a
spherical particle characterised by the hydrodynamic radius $a/2$ and
drag coefficient $\zeta_\sloc$, $\eta_\sloc = \zeta_\sloc/(6 \pi
a/2)$, where we anticipate that the viscosity generally varies from
location to location.  One may associate, by detailed balance, the
above drag coefficient to a diffusion constant.  Indeed, in the
presence of an external force ${\bf f}$ and in steady state, the
diffusive flux $-D_\sloc {\bm \nabla n}$ should be exactly compensated
by the force-induced flux $v n = (f/\zeta_\sloc) n$. Since $n \propto
e^{{\bm f} \br/k_B T}$, one obtains (locally) the venerable Einstein
relation $\zeta_\sloc = k_B T/D_\sloc$.  The resulting Stokes'
viscosity is thus $\eta_\sloc = k_B T/D_\sloc (6 \pi a/2)$.
Substituting Eq.~(\ref{DactivatedAv}) and averaging
Eq.~(\ref{viscDef}) with respect to the barrier distribution (and
dropping the ``local'' label) yields for the steady-state viscosity of
the actual heterogeneous liquid~\cite{Lionic}:
\begin{equation} \label{eta1} \eta = \frac{2 k_B T}{\pi a d_L^2} \la
  \tau \ra.
\end{equation}
That is, the viscosity scales {\em linearly} with the relaxation time,
in contrast with the mobility from Eq.~(\ref{DactivatedAv}), which
shows an inverse scaling. The relation in Eq.~(\ref{eta1}) hearkens
back to Maxwell's phenomenological argument~\cite{LLelast} about the
intrinsic connection between the elasticity, viscosity, and the
relaxation times in very viscous liquids: When $\tau$ is sufficiently
small, the structures do not live long enough to pass on momentum
appreciably. In the opposite extreme of very long-lived structures,
the response is purely elastic and so momentum is transferred
infinitely efficiently, via elastic waves.  At high frequencies
exceeding the inverse relaxation time, the response is largely elastic
so that the stress tensor goes as $K u$, where $u$ is the magnitude of
the deformation. On the other hand, at frequencies below $\tau^{-1}$,
the response should be purely liquid-like: $\eta \dot{u}$. At the
crossover between the largely viscous and elastic response, $\omega
\tau \simeq 1$, and so $\dot{u} \simeq (d/dt) [u(t=0) e^{-t/\tau}] =
u/\tau$. This yields $K \simeq \eta \tau^{-1}$.

It turns out that Eq.~(\ref{eta1}), which relates microscopic
characteristics of the structural reconfigurations, allows one to
derive a Maxwell-type expression in a constructive manner.  First we
recall that the quantity $(d_L/a)^2$, which is the typical strain
squared, is intrinsically related to the temperature and elastic
moduli, via the equipartition theorem: $K a^3 (d_L/a)^2 = k_B T$, up
to a coefficient of order one.  One obtains, as a result:
\begin{equation} \label{eta3} \eta \simeq K \la \tau \ra.
\end{equation}
The above relation works very well in actual liquids. The relaxation
times can be measured directly by dielectric response, while the
elastic constants can be measured, for instance, by Brillouin
scattering. This simple, but nevertheless, fully constructive result
thus serves as an important test for the microscopic picture advanced
by the RFOT theory. Again, we observe the quantity $\alpha \sim
1/d_L^2$ enters explicitly in a physical quantity characterising the
landscape regime, see Eq.~(\ref{eta1}).

An important corollary of Eqs.~(\ref{DactivatedAv}) and (\ref{eta1})
is that the particle and momentum transport become increasingly {\em
  decoupled} from each other for broader barrier distributions because
the mobility is dominated by the fastest regions while the viscosity
is determined by the typical relaxation time. The simplest
quantitative measure of the decoupling is the deviation of the
quantity $\la \tau^{-1} \ra \la \tau \ra$ from unity.  The amount of
decoupling increases with the width of the barrier distribution,
which, in turn, is greater for more fragile liquids, by
Eq.~(\ref{relf}). (We remind, a liquid is the more fragile, the larger
the fragility coefficient $m$ from Eq.~(\ref{mdef}) or the smaller the
fragility $D$ from Eq.~(\ref{VFT}).)  The amount of decoupling will
depend on the precise shape of the barrier distribution. Generically,
and according to Eq.~(\ref{relf}) and Fig.~\ref{XWbetaD}, $\delta
F^\ddagger/F^\ddagger_\smp \simeq 0.25$ while $F^\ddagger_\smp/k_B T
\simeq 37$ near $T_g$. The decoupling, which can be roughly estimated
as $\la \tau^{-1} \ra \la \tau \ra \sim e^{+ \delta F^\ddagger/k_B
  T}$, thus could be as large as a few orders of magnitude.

Because $\la \tau^{-1} \ra \la \tau \ra > 1$ in the activated regime,
the diffusivity predicted from the Stokes-Einstein relation using the
measured viscosity is lower than the actual diffusivity. This
violation of the Stokes-Einstein reflects the transient breaking of
ergodicity that takes place when the landscape sets in. Indeed, the
Einstein relation $D = k_B T/\zeta$ is a consequence of detailed
balance, which follows from the ergodic
assumption~\cite{PhysRev.37.405, PhysRev.38.2265}.  It is the breaking
of the translational symmetry on timescales shorter than the
structural relaxation time $\tau$ that leads to the breaking of
ergodicity on the correspondingly short timescales.

\begin{figure}[t]
\begin{tabular*}{\figurewidth} {ll}
\begin{minipage}{.48 \figurewidth} 
  \begin{center}
    \includegraphics[width=0.48
    \figurewidth]{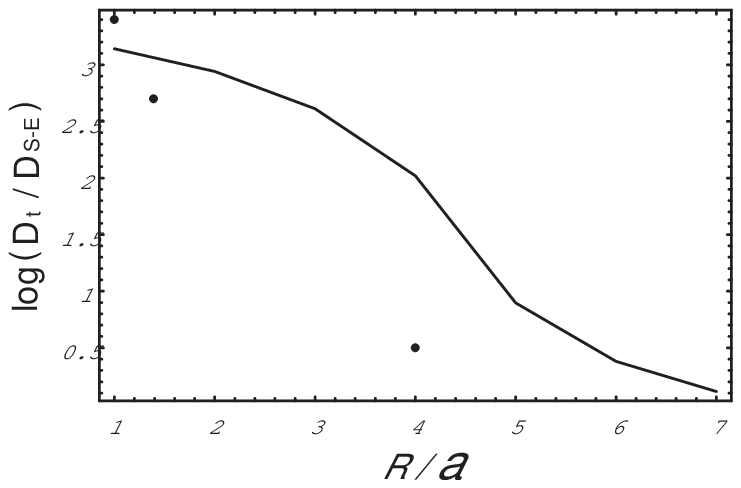} \\
    {\bf (a)}
   \end{center}
  \end{minipage}
&
\begin{minipage}{.48 \figurewidth} 
  \begin{center}
    \includegraphics[width=0.48 \figurewidth]{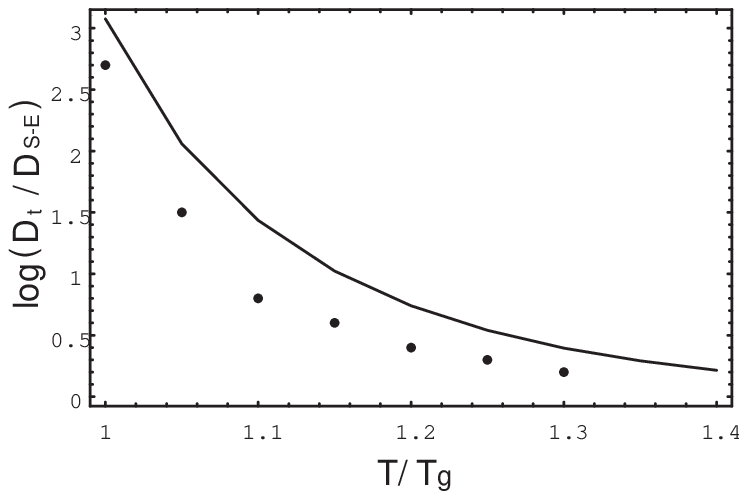}
    \\ 
    {\bf (b)}
  \end{center}
\end{minipage}
\end{tabular*}
\caption{\label{SEXW} RFOT-based predictions of the deviation from the
  Stokes-Einstein relation, due to Xia and Wolynes~\cite{XWhydro}
  along with experimental data of Cicerone and
  Ediger~\cite{CiceroneEdiger} for OTP. {\bf (a)} The ratio of the
  actual diffusivity to its value predicted by the Stokes-Einstein
  relation, as a function of the probe radius. {\bf (b)} Same as (a),
  but a function of temperature~\cite{XWhydro}.}
\end{figure}

Xia and Wolynes~\cite{XWhydro} have exploited these ideas to
quantitatively estimate the amount of decoupling between the diffusion
of a spherical probe of an arbitrary radius and momentum
transport. They have set up the problem as that of a spatially
inhomogeneous viscosity. Consistent with expectation, only probes that
are comparable to or smaller than the cooperativity size can ``sense''
the heterogeneity in the viscous response of the liquid and will
violate the Stokes-Einstein relation. Quantitative predictions of the
extent of the violation, due to Xia and Wolynes (XW), are shown in
Fig.~\ref{SEXW}(a), alongside the experimental data of Cicerone and
Ediger~\cite{CiceroneEdiger}, all near the glass transition. The
agreement is good, especially considering that no adjustable
parameters were used. As already mentioned, the decoupling goes
roughly as $e^{- \delta F^\ddagger/k_B T}$ and should be greater for
lower temperatures since the barrier $F^\ddagger$---and hence a
measure of the barrier distribution width $\delta F^\ddagger \propto
F^\ddagger$---increase with lowering temperature. This notion is borne
out both in the RFOT-based predictions and in observation, see
Fig.~\ref{SEXW}(b). (To avoid confusion we note that XW used the
``slip'' boundary conditions in their calculation, which engenders a
distinct value of the numerical coefficient in the Stokes formula from
the preceding qualitative argument, viz., $\eta = \zeta/4 \pi R$.)

\begin{figure}[t]
\begin{tabular*}{\figurewidth} {ll}
\begin{minipage}{.48 \figurewidth} 
  \begin{center}
    \includegraphics[width=0.32 \figurewidth]{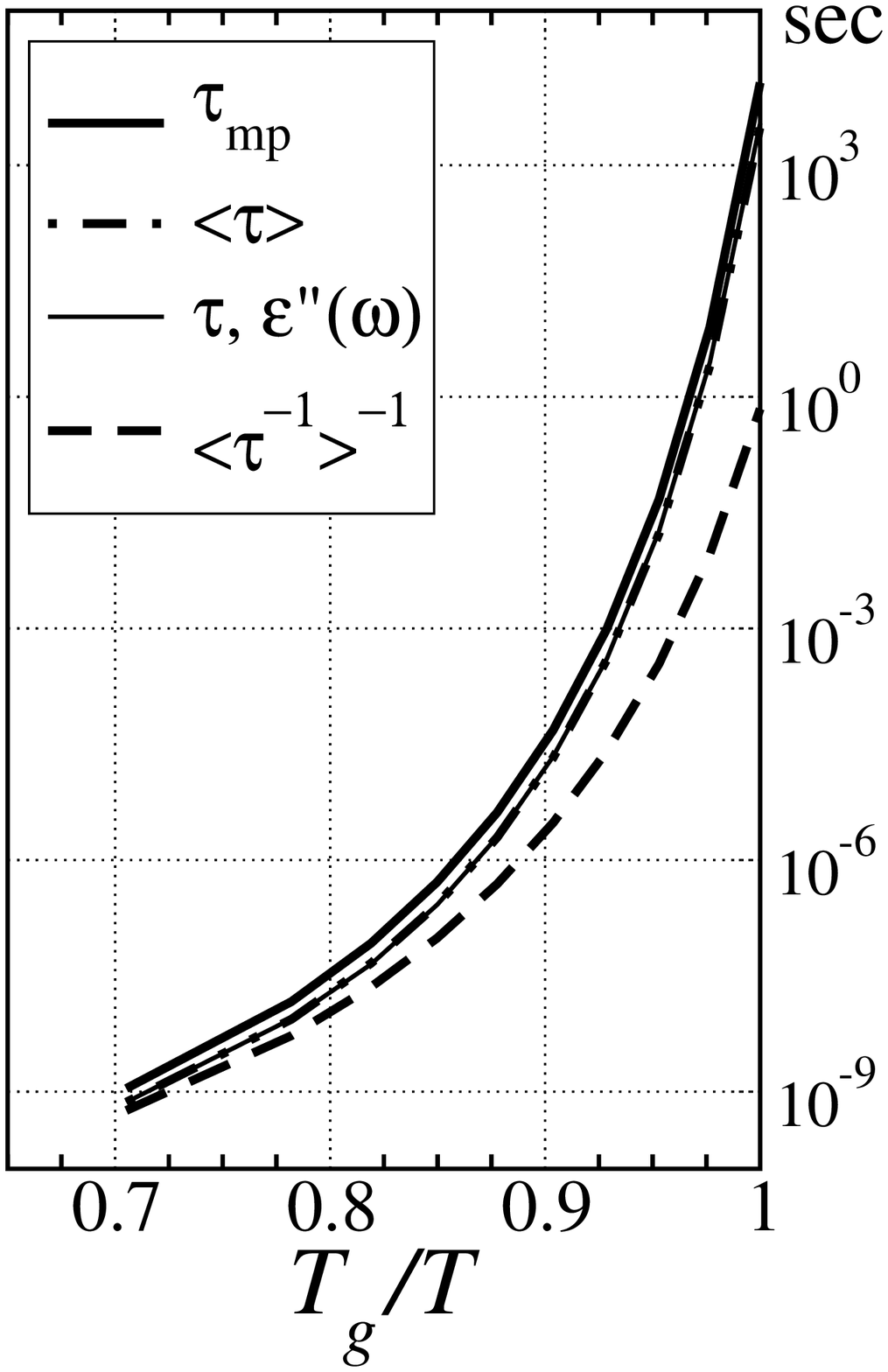} \\ \vspace{3mm}
    {\bf (a)}
   \end{center}
  \end{minipage}
&
\begin{minipage}{.48 \figurewidth} 
  \begin{center}
    \includegraphics[width=0.4 \figurewidth]{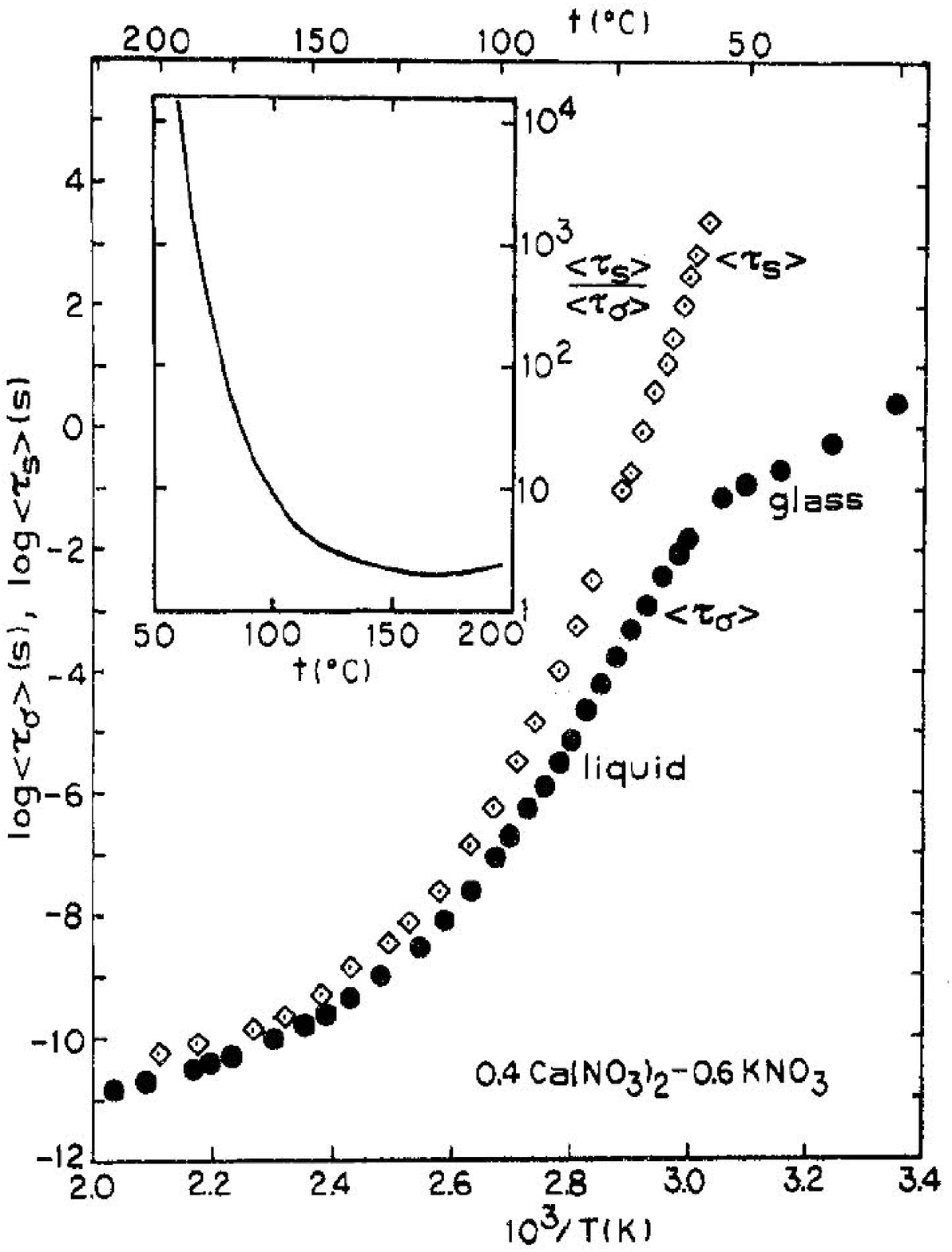}
    \\ 
    {\bf (b)}
  \end{center}
\end{minipage}
\end{tabular*}
\caption{\label{ionic} {\bf (a)} Different relaxation times derived
  from the barrier distribution in Eq.(\ref{pFm}), as functions of
  temperature. $\delta \wF = 0.25$, corresponding to fragility
  $D=4$~\cite{Lionic}. {\bf (b)} Relaxation time pertaining to the
  ionic conductivity, $\tau_\sigma$, and the structural relaxation
  time $\tau_s$. From Howell et al.~\cite{Howell1974}.}
\end{figure}

Lubchenko~\cite{Lionic} has addressed the decoupling between ionic
conductivity and momentum transfer in ionic liquids such as CKN (40\%
Ca(NO$_3$)$_2$-60\%KNO$_3$). In such liquids, every bead is
charged. In the presence of an electric field, there is a non-zero net
current density:
\begin{equation}
  \bj  = \la \bj' \ra = \frac{1}{\xi^3} \la \frac{\bmu_T}{\tau} \ra,
\end{equation}
because the transition barrier, and hence the relaxation time are
coupled to the electric field $\bE$: $\tau^{-1}(\bE) =
\tau^{-1}(\bE=0)(1+ \bmu_T^\ddagger \bE/k_B T)$, while the the
transition dipole moment $\bmu_T^\ddagger$ of a reconfiguring region
is correlated with the overall transition dipole moment $\bmu_T$. A
detailed calculation yields for the ionic conductivity tensor:
\begin{equation} \label{sij} \sigma_{ij} = \delta_{ij} \, \frac{\Delta
    \mu_\smol^2}{8 \, a^3 k_B T} \la \frac{1}{\tau} \ra,
\end{equation}
where $\Delta \mu_\smol$ is the transition dipole moment of a single
bead: $|\Delta \mu_\smol| \sim q d_L$, and $q$ is the bead charge.
Not surprisingly, the above expression for the mobility of charged
particles has the same structure as that for the diffusivity in
Eq.~(\ref{DactivatedAv}). The resulting temperature dependence of the
decoupling for a generic fragile substance is shown in
Fig.~\ref{ionic}(a), to be compared with the decoupling between ionic
transport and viscosity observed in KCN~\cite{Howell1974} shown in
Fig.~\ref{ionic}(b). Fig.~\ref{ionic}(a) also demonstrates that the
relaxation time, as determined by locating the peak in the imaginary
part of the dielectric susceptibility $\epsilon''(\omega)$ is
numerically close to the typical relaxation time, and so is the most
probable relaxation time.

Various other types of decoupling have been observed in viscous
liquids and, in fact, have been used to {\em detect} the crossover to
the activated transport. We turn to this next.

\section{At the crossover from collisional to activated transport}
\label{crossover}

Now that we have established a microscopic perspective on both the
thermodynamics and kinetics in the activated regime, we are ready to
discuss how the landscape regime sets in, in the first place. The
crossover spans a dynamical range of a couple of decades in
chemically-bonded glassformers, which is much less than the full
dynamical range of sixteen decades or so accessible during a
quench. Yet the crossover is a very important aspect of any
microscopic theory of the structural glass transition. It is the
transition that ultimately underlies the formation of the metastable
structures and the eventual, complete loss of ergodicity that takes
place at the temperature $T_g < T_\scr$. Unfortunately, the crossover
is much harder to describe quantitatively than the regime of
well-developed activated transport because the activated
reconfigurations are no longer decoupled from the vibrations. One may
say that the activated reconfigurations are strongly affected by
mode-coupling at the crossover. Conversely, one may say the
MCT-predicted slowing down becomes short-circuited by the activated
transition. Neither approach works quantitatively at the crossover, on
its own, nor is it clear at present how to marry the two in a formally
satisfactory way. (See however the earlier mentioned work in
Refs.~\cite{PhysRevE.72.031509, BBW2008} for phenomenological, hybrid
approaches.) Despite these complications, both quantitative and
qualitative characteristics of the crossover can be established by
building on the microscopic picture of the activated transport
reviewed above.

Already the recognition that particle transport occurs by activation
{\em below} the crossover yields a testable prediction that the
temperature dependence of the relaxation time for fragile substances
should not obey an Adam-Gibbs (AG) like expression in the full
temperature range between the melting temperature $T_m$ and glass
transition temperature $T_g$. This is because the crossover
temperature for sufficiently fragile liquids is below $T_m$. The
elegant analysis of Stickel et al.~\cite{Stickel} demonstrates that,
indeed, temperature dependences of $\alpha$-relaxation are well fitted
by {\em two} distinct AG forms, in the low-$T$ and high-$T$ portions
of the full temperature interval $T_g < T < T_m$.

To actually estimate $T_\scr$, one must first recognise that there
will be significant corrections to the RFOT-derived expression for the
reconfiguration barrier, Eq.~(\ref{F2}), upon approaching the
crossover from below.  We reiterate that the crossover, which is
centred at temperature $T_\scr$, is the finite-dimensional analog of
the mean-field transition at temperature $T_A$, at which the liquid
free energy $F(\alpha)$ from Eq.~(\ref{Far}) develops a metastable
minimum. As usual, the transition is lowered in finite dimensions,
compared with meanfield: $T_\scr < T_A$.  Lubchenko and
Wolynes~\cite{LW_soft} (LW) have elucidated the origin of this
lowering. They pointed out that as one approaches from below the
temperature $T_A$, at which the mean-field free energy $F(\alpha)$ has
a spinodal, fluctuations in the order parameter $\alpha$ become
increasingly strong.  The spinodal is illustrated by the thick solid
line in Fig.~\ref{RLFalpha}(a).
Thus the assumption that each metastable minimum is long-living
becomes progressively less reliable as the spinodal is approached
(from below); consequently, the expression (\ref{F2}) increasingly
{\em overestimates} the barrier.

In assessing the magnitude of the resulting barrier softening effects
LW noted that the mismatch penalty between distinct aperiodic minima
scales with a positive power of the barrier height $f^\ddagger$ in
Fig.~\ref{RLFalpha}(a) and thus must vanish at $T_A$. (The power is
$1/2$ in the standard thin interface limit.) This is because for a
transition between two aperiodic structures to take place, the regions
of small $\alpha$, such that $\alpha < \alpha^\ddagger$, are not
visited. We have seen this directly in Subsection~\ref{quant}. The
vanishing of the mismatch penalty and, hence, of the activation
barrier, at $T=T_A$, is in contrast with the simple prediction from
Eq.~(\ref{F2}), which ignores the effects of fluctuations by assuming
that the aperiodic minima are always separated by a finite barrier.

To quantitatively assess the barrier softening effects in actual
liquids, LW~\cite{LW_soft} utilised a simple $F(\alpha)$ curve that
can be parametrised so that $f^\ddagger(T=T_A) = 0$, while, at the
same time, $\Delta f(T=T_K) = 0$, to reflect the vanishing of the
configurational entropy at the (putative) Kauzmann temperature $T_K$.
Note such parametrisation is reasonable, in light of the data in
Fig.~\ref{RLFalpha}(b): In this Figure, we display results of Rabochiy
and Lubchenko's study~\cite{RL_LJ} of the free energy of a
Lennard-Jones liquid as a function of temperature, pressure, and
coordination. The latter can be controlled, within certain limits, by
varying the coefficient $\eta_\tRCP$ in the $g(R)$ ansatz from
Eq.~(\ref{gRBC}). One observes that given a fixed value of the barrier
$f^\ddagger$, the {\em shape} of the free energy curve $F(\alpha)$ is
nearly universal, not too far from the spinodal. This is an instance
of a law of corresponding states.

\begin{figure}[t] \centering
  \includegraphics[width= .7 \figurewidth]{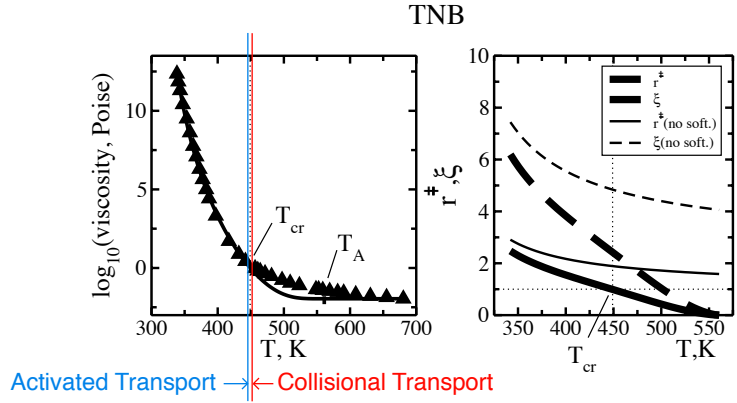}
  \caption{\label{LWcrossover} The left panel shows
    experimentally-determined viscosity of an organic glass-former TNB
    (symbols) alongside the theoretical prediction of an RFOT-based
    treatment that includes the barrier-softening
    effects~\cite{LW_soft} (thin line), Eq.~(\ref{F_r_interp}).  The
    temperature at which the landscape-based prediction of the RFOT
    theory and the experimental data diverge corresponds to the
    crossover between activated and collisional transport, above which
    mode-coupling effects dominate. The right panel compares
    RFOT-based predictions of the critical radius $r^\ddagger$ and
    cooperativity length $\xi$ with and without the barrier softening
    effects. Note the crossover happens to coincide with the
    temperature at which the critical radius for structural
    reconfigurations is numerically close to the molecular length
    scale, $r^\ddagger \approx a$.}
\end{figure}

Alongside the softening of the interface tension, one also expects
that there will be a reduction in interface renormalisation, for the
following reason: As the reconfiguration barrier becomes vanishingly
small, so does the critical nucleus. The latter, however, cannot be
smaller than the molecular size $a$. Since in the renormalisation
scenario, Subsection~\ref{mismatch}, the width of the interface is
tied to the droplet size, by Eq.~(\ref{zetar}), a renormalisation-like
scenario becomes internally inconsistent. LW the proceed to write down
a simple expression that interpolates in the simplest way between a
fully wetted interface at large $r$ and non-wetted interface in the $r
\to 0$ limit, without introducing an extra adjustable parameter:
\begin{equation} \label{F_r_interp} F(r)= \frac{\Sigma_K
    \Sigma_A}{\Sigma_K+\Sigma_A} - \frac{4 \pi}{3} (r/a)^3 T s_c,
\end{equation} 
where $\Sigma_K = 4 \pi \sigma_0 a^2 (r/a)^{3/2}$ and $\Sigma_A = 4
\pi f^\ddagger a^3 (r/a)^{2}$ are the wetted and non-wetted mismatch
penalties respectively.  Experimentally determined $T$-dependences of
relaxation time data can now be fitted to the barrier predicted by
Eq.~(\ref{F_r_interp}), using only the quantity $T_A$ and $\Delta
c_p(T_g)$ from Eq.~(\ref{scTRA}) as adjustable parameters. The latter
parameter can be used to estimate the bead count which, then, can be
independently checked against chemical intuition. ($T_K$ can be taken
from calorimetry while the prefactor $\tau_0$ can be fixed at a nearly
universal value of $1$~ps, which may introduce some numerical
uncertainty but removes an even greater source of uncertainty due to
underconstrainment in the fit.)  Using only a few low-temperature
experimental points produces excellent fits in {\em extended}
temperature ranges, at viscosities above $10$~Ps or so, see
Fig.~\ref{LWcrossover}. Interestingly, at the very same value of the
viscosity, the critical radius happens to numerically coincide with
the molecular size $a$.

The just mentioned value of the viscosity at which the RFOT-predicted
relaxation time (now including the softening effects) diverges from
experiment signifies the temperature at which collisional effects
become important. This temperature thus falls within the temperature
range of the crossover.  An important message of the work in
Ref.~\cite{LW_soft} is that the fluctuation effects due to the
spinodal are most significant in fragile substances, in which the
mean-field temperature $T_A$ and its finite-dimensional analog
$T_\scr$ and the Kauzmann temperature are numerically close. Thus in
fragile substances, the softening corrections likely modify the
barrier value predicted by Eq.~(\ref{F2}).  Conversely, in strong
substances, the simple expression for the barrier from Eq.~(\ref{F2})
should be adequate in a very broad temperature range.

\begin{figure}[t]
\begin{tabular*}{\figurewidth} {ll}
\begin{minipage}{.48 \figurewidth} 
  \begin{flushleft}
    \includegraphics[width=0.4 \figurewidth]{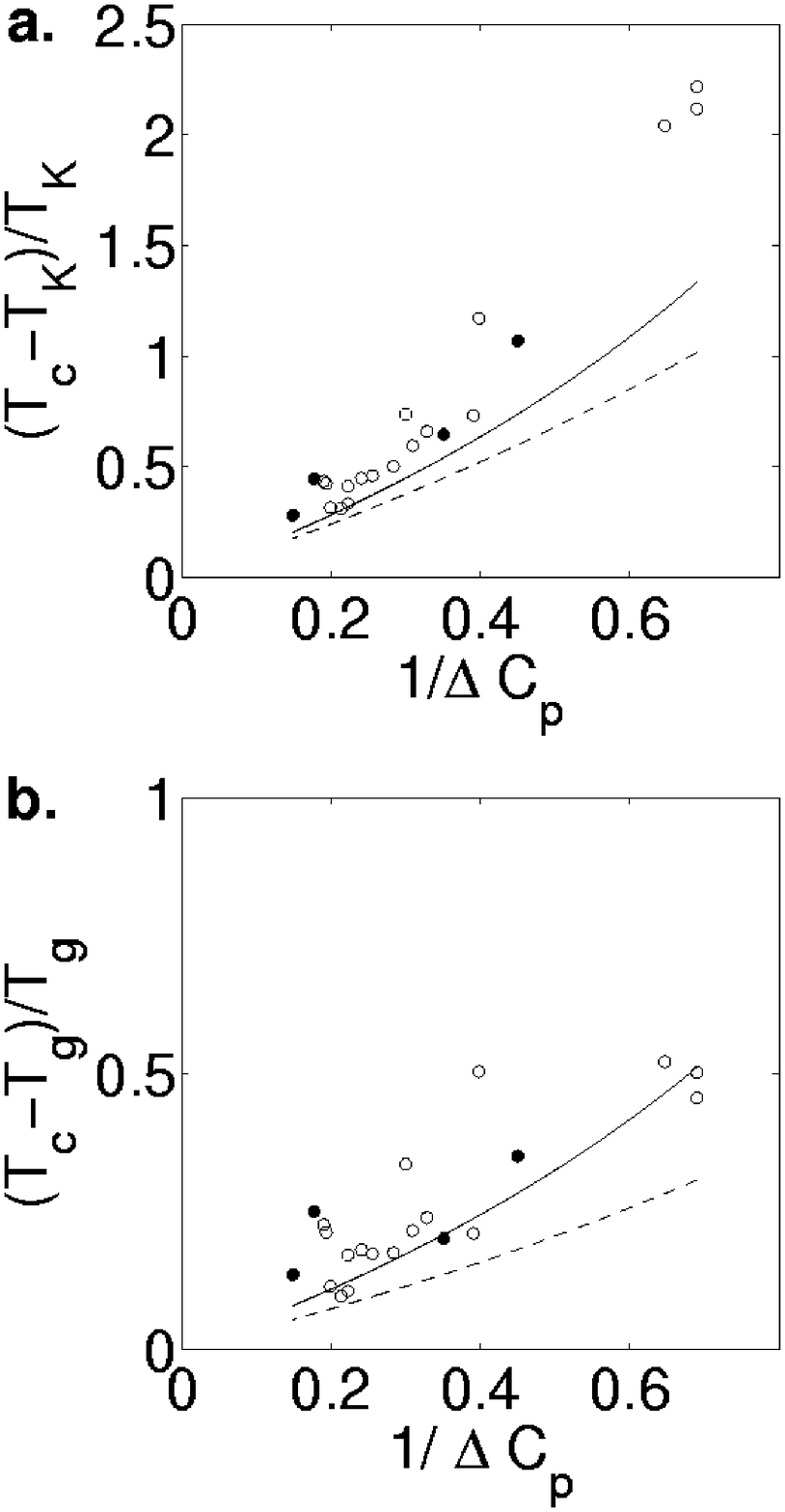}
    \\ 
  \end{flushleft}
\end{minipage}
&
\begin{minipage}{.48 \figurewidth} 
  \begin{center}
    \includegraphics[width=0.48 \figurewidth]{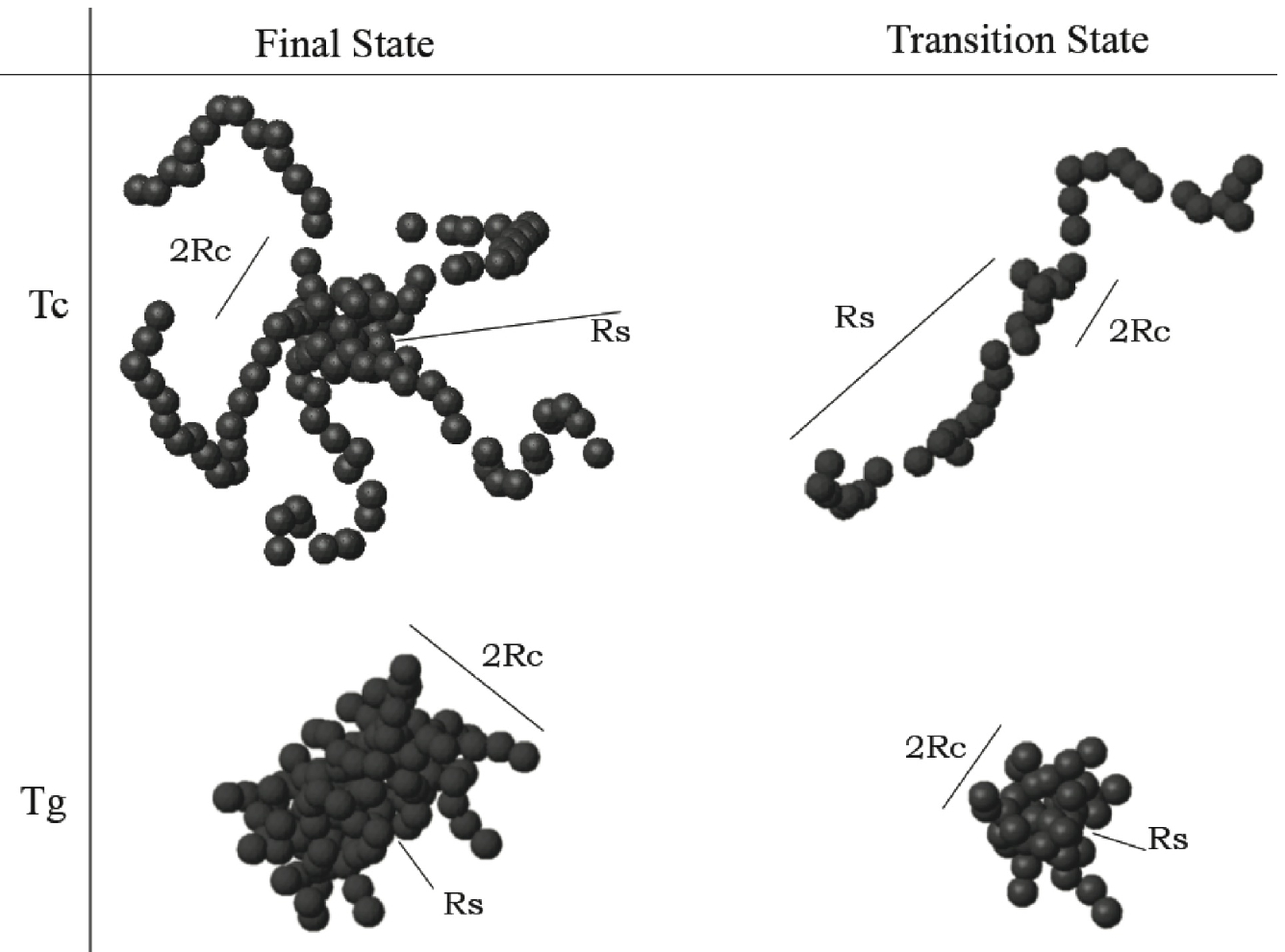} \\
    {\bf (c)}  \\ \vspace{5mm}
    \caption{\label{strings} {\bf (a)-(b) (left)} RFOT-based
      predictions for the crossover temperature, due to Stevenson et
      al.~\cite{SSW}, as a function of the inverse heat capacity jump
      at $T_g$. Solid and dashed lines correspond to percolation- and
      string-based estimates respectively. The circles show
      experimentally determined values, the filled circles
      corresponding to polymers. {\bf (c) (top)} Schematic of a
      reconfiguring regions at temperatures close to the crossover
      (upper portion) and far below the crossover (lower
      portion)~\cite{SSW}. The crossover is signalled by a large size
      $R_s$ of the non-compact halo, relative to the size $R_c$ of the
      compact core.}
   \end{center}
  \end{minipage}
\end{tabular*}
\end{figure}

In a more microscopic vein, Stevenson, Schmalian, and
Wolynes~\cite{SSW} have addressed the issue of how one must modify the
nucleation scenario from Section~\ref{ActTransport} near the crossover
to the mode-coupling-dominated transport. These authors argued that
viewing a cooperatively reconfiguring region as a compact object is
too simplistic close to $T_A$. Indeed, we have already mentioned that
the closer to $T_A$ the smaller the extent of the reconfiguration. In
other words, moving {\em individual} particles becomes less and less
costly.  Under these circumstances, the penalty for having non-compact
shapes is outweighed by the diversity of all possible non-compact
shapes even as the compact core of the reconfiguring region may be
still of appreciable size, see Fig.~\ref{strings}(c). The multiplicity
of all percolated, non-compact shapes is determined by the
connectivity of the lattice and does not involve adjustable
parameters. One then replaces the mismatch penalty in Eq.~(\ref{FN1})
with two terms: one is the free energy cost of moving $N$ particles
within a non-compact shape, the other the entropic gain due to the
multiplicity of all possible shapes. The cost is computed analogously
to how we estimated the molecular surface tension $\sigma_0$ in
Subsection~\ref{mismatch}, except for a general number of broken
contacts. In the XW approximation~\cite{XW}, Eq.~(\ref{sigmaXW}), this
cost is entirely entropic and scales linearly with temperature.  In
the large $N$ limit, the free energy cost of moving $N$ particles
within a non-compact shape turns out to scale {\em linearly} with $N$,
and so does the overall free energy profile as a result:
\begin{equation} \label{FSSperc} F(N) = (T s^\text{perc} - T s_c) N,
\end{equation} 
where the quantity $s^\text{perc} \simeq 1.28 k_B$ per particle, gives
the full free energy cost of excitations that percolate into a
non-compact cluster, divided by temperature. In the absence of
structural degeneracy, the cost of moving beads always outweighs the
entropic gain due to the multiplicity of the non-compact structures,
consistent with individual structures being mechanically (meta)stable.

The linear scaling of the free energy profile in Eq.~(\ref{FSSperc})
means the reconfiguration is either exclusively uphill or downhill. If
the configurational entropy exceeds a threshold value $s_c =
s^\text{perc}$ (which, note, is system-independent) the
reconfiguration is exclusively downhill, implying the activation
barrier is strictly zero. This, in turn, signals the crossover, thus
allowing one to combine system-specific values of $T_K$ and $\Delta
c_p$ with the universal $s_c \simeq 1.28 k_B$ to estimate the
crossover temperature, by Eq.~(\ref{scTRA}); the resulting prediction
for $T_\scr$ are shown with the solid line in Fig.~\ref{strings}(a)
and (b). (For lower temperatures, non-compactness is still allowed but
can be thought of as a perturbation to the relatively simple scenario
that led to Eq.~(\ref{FN1}).)

An instructive variation on the preceding argument is to think of the
various non-compact, percolated shapes as a small, compact core
dressed by string-like objects. This view is instructive since it
emphasises that to create a non-compact shape one must consider
contiguous chains (or ``strings'') of particle movements that
originate at the compact core but end {\em before} they return to the
core. The diversity of all such strings can {\em also} be estimated
without adjustable parameters and gives a similar, system-independent
estimate for the threshold value of the configurational entropy: $s_c
\simeq 1.13 k_B$.  Thus the ``percolation'' and ``string'' views are
mutually consistent. The string-based estimates of $T_\scr$ are shown
with the dashed line in Fig.~\ref{strings}(a)-(b).  One valuable
aspect of the string approach is that it allows one to make
connections between the RFOT theory and simulations, which show
string-like excitations as the viscous slowing-down sets
in~\cite{Glotzer_strings}. The presence of quasi-one dimensional,
string-like excitations is also consistent with our early,
symmetry-based discussion of the liquid-to-solid transition in
Subsection~\ref{driving}. There we observed how quasi-one dimensional
modes emerge during incipient solidification driven by steric effects,
see Eq.~(\ref{dF}).

The above notion of the universality of the magnitude of the
configurational entropy at the crossover predicted by Stevenson et
al.~\cite{SSW} was soon afterwards utilised by Hall and
Wolynes~\cite{HallWolynes_JPCB}, who used the density functional
theory to predict the crossover temperature for Lennard-Jones liquids,
in addition to the meanfied temperature $T_A$ and the putative
Kauzmann temperature. In particular, these quantities have been
computed as functions of pressure and show good agreement with
experiment.

\begin{figure}[t]
  \begin{tabular*}{\figurewidth} {ll}
    \begin{minipage}{.5 \figurewidth} 
      \begin{center}
        \includegraphics[width = .5 \figurewidth]{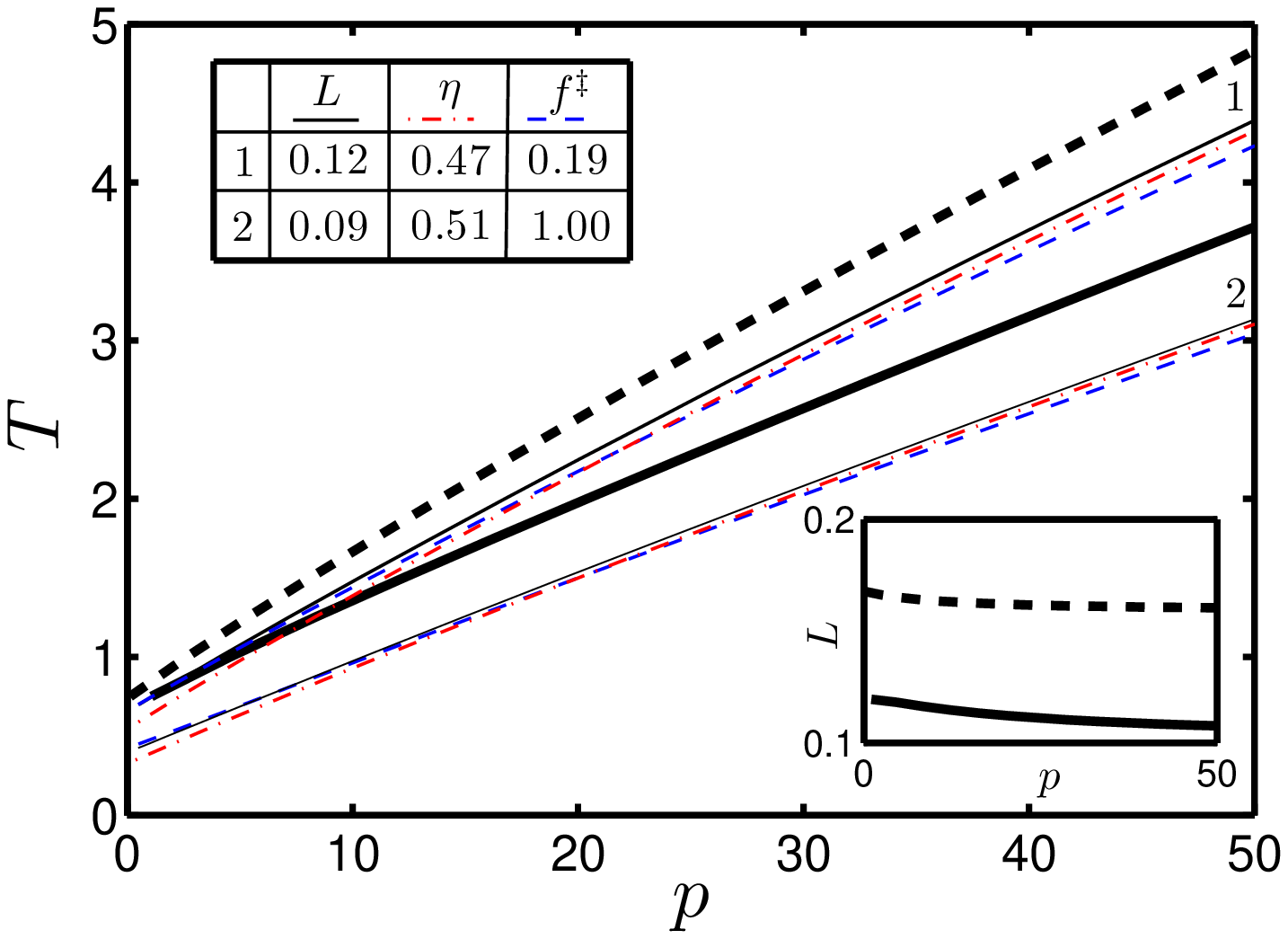} \\
        {\bf (a)}
      \end{center}
  \end{minipage}
  &
  \begin{minipage}{.46 \figurewidth} 
    \begin{center}
      \includegraphics[width=0.45 \figurewidth]{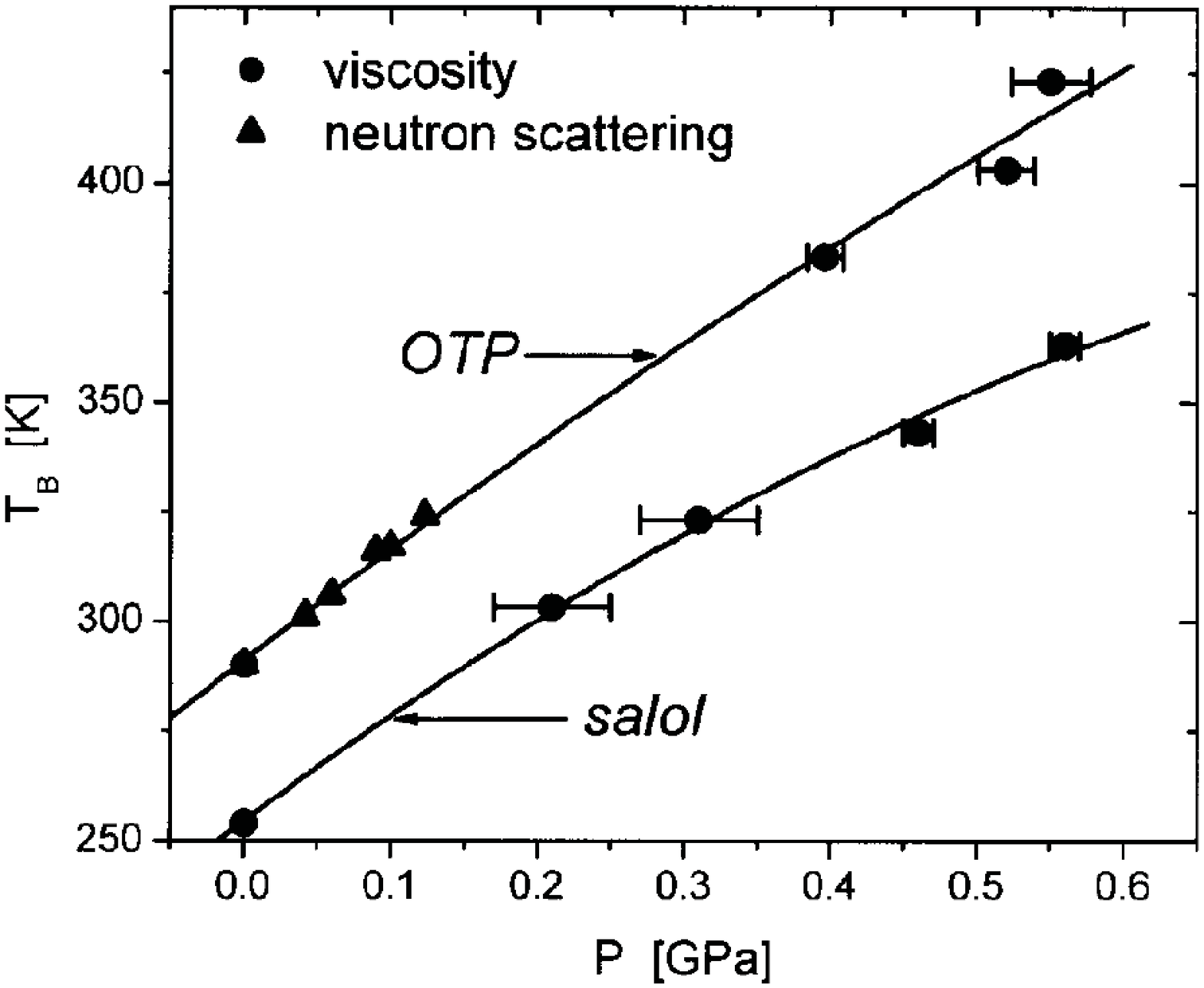} \\
    {\bf (b)} 
  \end{center}
\end{minipage}
\end{tabular*}
\caption{\label{lowP} {\bf (a)} The pressure and temperature
  dependence of the barrier height $f^\ddagger$, Lindemann ratio $L$,
  and the filling fraction $\eta$ shown as two sets of isolines of the
  corresponding surfaces. The thick dashed line shows the pressure
  dependence of the ``spinodal'' temperature $T_A$; it is an isoline
  for $f^\ddagger$ but not for $L$ and $\eta$. The thick solid line is
  the liquid-crystal coexistence line; its l.h.s. end is the triple
  point. The inset shows the pressure dependence of the Lindemann
  ratio along the ``spinodal'' and liquid-crystal coexistence
  lines. $\eta_\tRCP = 0.64$. From Ref.~\cite{RL_LJ}. {\bf (b)}
  Pressure dependence of the dynamic crossover temperature, after
  Casalini and Roland~\cite{PhysRevLett.92.245702}, to be compared
  with the thin solid line 1 in panel (a). Note that the solid lines
  are a guide to the eye and are not theoretical predictions. In any
  event, the slope of $\sim 100K/1GPa$ is consistent with the RFOT
  theory, see text. }
\end{figure}

We now return to the law of corresponding states illustrated in
Fig.~\ref{RLFalpha}(b). Rabochiy and Lubchenko~\cite{RL_LJ} used this
notion to argue that the crossover should take place at a universal
value of the farrier $f^\ddagger$. Indeed, activated reconfigurations
become essentially one-particle events at the crossover, as we
discussed earlier in this Section, implying fluctuations in the order
parameter $\alpha$ at different spots become mutually
uncorrelated. Thus the local value of the bulk density
$f(\alpha)$---corresponding to the free energy $F(\alpha)$---fully
characterises the stability of a local region, regardless of its
environment. In other words, the sample can be thought of as a
collection of uncorrelated, anharmonic degrees of freedom subject to a
(free) energy function $f(\alpha)$. Each such degree of freedom---and,
hence, the aperiodic structures themselves---become marginally stable
when at the typical thermal displacement away from the minimum,
$\alpha$ is at its saddle point value $\alpha^\ddagger$. In other
words, $\la (\alpha - \alpha_0)^2 \ra^{1/2} = (\alpha_0 -
\alpha^\ddagger)$, times a factor of order 1.  Combined with the near
universality of the shape of $f(\alpha)$, this implies that the
crossover is signalled by a system-independent value of $f^\ddagger$.
This criterion can be made more practical by noticing~\cite{RL_LJ}
that the (computed) $f^\ddagger=\text{const}$ lines in the $(p, T)$
plane nearly coincide with the (computed) lines $L=\text{const}$,
where $L$ is the Lindemann ratio from Eq.~(\ref{Lratio}), where the
lattice is now aperiodic. This is illustrated in Fig.~\ref{lowP}(a). We
thus arrive, semi-phenomenologically, at a criterion that the
crossover corresponds to a universal value of the ratio of the
vibrational displacement to particle spacing, irrespective of pressure
or temperature. This is in harmony with the systematic version of the
Lindemann criterion of melting~\cite{L_Lindemann}, under which the
vibrational displacement at the solid-liquid interface is a universal
quotient of the particle spacing, when the solid and liquid are in
equilibrium. The crossover {\em also} corresponds to an equilibrium
between liquid and solid, except the latter solid is now
aperiodic. RL~\cite{RL_LJ} took this notion further to determine the
actual value of $L$ that signals the crossover. These authors have
evaluated the values of the Lindemann ratio $L$ at the phase boundary
for the liquid and the {\em periodic} crystal as a function of
pressure (or $T)$, the totality of which form a line in the $(L, p,
T)$ space; this line is shown as the thick solid line in
Fig.~\ref{lowP}(a). The $L(p, T)$ plane for the Lindemann ratio in the
aperiodic crystal mostly lies above the line $L(p, T)$ for the
periodic crystal, but turns out to intersect it near the triple
point. Note the value of $L$ at the triple point is unique for each
substance. On the other hand, the Lindemann ratio at the crossover is
nearly unique, as just discussed. Based on these two notions and the
fact that the $(L, p, T)$ plane crosses the $L(p, T)$ line near the
triple point, RL surmised that the Lindemann ratio at the crossover
$L_\scr$ should be equal to its value at the triple point. The
classical DFT, in the modified weighted density approximation (MWDA),
yields the value of $0.12$ for $L_\scr$~\cite{RL_LJ}. The actual value
of the Lindemann ratio near the triple point of the Lennard-Jones
system is
\begin{equation} \label{L145} L_\scr = 0.145,
\end{equation}
according to simulations~\cite{Hansen}. This number will be of use in
short order.

The above findings on the universality of the Lindemann displacement
at the crossover can be used to make several testable predictions.
Several of these predictions rationalise the pressure dependence of
the fragility, depending on the degree of bonding
directionality~\cite{RL_LJ}. Here we limit ourselves to covering only
two of these predictions.

As pressure is increased, any substance will behave more and more like
that made of rigid particles, in the absence of a structural
transformation that would increase the coordination in a discontinuous
way. Under these circumstances, the external pressure becomes
increasingly close to the kinetic pressure, which corresponds to the
amount of momentum a particle transfers to its cage per unit time, per
unit area: $p \sim 2 (m v_\sth)(v_\sth/d_0)/a^2 \simeq 2 (a/d_0) n k_B
T \simeq 2 k_B T/d_0 a^2$, thus implying
$p \simeq 2 \frac{k_B T}{a^2} \frac{1}{d_0}$,
where $v_\sth$ is the average thermal speed and $d_0$ is the
confinement length of a particle in its cage. Since the confinement
length is equal to $d_L$ and, furthermore, is constant at the
crossover, we obtain a simple estimate for pressure dependence of 
the crossover temperature:
\begin{equation} \label{TcrP} k_B T_\scr \simeq (d_L/a) (a^3) \, p,
\end{equation}
up to an additive correction that reflect the softness of the local
molecular field acting on the individual particle~\cite{RL_LJ,
  LW_Wiley}. A quick estimate using a typical $a = 3$~\AA ~and $d_L/a
= 1/10$ yields that per each extra half-GPa in pressure, the critical
temperature will rise by 100 degrees or so. This is clearly consistent
with the experimental data in Fig.~\ref{lowP}(b).

\begin{figure}[t]
  \centering
  \includegraphics[width=0.48 \figurewidth]{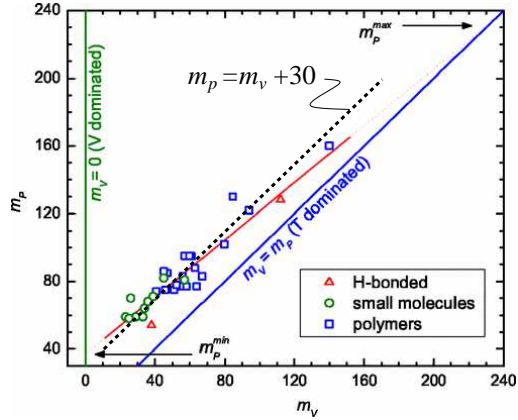}
  \caption{\label{mpmvFig} Experiment (symbols): Fragility indexes at
    constant pressure and volume for several substances, after
    Casalini and Roland.\cite{PhysRevE.72.031503}. Theory (black
    dashed line): Prediction from Eq.~(\ref{mpmv2})~\cite{RL_LJ}.}
\end{figure}

Consider now a simple identity: $\left(\prtl s_c/\prtl \ln T \right)_p
= \left(\prtl s_c/\prtl \ln T\right)_V + \left(\prtl s_c/\prtl \ln V
\right)_T \left(\prtl \ln V/\prtl \ln T\right)_p$. Combining this with
the RFOT-derived Eqs.~(\ref{XWbarrier}) and (\ref{mdef}), we obtain
a simple relation between the fragility coefficient at constant
pressure ($m_p$) and volume ($m_v$) \cite{RL_LJ}:
\begin{equation} \label{mpmv1} m_p \simeq m_V + 18 T \alpha_t (\prtl
  s_c/\prtl \ln V)_T, \text{ at } T=T_g.
\end{equation}
A dimensionless measure of the thermal expansivity $T \alpha_T \equiv
(\prtl \ln V/\prtl \ln T)$ is an important quantity that determines
the pressure dependence of the fragility~\cite{RL_LJ}.  It is also an
interesting quantity in that it varies remarkably little, 0.16 \ldots
0.19, between many substances, see Fig.~(12) of Ref.~\cite{RL_LJ},
which is often called the Boyer-Bondi rule~\cite{BondiBook,
  PhysRevE.72.031503}. Note that the thermal expansivity $\alpha_t$ is
entirely determined by the anharmonic response of the lattice.

Using a generic value $T \alpha_t = 0.17$ and the RL's prediction that
for the Lennard-Jones liquid, $(\prtl s_c/\prtl \ln V) \approx 10
k_B$ (Fig.~10 of Ref.~\cite{RL_LJ}), we obtain a simple relation:
\begin{equation} \label{mpmv2} m_p \simeq m_V + 30,
\end{equation}
This prediction matches well the data of Casalini and
Roland~\cite{PhysRevE.72.031503}, see Fig.~\ref{mpmvFig}.

The above prediction concerning the universality of the Lindemann
ratio $L$ near the crossover has been recently utilised by Rabochiy
and Lubchenko~\cite{RL_Tcr} to develop a simple way to estimate the
crossover temperature based on the elastic properties of the
substance. According to their formula in Eq.~(\ref{alpha1}), there is
an intrinsic relationship between the elastic constants, temperature,
and the typical vibrational displacement in a harmonic solid. Further,
the lattice spacing in a random-close packing is approximately given
by $r_\snn \approx 1.14/\rho^{1/3}$. Combined with
Eqs.~(\ref{alpha1}), (\ref{Lratio}), and (\ref{L145}), this yields
$\frac{\mu}{\rho k_B T_\scr}\frac{ 3K + 4 \mu }{ 6K +11 \mu } \simeq
5.8$. After expressing the elastic constants in terms of speeds of
longitudinal ($v_L$) and transverse ($v_T$) sounds, one obtains a
simple, testable relation:
\begin{equation} \label{main1} \frac{M}{N_b k_BT_\scr} \, \frac{ v_T^2
    v_L^2 } {2 v_L^2 + v_T^2}= 5.8
\end{equation}
where $N_b$ is the bead count from Eq.~(\ref{Nb}). Note the ratio made
of the speeds of sound can written out as $(2/v_L^2 + 1/v_T^2)$, which
reflects the contributions of the two transverse and one longitudinal
phonon branches to the phonon sums. As in Eq.~(\ref{c2kk}), we assume
the isothermal and adiabatic sound speeds are numerically close.

\begin{figure}[t]
\begin{tabular*}{\figurewidth} {ll}
\begin{minipage}{.48 \figurewidth} 
  \begin{center}
    \includegraphics[width=0.48 \figurewidth]{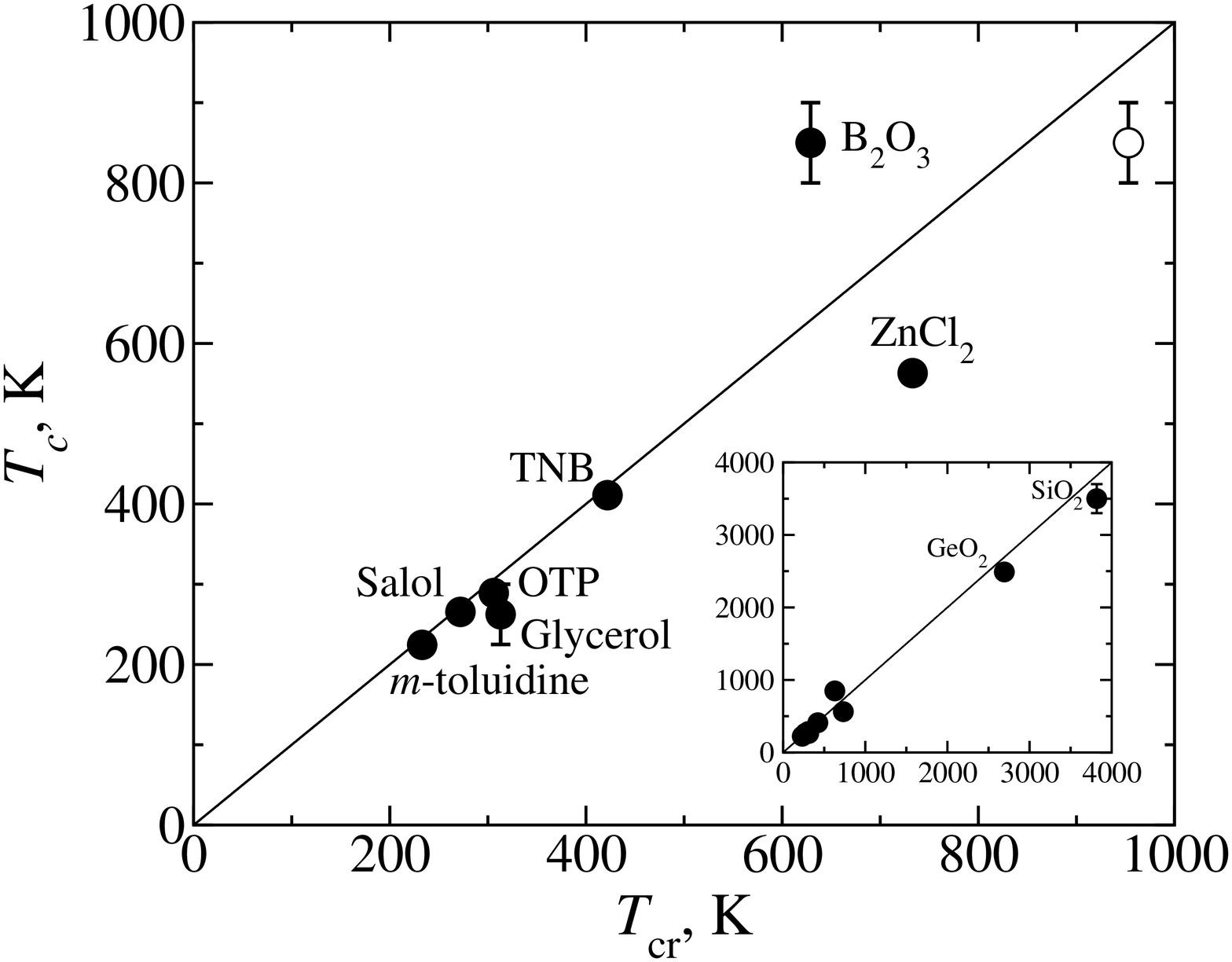} \\
    {\bf (a)}
   \end{center}
  \end{minipage}
&
\begin{minipage}{.48 \figurewidth} 
  \begin{center}
    \includegraphics[width=0.48 \figurewidth]{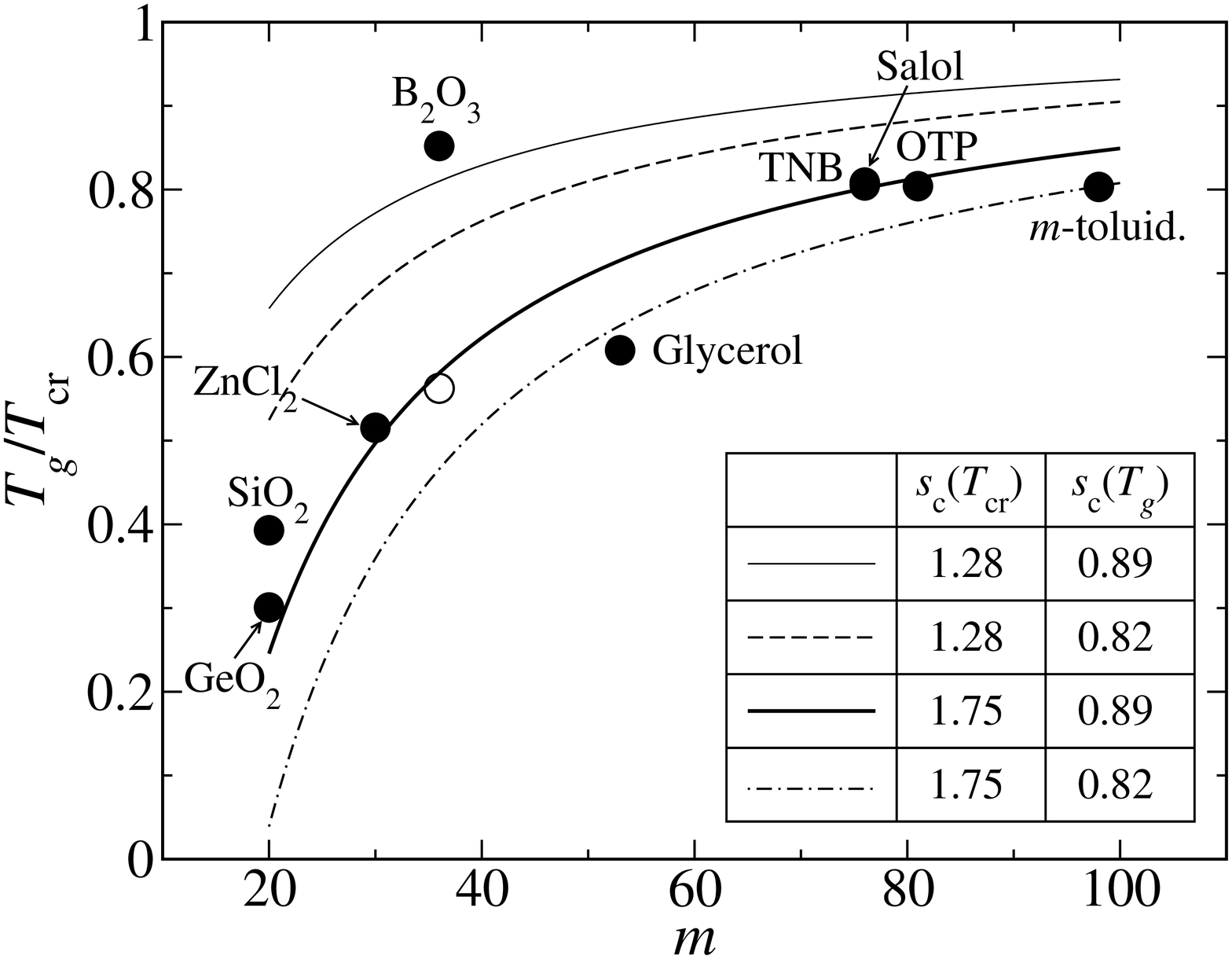}
    \\ 
    {\bf (b)} 
  \end{center}
\end{minipage}
\end{tabular*}
\caption{\label{RLTcrFig} {\bf (a)} Predicted values of the crossover
  temperature $T_\scr$ plotted against experimentally determined
  dynamic crossover temperature $T_c$. {\bf (b)}. Same values, in the
  form of the $T_g/T_\scr$ ratio, plotted against the experimentally
  determined fragility coefficient $m$. ($T_g$ is also experimentally
  determined.) Both panels are from Ref.~\cite{RL_Tcr}.}
\end{figure}

Eq.~(\ref{main1}) allows one to predict the crossover temperature for
actual substances, if data on the temperature dependent speeds of
sound are available. RL made such predictions for several specific
substances; the results of the calculation are shown in
Fig.~\ref{RLTcrFig}(a) alongside the experimentally determined values
of $T_\scr$. As already mentioned, Lubchenko and Wolynes had predicted
that the crossover temperature should be relatively close to the glass
transition temperature for fragile substances and vice versa for
strong substances. Fig.~\ref{RLTcrFig}(b) directly demonstrates this
notion. The ratio $T_g/T_\scr$ ($T_g$ measured, $T_\scr$ computed) is
shown as a function of experimentally determined fragility coefficient
$m$. The RFOT theory predicts~\cite{RL_Tcr, SSW}:
\begin{equation} \label{TcrVSm} \frac{T_g}{T_{cr}}= 1 -\frac{1}{m} \,
  \left(\frac{s_c\left(T_{cr}\right)}{s_c\left(T_g\right)}-1\right)
  \frac{32k_B}{\ln\left(10\right)s_c\left(T_g\right)}.
\end{equation}
This result is shown by smooth lines in Fig.~\ref{RLTcrFig}(b).  The
computed values of $T_\scr$ are clearly consistent with observation.
Note that predicted values of $T_\scr$ are consistently {\em above}
$T_g$.  This is reassuring as there is little {\em \`{a} priori}
reason for a combination of elastic constants and the bead
count---neither of which directly have to do with the RFOT or glass
transition---to produce a temperature that is consistently above
$T_g$. This robustness is consistent with the robustness of the
Lindemann criterion of melting in crystalline
materials~\cite{GrimvallSjodin}.

\begin{figure}[t]
\begin{tabular*}{\figurewidth} {ll}
\begin{minipage}{.5 \figurewidth} 
  \begin{center}
    \includegraphics[width= .46 \figurewidth]{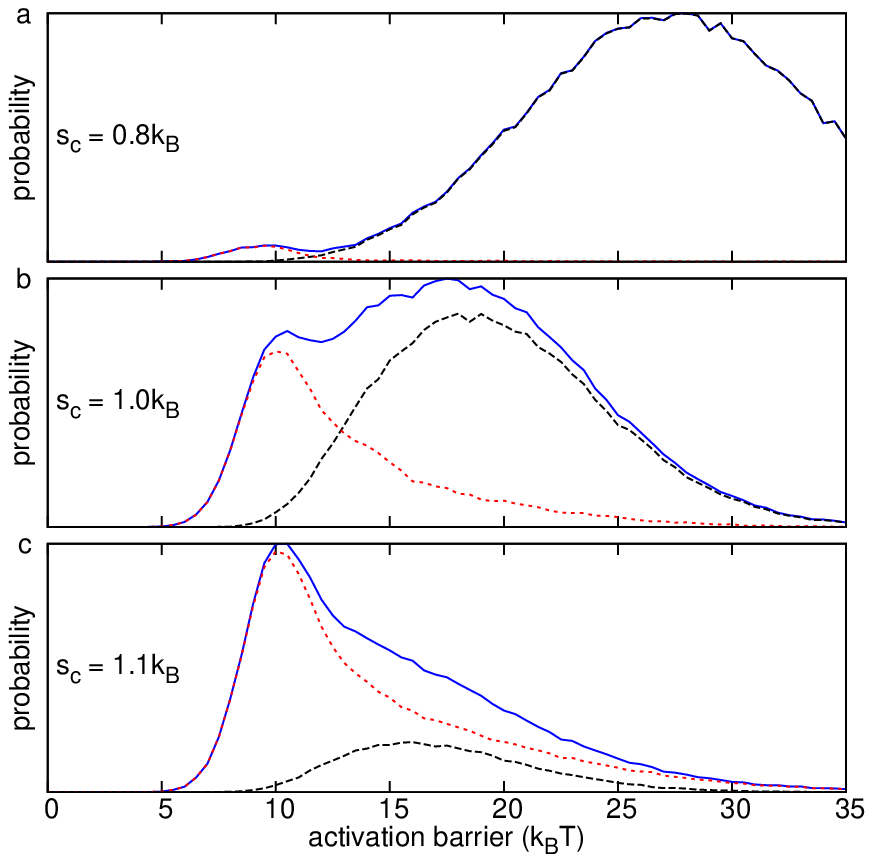}
    \\ {\bf (a)}
   \end{center}
  \end{minipage}
&
\begin{minipage}{.46 \figurewidth} 
  \begin{center}
    \includegraphics[width= .46 \figurewidth]{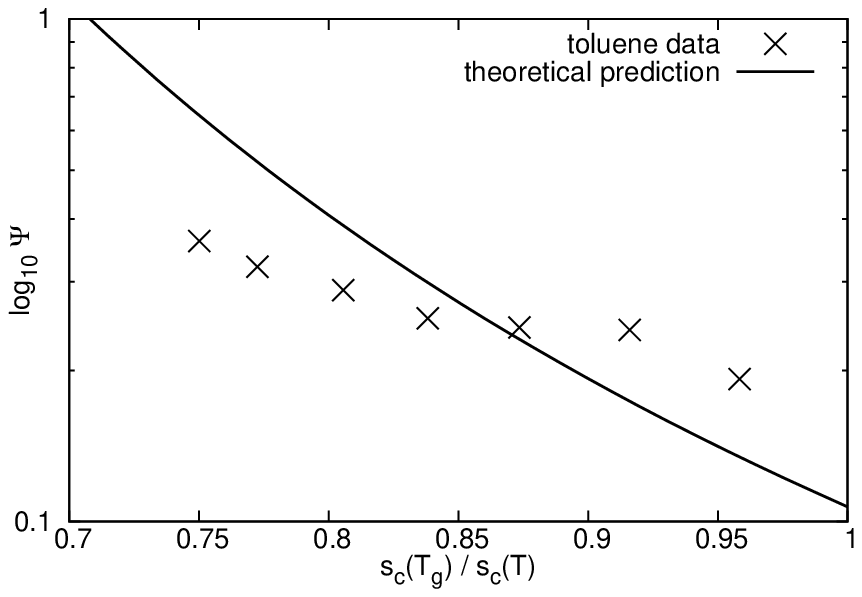} \vspace{3mm}
    \\ {\bf (b)}
  \end{center}
\end{minipage}
\end{tabular*}
\caption{ \label{betaRlxn} {\bf (a)} The apparent barrier distribution
  for the structural relaxations in the landscape regime, where the
  contribution of the non-compact, string-like excitations on top of
  the compact nucleation events shows up as a relatively distinct,
  low-barrier subset of motions.  As the temperature is decreased
  (bottom to top), the stringy subset separates in time from the
  compact excitations and becomes less important. This is illustrated
  in panel {\bf (b)}, where the partial contribution of the stringy
  modes to the overall excitation spectrum is plotted as a function of
  temperature~\cite{SWbeta}. The experimental data (symbols) are from
  Ref.~\cite{0953-8984-11-10A-010}.}
\end{figure}

The notion of the crossover as a regime in which the entropic and
enthalpic contributions to the free energy balance out, seems to be
consistent with findings of Biroli, Karmakar, and
Procaccia~\cite{PhysRevLett.111.165701} on the apparent coincidence
between the point-to-set length and (a fixed multiple of) the spatial
extent of marginally stable vibrational modes, in a finite temperature
range, see Fig.~\ref{szamel}(b).

We have seen that compact reconfigurations become increasingly
``dressed'' with string-like excitations~\cite{SSW}, upon approaching
the crossover from below. This microscopic picture may shed some light
on the poorly-understood beta-relaxations, see
Fig.~\ref{LL}. Similarly to the way local fluctuations in the
configurational-entropy content result in a distribution of the local
escape barrier from long-lived configurations, the very same
fluctuations will also affect the ease at which the stringy motions
can be excited. Indeed, the free energy cost of such motions is
directly connected with the local multiplicity of possible particle
arrangements. Stevenson and Wolynes~\cite{SWbeta} (SW) have studied
effects of local fluctuations of the landscape degeneracy on the ease
of string generation. Clearly, such strings will be more abundant in
regions characterised by relatively large values of $s_c$, even though
the compact core for the reconfiguration would be smaller than
average, by Eq.~(\ref{F2}).  SW have established that the most facile
subset of the string excitations engender the appearance of an
additional subset of structural relaxations on the low barrier side of
the distribution of the $\alpha$-relaxation barriers, see
Fig.~\ref{betaRlxn}(a). The latter figure demonstrates how the
low-cost stringy motions dominate the structural relaxation at high
temperatures near the crossover but become increasingly subdominant to
the nucleation events proper deeper in the landscape regime. This can
also be seen in Fig.~\ref{betaRlxn}(b), where the contribution of the
low-cost stringy excitations to the overall relaxation is plotted as a
function of temperature. The latter predictions are in qualitative
agreement with observation.  The just described relaxation mechanism
overlaps, frequency-wise, with and thus amounts to a universal
contribution to the set of excitations discussed under the umbrella of
beta-relaxations. It is likely that other, system-specific
contributions to the latter relaxations are present.

In concluding this Section, we reiterate the most important
qualitative features of the crossover, which is the finite-dimensional
realisation of the random first order transition. The crossover can be
thought of in two related ways: If approached from above, it is
signalled by an increase in collision-driven mode-coupling effects. In
the mean-field limit, this increase would lead to a complete kinetic
arrest inside a particular free energy minimum, at a temperature
$T_A$. The latter temperature is above the temperature $T_K$, thus
implying the liquid is completely frozen even as its configurational
entropy is perfectly finite.  This seeming paradox is however resolved
in finite dimensions, whereby the liquid is allowed to reconfigure by
activation. What would be a sharp transition in the mean-field limit,
at the temperature $T_A$, now becomes a soft crossover centred at a
temperature $T_\scr < T_A$. If approached from {\em below}, the
crossover is signalled by a rapid increase in the rate of structural
reconfiguration, which is made even more precipitous by the
barrier-softening effects stemming from the fluctuations of the order
parameter $\alpha$. The latter quantity can be thought of as a local
``stiffness,'' by Eq.~(\ref{alpha1}). As the barrier vanishes, the
viscous response of the liquid is now dominated by the collisional
effects. Importantly, the vanishing of the curvature of the
metastable, aperiodic minimum in $F(\alpha)$ at $\alpha_0$,
Fig.~\ref{RLFalpha}(a), does not lead to a diverging length scale as
it would do during ordinary transitions between phases that are each
characterised by a single free energy minimum.  The fluctuations in
the order parameter $\alpha$ play a triple role here: First, they
lower the transition temperature from its mean-field value $T_A$ to a
lower value $T_\scr$. Second, they also lower the barrier for the
activated reconfigurations. The latter then destroy the long-range
correlations that seem to be called for, at least superficially, by
the vanishing of $F''(\alpha_0)$ at $T_A$. Third, because the
activated events are low-barrier and ultra-local near what would be a
sharp ``spinodal'' at $T_A$ in meanfield, the latter spinodal becomes
a gradual crossover in finite dimensions. Alternatively said,
high-frequency modes freeze first, followed by the progressively
slower modes as temperature is lowered.  Consistent with our earlier
conclusions that a continuous liquid-to-solid transition would be
lowered by quasi-one-dimensional motions, string-like excitations are
predicted and appear to be observed in simulation near the crossover.

\section{Relaxations far from equilibrium: glass ageing and
  rejuvenation}

The emergence of the transient structures and, hence, activated
transport---as a prelude to the actual glass transition---seems
intuitive. Indeed, once exponentially many distinct minima have
formed, it is easy to imagine how the minima become progressively
deeper with lowering the temperature, since the enthalpy must decrease
with cooling. The glass transition then simply signifies a situation
in which the minima become so deep that the structure fails to
rearrange on the experimental timescale.

Yet, how do we know that transport involves activated events? This is
not entirely self-evident in view of the strongly non-Arrhenius
behaviour of the relaxation times in glassy liquids.  Given these
difficulties, can one name a {\em direct} experimental signature that
the transport is an activated process?  Such a direct signature is
actually provided by the relaxation in glasses themselves {\em below}
the glass transition. This relaxation, called ``ageing,'' is an
attempt for the glass to reach the structure that would be
representative of the liquid equilibrated at the ambient
temperature. To analyse ageing, the approach we took in
Subsection~\ref{ActTransport} must be modified to account for the
initial structure being different from an equilibrium one.

\subsection{Ageing}
\label{aging}

It is most straightforward to describe ageing using the free energy
formulation of the activated transport from Eq.~(\ref{FN1}).  Consider
an experiment in which an equilibrated liquid is rapidly cooled or
heated from temperature $T_1$ to temperature $T_2$, where both $T_1$
and $T_2$ are below the crossover temperature $T_\scr$.  Below, we
limit ourselves to temperature jumps that are faster than any
irreversible structural changes, but slower than the vibrational
relaxation. Lets call state 1 the configuration, in which the
vibrational degrees of freedom have already equilibrated at
temperature $T_2$, but where the other, anharmonic structural degrees
of freedom have not equilibrated. State 2 is the state {\em following}
a reconfiguration event whereby all degrees of freedom have
equilibrated at temperature $T_2$.  For simplicity, let us pretend for
now that structural reconfigurations are not accompanied by any volume
change, to be discussed later.

As during relaxation near equilibrium, the reconfiguration is subject
to the mismatch penalty, according to Lubchenko and
Wolynes~\cite{LW_aging}:
\begin{equation} \label{FNaging} F(N) = \gamma N^{1/2} + \Delta g(T_1,
  T_2) N,
\end{equation}
where the driving force $\Delta g(T_1, T_2) < 0$ for escape from state
1 to state 2,
\begin{equation} \Delta g(T_1, T_2) \equiv \Delta g(T_1 \to T_2),
\end{equation}
is given by the bulk free energy difference between the final and
initial state of the region. The reconfiguration barrier is given by
\begin{equation} \label{FDg} F^\ddagger(T_1, T_2) =
  \frac{\gamma^2}{4[-\Delta g(T_1, T_2)]}
\end{equation}

State 2 could be {\em any} structure representative of the liquid at
the ambient temperature, i.e., $T_2$. Thus the free energy
$\overline{G}^{(2)}$ of the final state is equal to free energy of the
liquid equilibrated at $T_2$:
\begin{equation} \label{qAging} \overline{G}^{(2)} = \overline{G}(T_2)
  = \overline{G}_i(T_2) - T_2 S_c(T_2),
\end{equation}
c.f. Eq.~(\ref{gequil}). The (average) free energy of an individual
state,
\begin{equation} \label{Gi2} \overline{G}_i(T_2) = \overline{H}_i(T_2)
  - T_2 \overline{S}_{\svibr, \, i}(T_2),
\end{equation}
accounts for the vibrational entropy of that state, as before.

If the temperatures $T_1$ and $T_2$ were equal, state 1 would
correspond to an {\em individual} free energy minimum at temperature
$T=T_1=T_2$.  Instead, the structure of state 1 only {\em
  approximately} corresponds to an individual minimum that would be
representative of a liquid equilibrated at $T_1$, since the vibrations
have already equilibrated at temperature $T_2$. The vibrations at
$T_1$ and $T_2$ are of different magnitude, if $T_1 \ne T_2$, and so
some shift in the average position of the particles is expected upon
vibrational relaxation, owing to the anharmonicity of the
lattice. Still, the structure in state 1 is completely isomorphic to
the structure at $T_1$.  We will thus assume approximately that the
configuration---and hence the enthalpy---of state 1 are the same as in
equilibrium at temperature $T_1$. On the other hand, the vibrational
entropy in state 1 is approximately equal to the equilibrium
vibrational entropy at temperature $T_2$:
\begin{equation} \label{Gi1} \overline{G}^{(1)} \approx
  \overline{H}_i(T_1) - T_2 \overline{S}_{\svibr, \, i}(T_2).
\end{equation}
Thus the bulk driving force $\Delta g$ for reconfiguration, per
particle, is given by~\cite{LW_aging}:
\begin{equation} \label{Deltag} \Delta g(T_1, T_2) =
  \overline{g}^{(2)} - \overline{g}^{(1)} \approx \overline{h}(T_2) -
  \overline{h}(T_1) - T_2 s_c(T_2).
\end{equation}
Using Eq.~(\ref{scTRA}), one obtains an explicit expression for the
driving force:
\begin{equation} \label{df1} - \Delta g(T_1, T_2) = \Delta c_p (T_g)
  T_g \left[ \left(\frac{T_2}{T_K}-1 \right) - \ln \left(
      \frac{T_2}{T_1} \right) \right].
\end{equation}

We now specifically focus on downward temperature quenches, $T_2 <
T_1$. Given the postulated quickness of the quench, we must identify
$T_1$ as the glass transition temperature: $T_1 = T_g$, while $T_2$ is
the ambient temperature $T$. Under these circumstances,
$\overline{h}(T) - \overline{h}(T_g) = \int_{T_g}^{T} \Delta c_p dT <
0$. Thus, the driving force for the escape from the structure quenched
at $T_g$, at the still lower temperature temperature $T < T_g$,
exceeds the driving force for reconfigurations in a liquid
equilibrated at the ambient temperature $T$:
\begin{equation} \label{drForceAging} -\Delta g(T_g, T) > - \Delta
  g(T, T), \text{ if } T < T_g.
\end{equation}

It is easy to see that the activation barrier for ageing is only weakly
temperature dependent~\cite{LW_aging}. Indeed, at $T=T_g$, $\Delta g =
- \Delta c_p(T_g) (T_g/T_K - 1)$. The most stable aperiodic structure
corresponds to the putative ideal-glass state that would be
equilibrium at the Kauzmann temperature $T_K$, where $s_c= 0$. Hereby,
the driving force is $\Delta g = \overline{h}(T_K) - \overline{h}(T_g)
= - \Delta c_p(T_g) \ln(T_g/T_K)$. To decide on the temperature
dependence of the mismatch penalty, if any, we revisit to
Eq.~(\ref{h2}), in which the energy prefactor $h$ now consists of the
contribution from the free energy fluctuations on the inside, $\delta
G_i(T_2)$, and that on the outside $\delta [H_i(T_1) - T_2
S_\svibr(T_2)]$. According to Eq.~(\ref{gamma2}), $\gamma \propto [K
(T_1 + T_2)/\bar{\rho}]^{1/2}$, considering that the bulk modulus and
density barely change with temperature in the frozen glass. Thus the
reconfiguration barrier goes as $(2T_g)^2/(T_g/T_K - 1)$ and
$(T_g+T_K)^2/\ln(T_g/T_K)$ at $T_g$ and $T_K$ respectively.  The two
quantities differ by at most 25\%, since empirically, the ratio
$T_g/T_K$ is numerically at most 2 (for very strong substances), but
is usually considerably less than 2.

\begin{figure}[t]
\begin{tabular*}{\figurewidth} {ll}
\begin{minipage}{.48 \figurewidth} 
  \begin{center}
    \includegraphics[width=0.35 \figurewidth]{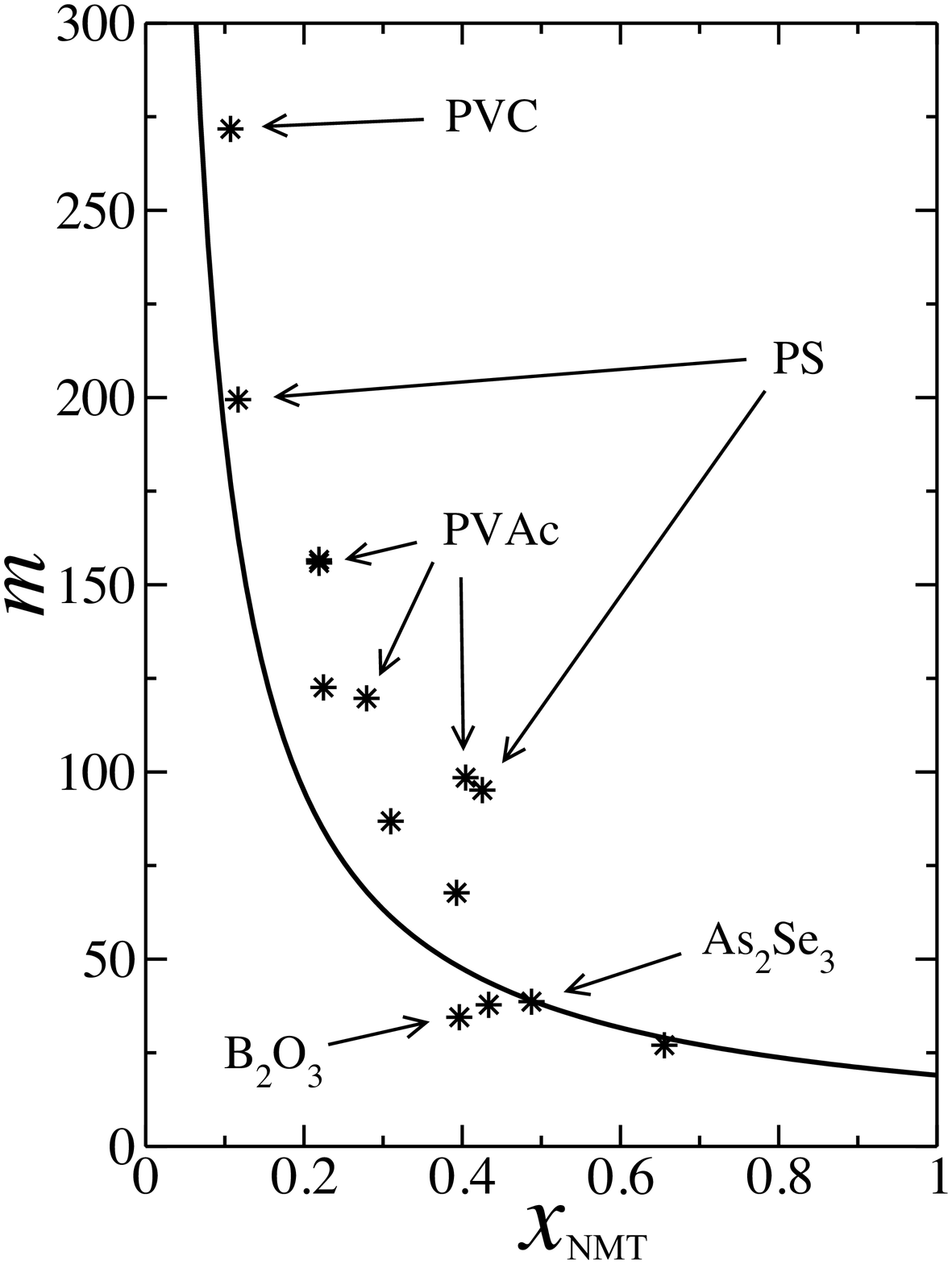} \\
    \vspace{-2mm} 
    {\bf (a)}
   \end{center}
  \end{minipage}
&
\begin{minipage}{.48 \figurewidth} 
  \begin{center}
    \includegraphics[width=0.4 \figurewidth]{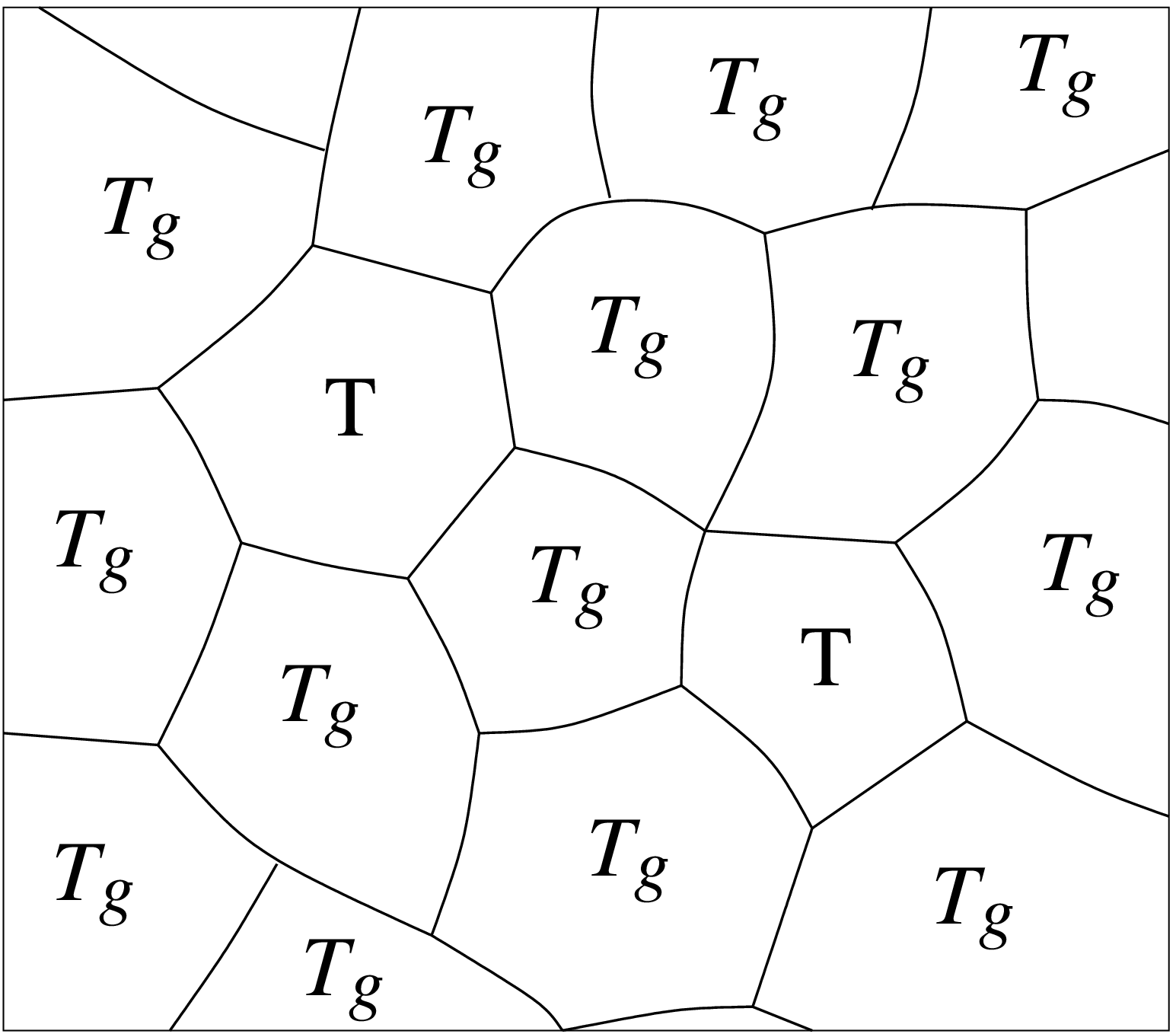}
    \\ \vspace{7mm}
    {\bf (b)}
  \end{center}
\end{minipage}
\end{tabular*}
\caption{\label{agingFigs} {\bf (a)} Smooth line: The RFOT-predicted,
  simple relation between the NMT coefficient $x$, which reflects the
  discontinuity of the apparent activation rate (\ref{actenergy}) and
  the fragility. Symbols: Experimental data. Note experimental data
  show significant scatter even for the same substance.  {\bf (b)}
  After a considerable period of ageing well below $T_g$ a patchwork of
  equilibrated and non-equilibrated mosaic cells will be found, leading
  to a distribution of fictive temperatures and emergence of
  ultra-slow relaxations.  After Ref.~\cite{LW_aging}}
\end{figure}

Because the activation barrier is only weakly temperature dependent
below $T_g$, the apparent activation energy (not the free energy!)
\begin{equation} \label{actenergy} E_\text{act} =
  \frac{\prtl(F^\ddagger/k_B T)}{\prtl (1/T)},
\end{equation}
which is generally not equal to $F^\ddagger$, is predicted to exhibit
a discontinuous {\em jump} following a rapid quench,
c.f. Fig.~\ref{LL}. (The activation {\em free} energy is continuous
through the glass transition.)  A convenient framework for
quantitative discussion of this jump is provided by the
Narayanaswamy-Moynihan-Tool
formalism~\cite{Tool,Narayanaswamy,Moynihan}, in which the relaxation
rate in quenched glasses is phenomenologically decomposed into a
purely activated part and a temperature-independent part reflecting a
{\em fictive} temperature $T_f$:
\begin{equation} \label{k_AG} k_\text{n.e.} = k_0
  \exp\left\{-x_\text{\tiny NMT}\frac{E_\text{act, eq}}{k_B T} -
    (1-x_\text{\tiny NMT}) \frac{E_\text{act, eq}}{k_B T_f}.
  \right\}, 
\end{equation}
Here $E_\text{act, eq}$ is the apparent activation barrier from
Eq.~(\ref{actenergy}) in an equilibrated liquid, i.e., just above the
glass transition: $E_\text{act, eq} = E_\text{act}(T_g^+)$. The
fictive temperature $T_f$ is an approximate concept. By construction,
it is chosen so that the structure equilibrated at $T_f$ would be as
similar as possible to the {\em actual} structure. Above $T_g$, $T_f =
T$, of course. Below $T_g$, one often adopts $T_f = T_g$ since the
structure of the frozen glass is quite similar to that of the liquid
at the glass transition, apart from some ageing and subtle changes
stemming from differences in the vibrational amplitude, which is an
anharmonic effect. For the NMT description to be
internally-consistent, one must disregard the temperature dependence
of the activation barrier below $T_g$, which we have seen is a good
approximation. This implies $F^\ddagger|_{T < T_g} \approx
F^\ddagger(T_g, T_K) $ leading to $ \prtl(F^\ddagger/k_B
T)/\prtl(1/T)|_{T<T_g} = F^\ddagger(T_g, T_K)$, where $F^\ddagger(T_1,
T_2)$ is from Eq.~(\ref{FDg}).  Finally, the quantity $x_\text{\tiny
  NMT}$, $0 < x_\text{\tiny NMT} < 1$, is a dimensionless measure of
the decrease in the apparent activation energy. Thus, $x_\text{\tiny
  MNT} E_\text{act, eq} = F^\ddagger(T_g, T_K) \Rightarrow
x^{-1}_\text{\tiny MNT} = E_\text{act, eq}/F^\ddagger(T_g, T_K)$.
Further using Eqs.~(\ref{mdef}), (\ref{FDg}), and (\ref{df1}), this
straightforwardly leads to the following estimate~\cite{LW_aging}:
\begin{equation} \label{DEapp} x_\text{\tiny MNT}^{-1} 
  = m \left\{ (\log_{10}e) \frac{F^\ddagger(T_g)}{k_B T_g}
    \left[\frac{\gamma(T_K)}{\gamma(T_g)} \right]^2
    \frac{(T_g/T_K-1)}{\ln(T_g/T_K)} \right\}^{-1}, 
\end{equation}
where $m$ stands for the fragility coefficient $m$ from
Eq.~(\ref{mdef}).  The last ratio on the r.h.s. depends on the
$T_g/T_K$ ratio only weakly, as already remarked. Using a specific,
generic value $T_g/T_K = 1.3$ and ignoring the temperature dependence
of the coefficient $\gamma$, one obtains a simpler yet relation
between the fragility coefficient and the discontinuity of the
apparent activation energy at the glass transition~\cite{LW_aging}:
\begin{equation} \label{m19x} m \simeq \frac{19}{x_\text{\tiny NMT}}.
\end{equation}
This simple relation agrees well with experiment, see
Fig.~\ref{agingFigs}(a), thus supporting the RFOT-advanced microscopic
picture on a very basic level.

We shall now discuss the effects of volume mismatch during ageing.  The
expression for the driving force in Eq.~(\ref{Deltag}) corresponds to
a process at constant pressure and thus is applicable for relatively
shallow quenches.  For considerable $T$-jumps, however, the situation
is more complicated since on the timescale of the nucleation event,
the sample is a mechanically stable solid with a non-zero shear
modulus.  And so, insofar as the equilibrium thermal expansivity
$\alpha_\text{eq} \equiv (1/V)(\prtl V/\prtl T)_p$ exceeds the
(largely vibrational) expansivity of a frozen glass
$\alpha_\text{vibr}$, downward $T$-jumps will be accompanied by some
stretching of the environment: A compact region of the material is
essentially replaced by a region with a smaller volume, following an
ageing event.  The expansivity of an equilibrated liquid usually does
significantly exceed that of the corresponding glass, see for instance
Ref.~\cite{10.1063/1.432870}. Despite this circumstance, the effects
of volume mismatch between the aged and unrelaxed glass do not
significantly affect the ageing rate, as we discuss in detail in
Appendix~\ref{volumemismatch}.

After a considerable period of ageing well below $T_g$, a patchwork of
equilibrated and non-equilibrated mosaic cells will
develop~\cite{LW_aging}, see Fig.~\ref{agingFigs}(b). If the
equilibrium energy at $T$ is further than a standard deviation from
the typical energy at $T_g$, the distribution of energies will be
noticeably bimodal and the idea of a single fictive temperature will
break down.  The typical magnitude of temperature fluctuations is
given by $(k_B T^2/\Delta c_v N^*)^{1/2}$~\cite{LLstat}. Thus,
significant deviations from a unimodal distribution of fictive
temperatures are not expected if $\Delta T = T_g -T < (k_B T^2/\Delta
c_v N^*)^{1/2} \equiv \delta T^*$. For $T_g$ relevant to 1
hr. quenches this gives $\delta T^*/T_g \simeq 0.07$. Most of the
Alegria et al.~\cite{Alegria} data lie in this modest quenching range,
while ``hyperquenched'' samples (with $\Delta T \gg \delta T^*$) will
often fall outside the allowed range of using a single fictive
temperature. When a sample has a two-peaked distribution of local
energies, the RFOT theory predicts an ultra-slow component of
relaxation will arise. Notice that an equilibrated region at the
temperature $T = T_g - \delta T^*$ will relax on the tens to hundreds
of hours scale, if $\tau_g$ is taken to be one hour. (The relaxation
barrier depends on temperature only weakly.)

\subsection{Rejuvenation}
\label{rejuvenation}

Let us now switch focus to upward temperature jumps, $T_2 > T_1$, which
is a different type of ageing experiment.  In this type of ageing, the
structure locally escapes from being relatively deep in the free
energy landscape to the region in the phase space where the free
energy minima are relatively shallow, by Eqs.~(\ref{XWbarrier}) and
(\ref{FKsc1}). Consistent with this notion, Eq.~(\ref{df1}) prescribes
that there is not as much driving force for structural relaxation when
$T_2 > T_1$ than in equilibrium at temperature $T=T_1=T_2$: 
\begin{equation} \label{drForceRejuv} - \Delta g(T_1, T_2) < - \Delta
  g(T_2, T_2), \text{ if } T_2 > T_1,
\end{equation}
c.f. Eq.~(\ref{drForceAging}).  For this reason, the process of glass
{\em rejuvenation}---i.e., warming up and subsequent equilibration of
a vitrified sample at the target temperature---is slower than the
relaxation time in equilibrium at that same temperature. On the other
hand, the relaxation barrier for a $T_2 > T_1$ process is lower than
the relaxation barrier in equilibrium at $T_1$, as follows from
Eq.~(\ref{df1}) (note $T_1 > T_K$ of course):
\begin{equation} \label{drForceRejuv2} - \Delta g(T_1, T_2) > - \Delta
  g(T_1, T_1), \text{ if } T_2 > T_1.
\end{equation}
Thus the sample will undergo irreversible relaxation to configurations
typical of equilibrium at temperature $T_2$, following the temperature
jump.  Furthermore, due to the relatively high mobility on the edge,
to be discussed in Section~\ref{ultrastable}, the sample will not melt
uniformly but will begin melting preferentially from the edge where it
is being heated. Once a region is melted, a {\em mobility front} will
propagate in the sample since the melted regions relax significantly
faster than the those regions not reached by the front. It is
convenient to define the mobility in direct relation to the fictive
temperature $T_f$:
\begin{equation} \label{mobTF} \bar\mu(T_F,T) = \mu_0 \exp \left\{-
    \frac{x_\text{\tiny NMT} E^\ddag}{k_B T} -
    \frac{\left(1-x_\text{\tiny NMT} \right) E^\ddag}{k_B T_F}
  \right\},
\end{equation}
c.f. Eq.~(\ref{k_AG}). The mobility $\mu$ (not to be confused with the
shear modulus) is a quantity of units inverse time that describes the
local relaxation rate for any quantity in question. The mobility can
be converted into specific transport coefficients of interest, if
desired. For instance, the diffusivity is equal to $ (d_L^2) \mu $,
c.f. Eq.~(\ref{Dactivated}). The parameter $x_\text{\tiny NMT}$ can be
computed using Eq.~(\ref{DEapp}) or (\ref{m19x}).

\begin{figure}[t]
\begin{tabular*}{\figurewidth} {ll}
\begin{minipage}{.46 \figurewidth} 
  \begin{center}
    \includegraphics[width=0.4 \figurewidth]{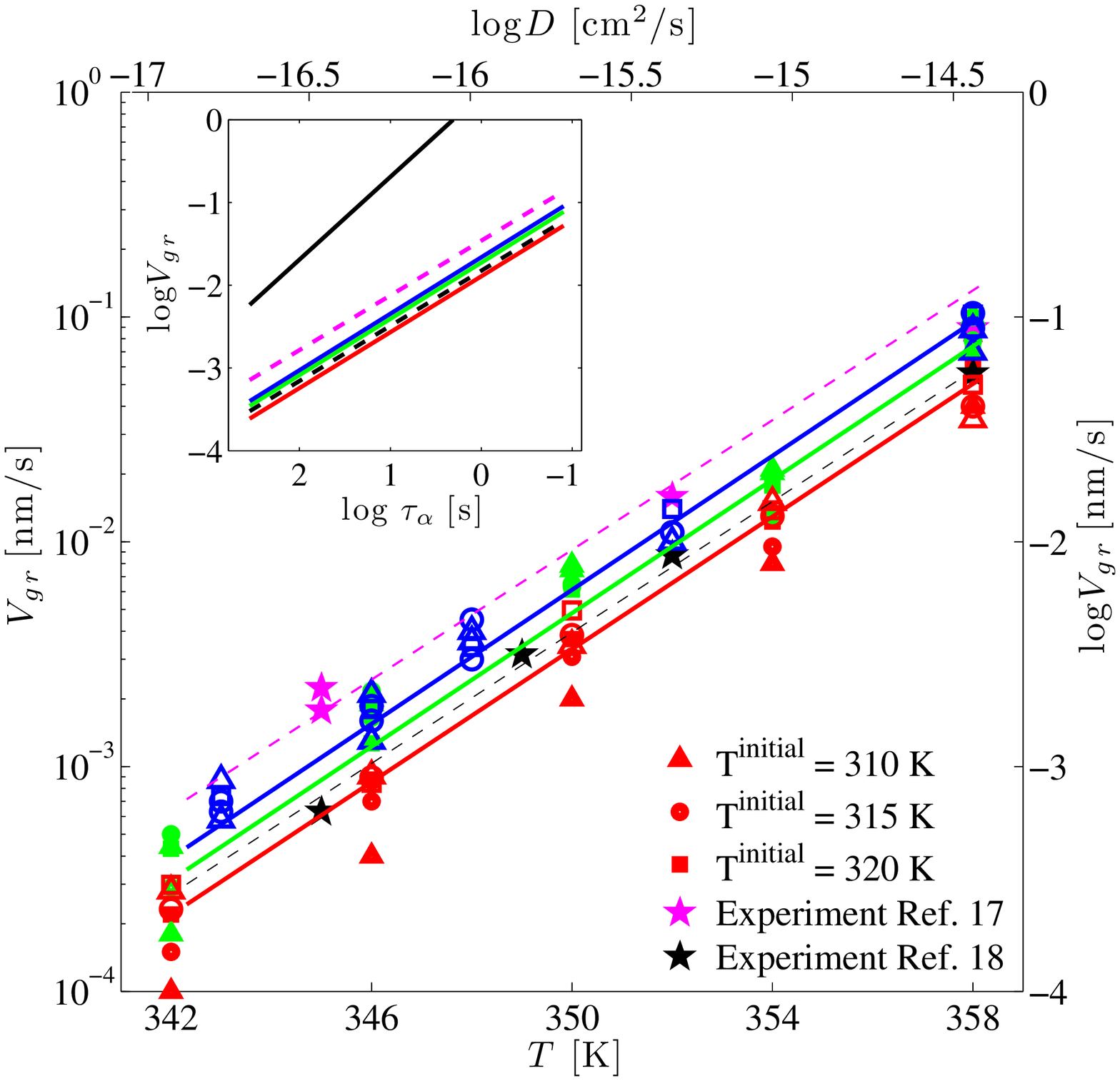} \\
    {\bf (a)}
   \end{center}
  \end{minipage}
&
\begin{minipage}{.56 \figurewidth} 
    \includegraphics[width=0.5 \figurewidth]{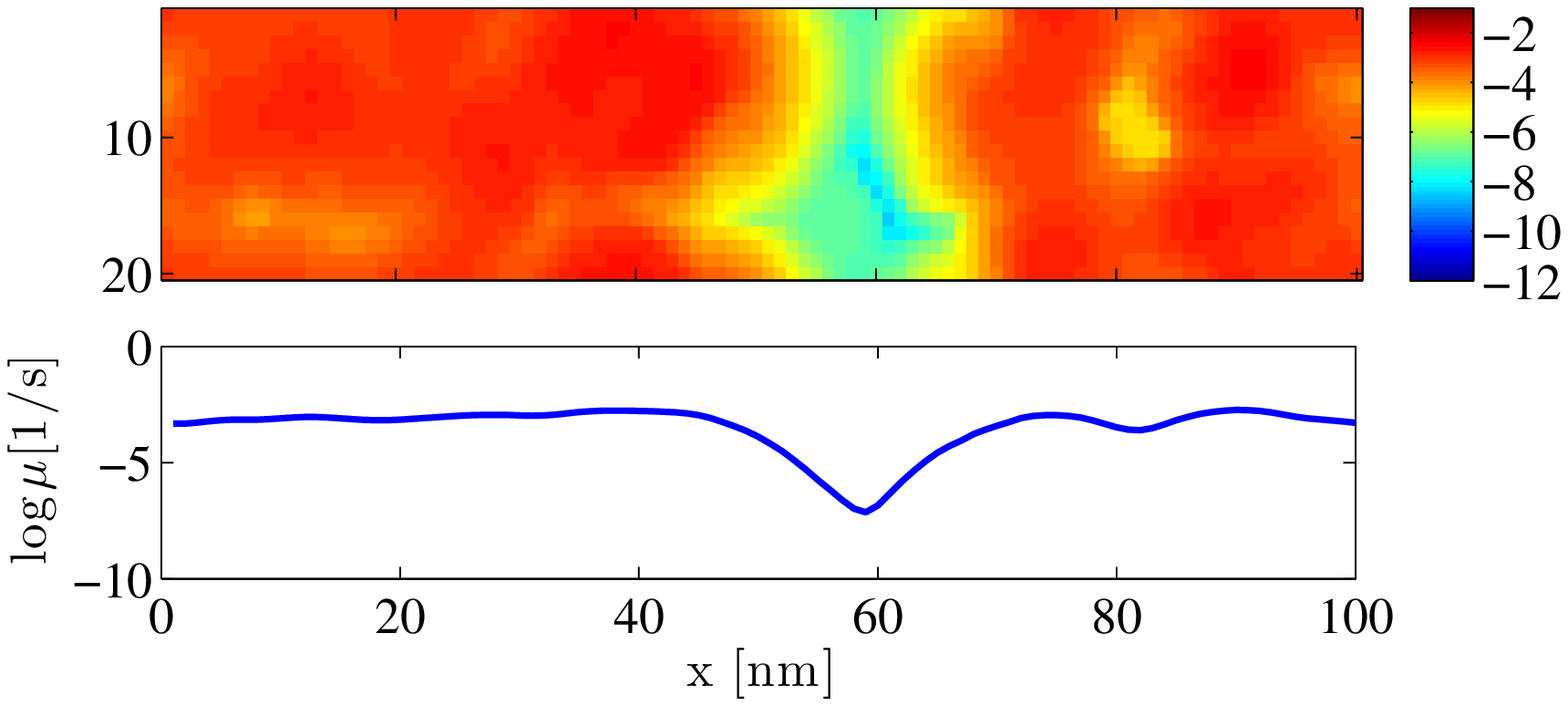}
    \\ \vspace{7mm}
  \begin{center}
    {\bf (b)}
  \end{center}
\end{minipage}
\end{tabular*}
\caption{\label{rejuvenationFig} {\bf (a)} The speed of the mobility
  front from the dynamics of the 1D deterministic linearised (solid
  red markers), the 1D deterministic non-linear Eq.~(\ref{NLFKPP})
  (opened red markers), the 1D stochastic linearised (solid green
  markers), the 1D stochastic non-linear (opened green markers), the
  2D stochastic linear (solid blue markers), and the 2D stochastic
  non-linear (opened blue
  markers)~\cite{PhysRevE.88.022308}. Different kinds of symbols
  indicate different initial fictive temperatures, the triangles are
  for $T^{\mathrm{initial}} = 310$~K, circles for
  $T^{\mathrm{initial}} = 315$~K, and squares for
  $T^{\mathrm{initial}} = 320$~K. The experimental measurements, as
  reported in Ref.~\cite{swallen2010transformation}, are shown as
  magenta stars and the black stars represent the more recent results
  from Ref.~\cite{sepulveda:204508}. All lines are drawn as guides for
  the eye. Inset: The front speed versus relaxation time. The
  continuum approximation for the front speed $(2/3)^{1/2} \xi
  \mu^{\mathrm{high}}$ is drawn as the black solid line. {\bf (b)}
  {\em Top}: Mobility field snapshot at $t = 4.6 \times 10^4$~s. The
  upper plot shows the mobility field from 2D stochastic linearised
  theory. The colour scheme denotes the mobility on a log-scale with
  corresponding col-or bars on the right. {\em Bottom}: Average values
  of the mobility fields along the y-axis.~\cite{PhysRevE.88.022308}}
\end{figure}

In estimating the speed of the mobility front, one must bear in mind
that the earlier derived expressions for the activation barrier only
apply to spatially homogeneous samples in the sense that the fictive
temperature is spatially uniform. (To avoid confusion we reiterate
that even if the fictive temperature is uniform, the dynamics are
still heterogeneous in the sense of Section~\ref{hetero}.) The actual
mobility should smoothly interpolate between slow and fast regions,
while the lengthscale of the variation is numerically close to the
cooperativity size $\xi$ from Eq.~(\ref{xi}).  A simple recipe to
determine such a interpolating mobility field is to solve the
differential equation $\xi^2 \nabla^2 \mu = \mu-\bar \mu$ with
appropriate boundary conditions~\cite{stevenson:234514}. Using $\xi$
in front of the Laplace operator automatically supplies the correct
length scale. (Note that since $\xi$ depends on the mobility only
logarithmically, using the typical value of $\xi$ characteristic of
the average mobility $\bar \mu$ suffices.) This phenomenological
equation suggests a simple way to go about constructing a hydrodynamic
continuum description for the time/space evolution of the mobility
field. Indeed, one may generally decompose the rate of change of local
mobility into a conserved and non-conserved part: $\dot \mu = -\bm
\nabla \bm j - \mu (\mu - \bar \mu)$. Thus in the absence of spatial
heterogeneity, $j = 0$, the mobility relaxes to its expectation value
$\bar \mu$ with the local relaxation rate, which is equal to the
mobility itself. Likewise, the flux $\bm j$ can be connected to the
magnitude of spatial variation in $\mu$ via a Fick-like law: $\bm j =
- \xi^2 \mu \bm \nabla \mu$. Putting these notions together yields the
following transport equation~\cite{PhysRevE.88.022308}:
\begin{equation}\label{NLFKPP} \dot \mu = \bm \nabla 
  \left( \mu \, \xi^2 \bm \nabla \mu \right) - \mu \left(\mu -
    \bar\mu \right) + \delta g + \bm \nabla \delta \bm j,
\end{equation}
where we have also included two terms corresponding to thermal noise,
$\bm \nabla \delta \bm j$ and $\delta g$, for the conserved and
non-conserved part of the mobility relaxation rate respectively. The
magnitude of these random-noise terms can be fixed using the
fluctuation-dissipation theorem~\cite{PhysRevE.88.022308}. The
linearised version of the above equation is easy to write down: $\dot
\mu = \bm \nabla \left(\bar \mu \, \xi^2 \bm \nabla \mu \right) - \bar
\mu \left(\mu - \bar\mu \right) + \delta g + \bm \nabla \delta \bm j$.
Note either this or the original non-linear equation (\ref{NLFKPP})
automatically insure that the mobility varies on the length scale
$\xi$.  The simplest way to quantify the approach of the local fictive
temperature to the ambient temperature is through the following,
ultralocal relaxation law
\begin{equation} \label{TFdot} \dot T_F = - \mu \left( T_F - T \right)
  + \delta \eta,
\end{equation}
where $\delta \eta$ is the corresponding noise-term. Note the fictive
temperature is a structural variable that has no dynamics of its own
other than through the mobility transport; thus there is no diffusive
term on the r.h.s. of Eq.~(\ref{TFdot}).

Aside from the mobility-dependent ``diffusivity,'' Eq.~(\ref{NLFKPP})
is the celebrated Fisher-Kolmogorov-Petrovsky-Piscounov~\cite{KPP,
  FisherKPP} equation that can be used to study migration of
organisms, as accompanied by death and procreation, and has also been
used to model flame propagation. This equation (in 1D) also governs
the generating function for the density distribution in the random
directed-polymer problem; curiously, it exhibits a singularity
intimately related to the glass transition in the random energy
model~\cite{DerridaSpohn}.

In Fig.~\ref{rejuvenationFig}(a), we show the results of the solution
of Eq.~(\ref{NLFKPP})---together with Eq.~(\ref{TFdot})---and its
linearised version. The solutions in the presence and absence of the
noise-terms, called ``stochastic'' and ``deterministic'' respectively,
are illustrated. One observes good agreement of the predictions with
the experimentally observed mobility front propagation. A snapshot of
the linearised, stochastic simulation in 2D is provided in
Fig.~\ref{rejuvenationFig}(b).

Note that a reasonable approximation for the speed of the mobility
front can be obtained without actually solving the differential
equations above, by using an analogy with flame
fronts~\cite{2009PNAS..106.1353P}. According to this approximation,
shown by the black solid line in the inset of
Fig.~\ref{rejuvenationFig}(a), the front speed is a fixed fraction of
the mobility $\mu^{\mathrm{high}}$ in the sample equilibrated at the
ambient temperature, viz., $(2/3)^{1/2} \xi
\mu^{\mathrm{high}}$. While an overestimation, this simple result
matches observation qualitatively.

\section{Rheological and Mechanical Anomalies}
\label{rheo}

\subsection{Shear thinning}

\begin{figure}[t]
\begin{tabular*}{\figurewidth} {ll}
\begin{minipage}{.50 \figurewidth} 
  \begin{center}
    \includegraphics[width=0.50
    \figurewidth]{WDvsTh.eps} \\
    {\bf (a)}
   \end{center}
  \end{minipage}
&
\begin{minipage}{.46 \figurewidth} 
  \begin{center}
    \includegraphics[width=0.46 \figurewidth]{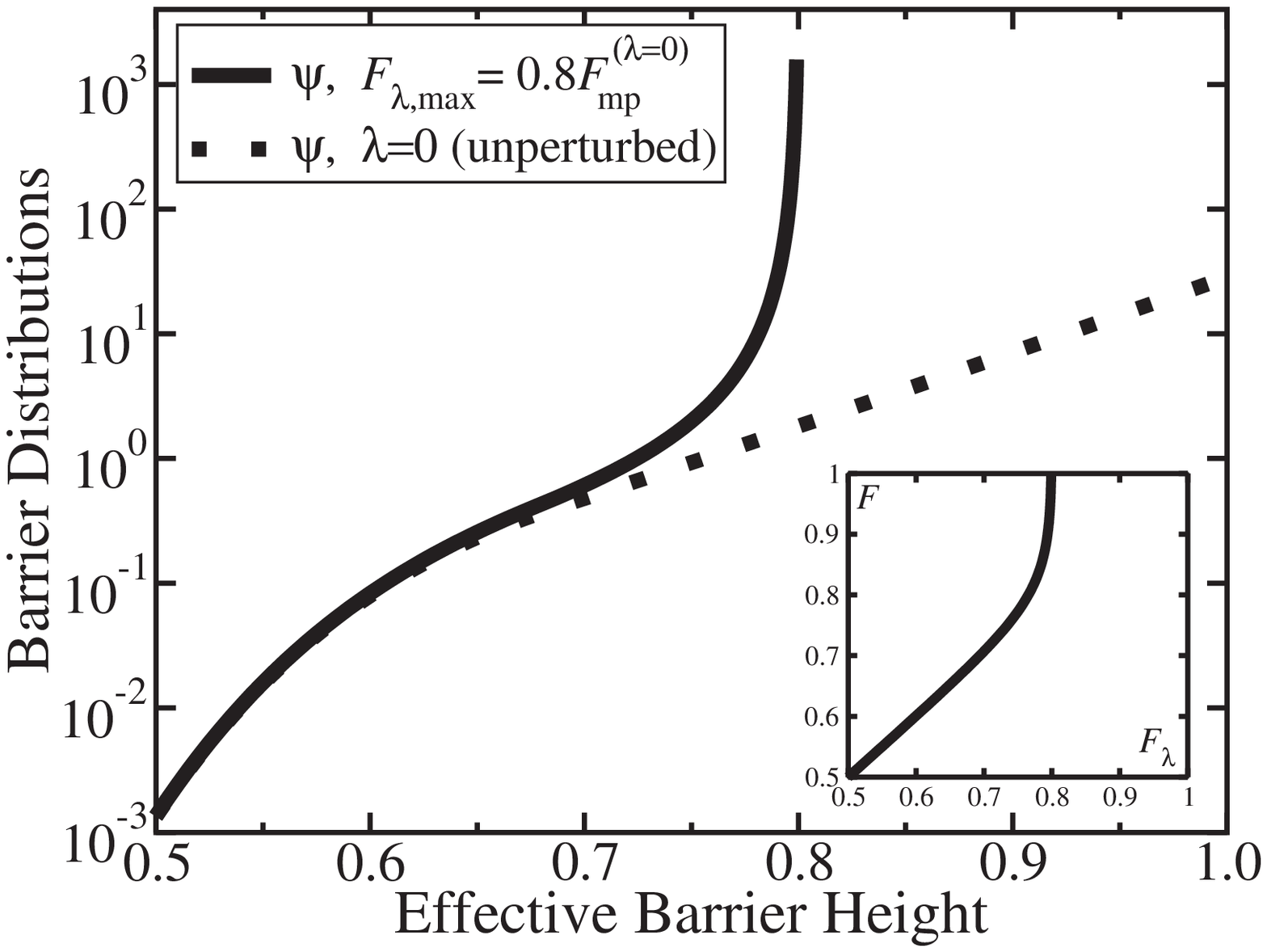}
    \\ 
    {\bf (b)}
  \end{center}
\end{minipage}
\end{tabular*}
\caption{\label{thinning} {\bf (a)} Normalised viscosity,
  $\eta/\eta(\de=0)$, vs. normalised strain rate $\de \la \tau
  \ra$. {\em Experiment}~\cite{WebbDingwellONSET}: Symbols, other than
  stars: data for Na$_2$Si$_4$O$_9$ at different temperatures; stars:
  Na$_2$Si$_2$O$_5$. Viscosities were determined by fibre
  elongation. {\em Theory}: Curves: theoretical predictions from
  Ref.~\cite{Lthinning}, the thin line corresponding to the
  approximate formula from Eq.~(\ref{lambdade}).  The bottom curve
  corresponds to the glass transition temperature, the top to the
  temperature $T_c$ of the crossover between activated and collisional
  transport. No adjustable parameters have been used. {\bf (b)} The
  solid line shows the effective distribution of barriers
  $\psi_\lambda$, in the presence of shear, the unperturbed
  distribution given by the dotted line. The unperturbed distribution
  is from Eq.(23) of Ref.\cite{Lionic}, with parameters corresponding
  to a fragile ($\beta \simeq 0.40$) substance near the glass
  transition on one-hour scale, so that the most probable barrier:
  $F_\smp = 37 k_B T$. The specific value of $\lambda$ was chosen to
  illustrate clearly how the motions at rates that would be slower
  than $\lambda$ in unperturbed fluid contribute to the high barrier
  peak in $\psi_\lambda$. These slow motions correspond to $F >
  F_{\lambda, \smax}$ in the inset. The energy units are chosen so
  that $F_\smp = 1$. From Ref.~\cite{Lthinning}}
\end{figure}

We have already seen that the microscopic picture advanced by the RFOT
theory can be used to estimate on a microscopic basis the viscosity,
which is the key quantity in the science of rheology. Next we shall
see how the very same picture allows one to answer an interesting
puzzle in the rheology of silicate melts. These melts exhibit
pronounced deviations from the Newtonian liquid response, in the form
of {\em shear thinning}, at shear rates that are, universally, about
three orders of magnitude {\em lower} than the inverse relaxation time
$\la \tau \ra^{-1}$. Shear thinning means the viscous response from
from Eq.~(\ref{viscDef}) is sublinear in the velocity gradient $\prtl
v_i/\prtl x_j$, implying an apparent viscosity that decreases with
$\prtl v_i/\prtl x_j$, see Fig.~\ref{thinning}(a). Viscosity
measurements can be imagined in a simple geometry in which two plates
move relative to each other and so only one component of the tensor
$\prtl v_i/\prtl x_j$ is non-zero, say $\prtl v_y/\prtl x$, which we
will denote as $\dot{\varepsilon}$. The latter quantity is of
dimensions inverse time, as is the mobility from Eq.~(\ref{mobTF}),
for instance. Thus, one would {\em \` a priori} expect that shear
thinning, if any at all, would set in at a rate $\dot{\varepsilon}$
comparable to the inverse of the typical relaxation time, or, perhaps,
typical relaxation rate.  (Because silicates are strong liquids, there
is not much difference between the two averages.) In contrast with
this expectation, the deviation from the Newtonian response in
Fig.~\ref{thinning}(a) takes place at shear rates that are much lower,
which is in the opposite direction from the decoupling of
mobility-like quantities we discussed in Subsection~\ref{decoupling}.

Lubchenko~\cite{Lthinning} has argued that the shear thinning is due
to a facilitation-like effect we have encountered earlier when
discussing the correlation between the stretching exponent $\beta$ and
the liquid's fragility $D$, in Subsection~\ref{betaDcorr}. There,
slower than the typical relaxations are facilitated because of nearby
regions that happen to be faster than typical. Here, there is
additional facilitation performed by the external shear itself. Denote
the rate of the additional relaxation of a region's environment with
$\lambda$. The survival probability for the region's relaxation is now
given by $p_{\lambda}(t) = \la e^{-t/\tau} e^{-\lambda t} \ra_{\tF, \,
  \tlambda}$. The resulting waiting time for the region's relaxation
is then
\begin{equation} \label{taul} \la \tau_\lambda \ra_{\tF, \, \tlambda}
  = \int_0^\infty dt \: p_\lambda(t) = \la \tau/(1 + \lambda \tau)
  \ra_{\tF, \, \tlambda} < \la \tau \ra_\tF,
\end{equation}
leading to a lowered viscosity:
\begin{equation} \label{eta2} \frac{\eta}{\eta(\de = 0)} = \frac{\la
    \tau_\lambda \ra_{\tF, \, \tlambda}}{\la \tau \ra_\tF} = \la
  \frac{ \tau}{1 + \lambda \tau} \ra_{\!\!\tF, \, \tlambda}
  \frac{1}{\la \tau \ra_\tF}.
\end{equation}

The rate $\lambda$ can be self-consistently
determined~\cite{Lthinning} at a given value of experimentally imposed
shear rate, by energy conservation: The energy flow through the domain
boundary, due to the external shear, must match the rate of energy
dissipation in the sample bulk, due to the shear. It turns out that
both at relatively high and low shear rate, $\lambda$ and $\de$ are
proportional to each other while the proportionality constant only
differs by a factor of two between the two extremes. Thus a simple
approximate expression for the relation between the quantities can be
written down, which does not depend on the detailed barrier
distribution:
\begin{equation} \label{lambdade} \lambda \approx [1.5 \sqrt{2} (a/d_L)
  (\xi/a)^{3/2}] \, \de.
\end{equation}
This equation indicates that the facilitation rate $\lambda$, as
sensed by an individual reconfiguring region, is significantly
enhanced compared to the experimentally imposed rate $\de$. The
enhancement is, in almost equal measure, is due to the smallness of
individual displacements $d_L$ from Eq.~(\ref{dLa10}) and to the size
of the cooperative region, which is about $(\xi/a)^3 \simeq 10^2$ near
$T_g$.  The numerical factor in the square brackets in
Eq.~(\ref{lambdade}) is thus about $10^3$ near $T_g$.  Conversely,
this enhancement implies that externally imposed shear at rate as low
as $10^{-3} \tau^{-1}$ will significantly increase the effective
relaxation rate of a region.

The theoretical prediction for the shear thinning, corresponding to
the simple form (\ref{lambdade}) and to a more accurate relation are
shown in Fig.~\ref{thinning}(a), alongside with experimental
data~\cite{WebbDingwellONSET}. The agreement between theory and
experiment is very good; note that no adjustable parameters were used.
The degree of shear thinning is predicted not to depend sensitively on
the width of the barrier distribution.

The argument in Ref.~\cite{Lthinning} self-consistently predicts the
magnitude of the shear rate at which the proposed picture breaks
down. In this picture, the external shear does not significantly
modify the mechanism of the relaxation itself but, instead, sets a
non-zero lower limit on the relaxation rate $\lambda$. At the same
time, this rate $\lambda$ may not exceed the intrinsic rate at which
the interface of an individual region relaxes, which can be estimated
geometrically. One can think of a region roughly as being in the
centre of a cube made of $3\times3\times3 = 27$ cooperative
regions. Thus the boundary of the region will relax, on average, once
per time $(\la \tau \ra/27)$ or so. If forced to relax at a higher
rate, the boundary must relax via other means, such as bond
breaking. This would occur at viscosities indicated by the horizontal
arrow in Fig.~\ref{thinning}(a). This estimate is quite consistent
with the shear rate at which the glassy fibres broke, whose elongation
was employed to measure the shear thinning.

Note that the shear-driven decrease in relaxation times,
Eq.~(\ref{taul}), implies lower effective barriers, by virtue of
Eq.(\ref{tauF1}):
\begin{equation} \label{Fl} F_\lambda \equiv F - k_B T \ln\left[1 +
    e^{(\tF-\tF_{\lambda, \tmax})/k_B T}\right],
\end{equation}
where $F_{\lambda, \smax}$ is the highest possible effective barrier
for a fixed $\lambda$:
\begin{equation}
  F_{\lambda, \smax} \equiv - k_B T \ln(\lambda \tau_0),
\end{equation}
see the inset of Fig.~\ref{thinning}(b). It is straightforward to show
that the distribution $\psi_\lambda$ of the shear-modified barriers
from Eq.~(\ref{Fl}) is related to the unperturbed barrier distribution
$\psi$ according to:
\begin{equation} \label{pFl} \psi_\lambda (F_\lambda) = \psi (F)
  \left[1 + e^{(\tF-\tF_{\lambda, \tmax})/k_B T}\right],
\end{equation}
where $F$ is understood as a function of $F_\lambda$ via
Eq.(\ref{Fl}). Both the unperturbed and shear-modified barrier
distributions (for a fragile substance near the glass transition) are
shown in Fig.\ref{thinning} for a particular value of $\lambda$.  We
thus directly observe how facilitation modifies the barrier
distribution. Presumably, one may use a similar approach to determine
the barrier distribution self-consistently and thus go beyond the
simple approximations made to derive the barrier distributions in
Eq.~(\ref{pF}) and (\ref{pFm}). To do so, one needs to better quantify
the coupling between nearby reconfiguring regions. Some progress along
these lines has been achieved recently in the context of glass
rejuvenation~\cite{2009PNAS..106.1353P, PhysRevE.88.022308}, as we saw
in Subsection~\ref{rejuvenation}.

\subsection{Mechanical Strength}

In the preceding Subsection, we used a purely kinetic argument to
estimate the threshold value of the shear rate beyond which a glassy
liquid will rupture. To develop a thermodynamic perspective on the
mechanical strength of a {\em frozen} glass, we will employ the
machinery developed earlier as part of the theory of ageing and of the
crossover phenomena, from Sections~\ref{aging} and \ref{crossover}
respectively.  The relevance of the crossover becomes clear after one
notes that on the verge of breakdown, the sample is near its
mechanical stability limit implying that the reconfiguration barriers
are low. This, in turn, means that one must include effects stemming
from the reconfiguring regions being not fully compact.  Wisitsorasak
and Wolynes~\cite{Wisitsorasak02102012} essentially repeat the steps
that led to the derivation of the free energy cost of reconfiguration
near the crossover, Eq.~(\ref{FSSperc}), except now the driving force
must also include two additional contributions:
\begin{equation} \label{FNstrength} F(N) = T \left[s_c^\text{perc} -
    \left( s_c + \frac{\Delta \Phi}{T} + \kappa \frac{\sigma^2 a^3}{2
        \mu} \right) \right] N.
\end{equation} 
Indeed, as we have seen in our analysis of ageing in
Subsection~\ref{aging}, the bulk driving force $\Delta g$ for
reconfiguration of a glass quenched to a temperature $T < T_g$ exceeds
the equilibrium configurational entropy at that temperature. By
Eq.~(\ref{df1}), together with $T_1 = T_g$ and $T_2 = T$, we get
$\Delta \Phi = \Delta c_p (T_g) T_g \ln(T_g/T)$. An additional
contribution to the driving force is the free energy of the elastic
stress that ends being released following the reconfiguration. In the
simplest, spherical geometry, this elastic energy is equal to $\kappa
\sigma^2/2 \mu$ per unit volume, where $\sigma$ is the magnitude of
stress, $\kappa \equiv 3 - 6/(7 - 5\nu)$, $\nu$ the Poisson ratio, and
$\mu$ the shear modulus.  This yields the final term on the r.h.s. of
Eq.~(\ref{FNstrength}).

As in the earlier analysis of the crossover, the free energy cost
$F(N)$ for reconfiguration, near the mechanical stability limit, is
either exclusively uphill or downhill, depending on the sign of the
expression in the square brackets in Eq.~(\ref{FNstrength}). The glass
becomes unstable when the expression vanishes. This notion, together
with some additional corrections, yields the following expression for
the threshold value of the stress beyond which the glass will break
catastrophically~\cite{Wisitsorasak02102012}:
\begin{equation} \label{limitingStrength} \sigma^*_{\text{pred}} =
  \sqrt{ \frac{2 \mu k_B T}{\kappa a^3} \left( \left[3.20
        \frac{T_K}{T} - 1.91 \right] - \frac{\Delta c_p(T_g)}{k_B}
      \frac{T_g}{T}\ln \frac{T_g}{T_K} \right)}
\end{equation}
As expected, the glass would be at its strongest at the (putative)
Kauzmann temperature $T_K$. Indeed, if it were possible to equilibrate
the liquid at $T_K$, this would correspond to the deepest valley in the
free energy landscape of the system.

\begin{figure}[t]
  \begin{tabular*}{\figurewidth} {cc}
    \begin{minipage}{.48 \figurewidth} 
      \flushleft
      \begin{center} 
        \includegraphics[width=0.46 \figurewidth]{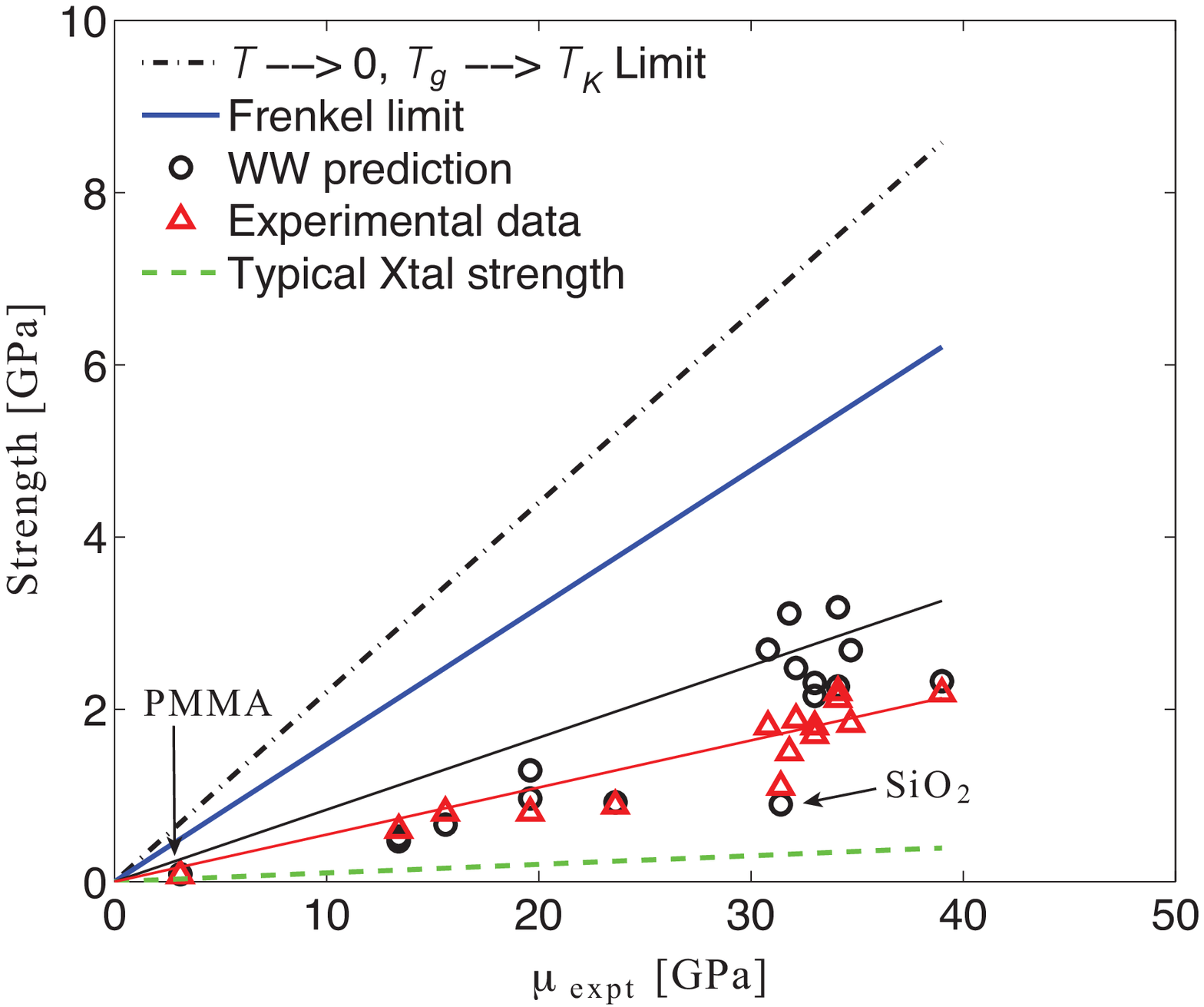}
        \\ {\bf (a)}
      \end{center}
    \end{minipage}
    &
    \begin{minipage}{.48 \figurewidth} 
      \begin{center}
        \includegraphics[width= .46 \figurewidth]{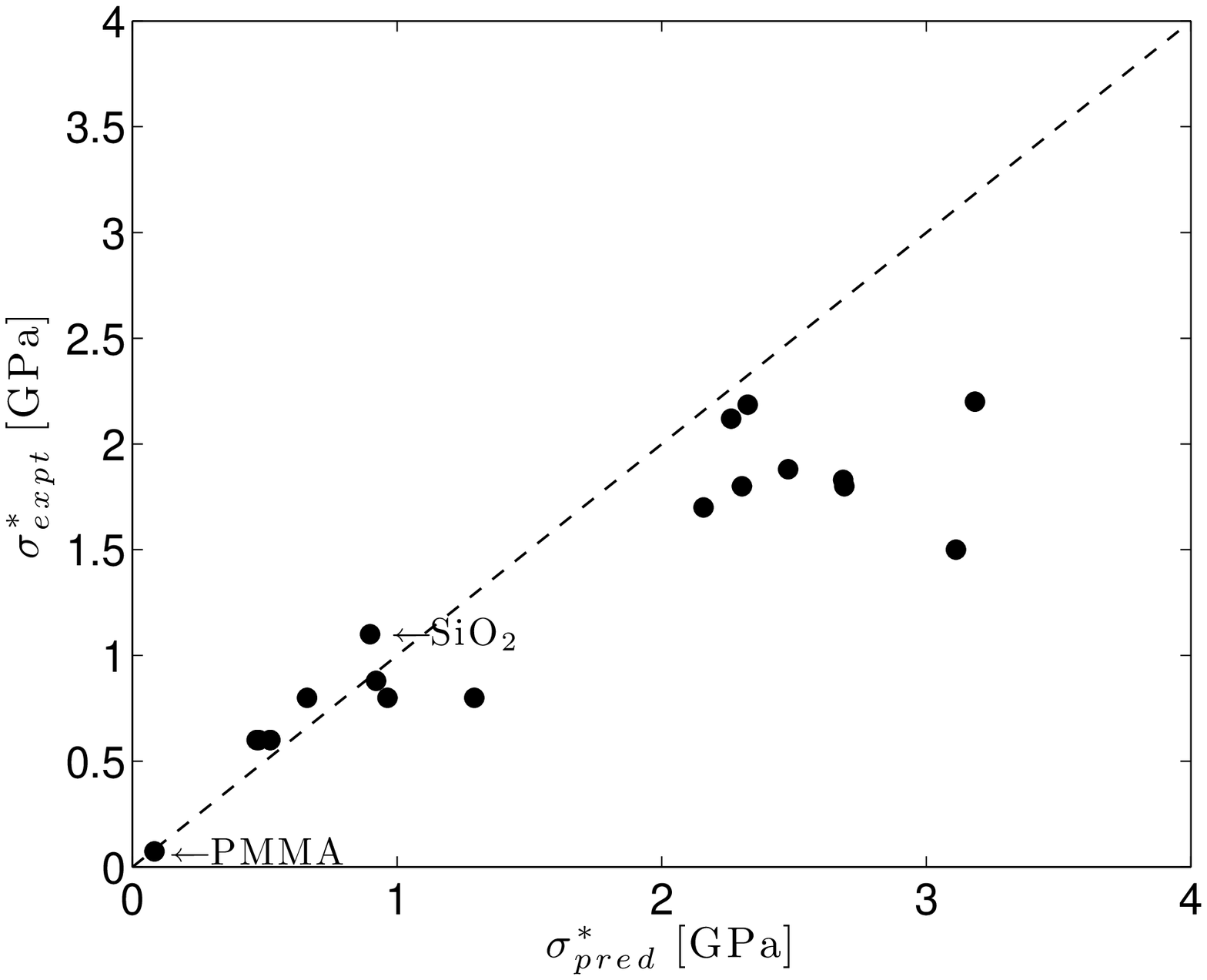} \\
        {\bf (b)}
      \end{center}
    \end{minipage}
  \end{tabular*}
  \caption{\label{strengthFig} {\bf (a)} Predicted values of the
    limiting strength for several glasses, due to Wisitsorasak and
    Wolynes~\cite{Wisitsorasak02102012} (black circles), plotted
    vs. the shear modulus for select materials, alongside their
    experimental values (red triangles). The black and red lines are
    best linear fits through the corresponding sets. {\bf (b)} Direct
    comparison of the theoretically-predicted and experimental values
    of the limiting strength, figure taken from
    Ref.~\cite{Wisitsorasak02102012}.}
\end{figure}

Predictions of the limiting strength, due to
Eq.~(\ref{limitingStrength}), are graphed in Fig.~\ref{strengthFig}(a)
against the shear modulus $\mu$, alongside the experimental values and
a number of other notable types of limiting strength, see the
legend. A direct comparison of the theoretically-predicted and
experimental values of the limiting strength is performed graphically
in Fig.\ref{strengthFig}(b). The agreement between theory and
experiment is quite satisfactory considering that no adjustable
parameters were used.

\section{Ultra-Stable Glasses}
\label{ultrastable}

Because of the dramatic increase in the relaxation time with lowering
temperature, there seems to be a natural limit to the depth one can
reach in the free energy landscape of a liquid. The depth simply goes
with the logarithm of the relaxation time times $k_B T$; the
relaxation time is limited by the duration of the experiment.  It is
quite rare that a controlled experiment lasts longer than a few hours
or days, a notable exception provided by the famous tar pitch
experiment in Australia~\cite{tarpitch}. Much, much longer {\em
  uncontrolled} experiments on glasses are also known, such as studies
of fossil amber~\cite{McKennaAmber}.

Given the intrinsic connection between the depth of the free energy
minima and their multiplicity, see Eq.~(\ref{F2}), there is a natural
lower bound on the enthalpy of a glass.  According to
Eq.~(\ref{scTRA}), the enthalpy of an equilibrated liquid at
temperature $T$ is:
\begin{equation} \label{hTglass} h(T) = \Delta c_p (T_g^\circ) \,
  T_g^\circ \, \ln \left( T/T_g^\circ \right) \ge \Delta c_p
  (T_g^\circ) \, T_g^\circ \, \ln \left( T_K/T_g^\circ \right),
\end{equation}
where we take as our reference temperature the conventional glass
transition temperature $T_g^\circ$, say, on the time scale of
$100$~sec.  The lower limit above is almost certainly an overestimate
because most liquids will undergo some sort of partial ordering before
the putative Kauzmann temperature could be reached, see
Section~\ref{fate} and also below.

\begin{figure}[t]
  \begin{tabular*}{\figurewidth} {cc}
    \begin{minipage}{.38 \figurewidth} 
      \flushleft
      \begin{center} 
        \includegraphics[width=0.34 \figurewidth]{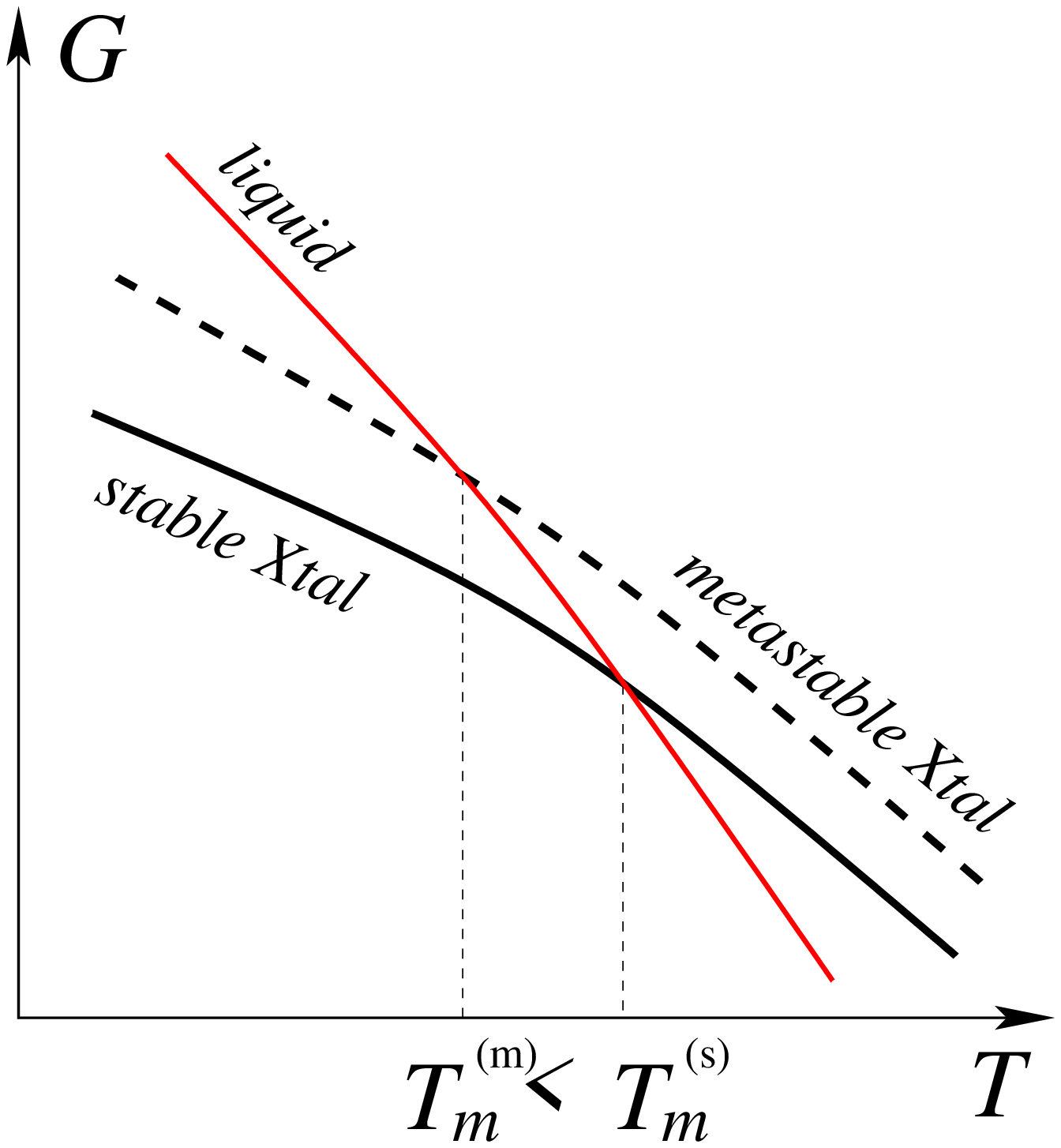} \vspace {6mm}
        \\  {\bf (a)}
      \end{center}
    \end{minipage}
    &
    \begin{minipage}{.58 \figurewidth} 
      \begin{center}
        \includegraphics[width= .58 \figurewidth]{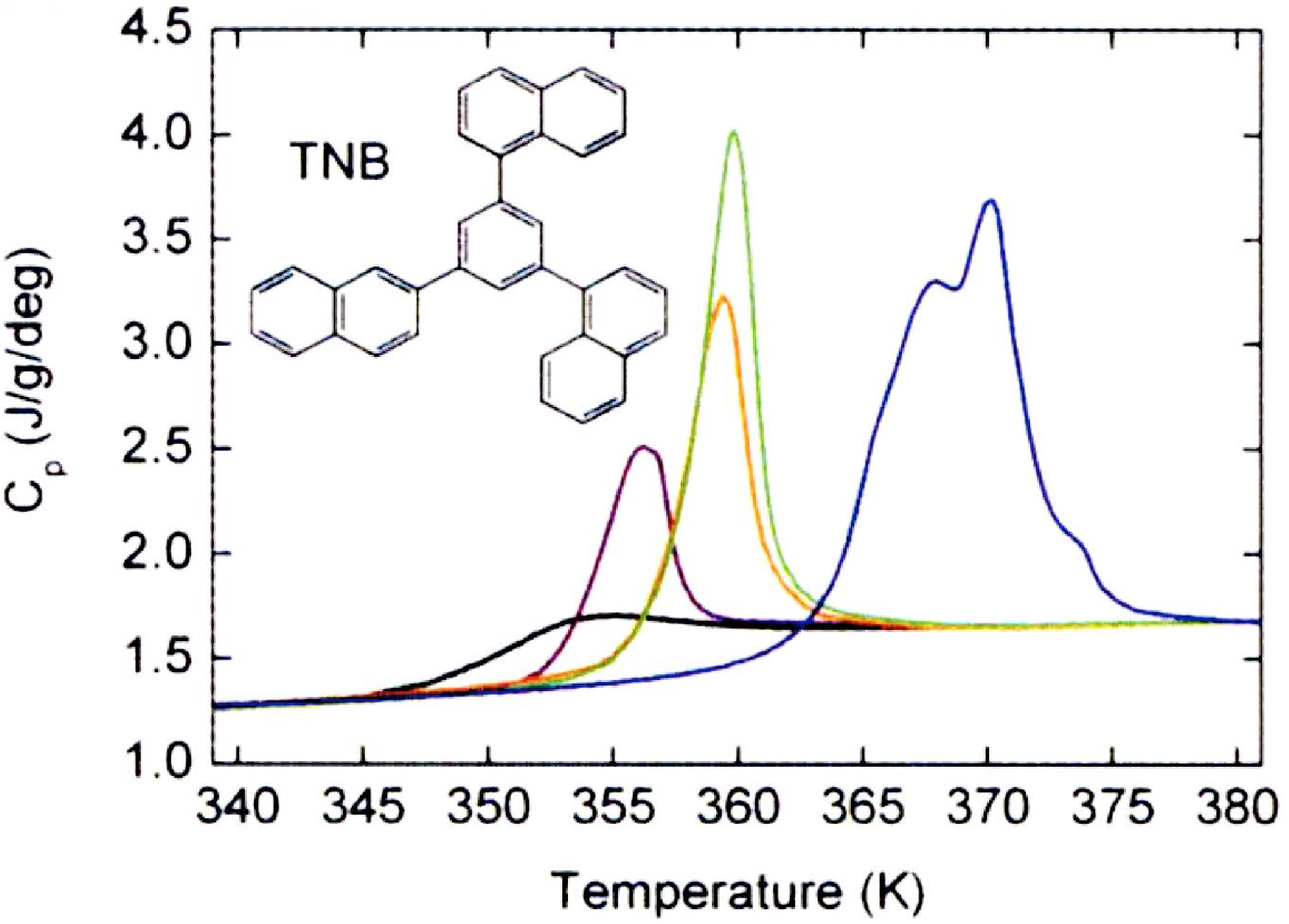} \\
        {\bf (b)}
      \end{center}
    \end{minipage}
  \end{tabular*}
  \caption{\label{metaStableLiq} {\bf (a)} A schematic of temperature
    dependences of the Gibbs free energy of a stable polymorph (black
    solid line), metastable polymorph of the same substance (black
    dashed line) and the corresponding liquid (red line). The less
    stable polymorph will melt at a lower temperature, $T_m^{(m)}$,
    than the stable one, $T_m^{(s)}$. (Melting is kinetically
    preferable to the nucleation of the stable crystal, at least at $T
    > T_m^{(m)}$, because the barrier for surface melting is only a
    $k_B T$ or so~\cite{L_Lindemann}. {\bf (b)} Heat capacity, $C_p$,
    of TNB samples: vapour deposited directly into a DSC pan at 296 K
    at a rate of ~5 nm/s (blue); ordinary glass produced by cooling
    the liquid at 40 K/min (black); ordinary glass annealed at 296 K
    for 174 days (violet), 328 K for 9 days (gold), and 328 K for 15
    days (green). (Inset) Structure of TNB. Figure from Swallen et
    al.~\cite{EdigerScience2007}.}
\end{figure}

The lower the temperature at which the sample has been equilibrated,
the deeper in the free energy landscape the liquid is. Consistent with
this notion, the escape barrier from a state in a liquid equilibrated
at temperature $T_1$ to a state equilibrated at temperature $T_2 >
T_1$, following a $T_1 \to T_2$ temperature jump, is the higher the
lower the temperature $T_1$ is, see Eq.~(\ref{drForceRejuv}).  In
turn, this means that given the same speed of heating, a sample
equilibrated at a lower temperature will melt at a higher temperature.
The latter situation is in a loose way similar to the
crystal-to-liquid transition, whereby a crystal made of molecules
bonded by stronger forces will melt a higher temperature. Suppose two
distinct polymorphs can be prepared at the same temperature, one of
the polymorphs must then be metastable. The metastable polymorph will
melt at a lower temperature that the stable one, because its free
energy is higher than that of the stable polymorph, see
Fig.~\ref{metaStableLiq}(a). For the same reason, different faces melt
at different temperatures because the bonding is generally differs in
strength depending on the specific face~\cite{Valenta, Lionic}. Note
this is consistent with the Lindemann criterion of
melting~\cite{Valenta, Lionic} since the vibrational displacement will
be longest along the direction of weakest bonding; this longest
displacement will satisfy the Lindemann criterion first as the crystal
is heated.  The corresponding face will thus be the first to melt.

In a fascinating contrast with crystals, the typical bonding strength
in glasses can be tuned {\em continuously}, simply by varying the
speed of quenching. The slow the speed, the lower the glass
temperature, the lower the enthalpy, the stronger the bonding, by
Eq.~(\ref{hTglass}). Another important distinction is that there is a
substantial barrier for melting a glass---as determined by the driving
force from Eq.~(\ref{drForceRejuv})---and so the melting of glass is
subject to kinetics as is the vitrification in the first place. In
contrast, the barrier for the melting of a {\em periodic} crystal is
very low, a $k_B T$ or so, and the melting temperature of the crystal
is largely determined by thermodynamics, not kinetics. To avoid
confusion, we reiterate that glasses melt by gradual softening, not by
sudden liquefaction, in contrast with periodic crystals.

Relatively recently, Ediger and coworkers~\cite{EdigerScience2007,
  doi:10.1021/jz3003266, doi:10.1021/jp405005n,
  PhysRevLett.113.045901} have generated glassy films by vapour
deposition at a temperature significantly below the glass
transition. These films melt at a significantly higher temperature
than conventionally-produced bulk glasses made by thermally quenching
a liquid at a generic rate, see Fig.~\ref{metaStableLiq}(b). In this
figure, the differential scanning calorimetry (DSC) curves other than
the blue curve correspond to conventional glasses, some of which have
also been subjected to additional thermal treatment. We clearly see
that the lower the enthalpy of the sample is---as could be determined
by integrating the heat capacity curves---the higher the temperature
at which the sample will melt. The blue curve, which describes the
vapour-deposited glass, clearly corresponds to a very stable glass. (It
appears that the most stable samples are obtained when the substrate
temperature is around 85\% of the conventional $T_g$.) Note that the
DSC peak corresponding to this stable glass has a structure; this
suggests the films are held together by distinct interactions that are
relatively well separated in terms of energy.  Consistent with these
notions, later measurements~\cite{doi:10.1021/jz3003266,
  doi:10.1021/jp405005n} confirmed that the packing in the stable
samples is indeed quite anisotropic, whereby the planar portions of
the constituent molecules seem to stack. Furthermore the direction of
stacking could be either perpendicular or orthogonal to the film.

\begin{figure}[t]
  \begin{tabular*}{\figurewidth} {cc}
    \begin{minipage}{.48 \figurewidth} 
      \flushleft
      \begin{center} 
        \includegraphics[width=0.46 \figurewidth]{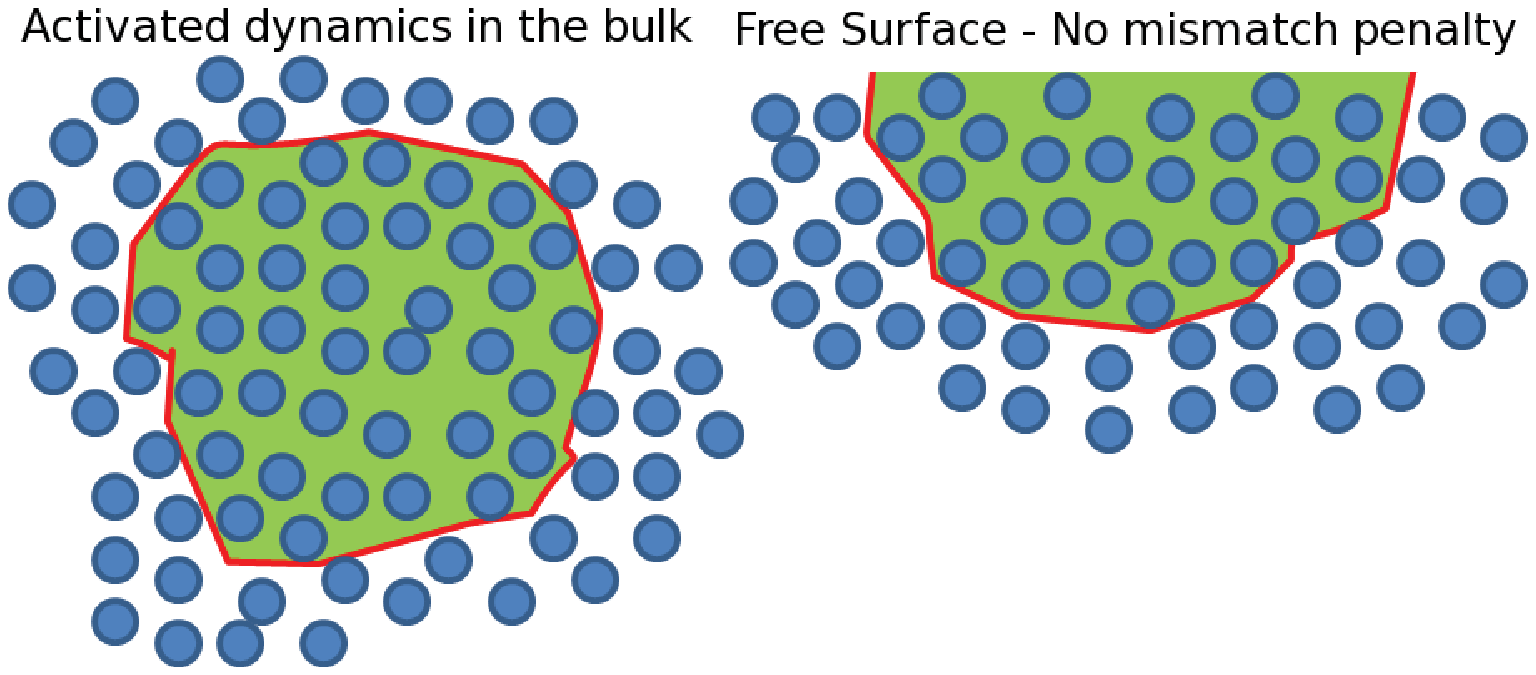}
        \\ {\bf (a)} \\ \vspace{13mm}
        \includegraphics[width=0.42
        \figurewidth]{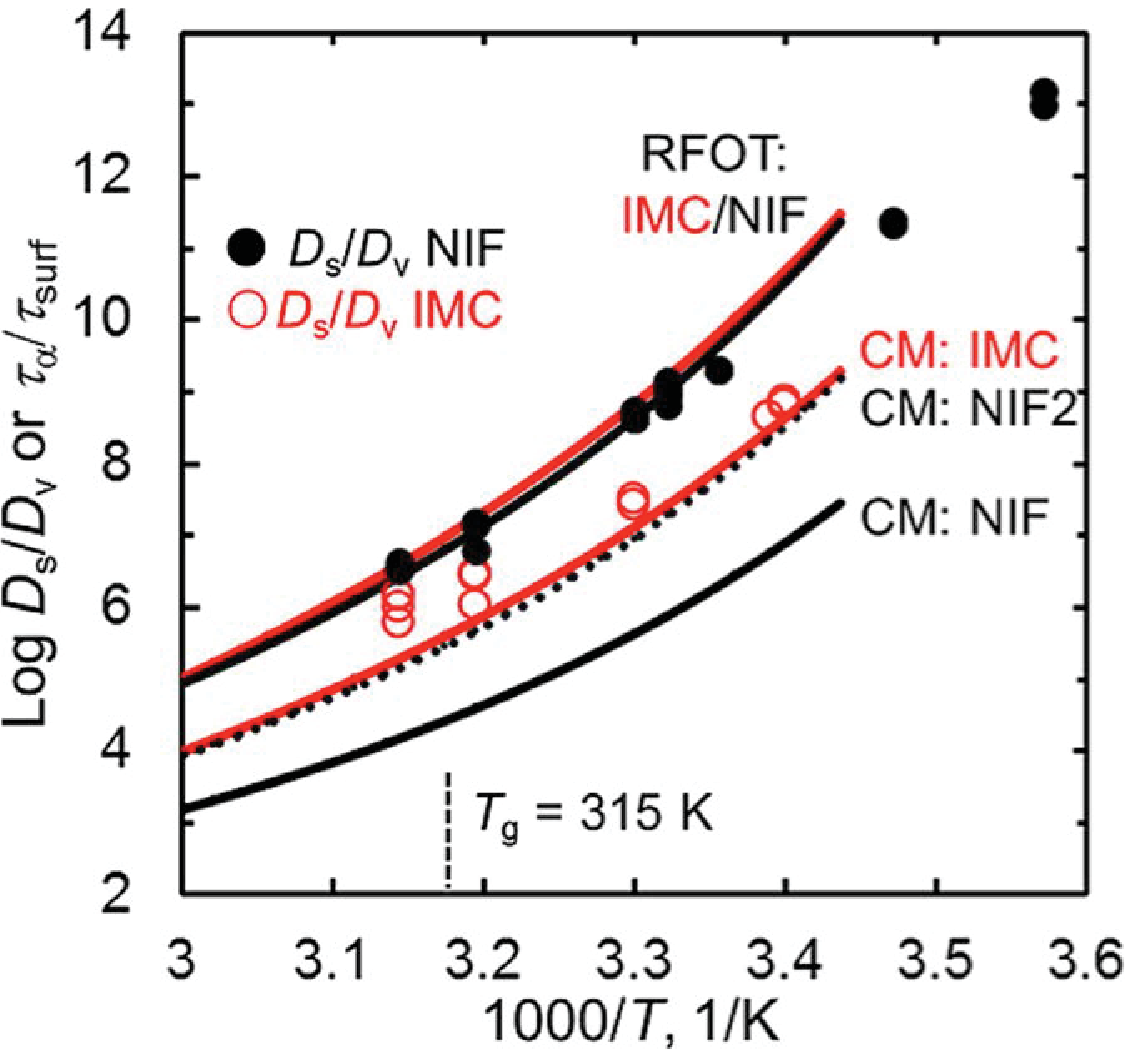}
        \\ {\bf (b)}

      \end{center}
    \end{minipage}
    &
    \begin{minipage}{.48 \figurewidth} 
      \begin{center}
        \includegraphics[width= .46 \figurewidth]{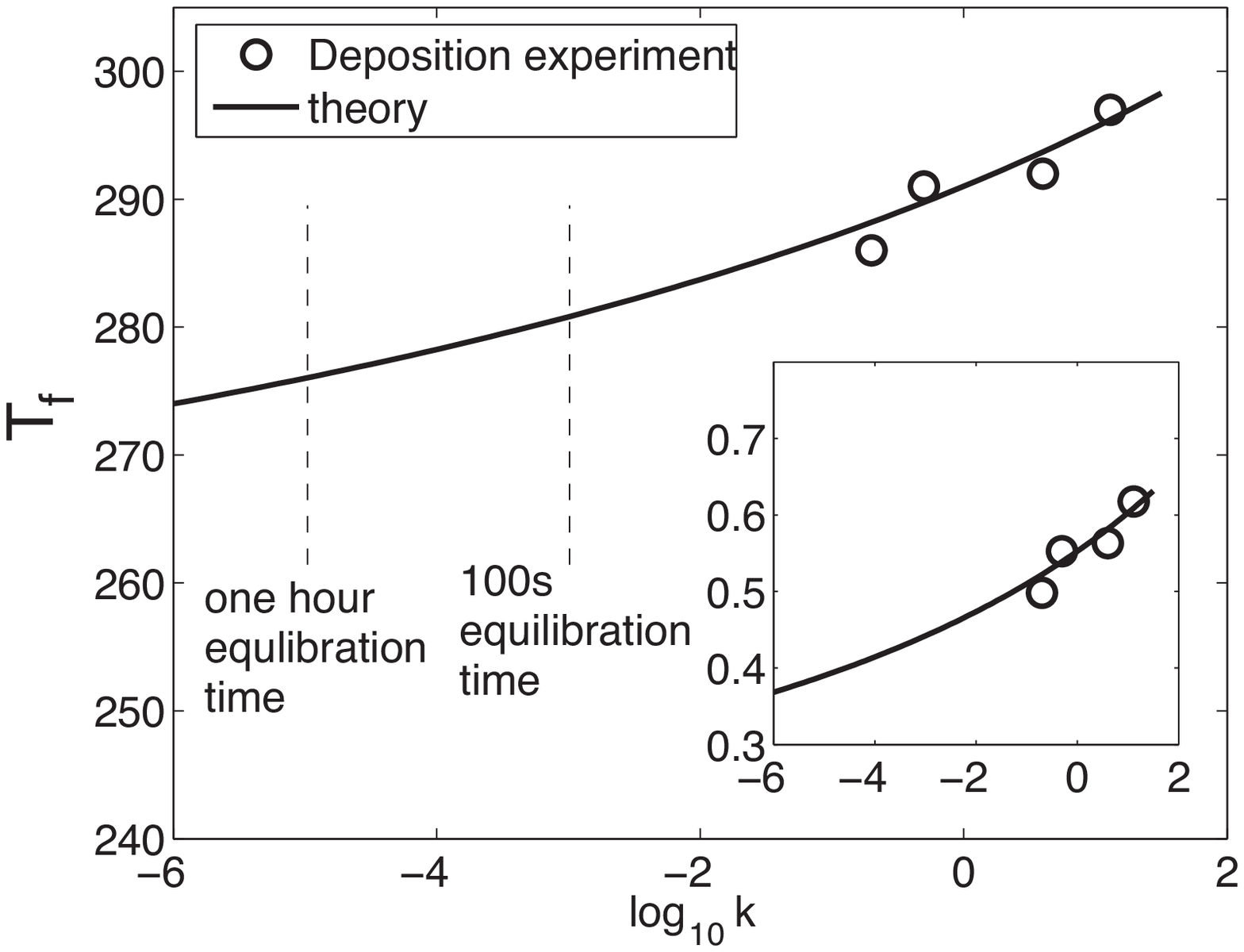} \\
        {\bf (c)}
      \end{center}
      \caption{\label{surfaceG} {\bf (a)} A schematic of a region
        reconfiguring near the surface of the sample. At the same
        curvature, both the volume and mismatch penalty are reduced
        resulting in a lowered reconfiguration barrier. {\bf (b)}
        Experimental data~\cite{doi:10.1021/jp404944s} for surface
        mobility for two glasses plotted against theoretical
        predictions, including those by the RFOT-based
        argument~\cite{stevenson:234514}. Figure from
        Ref.~\cite{doi:10.1021/jp404944s}. \\ {\bf (c)} Fictive
        temperatures vs deposition rate for the glass former IMC.
        Data for the deposition
        experiment~\cite{doi:10.1021/jp7113384} are shown with
        symbols. The experimental fictive temperature was determined
        by intersecting the apparent $T$-dependences of the enthalpy,
        as determined by integrating $\Delta c_p$ and the extrapolated
        enthalpy of the equilibrium liquid.  The same data are shown
        in the inset plotted as $s_c(T_f)/k_B$ per bead vs. deposition
        rate.}
    \end{minipage}
  \end{tabular*}
  
\end{figure}

Stevenson and Wolynes~\cite{stevenson:234514} have put forth an
argument as to why the {\em surface} of a glass---as it is being
deposited---could become more stable than the bulk. The main idea is
that at a fixed interface curvature, a reconfiguring region on the
surface has a smaller contact area with its environment. Indeed,
consider a roughly half-spherical region at the surface as in
Fig.~\ref{surfaceG}(a). A review of the argument leading to
Eq.~(\ref{sigmaXW}) indicates that there is no localisation penalty
for molecules at the surface. In turn, owing to the absence of the
mismatch penalty at the free surface, the overall nucleation profile
from Eq.~(\ref{Fr}) is modified to account both for the reduction in
the contact area of the region with its environment and the reduction
in the volume. For a strictly half-spherical region, the nucleation
profile is exactly a half of that for a full sphere, at the same value
of the curvature~\cite{stevenson:234514}:
\begin{equation} \label{Fr2} F(r) = 2 \pi r^2 \sigma_0 (a/r)^{1/2} -
  (2 \pi/3) (r/a)^3 T s_c.
\end{equation}
Thus the reconfiguration of a typically-sized region near the surface
is subject to a barrier that is about a half of its value in the bulk:
\begin{equation} \label{F2F} F^\ddagger_\text{surf} = \frac{1}{2}
  F^\ddagger_\text{bulk}, \text{ when  } r^\ddagger_\text{surf} =
  r^\ddagger_\text{bulk}.
\end{equation}
Consequently, the relaxation time at the surface is about the square
root of its value in the bulk~\cite{stevenson:234514}:
\begin{equation} \label{tauSurf} \tau_\text{surf} \approx \sqrt{\tau_0
    \tau_\text{bulk}},
\end{equation}
where the bulk relaxation time $\tau_\text{bulk}$ corresponds to the
relaxation time $\tau$ from Eq.~(\ref{tauF1}) and $\tau_0$ is the
prefactor from the same equation. Brian and
Yu~\cite{doi:10.1021/jp404944s} have tested the simple relation from
Eq.~(\ref{tauSurf}) and found it agrees well with their data on
surface mobility, see Fig.~\ref{surfaceG}(b).

The argument above can be also turned around: Suppose one prepares a
glass by surface deposition and controls the deposition rate so as to
give the molecules as much time to rearrange---before they get covered
by the next layer---as they would have in a {\em bulk} glass at the
same temperature. By equations (\ref{F2F}) and (\ref{F2}), the
configurational entropy of the so deposited glass is about twice lower
than in the bulk glass. In other words, the surface-deposited glass
corresponds to a bulk glass equilibrated at a significantly lower
temperature; one may thus say the surface glass has a significantly
lower fictive temperature than what would be available to a bulk glass
made by quenching.  This observation immediately explains the
remarkable stability of the surface glass in Fig.~\ref{metaStableLiq}.
Quantitatively, this notion can be expressed by relating the
deposition rate $k$ (of units length per unit time), the cooperative
size $\xi$, and the relaxation time $\tau$: $k = \xi/\tau$. Combined
with Eq.~(\ref{XWbarrier}), this quantity can be directly related to
the configurational entropy and the corresponding value of the fictive
temperature~\cite{stevenson:234514}:
\begin{equation}
  s_c(T_f) = \frac{32}{\ln(\xi / k \tau_0)}.
\end{equation}
The thus predicted values of the fictive temperature vs. the
deposition rate are plotted in Fig.~\ref{surfaceG}(c), alongside the
experimental values due to Kearns et al.~\cite{doi:10.1021/jp7113384}.
The agreement between theory and experiment is notable, especially in
view of the partial ordering that takes place in the ultrastable
glasses.  Indeed, it is not obvious that one would obtain quantitative
results by extrapolating configurational entropy from a regime in
which no apparent ordering takes place to a regime characterised by
some ordering.

\section{Ultimate Fate of Supercooled Liquids}
\label{fate}

Deeply supercooled liquids seem to be a great example of a
self-fulfilled prophecy: Once they fail to crystallise below the
fusion temperature, one can supercool them even more thus further
diminishing the probability to crystallise. This is expected since the
growth of the crystal nucleus is subject to viscous drag, which
ordinarily becomes increasingly strong with lowering the
temperature. Glycerol is an archetypal example of a system, in which
crystallisation is suppressed for a seemingly indefinite
period. Indeed, at normal conditions, glycerol is a supercooled liquid
with a seemingly indefinite shelf-life, unless it becomes ``infected''
by crystallites. (According to an anecdote told by Onsager, a whole
glycerol factory had to be shut down as a result of being infected by
such crystallites~\cite{PenroseGlycerineAnecdote}.)

\begin{figure}[t]
  \begin{tabular*}{\figurewidth} {cc}
    \begin{minipage}{.48 \figurewidth} 
      \flushleft
      \begin{center} 
        \includegraphics[width= .46 \figurewidth]{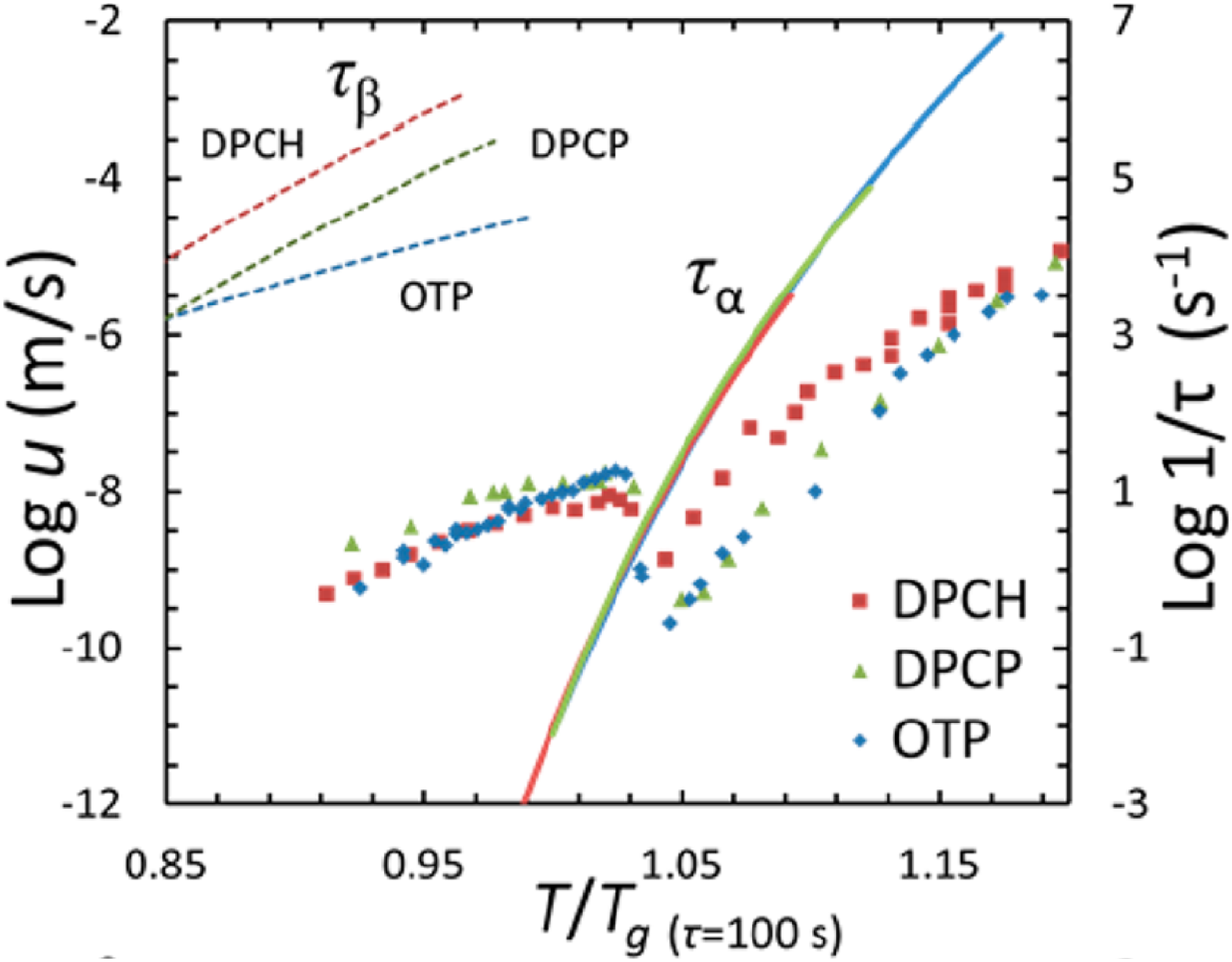} 
        \\ \vspace{10mm} {\bf (a)}
        \caption{\label{GCFig} {\bf (a)} Crystal growth rates
          (symbols) and dielectric relaxation times (curves) for three
          organic glassformers~\cite{doi:10.1021/jp501301y}. {\bf (b)}
          Excess scattering in an organic glass-former, BMMPC, at low
          values of the scattering vector, due to
          microcrystallites~\cite{Fischer1993183}. {\bf (c)} The
          annealing-time dependence of lengthscale of the density
          fluctuations leading to the extra scattering in panel (b),
          which confirms the crystalline nature of the scatterer. From
          Ref.~\cite{Fischer1991134}.}
      \end{center}
    \end{minipage}
    &
    \begin{minipage}{.48 \figurewidth} 
      \begin{center}
        \includegraphics[width= .44 \figurewidth]{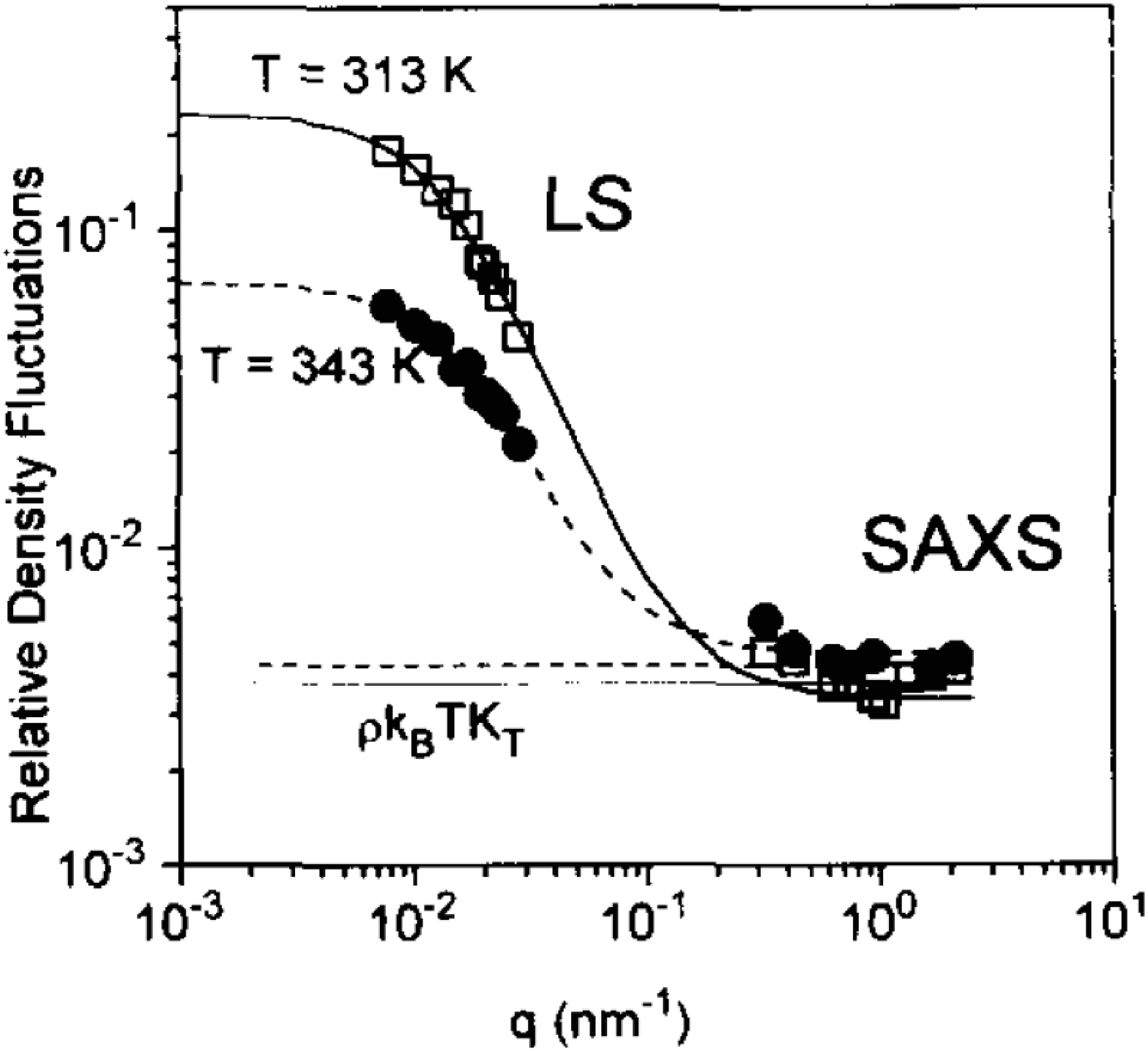} \\
        {\bf (b)} \\
        \includegraphics[width= .48 \figurewidth]{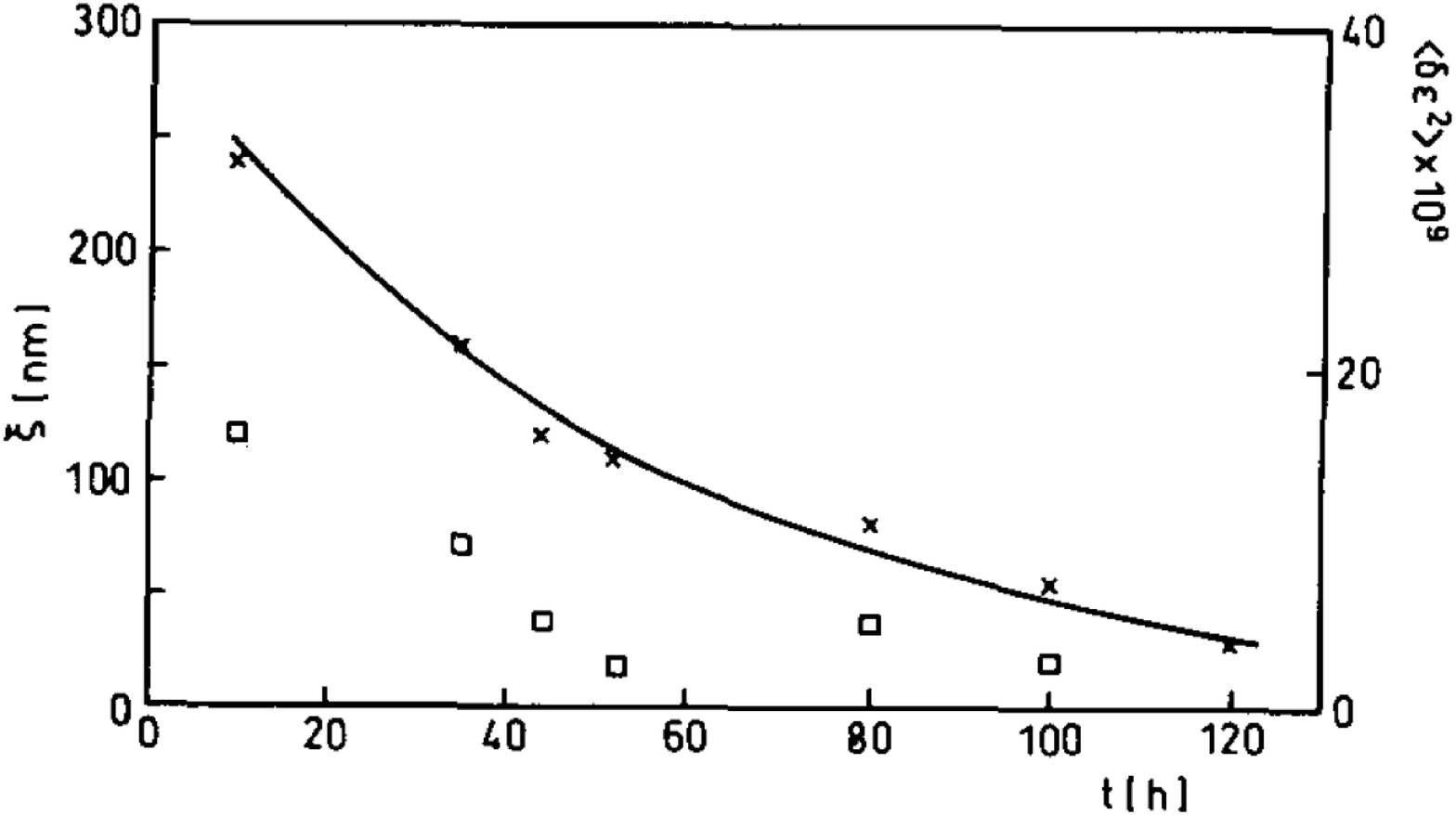} \\
        {\bf (c)}
      \end{center}
    \end{minipage}
  \end{tabular*}
\end{figure}

Yet it is not at all obvious that supercooled liquid should be harder
to crystallise with lowering the temperature, because crystal
nucleation and growth are subject to two competing factors. On the one
hand, the viscous drag makes it difficult for an incipient crystallite
to grow thus increasing the probability of its evaporation before it
ever reaches the critical size, as already mentioned. On the other
hand, the ordinary nucleation theory tells us that the activation
profile for nucleation is:
\begin{equation} \label{FXtal} F_\tX(r) = 4 \pi \sigma_\tX \: r^2 + (4
  \pi/3) \Delta g_\tX \: r^3,
\end{equation}
where $\Delta g_\tX < 0$ is the free energy difference between {\em
  equilibrated} crystal and liquid. Clearly, the equilibrium driving
force, $(-\Delta g_\tX)$, increases while the critical nucleus size
decreases upon cooling, see Fig.~\ref{metaStableLiq}(a). In fact, if
the liquid were uniform, it should presumably reach a mechanical
stability limit at some temperature $T < T_m$, implying crystal
nucleation is now a strictly downhill process, no matter how high the
viscosity is!  In line with the latter notions, the rate of crystal
growth in certain organic liquids displays a re-entrant behaviour as a
function of temperature, see Fig.~\ref{GCFig}(a).

One generally expects that the amount of crystallinity, if any, will
depend on the history of the sample. And so, for instance, the
crystallites' size and their growth in a glassy sample can be
controlled to some degree by varying the sample's temperature and,
hence, viscosity. Likewise, one can partially anneal out crystallites
by warming the liquid to a sufficient extent. To give a household
example of such a process, honey will begin to crystallise after
prolonged storage but can be made visibly crystal-free by
warming. Both the formation of crystallites and their melting in
organic glassformers is systematically illustrated with results of
Fischer et al. in Figs.~\ref{GCFig}(b) and (c).

\begin{figure}[t]
  \begin{center}
    \includegraphics[width=0.46 \figurewidth]{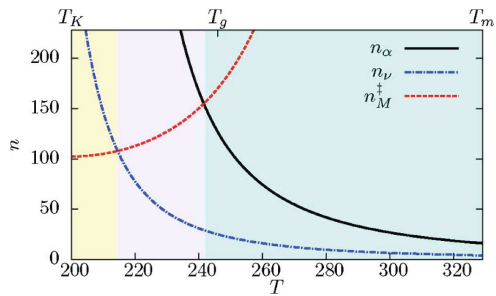}
  \end{center}
  \caption{\label{fateSize} Temperature dependence of the size
    $n_\alpha$ of a typical structural reconfiguration; the transition
    state size $n^\ddagger_M$ of classical crystal nucleation; and the
    number $n_\nu$ of particles involved in nanocrystallisation. The
    latter size indicates the size of a nucleus stable against
    evaporation in an {\em individual} free energy minimum of the
    liquid.  Within the shaded region on the right classical
    nucleation theory is valid.  In the shaded region on the left
    direct nanocrystallisation can take place.  Crystallisation in the
    centre region takes place through fluctuational, percolative
    nanocrystallisation. From Ref.~\cite{SWultimateFate}.}
\end{figure}

Stevenson and Wolynes~\cite{SWultimateFate} have pointed out that the
analysis of crystallisation in the landscape regime is complicated by
the fact that the liquid may not be spatially uniform or equilibrated
on the crystal-nucleation time. Qualitatively, if the critical nucleus
for crystallisation significantly exceeds the cooperativity size
$\xi$, crystal-nucleation occurs roughly as it would from the
uniform-liquid state. In the opposite case, where the crystal nucleus
is smaller than $\xi$, the crystal-nucleation would significantly
speed up, compared with the uniform liquid. This is because an
individual region is not configurationally equilibrated and is
significantly higher in free energy (by $T s_c$ per particle) than the
uniform liquid, see Eq.~(\ref{gequil}). Thus the driving force for
crystallisation becomes greater than its value for the
uniform-liquid-to-crystal transition. These notions are graphically
summarised in Fig.~\ref{fateSize}(a).  Thus SW have delineated
distinct regimes in which activated reconfiguration between aperiodic
structure kinetically competes with homogeneous crystallisation at
high temperatures and nano-crystallisation at lower temperatures, see
Ref.~\cite{SWultimateFate} for more detail. Regardless of
system-specific peculiarities, it is predicted that
nano-crystallisation will directly proceed at sufficiently low
temperatures. This, in effect, resolves the Kauzmann paradox in that
the putative ideal glass state would be always avoided owing to
partial ordering.

\section{Quantum Anomalies}
\label{Quantum}

So far, quantum mechanics has played no explicit role in our narrative
even though the inter-particle interactions, including in particular a
significant portion of the steric repulsion, are ultimately of
quantum-chemical origin. It is, in fact, fair to say that the
structural glass transition is an intrinsically classical
phenomenon. Indeed the very long lifetime of glassy, metastable states
is in contradiction with the Second Law and is predicated on the
system's ignorance of the actual stable state. Such ignorance is
guaranteed in classical mechanics since portions of the phase space
separated by a barrier are entirely ``unaware'' of each other. Barrier
crossing events are rare events that require activation.  In contrast,
the wavefunctions of the metastable and stable states have a non-zero
overlap, even if small. When the overlap is sufficiently small,
however, the ``self-awareness'' of glassy states as metastable
configurations should not affect the dynamics within those
states. Furthermore, quantum fluctuations in fact {\em augment} the
glass transition temperature when sufficiently weak, by making the
particles seem bigger than they are
classically~\cite{PhysRevB.68.134203, ReichmanBerne}.  Still, when the
wave-function overlap between distinct minima becomes sufficiently
large, the particles will eventually delocalise and the glass will
readily ``melt,'' in accordance with expectation.  In other words, the
glass transition is suppressed by quantum fluctuations of sufficient
magnitude, as any symmetry-lowering transition would be.

Some of the earliest quantum phenomena discussed in relation to
amorphous materials were of electronic nature, such as Anderson's
localisation~\cite{AndersonLoc} and Mott's variable range
hopping~\cite{Mott1993}.  These fascinating phenomena are generic
consequences of {\em static} disorder, largely regardless of the
history of the sample, and would not be unique to glasses, as opposed
to, for example, deposited amorphous films made of a poor
glass-former. (The real story is more complicated in that {\em glassy}
solids often exhibit stronger electron-phonon
interactions~\cite{Emin_rev, Emin_revII} and weaker scattering, owing
to lack of dangling bonds.)  Conversely, these electronic phenomena do
not induce nuclear dynamics other than vibrational displacement.

In contrast, here we discuss quantum phenomena that are largely unique
to glasses in that they rely on the existence of dynamics that
connects the many distinct aperiodic states available to a sample made
by quenching a liquid equilibrated below the crossover. We shall
observe that the very same quantities computed in
Section~\ref{ActTransport}, which set the length and time scales for
the classical phenomena near the glass transition temperature, also
determine the magnitude of excitations that operate down to sub-Kelvin
temperatures and are entirely of quantum nature.

\subsection{Two-Level Systems and the Boson Peak}
\label{TLS}

\begin{figure}[t]
  \begin{tabular*}{\figurewidth} {cc}
    \begin{minipage}{.50 \figurewidth} 
      \flushleft
      \begin{center} 
        \includegraphics[width=0.40 \figurewidth]{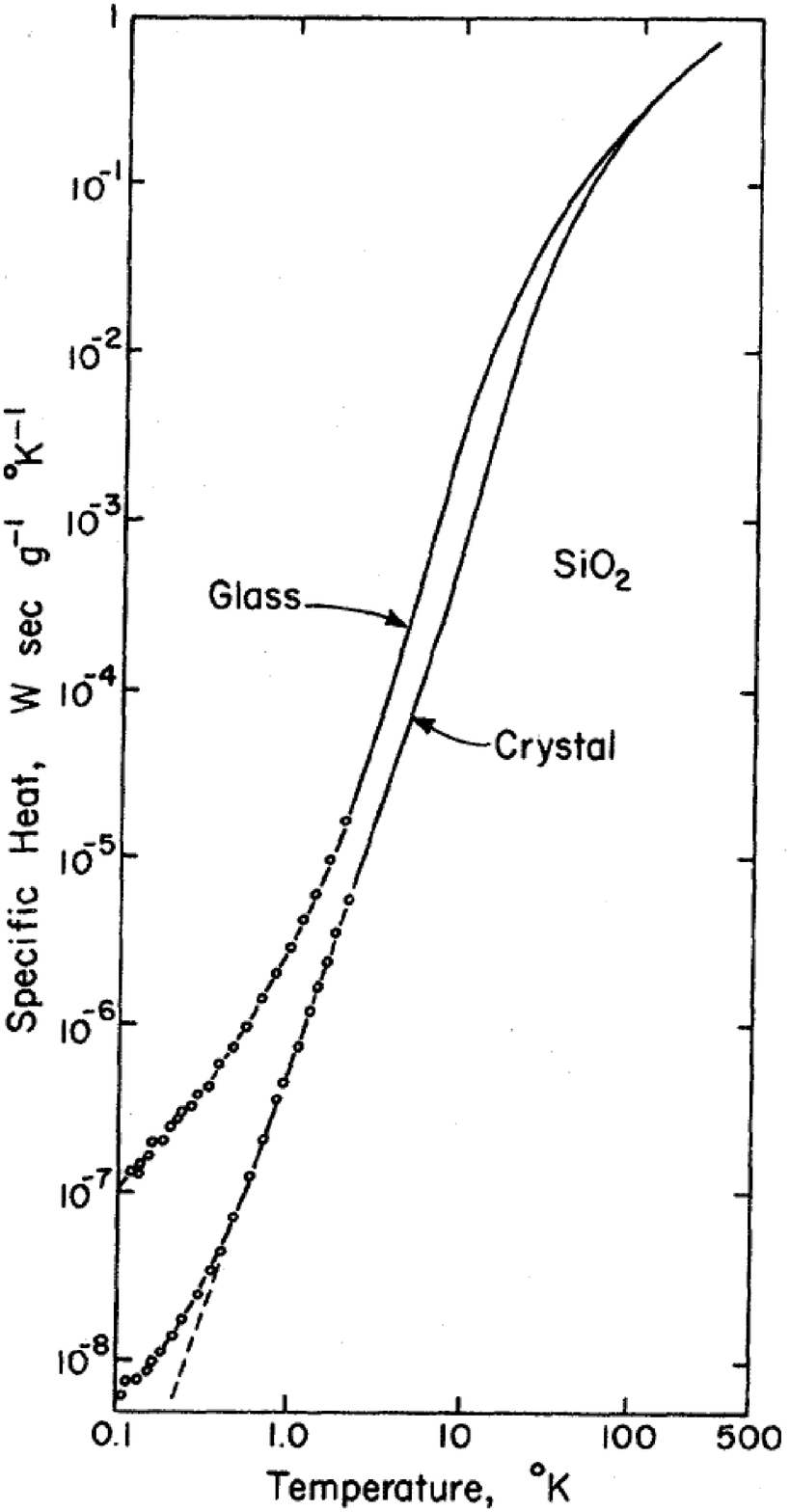}
        \\ {\bf (a)}
      \end{center}
    \end{minipage}
    &
    \begin{minipage}{.50 \figurewidth} 
      \begin{center}
        \includegraphics[width= .40 \figurewidth]{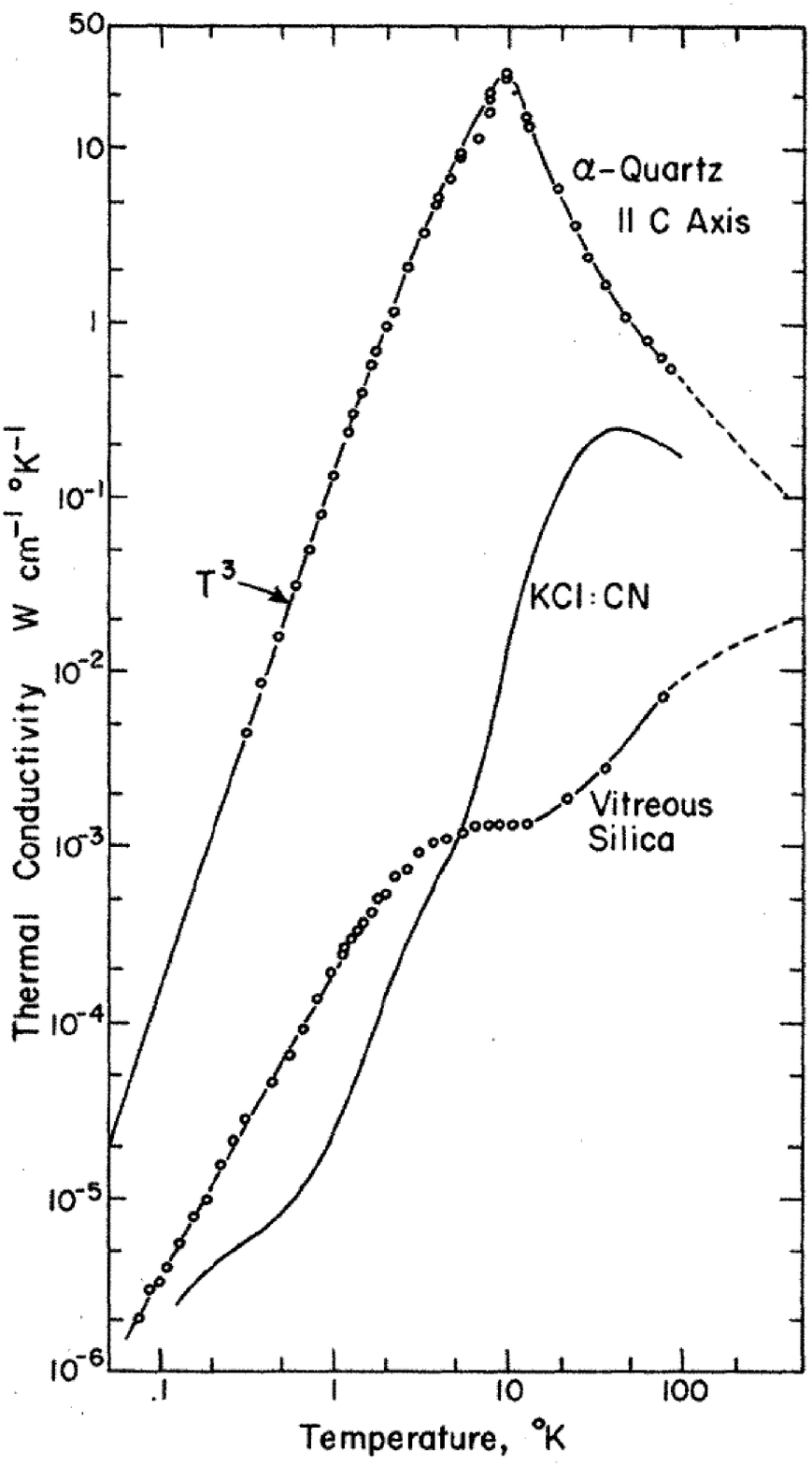} \\
        {\bf (b)}
      \end{center}
    \end{minipage}
  \end{tabular*}
  \caption{\label{ZP} {\bf (a)} Heat capacity of amorphous
    vs. crystalline SiO$_2$~\cite{ZellerPohl} {\bf (b)} Thermal
    conductivity of amorphous vs. crystalline SiO$_2$, samples'
    dimensions $5\times5\times40$~mm~\cite{ZellerPohl}.}
\end{figure}

A number of low temperature anomalies observed in cryogenic glasses
have certainly added to the mystique of glassy solids. Beginning from
the definitive measurements by Zeller and Pohl~\cite{ZellerPohl} some
four decades ago, it became clear that glasses are fundamentally
different in their thermal properties from periodic solids, see
Fig.~\ref{ZP}(a) and (b). As already discussed, glassy liquids
solidify in a gradual manner; they are aperiodic and vastly
structurally-degenerate. Yet this circumstance does not prevent
glasses from being macroscopically solid on very extended timescales
so long as the temperature is sufficiently below $T_g$. Indeed the
barrier for particle rearrangement in glasses is comparably high to or
often higher that the barrier for mechanical failure in crystalline
materials.

Given their macroscopic stability, one may reasonably expect that the
thermal properties of glasses at sufficiently low temperatures would
be identical to those of periodic crystals. This is because the
wavelength of thermal phonons can be made arbitrarily greater than the
correlation length characterising the structural inhomogeneity, if
any, in frozen glasses. Contrary to this expectation, the heat
capacity of cryogenic glasses is not cubic in temperature, but,
instead, is approximately {\em linear} down to lowest measured
temperatures and thus significantly exceeds the vibrational heat
capacity, Fig.~\ref{ZP}(a). (It is understood that equilibration could
become increasingly sluggish at sub-Kelvin temperatures which may
result in a transient low energy gap in the heat capacity.)  The
difference in phonon scattering between vitreous and crystalline
samples is no less dramatic at these temperatures: While the phonon
mean free path already exceeds the sample size in crystals---and so
one should properly speak of heat {\em conductance}---it is of
perfectly microscopic dimensions in glasses. In addition to the
significantly reduced magnitude, the heat conductivity in glasses also
has a distinct temperature dependence, approximately $\propto T^2$,
Fig.~\ref{ZP}(b), in contradistinction with the cubic law observed in
crystals.

\begin{figure}[t]
  \begin{tabular*}{\figurewidth} {ccc}
    \begin{minipage}{.5 \figurewidth} 
      \begin{center}
        \includegraphics[width= .5 \figurewidth]{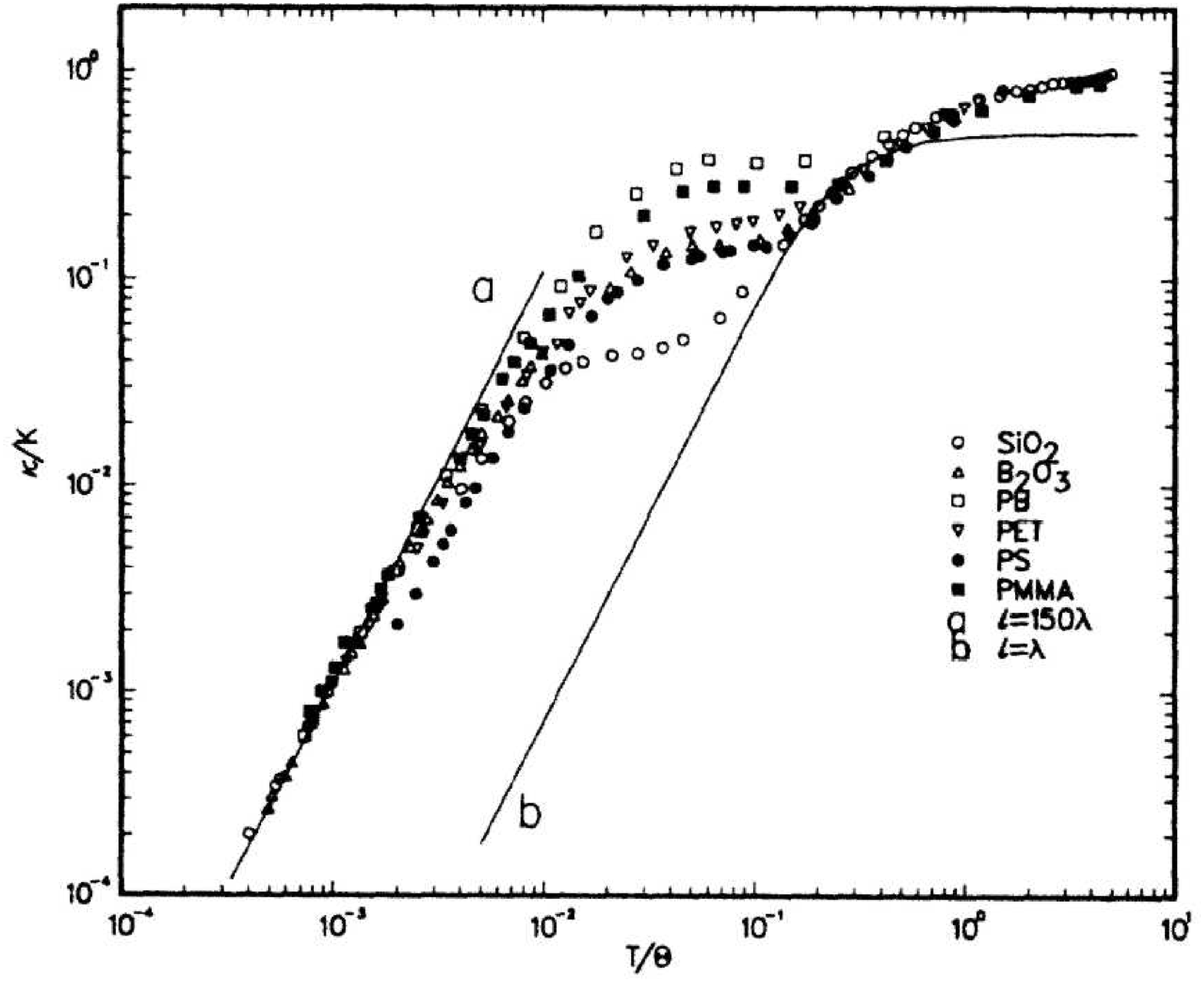}
        \\ {\bf (a)}
      \end{center}
    \end{minipage}
    &
    \begin{minipage}{.5 \figurewidth} 
      \flushleft
      \begin{center} 
        \includegraphics[width=0.45 \figurewidth]{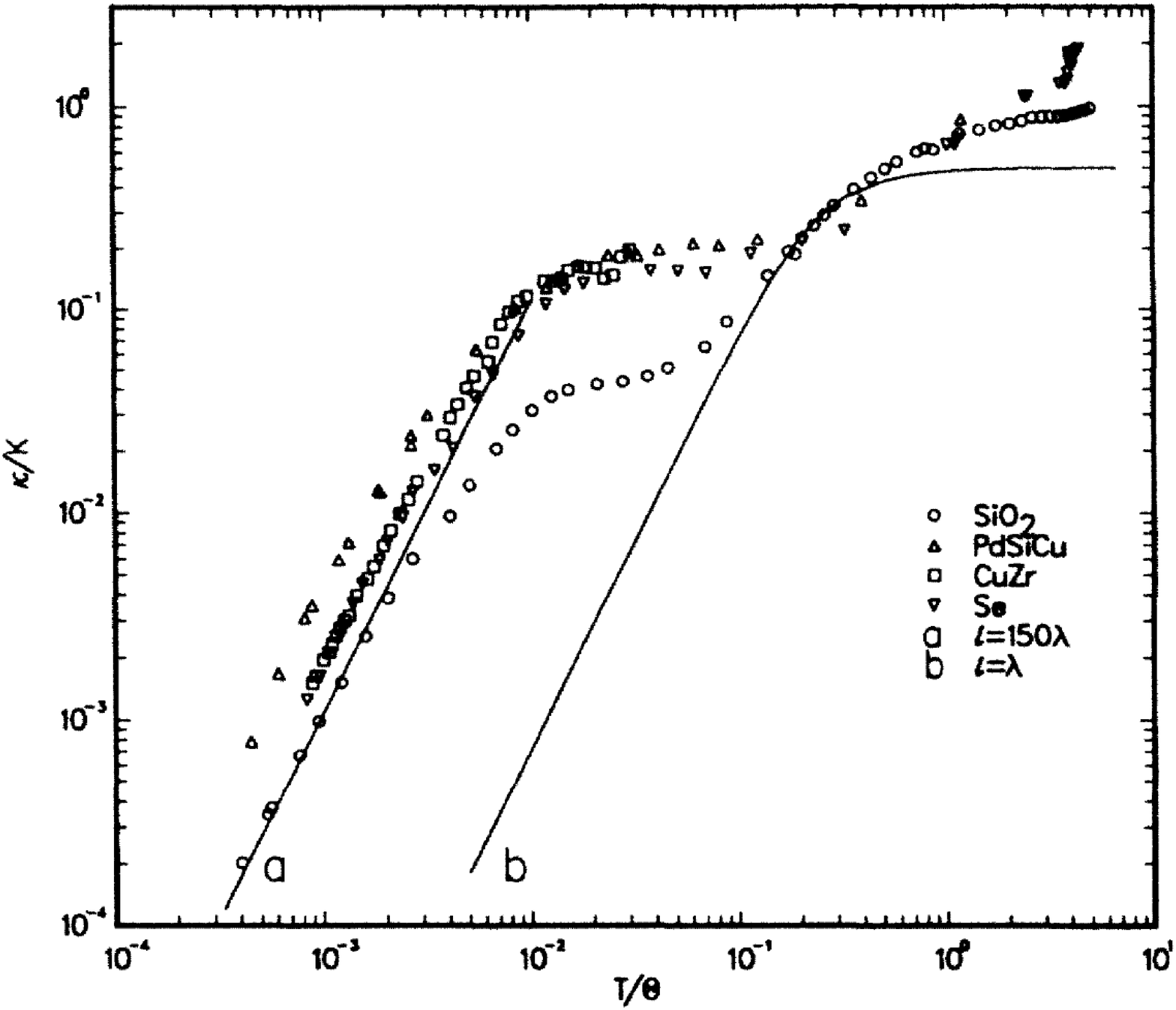}
        \\ {\bf (b)}
      \end{center}
    \end{minipage}
  \end{tabular*}
  \caption{\label{FA} {\bf (a)} Scaled thermal conductivity ($\kappa$)
    data for several amorphous materials. The horizontal axis is
    temperature in units the Debye temperature $T_D$. The vertical
    axis scale $K \equiv \frac{k_B^3 T_D^2}{\pi \hbar c_s}$.  The
    value of $T_D$ is somewhat uncertain, but its choice made in
    \cite{FreemanAnderson} is strongly supported by that it yields
    universality in the phonon localisation region. The solid lines
    are calculated using $\kappa \simeq \frac{1}{3} \sum_{\omega}
    C\sph(\omega) l_\smfp(\omega) c_S$ with $l_\smfp/\lambda = 150$
    and $l_\smfp/\lambda = 1$ respectively. {\bf (b)} Thermal
    conductivity of several metallic glasses plotted in the same
    fashion as in panel (a). The data for SiO$_2$ are given for
    comparison. Both figures from Freeman and
    Anderson~\cite{FreemanAnderson}.}
\end{figure}

Shortly after Zeller and Pohl's discovery, Anderson, Halperin, and
Varma~\cite{AHV}, and Phillips~\cite{Phillips} made an excellent
circumstantial case that both the excess heat capacity and phonon
scattering arise from the same microscopic entity. Clearly, the excess
heat capacity cannot be due to static inhomogeneities, as already
mentioned. Thus those workers concluded that cryogenic glasses must
host small groups of atoms performing strongly anharmonic motions that
result in local {\em resonances}; these can be approximated as
two-level systems at low temperatures: $E = -\frac{1}{2}(\epsilon
\sigma_z + \Delta \sigma_x)$. Expanding the diagonal component of the
energy splitting in terms of the local deformation $u$, $\epsilon
\approx \epsilon(u=0) + (\prtl \epsilon/\prtl u) u$ yields a coupling
to the sound waves: $g \equiv -\frac{1}{2} (\prtl \epsilon/\prtl u)$,
thus yielding the TLS energy function:
\begin{equation} \label{ETLS} E = -\frac{1}{2}(\epsilon \sigma_z +
  \Delta \sigma_x) + g u \sigma_z.
\end{equation}
A flat density of states, $n(E) = \bar{P} = \text{const}$,
approximately accounts for the $T$-dependence of {\em both} the heat
capacity and conductivity~\cite{AHV, Phillips, LowTProp}. Phonon-echo
and single-molecule spectroscopy have directly confirmed the resonant
nature of these mysterious excitations and even captured their
becoming multi-level at increasing temperatures~\cite{GG, Orrit,
  BauerKador}.

Already Zeller and Pohl noted the heat conductivity---a dimensional
quantity---seemed only mildly system-dependent, a point that was
forcefully brought home by Freeman and Anderson~\cite{FreemanAnderson}
some 15 years later, see Fig.~\ref{FA}(a).  The latter authors have
shown the ratio of the phonon mean-free path $\lambda_\smfp$ to the
thermal phonon wavelength $\lambda_\sth$ is near universally $\approx
150$ for all tested, insulating vitreous substances, some
polymeric. This ratio can be also expressed in terms of the parameters
of the two level systems:
\begin{equation} \label{Pg2} \frac{\l_\smfp}{\lambda_\sth} = \left(
    \frac{\bar{P} g^2}{\rho_m c_s^2} \right)^{-1} \simeq 150,
\end{equation}
where $\rho_m$ and $c_s$ are the mass density and speed of sound
respectively. One way to interpret the above figure is to say that the
vibrational plain waves---or phonons---are reasonably well-defined,
wave-like quasi-particles, despite the aperiodicity of the lattice. In
addition to the low-$T$ regime in Fig.~\ref{FA}(a), which corresponds
to the near universal line $\l_\smfp/\lambda_\sth \simeq 150$, there
is also the regime on the higher temperature flank on
Fig.~\ref{FA}(a), where this ratio is about one:
$\l_\smfp/\lambda_\sth \simeq 1$.  This regime is often referred to as
the Ioffe-Regel regime~\cite{Mott1990, Mott1993}, whereby the phonon
is no longer a well-defined quasi-particle. Instead, it is more
appropriate to speak of energy transport in the solid as hopping of
vibrational packets, a picture originally (and incorrectly) envisioned
by Einstein for {\em crystalline} solids~\cite{Einstein}. The
intermediate regime $1 < \l_\smfp/\lambda_\sth < 150$ is of
considerable interest, too: This region corresponds with a significant
rise in the heat capacity, Fig.~\ref{TLSPlateau}(a), often called the
heat-capacity ``bump,'' but also an increased rate of phonon
scattering, which leads to a ``plateau'' in the heat conductivity,
$10^1 \lesssim T/\theta \lesssim 10^2$ in Fig.~\ref{FA}(a). One should
not fail to notice that the plateau temperatures correspond to
terahertz frequencies and thus match, frequency-wise, the Boson Peak
excitations in Fig.~\ref{LL}.

Yu and Leggett~\cite{YuLeggett} (YL) have stressed that there is
little {\em \`{a} priori} reason for the ratio from Eq.~(\ref{Pg2}) to
be so consistent between different substances, considering that the
density of states $\bar{P}$ varies by at least two orders of
magnitude. (The magnitude of the variation, while considerable, is
still surprisingly small given the chemical variation among the tested
materials.) YL also noted that the phonon-mediated interaction between
two TLSs goes as $(g^2/\rho_m c_s^2)r^{-3}$---a fact we have already
encountered at the end of
Subsection~\ref{spinConnections}---consistent with Eq.~(\ref{Pg2}) and
the fact that $\bar{P}$ has dimensions inverse energy-volume.  Based
on this realisation and noting the long-range character of the $1/r^3$
interaction, YL proposed that the apparent density of states $\bar{P}$
pertains to some renormalised excitations which span a large number of
the local resonances. Thus this density of states could be a universal
quotient of the inverse of the elemental interaction $(g^2/\rho_m
c_s^2)^{-1}$ irrespective of the precise nature of those local
resonances. Showing why this quotient should be of order $10^2$ has
been difficult, but the ``interaction'' scenario is still being
pursued~\cite{BurinKagan, doi:10.1021/jp402222g, Esquinazi}. To avoid
confusion with the Bevzenko-Lubchenko model~\cite{BL_6Spin, BLelast},
we note that in the latter, the interaction is very strong, comparable
to the glass transition temperature, Fig.~\ref{BLfigure}(b), whereas
the interaction between the {\em apparent} two-level systems is closer
to sub-Kelvin energy scales~\cite{Silbey}, see also below.

Local resonances naturally arise in the RFOT scenario of the glass
transition. Hereby the near universality of the ratio in
Eq.~(\ref{Pg2}) can be traced to the near universality in the
cooperativity size $\xi$. To see this we first note that the activated
reconfigurations in a glassy liquid are perfectly reversible, so long
as the surrounding matrix has not re-arranged during the waiting time
for the reverse reconfiguration. Such reverse reconfigurations are
thus unlikely to happen at high temperatures, as follows from the
discussion of facilitation in Sections~\ref{hetero} and
\ref{rheo}. Now, Lubchenko and Wolynes~\cite{LW} (LW) have shown that
a certain fraction of the reconfigurations are essentially {\em
  zero-barrier}.  These reconfigurations---and their reverses!---would
be thus thermally active down to very low temperatures, while the
rest, slower reconfigurations are frozen. These reversible
reconfigurations underlie the structural resonances that give rise to
the TLS and the excess phonon scattering. What is the distribution of
the transition energy $\Omega(E)$ for such reversible transitions?
Setting aside the issue of the transition probability for a moment, we
note that save some ageing, the frozen matrix itself and its
excitations can thought of as pertaining exclusively to a {\em single}
temperature scale. The latter is the fictive temperature, of course;
in turn, it is approximately equal to the glass transition temperature
$T_g$. Thus $\prtl \ln \Omega(E)/\prtl E = 1/T_g$. Setting the ground
state energy at zero: $\int_{\infty}^0 dE \, \Omega(E) = 1$, one gets
$\Omega(E) = \frac{1}{T_g} e^{E/k_B T_g}$, per region that can
reconfigure on some specified time scale. Note this exponential
distribution can be explicitly obtained for spin glasses~\cite{MPV}
and the random energy model~\cite{Derrida}.

\begin{figure}[t]
  \begin{tabular*}{\figurewidth} {cc}
    \begin{minipage}{.54 \figurewidth} 
      \flushleft
      \begin{center} 
        \includegraphics[width=0.54 \figurewidth]{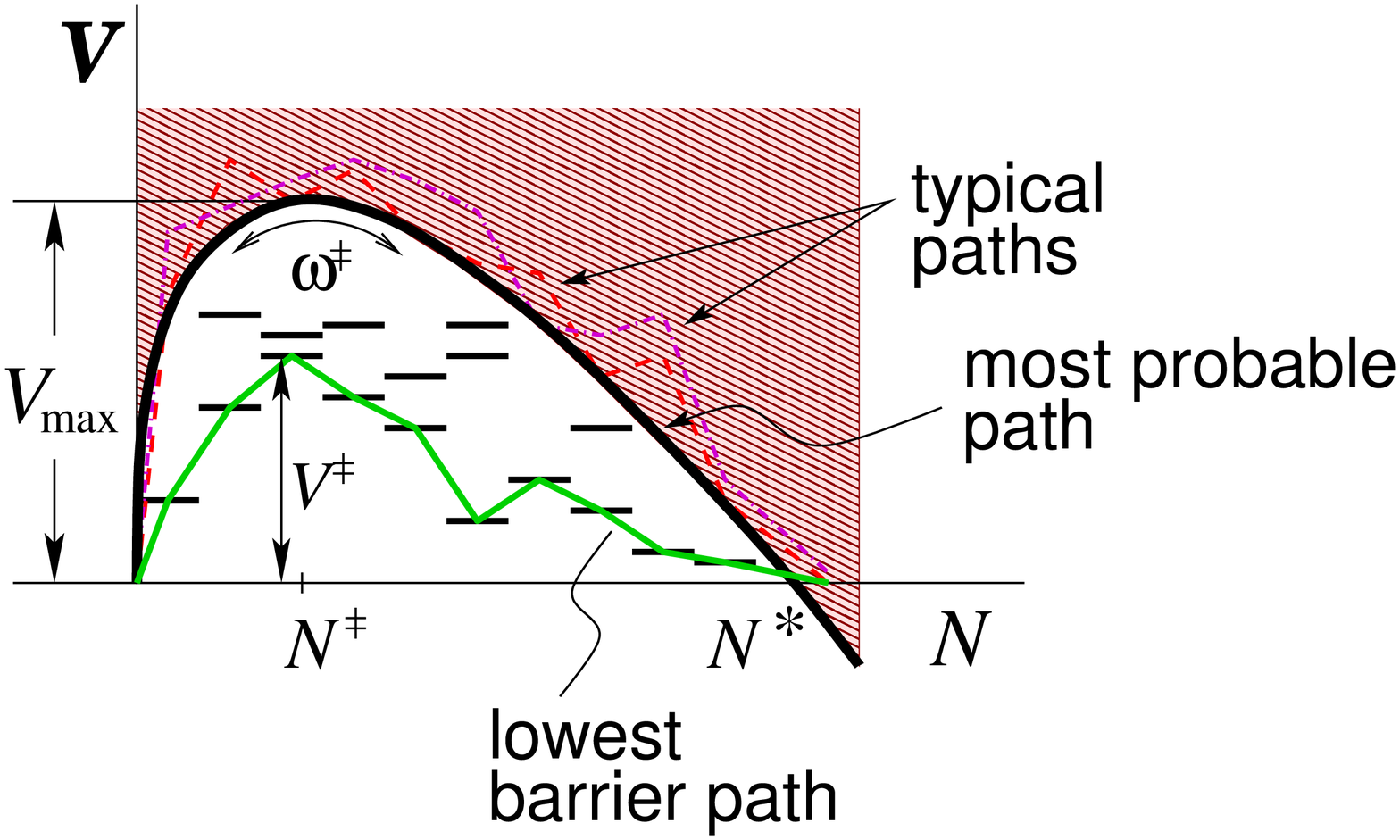}
        \\ {\bf (a)}
      \end{center}
    \end{minipage}
    &
    \begin{minipage}{.42 \figurewidth} 
      \begin{center}
        \includegraphics[width= .38 \figurewidth]{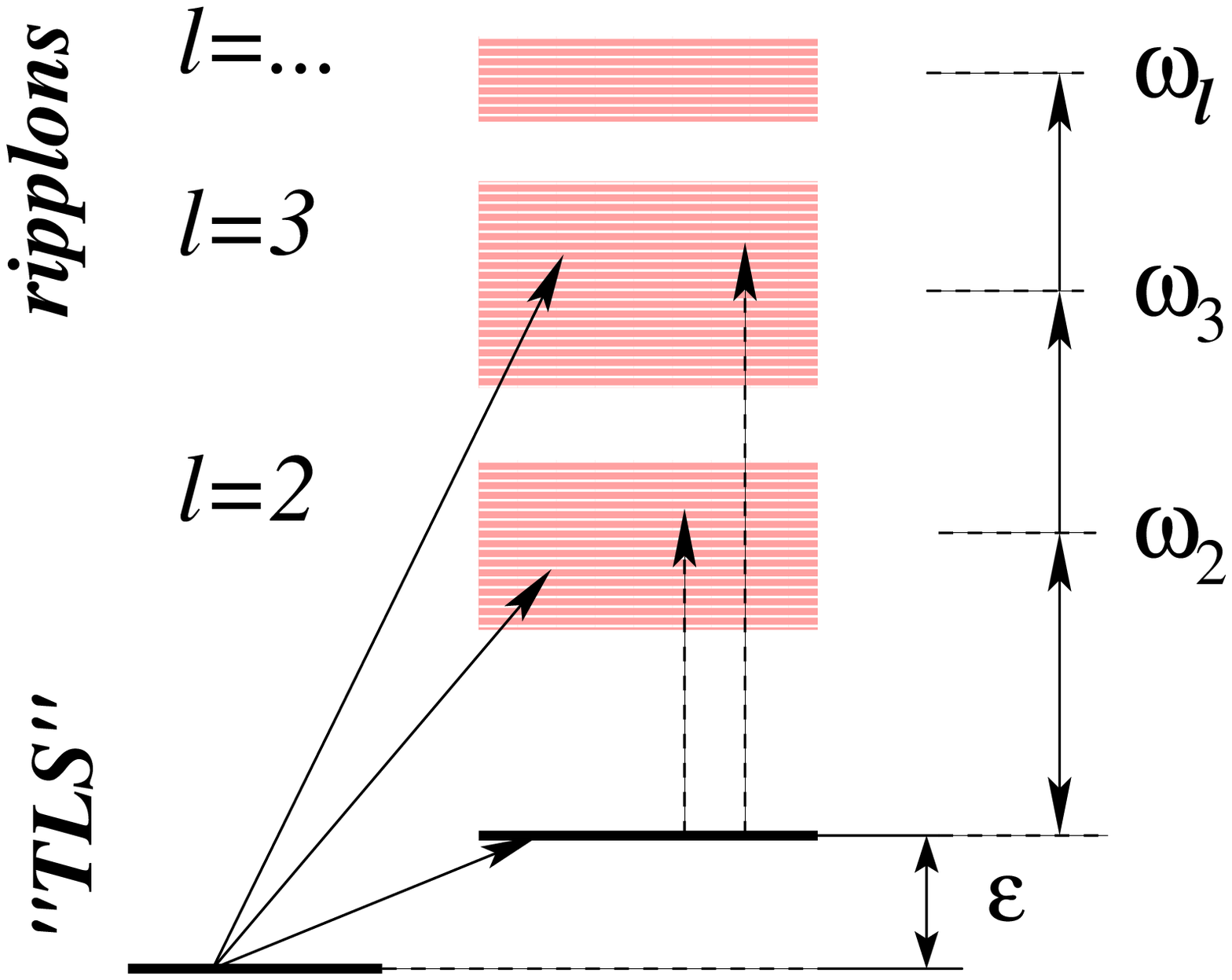} \vspace{2mm} \\
        {\bf (b)}
      \end{center}
    \end{minipage}
  \end{tabular*}
  \caption{\label{V} {\bf (a)} The black solid line shows the barrier
    along the most probable path. Thick horizontal lines at low
    energies and the shaded area at energies above the threshold
    represent energy levels available at size $N$.  The red and purple
    line demonstrate generic paths, green line shows the actual
    (lowest barrier) path that would be followed in the thermally
    activated regime, where $\hbar \omega^\ddagger < k_B T/2\pi$. {\bf
      (b)} Tunnelling to the alternative state at energy $\epsilon$
    can be accompanied by a distortion of the domain boundary and thus
    populating the vibrational states of the domain walls. All
    transitions exemplified by solid arrows involve tunnelling between
    the intrinsic states and are coupled linearly to the lattice
    distortion and contribute the strongest to the phonon
    scattering. The ``vertical'' transitions, denoted by the dashed
    arrows, are coupled to the higher order strain and contribute only
    to Rayleigh type scattering, which is much lower in strength than
    that due to the resonant transitions.}
\end{figure}

To convince ourselves in the existence of low barrier
reconfigurations, we may review the library construction in
Fig.~\ref{library} and recognise that a typical activation profile
gives the value of free energy at which the droplet$+$environment
system is guaranteed to have a state, however the transition state for
individual trajectories is {\em distributed}.  We reiterate this
notion in Fig.~\ref{V}(a). In contrast with the equilibrium
calculation in Subsection~\ref{mosaic}, the energies of both the
initial and transition state are randomly distributed variables, as
the surrounding matrix is now frozen and so one can no longer define a
typical, equilibrated initial state.  As a result, the barrier
distribution is exponential---similarly to that of the classical
density of states---but with a softer growth pattern, viz.,
$e^{V^\ddagger/\sqrt{2} k_B T_g}$.  Consequently, the barrier for
rearranging a region of size $\xi$ turns out to be lower than the
typical barrier for irreversible reconfigurations at $T_g$~\cite{LW,
  LW_RMP}, i.e., $\approx 26 k_B T_g$.  Though lower, this barrier is
still much much higher than what could be tunnelled through,
however. Conversely, the chances that a region of size $\xi$ will have
a zero reconfiguration barrier, $e^{- 26 k_B T_g/ \sqrt{2} k_B T_g}
\sim 10^{-8}$, are small, even if not astronomically small. LW go
further and consider the set of trajectories available to regions {\em
  larger} than $\xi$. To a region of size $N > N^*$, there are $e^{s_c
  (N-N^*)/k_B}$ more states available. And so given $s_c(T_g) = 0.82$,
at $(N-N^*) = 22$---only a 10\% increase relative to $N^*(T_g)$---one
{\em can} in fact find a trajectory with a zero barrier, since $e^{s_c
  (N-N^*)/k_B} \sim 10^8$.  Importantly, alternative configurations
are kinetically accessible only to regions larger than $N^*(T_g)$. The
energy of such a larger region is below that of the typically
reconfiguring region at $T_g$, implying only the negative tail of the
distribution $\Omega(E) = (1/T_g) e^{E/k_B T_g}$ is kinetically
accessible, by {\em tunnelling}. That the barrier is zero makes the
question of the tunnelling mass moot. Still, LW~\cite{LW} point out
the tunnelling coordinate is surprisingly low mass, in apparent
similarity to the domain-wall solution in the Su-Schrieffer-Heeger
hamiltonian~\cite{RevModPhys.60.781}. (This similarity will be
prominent in Subsection~\ref{midgap}.) We thus conclude that the
density of states of the resonances that would thermally active at
very low temperature is~\cite{LW, LW_BP, LW_RMP}:
\begin{equation} \label{nE} n(\epsilon) \simeq \frac{1}{T_g \xi^3}
  e^{-|\epsilon|/k_B T_g}.
\end{equation}
The resulting density of low energy excitations, $\epsilon \ll k_B
T_g$, is thus
\begin{equation} \label{Pbar} \bar{P} = 1/T_g \xi^3.
\end{equation}
Interestingly, it is determined by the characteristics of the solid
that were set at the temperature of preparation, which is two orders
of magnitude greater than the ambient, cryogenic temperature!  The
generic value of $\bar{P}$ is about $10^{45}$~J$^{-1}$m$^{-3}$,
consistent with observation~\cite{BerretMeissner}. It is easy to see
that he TLS density of states should decrease with lowering of the
glass transition temperature and, hence, of the rate of quench during
vitrification. Indeed, by Eqs.~(\ref{xiKsc}) and (\ref{xiXW}), we
obtain $\bar{P} \propto s_c^2$ and $\bar{P} \propto s_c^2/T$
respectively. Either quantity is an increasing function of
temperature, because $s_c \propto (T-T_K)$. We thus conclude that for
a material in the landscape regime, the density of states of the TLS
should be the lower the more stable the glass is. Generically, this
means that denser glasses have fewer two-level systems.

Let us return to a previous notion that near the crossover, the free
energy cost of particle localisation is comparable to that due to
elastic deformation of the newly formed elastic continuum. The latter
free energy is fixed by the equipartition theorem: $\la |g u| \ra
\simeq \rho_m c_s^2 a^3 \la u^2 \ra \simeq k_B T_\scr$, where $\rho_m$
is the mass density. (This simple estimate is consistent with a more
detailed argument~\cite{LW}.)  Thus we may estimate the TLS-phonon
coupling in Eq.~(\ref{ETLS}):
\begin{equation} \label{g} g \simeq \sqrt{\rho_m c_s^2 a^3 k_B T_g}.
\end{equation}
Eqs.~(\ref{Pbar}) and (\ref{g}) immediately yield that the mysterious
universality in the phonon scattering, as expressed in the ratio from
Eq.~(\ref{Pg2}) stems from the near universality of the cooperative
size near the preparation temperature of the sample:
\begin{equation} \label{llambda} \frac{\l_\smfp}{\lambda_\sth} =
  \left( \frac{\bar{P} g^2}{\rho_m c_s^2} \right)^{-1} \simeq \left(
    \frac{\xi(T_g)}{a} \right)^3 \simeq 10^2.
\end{equation}
According to the prediction above, the slower the quench used to
prepare the glass, the less intense phonon scattering will be at
cryogenic temperatures. This is consistent with our earlier statement
that glasses residing in deeper free energy minima are bonded more
strongly.  Conversely, glasses made by quicker quenches should exhibit
more phonon scattering. The latter notion, however, applies only so
long as the material remains in the landscape regime.  For instance,
we have seen that amorphous films prepared by vapour deposition on a
cold substrate, likely reside in relatively high energy states whereby
structural reconfigurations, if any, would involve bond
breaking. Consistent with this expectation, amorphous silicon films
show a significantly lower density of the two-level systems, according
to recent measurements of internal friction by Liu et
al.~\cite{PhysRevLett.113.025503}. These workers have shown that by
optimising the stability of such films---through varying the
deposition rate and temperature---one can further reduce the amount of
internal friction and the density of states of the
TLS.~\cite{QeenHellman2015} (The internal friction $Q^{-1}$ is equal
to the quantity in Eq.~(\ref{llambda}) times $\pi/2$.) We just saw
that a similar trend is expected in {\em quenched} glasses.

Another instructive example is provided by metallic glasses. According
to Fig.~\ref{FA}(b), the phonon scattering in the latter is similar to
insulating glasses but is weaker by a half-order of magnitude or
so. This strongly suggests that, while possibly present, the
cooperative rearrangements typical of the landscape regime are not as
abundant in (the rapidly quenched) metallic glasses as in their
insulating counterparts, which were made by leisurely cooling. This is
consistent with our earlier remarks in Subsection~\ref{MCT} that
metallic glasses made by rapid quenching may not be as deep in the
landscape regime as insulating glasses. Yet another interesting
example is the ultrastable glasses of indomethacine. These do not seem
to exhibit the two-level systems within the measured temperature
interval and the sensitivity of the
experiment~\cite{PerezCastaneda05082014}. According to the estimates
of the configurational entropy in ultrastable glasses, due to
Stevenson and Wolynes~\cite{SWultimateFate}, and Eq.~(\ref{NXW}), the
TLS density of states should be a half-order of magnitude lower in
such ultrastable glasses compared with those made by traditional
quenching. It will be interesting to see if these fascinating
materials do in fact exhibit the structural resonances at sufficiently
low temperatures.  It is, in principle, possible that the local
ordering in the deposited glassy films gives rise to new physics.

To obtain the basic estimates above, we did not have to discuss the
detailed distribution of the tunnelling amplitude $\Delta$ of the
TLS. This is because the distribution of the tunnelling barriers is
determined by a classical energy scale, $\sim k_B T_g$, which is much
greater than the thermal energy at the cryogenic temperatures in
question. This results in a nearly flat distribution of the barriers
$p(V^\ddagger) \approx \text{const}$, which leads to $p(\ln \Delta)
\approx \text{const}$ and, consequently, $p(\Delta) \propto
1/\Delta$. Semi-classical corrections to this simple result can be
obtained~\cite{LW_RMP}, which modify somewhat the simple
inverse-linear probability distribution: $p(\Delta) \propto
1/\Delta^{1+c}$, where $c \simeq \hbar \omega^\ddagger/\sqrt{2} k_B
T_g$ and $\omega^\ddagger$ is the typical under-barrier frequency, see
Fig.~\ref{V}(a). Simple estimates~\cite{LW_RMP} show that the
frequency $\omega^\ddagger$ scales with the Debye frequency
$\omega_D$, resulting in the numerical value of $c \simeq 0.1$. The
resulting correction to the $\Delta$ distribution partially accounts
for apparent deviations of the temperature dependences of the heat
capacity and conductivity from the simple linear and quadratic laws
respectively.

At sufficiently high temperatures, the low-energy, TLS approximation
becomes inadequate and one must confront the question of the full,
multi-level structure of the internal resonances. The transition to
such multilevel behaviour is expected to occur at a temperature $T'
\simeq \hbar \omega^\ddagger/2\pi$~\cite{LW, PGW_QTST}, even ignoring
damping. In the lowest order approximation, the higher excited states
can be imagined as the lowest excited state dressed with vibrations of
the domain wall that circumscribes the reconfiguring
region~\cite{LW_BP}. Indeed, the region itself is only defined within
the zero-point vibrations of the particles comprising its
boundary. Another way to look the vibrations of the domain walls is
that they are Goldstone particles that emerge when the mosaic of the
entropic droplets forms.

This multi-level structure of a tunnelling centre with added domain wall
vibrations is illustrated in Fig.~\ref{V}(b). Given the size of the
region, $(\xi/a) \simeq 6 \Rightarrow 2 \pi (r^*/a) \simeq 20$, and
the lowest wavelength $\simeq 2a$ for such vibrational excitations, we
obtain that harmonics order $2$ through $10$ or so can be
excited. (The zeroth harmonic corresponds to the uniform dilation,
while the first harmonic corresponds with the structural transition
itself.) Thus the basic frequency scale $\omega_\text{\tiny BP}$ for
these quasi-harmonic modes is given by the expression:
\begin{equation} \label{omegaBP} \omega_\text{\tiny BP} \simeq
  \frac{a}{\xi} \omega_D,
\end{equation}
where $\omega_D$ stands for the Debye frequency. (The actual predicted
scaling with $a/\xi$ is $(a/\xi)^{5/4}$~\cite{LW_BP, LW_RMP} which
would be hard to distinguish from the linear law in
Eq.~(\ref{omegaBP}).)  The identification of the frequency
$\omega_\text{\tiny BP}$ with the Boson peak is explained immediately
below.

\begin{figure}[t]
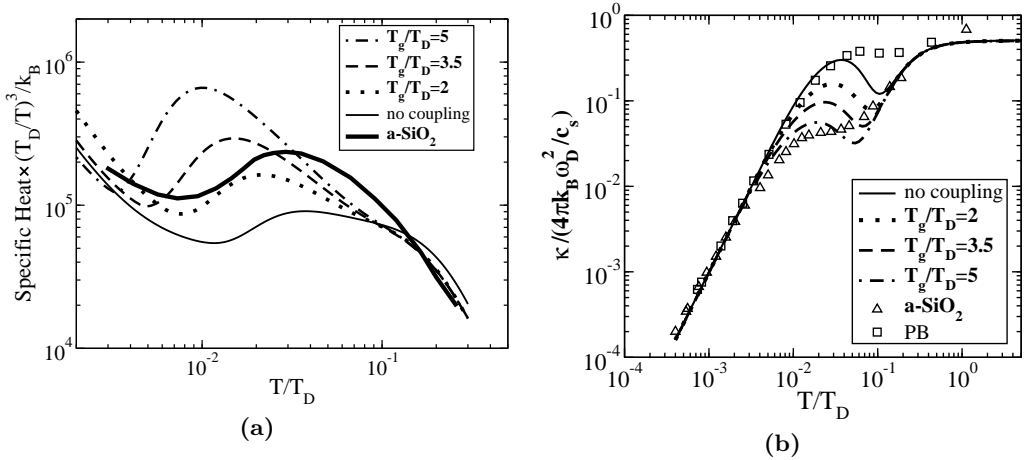

  \begin{tabular*}{\figurewidth} {cc}
    \begin{minipage}{.48 \figurewidth} 
      \flushleft
      \begin{center} 
        \includegraphics[width=0.48 \figurewidth]{bump_comp_Dless1.eps}
        \\ {\bf (a)}
      \end{center}
    \end{minipage}
    &
    \begin{minipage}{.5 \figurewidth} 
      \begin{center}
        \includegraphics[width= .46 \figurewidth]{plateau_comp4.eps} \\
        {\bf (b)}
      \end{center}
    \end{minipage}
  \end{tabular*}
  \caption{\label{TLSPlateau} {\bf (a)} The bump in the amorphous heat
    capacity, divided by $T^3$, as follows from the derived TLS $+$
    ripplon density of states illustrated in Fig.~\ref{V}(b),
    according to Refs.~\cite{LW_BP} and \cite{LW_RMP}. With the
    exception of the thin curve, the theoretical curves account for
    the shift in the ripplon frequency due to interaction with
    phonons. The amount of shift is determined by the ratio $T_D/T_g$
    of the Debye and glass transition temperatures. The thick solid
    line is experimental data for a-SiO$_2$ from \cite{Pohl}.  The
    experimental curve, originally given in J/gK$^4$, was brought to
    our scale by being multiplied by $\hbar^3 \rho c_s^3 (6 \pi^2)
    (\xi/a)^3/k_B^4$, where we used $\omega_D =
    (c_s/a)(6\pi^2)^{1/3}$, $(\xi/a)^3=200$, $\rho = 2.2
    \mbox{g/cm}^3, c_s=4100 \mbox{m/sec}$ and $T_D = 342 K$
    \protect\cite{FreemanAnderson}.  (The Debye contribution was
    included in the estimate of the total specific heat). {\bf (b)}
    Low temperature heat conductivity predicted in Refs.~\cite{LW_BP}
    and \cite{LW_RMP}.  The ``no coupling'' case neglects phonon
    coupling effects on the ripplon spectrum, as in panel (a). The
    (scaled) experimental data are taken from
    \protect\cite{smith_thesis} for a-SiO$_2$ ($k_B T_g/\hbar \omega_D
    \simeq 4.4$) and \protect\cite{FreemanAnderson} for polybutadiene
    ($k_B T_g/\hbar \omega_D \simeq 2.5$). The empirical universal
    lower $T$ ratio $l_\smfp/l \simeq 150$
    \protect\cite{FreemanAnderson}, is used explicitly here to
    superimpose theoretical predictions on the experiment. }
\end{figure}

The vibrations of the domain wall share many characteristics with
regular vibrational degrees of freedom such as the linear dependence
of their energy on the temperature. Still, these vibrational modes are
distinct from regular vibrational excitations on top of a unique
ground state in that they would not be activated in the absence of the
underlying anharmonic reconfiguration. To emphasise this distinction,
LW have called these compound excitations ``ripplons.''  Because the
excitation density of the ripplons is tied to the TLS density of
states, while exhibiting a similar coupling strength to the phonons,
it is straightforward to estimate the contribution of the ripplons to
both the heat capacity and phonon scattering. The resulting
predictions for these quantities are shown in Fig.~\ref{TLSPlateau}(a)
and (b). The agreement with experiment is satisfactory given that no
adjustable parameters are used and that damping of ripplon modes is
ignored. Such damping can be included in the treatment and results in
much better agreement with experiment~\cite{LW_BP, LW_RMP}.

Lubchenko and Wolynes~\cite{LW_Wiley} have pointed out the
identification of the ripplons with the modes responsible for the {\em
  high}-temperature Boson peak from Fig.~\ref{LL} is internally
consistent. In the Yoffe-Regel regime, vibrational energy is being
transferred by hopping of localised vibrations. The heat conductivity
is thus given by the standard kinetic theory expression: $\kappa =
(3N_a k_B/a^3) D$, where $N_a$ is the number of atoms per volume $a^3$
and $D$ is the diffusion constant for hopping of localised vibrations.
The length of the hop is determined by the lattice spacing $a$, while
the hop's waiting time is determined by the vibrational lifetime,
i.e. $\tau_\svibr$. The resulting diffusion constant is $D \simeq
a^2/6\tau_\svibr$, so that the heat conductivity:
\begin{equation} \label{kappaCond} \kappa \simeq \frac{k_B}{a
    \tau_\svibr} (N_a/2).
\end{equation}
Using $a = 3$\AA, $\tau_\svibr = 1$ psec, and $N_a = 3$ yields $\kappa
\simeq 0.1$ W/m$\cdot$K, in agreement with the typical experimental
value of the conductivity near $T_g$~\cite{FreemanAnderson}. In view
of Eq.~(\ref{kappaCond}) and the slow variation with temperature of
the thermal conductivity above the Debye
temperature,\cite{FreemanAnderson} the vibrational relaxation time
should indeed exhibit little temperature dependence well below $T_g$.
As a result, the corresponding peak in dielectric spectra should
exhibit little temperature dependence, as though a set of resonances
intrinsic to the lattice were responsible for the peak.  Further, we
recall Freeman and Anderson's empirical
observation~\cite{FreemanAnderson} that $\kappa$'s for several
different materials tend to saturate at a value numerically close to
$(k_B c_s/a^2) (4 \pi)^{1/3} (3 N_a)^{2/3}$. Combining this
observation with Eq.~(\ref{kappaCond}), we get $\tau_\svibr^{-1}
\simeq (c_s/a) 2(4\pi/3N_a)^{1/3}$, i.e. a universal fraction of the
Debye frequency.  As temperature is lowered, vibrational transfer
should become less overdamped, but so should the low-barrier
tunnelling motions discussed above. We thus expect that the motions
that give rise to the high-$T$ Boson peak will be mixed with the
vibrations of the domain walls, when both are present.

The relative unimportance of interaction between the structural
resonances is a key feature of the present microscopic picture that
distinguishes it from strong-interaction scenarios~\cite{BurinKagan,
  doi:10.1021/jp402222g, Esquinazi}. The weakness of the interaction
results from the low concentration of thermally active resonances, as
already mentioned. Still, each resonance is a multi-level system with
a rich structure. Transitions within an individual resonance are
coupled to transitions within the other resonances, via phonon
exchange. This sort of off-diagonal coupling between local resonances
lowers the energy~\cite{LLquantum} and is the mechanism of, for
instance, the dispersion interactions or the venerable Casimir effect.
Lubchenko and Wolynes~\cite{LW_RMP} (LW) have pointed out that since
the number of thermally active resonances in low temperature glasses
grows with temperature, so will the total amount of phonon-mediated
attraction between spatially separated portions of the
sample. Depending on the amount of the local anharmonicity of the
lattice, this heating-induced attraction may thus result in a negative
thermal expansivity! Negative thermal expansivity is, in fact,
observed in low temperature glasses~\cite{Ackerman}. The magnitude of
the ``Casimir'' effect in glasses depends sensitively on the number of
levels within an individual resonance. The results of LW analysis,
which explicitly included the ripplon states, are in quantitative
agreement with observation thus providing additional support for the
RFOT-based microscopic picture of the resonances depicted in
Fig.~\ref{V}(b).

As pointed out in the beginning of this Section, structural {\em
  degeneracy} is the key to the presently discussed glassy anomalies,
as opposed to effects of aperiodicity in a fully stable lattice per
se. We must be mindful that the vibrational response of stable
aperiodic lattices generally includes non-affine
displacements~\cite{PhysRevLett.97.055501}, which {\em also} violate
the Saint-Venant compatibility condition
(\ref{eq:SaintVenant})~\cite{0953-8984-26-1-015007}.  These modes,
which stem from a distribution in local elastic
response~\cite{6078032020110601, 0295-5075-104-5-56001,
  PhysRevB.79.060201}, have been proposed as the cause of the Boson
Peak, requiring however that the lattice be near its mechanical
stability limit~\cite{0295-5075-73-6-892, ParisiBP,
  0295-5075-104-5-56001}. In the absence of such marginal stability,
purely elastic scattering seems too weak to account for the apparent
magnitude of phonon scattering at Boson Peak
frequencies~\cite{ACAnd_Phi, Joshi, LW_RMP}.  While the lack of
periodicity in glasses undoubtedly contributes to the excess phonon
scattering, the presently discussed structural resonances, which are
inherently and strongly anharmonic processes, account {\em
  quantitatively} for the apparent magnitude of the heat capacity and
phonon scattering in a (logarithmically) broad temperature range that
covers both the two-level system and Boson peak dominated regimes.
The reader is referred to the detailed discussion of the combined
distribution of the TLS parameters $\epsilon$ and $\Delta$ from
Eq.~(\ref{ETLS}) by Lubchenko and Wolynes~\cite{LW_RMP}. In this
treatment, the deviations of both the heat capacity and conductivity
from strict linear and quadratic temperature-dependences,
respectively, are expected. While the deviation is in the correct
direction, its magnitude is somewhat below the observed value.

We note that only a relatively small number of regions host a
zero-barrier mode at cryogenic temperatures: $\xi^3 \int_0^T d\epsilon
\, n(\epsilon) \simeq T/T_g \ll 1$. This number, however, would be of
the order one near $T_g$ thus seemingly implying that the glass would
catastrophically liquefy upon approaching the glass transition from
below. This does not happen, of course.  In reality, the zero-barrier
modes become increasingly dampened with increasing temperature. This
damping stems from the interaction of each individual bead movement
along the trajectory in Fig.~\ref{V}(a) with the phonons. Quantitative
estimation of this damping does not appear to be
straightforward. Empirically, however, the phonons are entirely
dampened by the high-$T$ end of the plateau, and so should be
elemental tunnelling motions, as the two modes are completely
hybridised by that point. Another way of saying that the individual
bead motions are dampened to the fullest extent is that each bead is
subject to the regular viscous drag from the liquid. Under these
circumstances, the calculation for the reconfiguration rate from
Section~\ref{ActTransport} applies.

\subsection{The midgap electronic states}
\label{midgap}

The cryogenic anomalies in glasses we have just discussed were a part
of a thriving field of research on amorphous materials in the
1970s. Much of this effort seems to have been driven by potential
applications in renewable energy. For instance, amorphous silicon was
regarded to be a promising candidate material for photovoltaic
applications. Another family of amorphous semiconductors was also
studied, namely, the so called chalcogenide alloys, which contain
``chalcogens,'' i.e., elements from group XVI (S, Se, Te) and, often,
``pnictogens,'' i.e., elements from group XV (P, As, Sb, Bi) and also
elements from group XIV (Si, Ge). Archetypal representatives of this
group of compounds are As$_2$S$_3$ and As$_2$Se$_3$; both are good
glassformers. More recently, chalcogenides have gained prominence in
the context of applications in
opto-electronics~\cite{doi:10.1021/cr900040x, ISI:000250615400019,
  ISI:000261127100019, Hosseini2014, Kolobov2004, Steimer2008, AIST}
and are often called ``phase-change materials,'' since their optical
and electric properties can be readily
``switched''~\cite{PhysRevLett.21.1450} by converting the material
between its crystalline and amorphous form.

The chalcogenides were first brought into prominence by the Kolomiets
lab, including their remarkable property of being insensitive to
doping~\cite{Kolomiets1981, Mott1993}. This is in contrast with
traditional semiconductors, whose conductance can be manipulated by
adding elements that prefer to bond to fewer or more atoms than the
host species. Note that the prevailing view of amorphous insulators
seems that they are not too different from their crystalline
counterparts. Within the Anderson localisation
paradigm~\cite{AndersonLoc, RevModPhys.50.191}, these aperiodic
materials can still convey electricity within continuous {\em
  mobility} bands made of overlapping orbitals. In addition, there are
exponential (Urbach) tails of localised states extending away from the
mobility bands, which may serve as efficient traps for charge
carriers~\cite{Emin_rev, Emin_revII}. Since the tails decay
exponentially quickly, one may still speak of relatively well-defined
mobility {\em gaps}.

\begin{figure}[t]
  \begin{tabular*}{\figurewidth} {cc}
    \begin{minipage}{.46 \figurewidth} 
      \begin{center}
        \includegraphics[width= .46 \figurewidth]{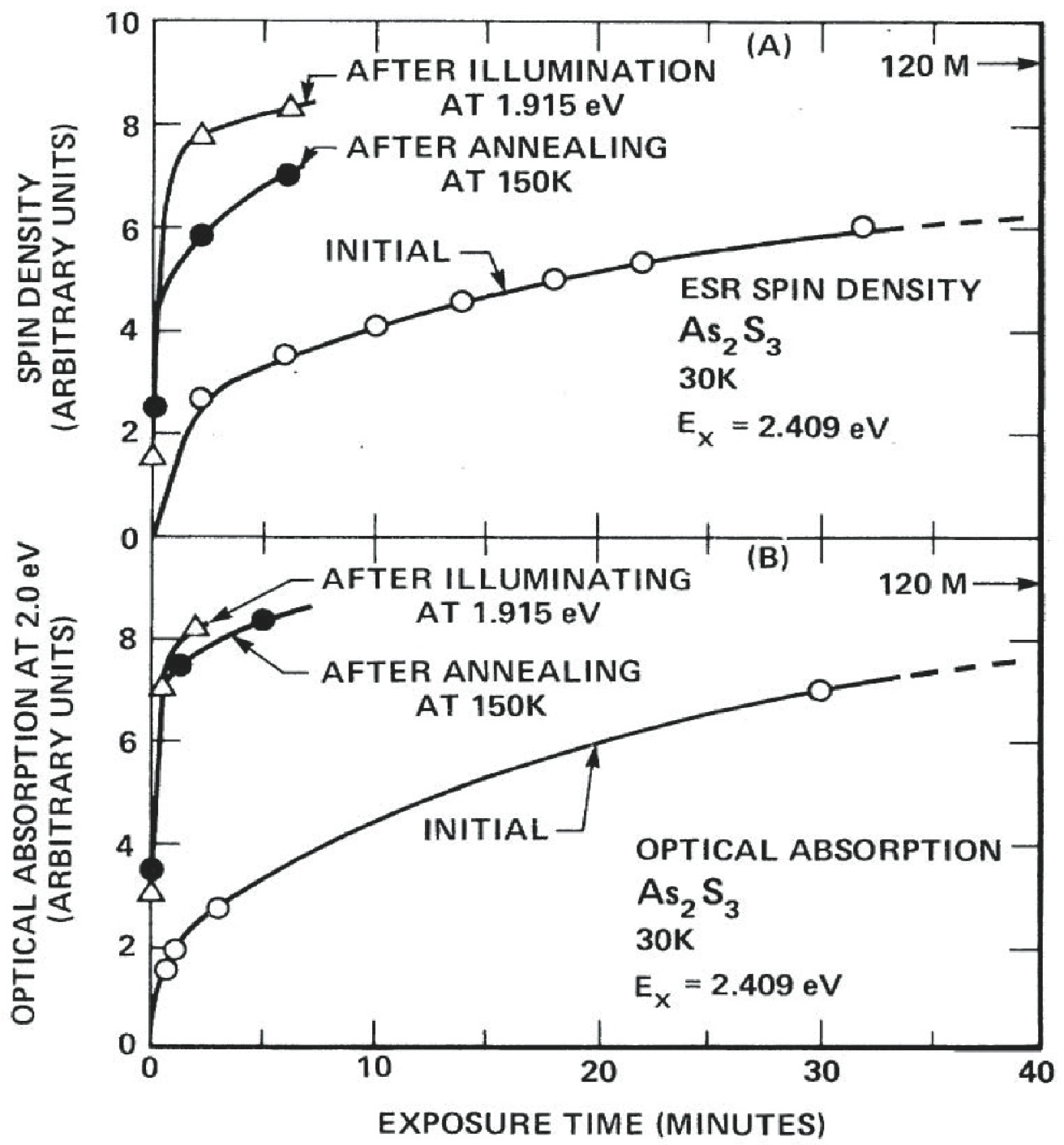}
         \\ {\bf (a)}
      \end{center}
    \end{minipage}    
    &
    \begin{minipage}{.48 \figurewidth} 
      \begin{center}
        \includegraphics[width= 0.46
        \figurewidth]{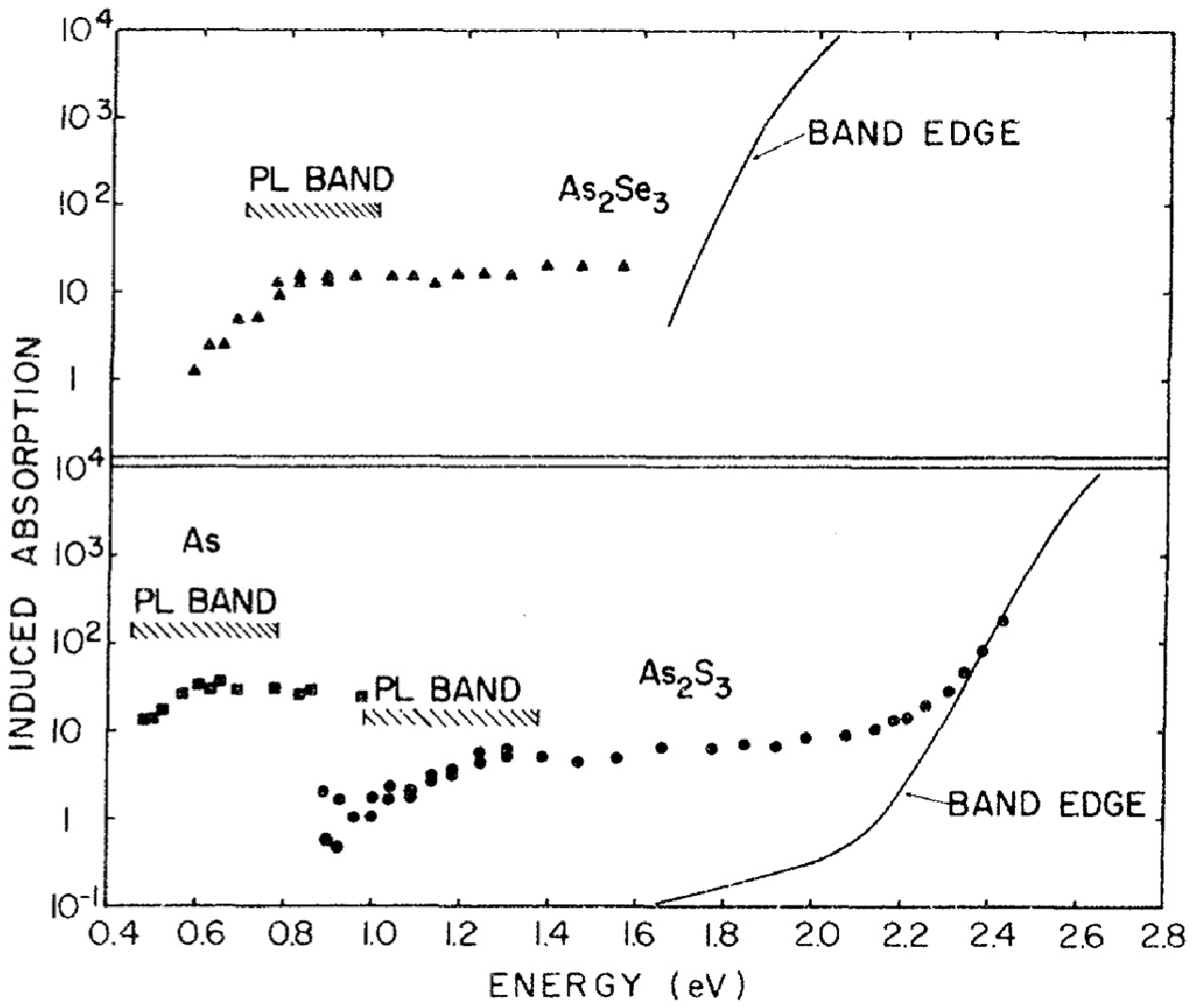}
         \vspace{12mm} 
        \\ {\bf (b)}
      \end{center}
    \end{minipage}
  \end{tabular*}
  \caption{\label{midgapExp} {\bf (a)} \small Time-dependence of
    photoinduced (a) ESR signal and (b) midgap absorption in amorphous
    As$_2$S$_3$~\cite{BiegelsenStreet}. {\bf (b)} Midgap absorption in
    amorphous As$_2$S$_3$ and As$_2$Se$_3$~\cite{PhysRevB.15.2278}.}
\end{figure}

While the ability of an equilibrated amorphous lattice to accommodate
a broad range of bonding preferences of individual atoms is perhaps
not too surprising, the apparent strict pinning of the Fermi level in
amorphous chalcogenides does beg for explanation. What is even more
puzzling, these materials cannot be doped extraneously yet, at the
same time, they seem to possess a robust, {\em intrinsic} number of
defect-like, midgap states. The latter states become apparent upon
irradiation of an amorphous chalcogenide with high intensity light at
gap frequencies, see Fig.~\ref{midgapExp}. Subsequently, the material
begins to absorb light at energies extending to or even below the
middle of the forbidden gap. At the same time, a population of
unpaired spins emerges whose number and time dependence match those of
the midgap light-absorbers. This number seems to saturate at $\sim
10^{20}$cm$^{-3}$ or one per several hundred atoms, nearly
universally.  None of these anomalies are seen in pristine samples
prior to irradiation, thus implying there are no ``dangling bonds.''
In contrast, amorphous silicon or germanium films always host
substantial numbers of unpaired spins hosted on such dangling
bonds~\cite{Mott1993}.

In 1975, Anderson~\cite{PWA_negU, PWA_negU2} proposed the so-called
negative-$U$ model that captures the salient features of these
mysterious midgap states. By the model's premise, there is effective
{\em attraction} between electrons on the same site, which can be
formally implemented using the Hubbard model with a negative $U$. The
mechanism of the effective attraction is similar to that giving rise
to conventional superconductivity---i.e., through vibrational modes of
the solid---except the electronic orbitals are {\em not} extended:
\begin{eqnarray} \label{negU} \cH &=& \epsilon(n_\uparrow +
  n_\downarrow) - g (n_\uparrow + n_\downarrow - 1) q + \frac{k
    q^2}{2} \nonumber \\ &+& U [ (n_\uparrow - 1/2) (n_\downarrow -
  1/2) +1/4],
\end{eqnarray}
where we have coupled the orbital to a single, harmonic
configurational coordinate $q$ and the electronic energies are
measured relative to the Fermi energy of the undistorted lattice, which
we set at zero. (Hamiltonian (\ref{negU}) is similar to the one used
in Ref.\cite{PWA_negU}, but modified so that the configurational
variable $q$ is unperturbed in the neutral, half-occupied state.) As
can be readily seen---after integrating out the configurational
coordinate---a sufficiently strong electron-phonon coupling leads to
an effective on-site attraction between two electrons (holes)
occupying the orbital:
\begin{equation} \label{Ueff} U/2 - g^2/2k = \eU < 0.
\end{equation}

The on-site energy and phonon-driven stabilisation are generally
distributed.  Despite the distribution, the Fermi level will be
strictly pinned, even if one attempts to dope the materials. This is
because the orbital is empty, when $2\epsilon + \eU > 0$, or doubly
occupied, when $2\epsilon + \eU < 0$. Further, optical excitations are
significantly faster than nuclear dynamics.  Thus a filled orbital is
not optically active in the midgap range when filled but {\em will}
give rise to midgap absorption when half-filled. In this picture of
phonon-driven electron pairing, the same lattice effects give rise to
the mobility gap, see also~\cite{ISI:A1972L738100004}. This is an
unwelcome feature---though probably not damning---since the amorphous
and crystalline chalcogenides are very much alike in terms of the gap
size and detailed bonding patterns, thus implying a similar, {\em
  chemical} origin of the forbidden gap.  It seems difficult to link
these detailed chemical interactions and the simple electron-phonon
hamiltonian in Eq.~(\ref{negU}). Likewise, it is difficult to see why
the concentration of the negative-$U$ centres would be, nearly
universally, at one per several hundred atoms. Despite these potential
issues, the model is quite elegant in its simplicity and has inspired
much further work~\cite{KAF, PhysRevLett.35.1293} on identifying
specific structural defects, in fiducial reference structures or
simulated systems, that could be responsible for both the generic
disorder-induced subgap states and negative-$U$ like
centres~\cite{KAF, PhysRevLett.35.1293, PhysRevB.23.2596,
  ShimakawaElliott, PhysRevLett.85.2785, Simdyankin, Dembovsky}.
While several of these detailed assignments are roughly consistent
with the shapes of ESR spectra in specific
compounds~\cite{PhysRevB.38.11048} they do not self-consistently
prescribe how the defects combine to form an actual lattice or address
the apparent near universal density of midgap states in many distinct
materials and stoichiometries.  Conversely, it is puzzling that
supercooled melts equilibrated on one hour times should carry a large
density of midgap electronic states that are very energetically costly
yet do not incur dangling bonds.

More recently, Zhugayevych and Lubchenko~\cite{ZL_JCP} took up the
notion from the library construction, Subsection~\ref{mosaic}, that
each structural reconfiguration is a set of dynamically connected
configurations. Two dynamically connected configurations differ only
by the position of one rigid molecular unit, or ``bead.'' This
statement can be made even stronger by noticing that consecutive bead
movements must be also nearby in space and thus form a quasi-one
dimensional chain in space. If such a chain were broken into two,
those two chains would have to be considered as distinct
reconfigurations if the gap between the chains is sufficiently large.
Indeed, the elastic interaction between two such movements decays as
$1/r^3$ with distance, implying reconfigurations of distinct regions
do not interact significantly. This picture can be refined to account
for the facilitation effects discussed in Subsection~\ref{betaDcorr},
but in any event, we will see shortly that even if a chain {\em is}
broken into several, shorter chains, the main conclusions are not
modified.

\begin{figure}[t]
  \begin{tabular*}{\figurewidth} {cc}
    \begin{minipage}{.52 \figurewidth} 
      \begin{center}
        \includegraphics[width= .52
        \figurewidth]{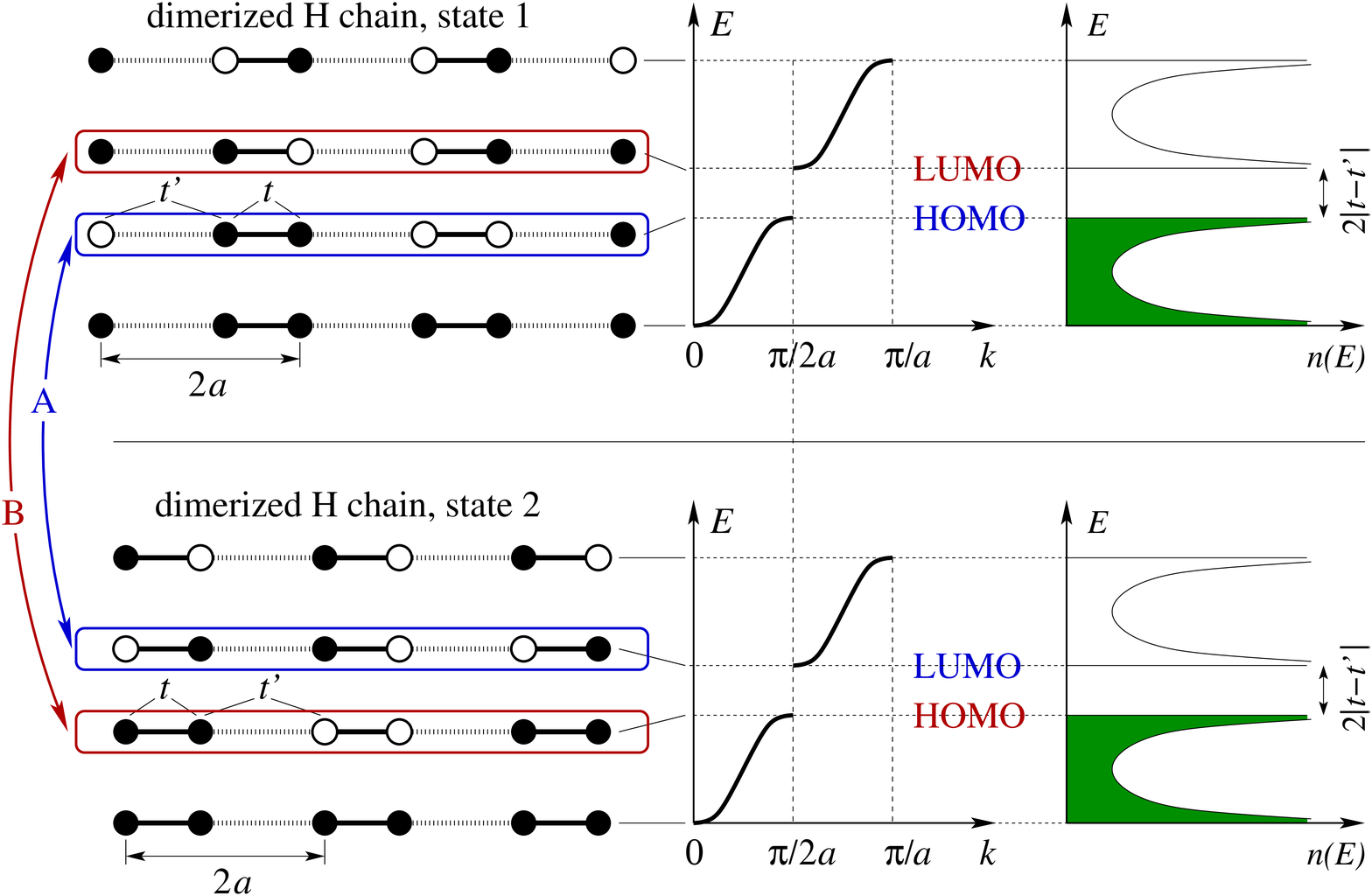}
         \\ {\bf (a)}
      \end{center}
    \end{minipage}    
    &
    \begin{minipage}{.44 \figurewidth} 
      \begin{center}
        \includegraphics[width= 0.44 \figurewidth]{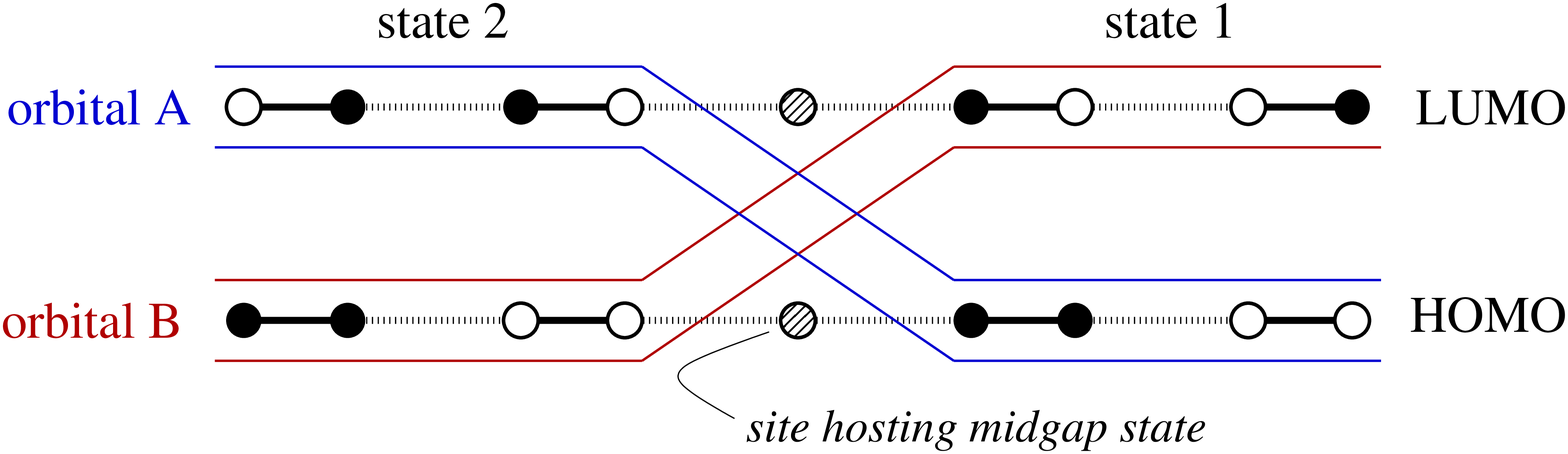}
        \\ {\bf (b)}
      \end{center}

      \caption{\label{MO} {\bf (a)} Molecular orbital view of a
        Peierls insulator. There are two alternative ways for a 1D
        metal do dimerize, call them states 1 and 2. Orbitals A and B
        represent the HOMO in LUMO in state 1 and vice versa in state
        2. {\bf (b)} Molecular orbital view of the topological midgap
        state centred at an under-coordinated atom. Hereby the
        orbitals A and B from panel (a) exchange their HOMO-LUMO
        identities at the malcoordinated atom. The effective gap
        vanishes and results in a mid-gap state, at that atom.}
    \end{minipage}
  \end{tabular*}
  
\end{figure}

One may write down the simplest one-electron Hamiltonian that directly
couples the electron density matrix to the mutual displacements of the
beads that eventually can reconfigure to a rearrange a region:
\begin{eqnarray} \label{SSH_E} \cH &=& \sum_n \sum_{s=\pm1/2} \left[
    t(x_{n}, x_{n+1}) (c^\dagger_{n,s} c_{n+1,s} + c^\dagger_{n+1,s}
    c_{n,s}) \right. + \left.  (-1)^n \epsilon_n c_{n,s}^\dagger
    c_{n,s} \right] \nonumber \\ &+& \cHlt (\{x_n\}),
\end{eqnarray}
where one ordinarily assumes that the lattice-mediated bead-bead
interaction $\cHlt (\{x_n\})$ depends quadratically on the their
mutual distance. The hopping matrix element $t(x_{n}, x_{n+1})$
depends exponentially on the distance $|x_n - x_{n+1}|$, but a linear
approximation already yields satisfactory accuracy. Eq.~(\ref{SSH_E})
is a generalisation of the
Su-Schrieffer-Heeger~\cite{PhysRevB.22.2099} Hamiltonian that
incorporates spatial variation in electronegativity $\epsilon_n =
\epsilon \ne 0$~\cite{PhysRevLett.49.1455}. If the latter is not too
large and the electron count is not too different from half-filling,
the energy function (\ref{SSH_E}) corresponds to a one-dimensional
metal. The latter are known to be Peierls-unstable toward partial
dimerization. Hereby, each dimer is held together by a two-centre
covalent bond while the dimers attract via a much weaker, closed-shell
interaction. This interaction is often called
``secondary''~\cite{Alcock1972, Pyykko, LandrumHoffmann1998,
  PapoianHoffmann2000} or ``donor-acceptor''~\cite{Bent1968}.  The
resulting bond-alternation pattern along the chain leads to the
formation of an insulating gap. There are two ways for the chain to
dimerize, call them state 1 and 2, see Fig.~\ref{MO}(a). (If the
electronegativity variation is sufficiently strong, the chain instead
becomes an ionic insulator with a gap largely determined by the
electronegativity differential $\epsilon$~\cite{PhysRevLett.49.1455}.)

But what happens if the very same chain happens to dimerize
inhomogeneously, so that different portions of the chain happen to
settle in distinct states? The atom at the interface between the two
states is either over-coordinated or under-coordinated. Heeger and
coworkers~\cite{PhysRevB.22.2099, RevModPhys.60.781} have shown, in
the context of trans-polyacetylene, that there appears a midgap state
centred on such a malcoordinated atom.  The origin of this midgap
state is topological and can be traced back to a rather general setup,
in which a fermionic degree of freedom is coupled to a classical
degree of freedom which is subject to a bistable
potential~\cite{PhysRevD.13.3398}. There is a heuristic way to see the
emergence of the midgap state using simple notions of the molecular
orbital theory, as we illustrate in Fig.~\ref{MO}. According to panel
(a), if the two orbitals at the edge of the forbidden gap of a Peierls
insulator---call them A and B---are HOMO and LUMO in state 1, then
they switch their roles in state 2.  In panel (b), we observe that at
the (undercoordinated) atom at the boundary between states 1 and 2,
the molecular terms corresponding to orbitals A and B, respectively,
must cross in the middle of the gap, implying there is midgap state at
zero energy, which is centred on the undercoordinated atom. This
molecular orbital-based picture demonstrates that not only will the
midgap state be in the middle of the gap---given the electron-hole
symmetry---but also that the orbitals from the valence and conductance
band contribute in equal measure to the wavefunction of the midgap
state. The state is thus robust with respect to effects of
electron-electron interactions, among others.

This robustness is of topological origin: First of all, the
malcoordination cannot be removed by elastic deformation, thus
automatically violating the Saint-Venant compatibility condition
(\ref{eq:SaintVenant}).  Another topological signature of the
malcoordination is explicitly seen using a continuum limit of the
Hamiltonian exhibited in Eq.~(\ref{SSH_E}), viz., $\cH = - i v
\sigma_3 \prtl_x + \Delta(x) \sigma_1 + \epsilon(x) \sigma_2$
\cite{PhysRevB.21.2388, PhysRevLett.49.1455}. Here, $\sigma_i$ are the
Pauli matrices, while $ -i v$, $\Delta(x)$ and $\epsilon(x)$
correspond, respectively, to the kinetic energy, local one-particle
gap and variation in electronegativity.  The local gap $\Delta(x)$ is
space dependent and, in fact, switches sign at the defect, thus
corresponding to a rotation of a vector $(\Delta, \epsilon)$. (This
space-dependent gap is the energy difference between A and B orbitals
in Fig.~\ref{MO}(b).) The orientation angle of this vector is the
topological phase associated with the defect; it follows a solitonic
profile across the interface.  This notion shows that in addition to
being robust with respect to local chemistry, the defects are stable
against mutual annihilation unless they travel along exactly the same
chain.  Indeed, the phases of defects travelling along different paths
can not cancel.

The midgap states are remarkable in several ways: First of all, they
exhibit a charge-spin relation opposite from that expected for a
regular fermion: A neutral defect is half-filled and thus has spin
$1/2$; a charged state is paramagnetic. From a chemist's viewpoint, a
neutral defect is merely a free radical solvated in a solid matrix.
Last but not least, the wave-function of the midgap state is
surprisingly extended. Indeed, a simple estimate shows that a generic
bound state at $\sim 1$eV below a continuum should be of atomic
dimensions. In contrast, both the wavefunction and the solitonic
deformation pattern of the 1D lattice are about 10 lattice spacings
across, for a semiconducting material~\cite{ZL_JCP}. (In
trans-polyacetylene, this extent is as large as 20 lattice spacings
because of the low effective mass of the electron in this material.)
Thus, the malcoordination is evenly distributed over a considerable
distance. Note that the bond length at the centre of the
malcoordination is about midway between that of the long and short
bond, also implying an intermediate bond strength.

In the ZL scenario, the effective attraction between two
electrons/holes occupying the midgap state, if any, is not directly
tied to the magnitude of the gap. Instead, it is a relatively subtle
effect that stems from the attraction between a charge and a neutral
radical embedded in a dielectric medium. While one expects such an
attraction when the radical-plus-matrix system is sufficiently
polarisable, its presence is not a given. We will return to this point
shortly.  In the most significant contrast with Anderson's
phenomenological negative-$U$ model and the subsequent ultralocal
defect views~\cite{KAF, PhysRevLett.35.1293, PhysRevB.23.2596,
  ShimakawaElliott, PhysRevLett.85.2785, Simdyankin, Dembovsky}, the
midgap states intrinsically stem from the structural degeneracy of the
lattice and are topologically stable.  The concentration of the midgap
states is determined by the concentration of the domain walls near the
glass transition, an {\em equilibrium} quantity:
\begin{equation} \label{nDW} n_\text{DW}(T_g) \simeq 1/\xi(T_g)^3
  \simeq 10^{20} \mbox{cm}^{-3},
\end{equation}
consistent with observation. For the ZL~\cite{ZL_JCP} scenario to
work, at least two basic requirements must be fulfilled by the
material: (a) the bonds along the chain should not be saturated, and
(b) the insulating gap should not be too large, since in this case the
chain is an ionic insulator, as opposed to a Peierls insulator. Only
one family of known glassformers seems to satisfy these restrictions,
namely, the chalcogenides. Already this can be used to argue
circumstantially for the presence of the midgap states in these
materials.


\begin{figure}[t]
  \begin{tabular*}{\figurewidth} {c|c}
    \begin{minipage}{.46 \figurewidth}  {\bf (a)} 
      \begin{center}
        \includegraphics[width= .46 \figurewidth]{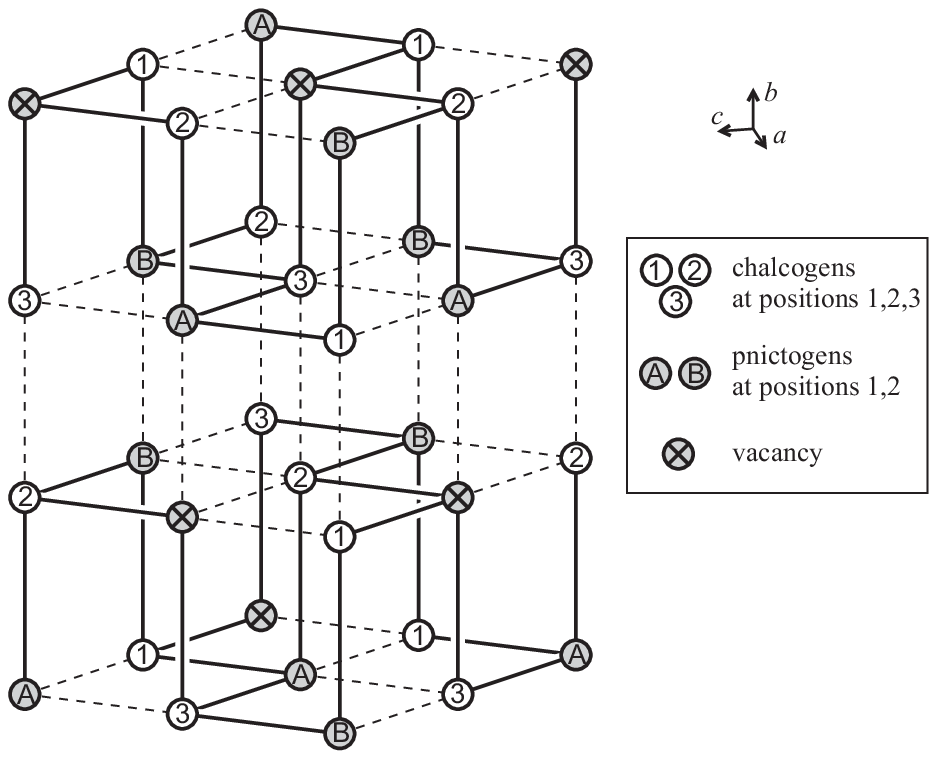}
      \end{center}
    \end{minipage}    
    &
    \begin{minipage}{.48 \figurewidth} {\bf (b)} 
      \begin{center}
        \includegraphics[width= 0.46
        \figurewidth]{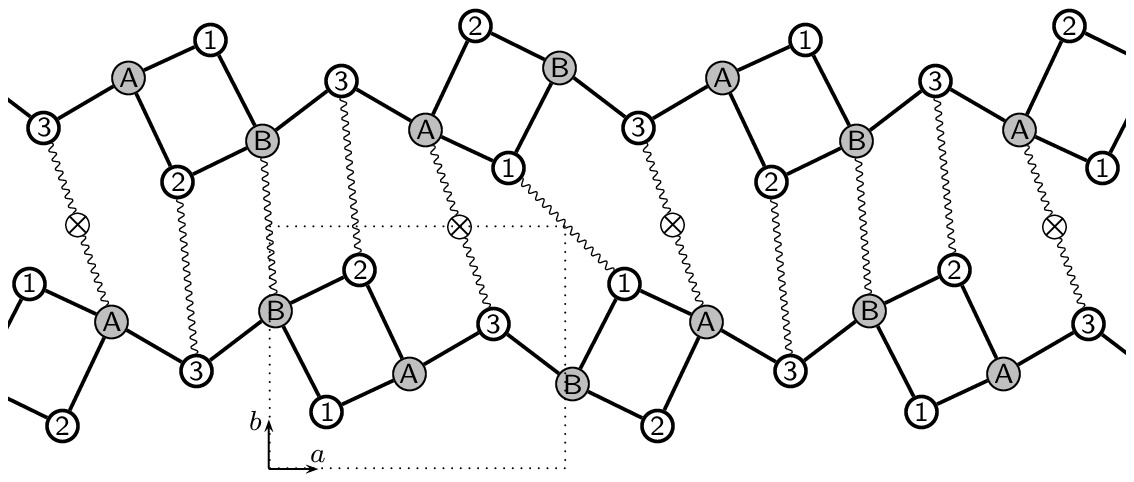}
      \end{center}
      \vspace{10mm} 

      {\bf (c)}
      \begin{center}
        \includegraphics[width= 0.36 \figurewidth]{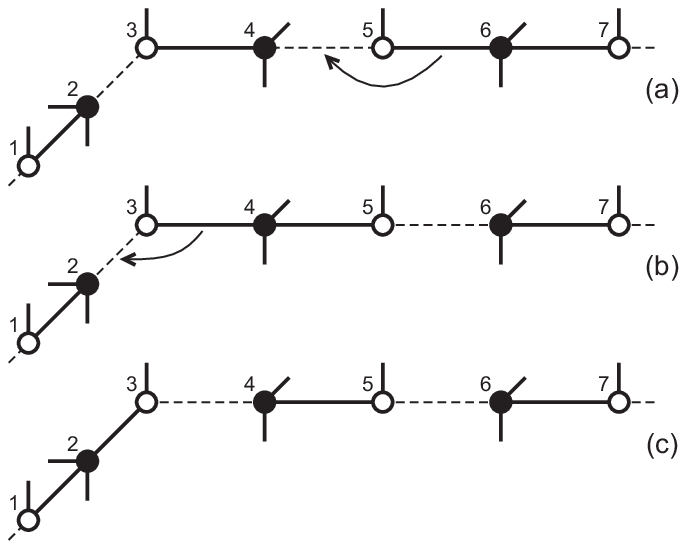}
      \end{center}
    \end{minipage}
  \end{tabular*}
  \caption{\label{As2Se3Xtal} {\bf (a)} A parent structure for
    Pn$_2$Ch$_3$ crystal (Pn=As, Ch=Se, S)~\cite{ZLMicro1}. The bond
    placement represents a particular symmetry breaking pattern. {\bf
      (b)} Side view of the actual As$_2$Se$_3$ structure. The wavy
    lines indicate inter-layer nearest neighbours in the parent
    structure, except the A-3 link, which is through a vacancy. The
    lengths of the links are only partially indicative of the actual
    bond length because the bonds are not parallel to the projection
    plane. From Ref.~\cite{ZLMicro1}. {\bf (c)} Illustration of the
    motion of an overcoordination defect by bond-switching along a
    linear chain (from atom 6 in (a) to atom 4 in (b)) or making a
    turn (from atom 4 in (b) to atom 2 in (c)). From
    Ref.~\cite{ZLMicro2}.}
\end{figure}

It turns out that in the specific case of chalcogenide glasses, the
specific atomic motifs leading to the midgap states can be directly
identified based on a relatively simple chemical yet constructive
picture. As the first step in their argument, Zhugayevych and
Lubchenko~\cite{ZLMicro1} (ZL) argue that these materials are symmetry
broken and distorted versions of much simpler, ``parent'' structures
defined on the simple cubic lattice.  This idea builds on the work of
Burdett and coworkers on the structure of solid arsenic and
electronically similar compounds~\cite{burdett5764,
  burdett5774}. Similarly, the crystal of As$_2$Se$_3$ can be thought
of as a distorted version of the parent structure in
Fig.~\ref{As2Se3Xtal}(a), in which some of the vertices are occupied
by vacancies. Each link in the parent structure corresponds with a
covalent bond in the actual, distorted structure, whereas a gap will
give rise to a secondary bond.  The actual, distorted structure is
much more complicated that its parent, see
Fig.~\ref{As2Se3Xtal}(b). The cause of the symmetry breaking in these
3D structures can no longer be traced with confidence to the Peierls
instability, especially in view of the significant $sp$-mixing in
these compounds~\cite{SeoHoffmann1999}, see, however,
Ref.~\cite{PhysRevB.76.052301}. Still, the very fact of symmetry
breaking between two adjacent covalent and secondary bonds that are
nearly co-linear is documented on thousands of
compounds~\cite{LandrumHoffmann1998}.

Much as the crystal of As$_2$Se$_3$ can be thought of as a distorted
version of the parent structure in Fig.~\ref{As2Se3Xtal}(a),
chalcogenide glasses can also be thought of as being distorted
versions of parent structures in which atoms and vacancies are placed
aperiodically. In either case, a pnictogen (chalcogen) forms three
(two) covalent bonds with its nearest neighbours and a few secondary
bonds~\cite{Alcock1972} with next nearest neighbours. The covalent
bonds must be approximately orthogonal.  The above notion directly
traces the thermodynamic stability of glasses to the same interactions
that underlie the stability of the crystal. This commonality betweem
periodic and aperiodic solids has been a recurring theme in this
article. In contrast with the crystal, however, the bonding in the
glass is typically somewhat weaker, consistent with its lower density
(4.57 vs. 4.81-5.01 g/cm$^3$ for crystalline and amorphous
As$_2$Se$_3$ respectively). More importantly, the entropy stemming
from the two alternative ways to perfectly dimerize the chains in the
parent structure scales with the sample's {\em area} and, hence, is
only sub-thermodynamic. Defects in the alternation pattern must be
present to account for the excess liquid entropy of a glassy
chalcogenide alloy. Such malcoordination defects are automatically
supplied by the domain walls of the mosaic!  Consistent with the
mobility of the domain walls, the malcoordination motifs can move and
make turns by switching bonds, very much like in the Grotthuss
mechanism of bond switching in water, see Fig.~\ref{As2Se3Xtal}(c).
This is an explicit demonstration of how a reconfiguration takes place
without breaking bonds. In addition, these observations enable one to
fix the bead count at, approximately, one bead per arsenic, consistent
with its value implied by the RFOT theory~\cite{ZLMicro2}. We remind
that the calorimetric bead count from Eq.~(\ref{Nb}) does not work
very well in the chalcogenides. Thus, we should expect that deciding
on the identity of the effective ultraviolet cutoff of the theory is
more complicated than counting the number of degrees of freedom
frozen-in at the liquid-to-crystal transition and may involve
system-specific microscopic considerations.

\begin{figure}[t]
  \begin{tabular*}{\figurewidth}{cc}
  \begin{minipage}{.48 \figurewidth}
    \flushleft 

    {\bf (a)} \\ \begin{center} \includegraphics[width= .48
      \figurewidth]{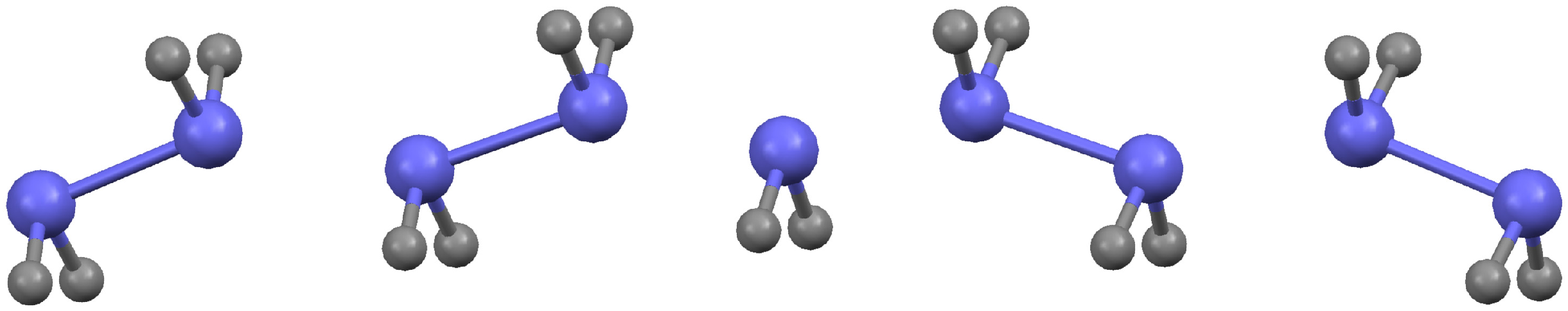}
    \end{center} 

    {\bf (b)} \\ \begin{center} \includegraphics[width=0.48
      \figurewidth]{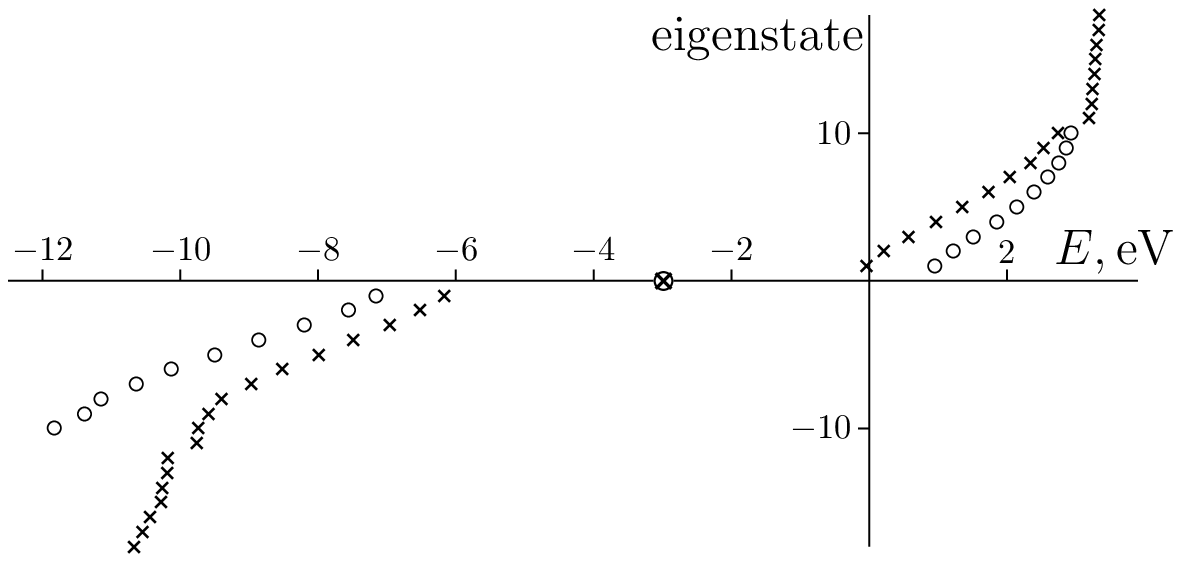}
    \end{center}
\end{minipage}
&
\begin{minipage}{.48 \figurewidth} 

  {\bf (c)} \\ \begin{center} \includegraphics[width=0.48
    \figurewidth]{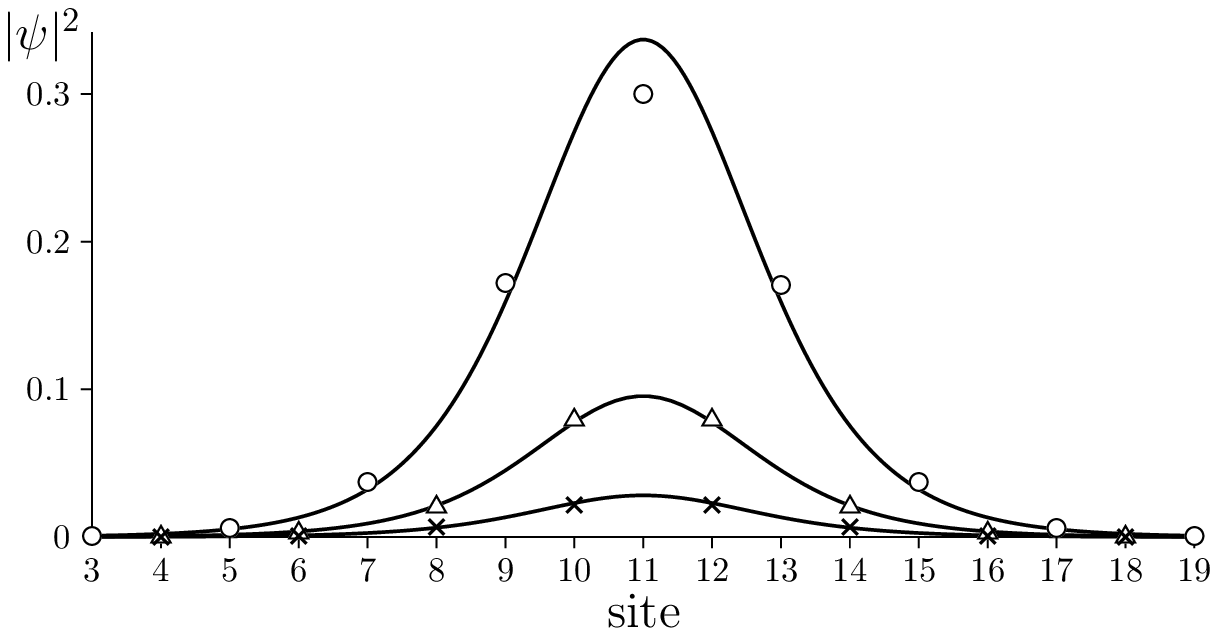}
  \end{center}

  \caption{\label{AsH2chain} {\bf (a)} Central part of neutral
    (AsH$_2$)$_{21}$ chain, whose ground state contains a coordination
    defect and the associated midgap level. {\bf (b)} Corresponding
    electronic energy levels: full MO calculation (crosses) vs.
    one-orbital model with renormalised $pp\sigma$ integrals
    (circles). States below the gap are filled; the midgap state is
    half-filled. {\bf (c)} The wave function squared of the midgap
    state: the circles correspond to the arsenics' $p_z$ atomic
    orbitals (AO) (total contribution 73\%), triangles As $s$-AO's
    (21\%), crosses -- the rest of AO's (6\%). Both figures from
    Ref.~\cite{ZLMicro2}.}
\end{minipage}
\end{tabular*}
\end{figure}

Thus in addition to being aperiodic, a parent structure for
chalcogenide glass must also contain under(over)-coordinated atoms.
On a pnictogen, for instance, such malcoordination corresponds to two
colinear secondary (covalent) bonds, along any of the three principal
axes. In actual distorted structures, the malcoordination is difficult
to detect because it is distributed over a large region, consistent
with the general difficulty in defining coordination in aperiodic
lattices. When $sp$-mixing is relatively weak, the distorted cubic
lattice can be presented as a collection of long stretches of
distorted one-dimensional chains. The rest of the lattice, to a good
approximation, renormalises the parameters of the chain, viz., the
on-site energies and hopping integrals~\cite{ZLMicro1,
  ZLMicro2}. Using these notions of quasi-one dimensionality, a
concrete example of a malcoordination pattern can be obtained with the
help of an actual 1D chain of $pp\sigma$-bonded molecules, such as the
hydrogen-passivated arsenics in Fig.~\ref{AsH2chain}.  As expected, a
neutral chain hosts an unpaired electron in its non-bonding
orbital. The latter orbital is indeed quite delocalized.

\begin{figure}[t]
  \begin{tabular*}{\figurewidth} {cc}
    \begin{minipage}{.52 \figurewidth} 
      \begin{center}
        \includegraphics[width= .52 \figurewidth]{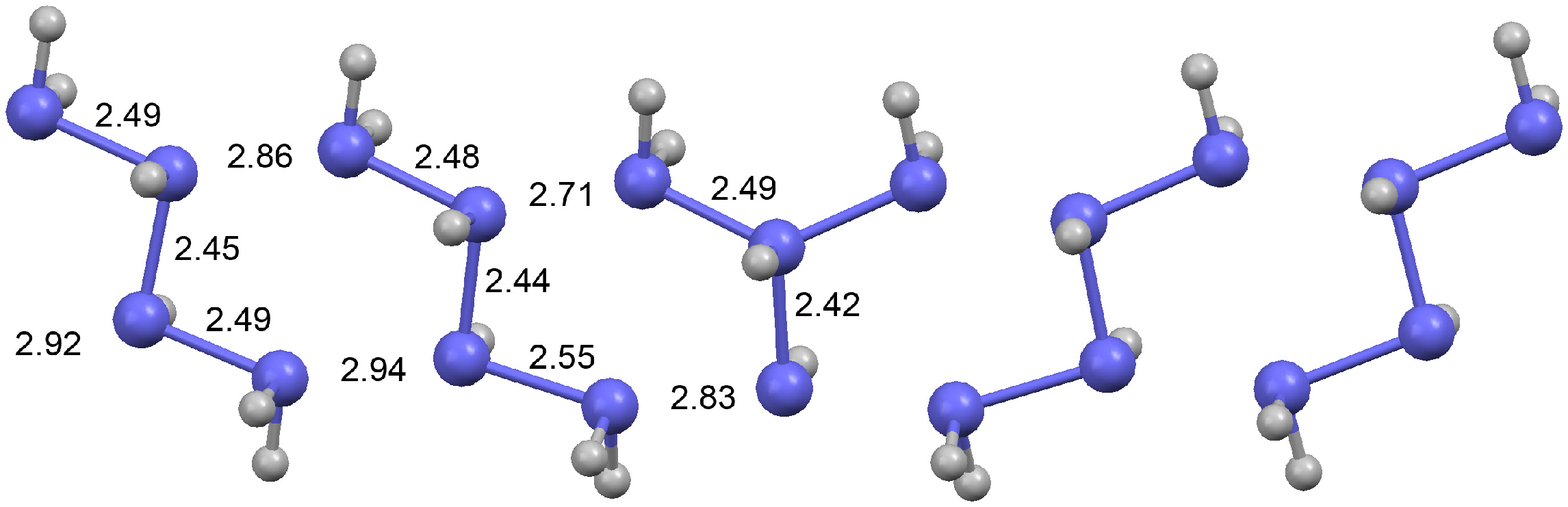}
         \\ {\bf (a)}
      \end{center}
      \caption{\label{defectTable} {\bf (a)} Central portion of a
        geometrically optimised double chain of hydrogen-passivated
        arsenics. The top and bottom chains, containing 19 and 17
        arsenics, host an over- and under-coordinated arsenic
        respectively.  The numbers denote the bond lengths in
        Angstroms. {\bf (b)} A compilation of the possible charged
        states of singly malcoordinated atoms. Here, ``Ch''=chalcogen,
        ``Pn''=pnictogen. Neutral states, not shown, imply dangling
        bonds and would be energetically costly.  Both graphics from
        Ref.~\cite{ZLMicro2}}
    \end{minipage}    
    &
    \begin{minipage}{.44 \figurewidth} 
      \begin{center}
        \includegraphics[width= 0.44 \figurewidth]{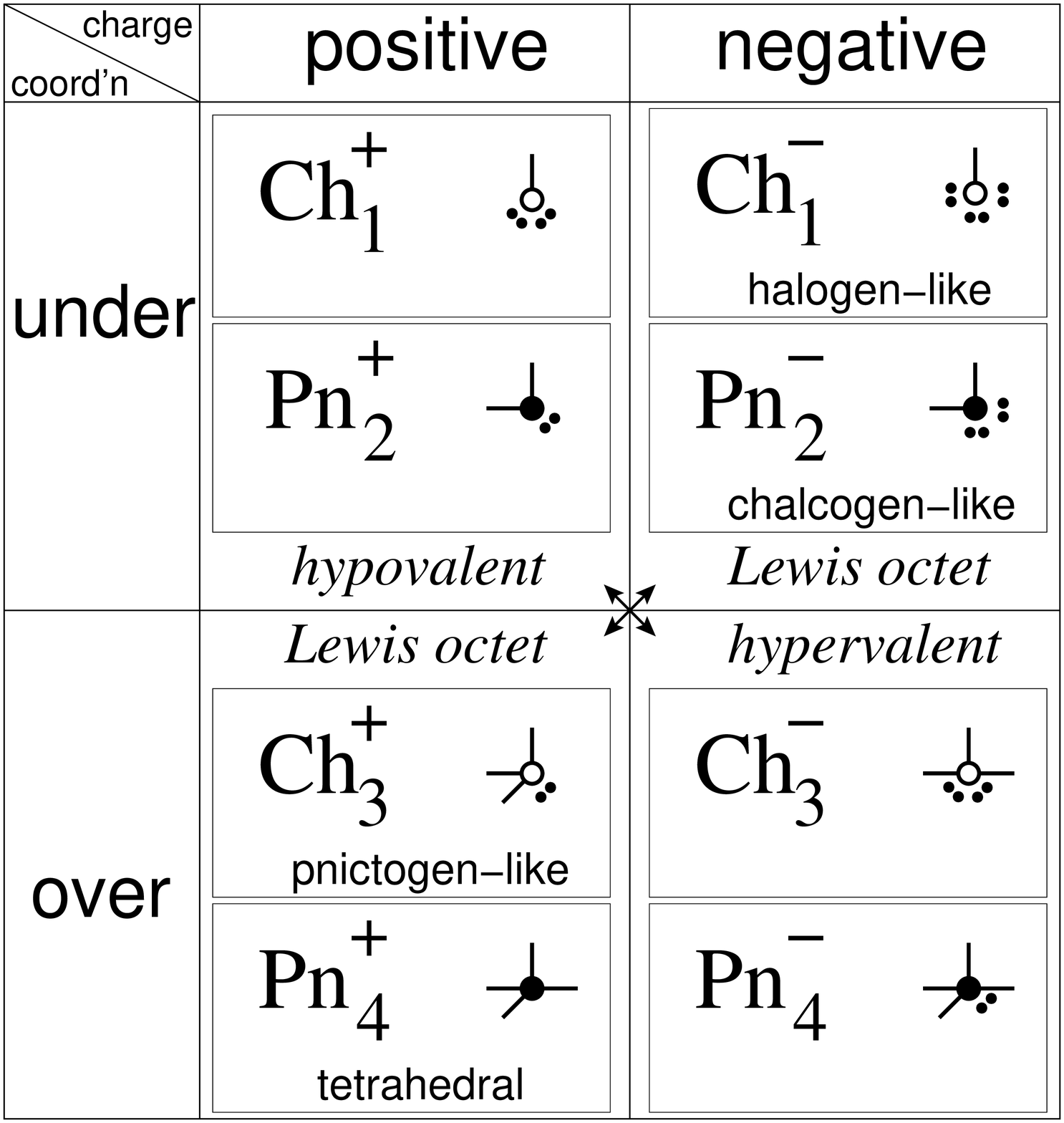}
        \\ {\bf (b)}
      \end{center}
    \end{minipage}
  \end{tabular*}
\end{figure}

It is interesting to note that interacting malcoordinated defects can
produce tetrahedral bonding patterns; the latter are not typical of
the crystalline chalcogenide alloys, in which the bonding is primarily
distorted-octahedral, see Fig.~\ref{defectTable}(a). It has been
suggested that such tetrahedral patterns emerge following the glass
transition in phase-change materials~\cite{ISI:000250615400019}.

As already mentioned, a half-filled defect is essentially a free
radical embedded in a dielectric medium and would be expected to be
stabilised by adding a charge, given sufficient polarisability. Such
stabilisation is imperative in order that the malcoordinated atom be
charged and ESR-inactive in its ground state. Consistent with these
notions, heteroatomic chalcogenides, which exhibit spatial variation
in electronegativity, are better glassformers than elemental arsenic,
for instance. The issue of the precise interaction between a radical
and a charge is, generally, a subtle quantum-chemical problem.  Still
there is an important limit in which this question can be resolved
using a very old chemical approximation, namely the Lewis octet rule.
In the ultra-local limit, the charged states of chalcogens and
pnictogens can be easily visualised as in Fig.~\ref{defectTable}(b).
According to this graphical table, there are several charged
configurations that exactly complete the Lewis octet and thus are
predicted to be stable. There are also two sets of configurations that
should be formally designated as hypo- or hypervalent. The double
arrows in the table indicate configurations that would be mutually
attractive and form a stable pair like that shown in
Fig.~\ref{defectTable}(a), in which a pair Pn$_4^+$Pn$_2^-$ is
shown. The ultralocal limit, although ultimately being an
oversimplification, is also instructive in that it allows one to
connect the present discussion with early, phenomenological proposals
on specific chemical motifs responsible for the midgap
states~\cite{KAF, PhysRevLett.35.1293}.

To summarise, the presence of the midgap states helps rationalise a
number of electronic anomalies in amorphous
chalcogenides~\cite{ShimakawaElliott} in a unified fashion.  Perhaps
the most important of these are the observation of a light-induced
paramagnetic signal and midgap absorption of light. The outcome of the
analysis~\cite{ZL_JCP} of the simple Hamiltonian (\ref{negU}) is also
consistent with the apparent existence of two distinct types of
photoluminescence and their fatigue in the
chalcogenides~\cite{TadaNinomiya, TadaNinomiya2, TadaNinomiya3}.  The
RFOT-based framework predicts a near universal density of the midgap
states; this density is intrinsic and originates from the inherent
density of cooperatively rearranging regions in an equilibrated
supercooled liquid.  The present microscopic picture and the much
earlier, phenomenological negative-$U$ model~\cite{PWA_negU,
  PWA_negU2} share the property of pinning the Fermi level close to
the middle of the gap thus explaining why conventional doping of
chalcogenides is inefficient~\cite{Kolomiets1981, Mott1993}.  Although
the states are not very close to the band edge, there is no subgap
absorption because of the Franck-Condon effect: The optical excitation
is faster than the atomic movements, implying the energy needed to
excite the electron from a filled level also includes the work needed
to deform the lattice to the configuration it had when the level is
half-filled to when it becomes completely filled. When half-filled,
the states {\em will} absorb light at subgap
frequencies~\cite{ZL_JCP}.

Finally, the extent of the midgap states in chalcogenides scales
inversely proportionally with the gap~\cite{RevModPhys.60.781,
  PhysRevB.21.2388, ZL_JCP}. In view of the effective attraction
between electrons, one expects that at high enough pressure, the
midgap orbitals will overlap leading to
superconductivity~\cite{ZL_JCP}. Chalcogenides are known to become
superconducting at high pressure~\cite{PhysRevB.15.973}. It will be
interesting to see whether this very quantum phenomenon originates in
things glassy.

\section{Summary and Connection with Jammed and Other Types of
  Aperiodic Solids}
\label{concl}

In an informal remark dating from 1995, P.~W.~Anderson
writes~\cite{PWA_lightly}: ``The deepest and most interesting unsolved
problem in solid state theory is probably the theory of the glass and
the glass transition. This could be the next breakthrough in the
coming decade.'' It is probably fair to say that the problem of the
structural glass transition has been as interesting and challenging as
it has been controversial. So much so that it found a way into mass
media some thirteen years after Anderson's remark was
published~\cite{NYT_glass}.

The reader has undoubtedly noticed that the present author feels
significantly more secure about the current status of the theory of
the glass transition than what the readers of the New York Times might
infer from several less-than-optimistic views expressed in
Ref.~\cite{NYT_glass}.  Although the RFOT theory uses approximations,
the physical picture of the structural glass transition presented bu
it is constructive and quantitative. It uses well-established concepts
of the classical density functional theory (DFT) complemented by more
recent insight about systems with relatively complicated free energy
landscapes; hereby the distinct free energy minima correspond to
long-lived, but not infinitely-lived states. Although it is often
difficult to treat finite-lifetime states using methods developed in
{\em equilibrium} thermodynamics, such methods become quantitative
(and physically transparent) given sufficient time scale separation
between motions of individual particles and the progress coordinate
that describes the escape from the long-lived states. Many examples of
successful quantitative descriptions of such long-lived states exist,
such as the transition state theory~\cite{Kramers, Hanggi_RMP, FW,
  PGW_QTST}, and, in particular, as applied to nucleation
phenomena. The RFOT description takes advantage of the accuracy of the
transition state theory to determine the distribution of the
relaxation rates and cooperativity sizes in glassy liquids, without
using adjustable parameters; in turn this allows one to predict the
glass transition temperature (on a given timescale) using
experimentally determined values of a specific quantity, viz., the
excess liquid entropy. This quantity is defined and measured without
reference to the glass transition itself. It should not be held
against the theory that it cannot (yet) predict the configurational
entropy for actual compounds, but only for model substances. Much the
same way, we do not hold against existing theories of
liquid-to-crystal transition that they cannot determine the melting
point for actual compounds, but only can predict melting points for
model substances with known interactions.

We have seen that the complicated free energy landscape of a glassy
liquid begins to form when a uniform liquid undergoes a breaking of
the translational symmetry, upon which the equilibrium density profile
is no longer flat but consists of disparate, narrow peaks; the
locations of the peaks form an aperiodic lattice. The breaking of the
symmetry is driven by steric effects, as can be shown using
independent arguments from the classical density functional theory,
the mode-coupling theory, and a symmetry-breaking perspective afforded
by the Landau-Ginzburg expansion.  The density-driven symmetry
breaking is entirely analogous to that during a density driven
transition from the uniform liquid to a periodic crystal, although
somewhat less so when the crystal structure is very open and mostly
determined by directional bonding. In contrast with the periodic
crystallisation, the peaks in the aperiodic structure are not entirely
stationary but move about in an intermittent, cooperative fashion; if
watched for a sufficiently long time, the liquid density profile
becomes uniform again. These motions correspond to transitions between
distinct metastable free energy minima and, hence, are driven by the
complexity of the free energy landscape. For this special mechanism to
work, the number of free energy minima must scale exponentially with
the region size; the log-number per particle is called the
configurational entropy $s_c > 0$.

This vast structural degeneracy introduces new physics in the problem,
along with new length and time scales. It is entropically unfavourable
for a glassy system to reside in any given free energy minimum
forever. The system thus undergoes local transitions to alternative
metastable states. In the mean-field limit, it is possible to think of
the liquid as a mosaic of regions, each of which is occupied by a
structure corresponding to one of the mean-field free energy
minima. The interfaces between the mosaic cells are made up of
configurations corresponding to the barrier in the bulk free energy
separating the minima. In finite dimensions, the picture is no longer
as simple, but the liquid can still be thought of as a mosaic of
regions characterised by relatively low free energy density separated
by regions at relatively high free energy density. Each stable region
can be thought of as being trapped in a state produced by {\em
  negative} free energy fluctuations, from a relatively high free
energy state typical of the liquid that just begins to enter the
landscape regime. Not every negative fluctuation would be long-lived,
however, but only such that {\em also} happens to coincide with a
transition to a specific free energy minimum. This is something that
would be impossible in a landscape-free system that has a
sub-thermodynamic number of alternative minima. Matching the magnitude
of the stabilisation due to the fluctuation with the entropic
stabilisation due to the complexity of the landscape, $\delta G = T
s_c N$, yields the size $N^*$ of the region that was stabilised by the
fluctuation. This size is finite because the fluctuation size scales
sublinearly with $N$: $\delta G = \gamma \sqrt{N}$. (Note the equation
$\gamma \sqrt{N} = T s_c N$ has no finite solutions when $s_c$ is
strictly zero.)  Conversely, $N^*$ is the region size that must be
reconfigured (in equilibrium) for the liquid to escape the free energy
minimum it is currently trapped in. It is the latter perspective that
is worked out in the microscopic picture presented here.  It is shown
that the escape from the current minimum can be thought of as an
activated, nucleation-like event driven by the multiplicity of the
free energy minima, $-T s_c N$, but is subject to a mismatch penalty
between the initial and final states.  If one takes the stabilised
state as the free energy reference, the mismatch penalty to flip a
region of size $N$ is the typical magnitude of the free energy
fluctuation: $\gamma \sqrt{N}$, Eq.~(\ref{FN12}).  The resulting
escape barrier is simply $F^\ddagger = \gamma^2/4 s_c T$.  This
formalism yields many quantitative predictions, including, in
particular, the $\alpha$-relaxation barrier and the cooperativity
length and those predictions can directly be extended to
non-equilibrium situations, such as ageing. A great deal of these
predictions have to do with the local {\em fluctuations} in the
structural degeneracy, which are directly reflected in the heat
capacity jump at the glass transition. Hereby the intrinsic
connections are revealed between the deviations from the Arrhenius
temperature dependence of $\alpha$-relaxation, relaxations'
exponentiality, and the fluctuation dissipation theorem.

Although transitions to both periodic and aperiodic crystal are
discontinuous---the restoring force in response to finite wavelength
deformation becomes finite via a discrete jump---the onset of measured
rigidity in glassy liquid is gradual. This is because the spatial
extent and lifetime of the long-lived structures increase continuously
with lowering temperature and/or increasing pressure. The material
becomes macroscopically rigid, if quenched sufficiently below the
glass transition temperature $T_g$. This is because the the barrier
for activated transitions, which is about $35 \ldots 39 k_B T_g$,
becomes very large compared with the ambient temperature, not because
the structure represents a unique energy minimum. (In fact, glasses
are vastly structurally degenerate, in contradistinction with periodic
crystals.) In this way, the rigidity of glasses and periodic crystals
are subtly different. Still, relics of the structural reconfigurations
that give rise to liquid flow above the glass transition survive down
to very low, sub-Kelvin temperatures.  These relics can be counted and
turn out to fully account for the mysterious structural resonances
people have phenomenologically associated with the so called two-level
systems. The physical boundaries of the reconfigurations---or domain
walls---are interesting objects in themselves: On the one hand, their
vibrations quantitatively account for the famous Boson peak. (These
vibrations can be thought of as the Goldstone modes that emerge when
the mosaic forms.) On the other hand, the domain walls can host very
special electronic states under certain circumstances that are
physically realised in amorphous chalcogenide alloys. The latter
alloys have come back into prominence owing to their potential as
phase-change memory materials.

As just said, the emergence of the landscape is the result of a
discontinuous transition and, in fact, could be thought of an avoided
critical point, see Fig.~\ref{ABphaseDiagram}. In the absence of
strongly-directional bonding, the criticality is avoided in {\em
  equilibrium} owing to steric effects, which stabilise locally dense
structures. This can be seen using already a Landau-Ginzburg expansion
and, quantitatively, the classical DFT. A quantitative criterion of
the proximity to the critical point is the ratio of the volumetric
particle size $a$ and the magnitude $d_L$ of the vibrational
displacement near the mechanical stability limit of the aperiodic
solid. In the case of aperiodic solids it is exactly the inverse of
the venerable Lindemann ratio.  By Eq.~(\ref{KWalpha1}), the $(a/d_L)$
ratio scales linearly with the dimensionality of space, up to a
logarithmic correction and factor of order 1. In 3D, both experiment
and calculation yield, nearly universally, $(a/d_L) \simeq 10^1$ for a
solid in equilibrium with the corresponding uniform liquid, be the
solid periodic or aperiodic.  For the liquid-to-solid transition to be
continuous, this ratio would have to be close to $10^0$. The $(a/d_L)$
ratio reappears under several guises during the analysis, because its
square corresponds with the elastic modulus of the substance in terms
of $k_B T$, by equipartition: $(a/d_L)^2 \simeq K a^3/k_B T$. ($K$ is
the bulk modulus.) We have seen that this ratio corresponds with the
excess relaxation time in a liquid compared to the corresponding
solid. This mismatch is behind the kinetic catastrophe of the
mode-coupling theory. The $(a/d_L)^2$ ratio also enters in the
expression for the escape barrier $F^\ddagger$ since, approximately,
$\delta G^2/N \equiv \gamma^2 \propto K$. Appropriately, the activated
reconfigurations do not involve breaking of individual bonds; in fact
$d_L \simeq a/10$ is comparable to the typical vibrational
magnitude. Despite the near harmonic nature of the motions of
individual atoms, the overall reconfiguration event is strongly
anharmonic, see Fig.~\ref{FNgraph}(a). Specific microscopic
realisations of such chemically-mild local motions have been
identified. For instance, a covalent bond in glassy chalcogenides can
gradually weaken into a secondary and, then, a van der Waals bond,
before finally rupturing~\cite{ZLMicro1, ZLMicro2}. This is quite
similar to the Grotthuss mechanism of bond switching in water. Another
often-cited example is the rotations of the SiO$_{4/2}$ tetrahedra in
glassy silica, during which covalent bonds are not broken but only
distorted following a reconfiguration event.  Yet another way to look
at this chemically-interesting aspect of the landscape regime is
provided by very very old samples of amber, which exhibit the
cryogenic two-level systems and Boson peak despite the significant
amount of cross-linking that took place over $10^8$ years
ago~\cite{PhysRevLett.112.165901}.

The $(d_L/a)$ ratio may thus be legitimately thought of as the small
parameter of the theory, non-withstanding the subtlety associated with
the fact that this small parameter is not fixed upfront but is
self-generated as a result of a {\em discontinuous} transition.
Consistent with the lack of critical-like fluctuations, the
correlation length for fluctuations in glassy liquids, has been shown
to be comparable to the molecular length scale~\cite{RL_sigma0}. As a
result, the cooperativity can be argued to be somewhat trivial, viz.,
due to ordinary Gaussian fluctuations. (A case for sub-dominant
fluctuations stemming from wandering of the domain walls has been made
recently~\cite{2014arXiv1411.4566B}.)

In contrast with substances in the landscape regime, amorphous films
made of a poor glass-former that favours open structures, such as
silicon or water at normal pressure, would rearrange by {\em
  bond-breaking}.  Accordingly, energy quantities reflecting bond
strength---as opposed to bond elasticity---would determine the
barrier. (Note bond strength and elasticity are correlated in actual
materials~\cite{GrimvallSjodin}, because the atomic size varies
relatively little across the periodic table and so the depth of a
bounding potential must correlate with its curvature at the bottom.)
Materials with open structures largely avoid entering the landscape
regime because of the relative unimportance of the steric effects and
thus are ``unaware'' of the aforementioned criticality in the first
place. The bond directionality could be so strong that the liquid
could expand upon freezing. Even in supercooled silicon, coordination
decreases with cooling~\cite{1.1631388}.  {\em Amorphous} films made
of such directionally bonding particles are even more open than the
crystal and exhibit dangling bonds.  In such films, which can be made
by rapid quench or by sputtering on a cold substrate, steric effects
are even less important than in corresponding crystals.  In a way, one
can think of such amorphous films as very cross-linked gels or
vulcanised rubbers. Although generally aperiodic, these materials may
be in the landscape regime only to a marginal extennt. The steric
interactions set the nearest neighbour distance but not the
coordination pattern itself; as a result the films reconfigure
predominantly by bond breaking. These notions are consistent with the
apparent scarcity of cryogenic excitations in silicon and germanium
films~\cite{Pohl_review, PhysRevLett.113.025503, QeenHellman2015}.
The latter excitations are predicted to be intrinsic (and quantum!)
features of landscape systems. To summarise, materials that favour
open structures avoid the landscape regime by either crystallising or
forming gel-like aperiodic structures. While such aperiodic open
structures may share some properties with glasses made by quenching an
equilibrated liquid, these properties are expected to be quite
system-specific, in contrast with glasses, whose key microscopic
properties are set by the cooperativity length at the glass transition
and the glass transition temperature itself.

To avoid confusion we note that materials with very directional
bonding may be poor glassformers at ordinary pressures but could be
vitrified more readily following pressurisation~\cite{Sharma1996,
  Brazhkin2003, Ponyatovsky1992, Hemley1988}, which would make steric
repulsion more prominent. This is not a general rule, however. For
instance, elemental calcium, which is close-packed at normal pressure,
becomes simple-cubic at sufficiently high pressures, supposedly due to
a density-induced hybridisation of electronic orbitals.  In any event,
it is requisite that an open structure undergo a pressure-induced
phase transition into a phase with higher coordination {\em before} it
can enter the landscape regime; the barrier for such a transition is
largely energetic because bond breaking must take place.  Polymers
seem to be a mixed case in that while exhibiting extremely anisotropic
bonding in one direction, they can usually pack quite well in the
other two dimensions. Indeed, polymers exhibit a broad distribution of
glassiness vs. local ordering. Many polymers do exhibit the cryogenic
two-level excitations~\cite{FreemanAnderson}, see Fig.~\ref{FA}(a).

In the opposite limit of completely isotropic interactions---as could
be realised, for instance, in Lennard-Jones liquids---steric effects
contribute prominently to the free energy, near melting, and so the
aperiodic-crystal phase would seem to be easily accessible from the
uniform liquid phase. However, the nucleation barrier for {\em
  periodic} crystallisation is very low in such systems because
monodisperse spheres line up in close-packed structures very
readily. Thus isotropically-interacting, monodisperse particles avoid
vitrification when quenched not too rapidly, owing to {\em periodic}
crystallisation, similarly to silicon. Yet in contrast with silicon,
crystallisation in rigid systems is induced by the very same forces
that cause the RFOT transition to the landscape regime. To appreciate
the distinction more clearly, we note that if forbidden to
crystallise---by controlling local orientational order, for
instance~\cite{C4SM02459A}---silicon would still fail to enter the
landscape regime but, instead, form a gel-like film even if cooled
leisurely. In contrast, hard spheres would undergo the RFOT transition
and eventually vitrify under such circumstances; we saw this
explicitly in Subsection~\ref{RFOTDFT}.

What should we think of an amorphous sample made of monodisperse
spheres? Such structures can be readily made with colloids, because of
particle-solvent friction, or by ultrafast quenching on a
computer. The resulting structure is caught in a relatively shallow,
high altitude minimum and could catastrophically relax into a denser
structure in the presence of thermal fluctuations. Conversely, at
zero-temperature or infinite pressure such structures would become
and, consequently, {\em jammed}. Consistent with these systems being
far away from equilibrium, there is no longer one-to-one
correspondence between density and pressure, at constant
temperature. In a mean-field limit, the least one may do is to use a
{\em triad} of intensive variables to describe the state of the
system, such as pressure-density-temperature or pressure-coordination
number-temperature, etc.

\begin{figure}[t]
  \centering
  \includegraphics[width= .6 \figurewidth]{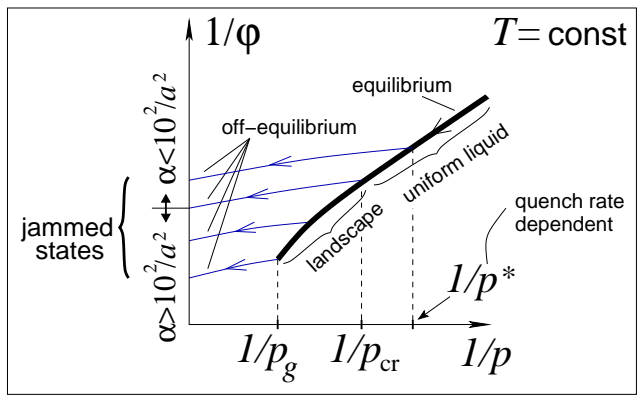}
  \caption{\label{Kurchan} The thick solid line shows the pressure
    dependence of the inverse filling fraction as resulting from a
    relatively (but not infinitely) slow quench. By varying the speed
    of quenching, one may control the density and pressure, $p^*$, at
    which the liquid is brought out of equilibrium, as shown by the
    thin solid lines. Apart from ageing, there is one-to-one
    correspondence between the state at which the system vitrified and
    the resulting jammed structure. Given a high enough quench rate,
    however, the uniform liquid can be brought out of equilibrium at a
    density {\em below} the density $\rho_\scr$, at which the
    landscape regime sets in in equilibrium. Under these
    circumstances, there is no longer direct correspondence between
    the $p = \infty$ jammed states and states equilibrated in a
    specific metastable free energy minimum. Roughly speaking, the
    jammed states will sample portions of the phase space
    characterised by a relatively small value of the force constant
    $\alpha$ of the effective Einstein oscillator. These high enthalpy
    states are vestiges of the avoided critical point from
    Fig.~\ref{ABphaseDiagram}.  This is a non-meanfield effect,
    c.f. the mean-field analysis of Mari et
    al.~\cite{PhysRevLett.103.025701}, see also Ref.~\cite{LW_Wiley}.}
\end{figure}

The jammed states could be thought of, generally, as corresponding to
states near the spinodal in $F(\alpha)$ in Fig.~\ref{RLFalpha}(a), but
with the quantity $\alpha^\ddagger a^2 \simeq \alpha_0 a^2 \simeq
(a/d_L)^2$ now distributed in a range $10^0 \ldots 10^2$, as opposed
to the equilibrium value of $10^2$ or so. This correspondence can be
seen using, as a starting point, results of the elegant mean-field
analysis of Mari, Krzakala, and Kurchan~\cite{PhysRevLett.103.025701}.
By solving a liquid model defined on a Bethe tree, Mari et al. have
explicitly shown that one may think of a jammed state as a liquid
first equilibrated in the landscape regime and then rapidly compressed
at constant temperature. According to this result, we may associate to
each jammed state an {\em equilibrated} state and the corresponding
values of $\alpha^\ddagger$ and $\alpha_0$, see Fig.~\ref{Kurchan}. In
the mean-field limit, the smallest possible value of $\alpha$ is that
at the density $\rho_A$ at which the metastable minimum in $F(\alpha)$
just begins to appear.  In finite dimensions, the landscape regime
sets in within a finite density interval centred at $\rho_\scr >
\rho_A$, as is reflected in Fig.~\ref{Kurchan}. (The crossover is
usually well separated from the sharply defined
$\rho_A$~\cite{LW_soft}.)  Yet given a fast enough speed of
compression, a liquid can be jammed starting also from a state {\em
  below} $\rho_\scr$, since metastable structures already appear but
quickly ``melt'' into the uniform liquid characterised by smaller
values of $\alpha$.  The faster the quench speed, the less stable the
resulting structure will be. The important thing is, if the density at
which the liquid fell out of equilibrium is below $\rho_\scr$, one
must associate the jammed structure not with a particular, equilibrium
$F(\alpha)$ but, instead, a set of location-specific $F(\alpha)$'s,
some of which are necessarily non-equilibrium. Because the system
undergoes significant fluctuations toward the uniform liquid state,
there will be many regions in which the typical value of $\alpha$
would be significantly less than the equilibrium $10^2/a^2$ and, in
fact, as small as $10^0/a^2$. The latter value is pertinent to
configurational equilibration in uniform liquids, which occurs by
particles exchanging their positions. Alternatively said, that
$\alpha^\ddagger a^2 \simeq \alpha_0 a^2$ could be as small as $1$
signifies that particles could move a distance comparable to their
size, after the liquid is unjammed, in a sort of a local avalanche. We
have explicitly seen in Fig.~\ref{Sciortino}(d) how small values of
$\alpha$ stem from fluctuations away from the landscape
regime. Alternatively, one may think of such fluctuations as arising
from local reduction in dimensionality, according to
Eq.~(\ref{KWalpha1}).

The smallness of the effective $\alpha$ is a key difference between a
jammed and vitrified system.  We reiterate that because the system is
off-equilibrium, there is no intrinsic relation between the stiffness
of the jammed system and the value of $\alpha_0$ of the type from
Eq.~(\ref{alpha1}); the stiffness becomes infinitely high in the $p
\to \infty$ limit whereas $\alpha_0$ remains finite. Said in simpler
terms, the spacing between (vibrating) particles in contact, in a
jammed structure, can be made arbitrarily small while the structure
remains practically the same. The particle displacement upon
unjamming, which is largely determined by the structure, is thus
decoupled from the vibrational motions.

Now take the zero temperature limit. Because the system is confined to
an {\em individual} free energy minimum at $T=0$, before it escapes to
a much lower enthalpy state, the correlations should be mean-field
like. This implies, among other things, that the relaxation time
should scale quadratically with the correlation length, $\tau \propto
\xi_A^2$, as follows from the mean-field Onsager-Landau ansatz for the
relaxation of the pertinent order parameter $\phi$: $\dot{\phi}
\propto - \delta F/\delta \phi$ and Eq.~(\ref{LG}). Given the
decoupling between thermal vibration and density, it is convenient to
choose the filling fraction as the order parameter.  Now, the scaling
$\xi_A \sim |\phi_A - \phi|^{-1/4}$ applies near a mean-field
spinodal~\cite{ISI:A1978GF11500008}, where $\phi_A$ is the location of
the metastable minimum.  Thus we obtain $\tau \propto |\phi_A -
\phi|^{-1/2}$. This is consistent with the $\omega^* \propto |\phi_A -
\phi|^{1/2}$ scaling discussed in Refs.~\cite{SilbertLiuNagel,
  LiuNagelSaarlosWyart}, where $\omega^*$ is the infra-red edge of
stable harmonic modes, and also with recent unjamming simulations of
soft spheres by Ikeda, Berthier, and Biroli~\cite{ikeda:12A507}, in
the low temperature limit. Although reasonable, the above discussion
is clearly qualitative and, certainly, is not a full-fledged field
theory. Such a field theory would have to contain exponentially many
order parameters and appears to be a complicated undertaking. Some
progress toward obtaining such a field-theoretical description of the
landscape regime in {\em finite} dimensions has been achieved recently
with the help of replica techniques, see Refs.~\cite{DSW_Wiley,
  2014arXiv1411.4566B} for instance.

In light of our earlier statements that the quantity $\alpha_0 a^2$
reflects the sharpness of the first order transition from liquid to
solid, we may conclude that, in fact, the lower density jammed states
are vestiges of an avoided critical point.  That the critical point is
destroyed in equilibrium, means in thermodynamic terms that the
critical fluctuations correspond to higher free energy
states. (Sometimes, such high free energy states may still correspond
to {\em conditional} equilibrium, such as the liquid-liquid separation
in protein solutions mentioned in Subsection~\ref{sc}.)  We have
argued in Subsection~\ref{driving} that the critical point is pushed
down to zero temperature, which implies the solid formed as a result
of the continuous transition would be only marginally stable. And
indeed, jammed systems are effectively zero temperature. Long-range,
critical correlations are destroyed by fluctuations in equilibrium,
but would be present in jammed systems which are not allowed to
equilibrate.  In fact, we just saw that a diverging lengthscale
appears during unjamming at low temperatures.

We thus conclude that there is a {\em continuum} of jammed states.
The states from the more stable, higher density part of this continuum
are generated by quenching a liquid starting at a density above
$\rho_\scr$, i.e., from a state that was fully equilibrated in the
landscape regime. In other words, the starting structure is trapped in
a relatively well-defined free energy minimum.  The resulting jammed
structure is hyperstatic.  It appears that while being jammed, such
hyperstatic structures could exhibit additional, replica-symmetry
breaking transitions~\cite{2015arXiv150107244C, doi:10.1021/jp402235d}
that are similar to the so called Gardner transitions in mean-field
spin glasses~\cite{Gardner1985747}. It is quite possible that in
finite dimensions, these additional symmetry-breaking transitions have
to do with freezing out the residual string-like excitations that
would be characteristic of the crossover to the landscape regime.  In
contrast, the lower density states from the continuum of jammed states
can be traced to the very critical point that is avoided when the
system undergoes an equilibrium, discontinuous transition to the
landscape regime.  These relatively low-density structures are
generated when the liquid falls out of equilibrium at a density below
$\rho_\scr$ and are {\em hypostatic}. Because the crossover has a
finite width, there is no sharp boundary between the hyperstatic and
hypostatic regimes, which seems to be consistent with recent
simulations of Morse and Corwin~\cite{MorseCorwin}. In any event, the
outcome of a jamming experiment likely depends on the precise
quenching protocol for {\em any} starting density, including the
putative $\rho_K$. This is probably true even in meanfield, owing to
the aforementioned Gardner transitions.  Also in meanfield, the
crossover is replaced by a sharply defined density $\rho_A$, implying
one can unambiguously define an {\em isostatic} jammed structure,
which then sharply demarcates the hyper- and hypostatic regimes.
According to the above discussion, an amorphous sample made of
isotropically-interacting particles would be similar to a quenched
glass, the degree of similarity depending on the speed of quenching.
The greater the similarity, the greater the number of the low-energy
resonances that give rise to the two-level systems.

The above discussion also implies that given a fixed quenching rate,
the glass-forming ability is optimised within a {\em window} of
bonding directionality, the low and high extremes corresponding to
relatively close-packed and open-structure crystal states
respectively.

\begin{figure}[t]
  \centering
  \includegraphics[width= .7 \figurewidth]{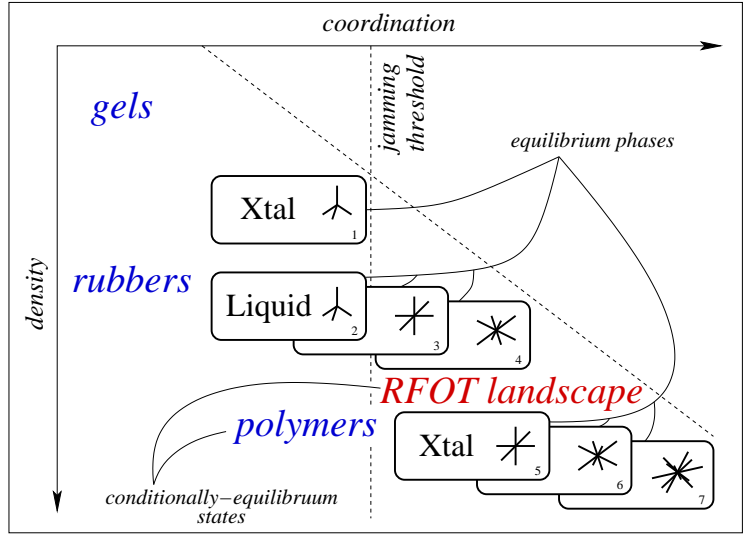}
  \caption{\label{densCoordDiagram} Qualitative structural map of
    condensed matter phases, periodic and aperiodic, in terms of
    density and average coordination. Boxes 1 through 7 correspond to
    equilibrium phases. Box 1 could be exemplified by crystalline
    silicon.  Boxes 2 through 4 represent uniform liquids in the order
    of increasing coordination/density. Boxes 5 through 7 represent
    crystals, also in the order of increasing
    coordination/density. The stick figures inside boxes are meant to
    depict coordination in a pictorial way and, obviously, do not
    cover all possible ways to arrange particles.  Direct transitions
    between some equilibrium phases are allowed. For instance, $1
    \leftrightarrow 2$ pertains to water, $1 \leftrightarrow 3$ to
    germanium. In the present scheme, colloidal suspensions and
    room-temperature ionic liquids would be classified as liquids in a
    pre-landscape regime, even if very viscous.  Materials to the left
    of the ``jamming threshold'' line would have to undergo a
    discontinuous transition before they could be jammed let alone
    enter the landscape regime.}
\end{figure}

It is hoped that in addition to a systematic exposition of technical
aspects of the RFOT theory of the glass transition, the present
article has clarified the limits of applicability of the theory and
succeeded, at least to some degree, to place the theory of glasses
made by quenching liquids in a broader context of other amorphous
materials prepared in a variety of ways. This broader context is
graphically summarised in Fig.~\ref{densCoordDiagram}.

Anderson concluded his informal remark in Science by stating: ``The
solution of the problem of spin glass in the late 1970s had broad
implications in unexpected fields like neural networks, computer
algorithms, evolution, and computational complexity. The solution of
the more important and puzzling glass problem may also have a
substantial intellectual spin-off.  Whether it will help make better
glass is questionable.'' Better glass has definitely been made since
1995, such as the famous Gorilla Glass. It is fair to say that
theoretical investigations had little to do with that progress.
Still, in view of the complicated interplay between the glass
transition and details of bonding in various glasses, such as the
chalcogenides, there is hope that the theory will help make better
glass by at least allowing one to narrow the scope of the
search. Perhaps more importantly, it may help one make better {\em
  crystalline} solids by explaining how to avoid vitrification
altogether.


{\em Acknowledgement.} The author gratefully acknowledges his
collaborators Peter G. Wolynes, Andriy Zhugayevych, Pyotr Rabochiy,
Dmytro Bevzenko, Jon Golden, and Ho Yin Chan. His work has been
supported by the National Science Foundation (Grants CHE-0956127,
MCB-0843726, and MCB-1244568), the Alfred P. Sloan Research
Fellowship, the Welch Foundation Grant E-1765, the Arnold and Mabel
Beckman Foundation Beckman Young Investigator Award, and the Donors of
the Petroleum Research Fund of the American Chemical Society.

\appendix

\section{Volume mismatch during ageing}
\label{volumemismatch}

\begin{figure}[t] \centering
  \includegraphics[width = .7 \figurewidth]{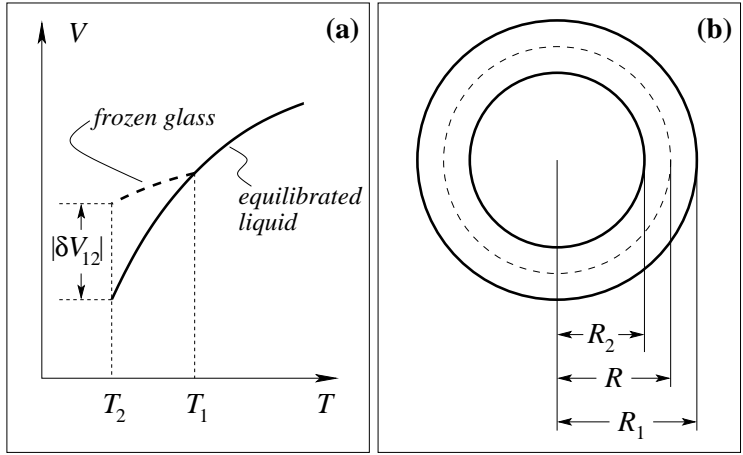}
  \caption{\label{contraction} {\bf (a)} The temperature dependence of
    the volume of an equilibrated sample (solid line) and of a sample
    rapidly quenched from temperature $T_1$ to temperature $T_2$. {\bf
      (b)}: Illustration of the volume mismatch due to ageing. A
    spherical region of radius $R_1$ of a sample rapidly quenched to
    temperature $T_2$ from temperature $T_1$ would be of radius $R_2$
    if fully equilibrated at the ambient pressure. To avoid rupture,
    both the chosen region and the environment must stretch, leading
    to a negative excess pressure. The boundary between the aged
    region and its environment is at radius $R$, $R_2 < R < R_1$.}
\end{figure}  

As before, consider a prompt thermal quench from equilibrium at
temperature $T_1$ to a different temperature $T_2$, at ambient
pressure.  The volume mismatch between a sample equilibrated at $T_2$
and the actual sample is the volume difference between a fully
equilibrated sample and the one that is only vibrationally relaxed:
\begin{equation} \label{DV} \DV = \int_{T_1}^{T_2} V (\alpha_\text{eq}
  - \alpha_\text{vibr}) \, dT.
\end{equation}
For concreteness, let us consider a downward $T$-jump: $T_2 < T_1
\Rightarrow \DV < 0$, see illustration in
Fig.~\ref{contraction}(a). Suppose a spherical region is of radius
$R_1$ before it structurally relaxed, {\em at the ambient pressure and
  temperature}, and would be of radius $R_2 < R_1$ after relaxation,
if {\em also} at the ambient pressure and temperature. In view of the
smallness of $\DV/V$ (usually not exceeding a few thousandth), $\delta
R/R \equiv (R_2 - R_1)/R \approx \DV/3V$, where $R$ stands for the
actual radius of the droplet upon reconfiguration. To avoid fracture,
both the inner and outer regions must stretch ($R_2 < R < R_1$), see
Fig.~\ref{contraction}(b), leading to an excess negative pressure
$p(r)$ throughout the sample. This pressure is constant within the
reconfigured region; it decreases away from the region and vanishes
strictly at the external boundary of the sample: $p(\infty)=0$. The
excess pressure can be straightforwardly related to the deformation
using a solved problem from Chapter 7 of Ref.~\cite{LLelast},
assuming, for simplicity, isotropic elasticity and a (macroscopic)
spherical sample. At the boundary of the reconfigured region, $R - R_2
= - p(R) R_2/3K_\sinn$ and $R - R_1 = p(R) R_1 /4 \mu_\sout$ where
$K_\sinn$ is the equilibrium bulk modulus at $T_2$ and the {\em
  ambient} pressure, while $\mu_\sout$ the shear modulus of an
individual aperiodic state, also at $T_2$ and ambient
pressure. Excluding $R$, we find:
\begin{equation} \label{p} p(R) = \frac{\DV}{V}
  \left(\frac{1}{K_\sinn} + \frac{3}{4\mu_\sout}\right)^{-1}.
\end{equation}
To estimate this excess pressure, $p \sim (\DV/V) \mu/2$, we note that
for a realistic quench, $|\DV/V| \lesssim 5 \cdot
10^{-3}$~\cite{Kovacs1977}, while the rule of thumb for elastic
constants is $\mu(T_\scr) a^3 \simeq 12 k_B T_\scr$, see
Eq.~(\ref{main1}). And so, generically, $\mu(T_g) a^3 \simeq 20 k_B
T_g$, since $T_g < T_\scr$, while $\mu$ usually increases with
lowering temperature. As a result, the excess pressure is comparable
to but probably less than $|\DV/V| 20 k_B T/a^3$. The latter does not
exceed $10^{-1} k_B T/a^3$, which is greater than the atmospheric
pressure by a couple of orders of magnitude, but is comparably less
than the {\em kinetic} pressure. The latter is the relevant pressure
reference here; its numerical value is about $(k_B T/a^3) (a/d_L) \sim
10 k_B T/a^3$~\cite{RL_LJ, LW_Wiley}, see Subsection~\ref{quant}.

The environmental deformation, as a function of the radius vector
$\br$, is given by~\cite{LLelast}
\begin{equation} \label{urA} \bu(r) = (\br/r^3) \, p(R) R^3/4 \mu_\sout
  \hspace{3mm} (r > R).
\end{equation}
The resulting volume change of the sample is $4 \pi \times p(R) R^3/4
\mu_\sout$, yielding for the relative volume change of the whole
sample:
\begin{equation} \label{dVV} \left. \frac{\delta V}{V}
  \right|_\text{sample} = \frac{\DV}{V} \left(1 + \frac{4\mu_\sout}{3
      K_\sinn} \right)^{-1} 
\end{equation}
Note this total volume change takes place even as the environmental
deformation from Eq.~(\ref{urA}) is pure shear: ${\bm \nabla} \bu
\equiv u_{ii} = 0$.

Also of potential interest is the free energy excess due to the
deformation of the environment, which can be estimated by integrating
the standard expression for the free energy density over space,
assuming again the isotropic elasticity and remembering that $u_{ii} =
0$ for $r > R$: $\int_R^\infty (\mu_\sout u_{ik}^2) 4 \pi r^2
dr$. This yields for the excess energy:
\begin{equation}\label{elastexcess}
  E_\text{exc} = \frac{\pi  p(R)^2 }{2 \mu_\sout} R^3, 
\end{equation}
where $p(R)$ is from Eq.~(\ref{p}). Note this energy scales with the
volume of the droplet.  As expected, there would not be any excess
pressure or stored energy, if the material around the nucleus could
flow, in which case $\mu_\sout =0$ ($p \propto \mu_\sout$ for
$\mu_\sout \rightarrow 0$, by Eq.~(\ref{p})). Finally note that
formulae (\ref{p})-(\ref{elastexcess}) apply for a $\DV$ of either
sign, i.e., for both downward and upward $T$-jumps. The physical
difference between the two cases is that the excess pressure $p$ will
be negative and positive for downward and upward jumps respectively.

It is apparent from Eq.~(\ref{dVV}) that, owing to the shear
resistance of the sample, the relative volume change of the sample is
always smaller than that of the reconfiguring (microscopic)
sub-regions, since the r.h.s. of Eq.~(\ref{dVV}) is less than $\DV/V$,
as long as $\mu_\sout > 0$. This would seemingly imply that the sample
will never reach its equilibrium volume! There is no contradiction
here, however. As the sample equilibrates, i.e., the ageing proceeds,
the assumption of the mechanical stability is no longer valid. Hereby
the material relaxes throughout, all stored excess energy from
Eq.~(\ref{elastexcess}) is released, as recently described in
Ref.~\cite{Wisitsorasak02102012}, while the shear modulus $\mu_\sout$
that enters in Eq.~(\ref{dVV}) effectively decreases and vanishes upon
complete equilibration. Note that to achieve the latter, regions that
originally relaxed at a non-zero excess pressure $p$, will also have
to relax until they are in equilibrium at $p=0$.

To quantify the significance of the volume-mismatch effect for the
thermodynamics of ageing we first note the stabilisation of state 2
from Eq.~(\ref{Deltag}), per particle, is approximately $\delta \Delta
g \simeq (V p)/N = p a^3$, where we have used the usual expression for
the Gibbs energy increment: $dG = -SdT + Vdp$. Together with
Eq.~(\ref{p}), this yields
\begin{equation} \label{dDf} \delta \Delta g = \frac{\DV}{V}
  \left(\frac{1}{K_\sinn} + \frac{3}{4\mu_\sout}\right)^{-1} a^3.
\end{equation}
We thus observe that {\em initial} ageing events always {\em overshoot}
in the following sense: For downward $T$-jumps, the final state free
energy is lower than the equilibrium free energy at the ambient
pressure ($\DV < 0$) and vice versa for upward $T$-jumps ($\DV > 0$).

To get a quantitative sense of this overshoot, let us adopt $\delta
\Delta g \simeq (\DV/V) \mu a^3/2$ for the sake of argument. As in the
above estimate for the excess pressure, the stabilisation for the
$\DV/V = - 1 \ldots 5 \cdot 10^{-3}$ range thus corresponds to $\delta
\Delta g \approx - 0.01 \ldots 0.05 k_B T$. This stabilisation will be
partially offset by the excess elastic energy stored in the
environment, from Eq.~(\ref{elastexcess}): $E_\text{exc}/N \simeq
(3/8) p^2 a^3/\mu_\sout \simeq +0.1 \cdot (\DV/V)^2 \mu a^3$, however
the offset is second order in the volume mismatch and sub-dominant by
several orders of magnitude to the main effect. 

To quantify the effects of the overshoot caused by the volume-mismatch
on the kinetics of ageing, we note the free energy difference $\Delta
g$ itself is about $-0.9 k_B T$ at equilibrium just above the glass
transition on 1 hr time scale, according to Eq.~(\ref{FXW}) and using
1 psec for the rate prefactor. For quenches below the glass
transition: $T_1 = T_g$, $T_2 < T_1$, $\Delta g$ is only weakly
$T$-dependent and predicted to decrease somewhat by absolute
value~\cite{LW_aging}, see Eq.~(\ref{df1}). Assuming for the sake of
argument $\Delta f = -0.8 k_B T$ (i.e., a relaxation time of $10^5$ s)
one gets, upon replacing $\Delta g = -0.8 k_B T$ by $\Delta g = -0.85
k_B T$, a speed up by a factor of ten or so. Note that the thermal
expansivity and the elastic moduli tend to
anti-correlate~\cite{GrimvallSjodin}, and so the combination $\DV
\mu$, for similar quench depths, should not vary wildly among
different substances.

We have thus established that as regards thermodynamics and kinetics
of ageing, the volume-mismatch effect is quantitatively, but not
qualitatively significant. One possible exception is melting of
ultrastable glasses, where the volume mismatch could exceed one
percent~\cite{doi:10.1021/jz3003266}, thus leading to a relatively
large $\delta \Delta g$.

\vspace{-3mm}

\bibliographystyle{tADP} 
\bibliography{/Users/vas/Documents/tex/ACP/lowT}

\end{document}